\immediate\write18{makeindex \jobname.nlo -s nomencl.ist -o \jobname.nls}
\documentclass[12pt]{caltech_thesis}

\pdfoutput=1

\usepackage{todonotes}

\usepackage[utf8]{inputenc}
\usepackage[T1]{fontenc}
\usepackage{newtxtext}

\usepackage{appendix}

\usepackage{amsfonts,amssymb,amsmath,amsthm}
\usepackage{mathtools}
\usepackage{tikz-cd}
\usepackage{enumitem}
\usepackage{algorithm}
\usepackage[noend]{algpseudocode}
\usepackage[hidelinks]{hyperref}
\usepackage{hhline}
\usepackage{tablefootnote}

\DeclareUnicodeCharacter{012D}{\u{\i}}

\algnewcommand\algorithmicinput{\textbf{Input:}}
\algnewcommand\INPUT{\item[\algorithmicinput] \strut}
\algnewcommand\algorithmicoutput{\textbf{Output:}}
\algnewcommand\OUTPUT{\item[\algorithmicoutput] \strut}

\algnewcommand\algorithmicto{\textbf{to}}
\algnewcommand\algorithmicforto{\textbf{for}}
\algdef{S}[FOR]{ForTo}[3]{\algorithmicforto\ #1 $\gets$ #2 \algorithmicto\ #3 \algorithmicdo}
   
\OnehalfSpacing*

\theoremstyle{plain}
\newtheorem{thm}{Theorem}[chapter]
\newtheorem{lem}{Lemma}[chapter]
\newtheorem{defi}{Definition}[chapter]
\newtheorem*{defi*}{Definition}
\newtheorem{cor}{Corollary}[chapter]
\newtheorem{conj}{Conjecture}[chapter]
\newtheorem*{conj*}{Conjecture}

\newtheorem*{prob*}{Problem}
\newtheorem{cond}{Condition}[chapter]

\theoremstyle{remark}
\newtheorem*{rem}{Remark}
\newtheorem{exmp}{Example}[chapter]
\newtheorem*{exmp*}{Example}
\newtheorem*{hyp}{Hypothesis}

\newcommand{\nospacepunct}[1]{\makebox[0pt][l]{\,#1}}

\newcommand{\Q}{\mathbb{Q}}
\newcommand{\Z}{\mathbb{Z}}
\newcommand{\N}{\mathbb{N}}
\newcommand{\C}{\mathbb{C}}

\newcommand{\F}{\mathbb{F}}
\newcommand{\ord}{\mathcal{O}}
\newcommand{\aut}{\mathrm{Aut}}

\newcommand{\sym}{\mathrm{Sym}}
\newcommand{\alt}{\mathrm{Alt}}

\newcommand{\gal}{\mathrm{Gal}}

\newcommand{\map}{\mathrm{Map}}
\renewcommand{\ker}{\mathrm{Ker}}

\newcommand{\tr}{\mathrm{Tr}}
\newcommand{\ind}{\mathrm{Ind}}

\newcommand{\twr}{\operatorname{twr}}

\DeclareMathOperator{\spec}{Spec}

\newcommand{\id}{\mathrm{id}}

\newcommand{\cha}{\mathrm{char}}
\newcommand{\rad}{\mathrm{Rad}}
\newcommand{\ann}{\mathrm{Ann}}

\newcommand{\gl}{\mathrm{GL}}
\newcommand{\gammal}{\mathrm{\Gamma L}}
\newcommand{\pgl}{\mathrm{PGL}}
\newcommand{\pgammal}{\mathrm{P\Gamma L}}
\newcommand{\agl}{\mathrm{AGL}}
\newcommand{\proj}{\mathbb{P}}
\newcommand{\soc}{\mathrm{soc}}
\newcommand{\inn}{\mathrm{Inn}}
\newcommand{\out}{\mathrm{Out}}
\newcommand{\hol}{\mathrm{Hol}}

\makeatletter

\newcommand{\Rmnum}[1]{\expandafter\@slowromancap\romannumeral #1@}
\makeatother

\usepackage{thmtools, thm-restate}

\usepackage[
    backend=biber,natbib,
    maxbibnames=9,
    maxcitenames=9,
    style=alphabetic,
    sorting=nyt,
]{biblatex}

\begin{document}

\title{$\mathcal{P}$-schemes and Deterministic Polynomial Factoring over Finite Fields}
\author{Zeyu Guo}

\degreeaward{Doctor of Philosophy}                 % Degree to be awarded
\university{California Institute of Technology}    % Institution name
\address{Pasadena, California}                     % Institution address
\unilogo{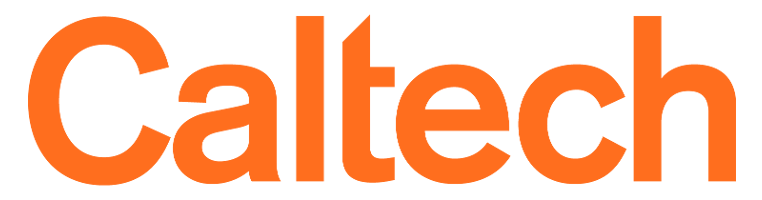}                                 % Institution logo
\copyyear{2017}  % Year (of graduation) on diploma
\defenddate{May 22}          % Date of defense

\orcid{0000-0001-7893-4346}

\rightsstatement{All rights reserved}

\maketitle[logo]

\begin{acknowledgements} 	
I would like to thank my advisor, Chris Umans, for his patient guidance and continual encouragement throughout my graduate studies.
His passion, enthusiasm, and dedication for research are truly  inspiring. I am very fortunate to have such a great teacher as my advisor.

I am indebted to Michael Aschbacher, who taught me a graduate algebra course, and Matthias Flach, who taught me a course on algebraic number theory. The knowledge I learned from their courses is crucial for the work presented in this thesis.
 
I also want to thank Leonard Schulman, Anand Kumar Narayanan, Manuel Arora, and Jenish Mehta for many helpful conversations. In particular, I am grateful to Manuel Arora for explaining to me his work on $m$-schemes. 
%I also want to thank Anand Kumar Narayanan for carefully reading part of the thesis and offering helpful feedbacks.

Finally, I want to thank my family and friends for their continual support and encouragement.
\end{acknowledgements}

\begin{abstract}
We introduce a family of mathematical objects called {\em $\mathcal{P}$-schemes}, where $\mathcal{P}$  is a poset of subgroups of a finite group $G$. A $\mathcal{P}$-scheme is a collection of partitions of the right coset spaces $H\backslash G$, indexed by $H\in \mathcal{P}$, that satisfies a list of axioms. These objects generalize the classical notion of association schemes   \citep{BI84} as well as the notion of $m$-schemes \cite{IKS09}. 

Based on  $\mathcal{P}$-schemes, we develop a unifying framework for the problem of deterministic factoring of  univariate polynomials over finite fields under the generalized Riemann hypothesis (GRH). 
More specifically, our results include the following:

\begin{itemize}
\item We show an equivalence between $m$-schemes as introduced in \citep{IKS09} and $\mathcal{P}$-schemes in the special setting that $G$ is a multiply transitive permutation group and $\mathcal{P}$ is a poset of pointwise stabilizers, and therefore realize the theory of $m$-schemes as part of the richer theory of $\mathcal{P}$-schemes. 

\item We give a generic deterministic   algorithm that computes the factorization of the input polynomial $f(X)\in\F_q[X]$ given a ``lifted polynomial'' $\tilde{f}(X)$ of $f(X)$ and a collection $\mathcal{F}$ of ``effectively constructible''  subfields of the splitting field of $\tilde{f}(X)$ over a certain base field. 
It is routine to compute $\tilde{f}(X)$ from $f(X)$ by lifting the coefficients of $f(X)$ to a number ring.
The algorithm  then successfully factorizes $f(X)$ under GRH in time polynomial in the size of $\tilde{f}(X)$ and $\mathcal{F}$, provided that a certain condition concerning  $\mathcal{P}$-schemes is satisfied, for $\mathcal{P}$ being the poset of subgroups of the Galois group $G$ of $\tilde{f}(X)$ defined by $\mathcal{F}$ via the Galois correspondence.
%The algorithmic problem of deterministic polynomial factoring over finite fields is therefore reduced to the combinatorial problem of proving a condition about $\mathcal{P}$-schemes. 
By considering various choices of $G$, $\mathcal{P}$ and verifying the condition, we are able to derive the main results of known  (GRH-based) deterministic factoring algorithms  \citep{Hua91-2, Hua91, Ron88, Ron92, Evd92, Evd94, IKS09} from our generic algorithm in a uniform way.
 
\item 
We investigate the {\em schemes conjecture} in \citep{IKS09}  and formulate analogous conjectures associated with various families of permutation groups, each of which has applications on deterministic polynomial factoring. Using a technique called induction of $\mathcal{P}$-schemes, we establish reductions among these conjectures and show that they form a hierarchy of relaxations of the original schemes conjecture.
%Moreover, I showed that the schemes conjecture in \citep{IKS09} can be viewed as the most difficult one of them, namely the one associated with symmetric groups.
%Finally, I showed that these new conjectures have a hierarchical structure by establishing reductions among them. 

\item We connect the complexity of deterministic polynomial factoring with the complexity of the Galois group $G$ of $\tilde{f}(X)$. Specifically, using techniques from permutation group theory, we obtain a (GRH-based) deterministic factoring algorithm  whose running time is bounded in terms of the noncyclic composition factors of $G$. In particular,  this algorithm runs in polynomial time if $G$ is in $\Gamma_k$ for some $k=2^{O(\sqrt{\log n})}$, where $\Gamma_k$ denotes the family of finite groups 
whose  noncyclic composition factors are all isomorphic of subgroups of the symmetric group of degree $k$. Previously, polynomial-time algorithms for $\Gamma_k$ were known only for bounded $k$.

\item We discuss various aspects of the theory of $\mathcal{P}$-schemes, including techniques of constructing new $\mathcal{P}$-schemes from old ones, $\mathcal{P}$-schemes for symmetric groups and linear groups, orbit $\mathcal{P}$-schemes, etc. For the closely related theory of $m$-schemes, we provide explicit constructions of strongly antisymmetric homogeneous $m$-schemes for $m\leq 3$. We also show that all antisymmetric homogeneous orbit $3$-schemes have a matching for $m\geq 3$, improving a result in \citep{IKS09}  that confirms the same statement for $m\geq 4$.
% and obtaining a polynomial-time algorithm for $d=n$ would fully resolve the problem of deterministic polynomial factoring over finite fields under the assumption of GRH.
\end{itemize}

In summary, our framework reduces the algorithmic problem of deterministic polynomial factoring over finite fields to a combinatorial problem concerning $\mathcal{P}$-schemes, allowing us to not only recover most of the known
results but also discover new ones. We believe progress in understanding $\mathcal{P}$-schemes associated with various families of permutation groups will shed some light on the ultimate goal of solving deterministic polynomial factoring over finite fields in polynomial time.
\end{abstract}

\tableofcontents

\mainmatter

\chapter{Introduction}\label{chap_intro}
%\addcontentsline{toc}{chapter}{Introduction}

We are interested in the problem of {\em deterministic univariate polynomial factoring} over finite fields: given a univariate polynomial $f$  of degree $n\in\N^+$ over a finite field $\F_q$, our goal is to {\em deterministically} compute a factorization of $f$ over $\F_q$
\[
f(X)=c\cdot\prod_{i=1}^k f_i(X),
\]
where $c\in\F_q$ is the leading coefficient of $f$ and each factor $f_i$ is {\em irreducible} over $\F_q$.
This is called the {\em complete factorization}\index{complete!factorization} of $f$ over $\F_q$. It is unique up to the  order of the factors $f_i$, since $\F_q[X]$ is a {\em unique factorization domain}\index{unique factorization domain}.
In addition, we are also interested in the more moderate goal of deterministically computing a {\em proper factorization}\index{proper!factorization} of $f$, i.e., factoring $f$ into more than one factors where each factor is allowed to be reducible.

\section{Previous work} 

Univariate polynomial factoring over finite fields has been extensively studied over the years as one of the most fundamental problems in computer algebra and  a common subroutine of many algorithms in coding theory, cryptography, computational number theory, etc.
%This problem has been extensively studied over the years. 
We review the previous work on this problem, with emphasis on {\em deterministic} factoring algorithms. For a detailed survey, see \citep{GP01}.

A truly polynomial-time factoring algorithm is required to factorize a  degree-$n$ polynomial $f(X)\in\F_q[X]$ in time $(n\log q)^{O(1)}$, since it takes $O(n\log q)$ bits to describe $f$.
 If randomness is allowed, such algorithms are well known: Berlekamp \citep{Ber70} described a randomized  algorithm that (completely) factorizes a univariate polynomial over $\F_q$ in polynomial time. The same paper also gave a deterministic reduction from the problem of factoring $f$ to the problem of finding the roots of certain other polynomials that split into $n$ linear factors over $\F_p$, where $p=\mathrm{char}(\F_q)$. More efficient randomized algorithms were discovered since then \citep{CZ81, VS92, KS98, U08, KU11}. The current best known running time has the exponent $3/2$ in  $n$, as achieved by \citep{KU11} based on the technique of {\em fast modular composition}.

On the other hand, despite much effort, factoring polynomials over finite fields in {\em deterministic} polynomial time remains a long-standing open problem. Berlekamp \citep{Ber67} gave the first deterministic algorithm for the general problem, whose running time is  polynomial in $n$ and $q$ (instead of $n$ and $\log q$).
His aforementioned paper \citep{Ber70} gave a deterministic algorithm that runs in time polynomial in $n$, $\log q$ and $p=\mathrm{char}(\F_q)$.
Deterministic algorithms with running time $(n\log q)^{O(1)}p^{1/2}$ were given in  \citep{Sho90, BKS15}.
Unfortunately, the  $p^{1/2}$-dependence on the characteristic $p$ of the field remains the best known for {\em unconditional} deterministic factoring algorithms, even if we only consider quadratic polynomials. 
Faster algorithms are known when $p-1$ is  assumed to be a {\em smooth} number \citep{Von87, Ron89, Sho96}. 
In addition, there are  deterministic algorithms for special polynomials based on the theory of elliptic curves or abelian varieties \citep{Sch85, Pil90}. Finally, the paper \citep{IKRS12} also unconditionally obtained some positive results on deterministic polynomial factoring in certain special cases.

A lot more is known if one accepts the generalized Riemann hypothesis (GRH): a deterministic polynomial-time algorithm that factorizes polynomials  of the form $X^n-a\in\F_p[X]$  under GRH was given in \citep{AMM77}. 
%which proceeds by   lifting $a\in\F_p$ to an integer and working over $\Z$.
Several GRH-based deterministic algorithms were proposed since then. These algorithms factorize a polynomial $f(X)\in \F_p[X]$ using the auxiliary information of a {\em lifted polynomial}, i.e., a polynomial $\tilde{f}(X)\in\Z[X]$ satisfying $\tilde{f}(X)\bmod p=f(X)$.
Huang  \citep{Hua91-2,Hua91} proved that a polynomial $f(X)\in\F_p[X]$ can be deterministically factorized  in polynomial time under GRH provided that the Galois group of the lifted polynomial is abelian.\footnote{In addition, $p$ is assumed to be a ``regular'' prime in \citep{Hua84, Hua91-2, Hua91} and also in \citep{Ron92}. This condition can be removed. See Section~\ref{sec_algreductiong} for a discussion.}  This was generalized in \citep{Evd92} to the case of solvable Galois groups.
 For a general Galois group $G$, the work \citep{Ron92} provided a deterministic algorithm that runs in time  polynomial in $|G|$ and the size of the input under GRH.
  In general, however, the cardinality of $G$ may be as large as $n!$, as attained by the symmetric group of degree $n$. Thus the algorithm in  \citep{Ron92} may take exponential time.
    
 In a different approach, R{\'o}nyai \citep{Ron88} showed that a polynomial $f(X)\in\F_q[X]$ of degree $n$ can be factorized deterministically in time  $(n^{n}\log q)^{O(1)}$ under GRH. The algorithm proceeds by manipulating tensor powers of the ring $\F_q[X]/(f(X))$, and does not need a lifted polynomial of $f$. Building on R{\'o}nyai's work, Evdokimov \citep{Evd94} showed that the problem can be solved in quasipolynomial time by presenting a deterministic $(n^{\log n}\log q)^{O(1)}$-time algorithm under GRH.
Evdokimov's algorithm remains the best known result on GRH-based deterministic polynomial factoring, although the $O(\log n)$ exponent of the running time was later improved by a certain constant factor \citep{CH00, IKS09, Gua09, Aro13}.

Efforts were made to understand the combinatorics behind R{\'o}nyai's and Evdokimov's algorithms \citep{CH00, Gao01}, culminating in the work \citep{IKS09} that proposed the notion of {\em $m$-schemes} together with an algorithm that subsumes those in \citep{Ron88, Evd94} (see also the follow-up work \citep{Aro13, AIKS14}). An $m$-scheme, parametrized by  $m\in\N^+$, is a collection of partitions of sets that satisfies a list of axioms. It was shown in \citep{IKS09} that whenever the algorithm fails to produce a proper factorization, there always exists an $m$-scheme satisfying strict combinatorial properties. Evdokimov's result can then be interpreted as the fact that such an $m$-scheme does not exist for sufficiently large $m=O(\log n)$. Finally, a conjecture on $m$-schemes, known as the {\em schemes conjecture}, was proposed in \citep{IKS09}, whose affirmative resolution would imply a polynomial-time factoring algorithm under GRH.

% Suppose $f(X)\in \F_p[X]$ is a nonzero polynomial of degree $n$ defined over a prime field $\F_p$. We could always choose a lifted polynomial $\tilde{f}$ of $f$ efficiently. Furthermore, we may assume $\tilde{f}$ is irreducible over $\Q$, otherwise we use the polynomial-time factoring algorithm for rational polynomials \citep{LLL82} to reduce to the subproblem for each factor. The polynomial $\tilde{f}$ is called an irreducible lifted polynomial of $f$. It turns out that the complexity of the Galois group of $\tilde{f}$ is related to the complexity of factoring $f$:\footnote{The Galois group of $\tilde{f}$ is the Galois group of the extension $L/\Q$, where $L$ is the splitting field of $\tilde{f}$ over $\Q$.} 

\paragraph{Role of GRH.} \index{GRH}\index{generalized Riemann hypothesis|see{GRH}}

GRH asserts that all nontrivial zeros of Dirichlet L-functions are on the line $\mathrm{Re}(z)=1/2$.
As noted in \citep{Ron92}, the known GRH-based algorithms (including our work) only need a consequence of GRH that finite fields can be efficiently constructed, and their $k$th power non-residues\footnote{For a prime factor $k$ of $q-1$, an element $x\in \F_q^\times$ is a {\em $k$th power residue}\index{power!residue} of $\F_q$  if $x\in (\F_q^\times)^k$. Otherwise it is a {\em $k$th power non-residue}\index{power!non-residue}.} can be efficiently found. Formally, for all the statements made under GRH throughout this thesis, we may use the following hypothesis instead.
\begin{hyp}[$*$]
 %Let $m$ be a positive integer and $p$ a prime number. 
 There exists a deterministic algorithm that given a prime number $p$ and an integer $d\in\N^+$, constructs\footnote{By constructing $\F_{p^d}$, we mean finding its {\em structure constants}\index{structure constants} in some $\F_p$-basis. See \citep{Len90}.} the finite field $\F_{p^d}$ in time polynomial in $d\log p$.
  In addition, given any prime factor $k$ dividing $p^d-1$, a $k$th power non-residue of $\F_{p^d}$ can be found deterministically in time polynomial in $k$ and $d\log p$.
\end{hyp}

See \citep{Hua91, LMO79} for the proof that Hypothesis ($*$) holds under GRH. By \citep{BIMS17}, it holds even under a weaker version of GRH, which asserts that all nontrivial zeros of Dirichlet L-functions are in the strip $\mathrm{Re}(z)\in [\frac{1}{2}-\epsilon, \frac{1}{2}+\epsilon]$ for some constant $\epsilon<1/2$.
 
\section{Main results}

In this thesis, we  introduce a family of mathematical objects called {\em $\mathcal{P}$-schemes}, generalizing the classical notion of association schemes   \citep{BI84} as well as the notion of $m$-schemes \citep{IKS09}. Based on  $\mathcal{P}$-schemes, we develop a unifying framework for  deterministic univariate polynomial factoring over finite fields under GRH.

\paragraph{$\mathcal{P}$-schemes.}
%We introduce the notion of {\em $\mathcal{P}$-schemes}, which is a generalization of the notion of $m$-schemes in \citep{IKS09}. 
Roughly speaking, given a finite group $G$ and a poset $\mathcal{P}$ of subgroups of $G$,  a $\mathcal{P}$-scheme is collection of partitions,
\[
\mathcal{C}=\{C_H: H\in\mathcal{P}\},
\]
satisfying certain constraints, where each $C_H$ is a partition of the right coset space $H\backslash G=\{Hg: g\in G\}$. 
The formal definition is given in Definition~\ref{defi_pscheme}. We also define various properties of $\mathcal{P}$-schemes, including {\em antisymmetry}, {\em strong antisymmetry}, {\em discreteness}, and {\em homogeneity}. These properties play important roles in our polynomial factoring algorithms.

When  $G$ is chosen to be a symmetric group and $\mathcal{P}$ is a poset of {\em stabilizer subgroups} (with respect to the natural action of $G$),  we  recover  the notion of $m$-schemes \citep{IKS09}:
\begin{thm}[informal]\label{thm_equivinformal}
Suppose $G=\sym(S)$ acts naturally on a finite set $S$ and $\mathcal{P}$ consists of the (pointwise) stabilizers $G_T$ for all subsets $T\subseteq S$ satisfying $1\leq |T|\leq m$. 
Then a $\mathcal{P}$-scheme $\mathcal{C}$ is equivalent to an $m$-scheme $\Pi$ on $S$. Moreover, $\mathcal{C}$ is antisymmetric (resp. strongly antisymmetric, discrete on $G_x$ for $x\in S$, homogeneous on $G_x$ for $x\in S$) iff $\Pi$ has the corresponding property.
\end{thm}
This result in fact holds as long as $G$ is $k$-transitive for sufficiently large $k$. See Theorem~\ref{thm_mandp} for the formal statement.

In this way, we regard the theory of $m$-schemes \citep{IKS09,Aro13, AIKS14} as part of the richer theory of $\mathcal{P}$-schemes. The advantage of adopting the notion of $\mathcal{P}$-schemes is that these objects capture not only the combinatorial structure of $m$-schemes but also the information provided by the group $G$ and the poset $\mathcal{P}$,  which allows  us to carry out both the Galois-theoretic/group-theoretic approach \citep{Hua91-2, Hua91, Evd92, Ron92} and the combinatorial approach \citep{Evd94, IKS09} of deterministic polynomial factoring in a uniform way.

\paragraph{A unifying framework for deterministic polynomial factoring.}

The theory of $\mathcal{P}$-schemes is applied to deterministic polynomial factoring as follows.
For simplicity, assume $f$ is a degree-$n$ polynomial  that is defined over a prime field $\F_p$ and factorizes into $n$ distinct linear factors over $\F_p$. Let $\tilde{f}(X)\in \Z[X]$ be an {\em irreducible lifted polynomial} of $f$, defined as follows: 
\begin{defi}[lifted polynomial]\label{defi_liftpoly}
A {\em lifted polynomial} of a degree-$n$ polynomial $f(X)\in\F_p[X]$ is a polynomial $\tilde{f}(X)\in \Z[X]$  of degree $n$ satisfying $\tilde{f}\bmod p=f$. An {\em irreducible lifted polynomial}\index{irreducible lifted polynomial} of $f$ is a lifted polynomial of $f$ that is irreducible over $\Q$. 
\end{defi}\index{lifted polynomial}
Let $L$ be the splitting field of $\tilde{f}(X)$ over $\Q$ and let $G=\gal(L/\Q)$. By Galois theory, we have a one-to-one correspondence between the subgroups of $G$ and the subfields of $L$
\[
H=\gal(L/K)\longleftrightarrow  K=L^H,
\]
where $L^H$ denotes the fixed subfield of $H$. 
%Thus any poset $\mathcal{P}$ of subgroups of $G$ corresponds to a poset $\mathcal{P}^\sharp:=\{L^H: H\in\mathcal{P}\}$ of subfields of $L$.

In  Chapter~\ref{chap_alg_prime}, we design a generic  algorithm, which we refer to as the {\em $\mathcal{P}$-scheme algorithm}, that deterministically factorizes $f$ under GRH given $f$ and $\tilde{f}$.  The generic part of the algorithm is a subroutine that uses $\tilde{f}$ to construct a poset of subfields of $L$, which in turn corresponds to a poset $\mathcal{P}$ of subgroups of $G$ by Galois theory. We then prove that the algorithm always produces the complete factorization (resp. a proper factorization) of $f$ under GRH, unless a combinatorial condition regarding $\mathcal{P}$-schemes fails to hold.\footnote{The condition requires all strongly antisymmetric $\mathcal{P}$-schemes to be discrete (resp. inhomogeneous) on $G_x$, where $x$ is a root of $\tilde{f}$ in $L$. See Theorem~\ref{thm_algmain2formal} for the formal statement.} Therefore the problem of deterministic polynomial factoring reduces to the problem of verifying this combinatorial condition about $\mathcal{P}$-schemes.

 By choosing various  posets $\mathcal{P}$ and verifying the condition, we  recover the main results of the previous work  \citep{Hua91-2, Hua91, Ron88, Ron92, Evd92, Evd94, IKS09} using the $\mathcal{P}$-scheme algorithm. 
Our algorithm thus provides a unifying framework for deterministic polynomial factoring over finite fields.

\paragraph{The generalized $\mathcal{P}$-scheme algorithm.}

The $\mathcal{P}$-scheme algorithm above is subject to the condition that the input polynomial is defined over a prime field $\F_p$ and factorizes into  distinct linear factors over $\F_p$. In Chapter~\ref{chap_alg_general}, we extend it to the {\em generalized $\mathcal{P}$-scheme algorithm}  that works for arbitrary polynomials $f(X)\in\F_q[X]$. The results obtained from the $\mathcal{P}$-scheme algorithm  are then proved in full generality. 

Several new ideas and a significant amount of work are required in the development of the generalized $\mathcal{P}$-scheme algorithm.  See Chapter~\ref{chap_alg_general} for the details.

\paragraph{Constructing new $\mathcal{P}$-schemes from old ones.}

We develop various techniques of constructing new $\mathcal{P}$-schemes from old ones, including {\em restriction}, {\em induction},  {\em extension}, etc. 
These techniques are useful for investigating the existence of certain $\mathcal{P}$-schemes, allowing us to reduce one case to another. 

In particular, using induction of  $\mathcal{P}$-schemes, we show that for finite groups $H\subseteq G$ and a poset  $\mathcal{P}$ of subgroups of $H$, a $\mathcal{P}$-scheme with various properties (antisymmetry, strong antisymmetry, etc.) can be used to construct  a $\mathcal{P}'$-scheme with the same properties, where $\mathcal{P}'$ is a certain poset of $G$. Intuitively, this means polynomial factoring ``becomes easier'' if the Galois group $G$ is replaced by a subgroup $H$. We make this intuition rigorous regarding the  {\em schemes conjecture} proposed in \citep{IKS09}. See below for a more detailed discussion.

In addition, we define the {\em direct product} and the {\em wreath product} of $\mathcal{P}$-schemes, generalizing the corresponding operations of permutation groups and association schemes \citep{SS98, Bai04}.  We also define the  direct product and the wreath product of $m$-schemes. A consequence of these operations is that either the schemes conjecture in \citep{IKS09} holds, or it has infinitely many counterexamples. 

\paragraph{Schemes conjectures for families of permutation groups.}

The work \citep{IKS09} proposed a combinatorial conjecture on $m$-schemes, called the {\em schemes conjecture}, whose positive resolution would imply a deterministic polynomial-time factoring algorithm under GRH.  
Proving this conjecture appears to be difficult.
However, as noted in Theorem~\ref{thm_equivinformal} above, an $m$-scheme is essentially a $\mathcal{P}$-scheme in the (worst) case of symmetric groups, with respect to a poset $\mathcal{P}$ of pointwise stabilizers.
This observation suggests that one should first formulate and attack the analogous conjectures for ``less complex'' Galois groups. 

For each family $\mathcal{G}$ of finite permutation groups, we formulate an analogous conjecture, called the {\em schemes conjecture for $\mathcal{G}$}. Like the original schemes conjecture, the  schemes conjecture for $\mathcal{G}$  also implies a deterministic polynomial-time factoring algorithm under GRH, provided that that Galois group of the lifted polynomial $\tilde{f}$, as a permutation group on the set of roots of $\tilde{f}$, is a member of $\mathcal{G}$.
Moreover, we show that these conjectures  form a hierarchy of relaxations of the original schemes conjecture in \citep{IKS09}. More specifically, for two families of finite permutation groups $\mathcal{G}$ and $\mathcal{G}'$ such that every member of $\mathcal{G}$ is (permutation isomorphic to) a subgroup of member in $\mathcal{G}'$,  the schemes conjecture for $\mathcal{G}$ is implied by that for $\mathcal{G}'$.
The worst case occurs when $\mathcal{G}$ is the family of symmetric groups, which yields (a slight relaxation of) the original schemes conjecture.
We hope  progress on this hierarchy of conjectures will shed some light on the original schemes conjecture  and pave the way for solving deterministic polynomial factoring over finite fields in polynomial time under GRH.

\paragraph{Galois groups with restricted noncyclic composition factors.}

% In particular, by studying $\mathcal{P}$-schemes in depth, we prove that a polynomial $f(X)\in \F_p[X]$ can be factorized in polynomial time given an irreducible polynomial $\tilde{f}(X)\in\Z[X]$ lifting $f(X)$ whose Galois group is in $\Gamma_d$ for $d=2^{O(\sqrt{\log n})}$, where $\Gamma_d$ denotes the family of finite groups whose noncyclic composition factors are isomorphic to subgroups of $\sym(d)$. Previously this was known only for bounded $d$. 

Using our framework of $\mathcal{P}$-schemes, we design a GRH-based deterministic  factoring algorithm that completely factorizes a polynomial $f$  using a lifted polynomial $\tilde{f}$, such that the running time of the algorithm is controlled by the {\em noncyclic composition factors}\footnote{Recall that a composition factor of a finite group is a {\em finite simple group}, and by the {\em classification of finite simple groups} (CFSG) it is isomorphic to one of the following groups: a cyclic group of prime order, an  alternating group, a classical group, an exceptional group of Lie type, or one of the 26 sporadic simple groups.} of the  Galois group  of $\tilde{f}$.
More specifically, we have
\begin{thm}[informal]\label{thm_compboundinf}
Under GRH, there exists a deterministic algorithm that 
given $f(X)\in\F_q[X]$ and a lifted polynomial\footnote{For a general (not necessarily prime) finite field $\F_q$, we use a more general definition of lifted polynomials (Definition~\ref{defi_genlpoly}) instead of Definition~\ref{defi_liftpoly}.} $\tilde{f}$ of $f$ with the Galois group $G$,
completely factorizes $f$ in time polynomial in $k(G)^{\log k(G)}$, $r(G)$ and the size of the input,  where $k(G)$ (resp. $r(G)$) is the maximum degree (resp. maximum order) of the alternating groups (resp. classical groups) among the composition factors of $G$.
\end{thm}
See Theorem~\ref{thm_compbound} for the formal statement.  Now fix $k\in\N^+$ and consider the family of finite groups whose noncyclic composition factors are all isomorphic to subgroups of $\sym(k)$. This family is commonly denoted by $\Gamma_k$ in the literature, and plays a significant role in graph isomorphism testing \cite{Luk82, Mi83}, asymptotic group theory \cite{BCP82, Pyb93, PS97} and computational group theory \cite{Luk93, Ser03}. 
It is known that a classical group of order $r$ lies in $\Gamma_k$ only if $r=k^{O(\log k)}$ \cite{Coo78}. So Theorem~\ref{thm_compboundinf} implies

\begin{thm}[informal]\label{thm_gammainf}
Under GRH, there exists a deterministic algorithm that 
given $f(X)\in\F_q[X]$ of degree $n$ and a lifted polynomial $\tilde{f}$ of $f$,
completely factorizes $f$ in time polynomial in $n$, $\log q$ and $k^{\log k}$, where $k$ is the smallest positive integer such that the Galois group of $\tilde{f}$ is in $\Gamma_k$.
\end{thm}
See Theorem~\ref{thm_gamma} for the formal statement. It refines and generalizes the main results of \citep{Hua91-2, Hua91, Evd92, Ron92, Evd94}.  Note that the algorithm runs in polynomial time under GRH provided that $k=2^{O(\sqrt{\log n})}$.
Previously,  polynomial-time factoring algorithms for $\Gamma_k$ were known only  for bounded $k$ under GRH \citep{Evd92, BCP82}. 
%: for $d\leq 4$ this follows directly from the deterministic polynomial-time factoring algorithm for solvable Galois groups \cite{Evd92}.  For $d=O(1)$, it follows from the proof in \cite{Evd92} together with the bound  in \cite{BCP82} for the orders of primitive permutation groups.
% Note that achieving $k=n$ would fully resolve the problem of deterministic polynomial factoring under GRH.

\paragraph{Other results.}
Finally, we list some other results obtained in this thesis.
\begin{enumerate}
\item The  schemes conjecture in \citep{IKS09} asserts that if a homogeneous antisymmetric orbit $m$-schemes on a set $S$   has no matching, then $m=O(1)$  (see Chapter~\ref{chap_pscheme} for the definition of matchings). Currently, the best known upper bound for $m$ is $m\leq c\log |S|+O(1)$, where $c= \frac{2}{\log 12}=0.5578\cdots$. We consider the analogous problem for a general linear group $\gl(V)$ over a finite field $\F_q$ acting naturally on $S=V-\{0\}$, and show that for this new problem, we have a slightly improved bound $m\leq c'\log |S|+O(1)$ where $c'=\frac{4}{4\log q + \log 12}\leq 0.5273\cdots$ (Theorem~\ref{thm_slightimp}). In addition, we  consider the analogous problems for the groups $\gl(V)$, $\gammal(V)$, $\pgl(V)$, and $\pgammal(V)$, and show that these problems are equivalent, in the sense that the optimal values of $m$ for them differ from each other by at most a constant (Theorem~\ref{thm_equivlin}).
\item We generalize the notion of orbit schemes in \citep{IKS09}, or what we call {\em orbit $m$-schemes}, to the notion of {\em orbit $\mathcal{P}$-schemes}. We also prove that an orbit $m$-scheme associated with a group $K$ is antisymmetric iff the order of $K$ is coprime to $1,2,\dots,m$ (Lemma~\ref{lem_antisym_orbitm}), which in turn shows that a result of \citep{Ron88, IKS09} on antisymmetric $m$-schemes is tight (cf. Lemma~\ref{lem_integrality_mscheme} and Example~\ref{exmp_antisym_orbit}).
\item The paper \citep{IKS09} showed that the schemes conjecture is true when restricted to orbit schemes, by proving that all antisymmetric homogeneous  orbit $m$-schemes on a set of cardinality greater than one have a matching for $m\geq 4$. We prove that the later statement in fact holds for $m\geq 3$ (Theorem~\ref{thm_nonexistence3sch}).
\end{enumerate}

\section{Outline of the thesis.} 

Basic notations and preliminaries are given in the next section, and additional preliminaries are given at the beginning of subsequent chapters.

Chapter~\ref{chap_pscheme} introduces definitions and develops basic results about $\mathcal{P}$-schemes: we first define  $\mathcal{P}$-schemes and their various properties. After reviewing the notion of $m$-schemes   in \citep{IKS09} and their connection with association schemes, we prove the formal version of Theorem~\ref{thm_equivinformal} above. Then we investigate the notion of {\em orbit schemes} in  \citep{IKS09}, and extend it  to our framework of $\mathcal{P}$-schemes. Finally, some concrete examples of {\em strongly antisymmetric homogeneous} $m$-schemes are given for small $m$.

\begin{figure}[htb]
\centering
\begin{tikzpicture}[>=stealth, mynode/.style={shape=rectangle,draw,rounded corners, thick,  inner sep=6pt}, node distance = 1.5cm and 0.8cm]
    % create the nodes
    \node (c1) [mynode] {Chapter~\ref{chap_pscheme}};
    \node (c2) [mynode, below= of c1]{Chapter~\ref{chap_alg_prime}};
    \node (c4) [mynode, below=of c2]{Chapter~\ref{chap_alg_general}};
    \node (c3) [mynode, left=of c4]{Chapter~\ref{chap_constructnum}};
    \node (c5) [mynode, right=of c2]{Chapter~\ref{chap_common}};
    \node (c6) [mynode, below=of c5, align=center]{Chapter~\ref{chap_sym}};
    \node (c7) [mynode, below=of c4]{Chapter~\ref{chap_primitive}};
    % connect the nodes
  \draw[->] (c1) to  (c2);
  \draw[->] (c1) to  (c5);
  \draw[->] (c2) to  (c4);
  \draw[->, dashed] (c3) to  (c4);
  \draw[->, dashed] (c3) to (c7);
  \draw[->, dashed] (c4) to  (c7);
  \draw[->] (c5) to  (c6);
  \draw[->] (c6) to  (c7);
\end{tikzpicture}
\caption{Dependencies among chapters}\label{fig_dependency}
\end{figure}
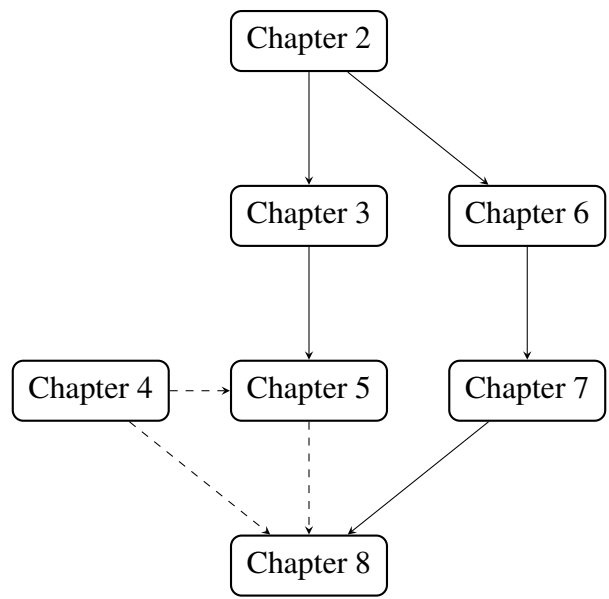

The rest of the thesis is divided into two parts: Chapters~\ref{chap_alg_prime}--\ref{chap_alg_general} constitute the algorithmic part of the thesis, whereas Chapters~\ref{chap_common}--\ref{chap_primitive} focus on further development of the theory of $\mathcal{P}$-schemes. The latter  is mostly algorithm-free, except that Section~\ref{lem_syspn} contains an algorithm  that depends on Section~\ref{sec_compositum}, Section~\ref{sec_consred}, and Theorem~\ref{thm_algmain2formalg}. 
The dependencies among chapters are roughly illustrated in Figure~\ref{fig_dependency}. 
%In addition, Section~\ref{sec_consred} uses facts about $\mathcal{P}$-schemes whose proofs are deferred to Chapter~\ref{chap_common}. Finally, Section~\ref{sec_symnatural} may be read right after Chapter~\ref{chap_pscheme}.

In Chapter~\ref{chap_alg_prime}, we develop the $\mathcal{P}$-scheme algorithm, and use it to reprove the main results of  \citep{Hua91-2, Hua91, Ron88, Ron92, Evd94, IKS09}.
As mentioned above, the results in Chapter~\ref{chap_alg_prime} are subject to the condition that the input polynomial is defined over a prime field $\F_p$ and factorizes into  distinct linear factors over $\F_p$.
% This assumption is later lifted by the {\em generalized $\mathcal{P}$-scheme algorithm} in Chapter~\ref{chap_alg_general}.

The $\mathcal{P}$-scheme algorithm  requires a subroutine that constructs a collection of number fields. In Chapter~\ref{chap_constructnum}, we discuss various ways of implementing this subroutine  and   survey techniques of constructing number fields  in the literature \citep{Len83, Lan84, Lan85, LM85, Evd92}. 
%In particular, we reprove the main result of \citep{Evd92} in the terminology of $\mathcal{P}$-schemes, which states that under GRH, there exists a deterministic algorithm factoring $f(X)\in \F_p[X]$ in polynomial time provided that a solvable lifted polynomial of $f$ is given.\footnote{In fact, we only prove it under the assumption that $f$ factorizes into  distinct linear factors over $\F_p$. This assumption is again lifted by the generalized $\mathcal{P}$-scheme algorithm in Chapter~\ref{chap_alg_general}.}

In Chapter~\ref{chap_alg_general}, we develop the generalized $\mathcal{P}$-scheme algorithm where the condition about the input polynomial  is no longer needed. The results in  Chapter~\ref{chap_alg_prime} are then proved in full generality.

Chapter~\ref{chap_common} develops various techniques of constructing new $\mathcal{P}$-schemes from old ones. 
%These techniques are useful for investigating the existence of certain $\mathcal{P}$-schemes, allowing us to reduce one case to another. 
 In Section~\ref{sec_schconj}, we formulate the schemes conjectures for families of finite permutation groups and show that these conjectures form a hierarchy of relaxations of the schemes conjecture proposed in \citep{IKS09}. Our result that an antisymmetric homogeneous orbit $m$-scheme on a set of cardinality $n>1$ has a matching for $m\geq 3$ is proved in Section~\ref{sec_primitiveorbit}, where we also discuss {\em primitivity} of $m$-schemes.

Chapter~\ref{chap_sym} discusses the (non-)existence of certain $\mathcal{P}$-schemes for symmetric groups and linear groups. In particular, we review the result in \citep{Aro13}  on $m$-schemes (based on the work of \citep{Evd94, IKS09}, and independently discovered in \citep{Gua09}), and interpret it as a result about $\mathcal{P}$-schemes with respect to the natural action of symmetric groups. We also extend it to a more general result  about $\mathcal{P}$-schemes with respect to {\em standard actions} of symmetric groups. The analysis employs a technical ``self-reduction lemma'' proven in Section~\ref{sec_selfreduction}, which is also heavily used in Chapter~\ref{chap_primitive}. Some results about $\mathcal{P}$-schemes for linear groups are also given.

Finally, in Chapter~\ref{chap_primitive}, we describe our deterministic factoring algorithm for  Galois groups with restricted noncyclic composition factors. More specifically, we give the algorithm and its analysis in Section~\ref{sec_partialstab}, assuming a statement about $\mathcal{P}$-schemes for primitive permutation groups (Theorem~\ref{thm_primcriterion}). The rest of Chapter~\ref{chap_primitive} then focuses on verifying this statement.

\section{Notations and preliminaries}
%\label{sec_pre}

Denote by $\N^+$ the set of positive integers.  \nomenclature[1a]{$\N^+$}{set of positive integers}
For $k\in \N^+$, we denote by $[k]$ the set $\{1,2,\dots,k\}$.   \nomenclature[2a]{$[k]$}{set $\{1,2,\dots,k\}$}
For two sets $A$ and $B$, write $A-B$ for the set difference $\{x: x\in A ~\text{and}~ x\not\in B\}$.\footnote{This is often denoted by $A\setminus B$. We use $A-B$ to avoid confusion with a right coset space $H\backslash G$.} \nomenclature[2b]{$A-B$}{set difference $\{x: x\in A ~\text{and}~ x\not\in B\}$}
The cardinality of a finite set $S$ is denoted by $|S|$. \nomenclature[2c]{$\lvert S\rvert$}{cardinality of $S$}
Denote by $\log$ the logarithmic function with base $2$. \nomenclature[2d]{$\log$}{logarithmic function with base $2$}

A partition\index{partition} of a finite set $S$ is a set $P$ of nonempty subsets of $S$ satisfying $S=\coprod_{B\in P} B$, where $\coprod$ denotes the disjoint union.   \nomenclature[3a]{$\coprod S_i$}{disjoint union of sets $S_i$}
Each $B\in P$ is called a {\em block} \index{block} of $P$.  
For two partitions $P$ and $P'$ of $S$, we say $P$ {\em refines }$P'$, or $P$ is a {\em refinement}\index{refinement} of $P'$, if every block in $P'$ is a disjoint union of blocks in $P$. We say the refinement is {\em proper}\index{proper!refinement} if $P\neq P'$.
Denote by $0_S$ the coarsest partition of $S$, i.e. the one consisting of a single block $S$.  \nomenclature[3b]{$0_S$}{coarsest partition of a set $S$}
Denote by $\infty_S$ the finest partition of $S$, i.e., $\infty_S=\{\{x\}:x\in S\}$.  \nomenclature[3c]{$\infty_S$}{finest partition of a set $S$}
For $T\subseteq S$ and a partition $P$ of $S$, define $P|_T:=\{B\cap T: B\in S\}-\{\emptyset\}$ which is a partition of $T$, called the {\em restriction} of $P$ to $T$.\index{restriction!of a partition}
\nomenclature[3d]{$P|_T$}{restriction of a partition $P$ to a subset $T$}
For a set $S$ and $k\in\N^+$, define the set $S^{(k)}:=\{(x_1,\dots,x_k)\in S^k: x_i\neq x_j ~\text{for}~ i\neq j\}$  consisting of $k$-tuples of distinct elements. \nomenclature[3e]{$S^{(k)}$}{set of $k$-tuples of distinct elements from $S$}

Write $f\circ g$ for the composition of two functions $f$ and $g$, from right to left. \nomenclature[3f]{$f\circ g$}{composition of functions $f$ and $g$, from right to left}
We note that this is the common convention, although group theorists often use the opposite convention $gf$.
For a function $f$ and a subset $T$ of the domain of $f$, denote by $f|_T$ the restriction of $f$ to $T$. 
\nomenclature[3g]{$f|_T$}{restriction of a function $f$ to a subset $T$ of its domain}
For a field $K$, denote the characteristic of $K$ by $\cha(K)$.
\nomenclature[3h]{$\cha(K)$}{characteristic of a field $K$}

A polynomial is {\em monic}\index{monic polynomial} if its leading coefficient is one. For two polynomials $f(X),g(X)\in\F_q[X]$ over a finite field $\F_q$ that are not both zero, define their {\em greatest common divisor} \index{great common divisor} $\gcd(f,g)$ to be the unique monic polynomial $h(X)\in\F_q[X]$ of the greatest degree that divides both $f$ and $g$. 
\nomenclature[3i]{$\gcd(f,g)$}{greatest common divisor of polynomials $f$ and $g$}
It is well defined since $\F_q[X]$ is a unique factorization domain,\index{unique factorization domain} and can be computed efficiently from $f$ and $g$ using the {\em Euclidean algorithm} \citep{VG13}.

\paragraph{Basic notations about groups.}

All groups in this thesis are finite. 
Write $e$ for the identity element of a group.
\nomenclature[4a]{$e$}{identity element of a group}\nomenclature[4b]{$gH$}{left coset $\{gh: h\in H\}$}\nomenclature[4c]{$Hg$}{right coset $\{hg: h\in H\}$} 
For a group $G$, a subgroup $H$ of $G$, and $g\in G$, write $gH$ for the {\em left coset} $\{gh: h\in H\}$\index{left coset} and $Hg$ for the {\em right coset} $\{hg: h\in H\}$\index{right coset}. 
Write $G/H$ for the {\em left coset space} $\{gH: g\in G\}$ and $H\backslash G$ for the {\em right coset space} $\{Hg: g\in G\}$
\nomenclature[4d]{$G/H$}{left coset space $\{gH: g\in G\}$}
\nomenclature[4e]{$H\backslash G$}{ right coset space $\{Hg: g\in G\}$}
For two subgroups $H,K$ of $G$ and $g\in G$, write $HgK$ for the {\em double coset} $\{hgh': h\in H, h'\in K\}$\index{double coset},
\nomenclature[4f]{$HgK$}{double coset $\{hgh': h\in H, h'\in K\}$}
and write $H\backslash G/K$ for the {\em double coset space} $\{HgK: g\in G\}$.
\nomenclature[4g]{$H\backslash G/K$}{double coset space $\{HgK: g\in G\}$}
Define $[G:H]:=|G|/|H|$, called the {\em index} of $H$ in $G$.\index{index of a subgroup}
\nomenclature[4h]{$[G:H]$}{index of a subgroup $H$ in $G$}
Write $\langle H_1,\dots,H_k\rangle$ for the {\em join}\index{join} of subgroups $H_1,\dots,H_k$, i.e., the subgroup generated by $H_1,\dots,H_k$. 
\nomenclature[4i]{$\langle H_1,\dots,H_k\rangle$}{join of subgroups $H_1,\dots,H_k$}
Write $\langle g_1,\dots,g_k\rangle$ for the subgroup generated by the group elements $g_1,\dots,g_k$.
\nomenclature[4j]{$\langle g_1,\dots,g_k\rangle$}{subgroup generated by $g_1,\dots,g_k$}

A {\em subquotient}\index{subquotient} of a group $G$ is a quotient group of a subgroup of $G$.\index{subquotient}
Two subgroups $H$ and $H'$ are said to be {\em conjugate} in $G$ if $H'=gHg^{-1}$ for some $g\in G$.\index{conjugate!subgroup}
A subgroup $H$ is said to be {\em normal} in $G$ or a {\em normal subgroup} of $G$ if $gHg^{-1}=H$ for all $g\in G$.\index{normal subgroup} 
Write $H\unlhd G$ for $H$ being normal in $G$.
\nomenclature[4k]{$H\unlhd G$}{$H$ is a normal subgroup of $G$}
Define the {\em normalizer} \index{normalizer} of $H$ in $G$ to be $N_G(H):=\{g\in G: gHg^{-1}=H\}$. \nomenclature[4l]{$N_G(H)$}{normalizer of $H$ in $G$}
We have $H\unlhd N_G(H)$, and indeed $N_G(H)$ is the unique maximal subgroup of $G$ with this property.
The {\em center}\index{center of a group} of $G$, denoted by $Z(G)$, is the subgroup $\{g\in G: gh=hg~\text{for all}~h\in G\}$.
\nomenclature[4m]{$Z(G)$}{center of $G$}
A subgroup $H$ of $G$ is {\em maximal}\index{maximal subgroup} if $H\neq G$ and there exists no subgroup $H'$ of $G$ satisfying $H\subsetneq H'\subsetneq G$.

For a finite set $S$, denote by $\sym(S)$ and $\alt(S)$ the symmetric group and the alternating group on $S$ respectively. We also write $\sym(n)$ and $\alt(n)$ when $S=[n]$.  Permutations are often written in the cycle notation, where  $(a_1~a_2~\cdots~a_n)$ denotes the cyclic permutation sending $a_i$ to $a_{i+1}$ for $1\leq i<n$ and $a_n$ to $a_1$.
\nomenclature[4n]{$(a_1~a_2~\cdots~a_n)$}{permutation  sending $a_i$ to $a_{i+1}$ for $1\leq i<n$ and $a_n$ to $a_1$}
\nomenclature[4o]{$\sym(S)$, $\sym(n)$}{symmetric group}
\nomenclature[4p]{$\alt(S)$, $\alt(n)$}{alternating group}
%\nomenclature[4q]{$\sym(n)$}{symmetric group on $[n]$}
%\nomenclature[4r]{$\alt(n)$}{alternating group on $[n]$}

For a group $G$, denote by $\aut(G)$ the automorphism group of $G$, i.e., the group of invertible homomorphisms $\rho:G\to G$ where the group operation is defined by composition. \index{automorphism group!of a group}
\nomenclature[4s]{$\aut(G)$}{automorphism group of a group $G$}
For $g\in G$, the map $\tau_g:G\to G$ sending $h\in G$ to $ghg^{-1}$ is an automorphism of $G$, called an {\em inner automorphism} of $G$. \index{inner automorphism}
Define $\inn(G):=\{\tau_g: g\in G\}$,  called the {\em inner automorphism group} of $G$, which is a normal subgroup of $\aut(G)$. 
\nomenclature[4t]{$\inn(G)$}{inner automorphism group of a group $G$}
Define $\out(G):=\aut(G)/\inn(G)$, called the {\em outer automorphism group} of $G$. 
\index{outer automorphism group}
\nomenclature[4u]{$\out(G)$}{outer automorphism group of a group $G$}

\paragraph{Group actions.} 
Let $G$ be a group and $S$ be a finite set. A {\em (left) group action}\index{group action|see{action}} or an {\em action}\index{action} of $G$ on $S$ is a function  $\varphi:G\times S\to S$ satisfying (1) $\varphi(e,x)=x$ for all $x\in S$ and (2) $\varphi(g,\varphi(h,x))=\varphi(gh,x)$ for all $x\in S$ and $g,h\in G$.
We also say $G$ {\em acts on} $S$ and $S$ is a {\em $G$-set}.\index{Gset@$G$-set} 
%The cardinality of $S$ is called the {\em degree} of $G$ (with respect to the action $\varphi$).\index{degree}
 We usually denote $\varphi(g,x)$ as $\prescript{g}{}{x}$ when $\varphi$ is clear from the context. 
\nomenclature[5a]{$\prescript{g}{}{x}$}{alias for the element $\varphi(g,x)$ where $\varphi$ is a group action}
 For $T\subseteq S$, write $\prescript{g}{}{T}$ for the set $\{\prescript{g}{}{x}: x\in T\}$.
\nomenclature[5b]{$\prescript{g}{}{T}$}{ set $\{\prescript{g}{}{x}: x\in T\}$}
Again, we note that group theorists commonly adopt the right action convention $xgh=(xg)h$ instead of our left action convention. One can switch between the two conventions by taking the inverse map $g\mapsto g^{-1}$.

Given a $G$-set $S$, the elements of $G$ act as permutations of $S$. This gives a group homomorphism $\rho: G\to \sym(S)$, called a {\em permutation representation}\index{permutation representation} of $G$ on $S$. The action of $G$ on $S$ is {\em faithful}\index{faithful action} if $\rho$ is injective.
The image $\rho(G)$ is called a {\em permutation group}\index{permutation group} on $S$. When the action is faithful and clear from the context, we usually just say $G$ is a permutation group on $S$.

\paragraph{Orbits and stabilizers.}
For a $G$-set $S$, the {\em orbit}\index{orbit} or {\em $G$-orbit}\index{Gorbit@$G$-orbit|see{orbit}} of an element $x\in S$ is $Gx:=\{\prescript{g}{}{x}:g\in G\}$. 
\nomenclature[6a]{$Gx$}{$G$-orbit $\{\prescript{g}{}{x}:g\in G\}$ of an element $x$}
The set $S$ is a disjoint union of its $G$-orbits.
The {\em stabilizer}\index{stabilizer} of $x\in S$ is $G_x:=\{g\in G:\prescript{g}{}{x}=x\}$. 
\nomenclature[6b]{$G_x$}{stabilizer of an element $x$}
For $T\subseteq S$, define the {\em pointwise stabilizer}\index{pointwise stabilizer} 
\[
G_T:=\{g\in G: \prescript{g}{}{x}=x \text{ for all } x\in T\}
\] 
\nomenclature[6c]{$G_T$}{pointwise stabilizer of a set $T$}
and the {\em setwise stabilizer}\index{setwise stabilizer} 
\[
G_{\{T\}}:=\{g\in G: \prescript{g}{}{T}=T\}.
\] 
For $T=\{x_1,\dots,x_k\}\subseteq S$ we also write $G_{x_1,\dots,x_k}$ for $G_T$. 
Let $S^G:=\{x\in S: \prescript{g}{}{x}=x~\text{for all } g\in G\}$ be the {\em set of fixed points} of $G$.\index{set of fixed points}
\nomenclature[6d]{$G_{\{T\}}$}{setwise stabilizer of a set $T$}
\nomenclature[6e]{$G_{x_1,\dots,x_k}$}{pointwise stabilizer of $\{x_1,\dots,x_k\}$}
\nomenclature[6f]{$S^G$}{set of fixed points of $G$ in a set $S$}

An action of $G$ on a set $S$ is {\em transitive}\index{transitive action} if it has only one orbit. 
It is {\em semiregular}\index{semiregular action} if $G_x$ is trivial for all $x\in S$. A group action is {\em regular}\index{regular action} if it is both transitive and semiregular. 
For $k\in \N^+$,  an action of $G$ on $S$ induces an action on $S^{(k)}$ via
\[
\prescript{g}{}{(x_1,\dots,x_k)}=(\prescript{g}{}{x_1},\dots,\prescript{g}{}{x_k}),
\]
called the {\em diagonal action} of $G$ on $S^{(k)}$.\index{diagonal action}
For $1\leq k\leq |S|$,
we say the action of $G$ on $S$ is {\em $k$-transitive}\index{ktransitive@$k$-transitive} if the corresponding diagonal action of $G$ on $S^{(k)}$ is transitive.
We say it is {\em $(k+1/2)$-transitive}\index{kztransitive@$(k+1/2)$-transitive} if it is $k$-transitive, and in addition for all $T\subseteq S$ of cardinality $k$, either the $G_T$-orbit of every $x\in S-T$ contains more than one element, or $|S-T|=1$.
A $(k+1)$-transitive action is also $(k+1/2)$-transitive. 
For more discussion about half transitivity, see \citep{Wie14}.

\paragraph{$G$-modules and $G$-invariant elements.}
Given a group $G$, an abelian group $A$ is called a {\em $G$-module}\index{Gmodule@$G$-module} if it has an action of $G$ compatible with its abelian group structure, i.e., $\prescript{g}{}{x}+\prescript{g}{}{y}=\prescript{g}{}{(x+y)}$ for $x,y\in A$ and $g\in G$. 
The set of fixed points $A^G$ is a subgroup of $A$, known as the  subgroup of {\em $G$-invariant} elements of $A$.\index{Ginvariant@$G$-invariant}
\nomenclature[7a]{$A^G$}{subgroup (resp. subring, subfield) of $G$-invariant elements of the abelian group (resp. ring, field) $A$}
Suppose in addition that $A$ is a ring (resp. field) and the action of $G$ respects the multiplication of $A$ as well, then $A^G$ is a subring (resp. subfield) of $A$, called the {\em fixed subring} (resp. {\em fixed subfield}) of $A$ corresponding to $G$.
\index{fixed!subring}\index{fixed!subfield}

\chapter{\texorpdfstring{$\mathcal{P}$-schemes}{P-schemes}}\label{chap_pscheme}

We introduce the notion of {\em $\mathcal{P}$-schemes} in this chapter, which plays a central  role throughout the thesis.
A $\mathcal{P}$-scheme is a combinatorial structure associated with a group $G$ and a conjugation-closed poset $\mathcal{P}$ of subgroups of $G$. Roughly speaking, it contains a collection of partitions of right coset spaces $H\backslash G$ for $H\in \mathcal{P}$, and these partitions satisfy various consistency properties.

For every permutation group $G$, we define the integers $d(G), d'(G)\in\N^+$ in terms of $\mathcal{P}$-schemes associated with $G$, and show that they are bounded by the {\em minimum base size} of $G$. We will see in Chapter~\ref{chap_alg_prime} that $d(G)$ and $d'(G)$ are closely related to deterministic polynomial factoring.

The work \citep{IKS09} proposed the notion of $m$-schemes as a ``higher-order'' generalization of {\em association schemes} that are central in the field of {\em algebraic combinatorics} \citep{BI84}. We show $\mathcal{P}$-schemes are further generalization of $m$-schemes: an $m$-scheme arises as a $\mathcal{P}$-scheme associated with a symmetric group, or more generally with a multiply transitive group action.

Other results in this chapter include:
\begin{itemize}
\item We define {\em orbit $\mathcal{P}$-schemes},
generalizing the notion of {\em orbit $m$-schemes}  in \citep{IKS09}.
We also provide a simple and exact criterion for antisymmetry of orbit $m$-schemes. 
Using this criterion, we give examples of antisymmetric homogeneous orbit $m$-schemes on finite sets $S$ for $m$ up to $\ell-1$, where $\ell$ is the least prime factor of $|S|$. This result matches the upper bound $m<\ell$ established by R{\'o}nyai \citep{Ron88} for {\em arbitrary} antisymmetric homogeneous $m$-schemes. We reproduce R{\'o}nyai's argument and extend it to $\mathcal{P}$-schemes.

\item We also provide examples of $m$-schemes for small values of $m$. In particular, for $m\leq 3$, we give explicit constructions of
$m$-schemes satisfying the properties of {\em strong antisymmetry} and {\em homogeneity} that are closely related to deterministic polynomial factoring.
\end{itemize}

\paragraph{Outline of the chapter.}

Preliminaries are given in Section~\ref{sec_pscheme_pre}. 
In Section~\ref{sec_pscheme_defp},  we define the notion of $\mathcal{P}$-schemes and its various properties. We also define $d(G)$ and $d'(G)$ in terms of $\mathcal{P}$-schemes.  
In Section~\ref{sec_pscheme_defm},  we review the notion of $m$-schemes and prove the equivalence between $m$-schemes and a certain kind of $\mathcal{P}$-schemes. We also discuss the connection between $m$-schemes and association schemes.
In Section~\ref{sec_pscheme_orbit}, we define orbit $m$-schemes as well as orbit $\mathcal{P}$-schemes.  
An exact criterion of antisymmetry is given for orbit $m$-schemes. Then we discuss R{\'o}nyai's upper bound for $m$ for antisymmetric homogeneous $m$-schemes and extend it to $\mathcal{P}$-schemes.
Finally,  in Section~\ref{sec_3anti}, we describe explicit constructions of strongly antisymmetric homogeneous $m$-schemes for $m\leq 3$.

\section{Preliminaries}\label{sec_pscheme_pre}

Let $G$ be a group.
A {\em partially ordered set}\index{partially ordered set|see{poset}} or {\em poset}\index{poset} of subgroups of $G$ is simply a set of subgroups of $G$, partially ordered by inclusion. All posets of subgroups in this thesis are assumed to be conjugation-closed, and we give the following definition for such posets.

\begin{defi}[subgroup system] 
 A  poset  $\mathcal{P}$ of subgroups of $G$ is called a  {\em subgroup system}\index{subgroup system} over $G$ if it is closed under 
conjugation in $G$, i.e., $gHg^{-1}\in\mathcal{P}$ for all $H\in\mathcal{P}$ and $g\in G$.
\end{defi}

We introduce $\mathcal{P}$-schemes in next section, each associated with a subgroup system $\mathcal{P}$. While the definitions are formulated for general subgroup systems, those arising from the factoring algorithms have special forms. In particular, the following kind of subgroup systems are frequently used in the algorithms.
\begin{defi}[system of stabilizers]\label{def_stab}
Suppose $G$ is a finite group acting on a finite set $S$. For $m\in \N$, let $\mathcal{P}_{m}$ be the set of pointwise stabilizers for nonempty subsets $T\subseteq S$ of cardinality up to $m$:
\[
\mathcal{P}_{m}:=\{G_T: T\subseteq S, 1\leq |T|\leq m\}.
\]
Then $\mathcal{P}_{m}$ is a subgroup system over $G$, called the {\em system of stabilizers}\index{system of stabilizers} of depth\index{depth} $m$ (with respect to the action of $G$ on $S$).
\end{defi}
\nomenclature[a1a]{$\mathcal{P}_{m}$}{system of stabilizers of depth $m$}

\paragraph{Left and inverse right translation.} Let $H$ be a subgroup of $G$. There is an action of $G$ on the right coset space $H\backslash G$  defined by 
\[
\prescript{g}{}{Hh}=Hhg^{-1} \quad \text{for}~ Hh\in H\backslash G ~\text{and}~ g\in G,
\]
called the action of $G$ on $H\backslash G$ by {\em inverse right translation}\index{inverse right translation}. More generally, for a subgroup $G'\subseteq G$, we have the action of $G'$ on $H\backslash G$ by inverse right translation, defined by restricting the previous action of $G$ to $G'$.

We also have an action of the normalizer $N_G(H)$ on $H\backslash G$ defined by 
\[
\prescript{g}{}{Hh}=Hgh \quad \text{for}~ Hh\in H\backslash G ~\text{and}~ g\in N_G(H),
\]
called the action of $N_G(H)$ on $H\backslash G$ by {\em left translation}\index{left translation}.

It is easy to see that they are indeed well defined group actions. For example, we check that for left translation, the coset $\prescript{g}{}{Hh}=Hgh$ is independent of the representative $h$ of $Hh$: Suppose a different representative $h'$ is chosen such that $Hh=Hh'$, then we have   $gh'(gh)^{-1}=gh'h^{-1}g^{-1}\in gHg^{-1}=H$ for  $g\in N_G(H)$ and hence $Hgh=Hgh'$.  

For any $h\in G$, it holds that $Hgh=Hh$ iff $g\in H$. So the action of $N_G(H)$ on $H\backslash G$ induces a semiregular action of $N_G(H)/H$ on $H\backslash G$, defined by $\prescript{gH}{}{Hh}=Hgh$, called the action of $N_G(H)/H$ on $H\backslash G$ by left translation.

\paragraph{Equivalent actions and permutation isomorphic actions.}

 Let $G$ be a group and let $S,T$ be $G$-sets.
 We say the actions of $G$ on $S$ and $T$ are {\em equivalent}\index{equivalent actions} if there exists a bijective map $\lambda: S\to T$ satisfying $\lambda(\prescript{g}{}{x})=\prescript{g}{}{(\lambda(x))}$ for all $x\in S$ and $g\in G$.
And $\lambda$ is said to be an {\em equivalence} between the two actions.
%\todo{Define equivalence for two different groups} 

More generally, suppose $\phi: G\to H$ is a group isomorphism, $S$ is a $G$-set, and $T$ is an $H$-set. We say the action of $G$ on $S$ is {\em permutation isomorphic}\index{permutation isomorphic actions} to the action of $H$ on $T$ (with respect to $\phi$) if there exists a bijective map $\lambda: S\to T$ satisfying $\lambda(\prescript{g}{}{x})=\prescript{\phi(g)}{}{(\lambda(x))}$ for all $x\in S$ and $g\in G$.

The following lemma states that any transitive group action is equivalent to the action on a right coset space by inverse right translation.
\begin{lem}\label{lem_equivaction}
Let $G$ be a group acting transitively on a set $S$. For any $x\in S$, the map $\lambda_x: S\to G_x\backslash G$ sending $\prescript{g}{}{x}$ to $G_xg^{-1}$ for $g\in G$ is well defined and is an equivalence between the action of $G$ on $S$ and that on $G_x\backslash G$ by inverse right translation.
\end{lem}
\nomenclature[a1b]{$\lambda_x$}{the map from a $G$-orbit $S$ containing $x$ to $G_x\backslash G$ sending $\prescript{g}{}{x}$ to $G_x g^{-1}$}

\begin{proof}
As the action of $G$ on $S$ is transitive, for any $y\in S$ we can choose $g\in G$ such that $y=\prescript{g}{}{x}$. Suppose $g,g'$ are two such choices.
We have $\prescript{g^{-1}g'}{}{x}=\prescript{g^{-1}}{}{y}=x$ and hence $g^{-1}g'\in G_x$. So $G_xg^{-1}=G_xg'^{-1}$. Therefore $\lambda_x$ is well defined. It is surjective since any coset $G_xg\in G_x\backslash G$ is the image of $\prescript{g^{-1}}{}{x}$ for a representative $g$ of $G_xg$. And it is injective since $G_xg^{-1}=G_xg'^{-1}$ implies $g^{-1}g'\in G_x$ and hence $\prescript{g'}{}{x}=\prescript{g(g^{-1}g')}{}{x}=\prescript{g}{}{x}$. Finally we check that for any $y=\prescript{g}{}{x}$ and $h\in G$, it holds that
\[
\lambda_x(\prescript{h}{}{y})=\lambda_x(\prescript{hg}{}{x})=G_x(hg)^{-1}=(G_xg^{-1})h^{-1}=\prescript{h}{}{(\lambda(y))}
\]
as desired.
\end{proof}

\begin{cor}[orbit-stabilizer theorem]\index{orbit-stabilizer theorem}
Let $S$ be a $G$-set for a finite group $G$. Then $|Gx|=|G|/|G_x|$ for any $x\in S$.
\end{cor}

\paragraph{Projections and conjugations.} 
We define the following two kinds of maps between right coset spaces $H\backslash G$ for various subgroups $H\subseteq G$:
\begin{itemize}
\item (projection) for $H\subseteq H'\subseteq G$, define the {\em projection}\index{projection} $\pi_{H, H'}: H\backslash G\to H'\backslash G$ to be the map sending $Hg\in H\backslash G$ to $H'g\in H'\backslash G$, and
\item (conjugation) for $H\subseteq G$ and $g\in G$, define the {\em conjugation}\index{conjugation} $c_{H,g}: H\backslash G\to gHg^{-1}\backslash G$ to be the map sending $Hh\in H\backslash G$ to $(gHg^{-1})gh \in gHg^{-1}\backslash G$.
\end{itemize}
\nomenclature[a1c]{$\pi_{H,H'}$}{projection from $H\backslash G$ to $H'\backslash G$}
\nomenclature[a1d]{$c_{H,g}$}{conjugation from $H\backslash G$ to $gHg^{-1}\backslash G$}

\begin{lem}\label{lem_maps}
The maps $\pi_{H, H'}$ and $c_{H,g}$ are well defined and satisfy the following properties: 
\begin{itemize}
\item The maps $\pi_{H, H'}$ are surjective and  $c_{H,g}$ are bijective.
\item $c_{H',g}\circ \pi_{H,H'}=\pi_{gHg^{-1},gH'g^{-1}}\circ c_{H,g}$.
\item (transitivity)  $\pi_{H',H''}\circ \pi_{H,H'}=\pi_{H,H''}$ and $c_{gHg^{-1}, g'}\circ c_{H,g}=c_{H, g'g}$.
\item ($G$-equivariance) $\pi_{H, H'}(\prescript{g}{}{Hh})=\prescript{g}{}{\pi_{H,H'}(Hh)}$ and $c_{H, g'}(\prescript{g}{}{Hh})=\prescript{g}{}{c_{H,g'}(Hh)}$ with respect to the action of $G$ on $H\backslash G$ by inverse right translation.
\end{itemize}
\end{lem}

\begin{proof}
The proof is straightforward from the definitions. We check $c_{H',g}\circ \pi_{H,H'}=\pi_{gHg^{-1},gH'g^{-1}}\circ c_{H,g}$ and leave the rest to the reader: For $Hh\in H\backslash G$, we have 
\[
c_{H',g}\circ \pi_{H,H'}(Hh)=c_{H',g}(H'h)=(gH'g^{-1})gh
\]
and
\[
\pi_{gHg^{-1},gH'g^{-1}}\circ c_{H,g}(Hh)=\pi_{gHg^{-1},gH'g^{-1}}((gHg^{-1})gh)=(gH'g^{-1})gh
\]
as desired.
\end{proof}

Note that for $g\in N_G(H)$, the map $c_{H,g}$ is the permutation of $H\backslash G$ sending each $Hh$ to $\prescript{g}{}{Hh}$ with respect to the action of $N_G(H)$ on $H\backslash G$ by left translation.

\section{\texorpdfstring{$\mathcal{P}$-schemes}{P-schemes}}\label{sec_pscheme_defp}

We start with the definition of a {\em $\mathcal{P}$-collection}\index{Pcollection@$\mathcal{P}$-collection}, which is a collection of partitions of right coset spaces.

\begin{defi}[$\mathcal{P}$-collection]
Let $\mathcal{P}$ be a subgroup system over a finite group $G$. A {\em $\mathcal{P}$-collection} $\mathcal{C}$ is a family $\{C_H: H\in \mathcal{P}\}$ indexed by $\mathcal{P}$ where each $C_H$ is a partition of $H\backslash G$. 
\end{defi}

We are now ready to define the central object of this thesis.

\begin{defi}[$\mathcal{P}$-scheme]\label{defi_pscheme}
A $\mathcal{P}$-collection $\mathcal{C}=\{C_H: H\in \mathcal{P}\}$ is a {\em $\mathcal{P}$-scheme}\index{Pscheme@$\mathcal{P}$-scheme} if it has the following properties:
\begin{itemize}
\item {\em (compatibility)}\index{compatibility!of a $\mathcal{P}$-collection}  for $H,H'\in \mathcal{P}$ with $H\subseteq H'$ and $x,x'\in H\backslash G$  in the same block of $C_H$, the images $\pi_{H,H'}(x)$ and $\pi_{H,H'}(x')$ are in the same block of $C_{H'}$.
\item {\em (invariance)}\index{invariance!of a $\mathcal{P}$-collection}  for $H\in\mathcal{P}$ and $g\in G$, the map $c_{H,g}: H\backslash G\to gHg^{-1}\backslash G$ maps any block of $C_H$ to a block of $C_{gHg^{-1}}$.
\item {\em (regularity)}\index{regularity!of a $\mathcal{P}$-collection} for $H,H'\in\mathcal{P}$ with $H\subseteq H'$, any block $B\in C_H$, $B'\in C_{H'}$, the number of $x\in B$ satisfying $\pi_{H,H'}(x)=y$ is a constant when $y$ ranges over the elements of $B'$.
\end{itemize}
\end{defi}

It is worth noting that in a $\mathcal{P}$-scheme, the partition of $H\backslash G$ for some $H\in\mathcal{P}$ determines the partitions of $H'\backslash G$ for all $H'\in\mathcal{P}$ containing $H$: 
\begin{lem}\label{lem_min}
Let $\mathcal{C}=\{C_H: H\in \mathcal{P}\}$ be a $\mathcal{P}$-scheme. For $H,H'\in\mathcal{P}$ with $H\subseteq H'$,  the blocks of $C_{H'}$ are exactly the images of the blocks of $C_H$ under $\pi_{H,H'}$.
\end{lem}
\begin{proof}
Let $B'$ be a block of $C_{H'}$.
By compatibility, $B'$ is a union of $\pi_{H,H'}(B)$ for one or more blocks $B\in C_H$. Assume $\pi_{H,H'}(B)\subsetneq B'$ for some $B\in C_H$ and choose $y\in \pi_{H,H'}(B)$, $y'\in B' - \pi_{H,H'}(B)$. Then we have $|\{x\in B: \pi_{H,H'}(x)=y\}|>0$ but  $|\{x\in B: \pi_{H,H'}(x)=y'\}|=0$, which  contradicts regularity. 
\end{proof}

In particular, if $\mathcal{P}$ has the property that all minimal subgroups in $\mathcal{P}$ are conjugate in $G$, then by invariance and Lemma~\ref{lem_min}, the partition for one of the minimal subgroups determines the whole $\mathcal{P}$-scheme. For instance, this holds if $\mathcal{P}$ is a system of stabilizers $\mathcal{P}_{m}$ with respect to an $m$-transitive group action. 
%The minimal subgroups in this case are the stabilizers $G_T$, $|T|=m$ which are conjugate in $G$.

\begin{rem}
Besides the set-theoretic definition of $\mathcal{P}$-schemes given in Definition~\ref{defi_pscheme}, there also exists an equivalent ``algebraic'' or ring-theoretic definition of $\mathcal{P}$-schemes. It formulates the three defining properties (compatibility, invariance, and regularity) in a unifying way as closedness of rings under three kinds of maps, respectively: inclusions, conjugations, and trace maps. The interested reader is referred to Appendix~\ref{chap_unify} for further discussion.
\end{rem}

Next we define some optional properties of $\mathcal{P}$-schemes.

\paragraph{Homogeneity and discreteness.} 

Recall that for a finite $S$, we denote by $0_S$ the coarsest partition of $S$ and $\infty_S$ the finest partition of $S$.

\begin{defi}
A $\mathcal{P}$-scheme $\mathcal{C}=\{C_H: H\in \mathcal{P}\}$ is {\em homogeneous}\index{homogeneity!of a $\mathcal{P}$-scheme} on a subgroup $H\in\mathcal{P}$ if $C_H=0_{H\backslash G}$, and otherwise {\em inhomogeneous} on $H$. It is {\em discrete}\index{discreteness!of a $\mathcal{P}$-scheme} on $H$ if $C_H=\infty_{H\backslash G}$, and otherwise {\em non-discrete} on $H$. 
\end{defi}

We will see in Chapter~\ref{chap_alg_prime} that homogeneity (resp. discreteness) of $\mathcal{P}$-schemes is closely related to whether or not the factoring algorithm always produces a proper factorization (resp. the complete factorization) of the input polynomial. 

\paragraph{Symmetry and antisymmetry.}

Invariance of $\mathcal{P}$-schemes states that maps $c_{H,g}: Hh\mapsto (gHg^{-1})gh$ always send blocks to blocks. When $g\in N_G(H)$, the map $c_{H,g}$ is a permutation of $H\backslash G$, and we can impose on a $\mathcal{P}$-scheme the constraint that $c_{H,g}$ always sends a block to itself. Alternatively, we may require $c_{H,g}$ to always send a block to a different block when it is not the trivial permutation. These two constraints are captured by {\em symmetry} and {\em antisymmetry} of $\mathcal{P}$-schemes, respectively.

\begin{defi}\label{defi_symantisym}
A $\mathcal{P}$-scheme $\mathcal{C}=\{C_H: H\in \mathcal{P}\}$ is {\em symmetric}\index{symmetry!of a $\mathcal{P}$-scheme} if  for $H\in\mathcal{P}$ and $g\in N_G(H)$, the permutation $c_{H,g}$ of $H\backslash G$ maps every block of $C_H$ to itself.
And  $\mathcal{C}$ is {\em antisymmetric}\index{antisymmetry!of a $\mathcal{P}$-scheme} if  for $H\in\mathcal{P}$ and $g$ in $N_G(H)$ but not in $H$,  the permutation $c_{H,g}$ maps every block of $C_H$ to a different block.
\end{defi}

Symmetry (resp. antisymmetry) is equivalent to the property that for all $H\in\mathcal{P}$,  elements in each $(N_G(H)/H)$-orbit of $H\backslash G$ belong to the same block (resp. distinct blocks) of $C_H$, where $N_G(H)/H$ acts on $H\backslash G$ by left translation.

As will be seen in Chapter~\ref{chap_alg_prime}, antisymmetry of $\mathcal{P}$-schemes is important for deterministic polynomial factoring \citep{Ron88, Ron92, Evd94, IKS09}. For now we show that an antisymmetric $\mathcal{P}$-scheme is discrete on $H$ for any $H\in\mathcal{P}$ provided that $\mathcal{P}$ contains the trivial subgroup of $G$.

\begin{lem}\label{lem_antidisc}
Suppose $\mathcal{P}$ is a subgroup system over a finite group $G$ that contains the trivial subgroup $\{e\}$.
For $H\in\mathcal{P}$, all antisymmetric $\mathcal{P}$-schemes are discrete on $H$.
\end{lem}
\begin{proof}
Let $\mathcal{C}=\{C_H: H\in \mathcal{P}\}$ be an antisymmetric $\mathcal{P}$-scheme. As $N_G(\{e\})=G$ acts transitively on $\{e\}\backslash G$ by left translation, we have $C_{\{e\}}=\infty_{\{e\}\backslash G}$ by antisymmetry. 
Now consider an arbitrary subgroup $H\in\mathcal{P}$. By Lemma~\ref{lem_min}, we have $C_H=\{\pi_{\{e\}, H}(B): B\in C_{\{e\}}\}=\infty_{H\backslash G}$. So $\mathcal{C}$ is discrete on $H$.
\end{proof}

On the other hand, symmetry of $\mathcal{P}$-schemes plays no role in polynomial factoring as far as we know, and we only discuss it within this chapter.

\paragraph{Strong antisymmetry.} We introduce another property called {\em strong antisymmetry}, which is a strengthening of antisymmetry define above. It is based on an idea introduced by Evdokimov \citep{Evd94} which leads to his quasipolynomial-time factoring algorithm.

Antisymmetry states that no nontrivial permutation of blocks arises from a conjugation $c_{H,g}$ where $g\in N_G(H)$: For such a map $c_{H,g}$ and a block $B\in C_H$, either the image $c_{H,g}(B)$ is a different block, or $c_{H,g}$ is the identity map.
We strengthen this property by considering permutations arising from compositions of not only conjugations, but also projections and their inverses.  Of course, a projection $\pi_{H,H'}$ is not invertible whenever $H\subsetneq H'$. Nevertheless, it is possible that the restriction of $\pi_{H,H'}$ to some block $B\in C_H$ maps $B$ bijectively to some block $B'\in C_{H'}$, in which case the inverse map  $(\pi_{H,H'}|_B)^{-1}$ is well defined.
 
\begin{defi}\label{defi_strongasym}
A $\mathcal{P}$-scheme $\mathcal{C}=\{C_H: H\in \mathcal{P}\}$ is {\em strongly antisymmetric}\index{strong antisymmetry!of a $\mathcal{P}$-scheme} if for any sequence of subgroups $H_0,\dots, H_k\in\mathcal{P}$, $B_0\in C_{H_0},\dots, B_k\in C_{H_k}$, and maps $\sigma_1,\dots, \sigma_k$ satisfying
\begin{itemize}
\item $\sigma_i$ is a bijective map from $B_{i-1}$ to $B_i$,
\item $\sigma_i$ is of the form $c_{H_{i-1},g}|_{B_{i-1}}$, $\pi_{H_{i-1},H_i}|_{B_{i-1}}$, or  $(\pi_{H_i,H_{i-1}}|_{B_i})^{-1}$,
\item $H_0=H_k$ and $B_0=B_k$,
\end{itemize}
the composition $\sigma_k\circ\cdots\circ \sigma_1$ is the identity map on $B_0=B_k$.
\end{defi}

In other words, no nontrivial permutation could be obtained by composing maps of the form  $c_{H_{i-1},g}|_{B_{i-1}}$, $\pi_{H_{i-1},H_i}|_{B_{i-1}}$, or  $(\pi_{H_i,H_{i-1}}|_{B_i})^{-1}$.

A strongly antisymmetric $\mathcal{P}$-scheme is indeed antisymmetric: 
Assume  $\mathcal{C}=\{C_H: H\in \mathcal{P}\}$ is not antisymmetric, then there exist $H\in\mathcal{P}$, $g\in N_G(H) -  H$ and $B\in C_H$ such that $c_{H,g}(B)=B$. Let $\sigma_1$ be the map $c_{H,g}|_B: B\to c_{H,g}(B)=B$. 
It sends $x\in B$ to $\prescript{gH}{}{x}$ with respect to the action of  $N_G(H)/H$ on $H\backslash G$ by left translation.
As this action is semiregular and $gH\in N_G(H)/H$ is not the identity element, the map $\sigma_1$ is a nontrivial permutation of $B$.
So $\mathcal{C}$ is not strongly antisymmetric.

\paragraph{$d(G)$ and $d'(G)$.} 

For every finite permutation group $G$, we define $d(G),d'(G)\in\N^+$  which are closely related to deterministic polynomial factoring, as will be seen in Chapter~\ref{chap_alg_prime}. 

\begin{defi}\label{defi_dg}
Let $G$ be a finite permutation group on a finite set $S$. For $m\in\N^+$,  let $\mathcal{P}_{m}$ be the system of stabilizers of depth $m$ with respect to this action. Define $d(G),d'(G)\in \N^+$ as follows.
\begin{itemize}
\item Define $d(G)$ to be the smallest integer $m\in\N^+$ such that all strongly antisymmetric $\mathcal{P}_{m}$-schemes are discrete on $G_x$  for all $x\in S$.
\item If $G$ acts transitively on $S$ and $|S|>1$, define $d'(G)$ to be the smallest integer $m\in\N^+$ such that all strongly antisymmetric $\mathcal{P}_{m}$-schemes are inhomogeneous on $G_x$  for all $x\in S$. Otherwise let $d'(G)=1$.
\end{itemize}
\end{defi}
\nomenclature[a1e]{$d(G)$, $d'(G)$}{See Definition~\ref{defi_dg}}
%\nomenclature[a1f]{$d'(G)$}{See Definition~\ref{defi_dg}}

  We  have $1\leq d'(G)\leq d(G)\leq \max\{|S|-1, 1\}$  for any finite permutation group $G$ on a finite set $S$. The first two inequalities are obvious  and the last one follows from  Lemma~\ref{lem_antidisc} and the fact that any $g\in G$ fixing $|S|-1$ elements of $S$ is the identity.
  A better upper bound for $d(G)$ is given by the {\em minimal base size} of $G$.

%\paragraph{Bases of permutation groups.} When $G$ is a finite permutation group and $\mathcal{P}$ is the corresponding system of stabilizers, there is a natural connection between antisymmetry of $\mathcal{P}$-schemes and the notion of {\em bases} of $G$, which we explain now.

\begin{defi}[base]
Let $G$ be a finite permutation group on a finite set $S$.
A {\em base}\index{base} of $G$ is a set $B\subseteq S$ for which $G_B$ equals the trivial subgroup $\{e\}$. The {\em minimal base size}\index{minimal base size} of $G$, denoted by $b(G)$, is the minimum cardinality of a base of $G$.
\end{defi}
\nomenclature[a1g]{$b(G)$}{minimal base size of a permutation group $G$}

By Lemma~\ref{lem_antidisc}, we have
\begin{lem}\label{lem_antibasebound}
$d(G)\leq \max\{b(G),1\}$ for any finite permutation group $G$.
\end{lem}

We also prove the following bound in latter chapters based on the work of \citep{Evd94, IKS09, Gua09, Aro13}.
\begin{lem}\label{lem_logbound}
There exists an absolute constant $c>0$ such that $d(G)\leq c\log n+O(1)$ for any finite permutation group $G$ on a set of cardinality $n\in\N^+$.
\end{lem}
The best known upper bound for $c$ is $\frac{2}{\log 12}=0.55788\dots$, proved by \citep{Gua09, Aro13}. See Section~\ref{sec_symnatural} for more details.

\section{\texorpdfstring{$m$-schemes}{m-schemes}}\label{sec_pscheme_defm}

The paper \citep{IKS09} proposed the notion of {\em $m$-schemes}. In this section, we present their definition and show that it is generalized by the notion of $\mathcal{P}$-schemes: roughly speaking, an $m$-scheme could be regarded as a $\mathcal{P}$-scheme where $\mathcal{P}$ is a system of stabilizers with respect to an $m$-transitive group action.

We use the following notations: 

Let $S$ be a finite set and let $m\in\N^+$. Define an {\em $m$-collection}\index{mcollection@$m$-collection} on $S$ to be a collection of partitions $P_1,\dots,P_m$ of $S^{(1)},\dots,S^{(m)}$ respectively.
 
For $k\in [m]$, the symmetric group $\sym(k)$ acts on the set $S^{(k)}$ by permuting the $k$ coordinates, i.e., for $g\in\sym(k)$ and $x=(x_1,\dots,x_k)\in S^{(k)}$, we have $\prescript{g}{}{x}=(y_1,\dots,y_k)$ where $y_{\prescript{g}{}{i}}=x_i$, or equivalently $y_i=x_{\prescript{g^{-1}}{}{i}}$.

For $1<k\leq m$ and $i\in [k]$, let $\pi^k_i: S^{(k)}\to S^{(k-1)}$ be the projection omitting the $k$th coordinate. 
More generally, for a proper subset $T$ of $[k]$, let $\pi^k_T: S^{(k)}\to S^{(k-r)}$ be the projection omitting the coordinates whose indices are in $T$. 
\nomenclature[a1h]{$\pi^k_i$}{projection from $S^{(k)}$ to $S^{(k-1)}$ omitting the $k$th coordinate}
\nomenclature[a1i]{$\pi^k_T$}{projection from $S^{(k)}$ to $S^{(k-1)}$ omitting the coordinates with indices in $T$}
\nomenclature[a1j]{$c^k_g$}{permutation of $S^{(k)}$ sending $x$ to $\prescript{g}{}{x}$}

For $k\in [m]$ and $g\in \sym(k)$, let $c_g^k$ be the permutation of $S^{(k)}$ sending $x$ to $\prescript{g}{}{x}$, with respect to the above action of $\sym(k)$ on $S^{(k)}$.

\begin{defi}[$m$-scheme \citep{IKS09}]
An $m$-collection $\Pi=\{P_1,\dots,P_m\}$ on $S$ is an {\em $m$-scheme}\index{mscheme@$m$-scheme} if it has the following properties:
\begin{itemize}
\item {\em (compatibility)}\index{compatibility!of an $m$-collection} for $1<k\leq m, i\in [k]$ and elements $x,x'\in S^{(k)}$ in the same block of $P_k$, the elements $\pi^k_i(x), \pi^k_i(x')$ are in the same block of $P_{k-1}$.
\item {\em (invariance)}\index{invariance!of an $m$-collection} for $k\in [m]$ and $g\in\sym(k)$, the permutation $c_g^k$ of $S^{(k)}$ sends blocks of $P_k$ to blocks.
\item {\em (regularity)}\index{regularity!of an $m$-collection} for $1<k\leq m$, $i\in [k]$ and blocks $B\in P_k$, $B'\in P_{k-1}$, the number of $x\in B$ satisfying $\pi^k_i(x)=y$ is a constant when $y$ ranges over the elements of $B'$.
\end{itemize}
Furthermore, we say $\Pi$ is  {\em symmetric}\index{symmetry!of an $m$-scheme} (resp. {\em antisymmetric}\index{antisymmetry!of an $m$-scheme}) if for all $k\in [m]$ and $g\in \sym(k)-\{e\}$,  the permutation $c_g^k$ of $S^{(k)}$ sends every block of $P_k$ to itself (resp. a different block). And  $\Pi$ is said to be {\em homogeneous}\index{homogeneity!of an $m$-scheme} if $P_1$ equals the coarsest partition $0_{S}$.
\end{defi}

We also introduce the following definitions which did not appear in \citep{IKS09}.
\begin{defi}
An $m$-scheme  $\Pi=\{P_1,\dots,P_m\}$ on $S$ is said to be {\em discrete}\index{discreteness!of an $m$-scheme} if $P_1$ equals the finest partition $\infty_{S}$. It is said to be {\em strongly antisymmetric}\index{strong antisymmetry!of an $m$-scheme} if no nontrivial permutation of any block of $P_k$ for any $k\in [m]$ can be obtained by composing maps of the form $c_g^i|_B$, $\pi^i_T|_B$, or $(\pi^i_T|_B)^{-1}$, where $B$ is a block of $P_i$.
\end{defi}

\begin{rem}
The parameter $m$ is allowed to be arbitrarily  large in our definition. Nevertheless, the sets $S^{(k)}$ for $k=|S|+1,\dots,m$ are empty and hence the corresponding partitions $P_k$ contain no information. By discarding these partitions and replacing $m$ with $\min\{m, |S|\}$, we may assume $m\leq |S|$. 
\end{rem}

\subsection{The connection of $m$-schemes with $\mathcal{P}$-schemes}

 Given a finite set $S$ and $m\in \N^+$, let $G$ be a group acting $m'$-transitively on $S$ where $m':=\min\{m,|S|\}$.\footnote{In particular, we can take $G=\sym(S)$ acting naturally on $S$, which is $|S|$-transitive.} 
Choose $\mathcal{P}=\mathcal{P}_{m}$ to be the system of stabilizers of depth $m$ with respect to this action (see Definition~\ref{def_stab}).
We prove that for such $G$ and $\mathcal{P}$, every $\mathcal{P}$-scheme gives rise to an $m$-scheme on $S$, and (under an additional assumption), there is a one-to-one correspondence between $m$-schemes  on $S$ and $\mathcal{P}$-schemes, with various properties (symmetry, antisymmetry, etc.) preserved.

For $k\in [m']$  and $x=(x_1,\dots,x_k)\in S^{(k)}$, let $T_x=\{x_1,\dots,x_k\}$. The stabilizer $G_x$ with respect to the diagonal action of $G$ on $S^{(k)}$ equals the pointwise stabilizer $G_{T_x}$ with respect to the action of $G$ on $S$, and therefore $G_x=G_{T_x}\in\mathcal{P}$.
As the action of $G$ on $S^{(k)}$ is transitive (which follows from $m'$-transitivity of $G$ on $S$),  by Lemma~\ref{lem_equivaction}, we have an equivalence of group actions 
\[
\lambda_x: S^{(k)}\to G_{x}\backslash G
\] 
between the diagonal action of $G$ on $S^{(k)}$ and the action on $G_{x}\backslash G$ by inverse right translation.
It sends $\prescript{g}{}{x}$ to $G_x g^{-1}$ for $g\in G$.
We use these maps $\lambda_x$ to  construct an $m$-scheme  on $S$ from a $\mathcal{P}$-scheme, and vice versa.

\paragraph{From a $\mathcal{P}$-scheme to an $m$-scheme.}

We construct an $m$-scheme  on $S$ from a $\mathcal{P}$-scheme as follows.

\begin{defi}\label{defi_ptom}
Given a  $\mathcal{P}$-scheme $\mathcal{C}=\{C_H: H\in\mathcal{P}\}$, define an $m$-collection $\Pi(\mathcal{C})=\{P_1,\dots, P_m\}$ on $S$ as follows:
for each $k\in [m']$ where $m'=\min\{m,|S|\}$, pick $x=(x_1,\dots,x_k)\in S^{(k)}$, and define $P_k=\{\lambda_x^{-1}(B): B\in C_{G_x}\}$. For $m'<k\leq m$, the partition $P_k$ is a partition of the empty set $S^{(k)}$ and is unique.
\end{defi}
\nomenclature[a1k]{$\Pi(\mathcal{C})$}{$m$-scheme constructed from a $\mathcal{P}$-scheme $\mathcal{C}$ (see Definition~\ref{defi_ptom})}

\begin{lem}\label{lem_ptom}
$\Pi(\mathcal{C})$ as defined above is independent of the choices of elements $x$ and is an $m$-scheme. 
It is symmetric (resp. antisymmetric, strongly antisymmetric) if $\mathcal{C}$ is symmetric (resp. antisymmetric, strongly antisymmetric).
And it is homogeneous (resp. discrete) iff $\mathcal{C}$ is homogeneous on $G_x$ (resp. discrete on $G_x$) for $x\in S$.
\end{lem}

\begin{proof}
We may assume $m\leq |S|$.
Fix $k\in [m]$ and we show that $P_k$ does not depend on the choice of $x\in S^{(k)}$.
Consider two elements $x,x'\in S^{(k)}$. Choose $h\in G$ such that $x'=\prescript{h}{}{x}$. Such $h$ exists since $G$ acts transitively on $S^{(k)}$.
Then $G_{x'}=hG_{x}h^{-1}$ and we have the conjugation $c_{G_x,h}: G_x\backslash G\to G_{x'}\backslash G$ sending $G_xg$ to $G_{x'}hg$.
We check that 
%$\lambda_x: S^{(k)}\to G_x\backslash G$ and  $\lambda_{x'}: S^{(k)}\to G_{x'}\backslash G$ satisfy
$\lambda_{x'}= c_{G_x,h} \circ \lambda_x$.
This holds since for $y=\prescript{g}{}{x'}\in S^{(k)}$, we have $\lambda_{x'}(y)=G_{x'}g^{-1}$ and 
\[
c_{G_x,h} \circ \lambda_x(y)=c_{G_x,h} \circ  \lambda_x(\prescript{g}{}{x'})=c_{G_x,h} \circ \lambda_x(\prescript{gh}{}{x})=c_{G_x,h}(G_{x}(gh)^{-1})=G_{x'}g^{-1}.
\]
So $\lambda_{x'}^{-1}=\lambda_{x}^{-1}\circ c_{G_x,h}^{-1}=\lambda_{x}^{-1}\circ c_{G_{x'},h^{-1}}$. As $\mathcal{C}$ is invariant, the conjugation $c_{G_{x'},h^{-1}}$ sends blocks of $C_{G_{x'}}$ to blocks of $C_{G_x}$. 
So the two partitions $\{\lambda_x^{-1}(B): B\in C_{G_x}\}$ and $\{\lambda_{x'}^{-1}(B): B\in C_{G_{x'}}\}$ are identical, i.e., the elements $x$ and $x'$ define the same partition $P_k$.

Next we check that $\Pi(\mathcal{C})$ is an $m$-scheme.
For $1<k\leq m$, consider the elements $x\in S^{(k)}$ and $x'\in S^{(k-1)}$ as picked in Definition~\ref{defi_ptom}.
Let $\bar{x}=\pi^k_i(x)\in S^{(k-1)}$ so that $G_x\subseteq G_{\bar{x}}$.
Choose $h\in G$, satisfying $x'=\prescript{h}{}{\bar{x}}$ so that $G_{x'}=hG_{\bar{x}}h^{-1}$.
% Such an element $h$ exists since $G$ acts transitively on $S^{(k-1)}$.
Then the following diagram commutes:
\[
\begin{tikzcd}[column sep=3cm]
S^{(k)} \arrow[r, "\pi^k_i"] \arrow[d, "\lambda_x"']
& {S^{(k-1)}}  \arrow[d, "\lambda_{x'}"] \\
G_x\backslash G  \arrow[r,  "c_{G_{\bar{x},h}}\circ \pi_{G_x,G_{\bar{x}}}"]
& G_{x'}\backslash G \nospacepunct{.}
\end{tikzcd}
\]
To see this, note that for any $y=\prescript{g}{}{x}\in S^{(k)}$ where $g\in G$, we have 
\[
c_{G_{\bar{x},h}}\circ \pi_{G_x,G_{\bar{x}}}\circ \lambda_x(y)=c_{G_{\bar{x},h}}\circ \pi_{G_x,G_{\bar{x}}}(G_xg^{-1})=c_{G_{\bar{x},h}}(G_{\bar{x}}g^{-1})=G_{x'}hg^{-1},
\]
and 
\[
\lambda_{x'}\circ \pi^k_i(y)=\lambda_{x'}\circ\pi^k_i(\prescript{g}{}{x})=\lambda_{x'}\left(\prescript{g}{}{(\pi^k_i(x))}\right)=\lambda_{x'}(\prescript{g}{}{\bar{x}})=\lambda_{x'}(\prescript{gh^{-1}}{}{x'})=G_{x'}hg^{-1},
\]
as desired.
Also note that the maps $\lambda_x$ and $\lambda_{x'}$ are bijections, sending blocks to blocks. 
Compatibility and regularity of $\Pi(\mathcal{C})$ then follow from compatibility, regularity, and invariance of $\mathcal{C}$. 

For $k\in [m]$, $\tau\in\sym(k)$ and $x=(x_1,\dots,x_k)\in  S^{(k)}$, let $x'=c^k_\tau(x)\in S^{(k)}$.
Choose $h\in G$ such that $x=\prescript{h}{}{x'}$. 
%Such $h$ exists since $G$ acts transitively on $S^{(k)}$.
Then $G_{x}=hG_{x'}h^{-1}$. We also have $G_{x}=G_{x'}$ since they are both the pointwise stabilizer $G_T$ with respect to the action of $G$ on $S$, where $T=\{x_1,\dots,x_k\}$.
So $h\in N_G(G_x)$. We claim that the following diagram commutes:
\[
\begin{tikzcd}[column sep=large]
S^{(k)} \arrow[r, "c^k_\tau"] \arrow[d, "\lambda_x"']
& {S^{(k)}}  \arrow[d, "\lambda_{x}"] \\
G_x\backslash G  \arrow[r,  "c_{G_x,h}"]
& G_{x}\backslash G \nospacepunct{.}
\end{tikzcd}
\]
To see this, note that for any $y=\prescript{g}{}{x}\in S^{(k)}$ where $g\in G$, we have $c_{G_x,h}\circ \lambda_x(y)=c_{G_x,h}(G_xg^{-1})=G_{x}hg^{-1}$,
and 
\[
\lambda_{x}\circ c^k_\tau(y)=\lambda_{x}\circ c^k_\tau(\prescript{g}{}{x})=\lambda_{x}\left(\prescript{g}{}{(c^k_\tau(x))}\right)=\lambda_{x}(\prescript{g}{}{x'})=\lambda_{x}(\prescript{gh^{-1}}{}{x})=G_{x}hg^{-1},
\] 
as desired.
Invariance of $\Pi(\mathcal{C})$ then follows from invariance of $\mathcal{C}$. So $\Pi(\mathcal{C})$ is an $m$-scheme.

The previous diagram also shows that if $\mathcal{C}$ is symmetric (resp. antisymmetric) then so is $\Pi(\mathcal{C})$. Suppose a nontrivial permutation of some block of $P_k$ for some $k\in [m]$ can be obtained by composing maps of the form $c_g^i|_B$, $\pi^i_T|_B$, or $(\pi^i_T|_B)^{-1}$, then using the two diagrams above, we also obtain a nontrivial permutation of some block of $G_x\backslash G$ (where $x\in S^{(k)}$ is as chosen in Definition~\ref{defi_ptom}) by composing conjugations, projections, and their inverses (restricted to blocks).
Therefore, if $\mathcal{C}$ is strongly symmetric, so is $\Pi(\mathcal{C})$.

Finally, for any $x\in S$, the partition $P_1$ of $S$ is constructed using the bijection $\lambda_x: S\to G_x\backslash G$ and the partition $C_{G_x}$ of  $G_x\backslash G$. Therefore $\Pi(\mathcal{C})$ is homogeneous (resp. discrete) iff $\mathcal{C}$ is homogeneous on $G_x$ (resp. discrete on $G_x$).
\end{proof}

\paragraph{From an $m$-scheme to a $\mathcal{P}$-scheme.}
Conversely, we could also construct a $\mathcal{P}$-scheme from an $m$-scheme on $S$. Here we need an additional assumption that $m\neq |S|-1$ and $G$ acts $\min\{|S|, m+1/2\}$-transitively on $S$.\footnote{Recall that a group action of $G$ on $S$ is $(k+1/2)$-transitive if it is $k$-transitive, and in addition for all $T\subseteq S$ of cardinality $k$, either the $G_T$-orbit of every $x\in S-T$ contains more than one element, or $|S-T|=1$.}

\begin{lem}\label{lem_uniqueT}
Assume $m\neq |S|-1$ and $G$ acts $\min\{|S|, m+1/2\}$-transitively on $S$. For $T,T'\subseteq S$ of cardinality at most $m$, we have $G_T=G_{T'}$ iff $T=T'$. 
% the set of fixed points $S^{G_T}$ equals $T$.
And the normalizer $N_G(G_T)$ of $G_T$ is the setwise stabilizer $G_{\{T\}}$.
\end{lem}

\begin{proof}
The assumption implies that set of elements in $S$ fixed by $G_T$ (resp. $G_{T'}$) is precisely $T$ (resp. $T'$).  So $G_T=G_{T'}$ implies $T=T'$. The other direction is trivial.

For $g\in G$, we have $gG_Tg^{-1}=G_{\prescript{g}{}{T}}$. So $g\in N_G(G_T)$ iff $G_T=G_{\prescript{g}{}{T}}$, which holds iff $T=\prescript{g}{}{T}$ by the first part. So $N_G(T)=G_{\{T\}}$.
\end{proof}

\begin{defi}\label{defi_mtop}
Assume $m\neq |S|-1$ and $G$ acts $\min\{|S|, m+1/2\}$-transitively on $S$. 
Given an $m$-scheme $\Pi=\{P_1,\dots, P_m\}$ on $S$, define a $\mathcal{P}$-collection $\mathcal{C}(\Pi)=\{C_H: H\in\mathcal{P}\}$ as follows:
For $H\in\mathcal{P}$, pick $T\subseteq S$ of cardinality $k\in [m]$ such that $H=G_T$. By Lemma~\ref{lem_uniqueT}, such a set $T$ is unique.
Pick $x=(x_1,\dots,x_k)\in S^{(k)}$ such that $T=\{x_1,\dots,x_k\}$. Then $G_x=G_T=H$ and we have the map $\lambda_x: S^{(k)}\to H\backslash G$.
Define $C_H=\{\lambda_x(B): B\in P_k\}$.
\end{defi}
\nomenclature[a1l]{$\mathcal{C}(\Pi)$}{$\mathcal{P}$-scheme constructed from an $m$-scheme $\Pi$ (see Definition~\ref{defi_mtop})}

\begin{lem}\label{lem_mtop}
$\mathcal{C}(\Pi)$ as defined above is independent of the choices of elements $x$ and is a $\mathcal{P}$-scheme. 
It is symmetric (resp. antisymmetric, strongly antisymmetric) if $\Pi$ is symmetric (resp. antisymmetric, strongly antisymmetric). And it is homogeneous on $G_x$ (resp. discrete on $G_x$) for $x\in S$ iff $\Pi$ is homogeneous (resp. discrete). 
\end{lem}
\begin{proof}
Fix $H=G_T\in\mathcal{P}$ and we show that $C_H$ does not depend on the choices of $x$. Consider two elements $x=(x_1,\dots,x_k), x'=(x'_1,\dots,x'_k)\in S^{(k)}$ such that $T=\{x_1,\dots,x_k\}=\{x'_1,\dots,x'_k\}$. Then there exists $\rho\in\sym(k)$ such that $c^k_{\rho}(x')=x$. We check that $\lambda_{x'}=\lambda_x\circ c^k_{\rho}$: For any $y=\prescript{g}{}{x'}\in S^{(k)}$ where $g\in G$, we have $\lambda_{x'}(y)=Hg^{-1}$ and 
\[
\lambda_x\circ c^k_{\rho}(y)=\lambda_x\circ c^k_{\rho}(\prescript{g}{}{x'})=\lambda_x\left(\prescript{g}{}{(c^k_\rho(x'))}\right)=\lambda_x(\prescript{g}{}{x})=Hg^{-1}
\]
as desired. As $\Pi$ is invariant, the map $c^k_\rho$ sends blocks to blocks. Therefore $\{\lambda_x(B): B\in P_k\}=\{\lambda_{x'}(B): B\in P_k\}$. So the two elements $x$ and $x'$ define the same partition $C_H$.

Next we check that $\mathcal{C}(\Pi)$  is a $\mathcal{P}$-scheme. 
Consider a projection $\pi_{H,H'}:H\backslash G\to H'\backslash G$ where $H,H'\in\mathcal{P}$ and $H\subseteq H'$. Then there exist $T'\subseteq T\subseteq S$ such that $H=G_T$, $H'=G_{T'}$.
We may assume $|T|=|T'|+1$ by decomposing  $\pi_{H,H'}$ into the composition of more projections if necessary. Let $k=|T|$ and pick $x=(x_1,\dots,x_k)$ such that $T=\{x_1,\dots,x_k\}$. Choose the unique $i\in [k]$ such that $x_i\not\in T'$.
Let $x'=\pi^k_i(x)$. Then $H=G_T=G_x$ and $H'=G_{T'}=G_{x'}$.
We claim that the following diagram commutes:
\[
\begin{tikzcd}[column sep=large]
S^{(k)} \arrow[r, "\pi^k_i"] \arrow[d, "\lambda_x"']
& {S^{(k-1)}}  \arrow[d, "\lambda_{x'}"] \\
H\backslash G  \arrow[r,  "\pi_{H,H'}"]
& H'\backslash G\nospacepunct{.}
\end{tikzcd}
\]
To see this, note that for any $y=\prescript{g}{}{x}\in S^{(k)}$ where $g\in G$, we have  $\pi_{H,H'}\circ \lambda_x(y)=\pi_{H,H'}(Hg^{-1})=H'g^{-1}$,
and 
\[
\lambda_{x'}\circ \pi^k_i(y)=\lambda_{x'}\circ\pi^k_i(\prescript{g}{}{x})=\lambda_{x'}\left(\prescript{g}{}{(\pi^k_i(x))}\right)=\lambda_{x'}(\prescript{g}{}{x'})=H'g^{-1},
\]
as desired.
And $\lambda_x, \lambda_{x'}$ are bijections that send blocks to blocks. 
Compatibility and regularity of $\mathcal{C}(\Pi)$ then follow from those of $\Pi$. 

Now consider a conjugation $c_{H,h}: H\backslash G\to H'\backslash G$ for $H\in\mathcal{P}$ and $h\in G$, where $H'=hHh^{-1}$. Choose $x\in S^{(k)}$ for some $k\in [m]$ such that $H=G_x$. Let $x'=\prescript{h}{}{x}$ so that $H'=G_{x'}$.
Then the following diagram commutes:
\[
\begin{tikzcd}
 & {S^{(k)}}  \arrow[dr, "\lambda_{x'}"] \arrow[dl, "\lambda_{x}"']  \\
H\backslash G \arrow[rr, "c_{H,h}"] && H'\backslash G\nospacepunct{.}
\end{tikzcd}
\]
To see this, note that for any  $y=\prescript{g}{}{x'}\in S^{(k)}$ where $g\in G$, we have $\lambda_{x'}(y)=H'g^{-1}$ and 
\[
c_{H,h}\circ \lambda_{x}(y)=c_{H,h}\circ \lambda_{x}(\prescript{gh}{}{x})=c_{H,h}(H(gh)^{-1})=H'g^{-1},
\] 
as desired. So $\mathcal{C}(\Pi)$ is invariant. Therefore $\mathcal{C}(\Pi)$ is a $\mathcal{P}$-scheme.

Now we prove the claim that strongly antisymmetry is preserved. Assume that a map $\tau:B_1\to B_2$ between blocks $B_1, B_2\in C_H$ for some $H=G_T\in \mathcal{P}$ is obtained by composing conjugations, projections and their inverses (restricted to blocks). Let $k=|T|$. By the two diagrams above, we can obtain a map $\tau':B'_1\to B'_2$ between blocks $B'_1,B'_2\in P_k$ by composing maps of the form $c_g^i|_B$, $\pi^i_T|_B$, or $(\pi^i_T|_B)^{-1}$, such that the following diagram commutes
\[
\begin{tikzcd}[column sep=large]
B'_1 \arrow[r, "\tau'"] \arrow[d, "\lambda_x"']
& B'_2  \arrow[d, "\lambda_{x'}"] \\
B_1  \arrow[r,  "\tau"]
& B_2 
\end{tikzcd}
\]
for some $x,x'\in S^{(k)}$. We showed in the beginning that there exists $\rho\in\sym(k)$ satisfying $\lambda_{x'}=\lambda_x\circ c^k_{\rho}$. By replacing $\tau'$ with $c^k_{\rho}\circ\tau'$, $B'_2$ with $c^k_{\rho}(B'_2)$, and $\lambda_x'$ with $\lambda_x$, we may assume $x=x'$. Then if $B_1=B_2$ and $\tau$ is a nontrivial permutation of $B_1$, we also know that $B'_1=B'_2$ and $\tau'$ is a nontrivial permutation of $B'_1$. 
Therefore, if $\Pi$  is strongly antisymmetric, so is $\mathcal{C}(\Pi)$. The claim for antisymmetry is proved in the same way, except that we only consider maps $\tau$ arising from conjugations but not projections. And if $B_1\neq B_2$ for such $\tau$, we also get a map $\tau':B_1'\to B_2'$ arising from $c^k_{\tau_0}$ for some $\tau_0\in\sym(k)$ such that $B_1'\neq B_2'$. So symmetry is also preserved.

Finally, for any $x\in S$, the partition $C_{G_x}$ of $G_x\backslash G$ is constructed using the bijection $\lambda_x: S\to G_x\backslash G$ and the partition $P_1$. Therefore $\mathcal{C}(\Pi)$ is homogeneous on $G_x$ (resp. discrete on $G_x$) iff $\Pi$ is homogeneous  (resp. discrete).
\end{proof}

The maps $\mathcal{C}\mapsto \Pi(\mathcal{C})$ and $\Pi\mapsto\mathcal{C}(\Pi)$ are inverse to each other by construction. So Lemma~\ref{lem_ptom} and Lemma~\ref{lem_mtop} together establish the one-to-one correspondence between $\mathcal{P}$-schemes and $m$-schemes on $S$.

\begin{thm}\label{thm_mandp}
Suppose $m\neq |S|-1$ and $G$ is a finite group acting $\min\{|S|, m+1/2\}$-transitively on $S$, and $\mathcal{P}=\mathcal{P}_{m}$ is the system of stabilizers of depth $m$ with respect to this action.
The map $\mathcal{C}\mapsto \Pi(\mathcal{C})$ in Definition~\ref{defi_ptom} is a one-to-one correspondence between $\mathcal{P}$-schemes and $m$-schemes on $S$, with the inverse map   $\Pi\mapsto\mathcal{C}(\Pi)$  as defined in Definition~\ref{defi_mtop}.
And $\Pi(\mathcal{C})$ is symmetric (resp. antisymmetric, strongly antisymmetric, homogeneous, discrete) iff $\mathcal{C}$ is symmetric (resp. antisymmetric, strongly antisymmetric, homogeneous on $G_x$ for $x\in S$, discrete on $G_x$ for $x\in S$).
\end{thm}

\begin{rem}
The unpleasant assumption $m\neq |S|-1$ in Theorem~\ref{thm_mandp} is due to the technical fact that when $T\subseteq S$ has cardinality $|S|-1$, the pointwise stabilizer $G_T$ fixes not only $T$ but also the whole set $S$. 
This assumption  is needed if we want the correspondence in Theorem~\ref{thm_mandp} to preserve antisymmetry and homogeneity: Suppose $G$ is a permutation group on $S$ and $|S|=\ell$ is a prime number. Then for $m=\ell-1$, there exists an antisymmetric homogeneous $m$-scheme on $S$ (see Example~\ref{exmp_antisym_orbit} in Section~\ref{sec_pscheme_orbit}). On the other hand, note that $\mathcal{P}=\mathcal{P}_{m}$ contains the trivial subgroup $G_S=\{e\}$. So by Lemma~\ref{lem_antidisc}, all antisymmetric $\mathcal{P}$-schemes are discrete on $G_x$ for any $x\in S$. 
\end{rem}

\paragraph{Matchings.} The papers \citep{IKS09, AIKS14} formulated the idea of \citep{Evd94} with a notion called  a {\em matching}. We use the more general definition in \citep{AIKS14} (where it is called a {\em generalized matching}).

\begin{defi}[matching]\label{defi_matching}
Let $\Pi=\{P_1,\dots, P_m\}$ be an  $m$-scheme on $S$.
A block $B\in P_k$ for some $k\in [m]$ is called a {\em matching}\index{matching} of $\Pi$ if there exist two distinct proper subsets $T, T'$ of $[k]$ of the same cardinality such that $\pi^k_T(B)=\pi^k_{T'}(B)$ and $|B|=|\pi^k_T(B)|$.
\end{defi}

The work \citep{IKS09, AIKS14}  designed algorithms leading to $m$-schemes with no matching.
Now we explain the connection between this property and  strong antisymmetry of $m$-schemes. 

Given a matching $B\in P_k$ of $\Pi$, let $T,T'\subseteq [k]$ be as in Definition~\ref{defi_matching} and let $k':=k-|T|$. Then $B':=\pi^k_T(B)=\pi^k_{T'}(B)$ is a block of $P_{k'}$.  
We have two maps $\pi^k_T|_B$ and $\pi^k_{T'}|_B$ from $B$ to $B'$, both of which are bijective by the condition $|B|=|\pi^k_T(B)|$.
Moreover $\pi^k_T|_B\neq \pi^k_{T'}|_B$ as they omit different subsets of coordinates and the $k$ coordinates of elements in $S^{(k)}$ are all distinct.
So $\pi^k_{T'}|_B \circ (\pi^k_T|_B)^{-1}$ is a nontrivial permutation of $B'$. We conclude:
\begin{lem}\label{lem_antmatching}
A strongly antisymmetric $m$-scheme has no matching.
\end{lem}

So our definition of strong antisymmetry of $m$-schemes (or that of $\mathcal{P}$-schemes by Lemma~\ref{lem_ptom}) subsumes the property that no matching exists.

We will use strong antisymmetry instead of (non-existence of) matchings throughout this thesis.
The advantage of this comes from transitivity: Suppose $x\in H\backslash G$ is sent to a different element $y\in H\backslash G$ by a map $\tau$ that is a composition of conjugations, projections and their inverses (restricted to blocks), then  $x$ and $y$ belong to different blocks by strong antisymmetry. Suppose we also separate $y$ from another element $z\in H\backslash G$ in the same way. Then since the set of maps we consider are closed under composition, we get a map sending $x$ to $z$ and hence are able to separate them as well. The analyses in Chapter~\ref{chap_sym} and Chapter~\ref{chap_primitive} crucially exploit this property.
% On the other hand, it is not clear if the set of permutations arising from matchings are closed under composition.

\subsection{The connection of $m$-schemes with association schemes}

As shown in \citep{IKS09, AIKS14}, $m$-schemes are closely related to the notion of {\em association schemes}  \citep{BI84}.

\begin{defi}\label{defi_asso}
An {\em association scheme}\index{association scheme} on a finite set $S$ is a partition $P$ of $S\times S$ such that
\begin{itemize}
\item $1_S:=\{(x,x): x\in S\}$ is a block of $P$,
\item for a block $g\in P$, the set $g^*:=\{(y,x):(x,y)\in B\}$ is also a block, and
\item for every triple of blocks $g,g',g''\in P$, there exists an integer $c_{g',g''}^{g}\geq 0$ such that for any $(x,y)\in g$, the number of $z\in S$ satisfying $(x,z)\in g'$ and $(z,y)\in g''$ is $c_{g',g''}^{g}$. 
\end{itemize}
An association scheme $P$ is {\em symmetric} if $g=g^*$ for all $g\in P$, and {\em antisymmetric} if $g\neq g^*$ for all $g\in P-\{1_S\}$. The integer $c^{1_S}_{g, g^*}$ is called the {\em valency}\index{valency} of $g$.
\end{defi}
\nomenclature[a1m]{$1_S$}{block $\{(x,x):x\in S\}$ of an association scheme on $S$}
\nomenclature[a1n]{$c^g_{g',g''}$}{See Definition~\ref{defi_asso}}

We can obtain a homogeneous $3$-scheme from an association scheme and vice versa using the following constructions.

\begin{defi}\label{defi_3sch}
For a finite set $S$ and a partition $P$ of $S\times S$ such that $1_S\in P$, define the partition $P'$ of $S^{(3)}$ such that two elements $(x_1,x_2,x_3),(x'_1,x'_2,x'_3) \in S^{(3)}$ are in the same block of $P'$ iff $(x_i, x_j)$ and $(x'_i, x'_j)$ are in the same block of $P$ for all $1\leq i,j\leq 3$. 
And define a $3$-collection $\Pi(P)=\{P_1,P_2,P_3\}$ on $S$ by choosing $P_1=S$, $P_2=P - \{1_S\}$, $P_3=P'$.
Conversely, given a 3-collection $\Pi=\{P_1,P_2,P_3\}$ on $S$, define a partition $P(\Pi)$ of $S\times S$ by $P(\Pi):=P_2\cup\{1_S\}$ 
\end{defi}
\nomenclature[a1o]{$\Pi(P)$}{$3$-collection constructed from a partition $P$ (see Definition~\ref{defi_3sch})}
\nomenclature[a1p]{$P(\Pi)$}{partition constructed from a $3$-collection $\Pi$ (see Definition~\ref{defi_3sch})}

\begin{lem}[\citep{IKS09, AIKS14}]\label{lem_3sch}
 If $P$ is an association scheme, then $\Pi(P)$ is a homogeneous $3$-scheme. Conversely,  if $\Pi$ is a homogeneous 3-scheme, then $P(\Pi)$ is an association scheme. 
\end{lem}

By construction, this gives a one-to-one correspondence between association schemes on $S$ and equivalent classes of homogeneous $3$-schemes on $S$, where  two  homogeneous $3$-schemes $\{P_1,P_2,P_3\}$ and  $\{P'_1,P'_2,P'_3\}$ on $S$ are said to be equivalent if $P_1=P'_1$ and $P_2=P'_2$.

In addition, we obviously have
\begin{lem}\label{lem_3schsym}
If $\Pi$ is symmetric (resp. antisymmetric), so is $P(\Pi)$.
\end{lem}

Next we discuss the relation between symmetry and antisymmetry of an association scheme $P$ and those of $\Pi(P)$.
Obviously, for $\Pi(P)$ to be symmetric (resp. antisymmetric), it is necessary that $P$ is also symmetric (resp. antisymmetric). The exact condition is given as follows.

\begin{lem}\label{lem_as_antisym}
The $3$-scheme $\Pi(P)$ is symmetric iff $P$ is the trivial association scheme $\{1_S, S\times S-1_S\}$. It is antisymmetric iff $P$ is antisymmetric and $c_{g,g}^{g^*}=0$ for all $g\in P-\{1_S\}$. 
\end{lem}

\begin{proof}
The trivial association scheme $P=\{1_S, S\times S-1_S\}$ gives rise to the $3$-scheme $\Pi(P)=\{0_S, 0_{S^{(2)}},0_{S^{(3)}}\}$ which is symmetric. 
Suppose $P\neq \{1_S, S\times S-1_S\}$.   Let $g_1$ and $g_2$ be two distinct blocks in $P-\{1_S\}$. 
If $g_1=g_2^*$ then $P$ is not symmetric and hence neither is $\Pi(P)$. So assume $g_1\neq g_2^*$.
Fix $x\in S$.
Then $(x,y)\in g_1$ and $(x,z)\in g_2$ for some $y,z\in S-\{x\}$, and $y\neq z$.  Consider the element $t=(x,y,z)\in S^{(3)}$.
Let $h=(1~2~3)\in \sym(3)$ so that $\prescript{h}{}{t}=(z,x,y)$.
We have $\pi^3_3(t)=(x,y)\in g_1$ and $\pi^3_3(\prescript{h}{}{t})=(z,x)\in g_2^*\neq g_1$. By compatibility of $\Pi(P)$, the elements $t$ and $\prescript{h}{}{t}$ are in different blocks. So $\Pi(P)$ is not symmetric.

Suppose $\Pi(P)$ is antisymmetric, then so is $P$. We check that $c_{g,g}^{g^*}=0$ for all $g\in P-\{1_S\}$. Assume to the contrary that $c_{g,g}^{g^*}>0$ for some $g\in P-\{1_S\}$. Fix $(x,y)\in g^*$. Then there exists $z\in S$ such that $(x,z), (z,y)\in g$. Then for $t=(x,y,z)\in S^{(3)}$ and $h=(1~2~3)\in \sym(3)$, we have $\pi^3_i(t),\pi^3_i(\prescript{h}{}{t})\in g^*$ for all $1\leq i\leq 3$. It follows by definition that $t$ and $\prescript{h}{}{t}$ are in the same block, contradicting antisymmetry of $\Pi(P)$.

Conversely, suppose  $P$ is antisymmetric and $c_{g,g}^{g^*}=0$ for all $g\in P-\{1_S\}$. To prove $\Pi(P)$ is antisymmetric, it suffices to show that for any $t=(x,y,z)\in S^{(3)}$ and $h\in \sym(3)$, the elements $t$ and $\prescript{h}{}{t}$ are in different blocks. First assume $h$ is a transposition, e.g., $(1~2)$ (the other cases are symmetric). Then $\pi^3_3(t)=(x,y)$ and $\pi^3_3(\prescript{h}{}{t})=(y,x)$ are in different blocks by antisymmetry of $P$, and the claims follows by compatibility of $\Pi(P)$. Next assume $h$ is a $3$-cycle, e.g., $(1~2~3)$ (the other case is symmetric), so that $\prescript{h}{}{t}=(z,x,y)$. Let $g$ be the block in $P-\{1_S\}$ containing $(y,x)$, so that $(x,y)\in g^*$. As $c_{g,g}^{g^*}=0$, either $(x,z)$ or $(z,y)$ is not in $g$. If $(x,z)\not\in g$, we have $\pi^3_3(t)=(x,y)\in g^*$ and $\pi^3_3(\prescript{h}{}{t})=(z,x)\not\in g^*$. If $(x,z)\in g$ but $(z,y)\not\in g$, we have  $\pi^3_2(t)=(x,z)\in g$ and $\pi^3_2(\prescript{h}{}{t})=(z,y)\not\in g$. In either case  $t$ and $\prescript{h}{}{t}$ are in different blocks by compatibility of $\Pi(P)$.
\end{proof}

\begin{exmp}\label{exmp_as_antisym}
Let $S$ be a finite dimensional vector space over a finite field $\F_q$ where $\cha(\F_q)\not\in\{2,3\}$. Let $P$ be the partition of $S\times S$ such that $(x,y)$ and $(x',y')$ are in the same block iff $x-y=x'-y'$, which is an association scheme  \citep{BI84}. We check that $P$ satisfies the condition of Lemma~\ref{lem_as_antisym}, and hence  $\Pi(P)$ is antisymmetric.
For any $(x,y)\not\in 1_S$, we have $x-y\neq y-x$ since $x\neq y$ and $\cha(\F_q)\neq 2$, and therefore $(x,y)$ and $(y,x)$ are in different blocks. So $P$ is antisymmetric. Then we check that  $c_{g,g}^{g^*}=0$ for all $g\in P-\{1_S\}$. Assume to the contrary that  $c_{g,g}^{g^*}>0$ for some $g\in P-\{1_S\}$.
Fix $(x,y)\in g^*$ and choose $z\in S$ such that $(x,z),(z,y)\in g$. Then $x-z=z-y=y-x$, implying $3(x-z)=0$. This is impossible since $x\neq z$ and $\cha(\F_q)\neq 3$. 
\end{exmp}

The antisymmetric $3$-scheme $\Pi(P)$ in Example~\ref{exmp_as_antisym} is not strongly antisymmetric: For any distinct $x,y\in S$, let $B\in P-\{1_S\}$ be the block containing $t=(x,y)$. Then $\pi^2_1|_B$ and $\pi^2_2|_B$ are bijections from $B$ to $S$ sending $t$ to $y$ and $x$, respectively.
So $\pi^2_1|_B\circ (\pi^2_2|_B)^{-1}$ is a permutation of the unique block $S\in 0_{S}$ sending $x$ to $y$.

We do not know any example of an association scheme $P$ for which $\Pi(P)$ is strongly antisymmetric. The following lemma gives a sufficient condition for the existence of such an association scheme.

\begin{lem} 
Suppose $P$ is an antisymmetric association scheme satisfying (1) $c_{g,g}^{g^*}=0$ for all $g\in P-\{1_S\}$, and (2) for all blocks $g\in P$ and $g',g''\in P-\{1_S\}$, either $c_{g',g''}^g=0$ or  $c_{g',g''}^g>1$. Then $\Pi(P)$ is strongly antisymmetric.
\end{lem}
\begin{proof}
By Lemma~\ref{lem_as_antisym}, the 3-scheme $\Pi(P)$  is antisymmetric. And (2) implies that none of the projections $\pi^2_i$ and $\pi^3_i$ are invertible even restricted to blocks of $S^{(2)}$ and $S^{(3)}$ respectively. Strong antisymmetry of $\Pi(P)$ then follows from antisymmetry.
\end{proof}

In general, strongly antisymmetric $3$-schemes do exist. See Example~\ref{exmp_3anti} in Section~\ref{sec_3anti}.

\section{Orbit \texorpdfstring{$\mathcal{P}$-schemes}{P-schemes} and \texorpdfstring{$m$-schemes}{m-schemes}}\label{sec_pscheme_orbit}

An important family of $m$-schemes called {\em orbit schemes}, or what we call {\em orbit $m$-schemes}, was proposed and studied in \citep{IKS09}. The blocks of such $m$-schemes are orbits of group actions.

\begin{defi}[orbit $m$-scheme \citep{IKS09}]\label{defi_orbitm}
Given a finite set $S$, $m\in \N^+$, and a group $K\subseteq \sym(S)$ acting naturally on $K$, for each $k\in [m]$, define the partition $P_k$ of $S^{(k)}$ to be the partition into $K$-orbits with respect to the diagonal action of $K$ on $S^{(k)}$.
%, i.e., two elements $(x_1,\dots, x_k)$ and $(y_1,\dots,y_k)$ of $S^{(k)}$ are in the same block iff there exists $g\in K$ such that $\prescript{g}{}{x_i}=y_i$ for all $i\in [k]$. 
The $m$-collection $\Pi=\{P_1,\dots,P_m\}$ is called the {\em orbit $m$-scheme}\index{orbit!$m$-scheme} on $S$ associated with the group $K$.
\end{defi}

This is indeed an $m$-scheme:
\begin{thm}[\citep{IKS09}]\label{thm_orbitm}
The $m$-collection $\Pi$ in Definition~\ref{defi_orbitm} is an $m$-scheme on $S$.
\end{thm}

%This result can be proven directly, but we will derive it from a similar statement for $\mathcal{P}$-schemes (Theorem~\ref{thm_orbitp}), using the connection between $m$-schemes and $\mathcal{P}$-schemes given in the previous section.

%\begin{proof}
%For $1<k\leq m$, $i\in [k]$, $g\in K$ and $x\in S^{(k)}$, we have $\pi^k_i(\prescript{g}{}{x})=\prescript{g}{}{(\pi^k_i(x))}$ with respect to the diagonal action of $K$ on $S^{(k)}$. Therefore  if $x,x'\in S^{(k)}$ are in the same block of $P_k$ (i.e., the same $K$-orbit of $S^{(k)}$), then $\pi^k_i(x)$ and $\pi^k_i(x')$ are in the same block of $P_{k-1}$ (i.e., the same $K$-orbit of $S^{(k-1)}$). So $\Pi$ is compatible. 
%
%Similarly, for $k\in [m]$, $\tau\in \sym(k)$, $g\in K$ and $x\in S^{(k)}$, we have $c^k_\tau(\prescript{g}{}{x})=\prescript{g}{}{(c^k_\tau(x))}$. Therefore  if $x,x'\in S^{(k)}$ are in the same block of $P_k$, so are $c^k_\tau(x)$ and $c^k_\tau(x')$. So $\Pi$ is invariant.
%
%Finally, for $1<k\leq m$ and $y,y'\in S^{(k-1)}$ lying in the same block $B$ of $P_{k-1}$, choose $g\in K$ such that $y'=\prescript{g}{}{y}$. Then for $i\in [k]$ and $x\in S^{(k)}$, we have $x\in B$ and $\pi^k_i(x)=y$ hold  iff $x'=\prescript{g}{}{x}\in \prescript{g}{}{B}=B$  and $\pi^k_i(\prescript{g}{}{x})=\prescript{g}{}{(\pi^k_i(x))}=\prescript{g}{}{y}=y'$. So the map $x\mapsto \prescript{g}{}{x}$ is a one-to-one correspondence between $B\cap (\pi^k_i)^{-1}(y)$ and  $B\cap (\pi^k_i)^{-1}(y')$, and hence the two sets have the same cardinality. So $\Pi$ is regular. 
%\end{proof}

We define orbit $\mathcal{P}$-schemes in a similar way, except that the subgroup $K$ of $\sym(S)$ is now replaced with a subgroup of $G$, and the diagonal actions on $S^{(k)}$, $k\in [m]$ are replaced with the  actions on right coset spaces by inverse right translation.

\begin{defi}[orbit $\mathcal{P}$-scheme]\label{defi_orbitp}
Let $\mathcal{P}$ be a subgroup system over a finite group $G$, and let $K$ be a subgroup of $G$.
For $H\in\mathcal{P}$, define the partition $C_H$ of $H\backslash G$ to be the partition into $K$-orbits, with respect to the action of $K$ on $H\backslash G$ by inverse right translation. The $\mathcal{P}$-collection $\mathcal{C}=\{C_H: H\in\mathcal{P}\}$ is called the {\em orbit $\mathcal{P}$-scheme}\index{orbit!$\mathcal{P}$-scheme} associated with the group $K$.
\end{defi}

This construction indeed yields a $\mathcal{P}$-scheme:
\begin{thm}\label{thm_orbitp}
The $\mathcal{P}$-collection $\mathcal{C}$ in Definition~\ref{defi_orbitp} is a $\mathcal{P}$-scheme.
\end{thm}
\begin{proof}
Let $K$ act on each right coset space $H\backslash G$ by inverse right translation.
For $H,H'\in\mathcal{P}$ with $H\subseteq H'$, $g\in K$ and $x\in H\backslash G$, we have $\pi_{H,H'}(\prescript{g}{}{x})=\prescript{g}{}{(\pi_{H,H'}(x))}$ by Lemma~\ref{lem_maps}.
Therefore  if $x,x'\in H\backslash G$ are in the same block of $C_H$ (i.e., the same $K$-orbit of $H\backslash G$), then $\pi_{H,H'}(x)$ and $\pi_{H,H'}(x')$ are in the same block of $C_{H'}$ (i.e., the same $K$-orbit of $H'\backslash G$). So $\mathcal{C}$ is compatible. 

Similarly, for $H\in \mathcal{P}$, $h\in G$, $g\in K$ and $x\in H\backslash G$, we have $c_{H,h}(\prescript{g}{}{x})=\prescript{g}{}{(c_{H,h}(x))}$ by Lemma~\ref{lem_maps}. Therefore  if $x,x'\in H\backslash G$ are in the same block  of $C_H$,
then $c_{H,h}(x)$ and $c_{H,h}(x')$ are in the same block  of $C_{gHg^{-1}}$. So $\mathcal{C}$ is invariant.

For $H'\in\mathcal{P}$ and $y,y'\in H'\backslash G$ in the same block $B$ of $C_{H'}$, choose $g\in K$ such that $y'=\prescript{g}{}{y}$. As $g\in K$, we have $\prescript{g}{}{B}=B$.
For $H\in\mathcal{P}$ with $H\subseteq H'$ and $x\in H\backslash G$, we have $x\in B$ and $\pi_{H,H'}(x)=y$  iff $\prescript{g}{}{x}\in \prescript{g}{}{B}=B$  and $\pi_{H,H'}(\prescript{g}{}{x})=\prescript{g}{}{(\pi_{H,H'}(x))}=\prescript{g}{}{y}=y'$.
So the map $x\mapsto \prescript{g}{}{x}$ is a one-to-one correspondence between  $B\cap \pi_{H,H'}^{-1}(y)$ and  $B\cap \pi_{H,H'}^{-1}(y')$, and hence the two sets have the same cardinality. So $\mathcal{C}$ is regular. 
\end{proof}

The connection between Definition~\ref{defi_orbitm}  and Definition~\ref{defi_orbitp} is given by the following lemma. 
\begin{lem}\label{lem_orbitpm}
For a finite set $S$,  $m\in \N^+$, and a subgroup $K\subseteq \sym(S)$, let $\mathcal{P}=\mathcal{P}_{m}$ be the system of stabilizers of depth $m$ with respect to the natural action of $\sym(S)$ on $S$,
and let $\mathcal{C}=\{C_H: H\in\mathcal{P}\}$ be the orbit $\mathcal{P}$-scheme associated with $K$. Then the orbit $m$-scheme associated with $K$ is exactly $\Pi(\mathcal{C})$ as defined in Definition~\ref{defi_ptom}.
\end{lem}

\begin{proof}
We may assume $m\leq |S|$.
Let $G$ be the symmetric group $\sym(S)$ acting naturally on $S$. Suppose  $\Pi(\mathcal{C})=\{P_1,\dots,P_m\}$ where $P_k$ is a partition of $S^{(k)}$ for $k\in [m]$.
By  Definition~\ref{defi_ptom}, each partition $P_k$ is given by $P_k=\{\lambda_x^{-1}(B): B\in C_{G_x}\}$ for some $x=(x_1,\dots,x_k)\in S^{(k)}$, where $\lambda_x: S^{(k)}\to G_x\backslash G$ is an equivalence  between the diagonal action of $G$ on $S^{(k)}$ and the action  on $G_x\backslash G$ by inverse right translation.
It follows that $P_k$ is the partition into $K$-orbits with respect to the diagonal action, since $C_{G_x}$ is the partition into $K$-orbits with respect to the action by inverse right translation.
\end{proof}

%Theorem~\ref{thm_orbitm} then follows from Lemma~\ref{lem_ptom}, Lemma~\ref{lem_orbitpm} and Theorem~\ref{thm_orbitp}.

\paragraph{Antisymmetry of orbit $m$-schemes.} We prove a simple and exact criterion for antisymmetry of orbit $m$-schemes.
\begin{lem}\label{lem_antisym_orbitm}
The orbit $m$-scheme on $S$ associated with $K\subseteq \sym(S)$ is antisymmetric iff the order of $K$ is coprime to $1, 2,\dots, m$.
\end{lem}

\begin{proof}
Let $\Pi=\{P_1,\dots,P_m\}$ be the orbit $m$-scheme on $S$ associated with $K$.
Suppose the order of $K$ is divisible by an integer $k$ satisfying $1<k\leq m$. We may assume that $k$ is a prime integer.
By Cauchy's theorem (see, e.g., \citep{Lan02}), the group $K$ contains an element $g$ of order $k$.
The element $g$, as a permutation of $S$, has at least one $k$-cycle $(x_1~x_2~\cdots~x_k)$.
Consider the element $x=(x_1,\dots,x_k)\in S^{(k)}$, and let $B$ be the block of $P_k$ containing $x$.
By definition, the element $\prescript{g}{}{x}=(\prescript{g}{}{x_1},\dots,\prescript{g}{}{x_k})=(x_2,\dots,x_k,x_1)$ is also in $B$.
On the other hand, let $h=(1~2~\cdots~k)^{-1}\in\sym(k)$. The permutation $c^k_h$ of $S^{(k)}$ sends $x=(x_1,\dots,x_k)$ to $y=(y_1,\dots,y_k)$ defined by $y_i=x_{\prescript{h^{-1}}{}{i}}$ for $i\in [k]$. So $c^k_h(x)=(x_2,\dots,x_k,x_1)\in B$. 
Therefore $\Pi$ is not antisymmetric.

Conversely, assume $\Pi$ is not antisymmetric. Then for some integer $k$ satisfying $1<k\leq \min\{m,|S|\}$, $h\in\sym(k)-\{e\}$, and some element $x=(x_1,\dots,x_k)\in S^{(k)}$ lying in a block $B$ of $P_k$, we have $c^k_h(x)\in B$, i.e., $c^k_h(x)=\prescript{g}{}{x}$ for some $g\in K$ with respect to the diagonal action of $K$ on $S^{(k)}$. As the permutation $c^k_h$ of $S^{(k)}$ sends $x$ to $y=(y_1,\dots,y_k)$ defined by $y_i=x_{\prescript{h^{-1}}{}{i}}$ for $i\in [k]$, we see $\prescript{g}{}{x_i}=x_{\prescript{h^{-1}}{}{i}}$ for $i\in [k]$. Then $g$ preserves the set $T:=\{x_1,\dots,x_k\}$ and restricts to a nontrivial permutation $g|_T\in \sym(T)$ of $T$. Let $e$ be the order of $g|_T$. Then $e$ is not coprime to some integer $t$ where $t\leq |T|\leq m$.
The order of $K$ is a multiple of the order of $g$, which is a multiple of $e$. So the order of $K$ is not coprime to $t$ either.
\end{proof}

\begin{exmp}\label{exmp_antisym_orbit}
Let $S$ be a finite set satisfying $|S|>1$.  Let $K$ be a subgroup of $\sym(S)$ generated by a single $|S|$-cycle so that it acts regularly on $S$. Denote by $\ell$ the least prime factor of $|S|$. 
Let $\Pi$ be the orbit $m$-scheme on $S$ associated with $K$ where $m$ is an integer satisfying $1\leq m<\ell$.
Then $\Pi$  is homogeneous since $K$ acts transitively on $S$. 
The order of $K$ is $|S|$, which is coprime to $1,\dots,\ell-1$. 
So $\Pi$ is also antisymmetric by Lemma~\ref{lem_antisym_orbitm} and the fact $m\leq \ell-1$.
\end{exmp}

\paragraph{Upper bound of $m$ for antisymmetric homogeneous $m$-schemes.}

Let $S$ be a finite set satisfying $|S|>1$, and let $\ell$ be the least prime factor of $|S|$.
For $m\geq \ell$, the orbit $m$-schemes on $S$ in Example~\ref{exmp_antisym_orbit} are still homogeneous but no longer antisymmetric. Indeed, an argument of R{\'o}nyai \citep{Ron88} shows that for $m\geq \ell$, even general $m$-schemes on $S$ cannot be both homogeneous and antisymmetric. This was reproduced in \citep{IKS09} and we present it here.

\begin{lem}[\citep{Ron88, IKS09}]\label{lem_integrality_mscheme}
Let $S$ be a finite set satisfying $|S|>1$, and let $\ell$ be the least prime factor of $|S|$. There exists no antisymmetric homogeneous $m$-scheme on $S$ for $m\geq \ell$.
\end{lem}

\begin{proof}
%Let $n=|S|$.
Assume to the contrary that such an $m$-scheme $\Pi=\{P_1,\dots,P_m\}$ exists. The group $\sym(\ell)$ acts on $S^{(\ell)}$ by $\prescript{g}{}{x}=c^k_g(x)$. By antisymmetry of $\Pi$, this action induces a semiregular action on the set of blocks in $P_\ell$.
Let $B_1,\dots,B_k\in P_\ell$ be a complete set of representatives for the $\sym(\ell)$-orbits, i.e., each orbit contains exactly one $B_i$. Then we have
\[
\sum_{i=1}^k |B_i|=\frac{|S^{(\ell)}|}{|\sym(\ell)|}=\frac{|S|(|S|-1)\cdots (|S|-\ell+1)}{\ell!}
\]
Let $\pi$ be the projection from $S^{(\ell)}$ to $S$ sending $(x_1,\dots,x_\ell)$ to $x_1$. 
%It can be decomposed into projections $\pi^{k}_i$ for various $k\in [\ell]$ and $i\in [k]$. 
By regularity and homogeneity of $\Pi$, for each $i\in [k]$, the cardinality of $B_i\cap \pi^{-1}(y)$ is a constant $d_i\in \N^+$ independent of $y\in S$. Then
\[
\sum_{i=1}^k d_i = \sum_{i=1}^k \frac{|B_i|}{|S|}=\frac{(|S|-1)\cdots (|S|-\ell+1)}{\ell!}.
\]
As $|S|$ is a multiple of $\ell$, none of the factors $|S|-1$, \dots, $|S|-\ell+1$ of the numerator is divisible by the prime number $\ell$ appeared in the denominator. This contradicts the integrality of $\sum_{i=1}^k d_i$.
\end{proof}

The condition $m\geq \ell$ in Lemma~\ref{lem_integrality_mscheme} is tight, since Example~\ref{exmp_antisym_orbit} shows that antisymmetric homogeneous $m$-schemes exist for $m=\ell-1$. 

R{\'o}nyai's result can be extended to $\mathcal{P}$-schemes in the case that $\mathcal{P}$ is a system of stabilizers with respect to a transitive group action.
 
\begin{lem}\label{lem_integrality_pscheme}
Let $G$ be a finite group acting transitively on a set $S$ of cardinality $n>1$. Let $\mathcal{P}=\mathcal{P}_{m}$ be the corresponding system of stabilizers of depth $m$ for some $m\geq \ell$, where $\ell$ is the least prime factor of $n$. 
Then for any $x\in S$, there exists no antisymmetric $\mathcal{P}$-scheme that is homogeneous on $G_x$. In particular, $d'(G)<\ell$.
\end{lem}
 
 Lemma~\ref{lem_integrality_pscheme} can be easily proven using a technique called the {\em induction of $\mathcal{P}$-schemes}, to be discussed in Chapter~\ref{chap_common}. It allows us to reduce to the case $G=\sym(S)$.  The claim then follows immediately, since by Lemma~\ref{lem_ptom}, for $G=\sym(S)$, the existence of an antisymmetric $\mathcal{P}$-scheme homogeneous on $G_x$  implies the existence of an antisymmetric homogeneous $m$-scheme on $S$, which contradicts Lemma~\ref{lem_integrality_mscheme}. For now, we just provide a direct proof.
 
 \begin{proof}[Proof of Lemma~\ref{lem_integrality_pscheme}]
 Assume to the contrary that $\mathcal{C}=\{C_H: H\in \mathcal{P}\}$ is an antisymmetric $\mathcal{P}$-scheme that is homogeneous on  $G_x$ for some $x\in S$. As $\mathcal{C}$ is invariant and $G$ acts transitively on $S$ (and hence all one-point stabilizers $G_x$ are conjugate in $G$), we know $\mathcal{C}$ is homogeneous on $G_x$ for all $x\in S$.
 
Consider the set $S^{(\ell)}$ equipped with two actions: the diagonal action of $G$ and the action of $\sym(\ell)$ permuting the $\ell$ coordinates. The latter action is defined by $\prescript{g}{}{(x_1,\dots,x_\ell)}=(x_{\prescript{g^{-1}}{}{1}},\dots x_{\prescript{g^{-1}}{}{\ell}})$ for $g\in\sym(\ell)$ and $(x_1,\dots,x_\ell)\in S^{(\ell)}$. Note that these two actions commute with each other and combine to an action of $G\times \sym(\ell)$ on $S^{(\ell)}$. 
For $z\in S^{(\ell)}$, we have $\prescript{g}{}{Gz}=G\prescript{g}{}{z}$ for all $g\in \sym(\ell)$ and hence the action of $\sym(\ell)$ permutes the $G$-orbits within the $(G\times \sym(\ell))$-orbit $(G\times \sym(\ell))z$.

Now fix $z\in S^{(\ell)}$. We have the bijection $\lambda_z: Gz\to G_z\backslash G$ which is an equivalence between the action of $G$ on the $G$-orbit $Gz$ and the action on $G_z\backslash G$ by inverse right translation. We also have a semiregular action of $N_G(G_z)/G_z$ on $G_z\backslash G$ by left translation. This gives a injective group homomorphism $\phi: N_G(G_z)/G_z\hookrightarrow \sym(G_z\backslash G)$, and we denote its image by $N$. Then $|N|=|N_G(G_z)/G_z|$.

Let $H$ be the subgroup of $\sym(\ell)$ fixing $Gz$ setwisely, i.e.,  $H=\{g\in\sym(\ell):\prescript{g}{}{Gz}=Gz\}$. The action of $H\subseteq \sym(\ell)$ on $S^{(\ell)}$ restricts to an action on $Gz$ and hence we have a group homomorphism $H\to \sym(Gz)$. It is injective since elements in $Gz\subseteq S^{(\ell)}$ have distinct coordinates. 
Now, identifying $Gz$ with $G_z\backslash G$ via $\lambda_z$, we have an action of $H$ on $G_z\backslash G$ as well, defined by $\prescript{g}{}{\lambda_z(x)}=\lambda_z(\prescript{g}{}{x})$ for $x\in Gz$. This gives an injective group homomorphism $\phi': H\hookrightarrow \sym(G_z\backslash G)$. 

We claim that $\phi'(H)\subseteq N$. To see this, pick any $g\in H$. We have $\prescript{g}{}{G_ze}=G_zh_0$ for some $h_0\in G$, or equivalently $\prescript{g}{}{z}=\prescript{h_0^{-1}}{}{z}$. 
Then for any  $h\in G$, we have 
\[
\prescript{g}{}{G_zh}=\prescript{g}{}{\left(\lambda_z(\prescript{h^{-1}}{}{z})\right)}
=\lambda_z\left(
\prescript{g}{}{(\prescript{h^{-1}}{}{z})}
\right)
=\lambda_z\left(
\prescript{h^{-1}}{}{(\prescript{g}{}{z})}
\right)
=\lambda_z(\prescript{(h_0h)^{-1}}{}{z})
=G_z h_0h.
\]
In particular, for any $h\in G_z$, we have $\prescript{g}{}{G_zh}=G_zh_0h$ and other other hand $\prescript{g}{}{G_zh}=\prescript{g}{}{G_ze}=G_zh_0$. So $h_0hh_0^{-1}\in G_z$. Therefore $h_0\in N_G(G_z)$.
Furthermore, note that $h_0G_z\in N_G(G_z)/G_z$ sends any $G_zh\in G_z\backslash G$ to $G_z h_0h=\prescript{g}{}{G_zh}$ by left translation.
So $\phi'(g)=\phi(h_0G_z)\in N$. Therefore $\phi'(H)\subseteq N$, as desired. 
%In particular, we know $|H|$ divides $|N|$.

By antisymmetry, the action of $N$ on $G_z\backslash G$ induces a semiregular action on the set of blocks of $C_{G_z}$, which induces a semiregular action of $\phi'(H)$ on the set of blocks of $C_{G_z}$.  Let $B_1,\dots,B_k\in C_{G_z}$ be a complete set of representatives for the $\phi'(H)$-orbits.
Then we have
\[
\sum_{i=1}^k |B_i|=\frac{|G_z\backslash G|}{|\phi'(H)|}=\frac{|Gz|}{|H|}.
\]
Choose $x\in S$ such that $G_z \subseteq G_x$. By regularity and homogeneity on $G_x$, for each $i\in [k]$, the cardinality of $B_i\cap \pi_{G_z,G_x}^{-1}(y)$ is a constant $d_i\in\N^+$ independent of $y\in G_x\backslash G$, and hence $|B_i|$ is a multiple of $|G_x\backslash G|=n$. Therefore $|Gz|$ is a multiple of $n\cdot |H|$. 

 By the orbit-stabilizer theorem, the number of $G$-orbits contained in $(G\times \sym(\ell))z$ is $|\sym(\ell)|/|H|$, and these $G$-orbits all have the same cardinality $|Gz|$. So 
 \[
 |(G\times \sym(\ell))z|=\frac{|\sym(\ell)|}{|H|}\cdot |Gz|,
 \]
 which is a multiple of $n\cdot |\sym(\ell)|=n \ell!$ since $|Gz|$ is a multiple of  $n\cdot |H|$. As this holds for arbitrary $z\in S^{(\ell)}$, we know $|S^{(\ell)}|=n(n-1)\cdots(n-\ell+1)$ is also a multiple of $n \ell!$. But this is not possible since $n-1,\dots, n-\ell+1$ are not divisible by the prime number $\ell$.
 \end{proof}
 
\section{Strongly antisymmetric homogeneous \texorpdfstring{$m$-schemes}{m-schemes} for \texorpdfstring{$m\leq 3$}{m>=3}}\label{sec_3anti}

In this section, we give examples of strongly antisymmetric homogeneous $m$-schemes on a finite set $S$ where $|S|>1$ and $m\in\{1,2,3\}$.

\paragraph{The case $m=1$.} For all finite sets $S$, there exists a unique homogeneous $1$-scheme $\Pi=\{P_1\}$ on $S$, given by $P_1=0_{S}$. It is obviously antisymmetric since $\sym(1)$ is the trivial group.  And it is also strongly antisymmetric since there exists no projection $\pi^k_i$ for $m=1$.

\paragraph{The case $m=2$.} We discuss the following explicit construction of orbit $2$-schemes.

\begin{exmp}\label{exmp_paley}
Let $q$ be a prime power of the form $q=4k+3$ for some $k\in\N$.\footnote{In particular, we may choose $q$ to be a prime number. By Dirichlet's theorem on arithmetic progressions \citep{Neu99}, there exist infinitely many prime numbers of the form $4k+3$.}
The multiplicative group $\F_q^\times$ is a cyclic group of order $4k+2$.  
Denote by $\chi_2:\F_q^\times\to \C$ the unique nontrivial quadratic character of $\F_q^\times$, which sends quadratic residues to $1$ and non-residues to $-1$.
Its kernel $\ker(\chi_2)$ is the unique subgroup of $\F_q^\times$ of index two.
For $u\in\F_q^\times$ and $v\in \F_q$, denote by $\phi_{u,v}$ the affine linear transformation of $\F_q$ sending $x\in\F_q$ to $ux+v$. Define $K$ by
\[
K:=\{\phi_{u,v}: u\in \ker(\chi_2), v\in\F_q\}.
\]
Then $K$ is a subgroup of  $\sym(\F_q)$.\footnote{The group $K$ is also a subgroup of the {\em general affine group} $\agl_1(q)$ and is isomorphic to a semidirect product $\F_q\rtimes \ker(\chi_2)$.}
Let $\Pi=\{P_1,P_2\}$ be the orbit $2$-scheme on $\F_q$ associated with the subgroup $K$.
\end{exmp}

The partitions $P_1$ and $P_2$ are given as follows:
as $K$ acts transitively on $\F_q$, we have $P_1=0_{\F_q}$ and $\Pi$ is homogeneous.
For $(a,b)\in \F_q^{(2)}$, we have $\prescript{\phi_{1,b}}{}{(a-b,0)}=(a,b)$ and $\phi_{1,b}\in K$, and hence $(a,b)$ and $(a-b,0)$ are in the same block of $P_2$. Two elements $(c,0),(d,0)\in  \F_q^{(2)}$ are in the same block iff $\prescript{g}{}{c}=d$ for some $g\in K_0$, where $K_0$ is the stabilizer of $0\in\F_q$. As $K_0=\{\phi_{u,0}: u\in \ker(\chi_2)\}$, we see that $(c,0)$ and $(d,0)$ are in the same block iff $\chi_2(c)=\chi_2(d)$. We conclude that $P_2$ contains two blocks $B_{+1}$ and $B_{-1}$, where
\[
B_s=\{(a,b)\in \F_q^{(2)}: \chi_2(a-b)=s\} 
\]
for $s=\pm 1$.

The order of $K$ is $q(q-1)/2=(4k+3)(2k+1)$ which is odd. So $\Pi$ is antisymmetric by Lemma~\ref{lem_antisym_orbitm}.
For every $y\in \F_q$, the number of elements in $B_{+1}$ (or $B_{-1}$) mapped to $y$ by the projection $\pi^2_1$ (or $\pi^2_2$) is $(q-1)/2$, which is greater than one when $q>3$. Therefore when $q>3$, the two projections $\pi^2_1$ and $\pi^2_2$ restricted to $B_1$ (or $B_2$) are not invertible, and hence $\Pi$ is strongly antisymmetric. We conclude:

\begin{lem}\label{lem_anti2ch}
The orbit $2$-scheme $\Pi$ in Example~\ref{exmp_paley} is homogeneous and  antisymmetric. It is strongly antisymmetric when $q>3$.
\end{lem}

We remark that the partition $P:=P_2\cup \{1_{\F_q} \}$ of $\F_q\times \F_q$ (where $1_{\F_q}=\{(a,a):a\in \F_q\}$) is actually an antisymmetric association scheme on $\F_q$. It is known as an association scheme of {\em Paley tournaments}\index{Paley tournament} \citep{ER63, BI84, BCN89}, or more generally a {\em cyclotomic scheme}\index{cyclotomic scheme} \citep{BCN89}.

Recall that for an association scheme $P$ on a set $S$, blocks $g,g',g''\in P$, and  $(x,y)\in g$, we use $c_{g',g''}^{g}$ to denote the number of $z\in S$ satisfying $(x,z)\in g'$ and $(z,y)\in g''$. When $P$ is antisymmetric and has only three blocks, the quantities  $c_{g',g''}^{g}$ only depend on $n$.\footnote{This is a folklore result. Such an association scheme is equivalent to a {\em doubly regular tournament}\index{doubly regular tournament}. See, e.g., \citep{RB72}.}  We state it formally for the cases $g,g',g''\neq 1_S$.
\begin{lem}\label{lem_antiasso}
Let $P$ be an antisymmetric association scheme on a set $S$ of cardinality $n$ containing only three blocks $1_S$, $g$ and $g^*$. Then for $u,v,w\in \{g,g^*\}$, we have
\[
c^{u}_{v, w}=\begin{cases} 
(n+1)/4 & \text{if}~ u^*=v=w,\\
(n-3)/4 & \text{otherwise}.
\end{cases}
\]
\end{lem}

\begin{proof}
From the basic properties of association schemes, we have $c^{g^*}_{g, g}=c^g_{g^*, g^*}$, $c^{g}_{g, g}=c^g_{g, g^*}=c^g_{g^*, g}=c^{g^*}_{g^*, g^*}=c^{g^*}_{g^*, g}=c^{g^*}_{g, g^*}$, and $\sum_{w\in P} c^u_{v, w}=(n-1)/2$ for $u,v\in\{g,g^*\}$.\footnote{See, e.g., \citep[Section \RN{2}.2, Proposition 2.2]{BI84} and note that $g,g^*$ have the same valency $(n-1)/2$.}
Also note that $c^{u}_{v, 1_S}$ equals one when $u=v$ and zero otherwise. The claim then follows by simple calculations.
\end{proof}

In particular, Lemma~\ref{lem_antiasso} applies to the association scheme $P=P_2\cup \{1_{\F_q}\}$ above. This is used in the next example for the proof of strong antisymmetry.

\paragraph{The case $m=3$.} We have noted that for $P_2$ as defined in Example~\ref{exmp_paley}, the partition $P=P_2\cup \{1_{\F_q}\}$  of $\F_q\times \F_q$  is an antisymmetric association scheme on $\F_q$. Thus by Lemma~\ref{lem_3sch}, we have a homogeneous $3$-scheme $\Pi(P)$. Unfortunately, $\Pi(P)$ is not necessarily  antisymmetric: there may exist distinct elements $a,b,c\in\F_q$ such that $\chi_2(a-b)=\chi_2(b-c)=\chi_2(c-a)$, and the block containing $(a,b,c)\in\F_q^{(3)}$ is preserved by the $3$-cycles in $\sym(3)$.

 However, it is possible to modify  $\Pi(P)$ to get an explicit construction of strongly antisymmetric homogeneous $3$-schemes. The idea is to use a nontrivial cubic character besides the quadratic character $\chi_2$.
 
\begin{exmp}\label{exmp_3anti}
 Let $q$ be a prime power of the form $36k+11$ or $36k+23$ for some $k\in\N$.\footnote{Again, by Dirichlet's theorem on arithmetic progressions \citep{Neu99}, there exist infinitely many such $q$.} The congruence is chosen so that $q-1$ is divisible by $2$ but not by $3$ or $4$, and $q^2-1$ is divisible by $3$ but not by $9$. In particular, the condition $q\equiv 3\bmod 4$ in Example~\ref{exmp_paley} still holds. 
Define a $3$-collection $\Pi=\{P_1, P_2, P_3\}$ on $\F_q$ as follows: $P_1$ and $P_2$ are constructed in the same way as in  Example~\ref{exmp_paley}, i.e., $P_1=0_{\F_q}$ and $P_2$  contains two blocks, $B_{+1}$ and $B_{-1}$, where
\[
B_s=\{(a,b)\in \F_q^{(2)}: \chi_2(a-b)=s\} 
\]
for $s=\pm 1$, and $\chi_2:\F_q^\times\to \C$ is the unique nontrivial quadratic character of $\F_q^\times$.

To construct $P_3$, we consider the quadratic extension $\F_{q^2}$ of $\F_q$. Its multiplicative group $\F_{q^2}^\times$ is a cyclic group of order $q^2-1$ which is divisible by $3$. 
Choose a nontrivial cubic character $\chi_3:\F_{q^2}^\times\to \C$.
%The image of $\chi_3$ is the subgroup of third roots of unity of $\C^\times$, which we denote by $\mu_3$.  
Let $\omega$ be a primitive third root of unity in $\F_{q^2}$ so that $1+\omega+\omega^2=0$. 
For $(a,b,c)\in \F_q^{(3)}$, we have $a+\omega b+\omega^2 c=(a-c)+\omega(b-c)$ which is nonzero since $\omega\not\in \F_q$. So $a+\omega b+\omega^2 c\in \F_{q^2}^\times$. We define a function $s$ on $\F_q^{(3)}$ by
\[
s(a,b,c):=\begin{cases}
\chi_3(a+\omega b+\omega^2 c) &\text{if}~ \chi_2(a-b)=\chi_2(b-c)=\chi_2(c-a),\\
1 &\text{otherwise.}
\end{cases}
\]
For $(a,b,c)\in  \F_q^{(3)}$, call the quadruple 
\[
(\chi_2(a-b),\chi_2(b-c),\chi_2(c-a), s(a,b,c))
\]
the {\em signature} of $(a,b,c)$.
Choose the partition $P_3$ of $\F_q^{(3)}$ such that two triples $(a,b,c), (a',b',c')\in \F_q^{(3)}$ are in the same block iff they have the same signature.
%So for each block $B\in P_3$, there exists $s_1,s_2,s_3\in\{\pm 1\}$ and $s_4\in \mu_3$ such that
%\[
%B=\left\{ 
%(a,b,c)\in \F_q^{(3)}:   
%\begin{aligned}
%        &\chi_2(a-b)=s_1, \chi_2(b-c)=s_2, \chi_2(c-a)=s_3,\\
%        &\chi_3(a+\omega b+\omega^2 c)=s_4
%\end{aligned}
%\right\}
%\]
%and we denote such a block by $B_{s_1,s_2,s_3,s_4}$ .
\end{exmp}

\begin{lem}
The $3$-collection $\Pi$ in Example~\ref{exmp_3anti} is an antisymmetric homogeneous $3$-scheme on $\F_q$.  It is strongly antisymmetric when $q>11$.
\end{lem}
\begin{proof}
We first check that $\Pi$ is an antisymmetric $3$-scheme.

For compatibility, we need to verify that if  $(a,b,c), (a',b',c')\in \F_q^{(3)}$ are in the same block of $P_3$, then their images under $\pi^3_i$ are  in the same block of $P_2$, $i=1,2,3$. This follows by construction.

For invariance and antisymmetry, we need to show that for any $g\in\sym(3)-\{e\}$, the signature of $(a,b,c)\in \F_q^{(3)}$ determines that of $\prescript{g}{}{(a,b,c)}$, and they are different. We note that $\chi_2(-1)=-1$ since  $|\F_{q}^\times|=q-1$ is not divisible by $4$, and $\chi_3(\omega)$ is a primitive third root of unity  in $\C$ since  $|\F_{q^2}^\times|=q^2-1$ is not divisible by $9$.

Suppose $g$ is a transposition, e.g., the one sending $(a,b,c)$ to $(b,a,c)$ (the other cases are symmetric). Then the signature of $\prescript{g}{}{(a,b,c)}$ is 
\begin{align*}
&(\chi_2(b-a),\chi_2(a-c),\chi_2(c-b), s(b,a,c))\\
&=(-\chi_2(a-b),-\chi_2(c-a),-\chi_2(b-c), s(b,a,c)).
\end{align*}
When $\chi_2(a-b),\chi_2(b-c),\chi_2(c-a)$ are not all equal, we have $s(b,a,c)=1$. Otherwise 
\[
s(b,a,c)=\chi_3(b+\omega a+ \omega^{2}c)=\chi_3(\omega)\chi_3(a+\omega^{-1}b+ \omega^{-2}c).
\]
The automorphism $x\mapsto x^q$ of $\F_{q^2}$ fixes $a,b,c\in\F_q$ and exchanges $\omega$ with $\omega^{-1}$.
So $\chi_3(a+\omega^{-1}b+ \omega^{-2}c)= \chi_3((a+\omega b+ \omega^2 c)^q)=\chi_3^q(a+\omega b+ \omega^2 c)$.
We see that in this case, the signature of $(a,b,c)\in \F_q^{(3)}$ determines that of $\prescript{g}{}{(a,b,c)}$. And they are different since $\chi_2(b-a)=-\chi_2(a-b)\neq \chi_2(a-b)$.

Suppose $g$ is a $3$-cycle, e.g., the one sending $(a,b,c)$ to $(b,c,a)$ (the other case is symmetric). Then the signature of $\prescript{g}{}{(a,b,c)}$ is 
$(\chi_2(b-c),\chi_2(c-a),\chi_2(a-b), s(b,c,a))$.
When $\chi_2(a-b),\chi_2(b-c),\chi_2(c-a)$ are not all equal, we have $s(b,c,a)=1$. Otherwise 
\[
s(b,c,a)=\chi_3(b+\omega c+\omega^2 a)=\chi_3(\omega^2)\chi_3(a+\omega b+\omega^2 c)\neq s(a,b,c).
\]
So again the signature of  $\prescript{g}{}{(a,b,c)}$  is determined by and different from that of $(a,b,c)$.

To prove regularity, let $K$ be the subgroup $\{\phi_{u,v}: u\in \ker(\chi_2), v\in\F_q\}$ of $\sym(\F_q)$ as in Example~\ref{exmp_paley}, and let $\Pi'=\{P_1', P_2', P_3'\}$ be the orbit $3$-scheme on $\F_q$ associated with $K$. Then $P_1=P_1'$ and $P_2=P_2'$. We claim that $P_3$ is a coarsening of $P_3'$, i.e., each block $B$ of $P_3$ is a disjoint union of a collection $I$ of blocks in $P_3'$. Assume the claim holds. Then for such a block $B$, an element $y\in \F_q^{(2)}$, and a projection $\pi^3_i$, we have 
\[
|B\cap (\pi^3_i)^{-1}(y)|=\sum_{B'\in I} |B'\cap (\pi^3_i)^{-1}(y)|.
\]
As $\Pi'$ is regular, it follows that $\Pi$ is also regular. So it remains to prove the claim.

The blocks of $P'_3$ are $K$-orbits. So it suffices to show that for $(a,b,c)\in\F_q^{(3)}$ and $\phi_{u,v}\in K$,  the elements $(a,b,c)$ and  $\prescript{\phi_{u,v}}{}{(a,b,c)}=(ua+v,ub+v,uc+v)$ have the same signature. We have 
\[
\chi_2((ua+v)-(ub+v))=\chi_2(u)\chi_2(a-b)=\chi_2(a-b)
\]
since $u\in(\F_q^\times)^2$. Similarly $\chi_2((ub+v)-(uc+v))=\chi_2(b-c)$ and  $\chi_2((uc+v)-(ua+v))=\chi_2(c-a)$. 
Also note that $\F_q^\times$ is contained in the kernel of $\chi_3$,\footnote{Otherwise the intersection of $\F_q^\times$ with the kernel has order $(q-1)/3$, which is impossible as $3$ does not divide $q-1$.} and  hence $\chi_3(u)=1$. Therefore
\begin{align*}
&\chi_3((ua+v)+\omega (ub+v)+ \omega^2 (uc+v))\\
&=\chi_3(u(a+\omega b+ \omega^2 c)+v(1+\omega+\omega^2))\\
&=\chi_3(u(a+\omega b+ \omega^2 c))\\
&=\chi_3(a+\omega b+ \omega^2 c)
\end{align*}
and hence $s(a,b,c)=s(ua+v,ub+v,uc+v)$, as desired.

Homogeneity holds since $P_1=0_{\F_q}$. Next we show that $\Pi$ is  strongly antisymmetric when $q>11$.
To prove this, it suffice to show that the projections $\pi^2_i$ and $\pi^3_i$ are not invertible even restricted to each block. For $\pi^2_i$ this holds when $q>3$, as shown in the proof of Lemma~\ref{lem_anti2ch}.
For $\pi^3_i$ we only need to check that the cardinalities of blocks of $P_3$ are greater than the cardinality $q(q-1)/2$ of blocks of  $P_2$.
Let $(a,b,c)$ be an element of $\F_q^{(3)}$ and let $B$ be the block of $P_3$ containing it. Let $(u,v,w,t)$ be the signature of $B$.
By Lemma~\ref{lem_antiasso}, if $u,v,w$ are not all equal, the cardinality of $B$ is $(q(q-1)/2)((q-3)/4)>q(q-1)/2$. If $u=v=w$, the block $B$ and two other blocks, whose signatures are $(u,v,w,\chi_3(\omega)t)$ and $(u,v,w,\chi_3^2(\omega)t)$ respectively, are permuted by $3$-cycles in $\sym(3)$, and their disjoint union has cardinality $(q(q-1)/2)((q+1)/4)$ by Lemma~\ref{lem_antiasso}. So
\[
|B|=\frac{1}{3}\cdot \frac{q(q-1)}{2}\cdot\frac{q+1}{4}>\frac{q(q-1)}{2}
\]
as desired.
\end{proof}

Unlike Example~\ref{lem_anti2ch}, the $3$-schemes constructed in Example~\ref{exmp_3anti} are not orbit $m$-schemes. In fact, we prove in Theorem~\ref{thm_nonexistence3sch} later that  no strongly antisymmetric homogeneous orbit $m$-schemes on $S$ exist if $|S|>1$ and $m\geq 3$.
It strengthens the result in \citep{IKS09} that no such  $m$-schemes exist for $m\geq 4$.
%We also prove a similar claim for orbit $\mathcal{P}$-schemes, where $\mathcal{P}$ is a lattice a stabilizer of depth $m$. 

For $m\geq 4$, there are no known examples of strongly antisymmetric homogeneous $m$-schemes on $S$ (where $|S|>1$), even for general $m$-schemes.
It is conjectured in \citep{IKS09} that such $m$-schemes do not exist for $m\geq C$ where $C$ is an absolute constant. 
An affirmative solution to this conjecture would imply a polynomial-time deterministic factoring algorithm under GRH. See Theorem~\ref{thm_polyalgconj}.
Currently the best known upper bound for $m$ is $O(\log |S|+1)$ \citep{Evd94, IKS09, Gua09, Aro13}. See Theorem~\ref{thm_symbound2}.

%One way of achieving this is to relate it to the number of solutions to some quadratic equation over finite fields: Take $s_1=s_2=s_3=1$ for example. We may fix $a,b$ satisfying $\chi_2(a-b)=1$, and there are $q(q-1)/2$ such pairs $(a,b)$. Then we count the number of $c\in\F_q-\{a,b\}$ satisfying $\chi_2(b-c)=1$ and $\chi_2(c-a)=1$. Each such $c$ corresponds to four solutions to the equation $X^2+Y^2=-1$ over $\F_q$, given by $X=\pm\sqrt{(b-c)/(a-b)}$, $Y=\pm\sqrt{(c-a)/(a-b)}$.\footnote{The square roots are taken in $\F_q$.}
%And these are all the solutions over $\F_q$.
%On the other hand, it is well known that when $-1$ is a quadratic non-residue of $\F_q$ (as in our case), the number of solutions to $X^2+Y^2=c$ over $\F_q$ for $c\neq 0$ is $q+1$.\footnote{This follows from the fact that the conic $X^2+Y^2=c Z^2$ has $q+1$ rational points over $\F_q$, and since $-1$ is a quadratic non-residue, points ``at the infinity'' are not rational over $\F_q$. (reference needed)}
%So the number of choices of $c$ is $((q+1)-4)/4=(q-3)/4$, and the cardinality of the corresponding block is $q(q-1)(q-3)/8$. The calculation for other values of $s_i$ can be carried out similarly.

\chapter{The \texorpdfstring{$\mathcal{P}$-scheme}{P-scheme} algorithm}\label{chap_alg_prime}

In this chapter, we present a generic deterministic factoring algorithm called the {\em $\mathcal{P}$-scheme algorithm}, based on the notion of $\mathcal{P}$-schemes introduced in Chapter~\ref{chap_pscheme}. 

A  univariate polynomial over a finite field is said to be {\em square-free}\index{square-free} if it has no repeated factors, and {\em completely reducible}\index{completely reducible} over $\F_q$ if it factorizes into linear factors over $\F_q$. 
For simplicity, the algorithm in this chapter assumes that the input polynomial satisfies the following condition:

\begin{cond}\label{cond_spoly}
The input polynomial is defined over a prime field $\F_p$. In addition, it is square-free and completely reducible over $\F_p$.
\end{cond}

This assumption is commonly made in the literature (see, e.g., \citep{Ron88,  Evd94, CH00, IKS09, Aro13, AIKS14}) and is justified by standard reductions \cite{Ber70, Yun76, Knu98}. Specifically, Berlekamp \citep{Ber70} reduced the problem of completely factoring an arbitrary polynomial over a finite field to the problem of finding roots of certain other polynomials in $\F_p$. The latter problem further reduces to the problem of  completely factoring polynomials satisfying Condition~\ref{cond_spoly} by the technique of  {\em square-free factorization}\index{square-free factorization} \citep{Yun76, Knu98}. Alternatively, we develop an algorithm that works for arbitrary polynomials over finite fields in Chapter~\ref{chap_alg_general} without using these reductions.

\subsection*{Overview of the $\mathcal{P}$-scheme algorithm} 

The $\mathcal{P}$-scheme algorithm consists of three parts: 
\begin{enumerate}
\item reducing to the problem of computing an `` idempotent decomposition'' of a certain ring,
\item computing  idempotent decompositions of rings associated with a poset of number fields, 
\item constructing the poset of number fields used in the previous part. 
\end{enumerate}
Now we elaborate on each part.

\paragraph{Reduction to computing an  idempotent decomposition.}
It is well known that computing a factorization of $f$ is equivalent to finding zero divisors of the ring $\F_p[X]/(f(X))$ \citep{Ron88, Evd94, IKS09}. We focus on special zero divisors called {\em idempotent elements} or simply {\em idempotents}\index{idempotent},\footnote{Strictly speaking, we need to exclude the unity of the ring which is the only idempotent element that is not a zero divisor.}  i.e., those elements $x$ satisfying $x^2=x$. Two idempotents $x,y$ are said to be {\em orthogonal}\index{orthogonal idempotents} if $xy=0$. 
It can be shown that the problem of factoring $f$ reduces to decomposing the unity of the ring $\F_p[X]/(f(X))$ into a sum of nonzero mutually orthogonal idempotent elements, called an {\em idempotent decomposition}.

\begin{defi}\label{defi_pou}
An {\em  idempotent decomposition}\index{idempotent decomposition} of a ring $R$ is a set $I$ of nonzero mutually orthogonal idempotent elements of $R$ satisfying $\sum_{x\in I} x=1$. 
\end{defi}

On the other hand, recall that our algorithm uses a lifted polynomial $\tilde{f}(X)\in\Z[X]$ of $f$, as mentioned in the  introduction.  
%We can always pick $\tilde{f}(X)\in\Z[X]$ satisfying $\tilde{f}\bmod p\equiv f$. 
Furthermore, we may assume $\tilde{f}$ is an irreducible lifted polynomial  (see Definition~\ref{defi_liftpoly}) by running the factoring algorithm for rational polynomials \citep{LLL82} to factorize $\tilde{f}$ into the irreducible factors over $\Q$. See Section~\ref{sec_algputtogether} for more discussion. 
%By Gauss Lemma (see \citep[Section~\RN{4}.2]{Lan02}), we may assume each factor $\tilde{f}_i(X)$ lies in $\Z[X]$. Then we reduce the problem to each subproblem of factoring $\tilde{f}_i\bmod p$ over $\F_p$ using the irreducible lifted polynomial $\tilde{f}_i$. }
The polynomial $\tilde{f}$ defines a number field $F:=\Q[X]/(\tilde{f}(X))$. We show that, since $f$ is square-free and completely reducible over $\F_p$, the ring  $\F_p[X]/(f(X))$ is naturally isomorphic to $\bar{\ord}_F:=\ord_F/p\ord_F$, where $\ord_F$ is the {\em ring of integers} of the field $F$.
  %\footnote{This is not necessarily true if $f$ is not square-free, but there still exists a ring homomorphism from $\F_p[X]/(f(X))$ to $\bar{\ord}_F$. We  use this fact to solve the general case in Chapter X.}
Therefore the problem reduces to that of computing an  idempotent decomposition of  the ring $\bar{\ord}_F$.

\paragraph{Computing  idempotent decompositions for a poset of number fields.}
Denote by $L$ the  splitting field of $\tilde{f}$ over $\Q$ and $G$ the Galois group of $\tilde{f}$ over $\Q$, i.e., $G=\gal(L/\Q)$.
%While 
Conceptually, replacing $\F_p[X]/(f(X))$ with $\bar{\ord}_F$ 
%essentially does nothing since the two rings are isomorphic, 
 allows us to use the information provided by  the Galois group $G$. By the work of R{\'o}nyai \citep{Ron92}, a zero divisor $a\neq 0$ (or, in our language, an  idempotent decomposition) of $\bar{\ord}_F$ can be found efficiently if an efficiently computable nontrivial automorphism of the ring $\bar{\ord}_F$ is given. 
The Galois group $G$ 
%(which is an automorphism of the splitting field of $\tilde{f}$) 
naturally provides automorphisms of  $\bar{\ord}_F$, at least when $F$ is Galois over $\Q$. Moreover, these automorphisms can be efficiently computed thanks to the efficient polynomial factoring algorithms for number fields  \citep{Len83, Lan85}.
Using this idea, R{\'o}nyai  \citep{Ron92} gave a polynomial-time factoring algorithm    for the case that $F$ is Galois over $\Q$.

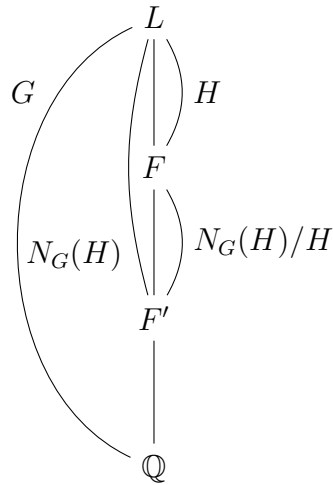
\begin{figure}
\centering
 \begin{tikzpicture}[node distance = 2cm, auto]
 \node (Q) {$\mathbb{Q}$};
 \node (K') [above of=Q] {$F'$};
 \node (K) [above of=K'] {$F$};
 \node (L) [above of=K] {$L$};
 \draw[-] (Q) to node {} (K');
 \draw[-] (K') to node {} (K);
 \draw[-] (K) to node {} (L);
 \draw[-] (K) to [bend right=30, swap] node {$H$} (L);
 \draw[-] (K') to [bend left=15, near start] node {$N_G(H)$} (L);
 \draw[-] (Q) to [bend left=65, near end] node {$G$} (L);
 \draw[-] (K') to [bend right=30, swap] node {$N_G(H)/H$} (K);
 \end{tikzpicture}
\caption{The tower of fields and Galois groups. Denote by $L$ the splitting field of $\tilde{f}$ over $\Q$ and regard $F$  as a subfield of $L$.}\label{fig_galois}
\end{figure}

When $F$ is not Galois over $\Q$, not every automorphism in $G$ restricts to an automorphism of $F$ or $\bar{\ord}_F$. One of our key observations is that $F$ may still admit a nontrivial automorphism group, from which we can compute a partial factorization of $f$. Indeed, we regard $F$ as a subfield of $L$ and let $H$ be the subgroup of $G$ fixing $F$. Then the automorphism group of $F$ is isomorphic to $N_G(H)/H$. 
The corresponding fixed subfield $F'=F^{N_G(H)/H}$ is the smallest subfield of $F$ such that $F/F'$ is Galois.
See Figure~\ref{fig_galois} for an illustration.

In the worst case, we may have $N_G(H)=H$ and then the automorphism group of $F$ is trivial. However,  an extension $K$ of $F$ may still have a nontrivial automorphism group, and hence a nontrivial  idempotent decomposition may be obtained for $\bar{\ord}_K:=\ord_K/p\ord_K$ instead of $\bar{\ord}_F$, where $\ord_K$ is the ring of integers of $K$.
For example, suppose $G$ is the symmetric group $\sym(n)$ permuting the $n$ roots of $\tilde{f}$. We identify $F$ with $\Q(\alpha)$ for some root $\alpha$ of $\tilde{f}$, and then $H$ is the stabilizer $G_\alpha$.  Let $\beta$ be a root of $\tilde{f}$ different from $\alpha$. Then the automorphism group of $K=F(\beta)=\Q(\alpha,\beta)$ is $N_G(G_{\alpha,\beta})/G_{\alpha,\beta}$, which is nontrivial as $N_G(G_{\alpha,\beta})$ contains the permutations swapping $\alpha$ and $\beta$. Another example is the case that $K$ equals the splitting field $L$ of $\tilde{f}$.  In this case, the automorphism group of $K$ is just $G$.

Motivated by the above observation, we design the algorithm so that it computes  idempotent decompositions not only for the number field $F$, but also simultaneously for a poset of subfields of $L$. Moreover, we compute homomorphisms between these fields,  which induce homomorphisms between the rings $\bar{\ord}_K$.
Using these homomorphisms, we show that the  idempotent decompositions can be properly refined, unless some consistency constraints  between them are satisfied. 

The connection with $\mathcal{P}$-schemes is as follows:
by Galois theory, the poset of subfields used by the algorithm corresponds to a poset $\mathcal{P}$ of subgroups of $G$. Suppose a field $K$ in the former poset is associated with a subgroup $H$. It can be shown that an  idempotent decomposition of $\bar{\ord}_K$ corresponds to a partition of the coset space $H\backslash G$. These partitions for various $H\in\mathcal{P}$ altogether form a $\mathcal{P}$-collection. Then the consistency constraints between the  idempotent decompositions are just the defining properties of $\mathcal{P}$-schemes in disguise, i.e. compatibility, regularity, and invariance. In addition, we incorporate in our algorithm R{\'o}nyai's technique \citep{Ron92} as mentioned above as well as its extension by Evdokimov \citep{Evd94}. They are characterized by  antisymmetry and strongly antisymmetry of $\mathcal{P}$-schemes respectively.

The main part of the algorithm has the following structure:  it constructs the rings $\bar{\ord}_K$ and the homomorphisms between them, and then maintains the  idempotent decompositions of these rings and iteratively refines them. Each time it calls a subroutine corresponding to some property of $\mathcal{P}$-schemes in attempt to obtain a refinement. Either the property is already satisfied, or strictly finer  idempotent decompositions are obtained by the subroutine.
The algorithm terminates when the  decompositions cannot be properly refined any more, in which case we are guaranteed to have a strongly antisymmetric $\mathcal{P}$-scheme.
%The running time of the algorithm depends on the total degree of the number fields. We define a complexity measure $c(P)$ called the {\em complexity of the poset $\mathcal{P}$} to capture it. 
This gives the following result.

\begin{thm}[informal]\label{thm_mainalginformal}
Under GRH, there exists a deterministic algorithm that given
%\begin{itemize}
%\item an irreducible polynomial $\tilde{f}\in\Z[X]$ with the Galois group $G$ and the splitting field $L$, such that $f=\tilde{f}\bmod p$ is square-free and completely reducible over $\F_p$, and 
a poset $\mathcal{P}^\sharp$ of subfields of $L$ corresponding to a poset $\mathcal{P}$ of subgroups of $G$,
%\end{itemize}
outputs  idempotent decompositions of $\bar{\ord}_K$ for $K\in\mathcal{P}^\sharp$ corresponding to a strongly antisymmetric $\mathcal{P}$-scheme. The running time is polynomial in the size of the input.
%$\tilde{f}$ and the complexity $c(\mathcal{P})$ of the poset $\mathcal{P}$.
\end{thm}

Suppose $F$ is in the poset $\mathcal{P}^\sharp$, corresponding to a group $H\in\mathcal{P}$. Then in the strongly antisymmetric $\mathcal{P}$-scheme produced by the algorithm, the partition of $H\backslash G$ translates into an  idempotent decomposition of the ring $\bar{\ord}_F$. In particular, it follows from the reduction in the first part of the algorithm that if all strongly antisymmetric $\mathcal{P}$-schemes are discrete (resp. inhomogeneous) on $H$, then we always obtain the complete factorization (resp. a proper factorization) of $f$.
 
\paragraph{Constructing a collection of number fields.} Theorem~\ref{thm_mainalginformal} is a generic result, as we may feed it any poset $\mathcal{P}^\sharp$ of subfields of $L$ and get a strongly antisymmetric $\mathcal{P}$-scheme, where $\mathcal{P}$ is the corresponding poset of subgroups of $G$. To obtain an actual factoring algorithm, we need to construct  such a poset. More precisely, we construct a collection $\mathcal{F}$ of number fields that are representatives of isomorphism classes of those in $\mathcal{P}^\sharp$, i.e., isomorphic fields in $\mathcal{P}^\sharp$ are represented by the same element in $\mathcal{F}$. The posets $\mathcal{P}$ and $\mathcal{P}^\sharp$ are determined once $\mathcal{F}$ is given. 

Let $H$ be the subgroup of $G$ fixing $F$. The collection $\mathcal{F}$ of number fields should satisfy the following two constraints: (1) $\mathcal{F}$ contains the field $F$,
 so that we can convert the partition on $H\backslash G$ in the $\mathcal{P}$-scheme into a factorization of $f$,
 and (2) all strongly antisymmetric $\mathcal{P}$-schemes are discrete (resp. inhomogeneous) on $H$, so that the algorithm always produces the complete factorization (resp. a proper factorization) of $f$. In addition, we want to bound the running time spent in constructing the fields in $\mathcal{P}^\sharp$, which controls the running time of the whole  algorithm. 
 
 We give various settings of $\mathcal{F}$ in which the two constraints above are satisfied. One of them  is to choose $\mathcal{F}=\{F,L\}$, where $L$ is the splitting field of $\tilde{f}$.
 % The running time is polynomial in $[L:\Q]=|G|$, $\log p$, and the length of $\tilde{f}$.
In another setting, we choose $\mathcal{F}$ so that $\mathcal{P}$ is a system of stabilizers of depth $m$ for sufficiently large $m\in\N^+$. They lead to factoring algorithms with various running time. 

For simplicity, we only state the results that $\mathcal{F}$ can be constructed in certain amount of time (see Section~\ref{sec_compnumflds}). The proofs are deferred  to Chapter~\ref{chap_constructnum}, where we give a more comprehensive investigation on the problem of constructing number fields.

\paragraph{Summary.}
The actual factoring algorithm combines the three parts above  in the opposite order: we first construct a collection $\mathcal{F}$ of number fields which determines the posets $\mathcal{P}$ and $\mathcal{P}^\sharp$. Then we run the algorithm in Theorem~\ref{thm_mainalginformal} to obtain a collection of  idempotent decompositions corresponding to a strongly antisymmetric $\mathcal{P}$-scheme. Finally we  extract a factorization of $f$ from the  idempotent decomposition of $\bar{\ord}_F$. This yields the main result of this chapter:

\begin{thm}[informal]\label{thm_algmain2informal}
Suppose
there exists a  deterministic  algorithm that given a   polynomial $g(X)\in\Z[X]$ irreducible over $\Q$, constructs in time $T(g)$ a collection $\mathcal{F}$ of subfields of the splitting field $L$ of $g$ over $\Q$ such that
\begin{itemize}
\item $F=\Q[X]/(g(X))$ is in $\mathcal{F}$, and
\item  all strongly antisymmetric $\mathcal{P}$-schemes are discrete (resp. inhomogeneous) on $\gal(L/F)\in \mathcal{P}$, where $\mathcal{P}$ is the subgroup system associated with $\mathcal{F}$.
\end{itemize}
Then under GRH, there exists a deterministic algorithm that given $f(X)\in\F_p[X]$ satisfying Condition~\ref{cond_spoly} and an irreducible lifted polynomial $\tilde{f}(X)\in\Z[X]$ of $f$, outputs the complete factorization (resp. a proper factorization) of $f$ over $\F_p$ in time polynomial in $T(\tilde{f})$ and the size of the input. 
\end{thm}

We show that many results achieved by known factoring algorithms \citep{Hua91-2, Hua91, Ron88, Ron92, Evd94, IKS09} can be derived from Theorem~\ref{thm_algmain2informal}. Thus the $\mathcal{P}$-scheme algorithm provides a unifying approach to  polynomial factoring over finite fields.

\paragraph{Outline of the chapter.}

Notations and mathematical preliminaries are given in Section~\ref{sec_algprelim}, and algorithmic preliminaries are given in Section~\ref{sec_algpre}.  We reduce the problem of factoring $f$ to that of computing an  idempotent decomposition of $\bar{\ord}_F$ in Section~\ref{sec_algreduction}. In Section~\ref{sec_algmain}, we  give the main body of the algorithm that computes  idempotent decompositions corresponding to a strongly antisymmetric $\mathcal{P}$-scheme, and use it to prove Theorem~\ref{thm_mainalginformal}. The next three sections (Section~\ref{sec_algci}, \ref{sec_algr} and \ref{sec_algsa}) describe three subroutines used by this algorithm. In Section~\ref{sec_compnumflds} we state some results on constructing a collection  $\mathcal{F}$ of number fields using $\tilde{f}$. Finally, in Section~\ref{sec_algputtogether}, we combine the results developed in the previous sections to prove Theorem~\ref{thm_algmain2informal}, and use it to derive the main results in \citep{Hua91-2, Hua91, Ron88, Ron92, Evd94, IKS09}.

\section{Preliminaries}\label{sec_algprelim}

We first review basic notations and facts in algebra. They are standard and can be found in various textbooks, e.g., \citep{Lan02, AM69, Mar77}. Then we discuss {\em splitting of prime ideals} in number field extensions. Finally, for the certain rings $\bar{\ord}_K$, we establish a one-to-one correspondence between their  idempotent decompositions and the partitions of certain right coset spaces.

All rings are assumed to be commutative rings with unity.

\paragraph{Ideals.}
 Recall that a subset $I$ of a ring $R$ is an {\em ideal}\index{ideal} of $R$ if (1) $I$ is a subgroup of the underlying additive abelian group of $R$, and (2) $R\cdot I=\{ra: r\in R, a\in  I\}\subseteq I$. 
For $x\in R$, denote by $(x)$, $xR$ or $Rx$ the ideal $\{rx:r\in R\}$ of $R$ generated by $x$.
\nomenclature[b1a]{$(x)$, $xR$, $Rx$}{ideal of a ring $R$ generated by $x$}
 
 An ideal of $R$ is {\em proper}\index{proper!ideal} if it is a proper subset of $R$. 
 Let $I$ be a proper ideal of $I$.
 We say $I$ is {\em prime}\index{prime ideal} if $I\neq R$ and $ab\in I$ implies $a\in I$ or $b\in I$ for any $a,b\in R$. 
And $I$ is {\em maximal}\index{maximal ideal} if $I\neq R$ and there exists no ideal $I'$ of $R$ satisfying $I\subsetneq I'\subsetneq R$.
A proper ideal $I$ is prime (resp. maximal) iff the quotient ring $R/I$ is an integral domain (resp. a field). 
In particular, maximal ideals are prime. 
For an ideal $I_0$ of $R$, the map $I\mapsto I/I_0$ is a one-to-one correspondence between the ideals of $R$ containing $I_0$ and the ideals of $R/I_0$, and it preserves primality and maximality.

If $\mathfrak{m}_1,\dots,\mathfrak{m}_k$ and $\mathfrak{m}$ are maximal ideals of $R$ and $\bigcap_{i=1}^k \mathfrak{m}_i\subseteq \mathfrak{m}$, then $\mathfrak{m}=\mathfrak{m}_i$ for some $i\in [k]$.\footnote{See \citep[Proposition~1.11]{AM69} for a more general statement for prime ideals.} 
In particular, if $\bigcap_{i=1}^k\mathfrak{m}_i=0$, then $\mathfrak{m}_1,\dots,\mathfrak{m}_k$ are the only maximal ideals of $R$. 

Two ideals $I,I'$ of $R$ are {\em coprime}\index{coprime ideals} if $I+I'=R$.
In particular, distinct maximal ideals are always coprime. For pairwise coprime ideals $I_1,\dots,I_k$, it holds that $\bigcap_{i=1}^k I_i=\prod_{i=1}^k I_i$.
We also have
\begin{lem}[Chinese remainder theorem]\index{Chinese remainder theorem} 
Suppose $I_1,\dots, I_k$ are pairwise coprime ideals of $R$. Then the ring homomorphism
\[
\phi: R/\bigcap_{i=1}^k I_i \to \prod_{i=1}^k R/I_i
\]
sending $x+\bigcap_{i=1}^k I_i$ to $(x+I_1,\dots,x+I_k)$ is an isomorphism.
\end{lem}

\paragraph{Semisimple rings.}

A (commutative) ring is {\em semisimple}\index{semisimple!ring} if it is isomorphic to a finite product of fields. 
The following lemma provides a characterization of semisimple rings.

\begin{lem}\label{lem_semisimple}
A ring $R$ is semisimple iff it has finitely many maximal ideals $\mathfrak{m}_1,\linebreak[0]\dots,\linebreak[0]\mathfrak{m}_k$ and $\bigcap_{i=1}^k \mathfrak{m}_i=0$, in which case $R$ is isomorphic to $\prod_{i=1}^k R/\mathfrak{m}_i$ via the map $x\mapsto (x+\mathfrak{m}_1,\dots,x+\mathfrak{m}_k)$.
\end{lem}
\begin{proof}
Suppose $R\cong \prod_{i=1}^k F_i$ is semisimple where each $F_i$ is a field. For $i\in [k]$, let $\pi_i:R\to F_i$ be the $i$th projection and $\mathfrak{m}_i$ be its kernel. Then $R/\mathfrak{m}_i\cong F_i$ and hence each $\mathfrak{m}_i$ is a maximal ideal of $R$. Moreover we have $\bigcap_{i=1}^k \mathfrak{m}_i=0$ and hence $\mathfrak{m}_1,\dots,\mathfrak{m}_k$ are the only maximal ideals. 
Conversely, suppose $R$ has finitely many maximal ideals $\mathfrak{m}_1,\dots,\mathfrak{m}_k$ and $\bigcap_{i=1}^k \mathfrak{m}_i=0$. Then by the Chinese remainder theorem, the map $R\to \prod_{i=1}^k R/\mathfrak{m}_i$ sending 
 $x\in R$ to $(x+\mathfrak{m}_1,\dots,x+\mathfrak{m}_k)$ is a ring isomorphism.
Each direct factor $R/\mathfrak{m}_i$ is a field, and hence $R$ is semisimple.
\end{proof}

The semisimple rings considered in this chapter are all  {\em semisimple $\F_p$-algebras}\index{semisimple!algebra}, i.e. semisimple rings that are also $\F_p$-algebras. 

\paragraph{Idempotent elements.} An element $x$ of a ring is an {\em idempotent element} (or just an {\em idempotent})\index{idempotent} if $x^2=x$. Two idempotents $x,y$ are {\em orthogonal}\index{orthogonal idempotents} if $xy=0$.
% If $x,y,z$ are idempotents and $x=y+z$, then $y$ and $z$ are orthogonal. 
A nonzero idempotent $x$ is {\em primitive}\index{primitive!idempotent} if it cannot be written as a sum of two nonzero orthogonal idempotents. As already stated in Definition~\ref{defi_pou}, an {\em idempotent decomposition}\index{idempotent decomposition} of a ring $R$ is a set $I$ of nonzero mutually orthogonal idempotents of $R$ satisfying $\sum_{x\in I} x=1$. We say such an idempotent decomposition is {\em proper}\index{proper!idempotent decomposition} if $|I|>1$ and {\em complete}\index{complete!idempotent decomposition} if all idempotents in $I$ are primitive. 

\begin{lem}\label{lem_idealidem}
Let $R$ be a semisimple ring. For every maximal ideal $\mathfrak{m}$ of $R$, there exists a unique primitive idempotent $\delta_{\mathfrak{m}}\in R$ satisfying $\delta_{\mathfrak{m}}\equiv 1 \pmod{\mathfrak{m}}$ and   $\delta_{\mathfrak{m}}\equiv 0 \pmod{\mathfrak{m}'}$ for all maximal ideals $\mathfrak{m}'\neq\mathfrak{m}$. 
Two elements $\delta_{\mathfrak{m}}$ and $\delta_{\mathfrak{m}'}$ are orthogonal iff $\mathfrak{m}\neq \mathfrak{m}'$.
Furthermore
\begin{itemize}
\item the map $\mathfrak{m}\mapsto \delta_{\mathfrak{m}}$ is a one-to-one correspondence between  the maximal ideals of $R$ and the primitive idempotents of $R$, and
\item the map $B\mapsto \sum_{\mathfrak{m}\in B} \delta_{\mathfrak{m}}$ is a one-to-one correspondence between the sets of maximal ideals of $R$ and the idempotents of $R$.
\end{itemize}
\end{lem}

\begin{proof}
This is clear from the isomorphism $R\cong \prod_{\mathfrak{m}\in S} R/\mathfrak{m}$, where $S$ denotes the set of all the maximal ideals of $R$.
\end{proof}

We also need the following lemma.
\begin{lem}\label{lem_quotienthom}
Suppose $\phi: R'\to R$ is a ring homomorphism between two semisimple rings $R,R'$. Let $\delta,\delta'$ be idempotents of $R$ and $R'$ respectively satisfying $\phi(\delta')\delta=\delta$. Then $\phi$ induces a ring homomorphism from $R'/(1-\delta')$ to $R/(1-\delta)$ sending $x+(1-\delta')$ to $\phi(x)+(1-\delta)$ for $x\in R'$.
\end{lem}
\begin{proof}
 It suffices to show that $\phi(1-\delta')$ is in the ideal $(1-\delta)$ of $R$, which holds since
$(1-\phi(\delta'))(1-\delta)=1-\phi(\delta')-\delta+\phi(\delta')\delta=1-\phi(\delta')=\phi(1-\delta').$
\end{proof}

\paragraph{Finitely generated modules and free modules.} A subset $S$ of an $R$-module $M$ {\em generates} $M$ if $\sum_{x\in S} Rx=M$.
And $M$ is {\em finitely generated}\index{finitely generated module} if it is generated by a finite subset $S$. A {\em basis}\index{basis of a free module} of $M$ over $R$, or an {\em $R$-basis} of $M$, is a subset $S\subseteq M$ generating $M$ for which the sum $M=\sum_{x\in S} Rx$ is a direct sum.
We say $M$ is {\em free}\index{free module} (over $R$) if it admits an $R$-basis. The {\em rank}\index{rank of a module} of a finitely generated free module over $R$ is the cardinality of any $R$-basis of it, which is finite and independent of the choice of the basis.

\paragraph{Number fields.} Elements in the algebraic closure $\bar{\Q}$ of $\Q$ are called {\em algebraic numbers}\index{algebraic number}.
An algebraic number is {\em integral}\index{integral} or an {\em algebraic integer}\index{algebraic integer} if it is a root of a monic polynomial in $\Z[X]$.
The set of algebraic integers is a subring of $\bar{\Q}$, denoted by $\mathbb{A}$. 
A {\em number field}\index{number field} is a finite degree field extension of $\Q$ in $\bar{\Q}$. 
For a number field $K$, the subring $\ord_K:=\mathbb{A}\cap  K$ is called the {\em ring of integers}\index{ring of integers} of $K$. It is embedded in the $\Q$-vector space $K$ as a lattice of rank $[K:\Q]$.
\nomenclature[b1b]{$\ord_K$}{ring of integers of a number field $K$}

Suppose $K/K_0$ is a number field extension. We say $\alpha\in K$ is a {\em primitive element}\index{primitive element} of $K$ over $K_0$ if $K=K_0(\alpha)$.  Primitive elements always exist for any number field extension by the {\em primitive element theorem}.\index{primitive element theorem}

\paragraph{Galois theory.}\index{Galois theory} Let $K/K_0$ be a field extension.
The set of automorphisms of $K$ fixing $K_0$ is a group, called the {\em automorphism group}\index{automorphism group!of a field extension} of $K$ over $K_0$, and is denoted by $\aut(K/K_0)$. We say $K$ is {\em Galois}\index{Galois extension} over $K_0$ if $|\aut(K/K_0)|=[K:K_0]$, in which case $\aut(K/K_0)$ is also called the {\em Galois group}\index{Galois group} of $K$ over $K_0$ and denoted by $\gal(K/K_0)$.
\nomenclature[b1c]{$\aut(K/K_0)$}{automorphism group of a field extension $K/K_0$}
\nomenclature[b1d]{$\gal(K/K_0)$}{Galois group of a Galois extension $K/K_0$}
\nomenclature[b1e]{$\gal(f/K_0)$}{Galois group of $K/K_0$ where $K$ is the splitting field of $f$ over $K_0$}

\begin{thm}[fundamental theorem of Galois theory]\label{thm_gal}\index{fundamental theorem of Galois theory}
Let $K/K_0$ be a Galois extension. Then for any intermediate field $K_0\subseteq E\subseteq K$, the extension $K/E$ is also a Galois extension. Furthermore, the map $E\mapsto \gal(K/E)$ is an inclusion-reversing one-to-one correspondence between the poset of intermediate fields $K_0\subseteq E\subseteq K$ and the poset of subgroups of $\gal(K/K_0)$, with the inverse map $H\mapsto K^H$.
%\begin{itemize}
%\item the map $E\mapsto \gal(K/E)$ is an inclusion-reversing correspondence between the poset of intermediate fields $K_0\subseteq E\subseteq K$ and the poset of subgroups of $\gal(K/K_0)$, with the inverse map $H\mapsto K^E$, and
%\item an intermediate field $E$ between $K$ and $K_0$ is Galois over $K_0$ iff $\gal(K/E)$ is a normal subgroup of $\gal(K/K_0)$, in which case $\gal(E/K_0)\cong \gal(K/K_0)/\gal(K/E)$.
%\end{itemize}
\end{thm}

Given a Galois extension $K/K_0$, two subfields $E, E'$ between $K$ and $K_0$ are {\em conjugate}\index{conjugate!subfield} over $K_0$ 
%(or just conjugate, if $K_0=\Q$) 
if there exists an isomorphism $\tau_0: E\to E'$ fixing $K_0$. Such an isomorphism always extends to an automorphism $\tau\in\gal(K/K_0)$ of $K$. The corresponding Galois groups $\gal(K/E)$ and $\gal(K/E')$  satisfy $\gal(K/E')=\tau\gal(K/E)\tau^{-1}$. So conjugate subfields of $K$ over $K_0$ correspond to conjugate subgroups in $\gal(K/K_0)$.

Now we restrict to number field extensions. Let $K/K_0$ be a number field extension. There exists a unique minimal number field that contains $K$ and is Galois over $K_0$, called the {\em Galois closure}\index{Galois closure} of $K/K_0$. For a polynomial $f(X)\in K_0[X]$ with roots $\alpha_1,\dots,\alpha_k\in\bar{\Q}$, the number field $K'=K_0(\alpha_1,\dots,\alpha_k)$ is called the {\em splitting field}\index{splitting field} of $f$ over $K_0$ and is Galois over $K_0$. We also write $\gal(f/K_0)$ for the corresponding Galois group $\gal(K'/K_0)$, called the Galois group of $f$ over $K_0$.
If $f$ is the minimal polynomial of a primitive element of $K$ over $K_0$, the splitting field of $f$ over $K_0$ is exactly the Galois closure of $K/K_0$.

Suppose $K/K_0$ is a Galois extension with the Galois group $G$. If $x\in K$ is an algebraic integer, so is $\prescript{g}{}{x}$ for any $g\in G$ since $\Z\subseteq K_0$ is fixed by $G$.
So the action of $G$ on $K$ restricts to an action on $\ord_K$. 

\paragraph{Splitting of prime ideals.}\index{splitting of prime ideals}
The ring of integers of a number field is an example of a {\em Dedekind domain}\index{Dedekind domain} \citep{AM69, Mar77}. An ideal of a Dedekind domain is a nonzero prime ideal iff it is a maximal ideal, and hence these two notions are  interchangeable. By convention, we use the notion of (nonzero) prime ideals instead of maximal ideals.

Let $K$ be a number field. It follows from the theory of Dedekind domains \citep{Mar77} that 
the ideal $p\ord_K$ of $\ord_K$ splits uniquely (up to the ordering) into a product
% and intersection
 of prime ideals of $\ord_K$:
\[
p\ord_K=\prod_{i=1}^k \mathfrak{P}_i.
%=\bigcap_{i=1}^k \mathfrak{P}_i.
\]
For $i\in [k]$, the quotient ring $\ord_K/\mathfrak{P}_i$ is a finite field extension of degree $d_i\in\N^+$ over $\F_p$, and $\sum_{i=1}^k d_i=[K:\Q]$.
We say $\mathfrak{P}_1,\dots,\mathfrak{P}_k$ are the prime ideals of $\ord_K$ {\em lying over}\index{lying over} $p$.
If $\mathfrak{P}_1\dots,\mathfrak{P}_k$ are distinct and $\ord_K/\mathfrak{P}_i\cong \F_p$ for all $i\in [k]$ (and hence $k=[K:\Q]$), we say $p$ {\em splits completely}\index{splitting of prime ideals!complete splitting} in $K$. 
It is known that if $p$ splits completely in $K$, then it also splits completely in any subfield of the Galois closure of $K/\Q$. See, e.g., \citep[Chapter~4]{Mar77}.
%It is easy to show that for a Galois extension $K/K_0$, the action of  $\gal(K/K_0)$ on $\ord_K$ induces an action on the set of nonzero prime ideals of $\ord_K$.
We also need the following result that identifies the set of prime ideals lying over $p$ with a right coset space in the case that $p$ splits completely in a Galois extension containing $K$.

\begin{thm}\label{thm_split}
Let $L$ be a Galois extension of $\Q$  such that $p$ splits completely in $L$, and let $G=\gal(L/\Q)$. Fix a prime ideal $\mathfrak{Q}_0$ of $\ord_L$ lying over $p$. For any subgroup $H\subseteq G$ and the corresponding fixed subfield $K=L^H$, the map $Hg\mapsto \prescript{g}{}{\mathfrak{Q}_0}\cap \ord_K$ is a one-to-one correspondence between the right cosets in $H\backslash G$ and the prime ideals of $\ord_K$ lying over $p$.\footnote{Note that this map is well defined: 
for another representative $hg\in G$ of $Hg$ where $h\in H$, we have  
$
\prescript{hg}{}{\mathfrak{Q}_0}\cap \ord_K=\prescript{h}{}{(\prescript{g}{}{\mathfrak{Q}_0}\cap \ord_K)}=\prescript{g}{}{\mathfrak{Q}_0}\cap \ord_K
$
since $\ord_K$  is fixed by $H$.}
\end{thm}
See, e.g., \citep[Theorem~33]{Mar77}. As the prime ideals of $\ord_K$ lying over $p$ are exactly those containing $p\ord_K$, we get the following correspondence by passing to the quotient ring $\bar{\ord}_K:=\ord_K/p\ord_K$.
\nomenclature[b1f]{$\bar{\ord}_K$}{quotient ring $\ord_K/p\ord_K$}

\begin{cor}\label{cor_idealcoset}
Let $L$, $G$, $\mathfrak{Q}_0$ be as in Theorem~\ref{thm_split}. For any subgroup $H\subseteq G$ and the corresponding fixed subfield $K=L^H$, the map $Hg\mapsto (\prescript{g}{}{\mathfrak{Q}_0}\cap \ord_K)/p\ord_K$ is a one-to-one correspondence between the right cosets in $H\backslash G$ and the prime (and maximal) ideals of $\bar{\ord}_K$.
\end{cor}

\paragraph{Idempotent decompositions vs. partitions of a right coset space.}

Suppose $p$ splits completely into a product of prime ideals $\mathfrak{P}_1,\dots,\mathfrak{P}_k$  in a number field $K$.
Then $\mathfrak{P}_1/p\ord_K,\dots,\mathfrak{P}_k/p\ord_K$ are the prime (and maximal) ideals of $\bar{\ord}_K$. As the intersection of these ideals equals $p\ord_K/p\ord_K=0$, the ring $\bar{\ord}_K$ is semisimple by Lemma~\ref{lem_semisimple}.
The prime ideals $\mathfrak{P}_i/p\ord_K$ correspond to the primitive idempotents of  $\bar{\ord}_K$  by Lemma~\ref{lem_idealidem} and also to the cosets in a right coset space by Corollary~\ref{cor_idealcoset}. We combine them and establish a correspondence between the idempotent decompositions of $\bar{\ord}_K$ and the partitions of a certain right coset space.

For a number field extension $L/K$, the inclusion $\ord_{K}\hookrightarrow \ord_{L}$ induces a map 
\[
i_{K,L}:\bar{\ord}_{K}\to \bar{\ord}_{L}
\]
with the kernel $(p\ord_L\cap \ord_{K})/p\ord_{K}$. As $p\ord_L\cap \ord_{K}=p\ord_{K}$,\footnote{To see this, note that if $x\in p\ord_L\cap \ord_{K}$, then $x/p\in \ord_L\cap K=\ord_{K}$.}
 this map is injective, which identifies  $\bar{\ord}_{K}$ with a subring of  $\bar{\ord}_{L}$. Also note that if $L/\Q$ is a Galois extension with the Galois group $G$, the action of $G$ on $\ord_L$ induces an action on $\bar{\ord}_L$ and permutes the maximal ideals of $\bar{\ord}_L$. These observations are used in Definition~\ref{defi_partitioncor} below.
\nomenclature[b1g]{$i_{K,L}$}{inclusion $\bar{\ord}_{K}\hookrightarrow \bar{\ord}_{L}$ in Chapter~\ref{chap_alg_prime}, or  $R_{K}\hookrightarrow R_{L}$ in Chapter~\ref{chap_alg_general}} 
 
Fix the following notations: let $L$ be a Galois extension of $\Q$ with $\gal(L/\Q)=G$ and suppose $p$ splits completely in $L$.
For a nonzero prime ideal $\mathfrak{Q}$ of $\ord_L$ lying over $p$, define $\bar{\mathfrak{Q}}:=\mathfrak{Q}/p\ord_L$ which is a prime (and hence maximal) ideal of $\bar{\ord}_L$, and let $\delta_{\bar{\mathfrak{Q}}}$ be the primitive idempotent of $\bar{\ord}_L$ satisfying $\delta_{\bar{\mathfrak{Q}}}\equiv 1\pmod{\bar{\mathfrak{Q}}}$ and $\delta_{\bar{\mathfrak{Q}}}\equiv 0\pmod{\bar{\mathfrak{Q}}'}$ for all maximal ideals $\bar{\mathfrak{Q}}'\neq \bar{\mathfrak{Q}}$ of $\bar{\ord}_L$ (cf. Lemma~\ref{lem_idealidem}). 
%Finally, fix a prime ideal $\mathfrak{Q}_0$ of $\ord_L$ lying over $p$.

\begin{defi}\label{defi_partitioncor}
 Suppose $H$ is a subgroup of $G$ and $K=L^H$.  Fix a prime ideal $\mathfrak{Q}_0$ of $\ord_L$ lying over $p$. Then
\begin{itemize}
\item for an idempotent decomposition $I$ of $\bar{\ord}_K$, define $P(I)$ to be the partition of $H\backslash G$ such that $Hg,Hg'\in H\backslash G$ are in the same block iff $\prescript{g^{-1}}{}{(i_{K,L}(\delta))}\equiv \prescript{g'^{-1}}{}{(i_{K,L}(\delta))}\pmod{\bar{\mathfrak{Q}}_0}$ holds for all $\delta\in I$, and
%\footnote{Here we regard $\delta\in I\subseteq\bar{\ord}_K$ as an element of $\bar{\ord}_L$ via the natural inclusion $\bar{\ord}_K\hookrightarrow \bar{\ord}_L$. }
\item for a partition $P$ of $H\backslash G$, define $I(P)$ to be the idempotent decomposition of $\bar{\ord}_K$ consisting of  the idempotents $\delta_B:=i_{K,L}^{-1}\left(\sum_{g\in G: Hg\in B}\prescript{g}{}{\delta_{\bar{\mathfrak{Q}}_0}}\right)$, where  $B$ ranges over the blocks in $P$.\footnote{We show in the proof of Lemma~\ref{lem_pandi} that $\sum_{g\in G: Hg\in B}\prescript{g}{}{\delta_{\bar{\mathfrak{Q}}_0}}$ does lie in the image of $i_{K,L}$, and hence $\delta_B$ is well defined.}
\end{itemize}
\end{defi}
\nomenclature[b1h]{$P(I)$}{See Definition~\ref{defi_partitioncor} and Definition~\ref{defi_partitioncorg}}
\nomenclature[b1i]{$I(P)$}{See Definition~\ref{defi_partitioncor} and Definition~\ref{defi_partitioncorg}}
\nomenclature[b1j]{$\delta_B$}{See Definition~\ref{defi_partitioncor} and Definition~\ref{defi_partitioncorg}}
\nomenclature[b1k]{$B_\delta$}{See Lemma~\ref{lem_pandi} and Lemma~\ref{lem_pandig}}

We have the following lemma, whose proof is routine and can be found  in Appendix~\ref{chap_omitted}. 
\begin{restatable}{lem}{lempi}\label{lem_pandi}
The partitions $P(I)$ and the idempotent decompositions $I(P)$ are well defined. 
And for any idempotent decomposition $I$  of $\bar{\ord}_K$, the idempotents $\delta\in I$ correspond one-to-one to  the blocks of $P(I)$ via the map
$\delta\mapsto B_\delta:=\{Hg\in H\backslash G: \prescript{g^{-1}}{}{(i_{K,L}(\delta))}\equiv 1\pmod{\bar{\mathfrak{Q}}_0}\}$ with the inverse map $B\mapsto \delta_B$.
\end{restatable}

Now we are ready to  establish the following correspondence.

\begin{lem}\label{lem_picorres}
The map $I\mapsto P(I)$ is a one-to-one correspondence between the idempotent decompositions of $\bar{\ord}_K$ and the partitions of $H\backslash G$, with the inverse map $P\mapsto I(P)$. 
\end{lem}

\begin{proof}
Note $I(P)=\{\delta_B: B\in P\}$ by definition and $P(I)=\{B_\delta: \delta\in I\}$ by Lemma~\ref{lem_pandi}. So $I=I(P(I))$ by Lemma~\ref{lem_pandi}.
Also note the map $B\mapsto \delta_B$ is injective, and hence the map $P\mapsto I(P)$ is also injective. So $P=I(P(I))$.
\end{proof}

\section{Algorithmic preliminaries}\label{sec_algpre}

In this section, we present some basic procedures used in the factoring algorithm, mostly related to number fields. Standard references include \citep{Len92, Coh13}.

Let $A$ be an $R$-algebra that is a free $R$-module of finite rank. In the factoring algorithm, we represent such an algebra by maintaining an $R$-basis $B=\{b_1,\dots,b_d\}$ of it.
The {\em structure constants}\index{structure constants} of $A$ in the basis $B$ are the constants $c_{ijk}\in R$ defined by $b_ib_j=\sum_{k=1}^d c_{ijk} b_k$.  Given these structure constants, arithmetic operations of $A$  can be performed in polynomial time, provided that the those of $R$ can also be performed in polynomial time.
In the discussion below, we use the phrase ``computing $A$'' for the task of computing the structure constants of $A$ in the $R$-basis $B$ associated with $A$.
And by ``computing $a$'' for $a\in A$ we mean computing the constants $c_i\in R$ satisfying $a=\sum_{i=1}^d c_i b_i$.
The interesting cases of $R$ to us are $\Z$, $\Q$, and $\F_p$.

Now let $R'$ be an $R$-algebra and let  $A'$ be an $R'$-algebra that is a free $R'$-module of finite rank.
Let $\phi: A\to A'$ be an $R$-linear map.
We use the phrase  ``computing $\phi$'' for the task of computing $\phi(b_i)\in A'$ for all $b_i\in B$, in terms of the coefficients of $\phi(b_i)$ in the $R'$-basis $B'$ associated with $A'$. The interesting cases to us are (1) $R=R'\in\{\Z,\Q,\F_p\}$, (2) $R=\Z$, $R'=\Q$ and $\phi$ is an inclusion that embeds a lattice in a vector space over $\Q$, and $(3)$ $R=\Z$, $R'=\Z/p\Z\cong \F_p$, and $\phi$ is a quotient map from a lattice to a vector space over $\F_p$.

The {\em size}\index{size} of an object used in the algorithm is the number of bits used to encode this object.

\paragraph{Encoding a number field.}

Let $K$ be a number field of degree $d\in\N^+$ over $\Q$.
We encode  $K$ using a primitive element $\alpha\in K$ over $\Q$, or more precisely, the minimal polynomial $g(X)\in\Q[X]$ of $\alpha$ over $\Q$. Given $g(X)$, we compute $\Q[X]/(g(X))$ in the standard $\Q$-basis $\{1+(g(X)),X+(g(X)),\dots,X^{d-1}+(g(X))\}$ and use it to represent $K$. This is justified by the isomorphism $\Q[X]/(g(X))\cong K$ sending $X+(g(X))$ to $\alpha$. 
%More specfically, we compute from $g(X)$ the {\em structure constants} of $\Q[X]/(g(X))$  in the standard $\Q$-basis $1,X,\dots,X^{d-1}$. Given these structure constants, arithmetic operations of $K$ can be performed in polynomial time in the length of $g(X)$.

%We say $K$ is {\em represented by $\alpha$} if it is represented this way. The length of this representation depends on $\alpha$. %Following \cite{WR76}, we bound the length in terms of the {\em size} of $\alpha$.

\paragraph{Computing $\bar{\ord}_K$.} Given $K$ and a prime number $p$, we want to compute the $\F_p$-algebra $\bar{\ord}_K=\ord_K/p\ord_K$.
It is natural to first compute the ring of integers $\ord_K$ and then pass to the quotient ring $\bar{\ord}_K$. Unfortunately, computing a $\Z$-basis of $\ord_K$ in $K$ is in general as hard as finding the largest square factor of a given integer \citep{Chi89, Len92}. We overcome the difficulty by working with a subring $\ord'_K\subseteq \ord_K$ instead of $\ord_K$ such that $[\ord_K: \ord'_K]$ is finite and coprime to $p$. Such a subring is called a {\em $p$-maximal order}\index{pmaximalorder@$p$-maximal order} of $K$, which can be efficiently computed:
\nomenclature[b1l]{$\ord'_K$}{a $p$-maximal order of $K$}

\begin{thm}\label{thm_zassenhaus}
There exists a polynomial-time algorithm that given $K$ and $p$, computes a $p$-maximal order $\ord'_K$ of $K$ together with the inclusion $\ord'_K\hookrightarrow K$.
\end{thm}

See, e.g., \citep[Chapter~6]{Coh13}. We may use $\ord_K'$ in place of $\ord_K$ thanks to the following lemma.
\begin{lem}\label{lem_ordisom}
For a $p$-maximal order $\ord'_K$ of $K$, the ring homomorphism $\ord'_K/p\ord'_K\to \ord_K/p\ord_K=\bar{\ord}_K$ induced from the inclusion $\ord'_K\hookrightarrow \ord_K$  is an  isomorphism.
\end{lem}
\begin{proof}
To show surjectivity, it suffices to show that $\ord'_K$ and $p\ord_K$ span $\ord_K$ over $\Z$. Note that $n_1:=[\ord_K: \ord'_K]$ is coprime to $p$ and $n_2:=[\ord_K:p\ord_K]$ is a power of $p$. The index of the lattice spanned by $\ord'_K$ and $p\ord_K$ in $\ord_K$ divides both $n_1$ and $n_2$ and hence equals one, as desired. 

On the other hand, note that $\ord_K$ and $\ord'_K$ are both lattices of rank $[K:\Q]$.
So $\bar{\ord}_K$ and $\ord'_K/p\ord'_K$ are both vector spaces of dimension $[K:\Q]$ over $\F_p$.
Therefore the map $\ord'_K/p\ord'_K\to \bar{\ord}_K$ is an isomorphism.
\end{proof}

This provides a method of computing the $\F_p$-algebra $\bar{\ord}_K$:
\begin{lem}\label{lem_residuering}
There exists a polynomial-time algorithm $\mathtt{ComputeQuotientRing}$ that given $K$ and $p$, computes the quotient ring $\bar{\ord}_K$,  a $p$-maximal order $\ord_K'$, the inclusion $\ord'_K\hookrightarrow K$, and the quotient map $\pi: \ord_K'\to \bar{\ord}_K$ sending $x\in  \ord_K'$ to $x+p\ord_K$.
\end{lem}
\begin{proof}
Compute  $\ord_K'$ and the inclusion $\ord'_K\hookrightarrow K$ using Theorem~\ref{thm_zassenhaus}. In particular the structure constants $c_{ijk}\in\Z$ of   $\ord_K'$  in some $\Z$-basis $\{b_1,\dots, b_d\}$ are computed, where  $d=[K:\Q]$. The structure constants of $\ord_K'/p\ord_K'$ in the $\F_p$-basis $\{b_1+p\ord_K',\dots, b_d+p\ord_K'\}$ are simply $c_{ijk}\bmod p$. By Lemma~\ref{lem_ordisom}, they are also the structure constants of $\bar{\ord}_K$ in the $\F_p$-basis $\{b_1+p\ord_K,\dots, b_d+p\ord_K\}$. The map $\pi$ is specified by the data $\pi(b_i)=b_i+p\ord_K$.
\end{proof}

Note that in addition to  $\bar{\ord}_K$, we also compute the auxiliary data of $\ord_K'$ and the maps from $\ord_K'$ to $\bar{\ord}_K$ and $K$. They are used  for the algorithms in Lemma~\ref{lem_computeresidue} and Lemma~\ref{lem_ringhom} below.

\paragraph{Computing the residue of an algebraic integer modulo $p$.} We need an algorithm computing the image of an algebraic integer $\alpha\in \ord_K$ in  $\bar{\ord}_K$, where $\alpha$ is given as an element of $K$.
\begin{restatable}{lem}{lemcomputeresidue}\label{lem_computeresidue}
There exists a polynomial-time algorithm $\mathtt{ComputeResidue}$ that  takes the following data as the input
\begin{itemize}
\item a number fields $K$,  a prime number $p$, and $\alpha\in \ord_K$ given as an element of $K$,
\item the outputs of $\mathtt{ComputeQuotientRing}$ (see Lemma~\ref{lem_residuering}) on the inputs $(K,p)$, i.e., the quotient ring $\bar{\ord}_K$, a maximal $p$-orders $\ord_K'$,  the inclusion $\ord_K'\hookrightarrow K$, and the quotient map $\ord_K'\to \bar{\ord}_K$,
\end{itemize}
and computes $\alpha+p\ord_K\in\bar{\ord}_K$.
\end{restatable}

The proof of Lemma~\ref{lem_computeresidue} can be found in Appendix~\ref{chap_omitted}. 

\paragraph{Computing embeddings of number fields.}
Embeddings of a number field in another can be computed efficiently, thanks to the polynomial-time factoring algorithms for number fields \citep{Len83, Lan85}. 
\begin{thm}[\citep{Len83, Lan85}]\label{thm_facnumfld}
There exists a polynomial-time algorithm that given a number field $K$ and a polynomial $g(X)\in K[X]$, factorizes $g(X)$ into irreducible factors over $K$.
\end{thm}

Let $K, K'$ be number fields and suppose $K$ is encoded with a primitive element $\alpha\in K$ whose minimal polynomial is $g(X)\in\Q[X]$. Each embedding $\phi$ of $K$ in $K'$ is determined by the image $\phi(\alpha)\in K'$ which is a root of $g(X)$. These roots can be enumerated by factoring $g(X)$ over $K'$ using Theorem~\ref{thm_facnumfld}. So we have:
\begin{lem}\label{lem_compembed}
There exists a polynomial-time algorithm $\mathtt{ComputeEmbeddings}$ that given number fields $K$ and $K'$, computes all the embeddings of $K$ in $K'$.
\end{lem}

\paragraph{Computing induced ring homomorphisms between $\bar{\ord}_K$.}

Let $\phi: K\hookrightarrow K'$ be an embedding of number fields, which restricts to an inclusion 
%$\phi|_{\ord_K}:$ 
$\ord_K\hookrightarrow \ord_{K'}$. By passing to the quotient rings $\bar{\ord}_K$ and $\bar{\ord}_{K'}$, we obtain a ring homomorphism $\bar{\phi}: \bar{\ord}_K\to \bar{\ord}_{K'}$. And we  say the map $\bar{\phi}$ is {\em induced from $\phi$}. 
 The following lemma states that $\bar{\phi}$ can be efficiently computed from $\phi$ and some auxiliary data.
% We present a more general statement below, with $\ord_K$ replaced by a subring $\ord$ of $\ord_K$.\footnote{As $\ord_K$ is a lattice, i.e., a free $\Z$-module of finite rank, so is its subring $\ord$.}
\nomenclature[b1m]{$\bar{\phi}$}{ring homomorphism $\bar{\ord}_K\to \bar{\ord}_{K'}$ induced from an embedding $\phi: K\hookrightarrow K'$}

\begin{restatable}{lem}{lemringhom}\label{lem_ringhom}
There exists a polynomial-time algorithm $\mathtt{ComputeRingHom}$ that takes the following data as the input
\begin{itemize}
\item number fields $K$, $K'$, an embedding $\phi: K\to K'$, and a prime number $p$,
%\item a subring $\ord$ of $\ord_K$, the quotient ring $\ord/p\ord$, the inclusion $\ord\hookrightarrow K$, and the quotient map $\ord\to \ord/p\ord$,
%\item a maximal $p$-orders $\ord_{K'}'$ of $K'$, the quotient ring $\bar{\ord}_{K'}$, the inclusion  $\ord_{K'}'\hookrightarrow K'$, and the quotient map   $\ord_{K'}'\to \bar{\ord}_{K'}$,
\item the outputs of $\mathtt{ComputeQuotientRing}$ (see Lemma~\ref{lem_residuering}) on the inputs $(K,p)$ and $(K',p)$ respectively,\footnote{That is, the quotient rings $\bar{\ord}_K$, $\bar{\ord}_{K'}$, the maximal $p$-orders $\ord_K'$, $\ord_{K'}'$, 
the inclusions $\ord_K'\hookrightarrow K$, $\ord_{K'}'\hookrightarrow K'$, and the quotient maps $\ord_K'\to \bar{\ord}_K$, $\ord_{K'}'\to \bar{\ord}_{K'}$.}
\end{itemize}
and computes the ring homomorphism $\bar{\phi}: \bar{\ord}_K\to \bar{\ord}_{K'}$ induced from $\phi$.
\end{restatable}

The proof of Lemma~\ref{lem_ringhom} can be found in Appendix~\ref{chap_omitted}.

\section{Reduction to computing an idempotent decomposition of \texorpdfstring{$\bar{\ord}_F$}{OF}} \label{sec_algreduction}

 Now we start describing the $\mathcal{P}$-scheme algorithm. Fix the following notations in the remaining sections:
 \begin{itemize}
\item $f(X)$: the input polynomial in $\F_p[X]$ to be factorized, which is square-free and completely reducible over $\F_p$,  
\item $\tilde{f}(X)$: an irreducible  lifted polynomial of $f(X)$  in $\Z[X]$,
\item $F$: the number field $\Q[X]/(\tilde{f}(X))$,
\item $L$: the splitting field of $\tilde{f}$ over $\Q$,
\item $G$: the Galois group $\gal(L/\Q)=\gal(\tilde{f}/\Q)$,
\item $\mathfrak{Q}_0$: a fixed prime ideal of $\ord_L$ lying over $p$.
 \end{itemize}

 In this section, we reduce the problem of factoring $f$ to that of computing an idempotent decomposition of $\bar{\ord}_F$. For simplicity, we first assume that  $\tilde{f}$ is a {\em monic} polynomial, and then remove the assumption at the end of this section.

 \paragraph{Ring isomorphism between $\F_p[X]/(f(X))$ and $\bar{\ord}_K$.} Let $\alpha:=X+(\tilde{f}(X))\in F$ which is a root of $\tilde{f}$.  As $\tilde{f}(X)\in\Z[X]$ is monic, we know $\alpha\in \ord_F$.
 Define the ring homomorphism $\tilde{\tau}:\F_p[X]\to \bar{\ord}_F$ by letting $\tilde{\tau}(X)=\alpha+p\ord_F$, which is well defined since $\bar{\ord}_F$ is an $\F_p$-algebra. Moreover, we have $\tilde{\tau}(f(X))=\tilde{f}(\alpha)+p\ord_F=0$.
 So $\tilde{\tau}$ induces a ring homomorphism $\tau:\F_p[X]/(f(X))\to  \bar{\ord}_K$ sending $X+(f(X))$ to $\alpha+p\ord_F$.
 
Let $f_1,\dots,f_n$ be the monic irreducible factors of $f$ over $\F_p$. As $f_i$ are irreducible and distinct, the ring $\F_p[X]/(f(X))$ is semisimple with the maximal ideals $(f_i(X))$, $i=1,\dots,n$. Then $\bar{\ord}_F$ is also semisimple. Indeed, we have the following lemma:
\begin{restatable}{lem}{lemringisom}\label{lem_ringisom}
The map $\tau:\F_p[X]/(f(X))\to  \bar{\ord}_F$ is a ring isomorphism, and $p$ splits completely in $F$.
\end{restatable}

\begin{proof}
The second claim follows from the first since $\F_p[X]/(f(X))$ has $n$ distinct maximal ideals.
To prove the first claim, note that the ring homomorphism $\F_p[X]/(f(X))\to \Z[\alpha]/p\Z[\alpha]$ sending $X$ to $\alpha+p\Z[\alpha]$ is an isomorphism.
So it suffices to show that the natural inclusion $\Z[\alpha]\hookrightarrow \ord_F$ induces an isomorphism $\Z[\alpha]/p\Z[\alpha]\to \bar{\ord}_F$. 
%Denote the latter map by $\upsilon$.

For $i\in [n]$, choose $\tilde{f}_i(X)\in\Z[X]$ that lifts the factor $f_i(X)\in\F_p[X]$ of $f$, and define the ideal $\mathfrak{P}_i$ of $\Z[\alpha]$ to be the one generated by $\tilde{f}_i(\alpha)$ and $p$. 
As $\Z[\alpha]/p\Z[\alpha]\cong \F_p[X]/(f(X))$ is semisimple, we have
$\bigcap_{i=1}^n \mathfrak{P}_i=p\Z[\alpha]$.
By \citep[Theorem~5.10]{AM69}, for each $i\in [n]$, we may choose a prime ideal $\mathfrak{Q}_i$ of $\ord_F$ lying over $p$ such that $\mathfrak{Q}_i\cap \Z[\alpha]=\mathfrak{P}_i$. Then we have
\[
p\ord_F \cap \Z[\alpha] \subseteq\left(\bigcap_{i=1}^n \mathfrak{Q}_i\right)\cap \Z[\alpha]=\bigcap_{i=1}^n \mathfrak{P}_i=p\Z[\alpha].
\]
So the map $\Z[\alpha]/p\Z[\alpha]\to \bar{\ord}_F$ is injective.
It is in fact an isomorphism since $\Z[\alpha]/p\Z[\alpha]$ and $\bar{\ord}_F$ are both vector spaces of dimension $n$ over $\F_p$.
\end{proof}

\paragraph{Extracting a factorization from an idempotent decomposition.} Let $I_F$ be an idempotent decomposition of $\bar{\ord}_F$. By Lemma~\ref{lem_ringisom}, the set $\tau^{-1}(I_F)=\{\tau^{-1}(\delta):\delta\in I_F\}$ is an idempotent decomposition of $\F_p[X]/(f(X))$. Given $\delta\in I_F$, 
% the element $\tau^{-1}(\delta)$ is a nonzero idempotent of $\F_p[X]/(f(X))$, and 
we can extract a factor $g_\delta(X)$ of $f(X)$ by
\[
g_\delta(X):=\mathrm{gcd}(f(X), h_\delta(X)),
\]
where $h_\delta(X)\in \F_p[X]$ is a nonzero polynomial of degree at most $n$  lifting $1-\tau^{-1}(\delta)\in \F_p[X]/(f(X))$. The factor $g_\delta(X)$ is the product of the monic irreducible factors $f_i(X)$ satisfying $\tau^{-1}(\delta)\equiv 1\pmod{f_i(X)}$. 
As $f(X)=\tilde{f}(X)\bmod p$ is monic and the elements  $\tau^{-1}(\delta)$ form an idempotent decomposition of  the ring $\F_p[X]/(f(X))$, we have the equality 
\[
f(X)=\prod_{\delta\in I} g_\delta(X).
\]
This gives the following algorithm that computes a factorization of $f$ from $I_F$: 

\begin{algorithm}[htbp]
\caption{$\mathtt{ExtractFactors}$}\label{alg_reduction}
\begin{algorithmic}[1]
\INPUT  \parbox[t]{.9\linewidth}{%
	 $p$, $f$, $\tilde{f}$,  $F$,  $\bar{\ord}_F$, idempotent decomposition $I_F$ of $\bar{\ord}_F$,\\
	  $p$-maximal order $\ord'_F$ of $F$ and maps $\ord'_F\hookrightarrow F$, $\ord'_F\to \bar{\ord}_F$ \strut
	  } 
\OUTPUT factorization of $f$
\State $\alpha\gets X+(\tilde{f}(X))\in F$
\State call $\mathtt{ComputeResidue}$ to compute $\alpha+p\ord_F\in  \bar{\ord}_F$
\State compute the ring homomorphism $\tau: \F_p[X]/(f(X))\to \bar{\ord}_F$ sending $X+(f(X))$ to $\alpha + p\ord_F$
 \For{$\delta\in I_F$}
    \State compute nonzero $h_\delta(X)\in \F_p[X]$ of degree at most $n$ lifting $1-\tau^{-1}(\delta)$
    \State $g_\delta(X)\gets\mathrm{gcd}(f(X), h_\delta(X))$
\EndFor
\State \Return the factorization $f(X)=\prod_{\delta\in I_F} g_\delta(X)$
\end{algorithmic}
\end{algorithm} 

For the purpose of computing the map $\tau$, the input contains some auxiliary data (e.g., a $p$-maximal order $\ord'_F$ and the related maps) other than the idempotent decomposition $I_F$. For now we note that the auxiliary data can be prepared in polynomial time using the subroutines in Section~\ref{sec_algpre}. Then we have:

\begin{thm}\label{thm_extract}
The algorithm $\mathtt{ExtractFactors}$ computes the factorization  $f(X)=\prod_{\delta\in I_F} g_\delta(X)$ in polynomial time. In particular, it computes the complete factorization (resp. a proper factorization) of $f(X)$ in polynomial time iff the idempotent decomposition $I_F$ of $\bar{\ord}_F$ is complete (resp. proper).
\end{thm}
\begin{proof}
 The algorithm clearly runs in polynomial time:  Line 1 is implemented by factoring $\tilde{f}$ over $F$ using Theorem~\ref{thm_facnumfld}. The loop in Lines 4--6 iterates $|I_F|\leq n$ times.  Line 5 is implemented by solving a system of linear equations over $\F_p$ and Line 6 by the Euclidean algorithm. 
 The fact that the factorization is complete (resp. proper) iff $I_F$ is complete (resp. proper) follows from the fact that  $\tau:\F_p[X]/(f(X))\to  \bar{\ord}_F$ is a ring isomorphism.
\end{proof}

Therefore the problem of computing the complete factorization (resp. a proper) factorization of $f$ reduces to the problem of computing the complete  (resp. a proper) idempotent decomposition of $\bar{\ord}_F$.

\paragraph{The reduction for non-monic polynomials.}

 After a slight change, the above reduction  also works for a possibly non-monic polynomial $\tilde{f}$. We explain it now.

Suppose $c\in\Z-\{0\}$ is the leading coefficient of $\tilde{f}$. Its residue $\bar{c}:=c\bmod p\in\F_p$ is nonzero since $\deg(\tilde{f})=\deg(f)=n$. 
Define $\tilde{f}'(X):=c^{n-1}\cdot \tilde{f}(X/c)\in \Z[X]$ and $f'(X):=\tilde{f}(X)\bmod p\in\F_p[X]$. The polynomials $\tilde{f}'$ and $f'$ are monic, and $f'(X)=\bar{c}^{n-1}\cdot f(X/\bar{c})$. Let $\alpha$ be a root of $\tilde{f}$ in $F$ as before. Then $\alpha':=c\alpha$ is a root of $\tilde{f}'$ and hence is in $\ord_F$.

Run the algorithm $\mathtt{ExtractFactors}$ above except that $f$, $\tilde{f}$ and $\alpha$ are replaced with $f'$, $\tilde{f}'$ and $\alpha'$ respectively. Then we obtain a factorization $f'(X)=\prod_{\delta\in I_F} g'_\delta(X)$ where  the factors $g'_\delta(X)\in\F_p[X]$ are monic. Substituting $X$ with $\bar{c}  X$, we obtain a factorization
\[
f(X)=\bar{c}\cdot \prod_{\delta\in I_F} g_\delta(X)
\]
with the monic factors $g_\delta(X):=\bar{c}^{-\deg(g'_\delta)}\cdot g'_\delta(\bar{c}  X)\in\F_p[X]$.
Theorem~\ref{thm_extract} then holds for $f$ and $\tilde{f}$.
% since it holds for $f'$ and $\tilde{f}'$.

\section{Main algorithm} \label{sec_algmain}

We present the main body of the $\mathcal{P}$-scheme algorithm in this section. Its input contains a collection of  number fields that are isomorphic to subfields of $L$.
% where $L$ is the splitting field of $\tilde{f}$ over $\Q$. 
In order to avoid duplicate data, we  assume that these number fields are mutually non-isomorphic.  This is formalized by the following %(algorithmic) 
definition:

\begin{defi}[$(\Q,g)$-subfield system]\label{defi_expfield}
Let $g(X)$ be a polynomial in $\Q[X]$ with the splitting field $L(g)$ over $\Q$.
Let $\mathcal{F}$ be a collection of number fields
% given to the algorithm (encoded by their structure constants) 
such that
(1) the fields in $\mathcal{F}$ are mutually non-isomorphic, and (2) each field $K'\in\mathcal{F}$ is isomorphic to a subfield of $L(g)$.
We say $\mathcal{F}$ is a {\em $(\Q,g)$-subfield system}.\index{subfield system}
\end{defi}

Given a $(\Q,g)$-subfield system, we define a subgroup system over $\gal(g/\Q)$ as follows.

\begin{defi}\label{defi_asssystem}
Let $g(X)$ be a polynomial in $\Q[X]$ with the splitting field $L(g)$ over $\Q$.
Let $\mathcal{F}$ be a {\em $(\Q,g)$-subfield system}. 
Define  $\mathcal{P}^\sharp$ to be the poset of subfields of $L(g)$ that includes all the fields isomorphic to those in $\mathcal{F}$:
\[
\mathcal{P}^\sharp:=\{K'\subseteq L(g): K'\cong K \text{ for some } K\in \mathcal{F}\}.
\]
By Galois theory, it corresponds to a poset $\mathcal{P}$ of subgroups of $\gal(g/\Q)$, given by
\[
\mathcal{P}:=\left\{H\subseteq\gal(g/\Q): (L(g))^H\in \mathcal{P}^\sharp \right\}
%=\left\{H\subseteq  \gal(g/\Q): (L(g))^H\cong K \text{ for some } K\in \mathcal{F}\right\}
\]
which is closed under conjugation in $\gal(g/\Q)$, and hence is a subgroup system over $\gal(g/\Q)$. We say $\mathcal{P}$ and $\mathcal{P}^\sharp$ are {\em associated with $\mathcal{F}$}.
\end{defi}
\nomenclature[b1n]{$\mathcal{P}^\sharp$}{poset of subfields corresponding to  $\mathcal{P}$  via Galois correspondence}

The pseudocode of the algorithm is given in Algorithm~\ref{alg_mainalg} below.  
Its input is the prime number $p$ and  a $(\Q,\tilde{f})$-subfield system $\mathcal{F}$. We fix $\mathcal{P}$ to be the subgroup system over $G=\gal(\tilde{f}/\Q)$ associated with $\mathcal{F}$.

\begin{algorithm}[htbp]
\caption{$\mathtt{ComputePscheme}$}\label{alg_mainalg}
\begin{algorithmic}[1]
\INPUT prime number $p$, $(\Q,\tilde{f})$-subfield system $\mathcal{F}$
\OUTPUT \parbox[t]{.9\linewidth}{%
	 for each $K\in\mathcal{F}$: $\bar{\ord}_K$, idempotent decomposition $I_K$ of $\bar{\ord}_K$,\\
	 $p$-maximal order $\ord'_K$ of $K$ and maps $\ord'_K\hookrightarrow K$, $\ord'_K\to \bar{\ord}_K$ \strut
	  }
\For{$K\in \mathcal{F}$} \strut
    \State  \parbox[t]{\dimexpr\linewidth-\algorithmicindent}{call $\mathtt{ComputeQuotientRing}$ to compute  $\bar{\ord}_F$, a $p$-maximal order $\ord'_K$ of $K$ and maps $\ord'_K\hookrightarrow K$, $\ord'_K\to \bar{\ord}_K$\strut} 
    \State  $I_K\gets \{1\}$, where $1$ denotes the unity of  $\bar{\ord}_K$ \strut
\EndFor 
 \For{$(K,K')\in \mathcal{F}^2$} \strut
    \State call $\mathtt{ComputeEmbeddings}$ to compute all the embeddings from $K$ to $K'$
    \For{embedding $\phi: K\hookrightarrow K'$}
        \State call $\mathtt{ComputeRingHom}$ to compute $\bar{\phi}:\bar{\ord}_K\to \bar{\ord}_{K'}$ induced from $\phi$
     \EndFor
\EndFor
\Repeat
    \State call  $\mathtt{CompatibilityAndInvarianceTest}$
    \State call  $\mathtt{RegularityTest}$
    \State call  $\mathtt{StrongAntisymmetryTest}$
\Until{$I_K$ remains the same in the last iteration for all $K\in\mathcal{F}$}
\State \Return $\bar{\ord}_K$, $I_K$, $\ord'_K$ and the maps $\ord'_K\hookrightarrow K$, $\ord'_K\to \bar{\ord}_K$ for $K\in\mathcal{F}$
\end{algorithmic}
\end{algorithm} 

The algorithm outputs, for every $K\in\mathcal{F}$, the ring $\bar{\ord}_K$ and an idempotent decomposition $I_K$ of $\bar{\ord}_K$, together with the auxiliary data of a $p$-maximal order $\ord'_K$ and the related maps $\ord'_K\hookrightarrow K$, $\ord'_K\to \bar{\ord}_K$.
%,as needed by the algorithm $\mathtt{ExtractFactors}$ in the previous section. 
We will see below that the idempotent decompositions $I_K$ altogether determine a $\mathcal{P}$-collection, which is guaranteed to be a strongly antisymmetric $\mathcal{P}$-scheme when the algorithm terminates.

 The first half (Lines 1--7) of the algorithm is the preprocessing  stage, where we  compute $\bar{\ord}_K$ for $K\in\mathcal{F}$ and the ring homomorphisms between them that are induced from the field embeddings.  For each $K\in\mathcal{F}$, we also initialize the idempotent decomposition $I_K$  of $\bar{\ord}_K$ to be the trivial one containing only the unity of $\bar{\ord}_K$.

The second half (Lines 8--12) is the ``refining'' stage. To understand it, we need to associate  a $\mathcal{P}$-collection $\mathcal{C}$ with the idempotent decompositions $I_K$.  By Lemma~\ref{lem_ringisom}, we know $p$ splits completely in $F$. So it also splits completely in  every subfield of $L$. In particular, for a field $K$ in $\mathcal{P}^\sharp$ or $\mathcal{F}$, the quotient ring $\bar{\ord}_K$ is semisimple.

%Fix a nonzero prime ideal $\mathfrak{Q}_0$ of $\ord_L$ lying over $p$. 
For each $H\in\mathcal{P}$, we define a partition $C_H$ of the coset space $H\backslash G$ as follows:
Let $K$ be the unique field in $\mathcal{F}$ isomorphic to $L^H$. Fix an isomorphism $\tau_H: K\to L^H$, which induces a ring isomorphism $\bar{\tau}_H: \bar{\ord}_K\to \bar{\ord}_{L^H}$.
Define $I_H:=\bar{\tau}_H(I_{K})$, which  is an idempotent decomposition of $\bar{\ord}_{L^H}$.
%From $\bar{\tau}_H$ and $I_{K}$, we obtain an idempotent decompostion $\bar{\tau}_H(I_{K})$ of $\bar{\ord}_{L^H}$. 
By  Definition~\ref{defi_partitioncor}, it corresponds to a partition $P(I_H)$ of $H\backslash G$.\footnote{Definition~\ref{defi_partitioncor} is made with respect to a fixed prime ideal $\mathfrak{Q}_0$ of $\ord_L$ lying over $p$. This ideal is chosen at the beginning of  Section~\ref{sec_algreduction}.}
And we define 
\[
C_H:=P(I_H).
\]
Finally, define the $\mathcal{P}$-collection $\mathcal{C}$ by 
\[
\mathcal{C}:=\{C_H: H\in\mathcal{P}\}.
\] 
%\todo{$\tau_H$ denotes a map between fields rather than quotient rings. Fix this.}
\nomenclature[b1o]{$\tau_H$}{fixed isomorphism $K\to L^H$ (which is $K_0$-linear in Chapter~\ref{chap_alg_general})}

%For $K\in\mathcal{F}$ and two idempotent decompositions $I,I'$ of $\bar{\ord}_K$, we say $I$ is a {\em refinement} of $I'$ if $P(\bar{\tau}_H(I))$ is a refinement of $P(\bar{\tau}_H(I'))$ for all $H\in\mathcal{P}$ such that $L^H\cong K$.

We call several subroutines to update $I_K$ in Lines 9--11, whose effects can be understood in terms of $\mathcal{C}$:

\begin{restatable}{lem}{lemcitest}\label{lem_citest}
There exists a subroutine  $\mathtt{CompatibilityAndInvarianceTest}$ that updates $I_K$ in time polynomial  in $\log p$ and the size of $\mathcal{F}$ so that the partitions $C_H\in \mathcal{C}$ are refined, and at least one partition $C_H$ is properly refined if $\mathcal{C}$ is not compatible or invariant.
\end{restatable}

\begin{restatable}{lem}{lemrtest}\label{lem_rtest}
There exists a subroutine $\mathtt{RegularityTest}$ that updates $I_K$ in  time polynomial in $\log p$ and the size of $\mathcal{F}$ so that the partitions $C_H\in \mathcal{C}$ are refined, and at least one partition $C_H$ is properly refined if $\mathcal{C}$ is compatible but not regular. 
\end{restatable}

\begin{restatable}{lem}{lemsatest}\label{lem_satest}
Under GRH, there exists a subroutine $\mathtt{StrongAntisymmetryTest}$ that updates $I_K$ in time polynomial  in $\log p$ and the size of $\mathcal{F}$ so that the partitions $C_H\in \mathcal{C}$ are refined, and at least one partition $C_H$ is properly refined  if $\mathcal{C}$ is  a $\mathcal{P}$-scheme, but not strongly antisymmetric.
\end{restatable}

We will describe these subroutines and prove the lemmas above in the next three sections. For now we just assume them and prove the main result of this section:

\begin{thm}[Theorem~\ref{thm_mainalginformal} restated]\label{thm_comppscheme}
Under GRH, the algorithm $\mathtt{ComputePscheme}$ runs in  time polynomial in the size of the input,
 and when it terminates, the $\mathcal{P}$-collection $\mathcal{C}$ is a strongly antisymmetric $\mathcal{P}$-scheme.
\end{thm}

\begin{proof}
We first analyze the running time.
As each field $K\in\mathcal{F}$ is encoded by a rational polynomial of degree $[K:\Q]$, the total degree $N:=\sum_{K\in\mathcal{F}} [K:\Q]$ is bounded by the size of $\mathcal{F}$. The loops in Lines 1--3  and Lines 4--7 iterate $|\mathcal{F}|\leq N$ and $|\mathcal{F}^2|\leq N^2$ times respectively. For each $(K,K')\in\mathcal{F}^2$, there are at most $[K:\Q]$ embeddings from $K$ to $K'$, and hence the inner loop in Lines 6--7 iterates at most $[K:\Q]$ times for each fixed $(K,K')$. 

For the loop in Lines 8--12, we consider $K\in\mathcal{F}$ and pick $H\in\mathcal{P}$ so that $L^H$ is isomorphic to $K$. By Lemma~\ref{lem_pandi}, the number of idempotents in $I_K$ equals the number of blocks in $C_H$, and this number increases every time $I_K$ is changed by the subroutines. On the other hand, the number of idempotents in $I_K$ is at most $[K:\Q]$. So the loop in Lines 8--12 iterates $O(N)$ times. The claim about the running time  easily follows.

Finally, note that the algorithm exits the loop in Lines 8--12 after an iteration iff all of the idempotent decompositions $I_K$ remain the same in that iteration, in which case $\mathcal{C}$ is already a strongly antisymmetric $\mathcal{P}$-scheme  by Lemma~\ref{lem_citest}, Lemma~\ref{lem_rtest} and Lemma~\ref{lem_satest}. 
\end{proof}

\begin{rem}
The input of the the algorithm contains $\mathcal{F}$  whose size may be much greater than that of $f$ and $\tilde{f}$. Therefore, the polynomiality of this algorithm in the size of its input does {\em not} imply that polynomial factoring over finite fields can be  solved in (deterministic) polynomial time. It does suggest, however, that the total degree of the fields in $\mathcal{F}$ over $\Q$ is the bottleneck of our factoring algorithm. 
%The problem of bounding the total degree  is the central topic of subsequent chapters. 
%Unfortunately, the number fields in $\mathcal{F}$ have super-polynomial total degrees in general . 
\end{rem}

\section{Compatibility and invariance test} \label{sec_algci}

The subroutine $\mathtt{CompatibilityAndInvarianceTest}$ is given in Algorithm~\ref{alg_citest}. It has the effect of properly refining at least one partition in $\mathcal{C}$, unless $\mathcal{C}$ is compatible and invariant.
\begin{algorithm}[htbp]
\caption{ $\mathtt{CompatibilityAndInvarianceTest}$}\label{alg_citest}
\begin{algorithmic}[1]
 \For{\strut $(K, K')\in\mathcal{F}^2$ and embedding $\phi: K'\hookrightarrow K$}
     \For{$(\delta,\delta')\in I_{K}\times I_{K'}$}
      \If{ $\bar{\phi}(\delta')\delta\not\in \{0,\delta\}$} \Comment{$\bar{\phi}: \bar{\ord}_{K'}\to \bar{\ord}_K$  is induced from $\phi$}
             \State $I_K\gets I_K - \{\delta\}$
             \State $I_K\gets I_K\cup \{\bar{\phi}(\delta')\delta, (1-\bar{\phi}(\delta')) \delta\}$
         \State \Return
      \EndIf
      \EndFor
  \EndFor
\end{algorithmic}
\end{algorithm}

This  subroutine  attempts to find a ring homomorphisms $\bar{\phi}: \bar{\ord}_{K'}\to \bar{\ord}_K$ (induced from a field embedding $\phi: K'\to K$) and idempotents $\delta\in I_K$, $\delta'\in I_{K'}$ such that $\bar{\phi}(\delta')\delta$ equals neither $\delta$ nor zero. If such $\delta$, $\delta'$, and $\bar{\phi}$ are found, the  subroutine  updates $I_K$ by replacing $\delta\in I_K$ with two new idempotents $\bar{\phi}(\delta')\delta$ and $(1-\bar{\phi}(\delta')) \delta$, neither of which is zero. It has the effect of splitting each block $B_{\bar{\tau}_H(\delta)}\in C_{H}=P(I_H)$ corresponding to $\bar{\tau}_H(\delta)\in I_H$ (see Lemma~\ref{lem_pandi}) into two blocks, where $H$ ranges over the subgroups in $\mathcal{P}$ satisfying $L^H\cong K$.  After the update,  the subroutine halts.

Now we prove Lemma~\ref{lem_citest} as promised before.

%\lemcitest*

\begin{proof}
[Proof of Lemma~\ref{lem_citest}]
Polynomiality of the running time is straightforward. To prove the rest of the claim,  we assume that no proper refinement is made, i.e. for all $K,K'\in\mathcal{F}$, $\delta\in I_K$, $\delta'\in I_{K'}$ and field embeddings $\phi: K'\hookrightarrow K$, we have $\bar{\phi}(\delta')\delta\in \{0,\delta\}$.  Then we show that $\mathcal{C}$ is compatible and invariant. 

For $H\in\mathcal{P}$, the isomorphism $\tau_H$ identifies $L^H\in\mathcal{P}$ with a field $K\in\mathcal{F}$. So the condition above can be reformulated as follows:
for all $H,H'\in\mathcal{P}$, $\delta\in I_H$, $\delta'\in I_{H'}$ and field embeddings $\phi: L^{H'}\hookrightarrow L^H$, we have $\bar{\phi}(\delta')\delta\in \{0,\delta\}$.

Now consider  $H,H'\in\mathcal{P}$  satisfying $H\subseteq H'$ and elements $Hg,Hg'\in H\backslash G$ in the same block $B\in C_H=P(I_H)$.
We want to show that $\pi_{H,H'}(Hg)=H'g$ and $\pi_{H,H'}(Hg')=H'g'$ are in the same block of $C_{H'}$.
  By Lemma~\ref{lem_pandi}, there exists an idempotent $\delta\in I_H$ for which 
 \begin{equation}\label{eq_blkb}
  B=\{Hh\in H\backslash G: \prescript{h^{-1}}{}{(i_{L^H,L}(\delta))}\equiv 1\pmod{\bar{\mathfrak{Q}}_0}\}
  \end{equation}
  holds, where $i_{L^H,L}:\bar{\ord}_{L^H}\hookrightarrow \bar{\ord}_{L}$ is induced from the natural inclusion $L^H\hookrightarrow L$. 
 Choose $\phi$ to be the natural inclusion $L^{H'}\hookrightarrow L^H$.
   As $\delta=\sum_{\delta'\in I_{H'}} \bar{\phi}(\delta')\delta$, there exists an idempotent $\delta'\in I_{H'}$ such that $\bar{\phi}(\delta')\delta\neq 0$. By assumption, we have $\bar{\phi}(\delta')\delta=\delta$. Again by Lemma~\ref{lem_pandi}, the set $B'$ given by
  \begin{equation}\label{eq_blkb2}
  B'=\{H'h\in H'\backslash G: \prescript{h^{-1}}{}{(i_{L^{H'},L}(\delta'))}\equiv 1\pmod{\bar{\mathfrak{Q}}_0}\}
  \end{equation}
  is a block of $C_{H'}=P(I_{H'})$. We claim that $H'g,H'g'\in B'$. To see this, note that as $Hg\in B$ , we have 
   \begin{equation}\label{eq_blkb3}
\prescript{g^{-1}}{}{(i_{L^{H},L}(\bar{\phi}(\delta')\delta))}=\prescript{g^{-1}}{}{(i_{L^{H},L}(\delta))}\equiv 1\pmod{\bar{\mathfrak{Q}}_0}.
\end{equation}
 It implies $\prescript{g^{-1}}{}{(i_{L^{H},L}(\bar{\phi}(\delta')))}\equiv 1\pmod{\bar{\mathfrak{Q}}_0}$.
 Note that $i_{L^{H'},L}=i_{L^{H},L}\circ \bar{\phi}$. So we have
 $\prescript{g^{-1}}{}{(i_{L^{H'},L}(\delta'))} \equiv 1\pmod{\bar{\mathfrak{Q}}_0}$ and hence $H'g\in B'$. Similarly, we have $H'g'\in B'$. So $H'g$ and $H'g'$ are in the same block of $C_{H'}$, as desired.
Therefore $\mathcal{C}$ is compatible.
 
Next consider  $H,H'\in\mathcal{P}$  satisfying $H'=hHh^{-1}$ for some $h\in G$ and elements $Hg,Hg'\in H\backslash G$ in the same block $B$ of $C_H$.
We want to show that $c_{H,h}(Hg)=H'hg$ and $c_{H,h}(Hg')=H'hg'$ are in the same block of $C_{H'}$. 
Again by Lemma~\ref{lem_pandi}, there exists an idempotent $\delta\in I_H$ for which \eqref{eq_blkb} holds.
Choose $\phi$ to be the isomorphism  $L^{H'}\to L^H$ sending $x\in L^{H'}$ to $\prescript{h^{-1}}{}{x}\in  L^H$. So $\bar{\phi}$ sends $x\in\bar{\ord}_{L^{H'}}$ to $\prescript{h^{-1}}{}{x}\in\bar{\ord}_{L^{H}}$, or more pedantically, to
\[
i_{L^H,L}^{-1}\left(\prescript{h^{-1}}{}{(i_{L^{H'},L}(x))}\right)\in L^H.
\]
 Again, as $\delta=\sum_{\delta'\in I_{H'}} \bar{\phi}(\delta')\delta$, there exists an idempotent $\delta'\in I_{H'}$ such that $\bar{\phi}(\delta')\delta\neq 0$. By assumption, we have $\bar{\phi}(\delta')\delta=\delta$. By Lemma~\ref{lem_pandi}, the set $B'$ given by \eqref{eq_blkb2} is a block of $C_{H'}=P(I_{H'})$.
 We claim that $H'hg,H'hg'\in B'$. To see this, note that \eqref{eq_blkb3} holds since $Hg\in B$. It implies that
 \[
\prescript{(hg)^{-1}}{}{(i_{L^{H'},L}(\delta'))}=\prescript{g^{-1}}{}{(i_{L^{H},L}(\bar{\phi}(\delta')))}\equiv 1\pmod{\bar{\mathfrak{Q}}_0}
 \]
 and hence $H'hg\in B'$. Similarly, we have $H'hg'\in B'$. So $H'hg$ and $H'hg'$ are in the same block of $C_{H'}$, as desired.
As $c_{H,h}$ is bijective, it maps blocks to blocks. Therefore $\mathcal{C}$ is invariant.
\end{proof}

\section{Regularity test} \label{sec_algr}

In this section we implement the subroutine $\mathtt{RegularityTest}$. It has the effect of properly refining at least one partition in $\mathcal{C}$ if $\mathcal{C}$ is compatible, invariant, but not regular.

A similar algorithm was proposed in \citep{Evd94, Gao01} based on generalizations of the Euclidean algorithm for polynomials over rings. We take an alternative approach developed in \citep{IKS09, IKRS12}  based on a ``free module test'':
\begin{restatable}[\citep{IKS09, IKRS12}]{lem}{lemfree}\label{lem_free}
There exists an algorithm $\mathtt{FreeModuleTest}$ that given a semisimple $\F_p$-algebra $A$ and a finitely generated $A$-module $M$, returns  a zero divisor $a$ of $A$  in polynomial time, such that $a$ is zero only  if $M$ is a free $A$-module.
%, i.e., isomorphic to a direct product of copies of $A$ as an $A$-module.
\end{restatable}

For completeness, we prove Lemma~\ref{lem_free}  in Appendix~\ref{chap_omitted}. 
In addition, we need the following subroutine.
\begin{restatable}{lem}{lemzerodivisor}\label{lem_zerodivisor}
There exists an algorithm  $\mathtt{SplitByZeroDivisor}$ that given 
\begin{itemize}
\item a semisimple $\F_p$-algebra $R$, an idempotent decomposition $I$ of $R$, and an idempotent $\gamma\in I$,
\item the ring $\bar{R}:=R/(1-\gamma)$, the quotient map $\pi: R\to \bar{R}$, and a zero divisor $a\neq 0$ of $\bar{R}$,
\end{itemize}
replaces $\gamma\in I$ with two nonzero idempotents $\gamma_1,\gamma_2$ satisfying $\gamma=\gamma_1+\gamma_2$ in polynomial time.
\end{restatable}

The proof of  Lemma~\ref{lem_zerodivisor}  can be found in Appendix~\ref{chap_omitted} as well. 
The subroutine $\mathtt{RegularityTest}$ is then implemented in Algorithm~\ref{alg_reg} below.

\begin{algorithm}[htbp]
\caption{$\mathtt{RegularityTest}$}\label{alg_reg}
\begin{algorithmic}[1]
 \For{\strut $(K, K')\in\mathcal{F}^2$ and embedding $\phi: K'\hookrightarrow K$}
     \For{$(\delta,\delta')\in I_{K}\times I_{K'}$ satisfying $\bar{\phi}(\delta')\delta=\delta$}
         \State compute $A=\bar{\ord}_{K'}/(1-\delta')$ and the quotient map $\bar{\ord}_{K'}\to A$
         \State compute $M=\bar{\ord}_K/(1-\delta)$  and the quotient map $\bar{\ord}_{K}\to M$
         \State \parbox[t]{.9\linewidth}{compute $\phi_{\delta,\delta'}: A\to M$ sending $u+(1-\delta')$ to $\bar{\phi}(u)+(1-\delta)$ for $u\in \bar{\ord}_{K'}$, making $M$ an $A$-algebra and hence an $A$-module\strut}
         \State \strut call $\mathtt{FreeModuleTest}$ with the input $A$ and $M$ to obtain $a\in A$
         \If{$a\neq 0$}
             \State call $\mathtt{SplitByZeroDivisor}$  to update $I_{K'}$ using the zero divisor $a$
             \State \Return
         \EndIf
\EndFor  
\EndFor
\end{algorithmic}
\end{algorithm}

The subroutine enumerates $(K,K')\in\mathcal{F}^2$, the  ring homomorphisms $\bar{\phi}: \bar{\ord}_{K'}\to \bar{\ord}_K$ (induced from the field embeddings $\phi: K'\to K$), and the idempotents $\delta\in I_K$, $\delta'\in I_{K'}$ satisfying $\bar{\phi}(\delta')\delta=\delta$. Line 3 and Line 4 compute the quotient rings $A=\bar{\ord}_{K'}/(1-\delta')$, $M=\bar{\ord}_K/(1-\delta)$ and the corresponding quotient maps. 
They are quotient rings of semisimple rings and hence also semisimple.
By Lemma~\ref{lem_quotienthom}, the map $\bar{\phi}$ induces a ring homomorphism $\phi_{\delta,\delta'}: A\to M$ sending $u+(1-\delta')$ to $\bar{\phi}(u)+(1-\delta)$ for $u\in \bar{\ord}_{K'}$, which we  compute at Line 5. It gives $M$ an $A$-algebra structure, and in particular an $A$-module structure. 
Then we call $\mathtt{FreeModuleTest}$ at Line 6 which returns a zero divisor $a$ of $A$ by Lemma~\ref{lem_free}.
If $a\neq 0$, we call $\mathtt{SplitByZeroDivisor}$ (with the input $R=\bar{\ord}_{K'}$, $I=I_{K'}$, $\gamma=\delta'$, $\bar{R}=A$, the quotient map $\bar{\ord}_{K'}\to A$, and the zero divisor $a$) to update $I_{K'}$, so that $\delta'$ is replaced with two nonzero idempotents by Lemma~\ref{lem_zerodivisor}.  After the update,  the subroutine halts.

\begin{proof}[Proof of Lemma~\ref{lem_rtest}]
The subroutine obviously runs in time polynomial in $\log p$ and the size of $\mathcal{F}$.
To prove the rest of the lemma, it suffices to show that a zero divisor $a\neq 0$ of $A$ is always found in Line 6 if $\mathcal{C}$ is compatible but not regular. 
%Lemma~\ref{lem_free} reduces it to the following lemma.

So assume $\mathcal{C}$ is compatible but not regular. Then there exist $H,H'\in\mathcal{P}$ satisfying $H\subseteq H'$, $B\in C_H$, $B'\in C_{H'}$ and $H'g,H'g'\in B'$ such that 
\begin{equation}\label{eq_cdegree}
|\pi_{H,H'}^{-1}(H'g)\cap B|\neq |\pi_{H,H'}^{-1}(H'g')\cap B|.
\end{equation}
By Lemma~\ref{lem_pandi}, there exist $\delta\in I_H$ and $\delta'\in I_{H'}$ such that
\[
B=\{Hh\in H\backslash G: \prescript{h^{-1}}{}{(i_{L^H,L}(\delta))}\equiv 1\pmod{\bar{\mathfrak{Q}}_0}\}
\]
and
\[
B'=\{H'h\in H'\backslash G: \prescript{h^{-1}}{}{(i_{L^{H'},L}(\delta'))}\equiv 1\pmod{\bar{\mathfrak{Q}}_0}\}.
\]
By \eqref{eq_cdegree} and compatibility of $\mathcal{C}$, we have $\pi_{H,H'}(B)\subseteq B'$. 
Let $\phi:L^{H'}\hookrightarrow L^H$ be the natural inclusion, which induces a ring homomorphism $\bar{\phi}:\bar{\ord}_{L^{H'}}\to\bar{\ord}_{L^H}$. 
We claim that $\bar{\phi}(\delta')\delta=\delta$ holds: assume to the contrary that it does not hold. Then there exists a maximal ideal $\mathfrak{m}$ of $\bar{\ord}_L$ such that 
\[
i_{L^H,L}(\delta)\equiv 1\pmod{\mathfrak{m}} \quad\text{and}\quad 
i_{L^H,L}(\bar{\phi}(\delta'))=i_{L^{H'},L}(\delta')\equiv 0\pmod{\mathfrak{m}}.
\]
Choose $h\in G$ such that $\mathfrak{m}=\prescript{h}{}{\bar{\mathfrak{Q}}_0}$.
Then we have 
\[
\prescript{h^{-1}}{}{(i_{L^H,L}(\delta))}\equiv 1\pmod{\bar{\mathfrak{Q}}_0}  \quad\text{and}\quad 
\prescript{h^{-1}}{}{(i_{L^{H'},L}(\delta'))}\equiv 0\pmod{\bar{\mathfrak{Q}}_0}.
\] 
It follows that $Hh \in B$ and $\pi_{H,H'}(Hh)=H'h\not\in B'$. But this contradicts  $\pi_{H,H'}(B)\subseteq B'$. So $\bar{\phi}(\delta')\delta=\delta$ holds.  

Define $A:=\bar{\ord}_{L^{H'}}/(1-\delta')$ and $M:=\bar{\ord}_{L^H}/(1-\delta)$. Let $\phi_{\delta,\delta'}:A\to M$ be the ring homomorphism sending $u+(1-\delta')$ to $\bar{\phi}(u)+(1-\delta)$ for $u\in\bar{\ord}_{L^{H'}}$, making $M$ an $A$-algebra and hence an $A$-module. We claim that $M$ is not free over $A$.  Assume to the contrary that $M$ is a free $A$-module. Denote its rank over $A$ by $k\in\N^+$. 
Define
\[
\mathfrak{P}:=(\prescript{g}{}{\mathfrak{Q}_0}\cap \ord_{L^{H'}})/p\ord_{L^{H'}}\subseteq \bar{\ord}_{L^{H'}}\quad\text{and}\quad \mathfrak{P}':=\mathfrak{P}/(1-\delta')\subseteq A,
\]
which are maximal ideals  of  $\bar{\ord}_{L^{H'}}$ and of $A$ respectively.
Then $M/\mathfrak{P}' M$ is a free $A/\mathfrak{P}'$-module of rank $k$.
On the other hand, we have the isomorphism
\[
M/\mathfrak{P}' M\cong \bar{\ord}_{L^H}/ (\bar{\phi}(\mathfrak{P})\bar{\ord}_{L^H}+(1-\delta)\bar{\ord}_{L^H}).
\]
It follows from the Chinese remainder theorem that $M/\mathfrak{P}'M$ is isomorphic to $\prod_{\mathfrak{m}\in S}  \bar{\ord}_{L^H}/\mathfrak{m}$ where $S$ denotes the set of the maximal ideals of  $\bar{\ord}_{L^H}$ containing  both $\bar{\phi}(\mathfrak{P})$ and $1-\delta$. 
As $p$ splits completely in $L^H$, each direct factor $\bar{\ord}_{L^H}/\mathfrak{m}$ is isomorphic to $\F_p$.
So $M/\mathfrak{P}'M$ is a vector space of dimension $|S|$ of $\F_p$.
On the other hand, as $p$ splits completely in $L^{H'}$, we have $A/\mathfrak{P}'\cong\F_p$.
So rank $k$ of $M/\mathfrak{P}'M$ over $A/\mathfrak{P}'$ equals $|S|$.

By Corollary~\ref{cor_idealcoset}, the  maximal ideals of  $\bar{\ord}_{L^H}$  are of the form by $\mathfrak{P}_{Hh}:=(\prescript{h}{}{\mathfrak{Q}_0}\cap \ord_{L^{H}})/p\ord_{L^{H}}$ which correspond one-to-one to the cosets $Hh\in H\backslash G$.
Each maximal ideal $\mathfrak{P}_{Hh}$ contains $\bar{\phi}(\mathfrak{P})$ iff $\mathfrak{P}$ is contained in
\[
\bar{\phi}^{-1}(\mathfrak{P}_{Hh})=(\prescript{h}{}{\mathfrak{Q}_0}\cap \ord_{L^{H'}})/p\ord_{L^{H'}},
\]
which, again by Corollary~\ref{cor_idealcoset}, holds iff  $H'g=H'h$. 
And $\mathfrak{P}_{Hh}$ contains $1-\delta$ iff $i_{L^H,L}(1-\delta)\in \prescript{h}{}{\mathfrak{Q}_0}$, which holds iff $Hh\in B$. So we have 
\[
k=|S|=|\{Hh\in B: H'g=H'h\}|=|\pi_{H,H'}^{-1}(H'g)\cap B|.
\]
But the same proof shows $k=|\pi_{H,H'}^{-1}(H'g')\cap B|$. This is a contradiction to \eqref{eq_cdegree}. Therefore $M$ is not free over $A$.

Identify $L^H$ (resp. $L^{H'}$) with a field in $\mathcal{F}$ using the isomorphism $\tau_H$ (resp. $\tau_{H'}$) chosen in Section~\ref{sec_algmain}. By Lemma~\ref{lem_free}, the subroutine is guaranteed to find a nonzero element $a\in A$ in Line 6.
It then updates an idempotent decomposition $I_{K'}$ and properly refines some  partition in $\mathcal{C}$ by Lemma~\ref{lem_zerodivisor}, as desired. 
\end{proof}

\section{Strong antisymmetry test} \label{sec_algsa}

In this section, we implement the subroutine $\mathtt{StrongAntisymmetryTest}$, which has the effect of properly refining at least one partition in $\mathcal{C}$ if $\mathcal{C}$ is a $\mathcal{P}$-scheme, but not a strongly antisymmetric $\mathcal{P}$-scheme.

This subroutine is based on an algorithm developed in \citep{Ron92}:

\begin{restatable}[\citep{Ron92}]{lem}{lemauto}\label{lem_auto}
Under GRH, there exists an algorithm $\mathtt{Automorphism}$ that given a ring $A$  isomorphic to a finite product of $\F_p$ and a nontrivial ring automorphism $\sigma$ of $A$, returns a zero divisor $a\neq 0$ of $A$  in polynomial time.
\end{restatable}

For completeness, we provide a proof of Lemma~\ref{lem_auto}  in Appendix~\ref{chap_omitted}.  

%Lemma~\ref{lem_auto}  was used in \citep{Ron92} to give a polynomial-time factoring algorithm in the case that $F/\Q$ is Galois. Even in non-Galois cases, it can be used to find partial factorizations, yielding an algorithm that performs an ``antisymmetry test'': the algorithm properly refines some partition in $\mathcal{C}$ if $\mathcal{C}$ is a $\mathcal{P}$-scheme but not an antisymmetric $\mathcal{P}$-scheme.
%The subroutine we give, however, is a further improvement of this algorithm, based an idea introduced in \citep{Evd94}. 

The subroutine $\mathtt{StrongAntisymmetryTest}$ is implemented in Algorithm~\ref{alg_satest} below. 

\begin{algorithm}[htbp]
\caption{$\mathtt{StrongAntisymmetryTest}$}\label{alg_satest}
\begin{algorithmic}[1]
\State \strut construct an edge-labeled directed graph $G=(V, E)$ where $V=\{(K,\delta): K\in\mathcal{F}, \delta\in I_K\}$ and $E=\emptyset$
\For{$(K,\delta)\in V$}
       \State compute $A_{K,\delta}:=\bar{\ord}_K/(1-\delta)$ and the quotient map $\bar{\ord}_K\to A_{K,\delta}$
%       \State add a self-loop of $(K,\delta)$ to $E$ with label $\id_{A_{K,\delta}}$, the identity map on $A_{K,\delta}$
\EndFor
 \For{$((K,\delta), (K',\delta'))\in V^2$ and $\phi: K'\hookrightarrow K$ satisfying $\bar{\phi}(\delta')\delta=\delta$}
         \State compute $\phi_{\delta,\delta'}: A_{K',\delta'}\to A_{K,\delta}$ sending $x+(1-\delta')$ to $\bar{\phi}(x)+(1-\delta)$
         \If{$\phi_{\delta,\delta'}$ is invertible}
             \State  \parbox[t]{0.9\linewidth}{$E\gets E\cup\{e,e'\}$, where the edge $e$ is from $(K',\delta')$ to $(K,\delta)$  with label $\phi_{\delta,\delta'}$, and  $e'$ is  from $(K,\delta)$ to $(K',\delta')$  with label $\phi_{\delta,\delta'}^{-1}$\strut}
%             \If{$\exists ~ e''\in E$ from $(K',\delta')$ \strut to $(K,\delta)$ with label $\psi\neq \phi_{\delta,\delta'}$}
%                 \State compute $\tau=\phi_{\delta,\delta'}\circ \psi^{-1}$
%                 \State  \parbox[t]{\dimexpr 0.9\linewidth-\algorithmicindent}{call $\mathtt{Automorphism}$ with the input $A_{K,\delta}$ and $\tau$ to obtain a zero divisor $a\neq 0$ of  $A_{K,\delta}$\strut}
%                 \State \strut call $\mathtt{SplitByZeroDivisor}$  to update $I_{K}$ using  $a$
%                 \State \Return
%             \EndIf
         \EndIf
\EndFor
\State search an nontrivial automorphism $\sigma$ of $A_{K,\delta}$ for some $(K,\delta)\in V$ that is a composition of maps in $\mathcal{L}:=\{\phi_{\delta,\delta'}: \text{there exists an edge $e\in E$ with label $\phi_{\delta,\delta'}$}\}$

\If{$\sigma$ is found at Line 8}
   \State {call $\mathtt{Automorphism}$ on $(A_{K,\delta},\sigma)$ to obtain a zero divisor $a\neq 0$ of  $A_{K,\delta}$\strut}
   \State \strut call $\mathtt{SplitByZeroDivisor}$  to update $I_{K}$ using  $a$
   \State \Return
\EndIf
\end{algorithmic}
\end{algorithm}

The subroutine first constructs an edge-labeled directed graph $G=(V, E)$, where the vertex set is 
\[
V:=\{(K,\delta): K\in\mathcal{F}, \delta\in I_K\}
\]
and each edge is labeled by a certain ring isomorphism to be determined later.
Initially the edge set $E$ is empty. 
For every vertex $(K,\delta)\in V$, we compute the ring  $A_{K,\delta}:=\bar{\ord}_K/(1-\delta)$ and the quotient map $\bar{\ord}_K\to A_{K,\delta}$ at Line 3.

Then we enumerate $((K,\delta), (K',\delta'))\in V^2$ and $\phi: K'\hookrightarrow K$ for which $\bar{\phi}(\delta')\delta=\delta$ holds, and for each of them, we compute a ring homomorphism 
\[
\phi_{\delta,\delta'}: A_{K',\delta'}\to A_{K,\delta}
\] 
that sends $x+(1-\delta')$ to $\bar{\phi}(x)+(1-\delta)$ for $x\in \bar{\ord}_{K'}$. The map $\phi_{\delta,\delta'}$ is well defined by Lemma~\ref{lem_quotienthom}. If $\phi_{\delta,\delta'}$ is an isomorphism (i.e., invertible), we add to $E$ an edge $e$ from $(K',\delta')$ to $(K,\delta)$ with label $\phi_{\delta,\delta'}$, and also an edge $e'$ from $(K,\delta)$ to $(K',\delta')$ with label $\phi_{\delta,\delta'}^{-1}$. 
%Here we only maintain one edge for each label, so adding an edge whose label already exists has no effect.

Next,  at  Line 8, we search a nontrivial automorphism $\sigma$ of $A_{K, \delta}$, $(K,\delta)\in V$,  such that $\sigma$ is a composition of  maps  in $\mathcal{L}$, where
\[
\mathcal{L}:=\{\phi_{\delta,\delta'}: \text{there exists an edge $e\in E$ with label $\phi_{\delta,\delta'}$}\}.
\]
 We sketch a way of implementing this step in time polynomial in $\log p$ and the size of $\mathcal{F}$: note that the edges whose labels compose into a nontrivial automorphism form a cycle of $G$.
 %, and hence are in the same strongly connected component. 
 So by  computing the strongly connected components of $G$ and restricting to each of them, we reduce to the case that $G$ is strongly connected. Fix a vertex $(K_0,\delta_0)\in V$. For every  $(K,\delta)\in V$, compute a ring isomorphism $\psi_{K,\delta}: A_{K_0,\delta_0}\to A_{K,\delta}$ that is a composition of maps in $\mathcal{L}$. These isomorphisms exist since we assume $G$ is strongly connected, and they can be computed by, e.g., the breadth-first search algorithm.
 %It is easy to see that a nontrivial automorphism $\sigma$ exists iff $\phi_{\delta,\delta'}\circ \psi_{K',\delta'}=\psi_{K,\delta}$ holds
 %for all $\phi_{\delta,\delta'}: A_{K',\delta'}\to A_{K,\delta}$ in $\mathcal{L}$.
Then we may find a nontrivial automorphism $\sigma$, if it exists, by enumerating the maps $\phi_{\delta,\delta'}: A_{K',\delta'}\to A_{K,\delta}$ in $\mathcal{L}$ and checking if the automorphism 
\[
\phi_{\delta,\delta'}\circ \psi_{K',\delta'}\circ\psi_{K,\delta}^{-1}: A_{K,\delta}\to A_{K,\delta}
\]
of $A_{K,\delta}$ is nontrivial.

Finally, if a nontrivial automorphism $\sigma$ of some ring $A_{K,\delta}$ is successfully found, we use it to update $I_K$ as follows: run the algorithm $\mathtt{Automorphism}$ on the input $(A_{K,\delta},\sigma)$ to obtain a zero divisor $a\neq 0$ of $\in A_{K,\delta}$.
Then call $\mathtt{SplitByZeroDivisor}$ (with the input $R=\bar{\ord}_{K}$, $I=I_{K}$, $\gamma=\delta$, $\bar{R}=A_{K,\delta}$, the quotient map $\bar{\ord}_{K}\to A_{K,\delta}$, and the zero divisor $a$) to update $I_{K}$, so that $\delta$ is replaced with two nonzero idempotents by Lemma~\ref{lem_zerodivisor}.

Now we analyze the subroutine. For $H\subseteq G$ and $B\in C_H$, there exists a unique idempotent $\delta=\delta_B\in I_H$ satisfying
\[
B=\{H h\in H\backslash G: \prescript{h^{-1}}{}{(i_{L^{H},L}(\delta))}\equiv 1\pmod{\bar{\mathfrak{Q}}_0}\}.
\]
See Definition~\ref{defi_partitioncor} and Lemma~\ref{lem_pandi}. Write $A_{L^H,\delta}$ for the ring $\bar{\ord}_{L^H}/(1-\delta)$.
The maximal ideals of $A_{L^H,\delta}$ are of the form $\mathfrak{m}/(1-\delta)$ where $\mathfrak{m}$ is a maximal ideal of $\bar{\ord}_{L^H}$ containing $1-\delta$. By Corollary~\ref{cor_idealcoset},  the map 
\[
Hg\mapsto \mathfrak{m}_{Hg}:=(\prescript{g}{}{\mathfrak{Q}_0}\cap \ord_{L^H})/p\ord_{L^H}
\] 
is a one-to-one correspondence between the right cosets in $H\backslash G$ and the  maximal ideals of $\bar{\ord}_K$.
And $\mathfrak{m}_{Hg}$ contains $1-\delta$ iff $i_{L^H,L}(1-\delta)\in\prescript{g}{}{\mathfrak{Q}_0}$, which holds iff $Hg\in B$. We conclude that the map 
\[
Hg\mapsto \mathfrak{m}_{Hg}/(1-\delta)
\]
 is a one-to-one correspondence between the right cosets in $B$ and the maximal ideals of $A_{L^H,\delta}$.
 
We also need the following technical lemma.
\begin{lem}\label{lem_pullback}
%Let $\mathcal{C}=\{C_H: H\in\mathcal{C}\}$ be a $\mathcal{P}$-scheme. 
Suppose $H,H'\in\mathcal{P}$, $B\in C_H$, $B'\in C_{H'}$, $\tau: B\to B'$, and $\phi: L^{H'}\to L^H$ are in one following cases:
\begin{enumerate}
\item $H\subseteq H'$, $\tau=\pi_{H,H'}|_B: B\to B'$, and $\phi: L^{H'}\to L^H$ is the natural inclusion.
\item $H'=hHh^{-1}$ for some $h\in G$, $\tau=c_{H,h}|_B: B\to B'$, and $\phi: L^{H'}\to L^H$ sends $x$ to $\prescript{h^{-1}}{}{x}$.
\end{enumerate}
Let $\delta:=\delta_B\in I_H$ and $\delta':=\delta_{B'}\in I_{H'}$ (see Definition~\ref{defi_partitioncor}). Let $\bar{\phi}:\bar{\ord}_{L^{H'}}\to\bar{\ord}_{L^H}$ be induced from $\phi$.
% Define $A_{L^H,\delta}:=\bar{\ord}_{L^H}/(1-\delta)$ and $A_{L^{H'},\delta'}:=\bar{\ord}_{L^{H'}}/(1-\delta')$.
 Then  $\bar{\phi}(\delta')\delta=\delta$ holds, so that the ring homomorphism 
\[
\phi_{\delta,\delta'}: A_{L^{H'},\delta'}\to A_{L^H,\delta}
\]  
sending $x+(1-\delta')$ to $\bar{\phi}(x)+(1-\delta)$ is well defined by Lemma~\ref{lem_quotienthom}. Moreover, for $Hg\in B$, we have 
\[
\phi_{\delta,\delta'}^{-1}(\mathfrak{m}_{Hg}/(1-\delta))=\mathfrak{m}_{\tau(Hg)}/(1-\delta').
\]
Finally, the map $\phi_{\delta,\delta'}$ is an isomorphism if $\tau$ is a bijection.
\end{lem}
\begin{proof}
We claim that for any $Hg\in H\backslash G$, it holds that  $\bar{\phi}^{-1}(\mathfrak{m}_{Hg})=\mathfrak{m}_{\tau(Hg)}$. Fix $Hg\in H\backslash G$. 
Note that $\bar{\phi}^{-1}(\mathfrak{m}_{Hg})$ is a prime (and hence maximal) ideal of $\bar{\ord}_{L^{H'}}$. Therefore to prove the claim, it suffices to show $\bar{\phi}(\mathfrak{m}_{\tau(Hg)})\subseteq \mathfrak{m}_{Hg}$. 
In the first case of the lemma, we have $\tau(Hg)=\pi_{H,H'}(Hg)=H'g$, and  
\[
\mathfrak{m}_{Hg}=(\prescript{g}{}{\mathfrak{Q}_0}\cap \ord_{L^H})/p\ord_{L^H} \quad\text{and}\quad \mathfrak{m}_{\tau(Hg)}=\mathfrak{m}_{H'g}=(\prescript{g}{}{\mathfrak{Q}_0}\cap \ord_{L^{H'}})/p\ord_{L^{H'}}.
\]
As $\phi: L^{H'}\to L^H$ is the natural inclusion, we have $\phi(\mathfrak{m}_{\tau(Hg)})\subseteq \mathfrak{m}_{Hg}$, as desired.

In the second case, we have $\tau(Hg)=c_{H,h}(Hg)=H'hg$, and
\[
\mathfrak{m}_{Hg}=(\prescript{g}{}{\mathfrak{Q}_0}\cap \ord_{L^H})/p\ord_{L^H} \quad\text{and}\quad \mathfrak{m}_{\tau(Hg)}=\mathfrak{m}_{H'hg}=(\prescript{hg}{}{\mathfrak{Q}_0}\cap \ord_{L^{H'}})/p\ord_{L^{H'}}.
\]
As $\phi: L^{H'}\to L^H$ sends $x$ to $\prescript{h^{-1}}{}{x}$, again we have $\phi(\mathfrak{m}_{\tau(Hg)})\subseteq \mathfrak{m}_{Hg}$. This proves the claim.

Next we prove $\bar{\phi}(\delta')\delta=\delta$. As $\bar{\ord}_{L^H}$ is semisimple, it suffices to show that for any maximal ideal $\mathfrak{m}_{Hg}$ containing $\bar{\phi}(\delta')$  also  contains $\delta$.  
Fix $Hg\in H\backslash G$ such that $\bar{\phi}(\delta')\in \mathfrak{m}_{Hg}$. Then $\delta'$ is contained in $\bar{\phi}^{-1}(\mathfrak{m}_{Hg})=\mathfrak{m}_{\tau(Hg)}$. As $\delta'=\delta_{B'}$, we have $\tau(Hg)\not\in B'$ and hence $Hg\not\in B$. Finally, as $\delta=\delta_B$, we have $\delta\in \mathfrak{m}_{Hg}$, as desired.

The next claim that $\phi_{\delta,\delta'}^{-1}(\mathfrak{m}_{Hg}/(1-\delta))=\mathfrak{m}_{\tau(Hg)}/(1-\delta')$ follows directly from $\bar{\phi}^{-1}(\mathfrak{m}_{Hg})=\mathfrak{m}_{\tau(Hg)}$. Now assume $\tau$ is a bijection. The kernel of $\phi_{\delta,\delta'}$ is
\[
\bigcap_{Hg\in B}\phi^{-1}_{\delta,\delta'}(\mathfrak{m}_{Hg}/(1-\delta))=\bigcap_{Hg\in B} \mathfrak{m}_{\tau(Hg)}/(1-\delta')=\bigcap_{H'g\in B'} \mathfrak{m}_{H'g}/(1-\delta')=0.
\]
So $\phi_{\delta,\delta'}$ is injective.
Also note that the dimension of $A_{L^H,\delta}$ (resp. $A_{L^{H'},\delta'}$) over $\F_p$ equals its number of maximal ideals, which is $|B|$ (resp. $|B'|$). As $\tau$ is bijective, we have $|B|=|B'|$. So $\tau$ is an isomorphism.
\end{proof}

Now we are ready to prove Lemma~\ref{lem_satest}, as promised.

\begin{proof}[Proof of  Lemma~\ref{lem_satest}]
Assume $\mathcal{C}$ is a $\mathcal{P}$-scheme but not a strongly antisymmetric $\mathcal{P}$-scheme.
By Lemma~\ref{lem_auto}, it suffices to show that some maps in $\mathcal{L}$ compose into a nontrivial automorphism of  $A_{K,\delta}$ for some $(K,\delta)\in V$.

As $\mathcal{C}$ is not strongly antisymmetric, there exist $k\in \N^+$, subgroups $H_0,\dots, H_k\in\mathcal{P}$, blocks $B_0\in C_{H_0},\dots, B_k\in C_{H_k}$, and maps $\sigma_1,\dots, \sigma_k$ satisfying
\begin{itemize}
\item $\sigma_i$ is a bijective map from $B_{i-1}$ to $B_i$,
\item $\sigma_i$ is of the form $c_{H_{i-1},g}|_{B_{i-1}}$, $\pi_{H_{i-1},H_i}|_{B_{i-1}}$, or  $(\pi_{H_i,H_{i-1}}|_{B_i})^{-1}$,
\item $H_0=H_k$ and $B_0=B_k$,
\end{itemize}
and the composition $\tau:=\sigma_k\circ\cdots\circ \sigma_1$ is a nontrivial permutation of $B_0=B_k$.

Let $\delta_i:=\delta_{B_i}\in I_{H_i}$ and $A_{L^{H_i}, \delta_i}:=\bar{\ord}_{L^{H_i}}/(1-\delta_i)$ for $0\leq i\leq k$. 
By Lemma~\ref{lem_pullback}, for $i\in [k]$, there exists a ring isomorphism $\psi_i: A_{L^{H_i},\delta_i}\to A_{L^{H_{i-1}},\delta_{i-1}}$
such that 
\[
\psi_i^{-1}(\mathfrak{m}_{H_{i-1}g}/(1-\delta_{i-1}))=\mathfrak{m}_{\sigma_i(H_{i-1}g)}/(1-\delta_i)
\]
holds for all $H_{i-1}g\in B_{i-1}$. Moreover, for $i\in [k]$, the map $\psi_i$ is in one of the following two cases:
\begin{itemize}
\item  $\psi_i$ sends $x+(1-\delta_i)$ to $\bar{\phi}_i(x)+(1-\delta_{i-1})$ for $x\in \bar{\ord}_{L^{H_i}}$, where  $\phi_i$ is an embedding of  $L^{H_i}$ in $L^{H_{i-1}}$.
\item $\psi_i^{-1}$ sends $x+(1-\delta_{i-1})$ to $\bar{\phi}_i(x)+(1-\delta_i)$ for $x\in \bar{\ord}_{L^{H_{i-1}}}$,  where  $\phi_i$ is an embedding of  $L^{H_{i-1}}$ in $L^{H_{i}}$.
\end{itemize}
Here the first case occurs when $\sigma_i$ is of the form $c_{H_{i-1},g}|_{B_{i-1}}$ or $\pi_{H_{i-1},H_i}|_{B_{i-1}}$, and the second one occurs  when $\sigma_i$ is of the form  $(\pi_{H_i,H_{i-1}}|_{B_i})^{-1}$.

Consider the automorphism $\sigma:=\psi_1\circ \cdots\circ\psi_k$ of $A_{L^{H_k},\delta_k}=A_{L^{H_0},\delta_0}$. We have
\[
\sigma^{-1}(\mathfrak{m}_{H_0g}/(1-\delta_0))=\mathfrak{m}_{\tau(H_0g)}/(1-\delta_0)
\]
for all $H_0g\in B_0$. As $\tau$ is a nontrivial permutation of $B_0$, there exists $H_0g\in B_0$ satisfying $\tau(H_0g)\neq H_0g$ and hence $\mathfrak{m}_{H_0g}/(1-\delta_0)\neq \mathfrak{m}_{\tau(H_0g)}/(1-\delta_0)$. So $\sigma$ is a nontrivial automorphism.

Finally, identifying each field $L^{H_i}$ with a field $K_i\in\mathcal{F}$ using the isomorphisms $\tau_{H_i}: K_i\to L^{H_i}$, we see that the ring isomorphisms $\psi_i$ are identified with maps in $\mathcal{L}$, and they compose into a nontrivial automorphism of $A_{K_0,\bar{\tau}_{H_i}^{-1}(\delta_0)}$. Here $K_0$ is the unique field in $\mathcal{F}$ isomorphic to $L^{H_0}$ and $\bar{\tau}_{H_i}^{-1}(\delta_0)\in I_{K_0}$. The lemma follows.
\end{proof}

\section{Constructing a collection of number fields}\label{sec_compnumflds}

The last ingredient of the $\mathcal{P}$-scheme algorithm is a subroutine that constructs a $(\Q,g)$-subfield system given a polynomial $g(X)\in\Q[X]$ irreducible over $\Q$. 
%In applications, we choose $g$ to be the irreducible lifted polynomial $\tilde{f}$ of $f$.

This subroutine can be implemented in various ways, leading to  algorithms with different running time. We mention two results of this kind: computing the splitting field of $g$, and computing a  $(\Q,g)$-subfield system whose associated subgroup system  is a system of stabilizers. For simplicity, we only state the results, deferring the proofs and the algorithms to Chapter~\ref{chap_constructnum} where we discuss the problem of constructing number fields in depth.

%Fix the following notations in this section: denote the splitting field of $g$ by $L$ and the degree of $g$ by $n$.  Let $F=\Q[X]/(g(X))$.
%Fix a subfield $F'$ of $L$ that is isomorphic to $F$. Let $G=\gal(L/\Q)$ and $H=\gal(L/F')$.

\paragraph{Computing the splitting field of a polynomial.} 
%Suppose $\mathcal{F}=\{F, L\}$. In this case we can compute $\mathcal{F}$ using the following lemma.
The splitting of a polynomial over $\Q$ can be effectively constructed by the following lemma.

\begin{lem}\label{lem_compsplit}
There exists a deterministic algorithm that given a polynomial $g(X)\in\Q[X]$ irreducible over $\Q$, computes its splitting field $L(g)$ over $\Q$ in time polynomial in $[L(g):\Q]$ and the size of $g$.
\end{lem}

The proof is deferred to Chapter~\ref{chap_constructnum}.
% We apply Lemma~\ref{lem_compsplit} with $g=\tilde{f}$.
%The associated subgroup system $\mathcal{P}$ contains the trivial subgroup $\{e\}$ and the conjugates of $H$ in $G$.
%Note that by Lemma~\ref{lem_antidisc}, all antisymmetric $\mathcal{P}$-schemes are discrete on $H$.

\paragraph{System of stabilizers.} 
%The group $G$ acts on $H\backslash G$ by inverse right translation. 
We also have an algorithm that computes a $(\Q,g)$-subfield system  whose associated subgroup system is a system of stabilizers:

\begin{lem}\label{lem_compstab}
There exists a deterministic algorithm that given a polynomial $g(X)\in\Q[X]$ irreducible over $\Q$ and a positive integer $m\leq \deg(g)$, computes a $(\Q,g)$-subfield system $\mathcal{F}$,  such that the subgroup system associated with $\mathcal{F}$ is the system of stabilizers of depth $m$ over $G(g/\Q)$ with respect to the action of $\gal(g/\Q)$ on the set of roots of $g$ in $L(g)$, where $L(g)$ denotes the splitting field of $g$ over $\Q$.
Moreover, the algorithm runs in time polynomial in $(\deg(g))^m$ and the size of $g$.
\end{lem}

The proof is again deferred to Chapter~\ref{chap_constructnum}. 

%We also need the following theorem.

%\begin{thm}\label{thm_discsys}
%There exists a constant $c>0$ such that 
%for any transitive action of a finite group $G$ on a set $S\neq \emptyset$, positive integer $m>c\log |S|$, and $x\in S$, all strongly antisymmetric $\mathcal{P}$-schemes are discrete on $G_x$ where $\mathcal{P}=\mathcal{P}_{S,m}$ is the system of stabilizers of depth $m$ with respect to the action of $G$ on $S$.
%\end{thm}
%
%We prove this theorem in Chapter X. It was essentially proven in \citep{Evd94} and reproved in \citep{IKS09} for the case that $G=\sym(n)$ acts naturally on $[n]$.

\section{Putting it together} \label{sec_algputtogether}

We combine the results in previous sections to obtain the $\mathcal{P}$-scheme algorithm.\index{Pschemealgorithm@$\mathcal{P}$-scheme algorithm}  The pseudocode is given in Algorithm~\ref{alg_pschalg} below.

\begin{algorithm}[htbp]
\caption{$\mathtt{PschemeAlgorithm}$}\label{alg_pschalg}
\begin{algorithmic}[1]
\INPUT $f(X)\in \F_p[X]$ and its irreducible lifted polynomial $\tilde{f}(X)\in\Z[X]$
\OUTPUT factorization of $f$
\State call $\mathtt{ComputeNumberFields}$ to compute a $(\Q,\tilde{f})$-subfield system $\mathcal{F}$ such that (1) $F=\Q[X]/(\tilde{f}(X))\in \mathcal{F}$, and  (2) for some $H\in\mathcal{P}$ satisfying $L^H\cong F$, all strongly antisymmetric $\mathcal{P}$-schemes are discrete (resp. inhomogeneous) on $H$, where $\mathcal{P}$ is the subgroup system over $G=\gal(\tilde{f}/\Q)$ associated with $\mathcal{F}$
\State call $\mathtt{ComputePscheme}$ on the input $(p, \mathcal{F})$ to obtain $I_K$ for $K\in\mathcal{F}$
\State call $\mathtt{ExtractFactors}$ to extract a factorization of $f$ from $I_F$, and output it
\end{algorithmic}
\end{algorithm}

The subroutine $\mathtt{ComputeNumberFields}$ at Line 1 is the generic part of the algorithm and can be implemented in various ways. It is supposed to compute a  $(\Q,\tilde{f})$-subfield system  $\mathcal{F}$ such that  $F\in\mathcal{F}$, and the associated subgroup system $\mathcal{P}$ over $G$ satisfies a certain combinatorial property (see Theorem~\ref{thm_algmain2formal} below).
The latter condition is used to show that the factoring algorithm always produces the complete factorization (resp. a proper factorization) of $f$.
%The subroutine $\mathtt{ComputeNumberFields}$ at Line 1 computes $\mathcal{F}$.

The algorithm $\mathtt{ComputePscheme}$ (see Section~\ref{sec_algmain}) at Line 2 takes the input $(p, \mathcal{F})$ and outputs data that includes the idempotent decompositions $I_K$ for $K\in\mathcal{F}$.
Finally, we call the subroutine $\mathtt{ExtractFactors}$ (see Section~\ref{sec_algreduction}) at Line 3 to extract a factorization of $f$ from $I_F$.

The following theorem is the main result of this chapter.

\begin{thm}[Theorem~\ref{thm_algmain2informal} restated]\label{thm_algmain2formal}
Suppose there exists a deterministic algorithm that given a polynomial $g(X)\in\Z[X]$ irreducible over $\Q$, constructs a  $(\Q,g)$-subfield system $\mathcal{F}$  in time $T(g)$  such that
\begin{itemize}
\item $\Q[X]/(g(X))$ is in $\mathcal{F}$, and
\item  for some $H\in\mathcal{P}$ satisfying $(L(g))^H\cong \Q[X]/(g(X))$, all strongly antisymmetric $\mathcal{P}$-schemes are discrete (resp. inhomogeneous) on $H$, where $\mathcal{P}$ is the subgroup system over $\gal(g/\Q)$ associated with $\mathcal{F}$, and $L(g)$ is the splitting field of $g$ over $\Q$.
\end{itemize}
Then under GRH, there exists a deterministic algorithm that given a  polynomial $f(X)\in\F_p[X]$ satisfying Condition~\ref{cond_spoly} and an irreducible lifted polynomial $\tilde{f}(X)\in\Z[X]$ of $f$, outputs the complete factorization (resp. a proper factorization) of $f$ over $\F_p$ in time polynomial in $T(\tilde{f})$ and the size of the input.%\todo{Make clear the dependence on GRH.}
\end{thm}

\begin{proof}
Consider the algorithm $\mathtt{PschemeAlgorithm}$ above and implement the subroutine $\mathtt{ComputeNumberFields}$ using the hypothetical algorithm in the theorem. Choose $g=\tilde{f}$.
 By Theorem~\ref{thm_comppscheme}, the $\mathcal{P}$-collection $\mathcal{C}=\{C_H: H\in\mathcal{P}\}$ defined by $C_H=P(\bar{\tau}_H(I_{K}))$ is a strongly antisymmetric $\mathcal{P}$-scheme. By the second condition in the theorem, we have $C_H=\infty_{H\backslash G}$ (resp. $C_H\neq 0_{H\backslash G}$) for some $H\in\mathcal{P}$ satisfying $L^H\cong F$. So the corresponding idempotent decomposition $I_F$ is complete (resp. proper). By Theorem~\ref{thm_extract}, the algorithm outputs the complete factorization (resp. a proper factorization) of $f$ over $\F_p$. 

The subroutine $\mathtt{ComputeNumberFields}$ runs in time $T(\tilde{f})$. In particular, the size of $\mathcal{F}$ is bounded by  $T(\tilde{f})$. The claim about the running time then follows from Theorem~\ref{thm_comppscheme} and Theorem~\ref{thm_extract}.
\end{proof}

By Theorem~\ref{thm_algmain2formal} and Lemma~\ref{lem_compstab}, we have a deterministic factoring algorithm whose running time is related to the notations $d(G)$ and $d'(G)$ introduced in Definition~\ref{defi_dg}: 
\begin{cor}\label{cor_algdgbound}
Under GRH, there exists a deterministic algorithm that, given a polynomial $f(X)\in\F_p[X]$ of degree $n\in\N^+$  satisfying Condition~\ref{cond_spoly} and an irreducible\footnote{The assumption that $\tilde{f}$ is irreducible is not necessary, and can be avoided by adapting Lemma~\ref{lem_compstab}. We omit the details.} lifted polynomial $\tilde{f}(X)\in\Z[X]$ of $f$, computes the complete factorization (resp. a proper factorization) of $f$ over $\F_p$ in time polynomial in $n^{d(G)}$ (resp. $n^{d'(G)}$) and the size of the input, where $G$ is the permutation group $\gal(f/\Q)$ acting on the set of roots of $\tilde{f}$.
\end{cor}

%Therefore, to prove that the algorithm always produces the complete factorization (resp. a proper factorization) of $f$, it suffices to show that any strongly antisymmetric $\mathcal{P}$-scheme is discrete (resp. inhomogeneous) on $H$. 

\paragraph{The unifying framework via the $\mathcal{P}$-scheme algorithm.}

The $\mathcal{P}$-scheme algorithm and the underlying notion of $\mathcal{P}$-schemes provide a unifying framework for deterministic polynomial factoring over finite fields. To illustrate this point, we show that  the main results achieved by known factoring algorithms \citep{Hua91-2, Hua91, Ron88, Ron92, Evd94, IKS09} can be easily derived from Theorem~\ref{thm_algmain2formal} or Corollary~\ref{cor_algdgbound} for the special case that the input polynomial  satisfies Condition~\ref{cond_spoly} (the  general case is solved  in Chapter~\ref{chap_alg_general}).

Suppose we want to factorize $f(X)\in\F_p[X]$ given  a (possibly reducible) lifted polynomial $\tilde{f}(X)\in\Z[X]$ of $f$. We reduce to the case that the lifted polynomial is irreducible as follows: first use the  factoring algorithm for rational polynomials \citep{LLL82} to factorize $\tilde{f}$ into its irreducible factors $f_1(X),\dots,f_k(X)\in\Q[X]$ over $\Q$ in polynomial time.
By Gauss Lemma (see \citep[Section~\RN{4}.2]{Lan02}), we may assume each factor $\tilde{f}_i(X)$ lies in $\Z[X]$.
Then the problem of factoring $f(X)$ is reduced to the problem of factoring each $f_i(X):=\tilde{f}(X)\bmod p\in\F_p[X]$ with the aid of its irreducible lifted polynomial $\tilde{f}_i(X)$. Moreover, for $i\in [k]$, the Galois group $\gal(\tilde{f}_i(X)/\Q)$ is a quotient group of $\gal(\tilde{f}/\Q)$, and hence $|\gal(\tilde{f}_i(X)/\Q)|\leq |\gal(\tilde{f}(X)/\Q)|$.

So assume $\tilde{f}$ is irreducible over $\Q$. Choose $\mathcal{F}=\{F,L\}$ where $F=\Q[X]/(\tilde{f}(X))$ and $L$ is the splitting field of $\tilde{f}$ over $\Q$. Compute $\mathcal{F}$ in time polynomial in $[L:\Q]=\gal(\tilde{f}(X)/\Q)$ and the size of $\tilde{f}$ using Lemma~\ref{lem_compsplit}. By Lemma~\ref{lem_antidisc}, all antisymmetric $\mathcal{P}$-schemes are discrete on $H$ for all $H\in\mathcal{P}$ since the trivial subgroup $\{e\}$ is in $\mathcal{P}$. 
%a fortiori, so are all strongly antisymmetric $\mathcal{P}$-schemes. 
Therefore by  Theorem~\ref{thm_algmain2formal} and the reduction above, we have

\begin{thm}[\citep{Ron92}]\label{thm_ron92}
Under GRH, there exists a deterministic algorithm that, given a polynomial $f(X)\in\F_p[X]$  satisfying Condition~\ref{cond_spoly} and a lifted polynomial $\tilde{f}(X)\in\Z[X]$ of $f$, computes the complete factorization of $f$ over $\F_p$ in time polynomial in $|\gal(\tilde{f}/\Q)|$ and the size of the input.
\end{thm}

Now assume  $\tilde{f}$ is irreducible over $\Q$ and  $\gal(\tilde{f}/\Q)$ is abelian. Then $\gal(\tilde{f}/\Q)$ acts regularly on the set of roots of  $\tilde{f}$. So we have $|\gal(\tilde{f}/\Q)|=\deg(f)$. Then Theorem~\ref{thm_ron92} gives

\begin{cor}[\citep{Hua91-2, Hua91}]\label{cor_huang}
Under GRH, there exists a deterministic algorithm that, given a polynomial $f(X)\in\F_p[X]$  satisfying Condition~\ref{cond_spoly} and a lifted polynomial of $f$ with an abelian Galois group, computes the complete factorization of $f$ over $\F_p$ in polynomial time.
\end{cor}

Suppose only the polynomial $f$ is known. Let $n=\deg(f)$. We may lift $f$ to a degree-$n$ polynomial $\tilde{f}(X)\in\Z[X]$ such that all coefficients of $\tilde{f}$ are in the interval $[0, p-1]$. So the size of $\tilde{f}$ is $O(n\log p)$. Reduce to the case that $\tilde{f}$ is irreducible over $\Q$ as above. As  $\gal(\tilde{f}/\Q)$ is a subgroup of $\sym(n)$, we derive the following theorem from Theorem~\ref{thm_ron92}.
\begin{thm}[\citep{Ron88, Ron92}]\label{thm_ron87}
Under GRH, there exists a deterministic algorithm that, given a  polynomial $f(X)\in\F_p[X]$ of degree $n\in\N^+$ that satisfies Condition~\ref{cond_spoly}, computes the complete factorization of $f$ in time polynomial in $n!$ and $\log p$.
\end{thm}

Alternatively, Theorem~\ref{thm_ron87} can be derived from Corollary~\ref{cor_algdgbound} by noting $d(G)\leq n-1$ (where $G=\gal(\tilde{f}/\Q)$). Similarly, using the bound $d(G)=O(\log n)$ in Lemma~\ref{lem_logbound}, we derive the following theorem from Corollary~\ref{cor_algdgbound}.

%Now suppose we lift $f$ to $\tilde{f}$, reduce to the case that $\tilde{f}$ is irreducible over $\Q$, but compute $\mathcal{F}$ using Lemma~\ref{lem_compstab} instead,
%% instead of  Lemma~\ref{lem_compsplit}
%so that the associated subgroup system $\mathcal{P}$ over $\gal(\tilde{f}/\Q)$ is the system of stabilizers of depth $m$ for some $m\in \N^+$ (with respect to the action of $\gal(\tilde{f}/\Q)$ on
%the set of roots of $\tilde{f}$ in $L$). Here $m$ is chosen to be large enough so that all strongly antisymmetric $\mathcal{P}$-schemes are discrete on $G_\alpha$ for any root $\alpha$ of $\tilde{f}$ in $L$. By Lemma~\ref{lem_logbound}, we may choose $m=O(\log n)$.
%Then by Theorem~\ref{thm_algmain2formal}, we have

\begin{thm}[\citep{Evd94, IKS09}]\label{thm_evd94}
Under GRH, there exists a deterministic algorithm that, given a polynomial $f(X)\in\F_p[X]$ of degree $n\in\N^+$ satisfying Condition~\ref{cond_spoly}, computes the complete factorization of $f$ over $\F_p$ in time polynomial in $n^{\log n}$ and $\log p$.
\end{thm}

%Suppose we choose $m=\ell$ instead, where $\ell$ is the least prime factor of $n$. By Lemma~\ref{lem_integrality_pscheme}, all antisymmetric $\mathcal{P}$-schemes are inhomogeneous on $G_\alpha$.
%So Theorem~\ref{thm_algmain2formal} implies:

By Corollary~\ref{cor_algdgbound} and Lemma~\ref{lem_integrality_pscheme}, we have

\begin{thm}[\citep{Ron88, IKS09}]\label{thm_boundbyell}
Under GRH, there exists a deterministic  algorithm that, given a polynomial $f(X)\in\F_p[X]$ of degree $n>1$  satisfying Condition~\ref{cond_spoly}, computes a proper factorization of $f$ over $\F_p$ in time polynomial in $n^\ell$ and $\log p$, where $\ell$ is the least prime factor of $n$.
\end{thm}

In latter chapters, we also prove (and generalize) the main result of \citep{Evd92} using the $\mathcal{P}$-scheme algorithm. It states that polynomial factoring over finite fields can be solved in deterministic polynomial time under GRH given  a lifted polynomial that has a solvable Galois group.
For more details, see Theorem~\ref{thm_algsolvable} and Theorem~\ref{thm_algsolvableg}.

\chapter{Constructing number fields}\label{chap_constructnum}

In this chapter, we discuss the problem of constructing number fields using a polynomial $g(X)\in \Q[X]$   irreducible over $\Q$. In particular, we prove Lemma~\ref{lem_compsplit} and Lemma~\ref{lem_compstab} as promised before.

In fact, we consider the more general problem of constructing {\em relative number fields}, which we explain now.

\paragraph{Relative number fields.} Recall that a number field $K$ is encoded using the minimal polynomial $h(X)\in\Q[X]$ of a primitive element $\alpha$ of $K$ over $\Q$, i.e., $K=\Q(\alpha)$. Suppose $K_0$ is a number field encoded in this way.
 A {\em relative number field}\index{relative number field} $K$ over $K_0$ is a number field containing $K_0$, encoded by the minimal polynomial $h(X)\in K_0[X]$ of a primitive element $\alpha$ of $K$ over $K_0$ (i.e. $K=K_0(\alpha)$).
We regard $K$ as a $K_0$-algebra by maintaining  its structure constants in the standard $K_0$-basis 
\[
\{1+(h(X)),X+(h(X)),\dots, X^{d-1}+(h(X))\},
\]
 where $d=[K:K_0]$. Note that when $K_0=\Q$, this this the usual way we encode a number field.

Given a number field $K_0$, we discuss various techniques of constructing relative number fields over $K_0$ given a polynomial $g(X)\in K_0[X]$ irreducible over $K_0$. 
%Choosing $K_0=\Q$ gives the original problem.
In particular, we discuss the technique of adjoining roots of polynomials and use it to prove Lemma~\ref{lem_compsplit} and Lemma~\ref{lem_compstab}. 

Motivated by the $\mathcal{P}$-scheme algorithm in Chapter~\ref{chap_pscheme}, we consider the problem of  constructing a collection of (relative) number fields using $g(X)$, such that for the associated subgroup system $\mathcal{P}$, all strongly antisymmetric $\mathcal{P}$-schemes are discrete (resp. inhomogeneous) on a distinguished subgroup $H\in\mathcal{P}$.
We describe a reduction of this problem  to the case that the Galois group of $g(X)$ is a {\em primitive} permutation group. The idea  was essentially introduced in \citep{LM85}, leading to a polynomial-time algorithm that determines if a given rational polynomial is solvable.\footnote{A rational polynomial $g(X)\in\Q[X]$ is solvable if its roots are expressible in the field operations and radicals. It is equivalent to the solvability of the Galois group $\gal(g/\Q)$.}
It was also used in \citep{Evd92} to obtain to a polynomial-time factoring algorithm for $f(X)\in \F_p[X]$, provided that a solvable polynomial $\tilde{f}(X)\in\Z[X]$ lifting $f(X)$ is given. We reproduce the main result of \citep{Evd92} for the case that $f$ satisfies Condition~\ref{cond_spoly}. For the general case, see Chapter~\ref{chap_alg_general}.

We note that most results in this chapter are essentially  known in the literature, except that we state them in a relative setting or in the terminology of $\mathcal{P}$-schemes.
In particular,  the discussion about algebraic numbers in Section~\ref{sec_conspre} follows \citep{WR76}, and the   techniques of constructing number fields  are mostly folklore or from \citep{Lan84, LM85, Evd92}.

\paragraph{Outline of the chapter.} Notations and preliminaries are given in Section~\ref{sec_conspre}. In particular, we define the {\em complexity} of a subgroup system, which is used to bound the size of a collection of (relative) number fields and the running time of the algorithms. This notion also plays a role in subsequent chapters.
In Section~\ref{sec_compositum}, we discuss the technique of constructing  (relative) number fields by adjoining roots of a polynomial, and use it to prove Lemma~\ref{lem_compsplit} and Lemma~\ref{lem_compstab}.
In Section~\ref{sec_consred}, we establish the reduction to primitive Galois groups and use it to prove  the main result of \citep{Evd92} for the special case that $f(X)\in\F_p[X]$ satisfies Condition~\ref{cond_spoly}.
Finally, we discuss some other techniques in Section~\ref{sec_consother}. These techniques are not directly used in the thesis, but may still have their own interest.

\section{Preliminaries}\label{sec_conspre}

%In this section, we first discuss two measures that are closely related to the length of a collection of number fields: the  complexity of a subgroup system and the size of an algebraic integer.  Then we show how to identify a relative number field with an ordinary number field. 

Let $K$ and  $K'$ be relative number fields over a number field $K_0$. We say an embedding (resp. isomorphism) $\tau: K\to K'$ is an embedding (resp. isomorphism) {\em over $K_0$} if $\tau$ is $K_0$-linear, i.e., $\tau(ax)=a\tau(x)$ for all $a\in K_0$ and $x\in K$. By choosing $x=1$, we see that this is equivalent to $\tau(a)=a$ for all $a\in K_0$. We write $K\cong_{K_0} K'$ for the statement that $K$ is isomorphic to $K'$ over $K_0$.
\nomenclature[c1a]{$\cong_{K_0}$}{isomorphism over a field $K_0$}

\paragraph{$(K_0,g)$-subfield systems and the associated subgroup systems.} 
We generalize the notion of $(\Q,g)$-subfield systems (Definition~\ref{defi_expfield}) and the associated subgroup systems (Definition~\ref{defi_asssystem}) as follows:

\begin{defi}[$(K_0,g)$-subfield system]\label{defi_kgcollection}
Let $K_0$ be a number field.
Let $g(X)$ be a polynomial in $K_0[X]$ with the splitting field $L$ over $K_0$.
Let $\mathcal{F}$ be a collection of relative number fields over $K_0$
% given to the algorithm (encoded by their structure constants) 
such that
(1) the fields in $\mathcal{F}$ are mutually non-isomorphic over $K_0$, and (2) each field $K'\in\mathcal{F}$ is isomorphic to a subfield of $L$ over $K_0$.
We say $\mathcal{F}$ is a {\em $(K_0,g)$-subfield system}.\index{subfield system}
\end{defi}

\begin{defi}
Let $g(X)$ be a polynomial in $K_0[X]$ with the splitting field $L$ over $K_0$.
 Let $\mathcal{F}$ be a {\em $(K_0,g)$-subfield system}. 
Define  $\mathcal{P}^\sharp$ to be the poset of subfields of $L$ that includes all the fields isomorphic to those in $\mathcal{F}$ over $K_0$:
\[
\mathcal{P}^\sharp:=\{K'\subseteq L: K'\cong_{K_0} K \text{ for some } K\in \mathcal{F}\}.
\]
By Galois theory, it corresponds to a poset $\mathcal{P}$ of subgroups of $\gal(g/K_0)$, given by
\[
\mathcal{P}:=\left\{H\subseteq\gal(g/K_0): L^H\in \mathcal{P}^\sharp \right\},
\]
which is closed under conjugation in $\gal(g/K_0)$, and hence is a subgroup system over $\gal(g/K_0)$. We say $\mathcal{P}$ and $\mathcal{P}^\sharp$ are {\em associated with $\mathcal{F}$}.
\end{defi}

\paragraph{The complexity of a subgroup system.} The size of a $(K_0,g)$-subfield system $\mathcal{F}$ is primarily controlled by the total degree of the fields in $\mathcal{F}$ over $K_0$, which is the number of coefficients in $K_0$ we need to maintain.
We relate this quantity to the {\em complexity} of a subgroup system, defined as follows.

\begin{defi}[complexity of a subgroup system]\label{defi_complexity}
Suppose $\mathcal{P}$ is a subgroup system over a finite group $G$. Then $G$ acts on $\mathcal{P}$ by conjugation, i.e., $g\in G$ sends $H\in\mathcal{P}$ to $gHg^{-1}\in\mathcal{P}$. Let $\mathcal{P}_0\subseteq \mathcal{P}$ be a complete set of representatives of the $G$-orbits under this action. Define the {\em complexity}\index{complexity of a subgroup system} of $\mathcal{P}$ to be
\[
c(\mathcal{P}):=\sum_{H\in\mathcal{P}_0} [G:H].
\]
\end{defi}
\nomenclature[c1b]{$c(\mathcal{P})$}{complexity of a subgroup system $\mathcal{P}$}

As conjugate subgroups have the same order, the complexity $c(\mathcal{P})$ is well defined. And we have
\begin{lem}
For a $(K_0,g)$-subfield system $\mathcal{F}$, the total degree of the fields in $\mathcal{F}$ over $K_0$ equals $c(\mathcal{P})$,
where  $\mathcal{P}$ is the  subgroup system associated with $\mathcal{F}$.
\end{lem}
\begin{proof}
Conjugate subgroups correspond to conjugate subfields under the Galois correspondence. So for $K\in\mathcal{F}$ there exists a unique subgroup $H\in\mathcal{P}_0$ satisfying $L^H\cong_{K_0} K$. And the map $K\mapsto H$ is a one-to-one correspondence between $\mathcal{F}$ and $\mathcal{P}_0$.
Finally note that $[K:K_0]=[G:H]$ for $H$ corresponding to $K$. 
\end{proof}

%For example, suppose $\mathcal{F}=\{F,L\}$ where $F, L$ are relative number fields over $K_0$, $F\subseteq L$, and $L/K_0$ is Galois with the Galois group $G$. Then the complexity $c(\mathcal{P})$ of the associated subgroup system $\mathcal{P}$ is bounded by $[F:K_0]+[L:K_0]\leq 2|G|$.

The following lemma bounds the complexity of a system of stabilizers.
\begin{lem}\label{lem_complexitystab}
Let $G$ be a finite group acting on a finite set $S$. Let $m\in\N^+$ and $m'=\min\{|S|, m\}$.  Let $\mathcal{P}$ be the system of stabilizers of depth $m'$ with respect to the action of $G$ on $S$. Then
\[
c(\mathcal{P})\leq \sum_{k=1}^{m'} \prod_{i=1}^k (|S|-i)=O\left(|S|^{m'}\right).
\]
\end{lem}
\begin{proof}
Replacing $m$ with $m'$ does not change $\mathcal{P}$. So we may assume $m=m'\leq |S|$.
When $|S|\geq 2$, we have 
\[
\sum_{k=1}^{m} \prod_{i=0}^{k-1} (|S|-i)\leq \sum_{k=1}^{m} |S|^k=O\left(|S|^{m}\right).
\]
The same holds trivially when $|S|=1$.

Next we prove $c(\mathcal{P})\leq \sum_{k=1}^{m} \prod_{i=1}^k (|S|-i)$.
Let $\mathcal{P}_0\subseteq \mathcal{P}$ be as in Definition~\ref{defi_complexity}.
It suffices to find  an injective map  
\[
\tau: \coprod_{H\in\mathcal{P}_0} H\backslash G\hookrightarrow \coprod_{k=1}^m S^{(k)},
\]
since the cardinality of  $\coprod_{H\in\mathcal{P}_0} H\backslash G$ is $c(\mathcal{P})$, whereas the cardinality of  $\coprod_{k=1}^m S^{(k)}$ is $\sum_{k=1}^{m} \prod_{i=1}^k (|S|-i)$.

For each $k\in [m]$, the group $G$ acts diagonally on $S^{(k)}$. For each $H\in\mathcal{P}_0$, we pick $k=k(H)\leq m$ and $x=x(H)\in S^{(k)}$ such that $H=G_x$ with respect to the diagonal action. 
By Lemma~\ref{lem_equivaction}, we have an injective map $H\backslash G\to S^{(k)}$ whose image is the $G$-orbit of $x$.
These maps altogether give the map $\tau$. To show $\tau$ is injective, it suffices to show that for different $H,H'\in\mathcal{P}_0$, the coset spaces $H\backslash G$ and $H'\backslash G$ are mapped to different $G$-orbits. Assume to the contrary that they are mapped to the the same $G$-orbit $O$. So $x(H), x(H')\in O$. Then $k(H)=k(H')$ and $x(H')=\prescript{g}{}{(x(H))}$ for some $g\in G$.  But then we have 
\[
H'=G_{x(H')}=G_{\prescript{g}{}{x(H)}}=gG_{x(H)}g^{-1}=gHg^{-1},
\]
which is a contradiction to the choice of $\mathcal{P}_0$. So $\tau$ is injective.
\end{proof}

\paragraph{Algebraic numbers.}  
The fields in a $(K_0,g)$-subfield system $\mathcal{F}$  are encoded by polynomials in $K_0[X]$. So to bound the size of $\mathcal{F}$, we also need to bound the size of the coefficients of these polynomials, which are algebraic numbers in $K_0$. This is closely related to the following definition, introduced in \citep{WR76}.
\begin{defi}
For an algebraic number $\alpha$, define $\|\alpha\|$ to be the greatest absolute value of $i(\alpha)\in\C$ where $i$ ranges over the embeddings of $\Q(\alpha)$ in $\C$.\footnote{$\|\alpha\|$ is called the size of $\alpha$ in  \citep{WR76}. We reserve the term {\em size} (of an object) for the number of bits used to encode an object in an algorithm.}
\end{defi}

\nomenclature[c1c]{$ \lVert \alpha\rVert$}{greatest absolute value of $i(\alpha)$ where $i$ ranges over embeddings $\Q(\alpha)\hookrightarrow\C$}

For algebraic numbers $\alpha,\beta$, we clearly have $\|\alpha+\beta\|\leq \|\alpha\|+\|\beta\|$ and  $\|\alpha\cdot \beta\|\leq \|\alpha\|\cdot\|\beta\|$. 
%since  embeddings of $\Q(\alpha+\beta)$ and $\Q(\alpha\cdot\beta)$ in $\C$ extend to those of $\Q(\alpha,\beta)$.

The following lemma relates the size of an algebraic number $\alpha\in K_0$ (i.e., the number of bits used to encode $\alpha$ in $K_0$) to $\|\alpha\|$.

%For a number field $K_0$ is a number field encoded by a polynomial $h(X)\in\Q[X]$ of degree $n$ that is irreducible over $\Q$. Let $\alpha$ be a root of $h(X)$. Then $K_0$ is identified with $\Q(\alpha)$, and elements of  $K_0$ are represented in the standard basis $\{1,\alpha,\dots,\alpha^{n-1}\}$. We want to show that elements with bounded coefficients in this representation also have bounded sizes, and vice versa. This is achieved by the following lemma.

\begin{lem}\label{lem_lengthsize}
Suppose $K_0$ is a number field encoded by a polynomial $h(X)\in \Q[X]$ irreducible over $\Q$ of degree $n$ and size $s_0$.
Let $\alpha$ be an algebraic number in $K_0$ of size $s$. Let $D$ be the smallest positive integer such that $D\alpha$ is an algebraic integer.
Then $s$ is polynomial in $\log\|\alpha\|$, $\log D$ and $s_0$. Conversely, $\log\|\alpha\|$ and $\log D$ are polynomial in $s$ and $s_0$.
%and  let $h(X)\in \Q[X]$ be the minimal polynomial of $\alpha$ over $\Q$ of degree $n$ and length $s_0$.
%Let $\beta=\sum_{i=0}^{n-1} c_i \alpha^i\in K_0$.
%\begin{itemize}
%\item Suppose the length of every coefficient $c_i\in\Q$ is bounded by $s$. Then $\log\|\beta\|$ is polynomial in $s$ and $s_0$. Moreover there exists $D\in\N^+$ such that $Dc_i\in\Z$ for $0\leq i<n$ and  $\log D$ is polynomial in $s$ and $n$.
%\item Conversely, the length of every coefficient $c_i\in\Q$ is polynomial in $\log \|\beta\|$, $\log D$ and $s_0$,  where $D$ is any positive integer such that $Dc_i\in\Z$ for $0\leq i<n$.
%\end{itemize}
\end{lem}
\begin{proof}
Suppose $h(X)=\sum_{i=0}^n c_i X_i$ where $n=\deg(h)$ and  $c_i\in\Q$ for all $i$. By substituting $X$ with $X/k$ for some large enough $k\in\N^+$ and clearing the denominators, we may assume $h(X)\in\Z[X]$ and  $c_n=1$. 
Both the encoding of $h$ and that of $\alpha$ use at least $n$ coefficients in $\Q$. So we have $s,s_0\geq n$.

The algebraic number $\alpha\in K_0$ is encoded by the constants $d_0,\dots,d_{n-1}\in\Q$ satisfying 
\begin{equation}\label{eq_expansion}
\alpha=\sum_{i=0}^{n-1} d_i \beta^i,
\end{equation}
where $\beta$ is a root of $h$ in $K_0$. So we have  $\|\alpha\|\leq \sum_{i=0}^{n-1} |d_i|\|\beta\|^i$.
It was   shown  in  \citep{WR76} that $\|\beta\|\leq \sum_{i=0}^{n-1} |c_i|$. And we clearly have $\log |c_i|\leq s_0$ and $\log |d_i|\leq s$ for $0\leq i\leq n-1$. It follows that $\log\|\alpha\|$ is polynomial in $s$ and $s_0$. 

Let $D'\in\N^+$ be the least common multiple of the denominators of $d_i$. As  $h(X)\in\Z[X]$ and  $c_n=1$, we know $\beta$ is an algebraic integer. Then $D'\alpha$ is also an algebraic integer by \eqref{eq_expansion}.
So $D$ is bounded by $D'$. It follows that $\log D$ is polynomial in $s$ and $s_0$. Then the second claim of the lemma is proved.

For the first claim, it suffices to show that the size of each $d_i$ is polynomial in $\log\|\alpha\|$, $\log D$ and $s_0$.
This follows from \citep[Section~7 and Lemma~8.3]{WR76}.
\end{proof}

%
%The following lemma states that for an algebraic number $\alpha$, a bound on $\log \|\alpha\|$ can be turned into a bound on the length of coefficients of the minimal polynomial of $\alpha$ over $\Q$, and vice versa.

%\begin{lem}\label{lem_coeffsize}
%Let $\alpha$ be an algebraic number and let $h(X)\in \Q[X]$ be its minimal polynomial over $\Q$ of degree $n$. 
%\begin{itemize}
%\item Suppose the length of every coefficient of $h$ is bounded by $s$. Then $\log \|\alpha\|=O(s+\log n)$. Moreover $D\alpha$ is an algebraic integer for some $D\in\N^+$ where $\log D$  is polynomial in $s$ and $n$.  
%\item Conversely, the length of every coefficient of $h$ is polynomial in $\log\|\alpha\|$, $\log D$ and $n$, where $D$ is any positive integer for which $D\alpha$ is an algebraic integer. 
%\end{itemize}
%\end{lem}
%\begin{proof}
%Suppose $h(X)=\sum_{i=0}^n c_i X^i$, where $c_i\in Q$ and $c_n=1$. The first claim follows from the bound $\|\alpha\|\leq \sum_{i=0}^{n-1} |c_i|$, as shown in  \citep{WR76}. The integer $D$ can be chosen the least common multiple of the denominators of $c_i$. For the second claim, we may assume $\alpha$ is an algebraic integer by replacing $\alpha$ with $D\alpha$ and $c_i$ with $D^{n-i} c_i$. Then the coefficients of $h$ are integers. The claim follows by noting that up to sign, each coefficient $c_i$ is given by the $i$th elementary symmetric polynomial in the roots of $h$, and these roots  have the same size $\log\|\alpha\|$.
%\end{proof}

The following lemma relates the size of the minimal polynomial of an algebraic number $\alpha$ over a number field $K_0$ to $\|\alpha\|$.

\begin{lem}\label{lem_coeffsizeg}
Suppose $K_0$ is a number field encoded by a rational polynomial irreducible over $\Q$ of size $s_0$ (let $s_0=1$ if $K_0=\Q$).  
Let $\alpha$ be an algebraic number,  and let $D$ be the smallest positive integer such that $D\alpha$ is an algebraic integer.
Let $h(X)\in K_0[X]$ be the minimal polynomial of $\alpha$ whose size is $s$ and degree is $n$. 
Then $s$ is polynomial in $\log\|\alpha\|$, $\log D$, $s_0$ and $n$. Conversely, $\log\|\alpha\|$ and $\log D$ are polynomial in $s$ and $s_0$.
%\begin{itemize}
%\item Suppose the length of every coefficient of $h$ is bounded by $s$. Then $\log \|\alpha\|$ is polynomial in $s$, $s_0$ and $\log n$. Moreover $D\alpha$ is an algebraic integer for some $D\in\N^+$ where $\log D$  is polynomial in $s$, $s_0$ and $n$.  
%\item Conversely, the length of every coefficient of $h$ is polynomial in $\log\|\alpha\|$, $\log D$, $s_0$ and $n$, where $D$ is any positive integer for which $D\alpha$ is an algebraic integer. 
%\end{itemize}
\end{lem}
\begin{proof}
We clearly have $n\leq s$.
Suppose $h(X)=\sum_{i=0}^n c_i X^i$,  where $c_i\in K_0$ and $c_n=1$. It was as shown in \citep{WR76} that $\|\alpha\|\leq \sum_{i=0}^{n-1} \|c_i\|$.
It follows from   Lemma~\ref{lem_lengthsize} that $\log\|\alpha\|$ is polynomial in $s$ and $s_0$. 

Note that for sufficiently large $k\in\N^+$ that is polynomial in $s$ and $s_0$, the coefficients of the polynomial $k^n h(X/k)$ are all algebraic integers. It follows that $k\alpha$ is an algebraic integer (cf. \citep[Corollary~5.4]{AM69}). So $D$ is bounded by $k$ and hence is polynomial in $s$ and $s_0$. Then the second claim of the lemma is proved.

For the first claim, we may assume $\alpha$ is an algebraic integer by replacing $\alpha$ with $D\alpha$ and $c_i$ with $D^{n-i} c_i$. Then any conjugate $\alpha'$ of $\alpha$ over $\Q$ is also an algebraic integer, and $\|\alpha'\|=\|\alpha\|$.
For $0\leq i\leq n-1$, the coefficient $c_i$ of $h$ is (up to sign) given  by the $i$th elementary symmetric polynomial in a subset of conjugates of $\alpha$ over $\Q$. It follows from Lemma $\ref{lem_lengthsize}$  that the size of each $c_i$ is polynomial in  $\log\|\alpha\|$, $\log D$, $s_0$ and $n$. So $s$ is  polynomial in  $\log\|\alpha\|$, $\log D$, $s_0$ and $n$ as well.
\end{proof}

\paragraph{Finding a primitive element over $\Q$.} Suppose $K_0=\Q(\alpha)$ is a number field encoded by
the minimal polynomial of a primitive element $\alpha$ over $\Q$, 
and $K=K_0(\beta)$ is a relative number field over $K_0$, encoded by the minimal polynomial of a primitive element $\beta$ over $K_0$. We would like to represent $K$ directly in the form $\Q(\gamma)$, encoded by the minimal polynomial of a primitive element $\gamma$ over $\Q$. The first step is to find such an element $\gamma$, which can be achieved using a constructive version of the primitive element theorem (see, e.g., \citep{VW91}). For completeness, we give the details as follows. 

\begin{lem}\label{lem_primitive}\index{primitive element theorem}
Suppose $K_0$ is a number field and $\alpha,\beta$ are algebraic numbers. Let $d=[K_0(\alpha,\beta): K_0]$. Then $k\alpha+\beta$ is a primitive element of $K_0(\alpha,\beta)$ over $K_0$ for some integer $k\in [1,d+1]$. 
\end{lem}
\begin{proof}
Consider a ``bad'' nonzero integer $k$ for which $K_0(k\alpha+\beta)$ is a proper subfield of $K_0(\alpha,\beta)$. Let $L$ be the Galois closure of $K_0(\alpha,\beta)/K_0$. Then by the fundamental theorem of Galois theory, there exists an automorphism $\phi$ of $L$ fixing $K_0(k\alpha+\beta)$ but not  $K_0(\alpha,\beta)$. Then either $\phi(\alpha)\neq \alpha$ or $\phi(\beta)\neq \beta$.
As $\phi$ fixes $k\alpha+\beta$, we have $k\phi(\alpha)+\phi(\beta)=\phi(k\alpha+\beta)=k\alpha+\beta$, from which we see that actually  $\phi(\alpha)\neq \alpha$ and $\phi(\beta)\neq \beta$ both hold.
Then $k$ is determined by $\phi(\alpha)$ and $\phi(\beta)$ via $k=(\phi(\beta)-\beta)/(\alpha-\phi(\alpha))$.
So the number of bad  choices of $k$ is bounded by the number of  $(\phi(\alpha),\phi(\beta))$ where $\phi$ ranges over the automorphisms of $L$ fixing $K_0$. The later is the cardinality of the orbit of $(\alpha, \beta)$ under the action of $\gal(L/K_0)$. By the orbit-stabilizer theorem, it equals 
\[
[\gal(L/K_0):\gal(L/K_0(\alpha,\beta))]=[K_0(\alpha,\beta): K_0]=d.
\]
So there are at most $d$  bad  choices of $k$.
The lemma follows since $[1,d+1]$ contains more than $d$ integers.
\end{proof}

This gives an efficient algorithm of finding a primitive element over $\Q$:

\begin{lem}\label{lem_compprimitive}
There exists a polynomial-time algorithm that given a number field $K_0$ and a relative number field $K$ over $K_0$, find  a primitive element $\gamma$ of $K$ over $\Q$ and  its minimal polynomial $h(X)\in\Q[X]$ over $\Q$.
%and compute the isomorphism $K\cong \Q[X]/(h(X))$ sending $\gamma$ to $X+(h(X))$.
\end{lem}
\begin{proof}
Suppose $K_0$ is encoded by a polynomial $g(X)\in\Q[X]$ irreducible over $\Q$, and $K$ is encoded by a polynomial $g'(X)\in K_0[X]$ irreducible over $K_0$. Then we are explicitly given a root $\alpha$ of $g(X)$ and a root $\beta$ of $g'(X)$ in $K$, and $K=\Q(\alpha,\beta)$. 

Enumerate the integers $k\in [1, d+1]$, where $d=[K:\Q]$. For each $k$, we compute $\gamma=k\alpha+\beta\in K$, and then compute its minimal polynomial  $h(X)\in\Q[X]$ over $\Q$ by solving linear equations over $\Q$. 
This step runs in polynomial time by  Lemma~\ref{lem_coeffsizeg}. 
Output $\gamma$ and $h$ whenever $\deg(h)=[K:\Q]$. By Lemma~\ref{lem_primitive}, a primitive element $\gamma$ is guaranteed to be found. 
\end{proof}

By computing a primitive element over $\Q$, we can  efficiently turn a relative number field into an ordinary number field:

\begin{cor}\label{lem_reltoord}
There exists a polynomial-time algorithm that given a number field $K_0$ and a relative number field $K$ over $K_0$,  computes  an ordinary number field $K'$, a $\Q$-basis $B$ of $K$, and an isomorphism $\phi: K\to K'$ encoded by $\phi(x)\in K'$ for $x\in B$.
\end{cor}
\begin{proof}
Find  a primitive element $\gamma$ of $K$ over $\Q$ and its minimal polynomial $h(X)\in\Q[X]$ over $\Q$ using Lemma~\ref{lem_compprimitive}.
Compute $K':=\Q[X]/(h(X))$ and $B=\{1,\gamma,\gamma^2,\dots,\gamma^{d-1}\}$, where $d=[K:\Q]$.
Then compute the isomorphism $\phi: K\to K'$, which sends $\gamma^i$ to $X^i+(h(X))$ for $i=0,1,\dots,d-1$.
\end{proof}

As an application, we generalize  Lemma~\ref{lem_compembed} to obtain an efficient algorithm that computes embeddings of relative number fields over a given number field. 

\begin{lem}\label{lem_comprelembed}
There exists a polynomial-time algorithm $\mathtt{ComputeRelEmbeddings}$ that given a number field $K_0$ and relative number fields $K$ and $K'$ over $K_0$, computes all the embeddings of $K$ in $K'$ over $K_0$.
\end{lem}
\begin{proof}
Identify $K$ and $K'$ with ordinary number fields using  Corollary~\ref{lem_reltoord}.
Run the algorithm $\mathtt{ComputeEmbeddings}$ in Lemma~\ref{lem_compembed} to compute all the embeddings of $K$ in $K'$, and ignore those not fixing $K_0$.  
%Suppose $K_0$, $K$ and $K'$ are given in the $\Q$-basis  $B_0=\{1, \alpha,\alpha^2,\dots\}$, $K_0$-basis  $B=\{1, \beta,\beta^2,\dots\}$ and $K_0$-basis  $B'=\{1, \beta',\beta'^2,\dots\}$ respectively.
%Run the algorithm in  Lemma~\ref{lem_compprimitive} to compute primitive elements $\gamma$, $\gamma'$ of $K$ and $K'$ over $\Q$ respectively together with their minimal polynomials $h(X), h'(X)\in \Q[X]$. Then we identify $K$ with $\Q[X]/(h(X))$ using the $\Q$-basis $\{1, \gamma,\gamma^2,\dots\}$  and similarly identify $K'$ with  $\Q[X]/(h'(X))$. Run the algorithm  $\mathtt{ComputeEmbeddings}$ in Lemma~\ref{lem_compembed} to compute all embeddings from $K$ to $K'$. For each embedding $\phi$, we check if $\phi$ is defined over $K_0$, which is achieved by checking if the elements of the  $\Q$-basis $B_0$ of $K_0$ is fixed by $\phi$. If so we get an embedding $\phi$ over $K_0$. Finally compute  $\phi(x)$ for $x\in B$ and express them in the $K_0$-basis $B'$ of $K'$, so that $\phi$ is represented as an embedding of the relative number field $K$ in $K'$ over $K_0$.
\end{proof}

\section{Adjoining roots of polynomials}\label{sec_compositum}

One of the most basic techniques of constructing number fields is adjoining roots of polynomials. It can be  efficiently performed by the following lemma.

\begin{lem}\label{lem_comp}
There exists a polynomial-time algorithm $\mathtt{AdjoinRoot}$ that given a number field $K_0$, a relative number field $K$ over $K_0$, and a polynomial $h(X)\in K[X]$  irreducible over $K$, computes the relative number field $K'=K(\alpha)$ over $K_0$ (up to isomorphism over $K_0$), where $\alpha$ is an arbitrary root of $h(X)$.
Moreover, suppose $K$ is encoded by the minimal polynomial of a primitive element $\beta\in K$ over $K_0$.
Then $K'$ is encoded  by the minimal polynomial of an element of the form $\beta+k\alpha$ over $K_0$, where $1\leq k\leq [K': K_0]+1$.
\end{lem}

\begin{proof}
Form the $K$-algebra $K'':=K[X]/(h(X))$ which is a field. 
We need to encode $K''$ as a relative number field over $K_0$.
Let $\alpha:=X+(h(X))\in K''$ which is a root of $h(X)$. 
Then $\alpha$ and $\beta$ are explicitly known in $K''$.
Let $d:=[K'':K_0]+1$.
By Lemma~\ref{lem_primitive}, there exists $k\in [1,d+1]$ such that $\gamma=\beta+k\alpha$ is a primitive element of $K''$ over $K_0$. Compute such an element $\gamma$ by enumerating $k$ and checking if the degree of the minimal polynomial  of $\gamma$ over $K_0$ equals $d$.
Once $\gamma$ is found, compute the relative number field $K':=K_0[X]/(g(X))$ over $K_0$, where $g(X)$ is the minimal polynomial of $\gamma$ over $K_0$.
It is isomorphic to $K''=K(\alpha)$ over $K_0$ via the $K_0$-linear map sending $X+(g(X))$ to $\gamma$.
\end{proof}

By repeatedly adjoining roots, we obtain an algorithm that computes the splitting field of a given irreducible polynomial over a number field $K_0$. See Algorithm~\ref{alg_splittingfield}.

\begin{algorithm}[htbp]
\caption{$\mathtt{SplittingField}$}\label{alg_splittingfield}
\begin{algorithmic}[1]
\INPUT  number field $K_0$ and $g(X)\in K_0[X]$ irreducible over $K_0$
\OUTPUT the splitting field of $g$ over $K_0$ as a relative number field over $K_0$
\State $K\gets K_0$, regarded as a relative number field over $K_0$
\State factorize $g$ over $K$
\While{$g$ has an irreducible non-linear factor over $K$}
    \State pick an irreducible non-linear factor $g_0$ of $g$ over $K$
    \State run $\mathtt{AdjoinRoot}$ on $(K_0, K, g_0)$ to obtain $K'$
    \State $K\gets K'$  \strut  
    \State factorize $g$ over $K$
\EndWhile
\State \Return $K$
\end{algorithmic}
\end{algorithm} 

Line 2 and Line 7 are implemented using the polynomial-time factoring algorithms for number fields \citep{Len83, Lan85}.\footnote{Here we factorize $g$ over the relative number field $K$. It can be reduced to the problem of factoring polynomials over an ordinary number field by Corollary~\ref{lem_reltoord}.} 
And we have

\begin{lem}\label{lem_compsplitgeneral}
Given a number field $K_0$ and a polynomial $g(X)\in K_0[X]$ irreducible over $K_0$, the algorithm $\mathtt{SplittingField}$ computes the splitting field $K$ of $g$ over $K_0$ in time polynomial in $[K:K_0]$ and the size of the input.
\end{lem}
\begin{proof}
The algorithm initializes $K$ to $K_0$ and  keeps adjoining roots of $g$ to $K$ until it contains all these roots. The resulting field $K$ is by definition the splitting field of $g$ over $K_0$. At most $t:=\log [K:K_0]$ intermediate fields are constructed other than $K_0$. By induction and Lemma~\ref{lem_comp}, each intermediate field is encoded by the minimal polynomial of a primitive element $k_1\alpha_1+\dots+k_s\alpha_s$ over $K_0$ where $s\leq t$, all $\alpha_i$ are roots of $g$ and $1\leq k_i\leq [K:K_0]+1$.
The claim about the running time then follows from Lemma~\ref{lem_coeffsizeg} and Lemma~\ref{lem_comp}.
\end{proof}

Choosing $K_0=\Q$ proves Lemma~\ref{lem_compsplit}. Similarly, we have an algorithm constructing a $(K_0,g)$-subfield system whose associated subgroup system is a system of stabilizers. See Algorithm~\ref{alg_stabilizers} below.
 
 \begin{algorithm}[htbp]
\caption{$\mathtt{Stabilizers}$}\label{alg_stabilizers}
\begin{algorithmic}[1]
\INPUT  number field $K_0$,  $m\in\N$, and $g(X)\in K_0[X]$ irreducible over $K_0$
\OUTPUT $(K_0,g)$-subfield system  $\mathcal{F}$ 
%\State compute $F=K_0[X]/(g(X))$
%\State $\mathcal{F}\gets \{F\}$
\If{$m=0$}
\State \Return $\emptyset$
\EndIf
\State $m\gets \min(\deg(g),m)$
\State $\mathcal{F}\gets \{K_0[X]/(g(X))\}$
\ForTo{$i$}{$2$}{$m$}
   \State $\mathcal{F}_{\mathrm{old}}\gets\mathcal{F}$
   \For{$K\in\mathcal{F}_{\mathrm{old}}$}
      \State factorize $g$ over $K$
         \For{irreducible non-linear factor $g_0$ of $g$ over $K$}
              \State  run $\mathtt{AdjoinRoot}$ on $(K_0, K, g_0)$  to obtain $K'$
              \If{$K'$ is non-isomorphic to all fields in $\mathcal{F}$ over $K_0$}
                 \State $\mathcal{F}\gets\mathcal{F}\cup\{K'\}$
              \EndIf
         \EndFor
   \EndFor
\EndFor
\State \Return $\mathcal{F}$
\end{algorithmic}
\end{algorithm} 

Again, Line 8 is implemented using the polynomial-time factoring algorithms for number fields \citep{Len83, Lan85}.
The condition at Line 11 is checked using the algorithm  $\mathtt{ComputeRelEmbeddings}$ in Lemma~\ref{lem_comprelembed}. 

We have the following lemma.

\begin{lem}\label{lem_compstabsysg}
Given a number field $K_0$, an integer $m\in\N$, and a polynomial $g(X)\in K_0[X]$ irreducible over $K_0$, the algorithm $\mathtt{Stabilizers}$ computes a $(K_0,g)$-subfield system $\mathcal{F}$,
such that the subgroup system $\mathcal{P}$ associated with $\mathcal{F}$ is the system of stabilizers of depth $m$ over $G(g/K_0)$ with respect to the action of $\gal(g/K_0)$ on the set of roots of $g$ in $L$,
where $L$ denotes the splitting field of $g$ over $K_0$.
Moreover, the algorithm runs in time polynomial in $c(\mathcal{P})$ and the size of the input.
\end{lem}

\begin{proof}
If $m=0$, the algorithm simply returns $\mathcal{F}=\emptyset$.
It replaces $m$ with $\min(\deg(g),m)$ at Line 3, which does not change the desired subgroup system. So we may assume $m\leq\deg(g)$.
The condition at Line 11 guarantees that the fields in $\mathcal{F}$ are mutually non-isomorphic over $K_0$.
For $k\in [m]$, let $\mathcal{P}_k$ be the the system of stabilizers of depth $k$ over $G(g/K_0)$ with respect to the action of $\gal(g/K_0)$ on the set of roots of $g$ in $L$, and let $\mathcal{P}_k^\sharp$ be the  corresponding poset of  subfields of $L$ determined by the Galois correspondence.
Then $\mathcal{P}_k^\sharp$ consists of the fields of the form $K_0(\alpha_1,\dots,\alpha_i)$, where  $i\in [k]$ and $\alpha_1,\dots, \alpha_i$ are roots of $g$ in $L$. 

We want to show that at the end of the algorithm, the subgroup system $\mathcal{P}$ associated with  $\mathcal{F}$ equals $\mathcal{P}_m$.
And it suffices to prove that for $k\in [m]$, after the $k$th iteration of the loop in Lines 5--12,  every field in $\mathcal{F}$ is isomorphic to some field in $\mathcal{P}_k^\sharp$ over $K_0$ and vice versa.  This  follows from a simple induction on $k$. 

%Note that $c(\mathcal{P}_{S,m})$ is the sum of the degrees of the fields in $\mathcal{F}$ over $K_0$.
Denote by $d$ the maximum degree of the fields in $\mathcal{F}$ over $K_0$. Then $d$ and $|\mathcal{F}|$ are bounded by $c(\mathcal{P})$. By  induction and Lemma~\ref{lem_comp}, each field in $\mathcal{F}$ is encoded by the minimal polynomial of a primitive element $k_1\alpha_1+\dots+k_s\alpha_s$ over $K_0$ where $s\leq m\leq \deg(g)$, all $\alpha_i$ are roots of $g$, and $1\leq k_i\leq d+1$.  The claim about the running time then follows from  Lemma~\ref{lem_coeffsizeg} and Lemma~\ref{lem_comp}.
%We prove this claim by induction on $k$. The claim is obvious for $k=1$. For $k>1$, assume the claim holds for $k-1$. Then any field $K'$ added to $\mathcal{F}$ is the output of the algorithm $\mathtt{AdjoinRoot}$ on  a triple $(K_0, K, g_0)$ where $K\in\mathcal{F}$ is isomorphic to some field $K_0(\alpha_1,\dots,\alpha_i)\in \mathcal{F}_{k-1}$  over $K_0$ by assumption, and $g_0$ is irreducible over $K$. Let $\phi: K\to K_0(\alpha_1,\dots,\alpha_i)$ be an isomorphism over $K_0$. Identifying $K$ with $K_0(\alpha_1,\dots,\alpha_i)$ via $\phi$ and applying Lemma~\ref{lem_comp}, we see that up to an isomorphism over $K_0$, the field $K'$ is obtained  by adjoining to $K_0(\alpha_1,\dots,\alpha_i)$ a root $\alpha$ of $\phi(g_0)$ in $L$ which is also a root of $\phi(g)=g$.\footnote{Here we extend $\phi$ to an isomorphism $K[X]\to K_0[\alpha_1,\dots,\alpha_i]$, and it fixes $g\in K_0[X]$ since $\phi$ is defined over $K_0$.} The resulting field $K_0(\alpha_1,\dots,\alpha_i,\alpha)$ is clearly in $\mathcal{F}_k$. Conversely, let $K'$ be a field in $\mathcal{F}_k$. Then $K'=K_0(\alpha_1,\dots,\alpha_i)$ where $i\in [k]$ and $\alpha_1,\dots,\alpha_i$ are roots of $g$ in $L$. The field $K_0(\alpha_1,\dots,$
\end{proof}

By Lemma~\ref{lem_complexitystab}, the complexity $c(\mathcal{P})$ of the subgroup system $\mathcal{P}$ in Lemma~\ref{lem_compstabsysg} is bounded by $(\deg(g))^{m'}$, where $m'=\min\{\deg(g), m\}$. Lemma~\ref{lem_compstab} then follows by choosing $K_0=\Q$.

\section{Reduction to primitive group actions}\label{sec_consred}

Suppose $K_0$ is a number field, $g(X)\in K_0[X]$ is irreducible over $K_0$, and $L$ is the splitting field of $g$ over $K_0$.
The Galois group $\gal(g/K_0)=\gal(L/K_0)$ acts faithfully and transitively on the set of roots of $g$ in $L$, 
%\footnote{To see the faithfulness, let $N$ be the kernel of  the corresponding permutation representation. Then $N$ fixes all roots of $g$ and hence also fixes $L$. So $N$ is trivial.} 
 and hence is a transitive permutation group on this set. 

Motivated by Theorem~\ref{thm_algmain2formal}, we are interested in the problem of  constructing a $(K_0, g)$-subfield system $\mathcal{F}$   such that  all strongly antisymmetric $\mathcal{P}$-schemes are discrete (resp. inhomogeneous) on $H$, where $\mathcal{P}$  is the subgroup system over $\gal(g/K_0)$ associated with $\mathcal{F}$, and $H$ is a subgroup in $\mathcal{P}$ satisfying $L^H\cong_{K_0} K_0[X]/(g(X))$.
In this section, we describe a reduction, based on the work  \citep{LM85, Evd92}, that reduces the problem to the special case that $\gal(g/K_0)$ is a {\em primitive} permutation group.

\begin{defi}[primitive permutation group]
Suppose $G$ is a permutation group on a finite set $S$. A nonempty subset $B$ of $S$ is called a {\em set of imprimitivity}\index{set of imprimitivity}\footnote{A set of imprimitivity is also called a {\em block} by some authors. We reserve the term {\em block} to denote a set in a partition instead.}  of $G$ if for all $g\in G$, either $\prescript{g}{}{B}=B$ or $B\cap\prescript{g}{}{B}=\emptyset$.
A set of imprimitivity is {\em trivial} if it is a singleton or the whole set $S$. We say $G$ is {\em primitive}\index{primitive!permutation group} if  it only has trivial sets of imprimitivity. Otherwise $G$ is {\em imprimitive}\index{imprimitive!permutation group}. 
\end{defi}

It is well known that for transitive permutation groups, primitivity is equivalent to maximality of stabilizers.
\begin{lem}\label{lem_maximal}
Let $S$ be a finite set where $|S|>1$, and let $x\in S$. A transitive permutation group $G$ on $S$ is primitive iff $G_x$ is maximal in $G$.
\end{lem}
See, e.g., \citep{Wie14} for the proof of Lemma~\ref{lem_maximal}.  We also need the following result, proved in \citep{LM85}.

\begin{thm}[\citep{LM85}]\label{thm_lm}
There exists a polynomial-time algorithm $\mathtt{Tower}$ that given a number field $K_0$ and a polynomial $g(X)\in K_0[X]$ irreducible over $K_0$,\footnote{The paper \citep{LM85} presented their algorithm only for $K_0=\Q$, but it easily extends to a general base field $K_0$.} computes a tower of relative number fields over $K_0$ 
\[
K_{0} \subseteq K_{1}\subseteq \dots \subseteq K_{k-1}\subseteq K_k
\]
together with the inclusions $K_{i-1}\hookrightarrow K_i$ and the polynomials $g_i(X)\in K_{i-1}[X]$ irreducible over $K_{i-1}$ for $i\in [k]$, such that $K_k\cong_{K_0} K_0[X]/(g(X))$, and the following conditions are satisfied for $i\in [k]$: 
\begin{enumerate}
\item $K_i$ is isomorphic to $K_{i-1}[X]/(g_i(X))$ over $K_{i-1}$, and
\item the Galois group $G_i:=\gal(L_i/K_{i-1})$ acts primitively on the set of roots of $g_i$ in $L_i$, where $L_i$ is the Galois closure of $K_i/K_{i-1}$.
\end{enumerate}
\end{thm}

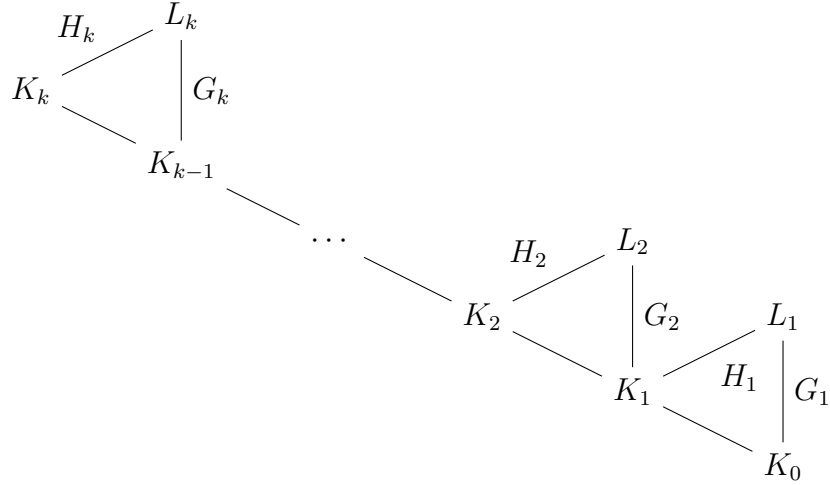
\begin{figure}[htb]
\centering
 \begin{tikzpicture}[auto]
 \node (Q) {$K_{0}$};
 \node (K1) [above of=Q, left of=Q, left of=Q] {$K_1$};
 \node (K2) [above of=K1, left of=K1,  left of=K1] {$K_2$};
 \node (C) [above of=K2, left of=K2,  left of=K2] {$\cdots$};
 \node (Kk2) [above of=C, left of=C,  left of=C] {$K_{k-1}$};
 \node (Kk) [above of=Kk2, left of=Kk2,  left of=Kk2] {$K_k$};
 \node (L1) [above of=Q, above of=Q] {$L_1$};
 \node (L2) [above of=K1, above of=K1] {$L_2$};
 \node (Lk) [above of=Kk2, above of=Kk2] {$L_k$};
 \draw[-] (Q) to node {} (K1);
 \draw[-] (K1) to node {} (K2);
 \draw[-] (K2) to node {} (C);
 \draw[-] (C) to node {} (Kk2);
 \draw[-] (Kk2) to node {} (Kk);
 \draw[-] (L1) to node {$G_1$} (Q);
 \draw[-] (L1) to node {$H_1$} (K1);
 \draw[-] (L2) to node {$G_2$} (K1);
 \draw[-] (K2) to node {$H_2$} (L2);
 \draw[-] (Lk) to node {$G_k$} (Kk2);
 \draw[-] (Kk) to node {$H_k$} (Lk);
 \end{tikzpicture}
\caption{The tower of fields and Galois groups  in Theorem~\ref{thm_lm}}
\label{fig_tower}
\end{figure}

For $i\in [k]$, let $H_i:=\gal(L_i/K_i)\subseteq G_i$.  See Figure~\ref{fig_tower} for an illustration. Note that the first condition above is equivalent to $K_i=K_{i-1}(\alpha)$ for some root $\alpha_i$ of $g_i$ in $L_i$. So $H_i$ is the stabilizer of $\alpha_i$. Then the second condition  is equivalent to maximality of $H_i$ in $G_i$.

The following theorem is the main result of this section.

 \begin{thm}\label{thm_primitivereduction}
Suppose there exists an algorithm $\mathtt{PrimitiveAction}$ that, given a number field $K_0$ and a polynomial $g(X)\in K_0[X]$ irreducible over $K_0$ with $\gal(g/K_0)$ acting primitively on the set of roots of $g$ in $L$, where $L$ is the splitting field of $g$ over $K_0$, computes a $(K_0,g)$-subfield system in time $T(K_0, g)$.
Then there exists an algorithm $\mathtt{GeneralAction}$ that given $K_0$ and  $g$ as above, but {\em without} the assumption that  $\gal(g/K_0)$  acts primitively on $S$, computes  
  \begin{itemize}
  \item  a  $(K_0,g)$-subfield system $\mathcal{F}$, and, 
 \item a tower of relative number fields $K_{0} \subseteq K_{1}\subseteq \dots \subseteq K_{k-1}\subseteq K_k$ over $K_0$ and $g_i(X)\in K_{i-1}[X]$ for $i\in [k]$ satisfying the conditions in Theorem~\ref{thm_lm}, such that $K_k\cong_{K_0} K_0[X]/(g(X))$ and the sizes of the polynomials $g_i$ are polynomial in the size of the input
 \end{itemize}
  in time polynomial in $\sum_{i=1}^k T(K_{i-1}, g_i)$ and the size of the input. 
  Moreover, if for each $i\in [k]$, the $(K_{i-1},g_i)$-subfield system $\mathcal{F}_i$ computed by  $\mathtt{PrimitiveAction}$ on the input $(K_{i-1}, g_i)$
  satisfies
  \begin{enumerate}
\item $K_{i-1}[X]/(g_i(X))\in \mathcal{F}_i$,
\item All strongly antisymmetric $\mathcal{P}$-schemes are discrete (resp. inhomogeneous) on $H$, where $\mathcal{P}$ is the subgroup system over $\gal(g_i/K_{i-1})$ associated with $\mathcal{F}_{i}$ and $H$ is a subgroup in $\mathcal{P}$ whose fixed subfield is isomorphic to $K_i$ over $K_{i-1}$.
\end{enumerate}
Then $\mathcal{F}$ satisfies
  \begin{enumerate}
\item $K_0[X]/(g(X))\in \mathcal{F}$,
\item All strongly antisymmetric $\mathcal{P}$-schemes are discrete (resp. inhomogeneous) on $H$, where $\mathcal{P}$ is the subgroup system over $\gal(g/K_0)$ associated with $\mathcal{F}$ and $H$ is a subgroup in $\mathcal{P}$ satisfying $L^H\cong_{K_0} K_0[X]/(g(X))$.
\end{enumerate}
\end{thm}

See Algorithm~\ref{alg_gaction} for the pseudocode of the algorithm $\mathtt{GeneralAction}$.  
It proceeds as follows:
maintain $\mathcal{F}$, which initially only contains $K_0[X]/(g(X))$. 
Then we call the algorithm $\mathtt{Tower}$  to compute a tower $K_{0} \subseteq K_{1}\subseteq \dots \subseteq K_{k-1}\subseteq K_k$ and  $g_i(X)\in K_{i-1}[X]$ for $i\in [k]$ as in Theorem~\ref{thm_lm}.
Next, run the hypothetical algorithm  $\mathtt{PrimitiveAction}$  in Theorem~\ref{thm_primitivereduction} on $(K_{i-1},g_i)$ for each $i\in [k]$ to obtain a $(K_{i-1},g_i)$-subfield system $\mathcal{F}_i$.
For $i\in [k]$, add the fields in $\mathcal{F}_i$ to $\mathcal{F}$, but encode them as relative number fields over $K_0$ (using  Lemma~\ref{lem_compprimitive}). In addition,  avoid adding fields to $\mathcal{F}$ that are isomorphic to some existent field $K\in \mathcal{F}$ over $K_0$, so that all the fields in $\mathcal{F}$ are mutually non-isomorphic over $K_0$. After all $\mathcal{F}_i$ are processed, output $\mathcal{F}$.

 \begin{algorithm}[htbp]
\caption{$\mathtt{GeneralAction}$}\label{alg_gaction}
\begin{algorithmic}[1]
\INPUT  number field $K_0$  and $g(X)\in K_0[X]$ irreducible over $K_0$
\OUTPUT $(K_0,g)$-subfield system $\mathcal{F}$
%\State compute $F=K_0[X]/(g(X))$
\State $\mathcal{F}\gets \{K_0[X]/(g(X))\}$
\State run  $\mathtt{Tower}$ on $(K_0, g)$ to obtain a tower $K_{0} \subseteq K_{1}\subseteq \dots \subseteq K_{k-1}\subseteq K_k$ and  $g_i(X)\in K_{i-1}[X]$ irreducible over $K_{i-1}$ for $i\in [k]$
\ForTo{$i$}{$1$}{$k$}
   \State run $\mathtt{PrimitiveAction}$ on $(K_{i-1}, g_i)$ to obtain   $\mathcal{F}_i$
   \For{$K\in\mathcal{F}_i$}
      \State compute a relative number field $K'$ over $K_0$ such that $K'\cong_{K_0} K$
      \If{$K'$ is non-isomorphic to all fields in $\mathcal{F}$ over $K_0$}
         \State $\mathcal{F}\gets\mathcal{F}\cup\{K'\}$
      \EndIf
   \EndFor
\EndFor
\State \Return $\mathcal{F}$
\end{algorithmic}
\end{algorithm}

The proof of Theorem~\ref{thm_primitivereduction} is based on the following lemma.

\begin{lem}\label{lem_aggregation}
Let $k\in \N^+$ and $G_k\subseteq G_{k-1}\subseteq\dots\subseteq G_1\subseteq G_0$ be a chain of finite groups.
For $i\in [k]$, let $N_i$ be a subgroup of $G_i$ that is normal in $G_{i-1}$, $\pi_i: G_{i-1}\to G_{i-1}/N_i$ be the corresponding quotient map, and $\mathcal{P}_i$ be a subgroup system over $G_{i-1}/N_i$ that contains $G_i/N_i$. Define
\[
\mathcal{P}=\{g\pi_i^{-1}(H)g^{-1}: 1\leq i\leq k, H\in \mathcal{P}_i, g\in G_0\},
\]
which is a subgroup system over $G_0$ and contains $\pi_i^{-1}(G_i/N_i)=G_i$ for all $i\in [k]$. 
%\footnote{Here $\pi_i^{-1}(H)\subseteq G_{i-1}$ is regarded as a subgroup of $G_0$.}
Then we have
\begin{enumerate}
\item If  for all $i\in [k]$, all strongly antisymmetric $\mathcal{P}_i$-schemes are discrete on $G_i/N_i$, then all strongly antisymmetric $\mathcal{P}$-schemes are discrete on $G_k$. 
\item  If  for some $i\in [k]$, all strongly antisymmetric $\mathcal{P}_i$-schemes are inhomogeneous on $G_i/N_i$, then all strongly antisymmetric $\mathcal{P}$-schemes are inhomogeneous on $G_k$. 
\end{enumerate} 
The same holds if strong antisymmetry is replaced by antisymmetry.
\end{lem}

We defer the proof of Lemma~\ref{lem_aggregation} to Section~\ref{sec_common_crq}.

%\begin{lem}\label{lem_complexityagg}
%Under the notations of Lemma~\ref{lem_aggregation}, it holds that 
%\[
%c(\mathcal{P})\leq [G_0:G_{k-1}]\cdot \left(\sum_{i=1}^k c(\mathcal{P}_i)\right).
%\]
%\end{lem}
%\begin{proof}
%For $i\in [k]$, let $\mathcal{P}_{i,0}$ be  a complete set of representatives of the orbits of $\mathcal{P}_i$ under the action of $G_{i-1}/N_i$ by conjugation. 
%%For $i\in [k]$ and  conjugate subgroups $H, H'=gHg^{-1}$ of $G_{i-1}/N_i$ where $g\in G_{i-1}/N_i$, we have 
%%$\pi_i^{-1}(H')=\tilde{g}\pi_i^{-1}(H)\tilde{g}^{-1}$ where $\tilde{g}\in G_{i-1}$ lifts $g$. So $\pi_i^{-1}(H)$ and $\pi_i^{-1}(H')$ are conjugate in $G_0$. Therefore 
%Every subgroup in $\mathcal{P}$ is conjugate to some subgroup of the form $\pi_i^{-1}(H)$ in $G_0$ where $i\in [k]$ and $H\in\mathcal{P}_{i,0}$, and we have 
%$[G_{i-1}: \pi_i^{-1}(H)]= [G_{i-1}/N_i : \pi_i^{-1}(H)/N_i]=[G_{i-1}/N_i: H]$.
%Therefore
%\begin{align*}
%c(\mathcal{P})&\leq \sum_{i=1}^k\sum_{H\in \mathcal{P}_{i,0}}[G_0: \pi_i^{-1}(H)]
%\leq  [G_0:G_{k-1}]\cdot\left(\sum_{i=1}^k\sum_{H\in \mathcal{P}_{i,0}} [G_{i-1}/N_i: H]\right)\\
%&= [G_0:G_{k-1}]\cdot \left(\sum_{i=1}^k c(\mathcal{P}_i)\right).
%\end{align*}
%\end{proof}

\begin{proof}[Proof of Theorem~\ref{thm_primitivereduction}]
The claims about $K_i$ and $g_i$ follow from Theorem~\ref{thm_lm}.
Use the following notations for $i\in [k]$:
\begin{itemize}
\item $L_i$: the splitting field of $g_i$ over $K_{i-1}$, which is a subfield of $L$.
\item $G_i:=\gal(L_i/K_{i-1})$ and $N_i:=\gal(L/L_i)$.
\item $\pi_i$: the natural projection $\gal(L/K_{i-1})\to \gal(L/K_{i-1})/N_i\cong G_i$.  
\item $\mathcal{P}_i$: the subgroup system over $G_i$ associated with $\mathcal{F}_i$. 
\end{itemize}
Then by construction, the subgroup system over $\gal(L/K_0)$  associated with  $\mathcal{F}$ is 
\[
\mathcal{P}:=\{g\pi_i^{-1}(H)g^{-1}: 1\leq i\leq k, H\in \mathcal{P}_i, g\in G\}.
\]
Assume the conditions on $\mathcal{F}_i$ in Theorem~\ref{thm_primitivereduction} are satisfied. Then for all $i\in [k]$, all strongly antisymmetric $\mathcal{P}_i$-schemes are discrete (resp. inhomogeneous) on $\gal(L_i/K_i)\in\mathcal{P}_i$.
Applying Lemma~\ref{lem_aggregation} to the chain 
\[
\gal(L/K_k)\subseteq \gal(L/K_{k-1})\subseteq \dots\subseteq \gal(L/K_1)\subseteq \gal(L/K_0)
\]
 and $N_i$, $\pi_i$, $\mathcal{P}_i$, we conclude that  all strongly antisymmetric $\mathcal{P}$-schemes are discrete (resp. inhomogeneous) on the subgroup $\gal(L/K_k)\in\mathcal{P}$. And the corresponding fixed subfield $K_k$ is isomorphic to $K_0[X]/(g(X))$ over $K_0$, as desired.

The total running time of the algorithm $\mathtt{PrimitiveAction}$  and the total size of $\mathcal{F}_i$  are both bounded by  $\sum_{i=1}^k T(K_{i-1}, g_i)$. The other operations take time polynomial in  the total size of $\mathcal{F}_i$ and the size of the input.  The claim about the running time follows.
\end{proof}

As an application, we prove the main result of  \citep{Evd92} for the special case that the input  polynomial satisfies Condition~\ref{cond_spoly} (i.e., it is defined over $\F_p$, square free, and complete reducible over $\F_p$).

\begin{thm}[\citep{Evd92}]\label{thm_algsolvable}
Under GRH, there exists a  deterministic polynomial-time  algorithm that, given a polynomial $f(X)\in\F_p[X]$ satisfying Condition~\ref{cond_spoly}  and a  lifted polynomial $\tilde{f}(X)\in\Z[X]$ of $f$ whose Galois group $\gal(\tilde{f}/\Q)$ is solvable, computes the complete factorization of $f$ over $\F_p$.
\end{thm}

The proof relies on the following bound for the orders of primitive solvable permutation groups, proved by P{\'a}lfy \citep{Pal82}.
\begin{thm}[\citep{Pal82}]\label{thm_solvablebound}
Let $G$ be a primitive solvable permutation group on a set of cardinality $n\in\N^+$. Then $|G|\leq 24^{-1/3} n^c$ for a constant $c=3.24399\dots$.
\end{thm}

\begin{proof}[Proof of Theorem~\ref{thm_algsolvable}]
As in Section~\ref{alg_pschalg}, we factorize $\tilde{f}$ into its irreducible factors $f_1(X),\dots,f_k(X)\in\Z[X]$ over $\Q$ in polynomial time using the factoring algorithm in \citep{LLL82}. The Galois groups $\gal(\tilde{f}_i(X)/\Q)$ are quotient groups of $\gal(\tilde{f}/\Q)$, and hence are solvable as well. By replacing $\tilde{f}(X)$ with $\tilde{f}_i(X)$ and $f(X)$ with $f_i(X):=\tilde{f}_i(X)\bmod p\in\F_p[X]$ for each $i\in [k]$, we reduce to the case that $\tilde{f}$ is irreducible over $\Q$.

Let $L$ be the splitting field of $\tilde{f}$ over $\Q$.
When $\gal(\tilde{f}/\Q)$ acts primitively on the set of roots of $\tilde{f}$  in $L$, its order is bounded by a polynomial in $\deg(f)$ by Theorem~\ref{thm_solvablebound}.
Then by  Theorem~\ref{lem_compsplitgeneral}, we can construct $\mathcal{F}$ in polynomial time such that $\Q[X]/(\tilde{f}(X))\in \mathcal{F}$ and all strongly antisymmetric $\mathcal{P}$-schemes are discrete on $H$, where $\mathcal{P}$ is the subgroup system  over $\gal(\tilde{f}/\Q)$ associated with $\mathcal{F}$ and $H$ is a subgroup in $\mathcal{P}$ satisfying $L^H\cong \Q[X]/(\tilde{f}(X))$.  By Theorem~\ref{thm_primitivereduction}, we also have a polynomial-time algorithm of constructing such $\mathcal{F}$ in the general case.
The theorem then follows from  Theorem~\ref{thm_algmain2formal}.
\end{proof}

In  Chapter~\ref{chap_alg_general}, we prove a generalization of Theorem~\ref{thm_algsolvable} (see Theorem~\ref{thm_algsolvableg}), which implies the main result of  \citep{Evd92} in its general form. In particular,  the assumption that $\tilde{f}$ satisfies Condition~\ref{cond_spoly} is no longer required.  

\section{Other techniques of constructing number fields}\label{sec_consother}

In this section, we survey some other techniques of constructing number fields. 
While we do not use these techniques directly in the thesis, they are worth mentioning because of their own interest  and their applications to other problems \citep{Lan84, LM85, Len92, Coh13}. 

\paragraph{Taking the compositum of number fields.}
Note that the fields computed in the two algorithms $\mathtt{SplittingField}$ and $\mathtt{Stabilizers}$ in Section~\ref{sec_compositum} are (up to isomorphism over $K_0$) compositums of conjugates of the field $K_0[X]/(g(X))$.  The general problem of constructing the compositum of (relative) number fields is solved  by the following lemma.

\begin{lem}\label{lem_composite}
There exists a polynomial-time algorithm that given a number field $K_0$ and relative number fields $K$, $L$ over $K_0$, constructs all the compositums $K' L'$ up to isomorphism over $K_0$ 
where  $K'$ (resp. $L'$) ranges over the conjugates of $K$ (resp. $L$) over $K_0$ in  the algebraic closure $\bar{K}_0$ of $K_0$.\footnote{Here $K$ and $L$ are embedded in $\bar{K}_0$ via some $K_0$-linear embeddings. The choices of these embeddings do not matter as we construct   $K'L'$ for all the conjugates $K'$ and $L'$ over $K_0$.}
\end{lem}
\begin{proof}
Take the irreducible polynomial $g(X)\in K_0[X]$ that encodes $L$, i.e., $L\cong_{K_0} K_0[X]/(g(X))$. Factorize $g(X)$ into irreducible polynomials $g_1(X),\dots,g_k(X)$ over $K$. Then compute and output the fields $K[X]/(g_1(X)),\dots, K[X]/(g_k(X))$.

To see that this gives the desired output, note that we may fix $K=K'$ as fields are constructed only up to isomorphism over $K_0$.
Let $\alpha_1,\dots,\alpha_n$ be the roots of $g$ in  $\bar{K}_0$, where $n=\deg(g)$. Then the conjugates of $L$ in $\bar{K}_0$ over $K_0$ are precisely $K_0(\alpha_1),\dots, K_0(\alpha_n)$. For $i\in [n]$, there exists a unique $j_i\in [k]$ such that $\alpha_i$ is the root of $g_{j_i}$, and the compositum of $K$ and $K_0(\alpha_i)$ is just $K(\alpha_i)\cong_{K_0} K[X]/(g_{j_i}(X))$.
\end{proof}

\paragraph{Taking the intersection of number fields.}

The intersection of two number fields can be computed efficiently, as shown in \citep{LM85}.

\begin{thm}[\citep{LM85}]
There exists a polynomial-time algorithm that given 
\begin{itemize}
\item number fields $K=\Q(\alpha)$, $K'=\Q(\beta)$ encoded by the minimal polynomials of  primitive elements $\alpha\in K$ and $\beta\in K'$ over $\Q$ respectively, and
\item the minimal polynomial $h_0(X)\in K[X]$ of $\beta$  over $K$,\footnote{The polynomial $h_0$ is needed for the problem to be well defined.} 
\end{itemize}
computes the number field $K\cap K'$ up to isomorphism.
\end{thm}

The algorithm in \citep{LM85} also extends to relative number fields. We omit the details.

\paragraph{Adjoining a square root of the discriminant.}

Suppose $K$ is a relative number field over $K_0$ encoded by the minimal polynomial $h(X)\in K_0[X]$ of a primitive element $\alpha\in K$ over $K_0$. Let $L$ be the Galois closure of $K/K_0$ and let $G=\gal(L/K_0)$. Then $G$ acts on the set $S$ of roots of $h$ in $L$ and hence can be identified with a subgroup of $\sym(S)$. 

Suppose $S=\{\alpha_1,\dots,\alpha_n\}$. Define  the {\em discriminant}\index{discriminant of a polynomial} of $h$ to be
\[
\Delta_h:=\prod_{1\leq i<j\leq n} (\alpha_i-\alpha_j)^2.
\]
We have $\prescript{g}{}{\Delta_h}=\Delta_h$ for all $g\in G$. So $\Delta_h\in L^G=K_0$.

Now consider the subfield $K_0':=K_0(\sqrt{\Delta_h})$ of $L$, where $\sqrt{\Delta_h}:=\prod_{1\leq i<j\leq n}  (\alpha_i-\alpha_j)$ is a square root of $\Delta_h$ in $L$. 
A permutation $g\in G$ fixes $\sqrt{\Delta_h}$ precisely when $g$ is an even permutation of $S$, which implies
\[
\gal(L/K_0')=G\cap \alt(S).
\]
With this observation, we have

\begin{lem}
There exists a polynomial-time algorithm that given a number field $K_0$ and a relative number field $K$ over $K_0$ encoded by $h(X)\in K_0[X]$, computes $L^{G\cap \alt(S)}$ up to isomorphism over $K_0$, where $L$ is the Galois closure of $K/K_0$, $G=\gal(L/K_0)$, and $S$ is the set of roots  of $h$ in $L$. 
\end{lem}
\begin{proof}
We have $L^{G\cap \alt(S)}=K_0(\sqrt{\Delta_h})$ by the above discussion. Let $n=\deg(h)$.
Then discriminant $\Delta_h$  satisfies the identity
\[
\Delta_h=(-1)^{n(n-1)/2}\mathrm{Res}(h,h'),
\]
where  $\mathrm{Res}(h,h')$ denotes the {\em resultant} of $h$ and its derivative $h'$. and is given by the determinant of the {\em Sylvester matrix} associated with $h$ and $h'$ \citep{Lan02}. Thus we can compute $\Delta_h$ in polynomial time. Then we test if $\Delta_h$ is a square in $K_0$ by factoring $X^2-\Delta_h$ over $K_0$. If $\Delta_h$ is a square, we have  $K_0(\sqrt{\Delta_h})=K_0$ and correspondingly $G\subseteq \alt(S)$. In this case we just output $K_0$. Otherwise we output $K_0[X]/(X^2-\Delta_h)$.
\end{proof}

\begin{rem}
The technique above was used in \cite{Lan84} for the determination of the Galois groups of number field extensions.
%, or more specifically to distinguish $\alt(S)$ from $\sym(S)$. 
It is not clear, however, if it helps for the problem of polynomial factoring over finite fields.
We note that replacing $K_0$ with $K_0'=K_0(\sqrt{\Delta_h})$ and $K$ with $K_0'K$ has the effect of reducing the Galois group $G$ to $G\cap \alt(S)$, but the order of $G$ is reduced by at most a factor of two. 
This does not help in the case that $G=\sym(S)$ and $\mathcal{P}$ is a system of stabilizers of depth $m\leq |S|-2$ (with respect to the natural action of $G$):
As both $\sym(S)$ and $\alt(S)$ are $k$-transitive for $k=|S|-2$, $\mathcal{P}$-schemes for $\sym(S)$ and those for $\alt(S)$ both correspond to $m$-schemes on $S$ (see Theorem~\ref{thm_mandp}), and hence they are the equivalent.
\end{rem}

\paragraph{Computing the fixed subfield of the automorphism group.}

The following lemma gives a characterization of the fixed subfield of an automorphism subgroup.

\begin{lem}\label{lem_autsym}
Suppose $K/K_0$ is a field extension and $\alpha$ is a primitive element of $K$ over $K_0$.
For a subgroup $U\subseteq\aut(K/K_0)$, the field $K^U$ is generated by elementary symmetric polynomials in the elements $\prescript{g}{}{\alpha}$  (indexed by $g\in U$) over $K_0$. 
\end{lem}

\begin{proof}
Let $K'$ be the subfield of $K$  generated by elementary symmetric polynomials in $\prescript{g}{}{\alpha}$, $g\in U$  over $K_0$. 
We obviously have $K'\subseteq K^U$. By Galois theory, it holds that $[K:K^U]=|U|$ (see, e.g., \citep[Section~\RN{6}.1, Theorem~1.8]{Lan02}). So it suffices to prove $[K:K']\leq |U|$.

Consider the polynomial $\phi(X)=\prod_{g\in U} (X-\prescript{g}{}{\alpha})$. The coefficients of $\phi(X)$ are, up to sign,  given by elementary symmetric polynomials in $\prescript{g}{}{\alpha}$, $g\in U$ and hence $\phi(X)\in K'[X]$.  As $\phi(\alpha)=0$, the minimal polynomial of $\alpha$ over $K'$ divides $\phi(X)$, and its degree is at most $\deg(\phi)=|U|$. So we have $[K'(\alpha): K']\leq |U|$. The claim follows by noting that $K'(\alpha)=K$.
\end{proof}

Lemma~\ref{lem_autsym} provides a method of computing the fixed subfield of the automorphism group $\aut(K/K_0)$:

 \begin{thm}\label{thm_normalizer}
There exists a polynomial-time algorithm that given a number field $K_0$ and a relative number field $K$ over $K_0$, computes the 
fixed subfield $K^{\aut(K/K_0)}\subseteq K$.
 \end{thm} 
 \begin{proof}
 Suppose $K$ is encoded by the minimal polynomial of a primitive element $\alpha$ over $K_0$.
 We compute all the automorphisms of $K$ in $\aut(K/K_0)$ using Lemma~\ref{lem_comprelembed}. Then we adjoining to $K_0$ the first $k$ elementary symmetric functions in $\prescript{g}{}{\alpha}$, $g\in \aut(K/K_0)$ where $k=|\aut(K/K_0)|$. The resulting field is exactly $K^{\aut(K/K_0)}$ by Lemma~\ref{lem_autsym}.
 \end{proof}
 
 More generally, given $K_0, K$ and a subgroup $U\subseteq \aut(K/K_0)$ of automorphisms of $K$, the same proof shows that $K^U$ can be constructed in polynomial time. 

Now suppose $L$ is a Galois extension of $K_0$ that contains $K$. Let $G=\gal(L/K_0)$ and $H=\gal(L/K)$. Then $\aut(K/K_0)$ is identified with $N_G(H)/H$, and we have $K^{\aut(K/K_0)}=K^{N_G(H)/H}=L^{N_G(H)}$. So Theorem~\ref{thm_normalizer} states that  $L^{N_G(H)}$ can be constructed in polynomial time given $K=L^H$ and $K_0$.
In the context of polynomial factoring using the $\mathcal{P}$-scheme algorithm, this means that we can efficiently enlarge a subgroup system $\mathcal{P}$ by including the normalizers $N_G(H)$ of  $H\in \mathcal{P}$.  

A natural question arising from this observation is whether adding the normalizers (or more generally subgroups between $N_G(H)$ and $H$) to the subgroup system helps a $\mathcal{P}$-scheme algorithm  obtain the complete factorization (resp. a proper factorization). By Theorem~\ref{thm_algmain2formal}, this reduces to the question  whether it helps for proving all strongly antisymmetric $\mathcal{P}$-schemes are discrete (resp. inhomogeneous) on a distinguished subgroup $H\in\mathcal{P}$.

For discreteness of strongly antisymmetric $\mathcal{P}$-schemes, we  give an affirmative answer in general: we show that for some subgroup system $\mathcal{P}$ and $H\in\mathcal{P}$, there exist strongly antisymmetric $\mathcal{P}$-schemes that are not discrete on $H$, but adding normalizers to the subgroup system rules out their existence.

\begin{exmp}\label{exmp_normalizer}
Choose a finite group  $G$ and a subgroup $H\subseteq G$ such that $N_G(H)$ is a proper normal subgroup of $G$.\footnote{For example, we may choose $G$ to be the semidirect product $(K\times K)\rtimes C_2$, where $K$ is a nontrivial finite group and $C_2$ permutes the two direct factors of $K\times K$. Let $H=K\times\{e\}$. Then $N_G(H)=K\times K\unlhd G$.}
Choose $\mathcal{P}=\{gHg^{-1}: g\in G\}$ which is a subgroup system over $G$. Define a $\mathcal{P}$-collection $\mathcal{C}=\{C_{H'}:H'\in\mathcal{P}\}$ as follows: 
the group $N_G(H)$ acts on $H\backslash G$ by left translation $\prescript{g}{}{Hh}=Hgh$ and $H\backslash G$ is partitioned into $N_G(H)$-orbits. Choose a complete set of representatives $B\subseteq H\backslash G$ for these orbits. Define $C_H=\{\prescript{g}{}{B}: g\in N_G(H)\}$. For any other subgroup $H'$ in $\mathcal{P}$, choose $g\in G$ such that $H'=gHg^{-1}$, and define $C_{H'}=\{c_{H,g}(B): B\in C_H\}$.
It is easy to see that $\mathcal{C}$ is a well defined strongly antisymmetric $\mathcal{P}$-scheme. 
Moreover, it is not discrete on $H$ since $N_G(H)$ does not act transitively on $H\backslash G$.

Now define $\mathcal{P}'=\mathcal{P}\cup \{N_G(H)\}$ which is also a subgroup system over $G$. We claim that any antisymmetric $\mathcal{P}'$-schemes $\mathcal{C}'$ must be discrete on any subgroup in $\mathcal{P}'$.
To see this, note that  $\mathcal{C}'$ is discrete on $N_G(H)\in\mathcal{P}'$ since $N_G(H)$ is normal in $G$.  Then $\mathcal{C}'$ is also discrete on all the other subgroups $H'\in\mathcal{P}'$ by compatibility, and the claim follows.   In particular, it is impossible to extend $\mathcal{C}$ to an antisymmetric $\mathcal{P}'$-scheme.
\end{exmp}

Despite the example above, adding normalizers to the subgroup system seems not helpful for attacking the most difficult cases in polynomial factoring:
for a subgroup system $\mathcal{P}$ over a finite group $G$, define 
\[
\mathcal{P}_+=\{U: H\subseteq U\subseteq N_G(H), H\in\mathcal{P}\},
\]
 which is also a subgroup system over $G$. For several important families of permutation groups, we show that if $\mathcal{P}$ is the corresponding system of stabilizers of certain depth $m$ (where $m$ is not too large), any $\mathcal{P}$-scheme $\mathcal{C}$ can be extended to a $\mathcal{P}_+$-scheme $\mathcal{C}'$ with antisymmetry and strong antisymmetry preserved. In particular, if $\mathcal{C}$ is not discrete or inhomogeneous on some subgroup $H\in\mathcal{P}$, then neither is $\mathcal{C}'$.
 \nomenclature[c1d]{$\mathcal{P}_+$}{subgroup system $\{U: H\subseteq U\subseteq N_G(H), H\in\mathcal{P}\}$}

\begin{lem}\label{lem_extsym}
Let $S$ be a finite set and let $G$ be $\sym(S)$ or $\alt(S)$ acting naturally on $S$. Let $\mathcal{P}$ be the system of stabilizers of depth $m$ over $G$ with respect to this action where $m<|S|/2$. Then any $\mathcal{P}$-scheme $\mathcal{C}$ can be extended to a $\mathcal{P}_+$-scheme $\mathcal{C}'$ such that $\mathcal{C}'$ is antisymmetric (resp. strongly antisymmetric) if so is $\mathcal{C}$.
 \end{lem} 

\begin{lem}\label{lem_extgl}
Let $V$ be a finite dimensional vector space over a finite field and let $G$ be $\gl(V)$ acting naturally on $S:=V-\{0\}$. Let $\mathcal{P}$ be the system of stabilizers of depth $m$ over $G$ with respect to this action where $m<\dim_F V$. Then any $\mathcal{P}$-scheme $\mathcal{C}$ can be extended to a $\mathcal{P}_+$-scheme $\mathcal{C}'$ such that $\mathcal{C}'$ is antisymmetric (resp. strongly antisymmetric) if so is $\mathcal{C}$.
 \end{lem} 

We defer the proofs of Lemma~\ref{lem_extsym} and Lemma~\ref{lem_extgl} to Section~\ref{sec_extension} . There we define the {\em closure}\index{closure} $\mathcal{P}_\mathrm{cl}$ of a subgroup system $\mathcal{P}$, and then show  that $\mathcal{P}$-schemes can always be extended to $\mathcal{P}_\mathrm{cl}$-schemes with antisymmetry and strong antisymmetry preserved. Lemma~\ref{lem_extsym} and Lemma~\ref{lem_extgl} then follow immediately once we verify that $\mathcal{P}_\mathrm{cl}=\mathcal{P}_+$ in these cases. 

%The cases in  Lemma~\ref{lem_extsym} and Lemma~\ref{lem_extgl} are the most difficult cases for polynomial factoring. This claim will be made rigorous in Chapter X.

\chapter{The generalized \texorpdfstring{$\mathcal{P}$-scheme}{P-scheme} algorithm}\label{chap_alg_general}

In Chapter~\ref{chap_alg_prime}, we developed the $\mathcal{P}$-scheme algorithm that  factorizes polynomials satisfying Condition~\ref{cond_spoly}, i.e., they are defined over a prime field $\F_p$, square-free, and completely reducible over $\F_p$.
In this chapter, we extend this algorithm to factorize general polynomials  $f(X)\in \F_q[X]$ over a finite field $\F_q$ of characteristic $p$.
The generality is reflected in the following three aspects: (1) $\F_q$ may be a non-prime field, (2) the degrees of the irreducible factors of $f$ may be greater than one, and (3) the multiplicities of the irreducible factors of $f$ may be greater than one.

\paragraph{Motivation.} Techniques like Berlekamp's reduction \citep{Ber70}, square-free factorization \citep{Yun76, Knu98} and distinct-degree factorization \citep{CZ81} were commonly used in literature to reduce the problem to the special case that the input polynomial satisfies Condition~\ref{cond_spoly}. However, these reductions do not preserve the information of the lifted polynomial $\tilde{f}$ employed by the $\mathcal{P}$-scheme algorithm.
%\footnote{Of course we could first apply the reductions and then pick a new lifting polynomial. But the information given by the original lifting polynomial is lost anyway.} 
Therefore, it is  desirable to avoid these reductions and extend the $\mathcal{P}$-scheme algorithm to the general setting  instead.

As a concrete example,  consider the following polynomial $\tilde{f}(X)\in\Z[X]$ irreducible over $\Q$, taken from \citep{KM00}:
 \begin{align*}
\tilde{f}(X)=&X^{14}+28 X^{11}+28 X^{10}-28 X^9+140 X^8+360 X^7+147 X^6\\&+196 X^5+336 X^4-546 X^3-532 X^2+896 X+823.
\end{align*}
For $p=43$, the reduced polynomial $f(X)=\tilde{f}(X) \bmod p$ has seven distinct linear factors and one irreducible factor of degree 7 over $\F_p$:
\begin{align*}
f(X)=&(X + 2)  (X + 4)  (X + 9) (X + 19) (X + 23) (X + 30) (X + 42)\\& (X^7 + 14 X^4 + 15 X^3 + 31 X^2 + 15 X + 38).
\end{align*}
The standard way of factoring $f$ over $\F_p$ is first applying distinct-degree factorization \citep{CZ81} to obtain a partial factorization $f=f_0f_1$, where
\[
f_0(X)=(X + 2)  (X + 4)  (X + 9) (X + 19) (X + 23) (X + 30) (X + 42)
\]
is the product of the linear factors and satisfies Condition~\ref{cond_spoly}. Then we factorize $f_0$ over $\F_p$. 
To achieve this goal deterministically, we pick a lifted polynomial $\tilde{f}_0(X)\in\Z[X]$ of $f$, which we may assume to be irreducible, and run the $\mathcal{P}$-scheme algorithm in Chapter~\ref{chap_alg_prime}. Suppose the $(\Q, \tilde{f})$-subfield system in the algorithm is constructed by Lemma~\ref{lem_compstab} and the associated subgroup system $\mathcal{P}$ is the system of stabilizers of depth $m$, where $m\in\N^+$ is sufficiently large. 
%The algorithm runs in time polynomial in $n^m$ and the length of the input. 
In the worst case, the action of $\gal(\tilde{f}_0/\Q)$ on the set of roots of $\tilde{f}$ is permutation isomorphic to the natural action of the symmetric group $\sym(7)$ on $[7]$. 
%\footnote{For example, this is the case when $\tilde{f}_0(X)=(X + 2)  (X + 4)  (X + 9) (x + 19) (X + 23) (X + 30) (X + 42)+43$.}  
Then we need $m\geq 3$ to obtain a proper factorization of $f$, since by Theorem~\ref{thm_mandp} and Lemma~\ref{lem_anti2ch}, there exists a strongly antisymmetric $\mathcal{P}$-scheme homogeneous on a stabilizer if $m\leq 2$.\footnote{For the same reason, one needs to choose $m\geq 3$ if the $m$-scheme algorithm \citep{IKS09} is used.}
%\footnote{However, the algorithm developed by Gao in \citep{Gao01} using ``balance tests'' does compute a proper factorization of $f$. Computer-aided search shows that Gao's algorithm fails if we take $p$ to be $170647$ instead of $43$. Gao's algorithm may be seen as an instantiation of the $m$-scheme algorithm for $m=2$ by making special choices. The situation here is similar to that of the simplex algorithm for linear programming, where no pivoting rules are known to work all the time but some rules may work well in some cases. Also see \citep{Sah08}.}

On the other hand, the action of the Galois group of $\tilde{f}$ on the  set of roots of $\tilde{f}$ is permutation isomorphic to the action of the wreath product\footnote{For the definition of the wreath product of groups, see Definition~\ref{defi_wrgroup}.}
 $C_7 \wr C_2$ on $[7]\times [2]$, where $C_7$ permutes $[7]$ cyclically and $C_2$ permutes the two copies of $[7]$. This action has a base of size two, which suggests that choosing $m=2$ is sufficient for  completely factoring $f$, provided that we have a generalization of Theorem~\ref{thm_algmain2informal} that employs the polynomial $\tilde{f}$.
The goal of this chapter is to establish such a generalization.

 The example above generalizes to an infinite family of instances: for every $k\in\N^+$, there exists $\tilde{f}(X)\in\Z[X]$ irreducible over $\Q$ of degree $2k$ such that the action of the Galois group on the set of roots of $\tilde{f}$ is permutation isomorphic to the action of $C_k \wr C_2$ on $[k]\times [2]$.\footnote{Shafarevich's theorem on solvable Galois groups \citep{Sha54, ILF97} implies that the existence of integral polynomials  realizing the family of groups $C_k \wr C_2$ as Galois groups. For an algorithm  explicitly computing such a polynomial, see \citep{KM00}.}
And for such $\tilde{f}$, there exists infinitely many prime numbers $p$ such that $f(X)=\tilde{f}(X)\bmod p$ has $k$ distinct linear factors and one irreducible factor of degree $k$.\footnote{This follows from Chebotar{\"e}v's density theorem. See, e.g.,  \citep{Neu99}.}
Using the generalized $\mathcal{P}$-scheme algorithm  developed in this chapter, it is sufficient to choose $m=2$ in order to completely factorize $\tilde{f}\bmod p$, leading to a polynomial-time factoring algorithm for such instances $(f,\tilde{f})$.  On the other hand, using distinct-degree factorization and the $\mathcal{P}$-scheme algorithm in Chapter~\ref{chap_alg_prime}, the best known general upper bound for $m$ is $O(\log k)$ (see Theorem~\ref{thm_evd94}), and the resulting algorithm takes superpolynomial time.

\paragraph{Lifted polynomial.} 

To formulate the main result of this chapter, we first need to generalize the notion of {\em lifted polynomials} (see Definition~\ref{defi_liftpoly}).
Recall that a lifted polynomial of $f(X)\in\F_p[X]$ is a polynomial $\tilde{f}(X)\in \Z[X]$ of degree $\deg(f)$ satisfying $\tilde{f}(X)\bmod p=f(X)$.
For the general case $\F_q=\F_{p^d}$, we fix the following notations: assume $\F_q$ is encoded by a monic irreducible polynomial $h(Y)\in\F_p[Y]$ of degree $d$, i.e., it is identified with $\F_p[Y]/(h(Y))$ via an isomorphism $\psi_0: \F_p[Y]/(h(Y))\to \F_q$ which we can efficiently compute. 
%\footnote{If $\F_q$ is encoded by the structure constants in some $\F_p$-basis, we can compute $h$ by first choosing $\alpha\in\F_q$ satisfying $\F_q=\F_p(\alpha)$ and then computing $h$ as the minimal polynomial of $\alpha$ over $\F_p$.}
Lift $h$ to a monic  polynomial $\tilde{h}(Y)\in\Z[Y]$ of degree $d$ which is necessarily irreducible over $\Q$.
Define $A_0:=\Z[Y]/(\tilde{h}(Y))$ and $K_0:=\Q[Y]/(\tilde{h}(Y))$.
Composing $\psi_0$ with the natural projection $A_0\to \F_p[Y]/(h(Y))$ sending $x$ to $x \bmod p$, 
we obtain a surjective ring homomorphism $\tilde{\psi}_0: A_0\to \F_q$. Finally extend $\tilde{\psi}_0$ to the ring $A_0[X]$  by applying it to each coefficient:
\[
\tilde{\psi}_0:  A_0[X]\to \F_q[X].
\]
With these notations, we generalize the definition of lifted polynomials as follows.

\begin{defi}[lifted polynomial]\label{defi_genlpoly}
Suppose $f(X)\in\F_q[X] $ is a polynomial  of degree $n\in\N^+$.
A {\em lifted polynomial}\index{lifted polynomial} of $f$ (with respect to $\tilde{h}$ and $\psi_0$) is a polynomial $\tilde{f}(X)\in A_0[X]$  of degree $n$ satisfying $\tilde{\psi}_0(\tilde{f})=f$. An {\em irreducible lifted polynomial}\index{irreducible lifted polynomial} of $f$ is a lifted polynomial of $f$ that is irreducible over $K_0$. 
\end{defi}

 Given $f(X)\in\F_q[X]$, we can choose a lifted polynomial $\tilde{f}$ of $f$ efficiently.
Furthermore, we argue that $\tilde{f}$ can be assumed to be irreducible over $K_0$. To see this, we need the following lemma.

\begin{restatable}{lem}{lemsubprob}\label{lem_subprob}
There exists a polynomial-time algorithm that given $p$ and  a polynomial $\tilde{f}(X)\in A_0[X]$ satisfying $\tilde{\psi}_0(\tilde{f})\neq 0$, computes an integer $D$ satisfying $D\equiv 1{\pmod p}$ and a factorization of $D\cdot \tilde{f}$ into irreducible factors $\tilde{f}_i$ over $K_0$. Furthermore all of the factors $\tilde{f}_i(X)$ are in $A_0[X]$.
\end{restatable}

The proof can be found in Appendix~\ref{chap_omitted2}. Compute $D$ and $f_i$ using the lemma above. We have $\tilde{\psi}_0(D\cdot \tilde{f})=\tilde{\psi}_0(\tilde{f})=f$ since $D\equiv 1{\pmod p}$. So the polynomials $\tilde{\psi}_0(\tilde{f}_i)$ are factors of $f$,
and we have reduced the problem to factoring each $\tilde{\psi}_0(\tilde{f}_i)\in\F_q[X]$ using its irreducible lifted polynomial $\tilde{f}_i$. 

The discussion above justifies the assumption that an irreducible lifted polynomial $\tilde{f}$ of $f$  is given, with respect to $\tilde{h}$ and $\psi_0$. The notations $\tilde{h}$, $\psi_0$, $A_0$, and $K_0$ are fixed throughout this chapter.

\nomenclature[d1a]{$\tilde{h}$, $\psi_0$, $A_0$, $K_0$}{See Chapter~\ref{chap_alg_general}}

\paragraph{Main result.} 

The main result of this chapter is a generalization of Theorem~\ref{thm_algmain2informal}:

\begin{thm}[informal]\label{thm_alginformalgeneral}
Suppose there exists a deterministic algorithm that given a polynomial $g(X)\in A_0[X]$ irreducible over $K_0$, constructs in time $T(g)$ a collection $\mathcal{F}$ of subfields of the splitting field $L$ of $g$ over $K_0$ such that
\begin{itemize}
\item $F=K_0[X]/(g(X))$ is in $\mathcal{F}$, and
\item  all strongly antisymmetric $\mathcal{P}$-schemes are discrete on $\gal(L/F)\in\mathcal{P}$, where $\mathcal{P}$ is the subgroup system associated with $\mathcal{F}$.
\end{itemize}
Then under GRH, there exists a  deterministic  algorithm that given $f(X)\in\F_q[X]$ and an irreducible lifted polynomial $\tilde{f}(X)\in A_0[X]$ of $f$, outputs the complete factorization of $f$ over $\F_q$ in time polynomial in $T(\tilde{f})$ and the size  of the input.
\end{thm}

See Theorem~\ref{thm_algmain2formalg} for the formal statement. For simplicity, here we only state the result for computing the complete factorization of $f$. The results for computing a proper factorization are slightly more complicated to state, and  we refer the reader to Section~\ref{sec_puttogetherg} for details.

%The condition on $\mathcal{P}$ can be weakened, but it involves the notion of  {\em $\mathcal{P}$-schemes of double cosets}.  See Theorem~\ref{thm_algmain2relax} for details.
 
\subsection*{Overview of the generalized $\mathcal{P}$-scheme algorithm} 

Recall that a $\mathcal{P}$-scheme algorithm in Chapter~\ref{chap_alg_prime} consists of three parts: (1) a reduction to the problem of computing an idempotent decomposition of the ring $\bar{\ord}_F$, where $F=\Q[X]/(\tilde{f}(X))$, (2) computing  idempotent decompositions for a collection  of number fields, and (3) constructing the collection of number fields used in the previous part. The factoring algorithm   in this chapter has the same structure but with some differences: we generalize the reduction in Part (1), where $F$ now denotes the number field $K_0[X]/(\tilde{f}(X))$. And in Part (3), we construct a collection of relative number fields over $K_0$ instead of ordinary number fields. The main difference is in Part (2), which  we now explain.

\paragraph{$\mathcal{P}$-schemes of double cosets.}
 In Chapter~\ref{chap_alg_prime}, we proved that for a subfield $K$ of the splitting field $L$ of $\tilde{f}$, $G$ the Galois group of $\tilde{f}$, and $H=\gal(L/K)$, an idempotent decomposition of the ring $\bar{\ord}_{K}$ corresponds to a partition of the right coset space $H\backslash G$.  The crucial condition for this claim to hold is that $p$ {\em splits completely} in the splitting field $L$ of $\tilde{f}$, which in turn relies on the assumption that $f$ is square-free and completely reducible over the field of definition. In general, one can prove that an idempotent decomposition of $\bar{\ord}_{K}$ corresponds to a partition of the {\em double coset space} $H\backslash G/\mathcal{D}_{\mathfrak{Q}_0}$ instead of the right coset space $H\backslash G$, where $\mathcal{D}_{\mathfrak{Q}_0}\subseteq G$ is known as the {\em decomposition group} (of a fixed prime ideal $\mathfrak{Q}_0$ of $\ord_L$ over $K_0$). 
For the special case studied in Chapter~\ref{chap_alg_prime}, the decomposition group $\mathcal{D}_{\mathfrak{Q}_0}$ is trivial, and hence the double coset space $H\backslash G/\mathcal{D}_{\mathfrak{Q}_0}$ coincides with the right coset space $H\backslash G$.

To address the general case, we define the notion of {\em $\mathcal{P}$-collections (resp. $\mathcal{P}$-schemes) of double cosets}, generalizing  (ordinary) $\mathcal{P}$-collections (resp. $\mathcal{P}$-schemes). Various properties including (strong) antisymmetry, discreteness and homogeneity can be extended to $\mathcal{P}$-schemes of double cosets.
In addition, as the rings $\bar{\ord}_K$ are not necessarily semisimple in general, we replace them with the rings $R_K$, defined by
\[
R_K:=\left\{x\in \bar{\ord}_K/\rad(\bar{\ord}_K): x^p=x\right\},
\]
where $\rad(\bar{\ord}_K)$ denotes the {\em radical} of $\bar{\ord}_K$. These rings have the advantage of being finite products of $\F_p$, so that we can directly use the results in Chapter~\ref{chap_alg_prime}.
Then we generalize the algorithm in Chapter~\ref{chap_alg_prime} to compute a collection of idempotent decompositions of the rings $R_K$ so that they correspond to a strongly antisymmetric $\mathcal{P}$-schemes of double cosets.

In addition, we introduce the following notations concerning partitions of double coset spaces: for every double coset $Hg\mathcal{D}_{\mathfrak{Q}_0}\in H\backslash G/\mathcal{D}_{\mathfrak{Q}_0}$ where $H\subseteq G$, we associate two positive integers $f(Hg\mathcal{D}_{\mathfrak{Q}_0})$ and $e(Hg\mathcal{D}_{\mathfrak{Q}_0})$, called the {\em inertia degree} and the {\em ramification index} of $Hg\mathcal{D}_{\mathfrak{Q}_0}$ respectively.\footnote{These names come from the fact that $f(Hg\mathcal{D}_{\mathfrak{Q}_0})$ (resp. $e(Hg\mathcal{D}_{\mathfrak{Q}_0})$) is the inertia degree (resp. ramification index) of the prime ideal of $\ord_{L^H}$ lying over $p$ corresponding to $Hg\mathcal{D}_{\mathfrak{Q}_0}$. See Definition~\ref{defi_indandeg} for details.} Then we say a partition $P$ of $H\backslash G/\mathcal{D}_{\mathfrak{Q}_0}$ has {\em locally constant inertia degrees (resp. ramification indices)} if for every block $B$ in $P$, all the double cosets in $B$ have the same inertia degree (resp. ramification index). 
%When $\mathcal{D}_{\mathfrak{Q}_0}$ is the trivial subgroup, we always have $f(Hg\mathcal{D}_{\mathfrak{Q}_0})=e(Hg\mathcal{D}_{\mathfrak{Q}_0})=1$ in which case these properties are automatically  satisfied, but in general this is not true. 
We design efficient algorithms that force the partitions in a $\mathcal{P}$-collection of double cosets to have locally constant inertia degrees and ramification indices. These algorithms may be regarded as the analogues of distinct-degree factorization \citep{CZ81} and square-free factorization \citep{Yun76, Knu98} that factorize a polynomial  according to the degrees and the multiplicities of the irreducible factors.

The discussion above is summarized by the following theorem, which generalizes Theorem~\ref{thm_mainalginformal} in Chapter~\ref{chap_alg_prime}.

\begin{thm}[informal]\label{thm_maindcinformal}
Under GRH, there exists a deterministic algorithm that given
a poset $\mathcal{P}^\sharp$ of number fields between $K_0$ and $L$ corresponding to a poset $\mathcal{P}$ of subgroups of $G$,
outputs   idempotent decompositions of $R_K$ for $K\in\mathcal{P}^\sharp$   corresponding to a strongly antisymmetric $\mathcal{P}$-scheme of double cosets $\mathcal{C}$ with respect to $\mathcal{D}_{\mathfrak{Q}_0}$.
%, where $L$ is the splitting field of $\tilde{f}$ over $K_0$ and $G=\gal(L/K_0)$.  
Moreover,  all the partitions in $\mathcal{C}$ have locally constant inertia degrees and ramification indices.
The running time is polynomial in the size of the input.
\end{thm}

\paragraph{From a $\mathcal{P}$-scheme of double cosets to an ordinary $\mathcal{P}$-scheme.}

Theorem~\ref{thm_maindcinformal} is still not enough for proving our main result (Theorem~\ref{thm_alginformalgeneral}), since the algorithm in Theorem~\ref{thm_maindcinformal}  only produces a strongly antisymmetric $\mathcal{P}$-scheme of double cosets rather than an (ordinary) $\mathcal{P}$-scheme. 
%This problem is due the fact that for $K=L^H$, the set of prime ideals that $p\ord_K$ splits into is identified with the double coset space $H\backslash G/\mathcal{D}_{\mathfrak{Q}_0}$ rather than the right coset space $H\backslash G$. 
While strongly antisymmetric $\mathcal{P}$-schemes of double cosets are interesting objects, we do not know if their existence implies the existence of  strongly antisymmetric (ordinary) $\mathcal{P}$-schemes.

To overcome this problem, we strengthen the algorithm by maintaining not only idempotent decompositions  of a collection of rings $R_K$, but also elements in rings of the form $\bar{\ord}_K$ or $(\bar{\ord}_K/\rad(\bar{\ord}_K))\otimes_{\F_q} \F_{q^i}$, $i\in\N^+$. 
More specifically, we compute auxiliary elements $s_\delta\in \bar{\ord}_K$ (resp. $t_\delta\in (\bar{\ord}_K/\rad(\bar{\ord}_K))\otimes_{\F_q} \F_{q^i}$) 
%\footnote{Here $i$ is the inertia degree of the corresponding block. See Lemma~\ref{lem_genset}.} 
 for number fields $K$ and idempotents $\delta$.
%whose corresponding ramification index (resp. inertia degree) is greater than one. 
Then we define a $\mathcal{P}$-collection $\tilde{\mathcal{C}}$ based on these auxiliary elements and the $\mathcal{P}$-scheme of double cosets $\mathcal{C}$ computed in Theorem~\ref{thm_maindcinformal}. Moreover, we describe subroutines that properly refines  the partitions in $\mathcal{C}$ unless $\tilde{\mathcal{C}}$  is a strongly antisymmetric $\mathcal{P}$-scheme. This allows us to strengthen  Theorem~\ref{thm_maindcinformal} so that the algorithm produces  a strongly antisymmetric (ordinary) $\mathcal{P}$-scheme $\tilde{\mathcal{C}}$ in addition to a $\mathcal{P}$-scheme of double cosets. See Theorem~\ref{thm_comppschemeg} for the formal statement. Our main result (Theorem~\ref{thm_alginformalgeneral}) then follows easily.

\paragraph{Outline of the chapter.}

Notations and mathematical preliminaries are given in Section~\ref{sec_prelimg}, and algorithmic preliminaries are given in Section~\ref{sec_algpreg}.  In Section~\ref{sec_algreductiong}, we reduce the problem of factoring $f$ to that of computing an idempotent decomposition of $R_F$. In Section~\ref{sec_algmaing}, we  give (a preliminary version of)  the main body of the algorithm that computes idempotent decompositions corresponding to a strongly antisymmetric $\mathcal{P}$-scheme of double cosets. This $\mathcal{P}$-scheme  also has the property that all of its partitions have locally constant inertia degrees and ramification indices, as guaranteed by the subroutines described in Section~\ref{sec_testlocalconst} and Section~\ref{sec_testlocalconst2}.

The next three sections address the problem of producing an (ordinary) $\mathcal{P}$-scheme from the above $\mathcal{P}$-scheme of double cosets.
More specifically, in Section~\ref{sec_defordschm}, we give a subroutine that computes the auxiliary elements $s_\delta$ and $t_\delta$, and use these elements to define a $\mathcal{P}$-collection $\tilde{\mathcal{C}}$.
In Section~\ref{sec_ordpschprop}, we introduce a property about $\mathcal{P}$-collections called $(\mathcal{C},\mathcal{D})$-separatedness, and use it to give a criterion for $\tilde{\mathcal{C}}$ being a strongly antisymmetric $\mathcal{P}$-scheme.
In Section~\ref{sec_compordpsch}, we modify the algorithm in Section~\ref{sec_algmaing} to produce a strongly antisymmetric $\mathcal{P}$-scheme, based on the results in Section~\ref{sec_defordschm} and Section~\ref{sec_ordpschprop}.

Finally, in Section~\ref{sec_puttogetherg}, we combine the results  in  previous sections to obtain the generalized $\mathcal{P}$-scheme algorithm, and use it to prove the main result of this chapter (Theorem~\ref{thm_alginformalgeneral}). Using the algorithm, we also obtain generalizations of the main results in \citep{Hua91-2, Hua91, Ron88, Ron92, Evd92, Evd94, IKS09}.

\section{Preliminaries}\label{sec_prelimg}

For a number field $K$, denote by $\bar{\ord}_{K}$ the quotient ring $\ord_{K}/p\ord_K$.
For $K_0=\Q[Y]/(\tilde{h}(Y))$, we have
\begin{lem}
The ideal $p\ord_{K_0}$ is a prime ideal of $\ord_{K_0}$. And $\bar{\ord}_{K_0}\cong \F_q$.
\end{lem}

\begin{proof}
Let $\bar{Y}:=Y+(\tilde{h}(Y))\in\ord_{K_0}$.
Consider the ring homomorphism $i: \F_p[Y]/(h(Y))\to \bar{\ord}_{K_0}$ sending $Y+(h(Y))$ to $\bar{Y}+p\ord_{K_0}$. Clearly $i$ is a nonzero map since $i(1)=1$. 
As  $\F_p[Y]/(h(Y))$ is a field,  the map $i$ is injective.  As  $\F_p[Y]/(h(Y))$ and $ \bar{\ord}_{K_0}$  both have dimension $\deg(h)$ over $\F_p$, the map $i$ is an isomorphism. So $\bar{\ord}_{K_0}\cong \F_p[Y]/(h(Y))\cong \F_q$ and $p\ord_{K_0}$ is  prime.
\end{proof}

In the following, we give some notations and facts from algebraic number theory. The proofs can be found in standard references like \citep{Neu99}.

\paragraph{Splitting of prime ideals.}\index{splitting of prime ideals}

Let $K$ be a finite extension of $K_0$.
The ideal $p\ord_K$ splits  in the unique way into a product  of prime ideals of $\ord_K$, up to the ordering:
\[
p\ord_K=\prod_{i=1}^k \mathfrak{P}_i^{e(\mathfrak{P}_i)}=\bigcap_{i=1}^k \mathfrak{P}_i^{e(\mathfrak{P}_i)},
\]
where $\mathfrak{P}_1,\dots, \mathfrak{P}_k$ are distinct and $e(\mathfrak{P}_i)\in \N^+$.
We say $\mathfrak{P}_1,\dots,\mathfrak{P}_k$ are the prime ideals of $\ord_K$ {\em lying over}\index{lying over}  $p$.
For $i\in [k]$,  define $\kappa_{\mathfrak{P}_i}:=\ord_K/\mathfrak{P}_i$ which is a finite field, called the {\em residue field}\index{residue field} of $\mathfrak{P}_i$. The inclusion $\ord_{K_0}\hookrightarrow \ord_K$ induces an embedding of $\bar{\ord}_{K_0}\cong \F_q$ in $\kappa_{\mathfrak{P}_i}$, making $\kappa_{\mathfrak{P}_i}$  an extension field of $\bar{\ord}_{K_0}$.
Let $f(\mathfrak{P}_i):=[\kappa_{\mathfrak{P}_i}: \bar{\ord}_{K_0}]$. We call  $e(\mathfrak{P}_i)$ and $f(\mathfrak{P}_i)$ the {\em ramification index}\index{ramification index!of a prime ideal} and the {\em inertia degree}\index{inertia degree!of a prime ideal} of $\mathfrak{P}_i$  (over $p\ord_{K_0}$) respectively.
It holds that
\[
\sum_{i=1}^k e(\mathfrak{P}_i)f(\mathfrak{P}_i)=[K:K_0].
\]
\nomenclature[d1b]{$\kappa_{\mathfrak{P}}$}{residue field of $\mathfrak{P}$}
\nomenclature[d1c]{$e(\mathfrak{P})$}{ramification index of $\mathfrak{P}$ over $p\ord_{K_0}$}
\nomenclature[d1d]{$f(\mathfrak{P})$}{inertia degree of $\mathfrak{P}$ over $p\ord_{K_0}$}

\paragraph{Vector spaces $\mathfrak{P}^i/\mathfrak{P}^{i+1}$.}
We also use the following facts implicitly:

 For a number field $K$, $i\in \N$ and a nonzero prime ideal $\mathfrak{P}$ of $\ord_K$,  the  abelian group $\mathfrak{P}^i/\mathfrak{P}^{i+1}$ is an one-dimensional vector space over the field $\kappa_{\mathfrak{P}}=\ord_K/\mathfrak{P}$, where the scalar multiplication is defined by
 \[
(u+\mathfrak{P})\cdot (v+\mathfrak{P}^{i+1})=uv+\mathfrak{P}^{i+1} \quad\text{for}~u\in \ord_K, v\in\mathfrak{P}^i.
 \]
 %\footnote{Here $\mathfrak{P}^i$ is defined to be $\ord_K$ when $i=0$.} 
 For $i,j\in\N$ and $u\in\mathfrak{P}^i-\mathfrak{P}^{i+1}$, the map 
 \[
 x+\mathfrak{P}^{j+1}\mapsto ux+\mathfrak{P}^{i+j+1}
 \]
 is an isomorphism from $\mathfrak{P}^j/\mathfrak{P}^{j+1}$ to $\mathfrak{P}^{i+j}/\mathfrak{P}^{i+j+1}$, both regarded as vector spaces over  $\kappa_{\mathfrak{P}}$. In particular, for $i,j\in\N$ and $u\in\mathfrak{P}^i-\mathfrak{P}^{i+1}$, we have $u^j\in \mathfrak{P}^{ij}-\mathfrak{P}^{ij+1}$.

Now suppose $K, K'$ are finite extensions of $K_0$ and $K\subseteq K'$. And $\mathfrak{P}$, $\mathfrak{Q}$ are prime ideals of $\ord_K$ and $\ord_{K'}$ respectively, both lying over $p$, such that $\mathfrak{Q}\cap \ord_K=\mathfrak{P}$. Then $e(\mathfrak{P})$  divides $e(\mathfrak{Q})$ and $f(\mathfrak{P})$ divides $f(\mathfrak{Q})$. 
%\footnote{They are known as the ramification index and the inertia degree of $\mathfrak{Q}$ over $\mathfrak{P}$ respectively.}
  And for $i\in\N$, the inclusion $\ord_K\hookrightarrow\ord_{K'}$ induces an inclusion $\mathfrak{P}^i/\mathfrak{P}^{i+1}\hookrightarrow \mathfrak{Q}^{i'}/\mathfrak{Q}^{i'+1}$ where $i'=i\cdot e(\mathfrak{Q})/e(\mathfrak{P})$.

\paragraph{The decomposition group and the inertia group.}

Let $L/K_0$ be a Galois extension of number fields with the Galois group $G=\gal(L/K_0)$.
Let $\mathfrak{P}$ be a prime ideal of $\ord_L$ lying over $p$.
The group
\[
\mathcal{D}_{\mathfrak{P}}:=\{g\in G:\prescript{g}{}{\mathfrak{P}}=\mathfrak{P}\}\subseteq G
\]
is called the {\em decomposition group}\index{decomposition group} of $\mathfrak{P}$ over $K_0$.
And the group
\[
\mathcal{I}_{\mathfrak{P}}:=\{g\in G:\prescript{g}{}{x}\equiv x{\pmod{\mathfrak{P}}} ~\text{for all}~ x\in\ord_L\}
\]
is a normal subgroup of $\mathcal{D}_{\mathfrak{P}}$, called the {\em inertia group}\index{inertia group} of $\mathfrak{P}$ over $K_0$.
Each automorphism $g\in \mathcal{D}_{\mathfrak{P}}$ of $L$ restricts to an automorphism of $\ord_L$  fixing $\ord_{K_0}$  and satisfying $\prescript{g}{}{\mathfrak{P}}=\mathfrak{P}$, and hence induces an automorphism $\bar{g}$ of the residue field $\kappa_{\mathfrak{P}}$ fixing $\bar{\ord}_{K_0}$, defined by
\[
\prescript{\bar{g}}{}{(x+\mathfrak{P})}=\prescript{g}{}{x}+\mathfrak{P}.
\]
The map $\pi:g\mapsto \bar{g}$ is a surjective group homomorphism from $\mathcal{D}_{\mathfrak{P}}$ to $\gal(\kappa_{\mathfrak{P}}/\bar{\ord}_{K_0})$ whose kernel is precisely $\mathcal{I}_{\mathfrak{P}}$, i.e, we have a short exact sequence
\[
1\to \mathcal{I}_{\mathfrak{P}}\to \mathcal{D}_{\mathfrak{P}}\xrightarrow{\pi} \gal(\kappa_{\mathfrak{P}}/\bar{\ord}_{K_0})\to 1.
\]
The Galois group $\gal(\kappa_{\mathfrak{P}}/\bar{\ord}_{K_0})$ is cyclic and is generated by the {\em Frobenius automorphism}\index{Frobenius automorphism} $x\mapsto x^q$ of $\kappa_{\mathfrak{P}}$ over $\bar{\ord}_{K_0}\cong \F_q$.
\nomenclature[d1e]{$\mathcal{D}_{\mathfrak{P}}$}{decomposition group of $\mathfrak{P}$ over $K_0$}
\nomenclature[d1f]{$\mathcal{I}_{\mathfrak{P}}$}{inertia group of $\mathfrak{P}$ over $K_0$}
\nomenclature[d1g]{$\mathcal{W}_{\mathfrak{P}}$}{wild inertia group of $\mathfrak{P}$ over $K_0$}

\paragraph{The wild inertia group.} Let $L$, $G$ and  $\mathfrak{P}$  be as above. The group
\[
\mathcal{W}_{\mathfrak{P}}:=\{g\in G:\prescript{g}{}{x}\equiv x{\pmod{\mathfrak{P}^2}} ~\text{for all}~ x\in\ord_L\}.
\]
is a normal subgroup of $\mathcal{I}_{\mathfrak{P}}$, called the {\em wild inertia group}\index{wild inertia group} of $\mathfrak{P}$ over $K_0$. 

Choose $\pi_L\in\mathfrak{P}-\mathfrak{P}^2$. We have a group homomorphism
$\mathcal{I}_{\mathfrak{P}}\to  \kappa_{\mathfrak{P}}^\times$
sending $g\in \mathcal{I}_{\mathfrak{P}}$ to the unique element $c_g\in \kappa_{\mathfrak{P}}^\times$ satisfying $\prescript{g}{}{\pi_L}+\mathfrak{P}^2= c_g(\pi_L+\mathfrak{P}^2)$.
This map is independent of the choice of $\pi_L$, and its kernel is precisely $\mathcal{W}_{\mathfrak{P}}$.
It is also known that $\mathcal{W}_{\mathfrak{P}}$ is a $p$-group.
See \cite[Section~\RN{2}.10]{Neu99}.  

In our factoring algorithm,  the group $G$ is a subgroup of $\sym(n)$ where $n$ is the degree of the input polynomial $f(X)\in\F_q[X]$. 
We can always assume $p>n$, since the case $p\leq n$ is solved in polynomial time by  Berlekamp's algorithm in \citep{Ber70}. Under this assumption, the $p$-subgroup $\mathcal{W}_{\mathfrak{P}}$ of $G$ is trivial, and hence the map $\mathcal{I}_{\mathfrak{P}}\to  \kappa_{\mathfrak{P}}^\times$ above is injective. In particular, the inertia group $\mathcal{I}_{\mathfrak{P}}$ is cyclic.

\paragraph{Prime ideals vs. double cosets.} We have the following generalization of Theorem~\ref{thm_split}, which gives a one-to-one correspondence between prime ideals  lying over $p$ and double cosets. See \citep{Neu99} for its proof.

\begin{thm}\label{thm_split_general}
Let $L/K_0$ be a Galois extension of number fields and let $G=\gal(L/K_0)$. Fix a prime ideal $\mathfrak{Q}_0$ of $\ord_L$ lying over $p$. For any subgroup $H\subseteq G$ and the corresponding fixed subfield $K=L^H$, the map $Hg\mathcal{D}_{\mathfrak{Q}_0}\mapsto \prescript{g}{}{\mathfrak{Q}_0}\cap \ord_K$ is a one-to-one correspondence between the double cosets in $H\backslash G/\mathcal{D}_{\mathfrak{Q}_0}$ and the prime ideals of $\ord_K$ lying over $p$.\footnote{Note that this map is well defined: 
for another representative $hgh'\in G$ of $Hg\mathcal{D}_{\mathfrak{Q}_0}$, where $h\in H$ and $h'\in \mathcal{D}_{\mathfrak{Q}_0}$, we have  
$
\prescript{hgh'}{}{\mathfrak{Q}_0}\cap \ord_K=\prescript{hg}{}{\mathfrak{Q}_0}\cap \ord_K=\prescript{h}{}{(\prescript{g}{}{\mathfrak{Q}_0}\cap \ord_K)}=\prescript{g}{}{\mathfrak{Q}_0}\cap \ord_K
$
since $\prescript{h'}{}{\mathfrak{Q}_0}=\mathfrak{Q}_0$ and $\ord_K$  is fixed by $H$.} Moreover, for $g\in G$ and the prime ideal $\mathfrak{P}=\prescript{g}{}{\mathfrak{Q}_0}\cap \ord_K$ corresponding to $Hg\mathcal{D}_{\mathfrak{Q}_0}$, define 
\[
n(\mathfrak{P}):=|\{ Hh\in H\backslash G: Hh\mathcal{D}_{\mathfrak{Q}_0}=Hg\mathcal{D}_{\mathfrak{Q}_0}\}|.
\] 
Then 
\[
e(\mathfrak{P})=|\{ Hh\in H\backslash G: Hh\mathcal{I}_{\mathfrak{Q}_0}=Hg\mathcal{I}_{\mathfrak{Q}_0}\}|
\quad\text{and}\quad
f(\mathfrak{P})=\frac{n(\mathfrak{P})}{e(\mathfrak{P})}.
\]
\end{thm}

%\paragraph{Ramification indices and inertia degrees of  double cosets.}

Motivated by Theorem~\ref{thm_split_general}, we  define the ramification index and the inertia degree of a double coset:
\begin{defi}\label{defi_indandeg}
Let $G$ be a finite group, $H, \mathcal{D}$ subgroups of $G$, and $\mathcal{I}$ a normal subgroup of $\mathcal{D}$.
Define the {\em ramification index}\index{ramification index!of a double coset} of a double coset $Hg\mathcal{D}\in H\backslash G/\mathcal{D}$ with respect to $(\mathcal{D}, \mathcal{I})$ to be
\[
e(Hg\mathcal{D}):=|\{ Hh\in H\backslash G: Hh\mathcal{I}=Hg\mathcal{I}\}|,
\]
which is well defined.\footnote{To see that $e(Hg\mathcal{D})$ is well defined, consider two representatives $g$ and $g'$ of $Hg\mathcal{D}$. Then $g'=sgt$ for some $s\in H$ and $t\in \mathcal{D}$. Note that $Hht\mathcal{I}=Hh\mathcal{I}t$ for all $h\in G$. It follows that the map $Hh\mapsto Hht$ is a bijection from $\{ Hh\in H\backslash G: Hh\mathcal{I}=Hg\mathcal{I}\}$ to $\{ Hh\in H\backslash G: Hh\mathcal{I}=Hg'\mathcal{I}\}$.}
And define the {\em inertia degree}\index{inertia degree!of a double coset} of $Hg\mathcal{D}$  with respect to $(\mathcal{D}, \mathcal{I})$ to be
 \[
f(Hg\mathcal{D}):=\frac{|\{Hh\in H\backslash G: Hh\mathcal{D}=Hg\mathcal{D}\}|}{e(Hg\mathcal{D})}.
\]
\end{defi}
\nomenclature[d1h]{$e(Hg\mathcal{D})$}{ramification index of a double coset $Hg\mathcal{D}$}
\nomenclature[d1i]{$f(Hg\mathcal{D})$}{inertia degree of a double coset $Hg\mathcal{D}$}

Suppose $L/K_0$ is a Galois extension of number fields with the Galois group $G$. Fix a prime ideal $\mathfrak{Q}_0$ of $\ord_L$ lying over $p$. Let $H$ be a subgroup of $G$ and $K=L^H$. Then by Theorem~\ref{thm_split_general}, the ramification index (resp. inertia degree) of a double coset $Hg\mathcal{D}_{\mathfrak{Q}_0}\in H\backslash G/\mathcal{D}_{\mathfrak{Q}_0}$ with respect to   $(\mathcal{D}_{\mathfrak{Q}_0}, \mathcal{I}_{\mathfrak{Q}_0})$  is precisely the ramification index (resp. inertia degree) of the corresponding prime ideal $\prescript{g}{}{\mathfrak{Q}_0}\cap \ord_K$ of $\ord_K$.

We also introduce the following notations concerning partitions of a double coset space.

\begin{defi}
Let $G, H, \mathcal{D}, \mathcal{I}$ be as in Definition~\ref{defi_indandeg}.
We say a partition $P$ of $H\backslash G/\mathcal{D}$ has {\em locally constant} ramification indices\index{locally constant!ramification indices} (resp. inertia degrees)\index{locally constant!inertia degrees} with respect to $(\mathcal{D},\mathcal{I})$ if for every $B\in P$, all the double cosets in $B$ have the same ramification index (resp. inertia degree) with respect to $(\mathcal{D},\mathcal{I})$. For such a partition $P$ and any $B\in P$, denote by $e(B)$ (resp. $f(B)$) the ramification index (resp. inertia degree) of any double coset in $B$.
\end{defi}

\paragraph{Radicals of rings and polynomials.} Let $A$ be a (commutative) ring. An element $x\in A$ is  {\em nilpotent}\index{nilpotent} if $x^k=0$ for some $k\in\N^+$. The {\em radical}\index{radical!of a ring} (or {\em nilradical})\index{nilradical|see{radical of a ring}} of $A$, denoted by $\rad(A)$, is the ideal consisting of the nilpotent elements of $A$. It equals the intersection of all the prime ideals of $A$ (see \citep{AM69}).

Let $g(X)\in\F_q[X]$ be a non-constant polynomial with the following factorization
\[
g(X)=c\cdot \prod_{i=1}^k (g_i(X))^{m_i}
\]
over $\F_q$, where $c\in\F_q$ is the leading coefficient of $g$ and $g_1,\dots,g_k$ are distinct monic irreducible polynomials over $\F_q$. 
%We call $m_i\in\N^+$ the {\em multiplicity} of $g_i$ in $g$.  
 Define the {\em radical}\index{radical!of a polynomial} $\rad(g)$ of $g$ to be the monic polynomial
$\prod_{i=1}^k g_i(X)\in \F_q[X]$.  For $A=\F_q[X]/(g(X))$, the ideal of $A$ generated by $\rad(g)+(g(X))\in A$ is precisely $\rad(A)$.
\nomenclature[d1j]{$\rad(A)$}{radical of a ring $A$}
\nomenclature[d1k]{$\rad(g)$}{radical of a polynomial $g$}

\paragraph{The ring $R_K$.} Suppose $K$ is a finite extension of $K_0$ and $p\ord_K$ splits into the product of prime ideals 
\[
p\ord_K=\prod_{i=1}^k  \mathfrak{P}_i^{e(\mathfrak{P}_i)},
\] where $\mathfrak{P}_1,\dots,\mathfrak{P}_k$ are distinct. The radical of $\bar{\ord}_K$ is given by
\[
\rad(\bar{\ord}_K)=\bigcap_{i=1}^k \mathfrak{P}_i /p\ord_K=\left(\bigcap_{i=1}^k \mathfrak{P}_i\right)/p\ord_K.
\]

By the Chinese remainder theorem, we have the isomorphism
\[
\bar{\ord}_K/\rad(\bar{\ord}_K) \to \prod_{i=1}^k \ord_K/\mathfrak{P}_i=\prod_{i=1}^k \kappa_{\mathfrak{P}_i},
\]
sending $x+\rad(\bar{\ord}_K)\in \bar{\ord}_K/\rad(\bar{\ord}_K)$ to $\left(\tilde{x}\bmod \mathfrak{P}_1,\dots,\tilde{x}\bmod \mathfrak{P}_k\right)$, where $\tilde{x}\in\ord_K$ is an arbitrary element lifting $x\in \bar{\ord}_K$. 
%Each factor $\ord_K/\mathfrak{P}_i$ is a finite extension of $\bar{\ord}_{K_0}\cong \F_q$. 
 In particular, the ring $\bar{\ord}_K/\rad(\bar{\ord}_K)$ is semisimple. 

Define  $R_K$ to be the subring of $\bar{\ord}_K/\rad(\bar{\ord}_K)$ consisting of elements fixed by the Frobenius automorphism $x\mapsto x^p$ over $\F_p$, i.e.,
\[
R_K:=\left\{x\in \bar{\ord}_K/\rad(\bar{\ord}_K): x^p=x\right\}.
\]
The isomorphism $\bar{\ord}_K/\rad(\bar{\ord}_K) \to \prod_{i=1}^k \kappa_{\mathfrak{P}_i}$  above identifies $R_K$ with the subring $\prod_{i=1}^k \F_p$ of $\prod_{i=1}^k \kappa_{\mathfrak{P}_i}$. So $R_K$ is a finite product of copies of $\F_p$ and in particular is semisimple.
\nomenclature[d1l]{$R_K$}{ring $\left\{x\in \bar{\ord}_K/\rad(\bar{\ord}_K): x^p=x\right\}$}

 Observe that the map $\mathfrak{m}\mapsto  (\mathfrak{m}/\rad(\bar{\ord}_K))\cap R_K$ is a one-to-one correspondence between the maximal ideals of $\bar{\ord}_K$ and those of $R_K$. Combining this fact with Theorem~\ref{thm_split_general}, we obtain

\begin{lem}\label{lem_idealdoublecosetg}
Let $L$, $G$, $\mathfrak{Q}_0$ be as in Theorem~\ref{thm_split_general}. For any subgroup $H\subseteq G$ and the corresponding fixed subfield $K=L^H$, the map 
\[
Hg\mathcal{D}_{\mathfrak{Q}_0}\mapsto  \frac{(\prescript{g}{}{\mathfrak{Q}_0}\cap \ord_K)/p\ord_K}{\rad(\bar{\ord}_K)}\cap R_K
\] is a one-to-one correspondence between the double cosets in $H\backslash G/\mathcal{D}_{\mathfrak{Q}_0}$ and the maximal ideals of $R_K$.
\end{lem}

\paragraph{Idempotent decompositions vs. partitions of a double coset space.}

In the following, we establish a one-to-one correspondence between the idempotent decompositions of $R_K$ and the partitions of a certain double coset space.

For a number field extension $L/K$, the inclusion $\ord_{K}\hookrightarrow \ord_{L}$ induces an inclusion $\bar{\ord}_{K}\hookrightarrow \bar{\ord}_{L}$. So we may regard $\bar{\ord}_{K}$ as a subring of  $\bar{\ord}_{L}$. Note that $\rad(\bar{\ord}_L)\cap \bar{\ord}_K=\rad(\bar{\ord}_K)$. Passing to the quotient rings yields an inclusion $\bar{\ord}_K/\rad(\bar{\ord}_K)\hookrightarrow \bar{\ord}_L/\rad(\bar{\ord}_L)$. Restricting to the subring $R_K$, we obtain an inclusion
\[
i_{K,L}:R_K\hookrightarrow R_L.
\]
Also note that if $L/K_0$ is a Galois extension with the Galois group $G$, the action of $G$ on $\ord_L$ induces an action on $R_L$ that permutes the maximal ideals of $R_L$. 
 
Fix the following notations: let $L$ be a Galois extension of $K_0$ with the Galois group $G=\gal(L/K_0)$.
For a (nonzero) prime ideal $\mathfrak{Q}$ of $\ord_L$ lying over $p$, define 
\[
\bar{\mathfrak{Q}}:= \frac{\mathfrak{Q}/p\ord_L}{\rad(\bar{\ord}_L)}\cap R_L,
\]
 which is a maximal ideal of $R_L$, and let $\delta_{\bar{\mathfrak{Q}}}$ be the primitive idempotent of $\bar{\ord}_L$ satisfying $\delta_{\bar{\mathfrak{Q}}}\equiv 1\pmod{\bar{\mathfrak{Q}}}$ and $\delta_{\bar{\mathfrak{Q}}}\equiv 0\pmod{\bar{\mathfrak{Q}}'}$ for all maximal ideal $\bar{\mathfrak{Q}}'\neq \bar{\mathfrak{Q}}$ of $\bar{\ord}_L$. Finally, fix a prime ideal $\mathfrak{Q}_0$ of $\ord_L$ lying over $p$.

\begin{defi}\label{defi_partitioncorg}
 Suppose $H$ is a subgroup of $G$ and $K=L^H$. Then
\begin{itemize}
\item for an idempotent decomposition $I$ of $R_K$, define $P(I)$ to be the partition of $H\backslash G/\mathcal{D}_{\mathfrak{Q}_0}$ where $Hg\mathcal{D}_{\mathfrak{Q}_0},Hg'\mathcal{D}_{\mathfrak{Q}_0}$ are in the same block iff $\prescript{g^{-1}}{}{(i_{K,L}(\delta))}\equiv \prescript{g'^{-1}}{}{(i_{K,L}(\delta))}\pmod{\bar{\mathfrak{Q}}_0}$ holds for all $\delta\in I$, and
%\footnote{Here we regard $\delta\in I\subseteq\bar{\ord}_K$ as an element of $\bar{\ord}_L$ via the natural inclusion $\bar{\ord}_K\hookrightarrow \bar{\ord}_L$. }
\item for a partition $P$ of $H\backslash G/\mathcal{D}_{\mathfrak{Q}_0}$, define $I(P)$ to be the idempotent decomposition of $R_K$ consisting of  the idempotents
\[
\delta_B:=i_{K,L}^{-1}\left(\sum_{g\mathcal{D}_{\mathfrak{Q}_0}\in G/\mathcal{D}_{\mathfrak{Q}_0}: Hg\mathcal{D}_{\mathfrak{Q}_0}\in B}\prescript{g}{}{\delta_{\bar{\mathfrak{Q}}_0}}\right),
\]
 where  $B$ ranges over the blocks in $P$.
\end{itemize}
\end{defi}

We have the following two lemma, whose proof is similar to that of  Lemma~\ref{lem_pandi} and and can be found in Appendix~\ref{chap_omitted2}. 

\begin{restatable}{lem}{lempig}\label{lem_pandig}
The partitions $P(I)$ and the idempotent decompositions $I(P)$ are well defined. 
And for any idempotent decomposition $I$  of $\bar{\ord}_K$, the idempotents $\delta\in I$ correspond one-to-one to  the blocks of $P(I)$ via the map 
\[
\delta\mapsto B_\delta:=\{Hg\mathcal{D}_{\mathfrak{Q}_0}\in H\backslash G/\mathcal{D}_{\mathfrak{Q}_0}: \prescript{g^{-1}}{}{(i_{K,L}(\delta))}\equiv 1\pmod{\bar{\mathfrak{Q}}_0}\}
\]
 with the inverse map $B\mapsto \delta_B$.
\end{restatable}

Now we are ready to establish the following correspondence.

\begin{lem}\label{lem_picorresg}
The map $I\mapsto P(I)$ is a one-to-one correspondence between the idempotent decompositions of $R_K$ and the partitions of $H\backslash G/\mathcal{D}_{\mathfrak{Q}_0}$, with the inverse map $P\mapsto I(P)$. 
\end{lem}

\begin{proof}
Note $I(P)=\{\delta_B: B\in P\}$ by definition and $P(I)=\{B_\delta: \delta\in I\}$ by Lemma~\ref{lem_pandig}. So $I=I(P(I))$ by Lemma~\ref{lem_pandig}.
Also note the map $B\mapsto \delta_B$ is injective, and hence the map $P\mapsto I(P)$ is also injective. So $P=I(P(I))$.
\end{proof}

\paragraph{$\mathcal{P}$-collections and $\mathcal{P}$-schemes of double cosets.}

Let $G$ be a finite group and $\mathcal{D}\subseteq G$ a subgroup.
We generalize projections and conjugations introduced in Chapter~\ref{chap_pscheme} so that they are defined between double coset spaces:
\begin{itemize}
\item (projection) for $H\subseteq H'\subseteq G$, define the {\em projection}\index{projection} $\pi_{H, H'}^\mathcal{D}: H\backslash G/\mathcal{D}\to H'\backslash G/\mathcal{D}$ to be the map sending $Hg\mathcal{D}\in H\backslash G/\mathcal{D}$ to $H'g\mathcal{D}\in  H'\backslash G/\mathcal{D}$, and
\item (conjugation) for $H\subseteq G$ and $g\in G$, define the {\em conjugation}\index{conjugation} $c_{H,g}^\mathcal{D}: H\backslash G/\mathcal{D}\to gHg^{-1}\backslash G/\mathcal{D}$ to be the map sending $Hh\mathcal{D}\in H\backslash G/\mathcal{D}$ to $(gHg^{-1})gh\mathcal{D}\in  gHg^{-1}\backslash G/\mathcal{D}$.
\end{itemize}
\nomenclature[d1m]{$\pi_{H,H'}^\mathcal{D}$}{projection from $H\backslash G/\mathcal{D}$ to $H'\backslash G/\mathcal{D}$}
\nomenclature[d1n]{$c_{H,g}^\mathcal{D}$}{conjugation from $H\backslash G/\mathcal{D}$ to $gHg^{-1}\backslash G/\mathcal{D}$}

Next we define $\mathcal{P}$-collections and $\mathcal{P}$-schemes  of double cosets. 

\begin{defi}\label{defi_double}
Let $\mathcal{P}$ be a subgroup system over a finite group $G$. Then a {\em $\mathcal{P}$-collection of double cosets}\index{Pcollection@$\mathcal{P}$-collection!of double cosets}  with respect to a subgroup $\mathcal{D}$ of $G$ is a family $\mathcal{C}=\{C_H: H\in \mathcal{P}\}$ indexed by $\mathcal{P}$ where each $C_H$ is a partition of $H\backslash G/\mathcal{D}$. 
Moreover, $\mathcal{C}$ is a {\em $\mathcal{P}$-scheme of double cosets}\index{Pscheme@$\mathcal{P}$-scheme!of double cosets} with respect to $\mathcal{D}$  if it has the following properties:
\begin{itemize}
\item (compatibility)\index{compatibility!of a $\mathcal{P}$-collection of double cosets}  for $H,H'\in \mathcal{P}$ with $H\subseteq H'$ and $x,x'\in H\backslash G/\mathcal{D}$  in the same block of $C_H$, the images $\pi_{H,H'}^\mathcal{D}(x)$ and $\pi_{H,H'}^\mathcal{D}(x')$ are in the same block of $C_{H'}$.
\item (invariance)\index{invariance!of a $\mathcal{P}$-collection of double cosets}  for $H\in\mathcal{P}$ and $g\in G$, the map $c_{H,g}^\mathcal{D}: H\backslash G/\mathcal{D}\to gHg^{-1}\backslash G/\mathcal{D}$ maps any block of $C_H$ to a block of $C_{gHg^{-1}}$.
\item (regularity)\index{regularity!of a $\mathcal{P}$-collection of double cosets} for $H,H'\in\mathcal{P}$ with $H\subseteq H'$, $B\in C_H$, $B'\in C_{H'}$, the number of $x\in B$ satisfying $\pi_{H,H'}^\mathcal{D}(x)=y$ is a constant when $y$ ranges over the elements of $B'$.
\end{itemize}
We also define the following optional properties for a $\mathcal{P}$-scheme of double cosets $\mathcal{C}$ with respect to $\mathcal{D}$:
\begin{itemize}
\item (homogeneity and discreteness) $\mathcal{C}$ is {\em homogeneous}\index{homogeneity!of a $\mathcal{P}$-scheme of double cosets} on $H\in\mathcal{P}$ if $C_H=0_{H\backslash G/\mathcal{D}}$, and otherwise {\em inhomogeneous} on $H$. It is {\em discrete}\index{discreteness!of a $\mathcal{P}$-scheme of double cosets} on $H$ if $C_H=\infty_{H\backslash G/\mathcal{D}}$, and otherwise {\em non-discrete}  on $H$. 
\item (antisymmetry) $\mathcal{C}$ is {\em antisymmetric}\index{antisymmetry!of a $\mathcal{P}$-scheme of double cosets} if  for $H\in\mathcal{P}$, $g\in N_G(H)$, $B\in C_H$ and $Hg\mathcal{D}\in B$,  either $c_{H,g}^\mathcal{D}(Hg\mathcal{D})=Hg\mathcal{D}$ or $c_{H,g}^\mathcal{D}(Hg\mathcal{D})\not\in B$.
\item (strong antisymmetry)  $\mathcal{C}$ is {\em strongly antisymmetric}\index{strong antisymmetry!of a $\mathcal{P}$-scheme of double cosets} if for any sequence of subgroups $H_0,\dots, H_k\in\mathcal{P}$, $B_0\in C_{H_0},\dots, B_k\in C_{H_k}$, and maps $\sigma_1,\dots, \sigma_k$ satisfying
\begin{itemize}
\item $\sigma_i$ is a bijective map from $B_{i-1}$ to $B_i$,
\item $\sigma_i$ is of the form $c_{H_{i-1},g}^\mathcal{D}|_{B_{i-1}}$, $\pi_{H_{i-1},H_i}^\mathcal{D}|_{B_{i-1}}$, or  $(\pi_{H_i,H_{i-1}}^\mathcal{D}|_{B_i})^{-1}$,
\item $H_0=H_k$ and $B_0=B_k$,
\end{itemize}
the composition $\sigma_k\circ\cdots\circ \sigma_1$ is the identity map on $B_0=B_k$.
\end{itemize}
\end{defi}

The notions of $\mathcal{P}$-collections and $\mathcal{P}$-schemes introduced in Chapter~\ref{chap_pscheme} correspond to the special case that $\mathcal{D}$ is trivial.

\paragraph{Extension of scalars of $\bar{\ord}_K/\rad(\bar{\ord}_K)$.} In Section~\ref{sec_defordschm}--\ref{sec_ordpschprop}, we need a family of rings $A_{K,i}$  that are obtained from $\bar{\ord}_K/\rad(\bar{\ord}_K)$ via ``extension of scalars'', whose definitions are given below.

Let $K$ be a finite extension of $K_0$. The inclusion $A_0\subseteq \ord_{K_0}\hookrightarrow \ord_K$ induces an embedding of $\F_q\cong A_0/pA_0$ in $\bar{\ord}_K/\rad(\bar{\ord}_K)$, endowing $\bar{\ord}_K/\rad(\bar{\ord}_K)$ the structure of an $\F_q$-algebra. For $i\in \N^+$, we define the tensor product
 \[
 A_{K,i}:=(\bar{\ord}_K/\rad(\bar{\ord}_K))\otimes_{\F_q} \F_{q^i},
 \]
 which is an $\F_{q^i}$-algebra and is spanned by tensors $a\otimes b$ over $\F_q$ where $a\in \bar{\ord}_K/\rad(\bar{\ord}_K)$ and $b\in  \F_{q^i}$ (see \citep{AM69} for the definition of tensor products of rings).
 Intuitively, the ring $A_{K,i}$ is obtained from $\bar{\ord}_K/\rad(\bar{\ord}_K)$ by extending the scalars from $\F_q$ to $\F_{q^i}$. 
And $\bar{\ord}_K/\rad(\bar{\ord}_K)$ is naturally identified with a subring of $A_{K, i}$ via $a\mapsto a\otimes 1$. 
As $\bar{\ord}_K/\rad(\bar{\ord}_K)$ is semisimple, so is $A_{K,i}$.\footnote{We use the fact that $\F_{q^d}\otimes_{\F_q} \F_{q^i}$ is semisimple for $d, i\in\N^+$: suppose $\F_{q^d}\cong \F_q[X]/(g(X))$ where $g(X)\in\F_q[X]$ is irreducible over $\F_q$. Then $\F_{q^d}\otimes_{\F_q} \F_{q^i}\cong \F_{q^i}[X]/(g(X))\cong\prod_{j=1}^k \F_{q^i}[X]/(g_j(X))$ where $g_1,\dots,g_k$ are the irreducible factors of $g$ over $\F_{q^i}$.}
The {Frobenius automorphism} $x\mapsto x^q$ of $\bar{\ord}_K/\rad(\bar{\ord}_K)$ over $\F_q$ induces an automorphism of $A_{K,i}$ over $\F_{q^i}$ sending $a\otimes b$ to $a^q\otimes b$. We denote this automorphism by $\sigma_{K,i}$.
\nomenclature[d1o]{$A\otimes_{\F_q} B$}{tensor product of $A$ and $B$ over $\F_q$}
\nomenclature[d1p]{$A_{K,i}$}{ring $(\bar{\ord}_K/\rad(\bar{\ord}_K))\otimes_{\F_q} \F_{q^i}$}
\nomenclature[d1q]{$\sigma_{K,i}$}{automorphism of $A_{K,i}$ sending $a\otimes b$ to $a^q\otimes b$}

  The following lemma is also needed,  whose proof is deferred to  Appendix~\ref{chap_omitted2}.

\begin{restatable}{lem}{lemtransitivitybc}\label{lem_transitivitybc}
For any maximal ideal $\mathfrak{m}$ of $\bar{\ord}_K/\rad(\bar{\ord}_K)$, the group $\langle \sigma_{K,i}\rangle$ generated by $\sigma_{K,i}$ acts transitively on the set of  the maximal ideal of $A_{K,i}$ containing $\mathfrak{m}$.
\end{restatable}

 Suppose $K,K'$ are extensions of $K_0$ and $K\subseteq K'$. Then the inclusion $\ord_K\hookrightarrow \ord_{K'}$ induces an embedding $\iota:\bar{\ord}_K/\rad(\bar{\ord}_K)\hookrightarrow \bar{\ord}_{K'}/\rad(\bar{\ord}_{K'})$, which in turn induces a ring homomorphism $\iota': A_{K,i}\hookrightarrow A_{K',i}$ sending $a\otimes b$ to $\iota(a)\otimes b$. 
 The map $\iota'$ is injective since $\F_{q^i}$ is a {\em flat} $\F_q$-module\index{flat module} (see, e.g., \citep[Proposition~2.19 and Exercise~2.4]{AM69}).  This allows us to regard $A_{K,i}$ as a subring of $A_{K',i}$. Note that $\iota'\circ\sigma_{K,i}=\sigma_{K',i}\circ \iota'$.

 Finally, suppose $L/K_0$ is a finite Galois extension with the Galois group $G$. The action of $G$ on $L$ induces an action on $\bar{\ord}_L/\rad(\bar{\ord}_L)$, which in turn induces an action on $A_{L,i}$ via $\prescript{g}{}{(a\otimes b)}:=\prescript{g}{}{a}\otimes b$. This action commutes with $\sigma_{L,i}$.\footnote{This follows from the fact that
 the action of $G$ on  $\bar{\ord}_L/\rad(\bar{\ord}_L)$ respects the multiplication and hence commutes with the automorphism $x\mapsto x^q$.} 

\section{Algorithmic preliminaries}\label{sec_algpreg}

In this section, we discuss some basic procedures used in the algorithm. 

\paragraph{Computation of radicals, and square-free factorization.}

We need to compute the radical of a finite dimensional (commutative) $\F_p$-algebra. This problem was  studied in \citep{FR85, Ron90} and solved in polynomial time in the more general setting of associative algebras. We state their result but restrict to the special case of commutative algebras.

\begin{thm}[\citep{FR85, Ron90}]\label{thm_comprad}
There exists a polynomial-time algorithm that given a finite dimensional (commutative) $\F_p$-algebra $A$, computes an $\F_p$-basis of $\rad(A)$ in $A$.
\end{thm}

See, e.g., \citep[Theorem~2.7]{Ron90}.

Next we discuss the problem of computing the radical of a nonzero polynomial $g(X)\in\F_q[X]$. This is solved via {\em square-free factorization}.

\begin{defi}
A {\em square-free factorization}\index{square-free factorization} of a nonzero polynomial $g(X)\in \F_q[X]$ over $\F_q$ is a factorization
\[
g(X)=c\cdot \prod_{i=1}^k (g_i(X))^{m_i},
\]
where $c\in\F_q$ is the leading coefficient of $g$ and the factors $g_1(X),\dots,g_k(X)\in\F_q[X]$ are monic, square-free, and pairwise coprime.
\end{defi}

\begin{thm}[\citep{Yun76, Knu98}]
There exists a polynomial-time algorithm that computes a square-free factorization of a given nonzero polynomial $g(X)\in \F_q[X]$. 
\end{thm}

Given the square-free factorization $g(X)=c\cdot \prod_{i=1}^k (g_i(X))^{m_i}$, the radical $\rad(g)$ is simply the product of $g_i(X)$.
So we have
\begin{cor}\label{cor_polyrad}
There exists a polynomial-time algorithm that given a nonzero polynomial $g(X)\in \F_q[X]$, computes its radical $\rad(g)$. 
\end{cor}

 Alternatively, we can compute $\rad(g)$ by computing the radical of $\F_q[X]/(g(X))$ and then its generator. The details are left to the reader.
 
 \paragraph{Computation of annihilators.}
 
 Let $R$ be a (commutative) ring. For a set $S\subseteq R$, define the {\em annihilator}\index{annihilator} $\ann_R(S)$ of $S$ to be the ideal
 \[
 \ann_R(S):=\{r\in R: rs=0 ~\text{for all}~s\in S\}
 \]
 of $R$. When $S$ is a singleton $\{s\}$, we also write $\ann_R(s)$ instead of $\ann_R(\{s\})$ and call it the annihilator of $s$.
 \nomenclature[d1r]{$\ann_R(S)$}{annihilator of $S$ in $R$}
 
When $R$ is an finite dimensional $\F_p$-algebra, we can efficiently compute the annihilator $\ann_R(s)$ of an element $s\in R$  by solving the system of $\F_p$-linear equations given by $x s=0$.  Similarly, when $S$ is an $\F_p$-subspace of $R$ (in particular, when $S$ is an ideal of $R$),  we can compute  $\ann_R(S)$  efficiently given $R$ and an $\F_p$-basis $B$ of $S$ by solving the system of $\F_p$-linear equations   $x s=0$, where $s$ ranges over the basis $B$. 

\paragraph{Computation of various rings and ring homomorphisms.}
The algorithm uses relative number fields over $K_0$ rather than ordinary number fields, i.e., every number field is an extension of $K_0$ and is encoded as a $K_0$-algebra $K_0[X]/(g(X))$ where $g(X)\in K_0[X]$ is irreducible over $K_0$.

Given a relative number field $K$ over $K_0$, we can identify $K_0$ with an ordinary number field $\tilde{K}$ by  Corollary~\ref{lem_reltoord}. It allows us to efficiently compute a $p$-maximal order $\ord'_K$ as well as the quotient ring $\bar{\ord}_K$ as in Chapter~\ref{chap_alg_prime}. We can also efficiently compute the rings $\bar{\ord}_K/\rad(\bar{\ord}_K)$ and $R_K$, which are  used in the generalized $\mathcal{P}$-scheme algorithm developed in this chapter. This is summarized by the following lemma, whose proof is deferred to  Appendix~\ref{chap_omitted2}.
\begin{restatable}{lem}{lemcomputerings}\label{lem_computerings}
There exists a polynomial-time algorithm $\mathtt{ComputeRings}$  that given $p$ and a relative number field $K$ over $K_0$, computes the following data
\begin{itemize}
\item a $p$-maximal order $\ord'_K$ of $K$ and the inclusion $\ord'_K\hookrightarrow K$,
\item $\bar{\ord}_K$ and the quotient map $\ord'_K\to \bar{\ord}_K$,
\item  $\bar{\ord}_K/\rad(\bar{\ord}_K)$ and the quotient map  $\bar{\ord}_K\to  \bar{\ord}_K/\rad(\bar{\ord}_K)$,
\item $R_K$ and the inclusion $R_K\hookrightarrow \bar{\ord}_K/\rad(\bar{\ord}_K)$,
\end{itemize}
where  $\bar{\ord}_K$,  $\bar{\ord}_K/\rad(\bar{\ord}_K)$, and $R_K$ are encoded as algebras over $\F_p$ and $\ord'_K$ is encoded as an algebra over $\Z$.
\end{restatable}

Suppose $K$ and $K'$ are relative number fields  over $K_0$ and $\phi:K\to K'$ is  a field embedding over $K_0$. The map $\phi$ induces a ring homomorphism $\bar{\phi}:\bar{\ord}_K\to \bar{\ord}_{K'}$ sending $x+p\ord_K\in \bar{\ord}_K$ to $\phi(x)+p\ord_{K'}$. As the image of an nilpotent element (resp. an element fixed by the automorphism $x\mapsto x^p$) under $\bar{\phi}$ is also nilpotent (resp. fixed by $x\mapsto x^p$), the map $\bar{\phi}$ induces a ring homomorphism $\bar{\ord}_K/\rad(\bar{\ord}_K)\to \bar{\ord}_{K'}/\rad(\bar{\ord}_{K'})$,  and we denote this map by $\hat{\phi}$.
Finally, the map $\hat{\phi}$ restricts to a ring homomorphism $R_K\to R_{K'}$, which we denote by $\tilde{\phi}$. The maps  $\bar{\phi}$, $\hat{\phi}$ and $\tilde{\phi}$ can be efficiently computed from $\phi$ (and some auxiliary data) by the following lemma.
 \nomenclature[d1s]{$\hat{\phi}$}{ring homomorphism $\bar{\ord}_K/\rad(\bar{\ord}_K)\to \bar{\ord}_{K'}/\rad(\bar{\ord}_{K'})$ induced from $\phi: K\hookrightarrow K'$}
 \nomenclature[d1t]{$\tilde{\phi}$}{ring homomorphism $R_K\to R_{K'}$ induced from $\phi: K\hookrightarrow K'$}
\begin{restatable}{lem}{lemringhomg}\label{lem_ringhomg}
There exists a polynomial-time algorithm $\mathtt{ComputeRingHoms}$ that given  $p$, relative number fields $K$, $K'$ over $K_0$, a field embedding $\phi: K\to K'$ over $K_0$, and the outputs of $\mathtt{ComputeRings}$ (see Lemma~\ref{lem_computerings}) on the inputs $(K,p)$ and $(K',p)$ respectively, computes the maps $\bar{\phi}:\bar{\ord}_K\to \bar{\ord}_{K'}$, $\hat{\phi}:\bar{\ord}_K/\rad(\bar{\ord}_K)\to \bar{\ord}_{K'}/\rad(\bar{\ord}_{K'})$ and $\tilde{\phi}:R_K\to R_{K'}$.
\end{restatable}

See Appendix~\ref{chap_omitted2} for its proof.

\section{Reduction to computing an idempotent decomposition of \texorpdfstring{$R_F$}{RF}} \label{sec_algreductiong}

 Now we start describing the generalized $\mathcal{P}$-scheme algorithm. It is always implicitly assumed that the prime number $p$, $\tilde{h}(Y)\in\Z[Y]$ and $h(Y)=\tilde{h}(Y)\bmod p\in\F_p[Y]$ are known to the algorithm, so that $\F_p[Y]/(h(Y))$, $A_0=\Z[Y]/(\tilde{h}(Y))$ and $K_0=\Q[X]/(\tilde{h}(Y))$ are also known. And $\F_p[Y]/(h(Y))$ is identified with a finite field $\F_q$ via an isomorphism $\psi_0:\F_p[Y]/(h(Y))\to \F_q$ that we can efficiently compute.
 
In addition, we fix the following notations in the remaining sections:
 \begin{itemize}
\item $f(X)$: the  input polynomial in $\F_q[X]$ to be factorized,
\item $\tilde{f}(X)$: an irreducible  lifted polynomial of $f(X)$  in $A_0[X]$,
\item $F$: the number field $K_0[X]/(\tilde{f}(X))$,
\item $L$: the splitting field of $\tilde{f}$ over $K_0$,
\item $G$: the Galois group $\gal(L/K_0)=\gal(\tilde{f}/K_0)$,
\item $\mathfrak{Q}_0$: a fixed prime ideal of $\ord_L$ lying over $p$.
 \end{itemize}
 
 In this section, we reduce the problem of factoring $f$ to computing an idempotent decomposition of $\bar{\ord}_F$, generalizing the result in Section~\ref{sec_algreduction}. For simplicity, we assume that  $\tilde{f}$ is a {\em monic} polynomial, and remove this assumption at the end of this section.

 \paragraph{Ring homomorphisms $\tau$ and $\bar{\tau}$.} 
 
 Let $\alpha:=X+(\tilde{f}(X))\in F$, which is a root of $\tilde{f}$ in $F$.  As $\tilde{f}(X)$ is a monic polynomial in $A_0[X]$ and $A_0\subseteq \ord_{K_0}$, we have $\alpha\in \ord_F$ (see \citep[Corollary~5.4]{AM69}). 
 %\footnote{See the discussion about irreducible lifting polynomials at the beginning of this chapter.}
  
  Consider the natural inclusion $A_0[X]/(\tilde{f}(X))=A_0[\alpha]\hookrightarrow \ord_F$. Taking the quotients of both sides of this map mod $p$ and identify $A_0/pA_0=\F_p[Y]/(h(Y))$ with $\F_q$ via $\psi_0$, we obtain a ring homomorphism
 \[
 \tau: \F_q[X]/(f(X))\to \bar{\ord}_F.
 \]
  Let $g:=\rad(f)$. Then the radical of $\F_q[X]/(f(X))$ is generated by $g(X)+(f(X))$. 
 we obtain a ring homomorphism
 \[
 \bar{\tau}:\F_q[X]/(g(X))\to \bar{\ord}_F/\rad(\bar{\ord}_F),
 \] 
 which  sends an element $h(X)+(g(X))$ to $\tau(h(X))+\rad(\bar{\ord}_F)$.
 Note that both $\F_q[X]/(g(X))$ and $\bar{\ord}_F/\rad(\bar{\ord}_F)$ are semisimple rings.
 
 We can efficiently compute $\bar{\tau}$ by the following lemma.
 \begin{lem}\label{lem_connectmap}
There exists a polynomial-time algorithm that given $f$, $\tilde{f}$, $F$ and  the outputs of $\mathtt{ComputeRings}$ (see Lemma~\ref{lem_computerings}) on the input $(F,p)$, computes
the $\F_q$-algebra $\F_q[X]/(g(X))$ (encoded in the standard $\F_q$-basis $\{1,X,\dots,X^{\deg(g)-1}\}$) and the map  $\bar{\tau}:\F_q[X]/(g(X))\to \bar{\ord}_F/\rad(\bar{\ord}_F)$.
\end{lem}
\begin{proof}
Compute $g$ using  Corollary~\ref{cor_polyrad} and form the $\F_q$-algebra $\F_q[X]/(g(X))$.
To compute $\bar{\tau}$, we first compute $\alpha=X+(\tilde{f}(X))\in F$ and $\bar{Y}:=Y+(\tilde{h}(Y))\in K_0\subseteq F$.
Then compute $\alpha+p\ord_F, \bar{Y}+p\ord_F\in\bar{\ord}_F$ by identifying $F$ with an ordinary number field (see Corollary~\ref{lem_reltoord}) and running the algorithm  $\mathtt{ComputeResidue}$ in Lemma~\ref{lem_computeresidue} on $\alpha,\bar{Y}\in F$. Next, compute $\tau: \F_q[X]/(f(X))\to \bar{\ord}_F$ as the  unique $\F_p$-linear map sending $X+(f(X))$ to  $\alpha+p\ord_F$ and $Y+(h(Y))\in\F_p[Y]/(h(Y))\cong\F_q$ to $\bar{Y}+p\ord_F$. Finally compute $\bar{\tau}$ from $\tau$ by passing to the quotients modulo radicals using the given map $\bar{\ord}_F\to \bar{\ord}_F/\rad(\bar{\ord}_F)$.
\end{proof}

  \paragraph{Extracting a factorization from an idempotent decomposition.} 
 
 We extract a factorization of $f$ from an idempotent decomposition of $R_F$. This is achieved by the algorithm $\mathtt{ExtractFactorsV2}$  below (see Algorithm~\ref{alg_reductiong}), extending the algorithm   in Section~\ref{sec_algreduction}.

 \begin{algorithm}[htbp]
\caption{$\mathtt{ExtractFactorsV2}$}\label{alg_reductiong}
\begin{algorithmic}[1]
\INPUT  \parbox[t]{\dimexpr\linewidth-\algorithmicindent}{%
	  $f$, $\tilde{f}$, $F$, the outputs of $\mathtt{ComputeRings}$ (see Lemma~\ref{lem_computerings}) on the input $(F,p)$,
	  and an idempotent decomposition $I_F$ of $R_F$\strut
	  } 
\OUTPUT factorization of $f$
\State compute $g=\rad(f)$, $\F_q[X]/(g(X))$ and $ \bar{\tau}:\F_q[X]/(g(X))\to \bar{\ord}_F/\rad(\bar{\ord}_F)$
 \State $I\gets \{1\}$, where $1$ denotes the unity of  $\F_q[X]/(g(X))$
 \For{$\delta'\in I_F$}
    \State $J\gets \bar{\tau}^{-1}((1-\delta')\bar{\ord}_F/\rad(\bar{\ord}_F))$
    \State compute $\delta_0\in J$ satisfying $(1-\delta_0)J=\{0\}$
    \For{$\delta\in I$ satisfying $\delta_0\delta\not\in\{0,\delta\}$}
        \State $I\gets I-\{\delta\}$
        \State $I\gets I\cup\{\delta_0\delta, (1-\delta_0)\delta\}$
    \EndFor
\EndFor
 \For{$\delta\in I$}
    \State compute nonzero $h_\delta(X)\in \F_q[X]$ of degree at most $\deg(g)$ lifting $1-\delta$
    \State $g_\delta(X)\gets\mathrm{gcd}(f(X), (h_\delta(X))^n)$ \Comment{$n=\deg(f)$}
    \State compute a square-free factorization $g_\delta(X)=\prod_{i=1}^{k_\delta} g_{\delta,i}(X)$
\EndFor
\State \Return the factorization $f(X)=\prod_{\delta\in I}\prod_{i=1}^{k_\delta} g_{\delta,i}(X)$
\end{algorithmic}
\end{algorithm} 
 
 The algorithm  first computes $g=\rad(f)$, the ring $\F_q[X]/(g(X))$, and the map $\bar{\tau}$  at Line 1 using Lemma~\ref{lem_connectmap}.
 It also maintains an idempotent decomposition $I$ of the ring $\F_q[X]/(g(X))$ which initially only contains the unity. 
 
 The loop in Lines 3--8 enumerates idempotents $\delta'\in I_F$. For each $\delta'$, we compute an ideal $J=\bar{\tau}^{-1}((1-\delta')\bar{\ord}_F/\rad(\bar{\ord}_F))$ of $\F_q[X]/(g(X))$ and an element $\delta_0\in J$ satisfying $(1-\delta_0)J=\{0\}$ by solving systems of linear equations. 
 As $\F_q[X]/(g(X))$ is semisimple, the element $\delta_0$ is the unique idempotent of $\F_q[X]/(g(X))$ that generates $J$.
 And we use it to refine $I$. 
 
The loop in Lines 9--12 extracts,  for each idempotent $\delta\in I$, a monic factor $g_\delta$ of $f$. Furthermore, we compute a square-free factorization  $g_\delta(X)=\prod_{i=1}^{k_\delta} g_{\delta,i}(X)$ for each factor $g_\delta$.
 Finally, the algorithm returns the factorization 
 \[
 f(X)=\prod_{\delta\in I}\prod_{i=1}^{k_\delta} g_{\delta,i}(X).
 \]

  The following theorem is the main result of this section. 
 \begin{thm}\label{thm_extfactorredg}
 The algorithm $\mathtt{ExtractFactorsV2}$ computes a factorization of $f$ over $\F_q$ in polynomial time, such that
 \begin{enumerate}
 \item the factorization is complete  if $I_F$ is a complete idempotent decomposition,
 \item the factorization is proper if $I_F$ is a proper idempotent decomposition, and
 \item at least one factor in the factorization is irreducible over $\F_q$ if $I_F$ contains a primitive idempotent.
 \end{enumerate}
 \end{thm}

 \paragraph{Analysis of the algorithm.}  To prove Theorem~\ref{thm_extfactorredg}, we introduce the following notations: let $S$ (resp. $S_F$) denote the set of the maximal ideals of $\F_q[X]/(g(X))$ (resp. $\bar{\ord}_F/\rad(\bar{\ord}_F)$). 
 For a maximal ideal $\mathfrak{m}$ of $\bar{\ord}_F/\rad(\bar{\ord}_F)$, the preimage $\bar{\tau}^{-1}(\mathfrak{m})$ is a prime (and hence maximal) ideal of $\F_q[X]/(g(X))$. So we obtain a map 
 \[
 \pi: S_F\to S,
 \]
 sending  $\mathfrak{m}$ to $\tau^{-1}(\mathfrak{m})$.   It can be shown that $\pi$ is surjective.\footnote{To prove this, it suffices to show that any prime ideal of  $A_0[X]/(\tilde{f}(X))=A_0[\alpha]\subseteq \ord_F$ is contained in a prime ideal of $\ord_F$, which follows from \citep[Theorem~5.10]{AM69}.}

 Suppose $f(X)=\prod_{i=1}^k (f_i(X))^{m_i}$ where $f_1,\dots,f_k$ are distinct monic irreducible factors of $f$ over $\F_q$.
 For $i\in [k]$, let $\mathfrak{m}_i$ be the (maximal) ideal of $\F_q[X]/(g(X))$ generated by $f_i(X)+(g(X))$.
 Then we have
 \[
 S=\{\mathfrak{m}_1,\dots,\mathfrak{m}_k\} \quad\text{and}\quad g(X)=\prod_{i=1}^k f_i(X).
 \]
The proof of Theorem~\ref{thm_extfactorredg} is based on the following lemma.
 \begin{lem}\label{lem_normmap}
 Let $I$ be the idempotent decomposition of $\F_q[X]/(g(X))$ given at the end of the algorithm $\mathtt{ExtractFactorsV2}$. 
 Define  the partition $P$ of $S$ by
    \[
  P:=\{B_{\delta}:\delta\in I\}, \quad\text{where}\quad B_\delta:=\{\mathfrak{m}\in S: \delta\equiv 1\pmod{\mathfrak{m}}\}
  \]
 and  the partition $P'$ of $S_F$ by
  \[
  P':=\{B'_{\delta}:\delta\in I_F\}, \quad\text{where}\quad B'_\delta:=\{\mathfrak{m}\in S_F: \delta\equiv 1\pmod{\mathfrak{m}}\}.
  \]
Then $P$ is the coarsest common refinement of the partitions $\{\pi(B), S-\pi(B)\}$, where $B$ ranges over the blocks in $P'$.
 Moreover, for each $\delta\in I$, the polynomial $g_\delta$ in the algorithm is given by
 \[
 g_\delta(X)=\prod_{i\in [k]: \mathfrak{m}_i\in B_{\delta}} (f_i(X))^{m_i}.
 \]
 \end{lem}
 \begin{proof}
 For the last claim, it suffices to prove, for all $i\in [k]$, that $h_\delta$ is divisible by $f_i$ iff $\mathfrak{m}_i\in B_\delta$.
By the choice of $h_\delta$,  it holds for all $i\in [k]$  that $h_\delta$ is divisible by $f_i$ iff $1-\delta\in \mathfrak{m}_i$. The claim then follows from the definition of $B_\delta$.

For the first claim, it suffices to show that for every $\delta'\in I_F$ enumerated at Line 3 and $\delta_0$ computed at Line 5 in the same iteration, it holds that $B_{\delta_0}\in \{\pi(B'_{\delta'}),S-\pi(B'_{\delta'})\}$. 
We claim that $B_{\delta_0}=S-\pi(B'_{\delta'})$. As the ideal $J$ computed at Line 4 is generated by $\delta_0$, this claim is equivalent to $J=\bigcap_{\mathfrak{m}\in \pi(B'_{\delta'})} \mathfrak{m}$. Note that for $\mathfrak{m}\in S_F$, it holds that $1-\delta'\in \mathfrak{m}$ iff $\mathfrak{m}\in B'_{\delta'}$ by the definition of $B'_{\delta'}$. So we have
\[
(1-\delta')\bar{\ord}_F/\rad(\bar{\ord}_F)=\bigcap_{\mathfrak{m}\in B'_{\delta'}}\mathfrak{m}
\]
and hence
\[
J= \bar{\tau}^{-1}\left(\bigcap_{\mathfrak{m}\in B'_{\delta'}}\mathfrak{m}\right)=\bigcap_{\mathfrak{m}\in B'_{\delta'}}\bar{\tau}^{-1}(\mathfrak{m})=\bigcap_{\mathfrak{m}\in B'_{\delta'}}\pi(\mathfrak{m})=\bigcap_{\mathfrak{m}\in \pi(B'_{\delta'})} \mathfrak{m}
\]
as desired.
\end{proof}
 
 We also need the following lemma.
 
 \begin{lem}\label{lem_muldeg}
$\pi: S_F\to S$ is bijective if $f$ is square-free, i.e., $m_i=1$ for $i\in [k]$.
 \end{lem}
 \begin{proof}
 Suppose $p\ord_F$ splits into the product of prime ideals by 
 \[
 p\ord_F=\prod_{i=1}^\ell\mathfrak{P}_i^{e(\mathfrak{P}_i)},
 \]
  where $\mathfrak{P}_1,\dots,\mathfrak{P}_\ell$ are distinct prime ideals lying over $p$. 
 For $j\in [\ell]$, let $\mathfrak{m}'_j:=\frac{\mathfrak{P}_j/p\ord_F}{\rad(\bar{\ord}_F)}$.
 Then $S_F=\{\mathfrak{m}'_1,\dots,\mathfrak{m}'_\ell\}$.
  Let $n=\deg(f)$.  Assume $f$ is square-free. Then we have
 \begin{equation}\label{eq_countindex}
 \sum_{i=1}^k \deg(f_i)= \sum_{i=1}^k m_i\deg(f_i)=n=\sum_{j=1}^\ell e(\mathfrak{P}_j) f(\mathfrak{P}_j).
 \end{equation}
% where the first equality uses the assumption that $f$ is square-free. 
Fix $i\in [k]$. We know $\pi^{-1}(i)\neq\emptyset$ since $\pi$ is surjective. Consider $j\in \pi^{-1}(i)$. As $\bar{\tau}(\mathfrak{m}_i)\subseteq \mathfrak{m}'_j$, the map $\bar{\tau}:\F_q[X]/(g(X))\to \bar{\ord}_F/\rad(\bar{\ord}_F)$ induces a field embedding
  \[
 \frac{\F_q[X]/(g(X))}{\mathfrak{m}_i}\hookrightarrow \frac{\bar{\ord}_F/\rad(\bar{\ord}_F)}{\mathfrak{m}'_j}.
 \]
 The left hand side is isomorphic to $\F_q[X]/(f_i(X))$ whereas the right hand side is isomorphic to $\ord_F/\mathfrak{P}_j=\kappa_{\mathfrak{P}_j}$.
 Therefore $\deg(f_i)$ divides $f(\mathfrak{P}_j)$. 
 
 Note that $e(\mathfrak{P}_j)\geq 1$ holds for all $j\in [\ell]$. It follows from \eqref{eq_countindex} that in fact  $e(\mathfrak{P}_j)=1$ holds for all $j\in [\ell]$. Moreover, for all $i\in [k]$, the set $\pi^{-1}(i)$ contains only one element $j_i\in [\ell]$, and $\deg(f_i)=f(\mathfrak{P}_{j_i})$.
 In particular, the map $\pi$ is bijective. 
 \end{proof}

Now we are ready to prove Theorem~\ref{thm_extfactorredg}. 

 \begin{proof}[Proof of Theorem~\ref{thm_extfactorredg}]
 Polynomiality of the algorithm is straightforward. Suppose $I_F$ is a complete idempotent decomposition of $R_F$. 
 It is also a complete idempotent decomposition of $\bar{\ord}_F/\rad(\bar{\ord}_F)$ since the maximal ideals of $\bar{\ord}_F/\rad(\bar{\ord}_F)$ correspond one-to-one to those of $R_F$ via $\mathfrak{m}\mapsto \mathfrak{m}\cap R_F$.
So the partition $P'$ in Lemma~\ref{lem_normmap} is $\infty_{S_F}$. By Lemma~\ref{lem_normmap} and surjectivity of $\pi$, the partition $P$ equals $\infty_{S}$, and the algorithm outputs the complete factorization $f(X)=\prod_{i\in [k]} (f_i(X))^{m_i}$. 
 
 Similarly, if $I_F$ contains a primitive idempotent $\delta$. Then $P'$ contains a singleton $B'_\delta$. By Lemma~\ref{lem_normmap}, the partition $P$ contains a singleton $\pi(B'_\delta)$, and algorithm outputs a factorization of $f(X)$ in which the irreducible factors $f_i(X)$ appear $m_i$ times, where $i$ is the unique index in $[k]$ satisfying $\pi(B'_\delta)=\{\mathfrak{m}_i\}$.

Finally, suppose $I_F$ is a proper idempotent decomposition of $R_F$, and hence a proper idempotent decomposition of $\bar{\ord}_F/\rad(\bar{\ord}_F)$. Then $P'\neq 0_{S_F}$.
If $\pi$ is bijective, then by Lemma~\ref{lem_normmap}, we have $P\neq 0_S$, and the algorithm outputs a proper factorization of $f$. Now suppose $\pi$ is not bijective. Then $f$ is not square-free by Lemma~\ref{lem_muldeg}. As we compute a square-free factorization for each $g_\delta$, the algorithm still outputs a proper factorization of $f$.
 \end{proof}

\paragraph{The reduction for non-monic polynomials.}

The same trick in Section~\ref{sec_algreduction} can be applied  to make  the above reduction work for a possibly non-monic polynomial $\tilde{f}$: let $c\in A_0$ be the leading coefficient of $\tilde{f}(X)\in A_0[X]$, and let $\bar{c}:=\tilde{\psi}_0(c)\in\F_q^\times$.
Compute the monic polynomials $\tilde{f}'(X):=c^{n-1}\cdot \tilde{f}(X/c)\in A_0[X]$ and $f'(X):=\bar{c}^{n-1} f(X/\bar{c}^{n-1})\in\F_q[X]$.
Run the algorithm $\mathtt{ExtractFactorsV2}$ on $f'$ and $\tilde{f}'$ instead of  $f$ and $\tilde{f}$, and obtain a factorization of $f'$. Finally, we recover a factorization of $f$ from that of $f'$ by substituting $X$ with $\bar{c} X$ in each factor.

\begin{rem} 
 The reduction in this section exploits the  well known connection between factorization of polynomials over finite fields and the splitting of prime ideals in number field extensions, which dates back to the  classical  work of Kummer and Dedekind (see, e.g., \citep[Proposition~\RN{1}.8.3]{Neu99}). The Kummer-Dedekind theorem, however, requires the map $\F_q[X]/(f(X))\to \bar{\ord}_F$ to be an isomorphism. For this reason, known factoring algorithms that use an irreducible lifted polynomial $\tilde{f}$ often assume $p$ is {\em regular}\index{regularity!of a prime} with respect to $\tilde{f}$.  See, e.g., \citep{Hua84, Hua91-2, Hua91, Ron92}.\footnote{We say $p$ is regular with respect to $\tilde{f}$ if $pA_0[\alpha]$ is coprime to the {\em conductor} of $A_0[\alpha]$. See \citep{Hua84} for the exact formulation of this condition. We remark that the journal version \citep{Hua91-2}  (and \citep{Hua91, Ron92}) assumes the stronger condition that $p$ is coprime to the discriminant of $\tilde{f}$.}
This assumption is {\em not} needed in our algorithm. The key observation is that we can always employ the surjective map $\pi$ from the set of prime ideals of $\bar{\ord}_F/\rad(\bar{\ord}_F)$ to that of $\F_q[X]/(g(X))$, where $g=\rad(f)$. 
%This map relates the splitting of $p$ in $F$ to the factorization of $f$.  
In algebro-geometric terminology, the map $\pi$ is interpreted as the morphism of reduced affine schemes\index{affine scheme} 
\[
\pi:\spec (\bar{\ord}_F)_{\mathrm{red}} \to \spec (A_0[\alpha]/p A_0[\alpha])_{\mathrm{red}}
\] 
induced from the morphism $\spec \ord_F\to \spec A_0[\alpha]$. The latter morphism is known as the {\em normalization}\index{normalization} of $\spec A_0[\alpha]$ (see \citep[Exercise~\RN{2}.3.8]{Har13}).
%We can strengthen Lemma~\ref{lem_muldeg} in the following way: for $i\in [k]$, let $S_i$ denote the set of $j\in [\ell]$ satisfying $\pi(\mathfrak{P}_j/p\ord_F)=(f_i(X))$. Then we have
%\begin{restatable}{lem}{lemcountindexadv}\label{lem_countindexadv}
%$m_i\deg(f_i)=\sum_{j\in S_i} e(\mathfrak{P}_j)f(\mathfrak{P}_j)$ for all $i\in [k]$.
%\end{restatable}
%In particular, as $\deg(f_i)\leq f(\mathfrak{P}_j)$ for $j\in S_i$ (see the proof of Lemma~\ref{lem_muldeg}), we see $m_i=1$ iff $S_i$  is a singleton and $e(\mathfrak{P}_j)=1$, $\deg(f_i)=f(\mathfrak{P}_j)$ for $j\in S_i$. 
% 
%A proof of  Lemma~\ref{lem_countindexadv} can be found in Chapter~\ref{chap_omitted2} of the appendix.
%%
%%\item Using the same idea of the regularity test (see Section X), the partition of unity $I$ can be further refined into a set of idempotents $\delta_{B'}$ such that $\mathfrak{m},\mathfrak{m}'\in S$ are in the same subset $B'$ only if $|\pi^{-1}(\mathfrak{m})\cap B|=|\pi^{-1}(\mathfrak{m}')\cap B|$ holds for all $B\subseteq S_F$ satisfying $\delta_B\in I_F$. We leave the details to the reader.
\end{rem}

\section{Producing a \texorpdfstring{$\mathcal{P}$-scheme}{P-scheme} of double cosets \texorpdfstring{$\mathcal{C}$}{C}} \label{sec_algmaing}
In this section, we present an algorithm that computes the idempotent decompositions of a  collection  of rings $R_K$ corresponding to a $\mathcal{P}$-scheme of double cosets. It extends the algorithm in Section~\ref{sec_algmain} and serves as (a preliminary version) of the main body of the generalized $\mathcal{P}$-scheme algorithm. 

The pseudocode of the algorithm is given in Algorithm~\ref{alg_mainalgg} below.
Its input is a $(K_0,\tilde{f})$-subfield system $\mathcal{F}$ (see Definition~\ref{defi_kgcollection}).
The algorithm outputs, for every $K\in\mathcal{F}$, an idempotent decomposition $I_K$ of the ring $R_K$, together with some auxiliary data.
%,as needed by the algorithm $\mathtt{ExtractFactorsV2}$ in the previous section. 
%We show that these idempotent decompositions altogether determine a $\mathcal{P}$-scheme of double cosets that satisfies strict properties.
%\todo{Run the tests for ramification indices and inertia degrees earlier}

\begin{algorithm}[htb]
\caption{$\mathtt{ComputeDoubleCosetPscheme}$}\label{alg_mainalgg}
\begin{algorithmic}[1]
\INPUT $(K_0,\tilde{f})$-subfield system  $\mathcal{F}$
\OUTPUT \parbox[t]{.9\linewidth}{%
	 for every $K\in\mathcal{F}$: the outputs of $\mathtt{ComputeRings}$ (see Lemma~\ref{lem_computerings}) on the input $(K,p)$, and an idempotent decomposition $I_K$ of $R_K$ \strut
	  }
\For{$K\in \mathcal{F}$} \strut
    \State  call $\mathtt{ComputeRings}$ on $(K,p)$ 
    \State  $I_K\gets \{1\}$, where $1$ denotes the unity of  $R_K$
 %   \State compute the inclusion $\F_q\hookrightarrow \bar{\ord}_K/\rad(\bar{\ord}_K)$
\EndFor 
 \For{$(K,K')\in \mathcal{F}^2$} \strut
    \State \parbox[t]{\dimexpr\linewidth-\algorithmicindent}{%
        call $\mathtt{ComputeRelEmbeddings}$ to compute all the embeddings from $K$ to $K'$ over $K_0$\strut
        }
    \For{\strut embedding $\phi: K\hookrightarrow K'$ over $K_0$}
        \State call $\mathtt{ComputeRingHoms}$ on $p$, $K$, $K'$ and $\phi$ to compute $\bar{\phi}$, $\hat{\phi}$ and $\tilde{\phi}$
     \EndFor
\EndFor
\Repeat
    \State call  $\mathtt{CompatibilityAndInvarianceTestV2}$
    \State call  $\mathtt{RegularityTestV2}$
    \State call  $\mathtt{StrongAntisymmetryTestV2}$
    \State call  $\mathtt{RamificationIndexTest}$
    \State call  $\mathtt{InertiaDegreeTest}$
\Until{$I_K$ remains the same in the last iteration for all $K\in\mathcal{F}$}
\State \Return  the outputs of $\mathtt{ComputeRings}$ on the input $(K,p)$ and   $I_K$ for $K\in\mathcal{F}$
\end{algorithmic}
\end{algorithm}

%\begin{algorithm}[htb]
%\caption{$\mathtt{Preprocess}$}
%\begin{algorithmic}[1]
%\For{$K\in \mathcal{F}$} \strut
%    \State  call $\mathtt{ComputeRings}$ on $(K,p)$ 
%    \State  $I_K\gets \{1_{R_K}\}$, where $1_{R_K}$ denotes the unity of  $R_K$
%    \State compute the inclusion $\F_q\hookrightarrow \bar{\ord}_K/\rad(\bar{\ord}_K)$
%        \ForTo{$i$}{$1$}{$[K:K_0]$}
%        \State compute  $A_{K,i}$ and the inclusions $\bar{\ord}_K/\rad(\bar{\ord}_K)\hookrightarrow A_{K,i}$, $\F_{q^i}\hookrightarrow A_{K,i}$
%         \EndFor
%\EndFor 
% \For{$(K,K')\in \mathcal{F}^2$} \strut
%    \State \parbox[t]{\dimexpr\linewidth-\algorithmicindent}{%
%        call $\mathtt{ComputeRelEmbeddings}$ to compute all the embeddings from $K$ to $K'$ over $K_0$\strut
%        }
%    \For{\strut embedding $\phi: K\hookrightarrow K'$ over $K_0$}
%        \State call $\mathtt{ComputeRingHoms}$ on $p$, $K$, $K'$ and $\phi$ compute $\bar{\phi}$, $\hat{\phi}$ and $\tilde{\phi}$
%     \EndFor
%\EndFor
%\end{algorithmic}
%\end{algorithm} 

We fix $\mathcal{P}$ to be the subgroup system over $G=\gal(\tilde{f}/K_0)$ associated with $\mathcal{F}$, i.e.,
\[
\mathcal{P}:=\left\{H\subseteq G: L^H\cong_{K_0} K \text{ for some } K\in \mathcal{F} \right\}.
\]

 The first half (Lines 1--7) of the algorithm is the preprocessing  stage:  for each $K\in\mathcal{F}$, we run  $\mathtt{ComputeRings}$ (see Lemma~\ref{lem_computerings}) on $(K,p)$ which returns the following data:
\begin{itemize}
\item a $p$-maximal order $\ord'_K$ of $K$ and the inclusion $\ord'_K\hookrightarrow K$,
\item $\bar{\ord}_K$ and the quotient map $\ord'_K\to \bar{\ord}_K$,
\item  $\bar{\ord}_K/\rad(\bar{\ord}_K)$ and the quotient map  $\bar{\ord}_K\to  \bar{\ord}_K/\rad(\bar{\ord}_K)$,
\item $R_K$ and the inclusion $R_K\hookrightarrow \bar{\ord}_K/\rad(\bar{\ord}_K)$.
\end{itemize}

%In addition, we compute the ring $A_{K,i}=\bar{\ord}_K/\rad(\bar{\ord}_K)\otimes_{\F_q} \F_{q^i}$ together with the inclusions  $\bar{\ord}_K/\rad(\bar{\ord}_K)\hookrightarrow A_{K,i}$, $\F_{q^i}\hookrightarrow A_{K,i}$ defined by $a\mapsto a\otimes 1$ and $b\mapsto 1\otimes b$ respectively.\footnote{To compute $A_{K,i}$, we first compute the inclusion $\F_q\hookrightarrow  \bar{\ord}_K/\rad(\bar{\ord}_K)$ at Line 4, endowing $\bar{\ord}_K/\rad(\bar{\ord}_K)$ the structure of an $\F_q$-algebra. To achieve this, we compute the image $\bar{Y}$ of $Y+(\tilde{h}(Y))\in\ord_{K_0}\subseteq\ord_K$ in $\bar{\ord}_K$ by Lemma~\ref{lem_computeresidue}. Then we compute the map $\F_p[Y]/(h(Y))\to \bar{\ord}_K$ sending $Y+(h(Y))$ to $\bar{Y}$, and compose it with the isomorphism $\psi_0^{-1}:\F_q\to \F_p[Y]/(h(Y))$ and the quotient map  $\bar{\ord}_K\to  \bar{\ord}_K/\rad(\bar{\ord}_K)$.}
% These objects are computed for the subroutines in Section X--X are not really needed in this section.

For $(K,K')\in\mathcal{F}$, we also compute all the embeddings $\phi$ from $K$ to $K'$ and the corresponding ring homomorphisms $\bar{\phi}:\bar{\ord}_K\to \bar{\ord}_{K'}$, $\hat{\phi}:\bar{\ord}_K/\rad(\bar{\ord}_K)\to \bar{\ord}_{K'}/\rad(\bar{\ord}_{K'})$ 
and $\tilde{\phi}: R_K\to R_{K'}$.
Moreover, for each $K\in\mathcal{F}$, we  initialize the idempotent decomposition $I_K$  of $R_K$ to be the trivial one containing only the unity of $R_K$.  

 The second half (Lines 8--14) of the algorithm refines the  idempotent decompositions $I_K$ for $K\in\mathcal{F}$. To analyze it, we associate a $\mathcal{P}$-collection $\mathcal{C}$ of double cosets with these idempotent decompositions.
For each $H\in\mathcal{P}$, define a partition $C_H$ of the coset space $H\backslash G/\mathcal{D}_{\mathfrak{Q}_0}$ as follows:
Let $K$ be the unique field in $\mathcal{F}$ isomorphic to $L^H$ over $K_0$. Fix an isomorphism $\tau_H: K\to L^H$ over $K_0$, which induces a ring isomorphism $\tilde{\tau}_H: R_K\to R_{L^H}$.
Define $I_H:=\tilde{\tau}_H(I_{K})$, which  is an idempotent decomposition of $R_{L^H}$.
By  Definition~\ref {defi_partitioncorg}, it corresponds to a partition $P(I_H)$ of $H\backslash G/\mathcal{D}_{\mathfrak{Q}_0}$.\footnote{Definition~\ref {defi_partitioncorg} is made with respect to a fixed prime ideal $\mathfrak{Q}_0$ of $\ord_L$ lying over $p$. This ideal is chosen at the beginning of  Section~\ref{sec_algreductiong}.}
And we define 
\[
C_H:=P(I_H).
\]
Finally, define 
\[
\mathcal{C}:=\{C_H: H\in\mathcal{P}\},
\] 
which is a $\mathcal{P}$-collection of double cosets  (with respect to $\mathcal{D}_{\mathfrak{Q}_0}$).

%We call several subroutines to update $I_K$ in Lines 3--7, whose effects can be understood in terms of $\mathcal{C}$.
The subroutines in Lines 9--11 extend those in  Section~\ref{sec_algci}, \ref{sec_algr}, and \ref{sec_algsa} respectively:

\begin{lem}\label{lem_citestg}
There exists a subroutine  $\mathtt{CompatibilityAndInvarianceTestV2}$ that updates $I_K$ in time  polynomial  in $\log p$ and the size of $\mathcal{F}$ so that the partitions $C_H\in \mathcal{C}$ are refined. Moreover, at least one partition $C_H$ is properly refined if $\mathcal{C}$ is not compatible or invariant.
\end{lem}

\begin{lem}\label{lem_rtestg}
There exists a subroutine $\mathtt{RegularityTestV2}$ that updates $I_K$ in time  polynomial  in $\log p$ and the size of $\mathcal{F}$ so that the partitions $C_H\in \mathcal{C}$ are refined. Moreover, at least one partition $C_H$ is properly refined if $\mathcal{C}$ is compatible but not regular. 
\end{lem}

\begin{lem}\label{lem_satestg}
Under GRH, there exists a subroutine $\mathtt{StrongAntisymmetryTestV2}$ that updates $I_K$ in time  polynomial  in $\log p$ and the size of $\mathcal{F}$ so that the partitions $C_H\in \mathcal{C}$ are refined. Moreover, at least one partition $C_H$ is properly refined  if $\mathcal{C}$ is  a $\mathcal{P}$-scheme of double cosets, but  not strongly antisymmetric.
\end{lem}

The proofs of Lemma~\ref{lem_citestg}--\ref{lem_satestg} (and the corresponding  subroutines) are almost the same as those of Lemma~\ref{lem_citest}--\ref{lem_satest}  in Chapter~\ref{chap_alg_prime}. For this reason, we only list the changes that need to be made  rather than describe the complete proofs and the subroutines.

\begin{proof}[Proof sketch of  Lemma~\ref{lem_citestg}--\ref{lem_satestg}]
We make the following changes to the proofs of Lemma~\ref{lem_citest}--\ref{lem_satest} and the corresponding subroutines: 

 each quotient ring $\bar{\ord}_K$  
% (as in the proofs in Chapter~\ref{chap_alg_prime}) 
is replaced with the ring $R_K$, which is still isomorphic to a finite product of copies of $\F_p$. 
A maximal ideal $\mathfrak{P}$ of $\bar{\ord}_K$ is replaced with the maximal ideal $(\mathfrak{P}/\rad(\bar{\ord}_K))\cap R_K$ of $R_K$. 
The subroutines enumerate field embeddings over $K_0$ instead of arbitrary field embeddings.
For each field embedding  $\phi: K\to K'$ over $K_0$, we use the ring homomorphism $\tilde{\phi}: R_K\to R_{K'}$ in place of $\bar{\phi}: \bar{\ord}_K\to\bar{\ord}_{K'}$. The ring isomorphisms $\bar{\tau}_H: \bar{\ord}_K\to \bar{\ord}_{L^H}$ are replaced with  $\tilde{\tau}_H: R_K\to R_{L^H}$.

A right coset $Hg$ is replaced with a double coset $Hg\mathcal{D}_{\mathfrak{Q}_0}$, and a right coset space $H\backslash G$ is replaced with $H\backslash G/\mathcal{D}_{\mathfrak{Q}_0}$. A projection $\pi_{H,H'}$ is replaced with $\pi_{H,H'}^{\mathcal{D}_{\mathfrak{Q}_0}}$, and a conjugation $c_{H,g}$ is replaced with $c_{H,g}^{\mathcal{D}_{\mathfrak{Q}_0}}$ (see Definition~\ref{defi_double}).
% And we use $\bar{\mathfrak{Q}}_0$ to denote the maximal ideal $\frac{\mathfrak{Q}_0/p\ord_L}{\rad(\bar{\ord}_L)}\cap R_L$ of $R_L$, rather than the maximal ideal $\mathfrak{Q}_0/p\ord_L$ of $\bar{\ord}_L$ as in Chapter~\ref{chap_alg_prime}.

Finally, instead of applying Corollary~\ref{cor_idealcoset}, Lemma~\ref{lem_pandi}, and Lemma~\ref{lem_picorres} from  Chapter~\ref{chap_alg_prime}, we apply Lemma~\ref{lem_idealdoublecosetg}, Lemma~\ref{lem_pandig}, and Lemma~\ref{lem_picorresg}, respectively. The details are left to the reader.
\end{proof}

In addition, the subroutines at Line 12 and Line 13  properly refine the partitions in $\mathcal{C}$ unless they all have locally constant ramification indices and inertia degrees:

\begin{lem}\label{lem_ramtest}
There exists a subroutine $\mathtt{RamificationIndexTest}$ that updates $I_K$ in time  polynomial  in $\log p$ and the size of $\mathcal{F}$ so that the partitions $C_H\in \mathcal{C}$ are refined. Moreover, at least one partition $C_H$ is properly refined unless all the partitions in $\mathcal{C}$  have locally constant ramification indices (with respect to $(\mathcal{D}_{\mathfrak{Q}_0}, \mathcal{I}_{\mathfrak{Q}_0})$).
\end{lem}

\begin{lem}\label{lem_inetest}
There exists a subroutine $\mathtt{InertiaDegreeTest}$ that updates $I_K$ in time  polynomial  in $\log p$ and the size of $\mathcal{F}$ so that the partitions $C_H\in \mathcal{C}$ are refined. Moreover, at least one partition $C_H$ is properly refined unless all the partitions in $\mathcal{C}$   have locally constant inertia degrees (with respect to $(\mathcal{D}_{\mathfrak{Q}_0}, \mathcal{I}_{\mathfrak{Q}_0})$).
\end{lem} 
 
  Lemma~\ref{lem_ramtest} and Lemma~\ref{lem_inetest} are proved in Section~\ref{sec_testlocalconst} and Section~\ref{sec_testlocalconst2}, respectively.

 Combining Lemma~\ref{lem_citestg}--\ref{lem_inetest} yields the main result of this section:

\begin{thm}[Theorem~\ref{thm_maindcinformal} restated]\label{thm_maindc}
Under the assumption of GRH, the  algorithm $\mathtt{ComputeDoubleCosetPscheme}$ runs in time polynomial in $\log p$ and the size of $\mathcal{F}$, and when it terminates,   $\mathcal{C}$ is a strongly antisymmetric $\mathcal{P}$-scheme of double cosets (with respect to $\mathcal{D}_{\mathfrak{Q}_0}$). Moreover, all the partitions in $\mathcal{C}$ have locally constant ramification indices and inertia degrees (with respect to $(\mathcal{D}_{\mathfrak{Q}_0}, \mathcal{I}_{\mathfrak{Q}_0})$).
\end{thm}

%\begin{proof}
%By Lemma~\ref{lem_ramtest} and Lemma~\ref{lem_inetest}, all the partitions in  $\mathcal{C}$ have locally constant ramification indices and inertia degrees after the execution of   Line 8 and Line 9. 
%%Note that these properties are preserved by further refinements.
%Then we run the loop in Lines 10--14 which, by Lemma~\ref{lem_citestg}--\ref{lem_satestg}, properly refines the idempotent decompositions $I_K$ unless $\mathcal{C}$ is a strongly antisymmetric  $\mathcal{P}$-scheme of double cosets.
%The number this loop is executed is bounded by the total degree of the fields $K\in\mathcal{F}$ over $K_0$. It follows that the algorithm runs in time polynomial in  $\log p$ and the size of $\mathcal{F}$.
%\end{proof}

\section{Testing local constantness of ramification indices}\label{sec_testlocalconst}

In this section, we describe the subroutine  $\mathtt{RamificationIndexTest}$ that properly refines at least one partition in $\mathcal{C}$ unless all the partition have locally constant ramification indices.

\begin{algorithm}[htbp]
\caption{$\mathtt{RamificationIndexTest}$}\label{alg_ramtest}
\begin{algorithmic}[1]
 \For{\strut $K\in\mathcal{F}$}
         \ForTo{$i$}{$1$}{$[K:K_0]$}  
%              \State compute $\ann_{\bar{\ord}_K}(\rad(\bar{\ord}_K)^i)$     
              \State $J\gets$ the image of  $\ann_{\bar{\ord}_K}(\rad(\bar{\ord}_K)^i)\subseteq \bar{\ord}_K$ in $ \bar{\ord}_K/\rad(\bar{\ord}_K)$
              \State find $\delta_0\in J\cap R_K$ satisfying $(1-\delta_0)J=\{0\}$
              \For{$\delta\in I_K$ satisfying $\delta_0\delta\not\in\{0,\delta\}$}
                   \State $I_K\gets I_K-\{\delta\}$
                   \State $I_K\gets I_K\cup\{\delta_0\delta, (1-\delta_0)\delta\}$
              \EndFor
          \EndFor  
\EndFor
\end{algorithmic}
\end{algorithm}

The pseudocode of the subroutine  is given in Algorithm~\ref{alg_ramtest} above.  We enumerate $K\in\mathcal{F}$ and  $i=1,2,\dots, [K:K_0]$. For each $K$ and $i$, we compute an ideal $J$ of $\bar{\ord}_K/\rad(\bar{\ord}_K)$, defined to be the image of  $\ann_{\bar{\ord}_K}(\rad(\bar{\ord}_K)^i)$ under the quotient map $\bar{\ord}_K\to \bar{\ord}_K/\rad(\bar{\ord}_K)$.
We also compute an element $\delta_0\in J\cap R_K\subseteq R_K$, satisfying $(1-\delta_0)J=\{0\}$. 
 As $\bar{\ord}_K/\rad(\bar{\ord}_K)$ and $R_K$ are semisimple, and  $\mathfrak{m}\mapsto \mathfrak{m}\cap R_K$ is a one-to-one correspondence between the maximal ideals of $\bar{\ord}_K/\rad(\bar{\ord}_K)$ and those of $R_K$, we know    $\delta_0$ is the unique idempotent of $\bar{\ord}_K/\rad(\bar{\ord}_K)$ (resp. $R_K$) that generates $J$ (resp. $J\cap R_K$).
Then we use $\delta_0$ to refine $I_K$.

Next we prove Lemma~\ref{lem_ramtest}.

\begin{proof}[Proof of Lemma~\ref{lem_ramtest}] 
The claim about the running time is straightforward. Suppose there exists $H\in\mathcal{P}$ such that $C_H$ does not have locally constant ramification indices. Choose $B\in C_H$ and $g,g'\in G$ such that $Hg\mathcal{D}_{\mathfrak{Q}_0}, Hg'\mathcal{D}_{\mathfrak{Q}_0}\in B$ and $e(Hg\mathcal{D}_{\mathfrak{Q}_0})<e(Hg'\mathcal{D}_{\mathfrak{Q}_0})$.

By Theorem~\ref{thm_split_general} and Definition~\ref{defi_indandeg}, the ideal $p\ord_{L^H}$ splits into the product of prime ideals $\prescript{h}{}{\mathfrak{Q}_0}\cap \ord_{L^H}$ by 
 \[
 p\ord_{L^H}=\prod_{Hh\mathcal{D}_{\mathfrak{Q}_0}\in H\backslash G/\mathcal{D}_{\mathfrak{Q}_0}} \left(\prescript{h}{}{\mathfrak{Q}_0}\cap \ord_{L^H}\right)^{e(Hh\mathcal{D}_{\mathfrak{Q}_0})}.
 \]
 For $Hh\mathcal{D}_{\mathfrak{Q}_0}\in H\backslash G/\mathcal{D}_{\mathfrak{Q}_0}$, define  
$\mathfrak{P}_{Hh\mathcal{D}_{\mathfrak{Q}_0}}:=\left(\prescript{h}{}{\mathfrak{Q}_0}\cap \ord_{L^H}\right)/p\ord_{L^H}$,
 which is a maximal ideal of $\bar{\ord}_{L^H}$.
By the Chinese remainder theorem, we have
\[
\bar{\ord}_{L^H}\cong 
\prod_{x\in  H\backslash G/\mathcal{D}_{\mathfrak{Q}_0}} \bar{\ord}_{L^H}/\mathfrak{P}_{x}^{e(x)}.
\]
And $\rad(\bar{\ord}_{L^H})=\prod_{x\in  H\backslash G/\mathcal{D}_{\mathfrak{Q}_0}} \mathfrak{P}_{x}$. So for $i\in\N$, we have
\begin{equation}\label{eq_annchar}
\ann_{\bar{\ord}_{L^H}}(\rad(\bar{\ord}_{L^H})^i)=\prod_{x\in  H\backslash G/\mathcal{D}_{\mathfrak{Q}_0}}  \mathfrak{P}_{x}^{\max\{0,e(x)-i\}}.
\end{equation}
Choose $i=e(Hg\mathcal{D}_{\mathfrak{Q}_0})$ and let $J$ be the image of $\ann_{\bar{\ord}_{L^H}}(\rad(\bar{\ord}_{L^H})^i)$ in the quotient ring $\bar{\ord}_{L^H}/\rad(\bar{\ord}_{L^H})$.
Let $\delta_0$ be the unique idempotent of $\bar{\ord}_{L^H}/\rad(\bar{\ord}_{L^H})$ that generates $J$.
It follows from \eqref{eq_annchar} that 
\[
\delta_0\equiv 1 \pmod{\mathfrak{P}_{Hg\mathcal{D}_{\mathfrak{Q}_0}}/\rad(\bar{\ord}_{L^H})} 
~~\text{and}~~ 
\delta_0\equiv 0 \pmod{\mathfrak{P}_{Hg'\mathcal{D}_{\mathfrak{Q}_0}}/\rad(\bar{\ord}_{L^H})}.
\]
Therefore
\begin{equation}\label{eq_remresidue}
\prescript{g^{-1}}{}{(i_{L^H,L}(\delta_0))}\equiv 1\pmod{\bar{\mathfrak{Q}}_0} \quad\text{and}\quad \prescript{g'^{-1}}{}{(i_{L^H,L}(\delta_0))}\equiv 0\pmod{\bar{\mathfrak{Q}}_0},
\end{equation}
where $i_{L^H,L}: R_{L^H}\hookrightarrow R_L$ is the inclusion induced from the natural inclusion $\ord_{L^H}\hookrightarrow \ord_L$.

On the other hand, by Lemma~\ref{lem_pandig}, the block $B\in C_H$ corresponds to an idempotent $\delta=\delta_B\in I_H$. And
\[
\prescript{h^{-1}}{}{(i_{L^H,L}(\delta))}\equiv 1\pmod{\bar{\mathfrak{Q}}_0}
\]
holds for all $h\in G$ satisfying $Hh\mathcal{D}_{\mathfrak{Q}_0}\in B$. In particular, it holds for $h=g$ and $h=g'$. It follows from \eqref{eq_remresidue} that $\delta_0\delta\not\in\{0,\delta\}$. 

Identifying $L^H$  with a field  $K\in\mathcal{F}$ using the isomorphism  $\tau_H: K\to L^H$ over $K_0$ chosen in Section~\ref{sec_algmaing}, we see that the subroutine is guaranteed to find an idempotent $\delta\in I_K$ satisfying $\delta_0\delta\not\in\{0,\delta\}$ at Line 5. The lemma follows.
\end{proof}

\section{Testing local constantness of inertia degrees}\label{sec_testlocalconst2}

In this section, we describe the subroutine  $\mathtt{InertiaDegreeTest}$ that properly refines at least one partition in $\mathcal{C}$ unless all the partition have locally constant inertia degrees.

\begin{algorithm}[htbp]
\caption{$\mathtt{InertiaDegreeTest}$}\label{alg_inetest}
\begin{algorithmic}[1]
 \For{\strut $K\in\mathcal{F}$}
         \ForTo{$i$}{$1$}{$[K:K_0]$}  
             \State \parbox[t]{0.9\dimexpr\linewidth}{  $J\gets$ the ideal of $\bar{\ord}_K/\rad(\bar{\ord}_K)$ generated by $\{x^{p^i}-x:x\in \bar{\ord}_K/\rad(\bar{\ord}_K)\}$ \strut}
              \State \strut find $\delta_0\in J\cap R_K$ satisfying $(1-\delta_0)J=\{0\}$
              \For{$\delta\in I_K$ satisfying $\delta_0\delta\not\in\{0,\delta\}$}
                   \State $I_K\gets I_K-\{\delta\}$
                   \State $I_K\gets I_K\cup\{\delta_0\delta, (1-\delta_0)\delta\}$
              \EndFor
          \EndFor  
\EndFor
\end{algorithmic}
\end{algorithm}

The pseudocode of the subroutine  is given in Algorithm~\ref{alg_inetest}.  We enumerate $K\in\mathcal{F}$ and  $i=1,2,\dots, [K:K_0]$. For each $K$ and $i$, we compute an ideal $J$ of $\bar{\ord}_K/\rad(\bar{\ord}_K)$, generated by the elements $x^{p^i}-x$, where $x$ ranges over $\bar{\ord}_K/\rad(\bar{\ord}_K)$.  Note that $J$ is just the $\F_p$-linear subspace of $\bar{\ord}_K/\rad(\bar{\ord}_K)$ spanned by $x^{p^i}-x$ where $x$ ranges over an $\F_p$-basis of  $\bar{\ord}_K/\rad(\bar{\ord}_K)$. So it can be efficiently computed.
We also compute an element $\delta_0\in J\cap R_K\subseteq R_K$ satisfying $(1-\delta_0)J=\{0\}$. 
 As  in  Algorithm~\ref{alg_ramtest}, it is the unique idempotent of $\bar{\ord}_K/\rad(\bar{\ord}_K)$ (resp. $R_K$) that generates $J$ (resp. $J\cap R_K$). Then we use $\delta_0$ to refine $I_K$. 

 Next we prove Lemma~\ref{lem_inetest}.

\begin{proof}[Proof of Lemma~\ref{lem_inetest}]
The claim about the running time is straightforward. Suppose there exists $H\in\mathcal{P}$ such that $C_H$ does not have locally constant inertia degrees. Choose $B\in C_H$ and $g,g'\in G$ such that $Hg\mathcal{D}_{\mathfrak{Q}_0}, Hg'\mathcal{D}_{\mathfrak{Q}_0}\in B$ and $f(Hg\mathcal{D}_{\mathfrak{Q}_0})>f(Hg'\mathcal{D}_{\mathfrak{Q}_0})$.

 For $Hh\mathcal{D}_{\mathfrak{Q}_0}\in H\backslash G/\mathcal{D}_{\mathfrak{Q}_0}$, define  
$\mathfrak{P}_{Hh\mathcal{D}_{\mathfrak{Q}_0}}:=\left(\prescript{h}{}{\mathfrak{Q}_0}\cap \ord_{L^H}\right)/p\ord_{L^H}$,
 which is a maximal ideal of $\bar{\ord}_{L^H}$.
By Theorem~\ref{thm_split_general}, Definition~\ref{defi_indandeg}, and the Chinese remainder theorem, we have
\[
\bar{\ord}_{L^H}/\rad(\bar{\ord}_{L^H})\cong 
\prod_{Hh\mathcal{D}_{\mathfrak{Q}_0}\in  H\backslash G/\mathcal{D}_{\mathfrak{Q}_0}} \bar{\ord}_{L^H}/\mathfrak{P}_{Hh\mathcal{D}_{\mathfrak{Q}_0}}
\]
and each factor $ \bar{\ord}_{L^H}/\mathfrak{P}_{Hh\mathcal{D}_{\mathfrak{Q}_0}}$ is an extension field of $\F_p$ of degree $f(Hh\mathcal{D}_{\mathfrak{Q}_0})$. Choose $i=f(Hg'\mathcal{D}_{\mathfrak{Q}_0})$ and let $J$ be the ideal of $\bar{\ord}_{L^H}/\rad(\bar{\ord}_{L^H})$ generated by $x^{p^i}-x$  where $x$ ranges over $\bar{\ord}_{L^H}/\rad(\bar{\ord}_{L^H})$.  Let $\delta_0$ be the unique idempotent of $\bar{\ord}_{L^H}/\rad(\bar{\ord}_{L^H})$ that generates $J$.
Note that we have 
\[
x^{p^i}\not\equiv x\pmod{\mathfrak{P}_{Hg\mathcal{D}_{\mathfrak{Q}_0}}/\rad(\bar{\ord}_{L^H})}\quad\text{for some}~x\in \bar{\ord}_{L^H}/\rad(\bar{\ord}_{L^H}), 
\] 
and
\[
x^{p^i}\equiv x\pmod{\mathfrak{P}_{Hg'\mathcal{D}_{\mathfrak{Q}_0}}/\rad(\bar{\ord}_{L^H})}\quad\text{for all}~x\in \bar{\ord}_{L^H}/\rad(\bar{\ord}_{L^H}).
\] 
So $J$ is contained in $\mathfrak{P}_{Hg'\mathcal{D}_{\mathfrak{Q}_0}}$ but not in $\mathfrak{P}_{Hg\mathcal{D}_{\mathfrak{Q}_0}}$.
It follows that 
\[
\delta_0\equiv 1 \pmod{\mathfrak{P}_{Hg\mathcal{D}_{\mathfrak{Q}_0}}/\rad(\bar{\ord}_{L^H})} 
~~\text{and}~~ 
\delta_0\equiv 0 \pmod{\mathfrak{P}_{Hg'\mathcal{D}_{\mathfrak{Q}_0}}/\rad(\bar{\ord}_{L^H})}.
\]
Then \eqref{eq_remresidue} in the proof of Lemma~\ref{lem_ramtest} holds. The rest of the proof follows the proof of Lemma~\ref{lem_ramtest}.
\end{proof}

\section{A \texorpdfstring{$\mathcal{P}$-collection}{P-collection} \texorpdfstring{$\tilde{\mathcal{C}}$}{~C} induced from \texorpdfstring{$\mathcal{C}$}{C} and auxiliary elements}\label{sec_defordschm}

The  idempotent decompositions $I_K$ produced in Section~\ref{sec_algmaing} define a $\mathcal{P}$-scheme of double cosets $\mathcal{C}$ rather than an (ordinary) $\mathcal{P}$-scheme. 
Section~\ref{sec_defordschm}--\ref{sec_compordpsch} are devoted to turning it to  a $\mathcal{P}$-scheme $\tilde{\mathcal{C}}$.
In particular, this section focuses on the definition of $\tilde{\mathcal{C}}$ as a $\mathcal{P}$-collection. 

We assume $p>\deg(f)$  in Section~\ref{sec_defordschm}--\ref{sec_compordpsch}. 
As mentioned in Section~\ref{sec_prelimg}, this assumption implies that the wild inertia group $\mathcal{W}_{\mathfrak{Q}_0}\subseteq G$ of $\mathfrak{Q}_0$ over $K_0$ is trivial.
%In addition, by Theorem~\ref{thm_split_general}, the ramification index $e(\mathfrak{P})$ and the inertia degree $f(\mathfrak{P})$ of a prime ideal $\mathfrak{P}$ of a field between 

Suppose the partitions in $\mathcal{C}$ all have locally constant ramification indices and inertia degrees (with respect to $(\mathcal{D}_{\mathfrak{Q}_0}, \mathcal{I}_{\mathfrak{Q}_0})$).
Then for  $K\in\mathcal{F}$ and $\delta\in I_K$, the (nonempty) set of maximal ideals $\mathfrak{P}$ of $\ord_K$ satisfying 
\[
\delta\equiv 1\pmod{\bar{\mathfrak{P}}} \quad\text{where}\quad \bar{\mathfrak{P}}:=\frac{\mathfrak{P}/p\ord_K}{\rad(\bar{\ord}_K)}\cap R_K
\]
all have the same ramification index $e(\mathfrak{P})$ and the same inertia  degree $f(\mathfrak{P})$.
We denote $e(\mathfrak{P})$ by $e_\delta$ and $f(\mathfrak{P})$ by $f_\delta$.
Note that $e_\delta$ and $f_\delta$ are coprime to $p$ by Theorem~\ref{thm_split_general} and the assumption $p>\deg(f)$.
 \nomenclature[d1u]{$e_\delta$, $f_\delta$}{See Section~\ref{sec_defordschm}}

Recall that for a finite extension $K$ of $K_0$ and $i\in\N^+$, we denote by $A_{K,i}$ the ring $(\bar{\ord}_K/\rad(\bar{\ord}_K))\otimes_{\F_q} \F_{q^i}$.
To define $\tilde{\mathcal{C}}$, we need an auxiliary collection of elements in rings $\bar{\ord}_K$ or $A_{K,i}$. We call such a collection of elements an {\em $\mathcal{I}$-advice}:

\begin{defi}\label{defi_auxielem}
Suppose $\mathcal{I}=\{I_K: K\in \mathcal{F}\}$ is a collection of idempotent decompositions of the rings $R_K$, $K\in\mathcal{F}$, that defines to a $\mathcal{P}$-collection of double cosets $\mathcal{C}$ (with respect to $\mathcal{D}_{\mathfrak{Q}_0}$), such that all the partitions in $\mathcal{C}$ have locally constant ramification indices and inertia degrees (with respect to $(\mathcal{D}_{\mathfrak{Q}_0}, \mathcal{I}_{\mathfrak{Q}_0})$). An {\em $\mathcal{I}$-advice}\index{Iadvice@$\mathcal{I}$-advice} $\{\mathcal{S},\mathcal{T}\}$ consists of the following data:
\begin{itemize}
\item $\mathcal{S}=\{s_\delta: \delta\in I_K, e_\delta>1\}$, where each $s_\delta\in\mathcal{S}$ is an element of $\bar{\ord}_K$ such that    $s_\delta\in\mathfrak{m}-\mathfrak{m}^2$ for all the maximal ideals $\mathfrak{m}$ of $\bar{\ord}_K$ satisfying $\delta\equiv 1\pmod{\mathfrak{m}/\rad(\bar{\ord}_K)}$.
%\begin{align*}
%s_\delta&\in\mathfrak{m}-\mathfrak{m}^2 & \text{if}\quad\delta&\equiv 1\pmod{\mathfrak{m}/\rad(\bar{\ord}_K)},~\text{and}\\
%s_\delta&\in \mathfrak{m}^k~\text{for all}~k\in\N^+   &\text{if}\quad\delta&\equiv 0\pmod{\mathfrak{m}/\rad(\bar{\ord}_K)}.
%\end{align*}

\item $\mathcal{T}=\{t_\delta: \delta\in I_K, f_\delta>1\}$, where each $t_\delta\in \mathcal{T}$ is an element of $A_{K,f_\delta}$ such that  $t_{\delta}\not\in\mathfrak{m}$  for all the maximal ideals $\mathfrak{m}$ of $A_{K,f_\delta}$ satisfying $\delta\equiv 1\pmod{\mathfrak{m}}$, and $\sigma_{K,f_\delta}(t_{\delta})=\xi\cdot t_\delta$, where $\xi\in \F_{q^{f_\delta}}$ is a primitive $f_\delta$th root of unity.\footnote{We regard $\delta\in R_K\subseteq \bar{\ord}_K/\rad(\bar{\ord}_K)$ as an element of $A_{K,f_\delta}$ via $\delta\mapsto \delta\otimes 1$,  and $\xi\in\F_{q^{f_\delta}}$ as an element of  $A_{K,f_\delta}$ via $\xi\mapsto 1\otimes \xi$. 
% Also recall that $\sigma_{K,f_\delta}$ is the automorphism of $A_{K,f_\delta}$ sending $a\otimes b\in A_{K,f_\delta}$ to $a^q\otimes b$.
}
\end{itemize}
\end{defi}

An $\mathcal{I}$-advice can be computed from $\mathcal{I}$ by the following lemma. Its proof is deferred to Appendix~\ref{chap_omitted2}. 

\begin{restatable}{lem}{lemgenset}\label{lem_genset}
Under GRH, there exists a subroutine $\mathtt{ComputeAdvice}$ that given $\mathcal{I}=\{I_K: K\in \mathcal{F}\}$ as in Definition~\ref{defi_auxielem}, either properly refines some idempotent decomposition  $I_K\in\mathcal{I}$, or computes $e_\delta,f_\delta$ for  $K\in\mathcal{F}$, $\delta\in I_K$ and an $\mathcal{I}$-advice.\footnote{We need to compute the rings $A_{K,f_\delta}$ before computing  the elements $t_\delta\in A_{K,f_\delta}$. These rings will be computed before the call of the subroutine  $\mathtt{ComputeAdvice}$. See Section~\ref{sec_compordpsch}.}
Moreover, the subroutine runs in time polynomial in $\log p$ and the size of $\mathcal{F}$.
\end{restatable}

We also need the following notations: recall that for $H\in\mathcal{P}$, we chose an isomorphism $\tau_H: K\to L^H$ over $K_0$ where $K$ is the unique field in $\mathcal{F}$ isomorphic to $L^H$ over $K_0$. 
The induced isomorphism $\bar{\ord}_K\cong \bar{\ord}_{L^H}$ identifies each $s_\delta\in \mathcal{S}$ (where $\mathcal{S}$ is as in Definition~\ref{defi_auxielem}) with an element in $\bar{\ord}_{L^H}$, which we denote by $s_{\delta,H}$. Similarly, we identify each $t_\delta\in \mathcal{T}$ with an element in $A_{L^H,f_\delta}$,  denoted by $t_{\delta,H}$.

Next we define a $\mathcal{P}$-collection $\tilde{\mathcal{C}}$  using $\mathcal{I}$    and an  $\mathcal{I}$-advice:
 
 \begin{defi}\label{defi_ordpcollection}
Let  $\mathcal{I}=\{I_K: K\in \mathcal{F}\}$ be as in Definition~\ref{defi_auxielem} and  $\{\mathcal{S},\mathcal{T}\}$ be an  $\mathcal{I}$-advice. Let $\mathcal{C}=\{C_H: H\in\mathcal{P}\}$ be the   $\mathcal{P}$-collection of double cosets with respect to $\mathcal{D}_{\mathfrak{Q}_0}$ associated with $\mathcal{I}$ (see Section~\ref{sec_algmaing}).
For $H\in\mathcal{P}$, let $K$ be the unique field in $\mathcal{F}$ isomorphic to $L^H$ over $K_0$, and   define the partition $\tilde{C}_H$ of $H\backslash G$ so that $Hg,Hg'\in H\backslash G$ are in the same block of $\tilde{C}_H$ iff the following conditions are satisfied:
\begin{enumerate}
\item $Hg\mathcal{D}_{\mathfrak{Q}_0}$ and $Hg'\mathcal{D}_{\mathfrak{Q}_0}$ are in the same block $B$ of $C_H$.
\item Let $\delta$ be the unique idempotent in $I_K$ such that  $\tilde{\tau}_H(\delta)=\delta_B$ (see Definition~\ref{defi_partitioncorg}), where $B\in C_H$ is as in the previous condition. If $e_\delta>1$, the order of the unique element $c$ in $\kappa_{\mathfrak{Q}_{0}}^\times$ satisfying
\[
\prescript{g^{-1}}{}{s_{\delta,H}}+I= c\cdot  (\prescript{g'^{-1}}{}{s_{\delta,H}} +I)
\]
is coprime to $e_\delta$, where $I=(\mathfrak{Q}_{0}/p\ord_{L})^{e(\mathfrak{Q}_0)/e_\delta+1}$.
\item  Let $\delta\in I_K$ be as in the previous condition. Let  $\mathfrak{m}_0$  be an arbitrary  maximal ideal of $A_{L, f_\delta}$ containing $\frac{\mathfrak{Q}_{0}/p\ord_{L}}{\rad(\bar{\ord}_{L})}$. If $f_\delta>1$, the order of the unique element $c$ in $(A_{L,f_\delta}/\mathfrak{m}_0)^\times$ satisfying
\[
\prescript{g^{-1}}{}{t_{\delta,H}}+\mathfrak{m}_0=c \cdot (\prescript{g'^{-1}}{}{t_{\delta,H}}+\mathfrak{m}_0) 
\]
 is coprime to $f_\delta$.
\end{enumerate}
Define   $\tilde{\mathcal{C}}=\{\tilde{C}_H: H\in\mathcal{P}\}$, which is a $\mathcal{P}$-collection.
We say $\tilde{\mathcal{C}}$ is the {\em  $\mathcal{P}$-collection associated with $\mathcal{I}$ and   $\{\mathcal{S},\mathcal{T}\}$}. 
 \end{defi}

We check that  $\tilde{\mathcal{C}}$ is well defined:

\begin{restatable}{lem}{lemordpcolwd}\label{lem_ordpcolwd}
The  $\mathcal{P}$-collection $\tilde{\mathcal{C}}$ in Definition~\ref{defi_ordpcollection} is well defined. 
\end{restatable}

 The proof of Lemma~\ref{lem_ordpcolwd}  is routine and can be found in Appendix~\ref{chap_omitted2}.

\section{\texorpdfstring{$(\mathcal{C},\mathcal{D})$-separated}{(C, D)-separated} \texorpdfstring{$\mathcal{P}$-collections}{P-collections}}\label{sec_ordpschprop}

We continue the discussion in the previous section.
Our goal is to compute $\mathcal{I}=\{I_K: K\in\mathcal{F}\}$ and an $\mathcal{I}$-advice $\{\mathcal{S},\mathcal{T}\}$ such that the associated $\mathcal{P}$-collection $\tilde{\mathcal{C}}$ is a strongly antisymmetric $\mathcal{P}$-scheme.
To achieve this goal, we introduce another property of $\mathcal{P}$-collections called {\em $(\mathcal{C},\mathcal{D})$-separatedness}:

\begin{defi} \label{defi_cdseparate}
Let $\mathcal{P}$ be a subgroup system over a finite group $G$, and let $\mathcal{C}=\{C_H: H\in\mathcal{P}\}$ be a $\mathcal{P}$-collection of double cosets with respect to a subgroup $\mathcal{D}$ of $G$.
We say a $\mathcal{P}$-collection $\tilde{\mathcal{C}}=\{\tilde{C}_H: H\in\mathcal{P}\}$ is {\em $(\mathcal{C}, \mathcal{D})$-separated}\index{CDseparatedness@$(\mathcal{C},\mathcal{D})$-separatedness} if the  following conditions are satisfied:
\begin{enumerate}
\item All the partitions $\tilde{C}_H\in\tilde{\mathcal{C}}$ are invariant under the action of $\mathcal{D}$ by inverse right translation, i.e. for all $B\in\tilde{C}_H$ and $g\in \mathcal{D}$, the set $\prescript{g}{}{B}=\{Hhg^{-1}: Hh\in B\}$ is also in $\tilde{C}_H$. 
\item For $H\in\mathcal{P}$, the map $\pi_H: H\backslash G\to H\backslash G/\mathcal{D}$ sending $Hg\in H\backslash G$ to $Hg\mathcal{D}$ maps each block of $\tilde{C}_H$ bijectively to a block of $C_{H}$.
\end{enumerate}
\end{defi}

It is worth noting that if $\tilde{\mathcal{C}}$ is $(\mathcal{C},\mathcal{D})$-separated, then all the partitions in  $\mathcal{C}$ automatically have locally constant ramification indices and inertia degrees:
% This observation is not needed for analyzing the algorithm, though.
\begin{lem}
Suppose   $\tilde{\mathcal{C}}=\{\tilde{C}_H: H\in\mathcal{P}\}$ is a $(\mathcal{C},\mathcal{D})$-separated $\mathcal{P}$-collection
where $\mathcal{P}$, $\mathcal{C}$, $\mathcal{D}$ are as in Definition~\ref{defi_cdseparate}.
Let $\mathcal{I}$ be a normal subgroup of $\mathcal{D}$.
Then all the partitions in  $\mathcal{C}$ have locally constant ramification indices and inertia degrees with respect to  $(\mathcal{D}, \mathcal{I})$.
\end{lem}

\begin{proof}
 Fix $H\in\mathcal{P}$, $B\in C_H$, and $\tilde{B}\in \tilde{C}_H$ such that $\pi_H(\tilde{B})=B$, where $\pi_H$ is as in Definition~\ref{defi_cdseparate}. Let $\mathcal{D}'$ be a subgroup of $\mathcal{D}$.
 % and consider the action of $\mathcal{D}'$ on $H\backslash G$ by inverse right translation. 
 Consider arbitrary $Hg\mathcal{D}', Hg'\mathcal{D}'\in B$ and lift them to $Hg,Hg'\in\tilde{B}$ respectively. Choose $h_1,\dots,h_k\in\mathcal{D}'$ such that  the $\mathcal{D}'$-orbit of $Hg$ is $\{Hgh_1,\dots,Hgh_k\}$ and the cosets $Hgh_i$ are all distinct. We claim $Hg'h_1,\dots, Hg'h_k$ are also distinct. Assume to the contrary that $Hg'h_{i_1}=Hg'h_{i_2}$ holds for distinct $i_1, i_2\in [k]$. Then $Hg'h_{i_1}$ and $Hg'h_{i_2}$ are in the same block of $\tilde{C}_H$. It follows by the first condition in Definition~\ref{defi_cdseparate} that $Hgh_{i_1}$ and $Hgh_{i_2}$ are also in the same block. But $Hgh_{i_1}\neq Hgh_{i_2}$ and they are both mapped to $Hg\mathcal{D}$ by $\pi_H$, contradicting the second condition in Definition~\ref{defi_cdseparate}. This proves the claim. So the cardinality of the $\mathcal{D}'$-orbit of any $Hg\in H\backslash G$ only depends on the block in $C_H$ containing $Hg\mathcal{D}$.
 In particular, this holds for $\mathcal{D}'=\mathcal{D}$ and $\mathcal{D}'=\mathcal{I}$.
The lemma then follows from  Definition~\ref{defi_indandeg}.
\end{proof}

The following lemma provides a criterion for a $(\mathcal{C},\mathcal{D})$-separated $\mathcal{P}$-collection to be a strongly antisymmetric $\mathcal{P}$-scheme.

\begin{lem}\label{lem_pcoltopsch}
Let $\mathcal{P}$ be a subgroup system over a finite group $G$, and let $\mathcal{C}=\{C_H: H\in\mathcal{P}\}$ be a $\mathcal{P}$-scheme of double cosets with respect to $\mathcal{D}\subseteq G$.
Suppose  $\tilde{\mathcal{C}}=\{\tilde{C}_H: H\in\mathcal{P}\}$ is a compatible, invariant, $(\mathcal{C},\mathcal{D})$-separated
%\footnote{Actually, only the second condition of Definition~\ref{defi_cdseparate} is needed. Our algorithm produces a $\mathcal{P}$-collection $\tilde{C}$ that satisfies both conditions, though.} 
$\mathcal{P}$-collection. Then it is actually a $\mathcal{P}$-scheme. Moreover, if $\mathcal{C}$ is antisymmetric (resp. strongly antisymmetric), so is $\tilde{\mathcal{C}}$.
\end{lem}

\begin{proof}
For the first claim, we just need to show $\tilde{\mathcal{C}}$ is regular. 
Consider $H,H'\in\mathcal{P}$ with $H\subseteq H'$.
Let $\pi_H: H\backslash G\to H\backslash G/\mathcal{D}$ be the map sending $Hg\in H\backslash G$ to $Hg\mathcal{D}$, and define $\pi_{H'}$ similarly.
Then the following diagram commutes.
\[
\begin{tikzcd}[column sep=large]
H\backslash G \arrow[r, "\pi_{H,H'}"] \arrow[d, "\pi_{H}"']
& H'\backslash G  \arrow[d, "\pi_{H'}"] \\
H\backslash G/\mathcal{D}  \arrow[r,  "\pi_{H,H'}^{\mathcal{D}}"]
& H'\backslash G/\mathcal{D}
\end{tikzcd}
\]
For $B\in \tilde{C}_H$ and $B'\in \tilde{C}_{H'}$ containing $\pi_{H,H'}(B)$, we need to show the map $\pi_{H,H'}|_{B}: B\to B'$ has the constant degree, i.e., the cardinality of $\pi_{H,H'}^{-1}(y)\cap B$ is independent of $y\in B'$. As $\tilde{\mathcal{C}}$ is $(\mathcal{C},\mathcal{D})$-separated, the map $\pi_H$ sends $B$ bijectively to $\pi_H(B)\in C_H$, and similarly $\pi_{H'}$ sends $B'$ bijectively to $\pi_{H'}(B')\in C_{H'}$. The claim then follows from regularity of $\mathcal{C}$.

Note that the conjugations also commute with the maps $\pi_H$, i.e., $\pi_{hHh^{-1}}\circ c_{H,h}=c_{H,h}^{\mathcal{D}}\circ\pi_H$ for $H\in\mathcal{P}$ and $h\in G$.
Assume $\tilde{\mathcal{C}}$ is not strongly antisymmetric. Then there exists a nontrivial permutation $\tau$ of a block $B\in \tilde{C}_{H}$ for some $H\in \mathcal{P}$ that arises as a composition of maps $\sigma_i: B_{i-1}\to B_i$, $i=1\dots,k$ where $B_i$ is a block of $\tilde{C}_{H_i}$, $H_i\in \mathcal{P}$, and $\sigma_i$ is of the form $c_{H_{i-1}, h}|_{B_{i-1}}$ (where $h\in G$), $\pi_{H_{i-1}, H_i}|_{B_{i-1}}$, or  $(\pi_{H_{i}, H_{i-1}}|_{B_{i}})^{-1}$ (see Definition~\ref{defi_strongasym}). 
As the maps $\pi_{H_i}|_{B_i}: B_i\to \pi_{H_i}(B_i)$ are bijective and commute with projections and conjugations, we see
$\tau':=\sigma'_k\circ\cdots\circ \sigma'_1$ is a nontrivial permutation of $\pi_H(B)\in C_{H}$, where each map $\sigma'_i:=\pi_{H_i}|_{B_i}\circ \sigma_i\circ(\pi_{H_{i-1}}|_{B_{i-1}})^{-1}$ is of the form $c_{H_{i-1}, h}^{\mathcal{D}}|_{B_{i-1}}$, $\pi_{H_{i-1}, H_i}^{\mathcal{D}}|_{B_{i-1}}$, or  $(\pi_{H_{i}, H_{i-1}}^{\mathcal{D}}|_{B_{i}})^{-1}$. So $\mathcal{C}$ is not strongly antisymmetric. 
The proof of antisymmetry is the same except that we only consider maps $\sigma_i$ that are conjugations.
\end{proof}

We need to compute $\mathcal{I}=\{I_K: K\in\mathcal{F}\}$ and an $\mathcal{I}$-advice $\{\mathcal{S},\mathcal{T}\}$ such that the associated $\mathcal{P}$-collection $\tilde{\mathcal{C}}$ is $(\mathcal{C},\mathcal{D}_{\mathfrak{Q}_0})$-separated. The following lemma states that for $\mathcal{P}$-collections arising from  Definition~\ref{defi_ordpcollection}, the first condition of $(\mathcal{C},\mathcal{D}_{\mathfrak{Q}_0})$-separatedness is in fact automatic.

\begin{lem}\label{lem_fixirt}
Let $\mathcal{I}$, $\{\mathcal{S},\mathcal{T}\}$, $\mathcal{C}$ and $\tilde{\mathcal{C}}$ be as in Definition~\ref{defi_ordpcollection}.
Then all the partitions in $\tilde{\mathcal{C}}$ are invariant under the action of $\mathcal{D}_{\mathfrak{Q}_0}$ by inverse right translation. 
%In other words, for $H\in\mathcal{P}$, $B\in \tilde{C}_H$ and $h\in \mathcal{D}_{\mathfrak{Q}_0}$, the set $B'=\{Hgh^{-1}: Hg\in B\}$ is also a block in $\tilde{C}_H$.
\end{lem}

To prove it, we need the following observation.

\begin{lem}\label{lem_congruencedi}
Let  $\mathfrak{m}_0$  be a maximal ideal of $A_{L, f_\delta}$ containing  $\frac{\mathfrak{Q}_{0}/p\ord_{L}}{\rad(\bar{\ord}_{L})}$.
For all $x\in A_{L,f_\delta}$, $\omega\in \mathcal{I}_{\mathfrak{P}}$, and $\sigma\in \mathcal{D}_{\mathfrak{Q}_0}$ such that the image of $\sigma$ in  $\gal(\kappa_{\mathfrak{Q}_0}/\bar{\ord}_{K_0})$ is the Frobenius automorphism $x\mapsto x^q$ over $\F_q$, it holds that $\prescript{\omega}{}{x}\equiv x\pmod{\mathfrak{m}_0}$ and  $\prescript{\sigma}{}{x}\equiv \sigma_{L,f_\delta}(x) \pmod{\mathfrak{m}_0}$. 
%In particular, we have $\prescript{\omega}{}{\mathfrak{m}_0}=\mathfrak{m}_0$.
\end{lem}

\begin{proof}
By bilinearity, we may assume $x=a\otimes b$ where $a\in \bar{\ord}_{L}/\rad(\bar{\ord}_{L})$ and 
$b\in \F_{q^{f_\delta}}$. 
As $\omega\in  \mathcal{I}_{\mathfrak{P}}$, it holds that $\prescript{\omega}{}{a}\equiv a\pmod{\frac{\mathfrak{Q}_{0}/p\ord_{L}}{\rad(\bar{\ord}_{L})}}$ and hence $\prescript{\omega}{}{(a\otimes b)}\equiv\prescript{\omega}{}{a}\otimes b\equiv a\otimes b\pmod{\mathfrak{m}_0}$.  
Similarly, we have $\prescript{\sigma}{}{a}\equiv a^q \pmod{\frac{\mathfrak{Q}_{0}/p\ord_{L}}{\rad(\bar{\ord}_{L})}}$  by definition and hence $\prescript{\sigma}{}{(a\otimes b)}\equiv\prescript{\sigma}{}{a}\otimes b\equiv a^q\otimes b\equiv\sigma_{L,f_\delta}(a\otimes b)\pmod{\mathfrak{m}_0}$.  
\end{proof}

Now we are ready to prove Lemma~\ref{lem_fixirt}.

\begin{proof}[Proof of Lemma~\ref{lem_fixirt}]
Consider $H\in\mathcal{P}$ and $Hg, Hg'$ in the same block of  $\tilde{C}_H$.
Fix $h\in \mathcal{D}_{\mathfrak{Q}_0}$.  
We prove $Hgh^{-1}, Hg'h^{-1}$ are also in the same block by verifying the three conditions in  Definition~\ref{defi_ordpcollection}. 
Let $B$ be the block of $C_H$ containing both $Hg\mathcal{D}_{\mathfrak{Q}_0}$ and $Hg'\mathcal{D}_{\mathfrak{Q}_0}$.
The first condition  in  Definition~\ref{defi_ordpcollection} obviously holds for $Hgh^{-1}$ and $Hg'h^{-1}$ since $Hgh^{-1}\mathcal{D}_{\mathfrak{Q}_0}=Hg\mathcal{D}_{\mathfrak{Q}_0}\in B$ and $Hg'h^{-1}\mathcal{D}_{\mathfrak{Q}_0}=Hg'\mathcal{D}_{\mathfrak{Q}_0}\in B$. 

Let $K$ be the field in $\mathcal{F}$ isomorphic to $L^H$ over $K_0$.
Let $\delta$ be the idempotent in $I_K$ satisfying $\tilde{\tau}_H(\delta)=\delta_B$ (see Definition~\ref{defi_partitioncorg}).
Suppose $e_{\delta}>1$. By  Definition~\ref{defi_ordpcollection}, the order of the unique element $c$ in $\kappa_{\mathfrak{Q}_{0}}^\times$ satisfying
\[
\prescript{g^{-1}}{}{s_{\delta,H}}+I= c\cdot  (\prescript{g'^{-1}}{}{s_{\delta,H}} +I)
\]
is coprime to $e_\delta$, where $I=(\mathfrak{Q}_{0}/p\ord_{L})^{e(\mathfrak{Q}_0)/e_\delta+1}$. We have $\prescript{h}{}{I}=I$ since $h\in  \mathcal{D}_{\mathfrak{Q}_0}$. Therefore
\[
\prescript{hg^{-1}}{}{s_{\delta,H}}+I= \prescript{h}{}{c}\cdot  (\prescript{hg'^{-1}}{}{s_{\delta,H}} +I),
\]
where $\prescript{h}{}{c}\in \kappa_{\mathfrak{Q}_{0}}^\times$ has the same order as  $c$. 
 So the second condition in Definition~\ref{defi_ordpcollection} is  satisfied by $Hgh^{-1}$ and $Hg'h^{-1}$.

Now suppose $f_{\delta}>1$.
Let  $\mathfrak{m}_0$  be  a maximal ideal of $A_{L, f_\delta}$ containing  $\frac{\mathfrak{Q}_{0}/p\ord_{L}}{\rad(\bar{\ord}_{L})}$.
By Definition~\ref{defi_ordpcollection}, the order of the unique element $c$ in $(A_{L,f_\delta}/\mathfrak{m}_0)^\times$ satisfying
\[
\prescript{g^{-1}}{}{t_{\delta,H}}+\mathfrak{m}_0=c \cdot (\prescript{g'^{-1}}{}{t_{\delta,H}}+\mathfrak{m}_0) 
\]
 is coprime to $f_\delta$.
Fix $\sigma\in \mathcal{D}_{\mathfrak{Q}_0}$ whose image   in  $\gal(\kappa_{\mathfrak{Q}_0}/\bar{\ord}_{K_0})$ is the Frobenius automorphism $x\mapsto x^q$ over $\F_q$.
Choose   $\omega\in \mathcal{I}_{\mathfrak{Q}_0}$ and $i\in\Z$ such that $h=\omega\sigma^i$. 
By Lemma~\ref{lem_congruencedi}, we have
\begin{align*}
\prescript{hg^{-1}}{}{t_{\delta,H}}
&\equiv\prescript{\omega\sigma^i}{}{(\prescript{g^{-1}}{}{t_{\delta,H}})}
\equiv\prescript{\sigma^i}{}{(\prescript{g^{-1}}{}{t_{\delta,H}})}
\equiv\sigma_{L,f_\delta}^i(\prescript{g^{-1}}{}{t_{\delta,H}})
\equiv \prescript{g^{-1}}{}{\left(\sigma_{L,f_\delta}^i(t_{\delta,H})\right)}\\
&\equiv \prescript{g^{-1}}{}{(\xi^i\cdot t_{\delta,H})}
\equiv\xi^i\cdot \prescript{g^{-1}}{}{t_{\delta,H}}\pmod{\mathfrak{m}_0},
\end{align*}
where $\xi$ is the primitive $f_\delta$th root of unity satisfying $\sigma_{K,f_\delta}(t_{\delta})=\xi\cdot t_\delta$ as in  Definition~\ref{defi_auxielem}.
%Here the second and the third congruence relations follow from Lemma~\ref{lem_congruencedi}. The fourth one holds since the action of $G$ on $A_{L,f_\delta}$ commutes with $\sigma_{L,f_\delta}$. The fifth one follows from  $\sigma_{L,f_\delta}(t_{\delta,H})=\xi\cdot t_{\delta,H}$, which in turn follows from $\sigma_{K,f_\delta}(t_{\delta})=\xi\cdot t_\delta$. The last one holds since $\xi^j\in\F_{q^{f_\delta}}\subseteq A_{L,f_\delta}$ is fixed by $G$. 
The same argument  shows $\prescript{hg'^{-1}}{}{t_{\delta,H}}\equiv \xi^i\cdot \prescript{g'^{-1}}{}{t_{\delta,H}}\pmod{\mathfrak{m}_0}$. It follows that
\[
\prescript{hg^{-1}}{}{t_{\delta,H}}+\mathfrak{m}_0=c \cdot (\prescript{hg'^{-1}}{}{t_{\delta,H}}+\mathfrak{m}_0).
\]
So the third condition in Definition~\ref{defi_ordpcollection} is  also satisfied by $Hgh^{-1}$ and $Hg'h^{-1}$.
\end{proof}

We also  show that $\mathcal{P}$-collections arising from  Definition~\ref{defi_ordpcollection} always satisfy a weakening of the second condition of  $(\mathcal{C},\mathcal{D}_{\mathfrak{Q}_0})$-separatedness, where bijectivity is replaced by injectivity:

\begin{lem}\label{lem_injsep}
Let $\mathcal{I}$, $\{\mathcal{S},\mathcal{T}\}$, $\mathcal{C}$ and $\tilde{\mathcal{C}}$ be as in Definition~\ref{defi_ordpcollection}.
Then for $H\in\mathcal{P}$, the map $\pi_H: H\backslash G\to H\backslash G/\mathcal{D}_{\mathfrak{Q}_0}$ sending $Hg\in H\backslash G$ to $Hg\mathcal{D}_{\mathfrak{Q}_0}$ maps each block of $\tilde{C}_H$ injectively to a block of $C_{H}$.
\end{lem}
\begin{proof}
Consider $H\in\mathcal{P}$ and $g\in G$ and $h\in\mathcal{D}_{\mathfrak{Q}_0}$ such that $Hg\neq Hgh^{-1}$.
We want to prove that $Hg$ and $Hgh^{-1}$ are in different blocks of $\tilde{C}_H$. 

Let $B$ be the block of $C_H$ containing $Hg\mathcal{D}_{\mathfrak{Q}_0}=Hgh^{-1}\mathcal{D}_{\mathfrak{Q}_0}$.
Let $K$ be the field in $\mathcal{F}$ isomorphic to $L^H$ over $K_0$.
Let $\delta$ be the idempotent in $I_K$ satisfying $\tilde{\tau}_H(\delta)=\delta_B$ (see Definition~\ref{defi_partitioncorg}).
Fix $\sigma\in \mathcal{D}_{\mathfrak{Q}_0}$ whose image   in  $\gal(\kappa_{\mathfrak{Q}_0}/\bar{\ord}_{K_0})$ is the Frobenius automorphism $x\mapsto x^q$ over $\F_q$.

As we assume $p>\deg(f)$, the wild inertia group $\mathcal{W}_{\mathfrak{Q}_0}\subseteq G$ of $\mathfrak{Q}_0$ over $K_0$ is trivial. So $\mathcal{I}_{\mathfrak{Q}_0}$ is a cyclic group of order $e(\mathfrak{Q}_0)$.
Fix a generator $\omega$ of $\mathcal{I}_{\mathfrak{Q}_0}$. 
By Theorem~\ref{thm_split_general} and Definition~\ref{defi_indandeg}, we know $e_\delta$ is the smallest positive integer $k$ satisfying $Hg \omega^{-k}=Hg$, and $f_\delta$ is the smallest positive integer $k$ satisfying $Hg \sigma^{-k}\mathcal{I}_{\mathfrak{Q}_0}=Hg \mathcal{I}_{\mathfrak{Q}_0}$.
So there exist unique  $i\in\{0,\dots,f_\delta-1\}$ and $j\in\{0,\dots,e_\delta-1\}$ such that $Hgh^{-1}=Hg\sigma^{-i}\omega^{-j}$.
As $Hg\neq Hgh^{-1}$, we have $(i,j)\neq (0,0)$.
By replacing $h$ with $\omega^j\sigma^i$ if necessary, we may assume $h=\omega^j\sigma^i$.

First assume $i\neq 0$. Then $f_\delta>1$. Let  $\mathfrak{m}_0$  be  a maximal ideal of $A_{L, f_\delta}$ containing  $\frac{\mathfrak{Q}_{0}/p\ord_{L}}{\rad(\bar{\ord}_{L})}$. 
As shown in the proof of Lemma~\ref{lem_fixirt}, we have 
\[
\prescript{hg^{-1}}{}{t_{\delta,H}} \equiv\xi^i\cdot \prescript{g^{-1}}{}{t_{\delta,H}}\pmod{\mathfrak{m}_0},
\]
 where $\xi$ is a primitive $f_\delta$th root of unity. The order of $\xi^i$ is $f_\delta/\gcd(f_\delta,i)>1$ and is a divisor of $f_\delta$. So the third condition  in Definition~\ref{defi_ordpcollection} is  not satisfied by $Hg$ and $Hgh^{-1}$. It follows that $Hg$ and $Hgh^{-1}$ are in different blocks of $\tilde{C}_H$, as desired.
 
 Now assume $i=0$ and $j\neq 0$.  Then $e_\delta>1$.
 Let $\mathfrak{m}_e=\mathfrak{Q}_0/p\ord_L$ and $k=e(\mathfrak{Q}_0)/e_\delta$.
 As shown in the proof of Lemma~\ref{lem_ordpcolwd}, we have $\prescript{g^{-1}}{}{s_{\delta,H}}\in \mathfrak{m}_{e}^k-\mathfrak{m}_{e}^{k+1}$.
 Choose $\pi_L\in \mathfrak{m}_e-\mathfrak{m}_e^2$.
 We have a group homomorphism
$\mathcal{I}_{\mathfrak{Q}_0}\to  \kappa_{\mathfrak{Q}_0}^\times$
sending $g\in \mathcal{I}_{\mathfrak{Q}_0}$ to the unique element $c_g\in \kappa_{\mathfrak{Q}_0}^\times$ satisfying $\prescript{g}{}{\pi_L}+\mathfrak{m}_e^2=c_g(\pi_L+\mathfrak{m}_e^2)$.
 This map is injective since its kernel  is $\mathcal{W}_{\mathfrak{Q}_0}=\{e\}$. In particular, we know $c_\omega$ is a primitive $e(\mathfrak{Q}_0)$th root of unity in $\kappa_{\mathfrak{Q}_0}^\times$.
 Choose $c\in\kappa_{\mathfrak{Q}_0}^\times$ such that 
 \[
 \prescript{g^{-1}}{}{s_{\delta,H}}+\mathfrak{m}_e^{k+1}=c(\pi_L^k+\mathfrak{m}_e^{k+1}),
 \]
 which exists since $\prescript{g^{-1}}{}{s_{\delta,H}}$ and $\pi_L^k$ are both  in $\mathfrak{m}_{e}^k-\mathfrak{m}_{e}^{k+1}$. Then we have
 \begin{align*}
 \prescript{hg^{-1}}{}{s_{\delta,H}}+\mathfrak{m}_e^{k+1}&=\prescript{\omega^j}{}{(\prescript{ g^{-1}}{}{s_{\delta,H}}+\mathfrak{m}_e^{k+1})}=\prescript{\omega^j}{}{(c(\pi_L^k+\mathfrak{m}_e^{k+1}))}
 \\
 &=c\cdot c_\omega^{jk}\cdot(\pi_L^k+\mathfrak{m}_e^{k+1})= c_\omega^{jk}\cdot (\prescript{g^{-1}}{}{s_{\delta,H}}+\mathfrak{m}_e^{k+1}).
 \end{align*}
 The order of $ c_\omega^{jk}\in \kappa_{\mathfrak{Q}_0}^\times$ is $e(\mathfrak{Q}_0)/\gcd(e(\mathfrak{Q}_0),jk)=e_\delta/\gcd(e_\delta,j)>1$, which is a divisor of $e_\delta$.  So the second condition  in Definition~\ref{defi_ordpcollection} is  not satisfied by $Hg$ and $Hgh^{-1}$. It follows that $Hg$ and $Hgh^{-1}$ are in different blocks of $\tilde{C}_H$, as desired.
\end{proof}

In the next section, we give subroutines that refine  the idempotent decompositions $I_K$  so that $\tilde{\mathcal{C}}$ is eventually a compatible, invariant, $(\mathcal{C},\mathcal{D})$-separated $\mathcal{P}$-collection, and hence a strongly antisymmetric $\mathcal{P}$-scheme.

\section{Producing an ordinary \texorpdfstring{$\mathcal{P}$-scheme}{P-scheme}}\label{sec_compordpsch}

We modify the algorithm $\mathtt{ComputeDoubleCosetPscheme}$ in Section~\ref{sec_algmaing} so that a $(\mathcal{C},\mathcal{D}_{\mathfrak{Q}_0})$-separated strongly antisymmetric $\mathcal{P}$-scheme is produced.
%More specifically, the new algorithm, called $\mathtt{ComputeOrdinaryPscheme}$, computes partitions of unity $I_K$ of the rings $R_K$ and the elements $s_\delta$, $t_\delta$, such that the induced  $\mathcal{P}$-scheme $\tilde{C}$ is a $(\mathcal{C},\mathcal{D}_{\mathfrak{Q}_0})$-separated strongly antisymmetric $\mathcal{P}$-scheme, where $\mathcal{C}$ is a  strongly antisymmetric $\mathcal{P}$-scheme of double cosets as in Section~\ref{sec_algmaing}.

The pseudocode of the modified algorithm is given in Algorithm~\ref{alg_mainalggv2}. 
Again, the algorithm takes a $(K_0,\tilde{f})$-subfield system $\mathcal{F}$ as the input, and outputs for every $K\in\mathcal{F}$ an idempotent decomposition $I_K$ of the ring $R_K$, together with some auxiliary data.

\begin{algorithm}[htbp]
\caption{$\mathtt{ComputeOrdinaryPscheme}$}\label{alg_mainalggv2}
\begin{algorithmic}[1]
\INPUT $(K_0,\tilde{f})$-subfield system $\mathcal{F}$
\OUTPUT \parbox[t]{.9\linewidth}{%
	 for every $K\in\mathcal{F}$: the outputs of $\mathtt{ComputeRings}$ (see Lemma~\ref{lem_computerings}) on the input $(K,p)$, and an idempotent decomposition $I_K$ of $R_K$ \strut
	  }
\For{$K\in \mathcal{F}$} \strut
    \State  call $\mathtt{ComputeRings}$ on $(K,p)$ 
    \State  $I_K\gets \{1\}$, where $1$ denotes the unity of  $R_K$
    \State compute the inclusion $\F_q\hookrightarrow \bar{\ord}_K/\rad(\bar{\ord}_K)$
        \ForTo{$i$}{$1$}{$[K:K_0]$}
        \State compute  $A_{K,i}$ and the inclusions $\bar{\ord}_K/\rad(\bar{\ord}_K)\hookrightarrow A_{K,i}$, $\F_{q^i}\hookrightarrow A_{K,i}$
         \EndFor
\EndFor 
 \For{$(K,K')\in \mathcal{F}^2$} \strut
    \State \parbox[t]{\dimexpr\linewidth-\algorithmicindent}{%
        call $\mathtt{ComputeRelEmbeddings}$ to compute all the embeddings from $K$ to $K'$ over $K_0$\strut
        }
    \For{\strut embedding $\phi: K\hookrightarrow K'$ over $K_0$}
        \State call $\mathtt{ComputeRingHoms}$ on $p$, $K$, $K'$ and $\phi$ compute $\bar{\phi}$, $\hat{\phi}$ and $\tilde{\phi}$
     \EndFor
\EndFor
\Repeat
\Repeat
\Repeat
    \State call  $\mathtt{CompatibilityAndInvarianceTestV2}$
    \State call  $\mathtt{RegularityTestV2}$
    \State call  $\mathtt{StrongAntisymmetryTestV2}$
    \State call  $\mathtt{RamificationIndexTest}$
    \State call  $\mathtt{InertiaDegreeTest}$
\Until{$I_K$ remains the same in the last iteration for all $K\in\mathcal{F}$}
\State call $\mathtt{ComputeAdvice}$ on $\mathcal{I}=\{I_K:K\in\mathcal{F}\}$
\Until{$I_K$ remains the same in the last iteration for all $K\in\mathcal{F}$}
\State call $\mathtt{SurjectivityTest}$
\State call $\mathtt{RingHomTest}$
\Until{$I_K$ remains the same in the last iteration for all $K\in\mathcal{F}$}
\State \Return  the outputs of $\mathtt{ComputeRings}$ on the input $(K,p)$ and $I_K$ for $K\in\mathcal{F}$
%, and the elements $s_\delta$ and $t_\delta$
\end{algorithmic}
\end{algorithm} 

 The first half (Lines 1--10) of the algorithm is the preprocessing  stage:  we compute the same data as in the algorithm $\mathtt{ComputeDoubleCosetPscheme}$.
In addition,  for $K\in\mathcal{F}$, we compute the inclusion $\F_q\hookrightarrow  \bar{\ord}_K/\rad(\bar{\ord}_K)$ at Line 4, endowing $\bar{\ord}_K/\rad(\bar{\ord}_K)$ the structure of an $\F_q$-algebra.\footnote{To achieve this, we compute the image $\bar{Y}$ of $Y+(\tilde{h}(Y))\in\ord_{K_0}\subseteq\ord_K$ in $\bar{\ord}_K$ by Lemma~\ref{lem_computeresidue}. Then   compute the map $\F_p[Y]/(h(Y))\to \bar{\ord}_K$ sending $Y+(h(Y))$ to $\bar{Y}$, and compose it with the isomorphism $\psi_0^{-1}:\F_q\to \F_p[Y]/(h(Y))$ and the quotient map  $\bar{\ord}_K\to  \bar{\ord}_K/\rad(\bar{\ord}_K)$.}
And for $1\leq i\leq [K:K_0]$, we compute the ring $A_{K,i}=\bar{\ord}_K/\rad(\bar{\ord}_K)\otimes_{\F_q} \F_{q^i}$ together with the inclusions  $\bar{\ord}_K/\rad(\bar{\ord}_K)\hookrightarrow A_{K,i}$, $\F_{q^i}\hookrightarrow A_{K,i}$ defined by $a\mapsto a\otimes 1$ and $b\mapsto 1\otimes b$ respectively.  

The second half (Lines 11--24) of the algorithm refines the  idempotent decompositions $I_K$ for $K\in\mathcal{F}$.
The loop in Lines 13--19 is the same as in the algorithm $\mathtt{ComputeDoubleCosetPscheme}$. It produces idempotent decompositions $I_K$ that define a strongly antisymmetric $\mathcal{P}$-scheme of double cosets $\mathcal{C}=\{C_H:H\in\mathcal{P}\}$ with respect to $\mathcal{D}_{\mathfrak{Q}_0}$, in which all the partitions have locally constant ramification indices and inertia degrees (with respect to $(\mathcal{D}_{\mathfrak{Q}_0}, \mathcal{I}_{\mathfrak{Q}_0})$).
After that, we call the subroutine $\mathtt{ComputeAdvice}$ in Lemma~\ref{lem_genset} on $\mathcal{I}=\{I_K: K\in\mathcal{F}\}$ at Line 20. It either properly refines some  $I_K$ or returns an $\mathcal{I}$-advice. 
In the former case, we start over from Line 13.  

So assume an $\mathcal{I}$-advice $\{\mathcal{S},\mathcal{T}\}$ is returned at Line 20. Let $\tilde{\mathcal{C}}=\{\tilde{C}_H: H\in\mathcal{P}\}$ be  the  $\mathcal{P}$-collection associated with $\mathcal{I}$ and   $\{\mathcal{S},\mathcal{T}\}$.
Next we need two new subroutines, $\mathtt{SurjectivityTest}$ and $\mathtt{RingHomTest}$:

\begin{restatable}{lem}{lembijtest}\label{lem_bijtest}
Under GRH, there exists a subroutine $\mathtt{SurjectivityTest}$ that updates $I_K$ in time  polynomial  in $\log p$ and the size of $\mathcal{F}$ so that the partitions $C_H\in \mathcal{C}$ are refined. Moreover, at least one partition $C_H$ is properly refined unless 
for all $H\in\mathcal{P}$, the map $\pi_H: H\backslash G\to H\backslash G/\mathcal{D}_{\mathfrak{Q}_0}$ sending $Hg\in H\backslash G$ to $Hg\mathcal{D}_{\mathfrak{Q}_0}$ maps each block of  $\tilde{C}_H$ surjectively to a block of $C_{H}$.
\end{restatable}

\begin{restatable}{lem}{lemringhomtestg}\label{lem_ringhomtestg}
Under GRH, there exists a subroutine $\mathtt{RingHomTest}$ that updates $I_K$ in time  polynomial  in $\log p$ and the size of $\mathcal{F}$ so that the partitions $C_H\in \mathcal{C}$ are refined. Moreover, at least one partition $C_H$ is properly refined unless $\tilde{\mathcal{C}}$ is compatible and invariant.
\end{restatable}

The proofs of the above two lemmas  are the most technical part of this chapter. We defer them to Appendix~\ref{chap_omitted2}.  
%We assume the wild inertia group $\mathcal{W}_{\mathfrak{Q}_0}\subseteq G$ of $\mathfrak{Q}_0$ over $K_0$ is trivial. As mentioned in Section~\ref{sec_prelimg}, this assumption always holds if $p>n=\deg(f)$. 

We run these two subroutines and repeat, until no idempotent decomposition $I_K$ is properly refined in the last iteration. By Lemma~\ref{lem_fixirt},  Lemma~\ref{lem_injsep}, Lemma~\ref{lem_bijtest}, and Lemma~\ref{lem_ringhomtestg}, the resulting $\mathcal{P}$-collection $\tilde{\mathcal{C}}$ is a  compatible, invariant, $(\mathcal{C},\mathcal{D}_{\mathfrak{Q}_0})$-separated $\mathcal{P}$-collection. Also note that $\mathcal{C}$ is a strongly antisymmetric $\mathcal{P}$-scheme of double cosets with respect to $\mathcal{D}_{\mathfrak{Q}_0}$. It follows from Lemma~\ref{lem_pcoltopsch} that $\tilde{\mathcal{C}}$ is a strongly antisymmetric $(\mathcal{C},\mathcal{D}_{\mathfrak{Q}_0})$-separated $\mathcal{P}$-scheme. We conclude

\begin{thm}\label{thm_comppschemeg}
Under GRH, the algorithm $\mathtt{ComputeOrdinaryPscheme}$ runs in time polynomial in $\log p$ and the size of $\mathcal{F}$,
and when it terminates, $\mathcal{C}$ is a strongly antisymmetric $\mathcal{P}$-scheme of double cosets (with respect to $\mathcal{D}_{\mathfrak{Q}_0}$),
and  $\tilde{\mathcal{C}}$  is a $(\mathcal{C},\mathcal{D}_{\mathfrak{Q}_0})$-separated strongly antisymmetric $\mathcal{P}$-scheme.
\end{thm}

\section{Putting it together}\label{sec_puttogetherg}\index{generalized Pscheme algorithm@generalized $\mathcal{P}$-scheme algorithm}

We combine the results in previous sections to obtain the generalized $\mathcal{P}$-scheme algorithm. 
%It generalizes the algorithm $\mathtt{PschemeAlgorithm}$ in Chapter~\ref{chap_alg_prime}.
For simplicity, we first focus on computing the complete factorization of the input polynomial $f$. The problem of computing a proper factorization of $f$ is discussed later in this section.

The algorithm takes a polynomial $f(X)\in\F_q[X]$ and an irreducible lifted polynomial $\tilde{f}(X)\in A_0[X]$ as the input, and outputs the complete factorization of $f$. Its pseudocode is given in Algorithm~\ref{alg_genpscheme} below. 

\begin{algorithm}[htbp]
\caption{$\mathtt{GeneralizedPschemeAlgorithm}$}\label{alg_genpscheme}
\begin{algorithmic}[1]
\INPUT $f(X)\in \F_q[X]$ and its irreducible lifted polynomial $\tilde{f}(X)\in A_0[X]$
\OUTPUT factorization of $f$
\If{$p\leq \deg(f)$} 
run Berlekamp's algorithm in \citep{Ber70} to compute the complete factorization of $f$, output it and halt
\EndIf
\State call $\mathtt{ComputeRelNumberFields}$ to compute a $(K_0,\tilde{f})$-subfield system $\mathcal{F}$ such that (1) $F=K_0[X]/(\tilde{f}(X))\in \mathcal{F}$, and  (2) for some $H\in\mathcal{P}$ satisfying $L^H\cong_{K_0} F$, all strongly antisymmetric $\mathcal{P}$-schemes are discrete on $H$, where $\mathcal{P}$ is the subgroup system over $G=\gal(\tilde{f}/K_0)$ associated with $\mathcal{F}$  
\State call $\mathtt{ComputeOrdinaryPscheme}$ on  $\mathcal{F}$ to obtain $I_K$ for $K\in\mathcal{F}$
\State call $\mathtt{ExtractFactorsV2}$ to extract a factorization of $f$ from $I_F$, and output it
\end{algorithmic}
\end{algorithm}

Line 1 checks whether $p> \deg(f)$ holds. If $p\leq \deg(f)$, we just run Berlekamp's algorithm in \citep{Ber70} to compute  the complete factorization of $f$ in time polynomial in $p$ and $\deg(f)$, output it, and halt.
This step justifies the assumption $p>\deg(f)$ made in Section~\ref{sec_defordschm}--\ref{sec_compordpsch}.

The subroutine $\mathtt{ComputeRelNumberFields}$ at Line 2 is the generic part of the algorithm. It is supposed to  compute a  $(K_0,\tilde{f})$-subfield system  $\mathcal{F}$ such that  $F\in\mathcal{F}$, and the associated subgroup system $\mathcal{P}$ over $G$ satisfies a certain combinatorial property (see Theorem~\ref{thm_algmain2formalg} below).
%The latter condition is used to show that the factoring algorithm always produces the complete factorization  of $f$.
The algorithm $\mathtt{ComputeOrdinaryPscheme}$ (see Section~\ref{sec_algmaing}) at Line 3 takes the input $\mathcal{F}$ and outputs data that includes the idempotent decomposition $I_F$ of $R_F$.
Finally, we call the subroutine $\mathtt{ExtractFactorsV2}$ (see Section~\ref{sec_algreductiong}) at Line 4 to extract a factorization of $f$ from $I_F$.

The following theorem is the main result of this chapter.

\begin{thm}\label{thm_algmain2formalg}
Suppose there exists a deterministic algorithm that given a polynomial $g(X)\in K_0[X]$ irreducible over $K_0$, constructs  a  $(K_0,g)$-subfield system $\mathcal{F}$ in time $T(g)$ such that
\begin{itemize}
\item $K_0[X]/(g(X))$ is in $\mathcal{F}$, and
\item  for some $H\in\mathcal{P}$ satisfying $(L(g))^H\cong_{K_0} K_0[X]/(g(X))$, all strongly antisymmetric $\mathcal{P}$-schemes are discrete on $H$, where $\mathcal{P}$ is the subgroup system over $\gal(g/K_0)$ associated with $\mathcal{F}$, and $L(g)$ is the splitting field of $g$ over $K_0$.
\end{itemize}
Then under GRH, there exists a deterministic algorithm that given a polynomial $f(X)\in\F_q[X]$ and an irreducible lifted polynomial $\tilde{f}(X)\in A_0[X]$ of $f$, outputs the complete factorization  of $f$ over $\F_q$ in time polynomial in $T(\tilde{f})$ and the size of the input.
\end{thm}

\begin{proof}
Consider the algorithm $\mathtt{GeneralizedPschemeAlgorithm}$ above and implement the subroutine $\mathtt{ComputeRelNumberFields}$ using the hypothetical algorithm in the theorem. 
The case $p\leq \deg(f)$ is solved by Berlekamp's algorithm in \citep{Ber70}. So assume $p>\deg(f)$.
Choose $g=\tilde{f}$.
%By assumption, for some $H\in\mathcal{P}$ satisfying $L^H\cong F$, all strongly antisymmetric $\mathcal{P}$-schemes are discrete on $H$. 
By Theorem~\ref{thm_comppschemeg}, 
the $\mathcal{P}$-collection $\mathcal{C}=\{C_H: H\in\mathcal{P}\}$ defined by $C_H=P(\tilde{\tau}_H(I_{K}))$ is a strongly antisymmetric $\mathcal{P}$-scheme of double cosets with respect to $\mathcal{D}_{\mathfrak{Q}_0}$, and
the $\mathcal{P}$-collection $\tilde{\mathcal{C}}=\{\tilde{C}_H: H\in\mathcal{P}\}$ associated with the collection of idempotent decompositions $\mathcal{I}=\{I_K:K\in \mathcal{F}\}$ and  the $\mathcal{I}$-advice produced by the algorithm $\mathtt{ComputeOrdinaryPscheme}$ is a $(\mathcal{C},\mathcal{D}_{\mathfrak{Q}_0})$-separated strongly antisymmetric $\mathcal{P}$-scheme.
By the second condition in the theorem, we have $\tilde{C}_H=\infty_{H\backslash G}$ (and hence $C_H=\infty_{H\backslash G/\mathcal{D}_{\mathfrak{Q}_0}}$) for some $H\in\mathcal{P}$ satisfying $L^H\cong_{K_0} F$.   So the  idempotent decomposition $I_F$ is complete. 
By Theorem~\ref{thm_extfactorredg}, the algorithm outputs  the complete factorization of $f$ over $\F_q$. 

The subroutine $\mathtt{ComputeRelNumberFields}$ runs in time $T(\tilde{f})$.
In particular, the size of $\mathcal{F}$ is bounded by  $T(\tilde{f})$. The claim about the running time then follows from Theorem~\ref{thm_comppschemeg} and Theorem~\ref{thm_extfactorredg}.
\end{proof}

By Theorem~\ref{thm_algmain2formalg} and Lemma~\ref{lem_compstabsysg}, we have the following partial generalization of Corollary~\ref{cor_algdgbound}.
\begin{cor}\label{cor_algdgboundg}
Under GRH, there exists a deterministic algorithm that, given a polynomial $f(X)\in\F_q[X]$ of degree $n\in\N^+$ and an irreducible\footnote{The assumption that $\tilde{f}$ is irreducible is not necessary, and can be avoided by adapting Lemma~\ref{lem_compstabsysg}. We omit the details.} lifted polynomial $\tilde{f}(X)\in A_0[X]$ of $f$, computes the complete factorization of $f$ over $\F_q$ in time polynomial in $n^{d(G)}$ and the size of the input, where $G$ is the permutation group $\gal(f/K_0)$ acting on the set of roots of $\tilde{f}$.
\end{cor}

%Therefore, to prove that the algorithm always produces the complete factorization (resp. a proper factorization) of $f$, it suffices to show that any strongly antisymmetric $\mathcal{P}$-scheme is discrete (resp. inhomogeneous) on $H$. 

\paragraph{The unifying framework via the generalized $\mathcal{P}$-scheme algorithm.}

In the following, we use  Theorem~\ref{thm_algmain2formalg} and Corollary~\ref{cor_algdgboundg} to derive generalizations of the main results in \citep{Hua91-2, Hua91, Ron88, Ron92, Evd92, Evd94, IKS09} in a uniform way.

Given a polynomial $f(X)\in\F_q[X]$ and  a (possibly reducible) lifted polynomial $\tilde{f}(X)\in A_0[X]$ of $f$. We reduce to the case that the lifted polynomial is irreducible as follows: 
using Lemma~\ref{lem_subprob}, we compute an integer $D$ satisfying $D\equiv 1\pmod{p}$ and a factorization of $D\cdot\tilde{f}$ into irreducible factors $\tilde{f}_1,\dots,\tilde{f}_k\in A_0[X]$ over $K_0$. 
Then we have 
\[
f(X)=\prod_{i=1}^k \tilde{\psi}_0(f_i)(X)
\]
 and the problem of factoring $f(X)$ is reduced to the problem of factoring each $\tilde{\psi}_0(f_i)\in\F_q[X]$ with the aid of its irreducible lifted polynomial $\tilde{f}_i(X)$ (see the discussion after  Lemma~\ref{lem_subprob}). Moreover, for $i\in [k]$, the Galois group $\gal(\tilde{f}_i(X)/K_0)$ is a quotient group of $\gal(\tilde{f}/K_0)$, and hence $|\gal(\tilde{f}_i(X)/K_0)|\leq |\gal(\tilde{f}(X)/K_0)|$.

So assume $\tilde{f}$ is irreducible over $K_0$. We choose $\mathcal{F}=\{F,L\}$ where $F=K_0[X]/(\tilde{f}(X))$ and $L$ is the splitting field of $\tilde{f}$ over $K_0$. Compute $\mathcal{F}$ in time polynomial in $[L:K_0]=\gal(\tilde{f}(X)/K_0)$ and the size of $\tilde{f}$ using  Lemma~\ref{lem_compsplitgeneral}. By Lemma~\ref{lem_antidisc}, all antisymmetric $\mathcal{P}$-schemes are discrete on $H$ for all $H\in\mathcal{P}$ since the trivial subgroup $\{e\}$ is in $\mathcal{P}$. 
%a fortiori, so are all strongly antisymmetric $\mathcal{P}$-schemes. 
Therefore by  Theorem~\ref{thm_algmain2formalg} and the reduction above, we have the following generalization of Theorem~\ref{thm_ron92}.

\begin{thm}
Under GRH, there exists a deterministic algorithm that, given a polynomial $f(X)\in\F_q[X]$ and a lifted polynomial $\tilde{f}(X)\in A_0[X]$ of $f$, computes the complete factorization of $f$ over $\F_q$ in time polynomial in $|\gal(\tilde{f}/K_0)|$ and the size of the input.
\end{thm}

Note $|\gal(\tilde{f}/K_0)|= \deg(f)$ when $\tilde{f}$ is irreducible over $K_0$ and $\gal(\tilde{f}/K_0)$ is abelian.  So we have the following generalization of Corollary~\ref{cor_huang}.

\begin{cor} 
Under GRH, there exists a deterministic algorithm that, given a polynomial $f(X)\in\F_q[X]$    and a lifted polynomial $\tilde{f}(X)\in A_0[X]$ of $f$ such that $\gal(\tilde{f}/K_0)$ is abelian, computes the complete factorization of $f$ over $\F_q$ in polynomial time.
\end{cor}

Suppose only the polynomial $f$ is known. Let $n=\deg(f)$. We can  efficiently compute a lifted polynomial $\tilde{f}(X)\in A_0[X]$ of $f$ whose size is polynomial in $n$ and $\log q$.\footnote{We also need to   choose $\tilde{h}(Y)\in\Z[Y]$, $h(Y)=\tilde{h}(Y)\bmod p\in\F_p[Y]$ and $\psi_0:\F_p[Y]/(h(Y))\to \F_q$ first, so that  $A_0=\Z[Y]/(\tilde{h}(Y))$, $K_0=\Q[Y]/(\tilde{h}(Y))$ and the notion of lifted polynomials are defined. The isomorphism $\psi_0$ can be efficiently computed  by \citep{Len91}.}
Reduce to the case that $\tilde{f}$ is irreducible over $K_0$ as above. As   $\gal(\tilde{f}/K_0)$ is a subgroup of $\sym(n)$, we have the following generalization of Theorem~\ref{thm_ron87}.

\begin{thm}\label{thm_ron87g}
Under GRH, there exists a deterministic algorithm that, given a polynomial $f(X)\in\F_q[X]$ of degree $n\in\N^+$, computes the complete factorization of $f$ in time polynomial in $n!$ and $\log q$.
\end{thm}

Now suppose we lift $f$ to $\tilde{f}$, reduce to the case that $\tilde{f}$ is irreducible over $K_0$, but compute $\mathcal{F}$ using  Lemma~\ref{lem_compstabsysg} instead. By Corollary~\ref{cor_algdgboundg} and Lemma~\ref{lem_logbound}, we have the following generalization of Theorem~\ref{thm_evd94}.
%% instead of  Lemma~\ref{lem_compsplit}
%so that the associated subgroup system $\mathcal{P}$ over $\gal(\tilde{f}/K_0)$ is the system of stabilizers of depth $m$ for some $m\in \N^+$ (with respect to the action of $\gal(\tilde{f}/K_0)$ on
%the set of roots of $\tilde{f}$ in $L$). Here $m$ is chosen to be large enough so that all strongly antisymmetric $\mathcal{P}$-schemes are discrete on $G_\alpha$ for any root $\alpha$ of $\tilde{f}$ in $L$. By Lemma~\ref{lem_logbound}, we may choose $m=O(\log n)$. Then by Theorem~\ref{thm_algmain2formalg}, we have following generalization of Theorem~\ref{thm_evd94}.

\begin{thm}
Under GRH, there exists a deterministic algorithm that, given a polynomial $f(X)\in\F_q[X]$ of degree $n\in\N^+$, computes the complete factorization of $f$ over $\F_q$ in time polynomial in $n^{\log n}$ and $\log q$.
\end{thm}

We also have the following theorem that  generalizes Theorem~\ref{thm_algsolvable} and the main result of  \citep{Evd92}. The proof is the same as that of  Theorem~\ref{thm_algsolvable}, except that Theorem~\ref{thm_algmain2formalg} is used instead of Theorem~\ref{thm_algmain2formal}, and the base field  $\Q$ is replaced by $K_0$.

\begin{thm} \label{thm_algsolvableg}
Under GRH, there exists a  deterministic polynomial-time  algorithm that, given a polynomial $f(X)\in\F_q[X]$   and a  lifted polynomial $\tilde{f}(X)\in A_0[X]$ of $f$ whose Galois group $\gal(\tilde{f}/K_0)$ is solvable, computes the complete factorization of $f$ over $\F_q$.
\end{thm}

\paragraph{Computing a proper factorization of $f$.}\index{proper!factorization}

Unlike the special case considered in Chapter~\ref{chap_alg_prime},    replacing discreteness by inhomogeneity in the second condition of Theorem~\ref{thm_algmain2formalg} does not automatically yield an algorithm computing a proper factorization of $f$. The reason is that even if a $(\mathcal{C},\mathcal{D}_{\mathfrak{Q}_0})$-separated $\mathcal{P}$-scheme $\tilde{\mathcal{C}}$ is inhomogeneous on a subgroup $H\in\mathcal{P}$, the corresponding $\mathcal{P}$-scheme of double cosets $\mathcal{C}$ may still be homogeneous on $H$. In fact, this is always the case  
when $H\backslash G/\mathcal{D}_{\mathfrak{Q}_0}$ is a singleton, or equivalently, when $\mathcal{D}_{\mathfrak{Q}_0}$ acts transitively on 
$H\backslash G$ by inverse right translation.

Still, by adapting the condition, we obtain some results on computing a proper factorization of $f$:
\begin{itemize}
\item We formulate a new condition on $\mathcal{P}$-schemes and use it to obtain
 algorithms computing one irreducible factor of $f$. See Theorem~\ref{thm_algmain2relax} and Corollary~\ref{cor_algmain2singleton}.
%They can be seen as the analogues of Berlekamp's reduction from polynomial factoring to root-finding of special polynomials \citep{Ber70}.

\item  We formulate conditions on $\mathcal{P}$ that involve not only ordinary $\mathcal{P}$-schemes but also $\mathcal{P}$-schemes of double cosets, and these conditions  can be used for computing the complete factorization as well as a proper factorization. See Theorem~\ref{thm_algmain2relax}. 

\item Finally,  we prove a generalization of  Lemma~\ref{lem_integrality_pscheme} for $\mathcal{P}$-schemes of double cosets, and use it to prove a generalization of Theorem~\ref{thm_boundbyell}. See Theorem~\ref{thm_boundbyellg}.
\end{itemize}

First, we introduce the following definition.
\begin{defi}\label{defi_havesingleton}
For a $\mathcal{P}$-scheme (resp. $\mathcal{P}$-scheme of double cosets) $\mathcal{C}=\{C_H: H\in\mathcal{P}\}$ and $H\in\mathcal{P}$, we say $\mathcal{C}$ {\em has a singleton}\index{having a singleton} on $H$ if the partition $C_H$  has a block that is a singleton.
\end{defi}

The following theorem is a variant of Theorem~\ref{thm_algmain2formalg} with weakened conditions on the subgroup system $\mathcal{P}$.

\begin{thm}\label{thm_algmain2relax}
Suppose there exists a deterministic algorithm that, given a polynomial $g(X)\in K_0[X]$ irreducible over $K_0$, constructs  a $(K_0,g)$-subfield system $\mathcal{F}$ in time $T(g)$ such that
\begin{itemize}
\item $K_0[X]/(g(X))$ is in $\mathcal{F}$, and
\item for some $H\in\mathcal{P}$ satisfying $(L(g))^H\cong_{K_0} K_0[X]/(g(X))$, all strongly antisymmetric $\mathcal{P}$-schemes of double cosets $\mathcal{C}$ with respect to $\mathcal{D}_{\mathfrak{Q}_0}$ 
that admit a $(\mathcal{C},\mathcal{D}_{\mathfrak{Q}_0})$-separated strongly antisymmetric $\mathcal{P}$-scheme
are discrete (resp. are inhomogeneous, have a singleton) on $H$,\footnote{Here $\mathcal{D}_{\mathfrak{Q}_0}$ is the decomposition group of a fixed prime ideal $\mathfrak{Q}_0$ of $\ord_{L(g)}$ lying over $p$. Different choices of $\mathfrak{Q}_0$  lead to conjugate subgroups and hence do not matter.} where $\mathcal{P}$ is the subgroup system over $\gal(g/K_0)$ associated with $\mathcal{F}$, and $L(g)$ is the splitting field of $g$ over $K_0$.  
\end{itemize}
Then under GRH, there exists a deterministic algorithm that, given a polynomial $f(X)\in\F_q[X]$ and an irreducible  lifted polynomial $\tilde{f}(X)\in A_0[X]$ of $f$,  outputs the complete factorization (resp. a proper factorization, an irreducible factor) of $f$ over $\F_q$ in time polynomial in $T(\tilde{f})$ and the size of the input.
\end{thm}

\begin{proof}
The proof is the almost same as that of Theorem~\ref{thm_algmain2formalg}. The second condition in the theorem are used to show $C_H=\infty_{\mathcal{D}_{\mathfrak{Q}_0}}$ (resp. $C_H\neq 0_{\mathcal{D}_{\mathfrak{Q}_0}}$, $C_H$ has a singleton) for some $H\in\mathcal{P}$ satisfying $L^H\cong_{K_0} K_0[X]/(\tilde{f}(X))$, and hence the corresponding idempotent decomposition is complete (resp. is proper, has a singleton). Then apply Theorem~\ref{thm_extfactorredg}. The details are left to the reader.
\end{proof}

Observe that  if a $\mathcal{P}$-scheme of double cosets $\mathcal{C}$ has a singleton on $H$, then a $(\mathcal{C},\mathcal{D}_{\mathfrak{Q}_0})$-separated $\mathcal{P}$-scheme also has a singleton on $H$. So we have the following corollary, which is an analogue of Theorem~\ref{thm_algmain2formalg}.

\begin{cor}\label{cor_algmain2singleton}
Suppose there exists a deterministic algorithm that, given a polynomial $g(X)\in K_0[X]$ irreducible over $K_0$, constructs  a  $(K_0,g)$-subfield system $\mathcal{F}$ in time $T(g)$ such that
\begin{itemize}
\item $K_0[X]/(g(X))$ is in $\mathcal{F}$, and
\item  for some $H\in\mathcal{P}$ satisfying $(L(g))^H\cong_{K_0} K_0[X]/(g(X))$, all strongly antisymmetric $\mathcal{P}$-schemes have a singleton on $H$, where $\mathcal{P}$ is the subgroup system over $\gal(g/K_0)$ associated with $\mathcal{F}$ and $L(g)$ is the splitting field of $g$ over $K_0$.
\end{itemize}
Then under GRH, there exists a deterministic algorithm that, given a polynomial $f(X)\in\F_q[X]$ and an irreducible lifted polynomial $\tilde{f}(X)\in A_0[X]$ of $f$, outputs  an irreducible factor of $f$ over $\F_q$ in time polynomial in $T(\tilde{f})$ and the size of the input.
\end{cor}

Finally, we give a generalization of Theorem~\ref{thm_boundbyell}:

\begin{thm}\label{thm_boundbyellg}
Under GRH, there exists a deterministic algorithm that, given a polynomial $f(X)\in\F_q[X]$ of degree $n\in\N^+$ that has $k>1$ irreducible factors over $\F_q$, computes a proper factorization of $f$ in time polynomial in $n^\ell$ and $\log q$, where $\ell$ is the least prime factor of $k$.
\end{thm}

To prove Theorem~\ref{thm_boundbyellg}, we need the following generalization of  Lemma~\ref{lem_integrality_pscheme}, whose proof is deferred to Appendix~\ref{chap_omitted2}.

\begin{restatable}{lem}{lemintegralitypschemeg}\label{lem_integrality_pschemeg}
Let $G$ be a finite group acting transitively on a set $S$. Let $\mathcal{D}$ be a subgroup of $G$  and let $k$ be the number of $\mathcal{D}$-orbits in $S$. 
Suppose $k>1$. Let $\ell\in\N^+$ be the least prime factor of $k$.
Let $\mathcal{P}=\mathcal{P}_{m}$ be the system of stabilizers of depth $m$ for some $m\geq \ell$ (with respect to the action of $G$ on $S$).
Then for any   $x\in S$ and any $\mathcal{P}$-scheme of double cosets $\mathcal{C}$ with respect to  $\mathcal{D}$ that is homogeneous on $G_x$, there exists no  antisymmetric $(\mathcal{C},\mathcal{D})$-separated $\mathcal{P}$-scheme.
\end{restatable}

\begin{proof}[Proof of Theorem~\ref{thm_boundbyellg}]
We may assume the irreducible factors of $f$ over $\F_q$ are all distinct and have the same degree $d$, since otherwise a proper factorization of $f$ can be found by square-free factorization \citep{Yun76, Knu98} or distinct-degree factorization \citep{CZ81}.
Compute $d$ as the smallest positive integer for which the automorphism $x\mapsto x^{q^d}$ fixes $\F_q[X]/(f(X))$. Then compute $k=n/d$ and $\ell$.

As in the proof of Theorem~\ref{thm_ron87g}, we choose a lifted polynomial $\tilde{f}\in A_0[X]$ of $f$ whose size is polynomial in $n$ and $\log q$, and reduce to the case that $\tilde{f}$ is irreducible over $K_0$.
Use  Lemma~\ref{lem_compstabsysg} to compute $\mathcal{F}$  so that the associated subgroup system $\mathcal{P}$ is the system of stabilizers  of depth $\ell$ with respect to the action of $\gal(\tilde{f}/K_0)$ on the set of roots of $\tilde{f}$ in $L$. 
This step takes time polynomial in $c(\mathcal{P})$ and the size of $\tilde{f}$, which is polynomial in $n^{\ell}$ and $\log q$.
The theorem then follows from  Theorem~\ref{thm_algmain2relax} and Lemma~\ref{lem_integrality_pschemeg}.
\end{proof}

\begin{rem}
We  may also derive  Theorem~\ref{thm_boundbyellg} from Theorem~\ref{thm_boundbyell} by reducing to the case that $f$ satisfies Condition~\ref{cond_spoly}:  by square-free factorization, we may assume $f$ is square-free. Compute the subring $R$ of $\F_q[X]/(f(X))$ fixed by the Frobenius automorphism $x\mapsto x^p$ over $\F_p$.
Then find an element $z\in R$ such that the minimal polynomial $g$ of $z$ over $\F_p$ is a degree-$k$ polynomial satisfying Condition~\ref{cond_spoly}. Such an element $z$ exists if $p\geq k$.
Then reduce to the problem of computing a proper factorization of $g$ over $\F_p$. We leave the details to the reader.
\end{rem}

\chapter{Constructing new \texorpdfstring{$\mathcal{P}$-schemes}{P-schemes} from old ones}\label{chap_common}

In the previous chapters, we developed a framework for polynomial factoring whose correctness relies on combinatorial properties of $\mathcal{P}$-schemes. Motivated by it, we continue our study on $\mathcal{P}$-schemes in this chapter and also in subsequent chapters. Techniques introduced in this chapter have a common theme, namely constructing  new $\mathcal{P}$-schemes from old ones. Such techniques include
\begin{itemize}
\item Inverse right translation on the set of $\mathcal{P}$-schemes. 
%We define the action of $G$ on the set of  $\mathcal{P}$-schemes by inverse right translation.
\item Restriction of $\mathcal{P}$-schemes to a subgroup, and its analogue for $m$-schemes.
%Given a subgroup system $\mathcal{P}$ over a group $G$, we define the restriction of $\mathcal{P}$ to a subgroup $G'\subseteq G$, which is a $\mathcal{P}$-scheme for some subgroup system $\mathcal{P}'$ over $G'$.
\item Passing to quotient groups. 
\item Induction of  $\mathcal{P}$-schemes.
\item Extension of $\mathcal{P}$-schemes to the closure of $\mathcal{P}$.
\item Restriction of $m$-schemes to a subset, and its generalization for $\mathcal{P}$-schemes.
\item Constructing primitive $m$-schemes from a general one.
\item Direct products and wreath products.
%Given a group $G$ and a subgroup system $\mathcal{P}$ over a quotient group $\bar{G}$ of $G$, we show that a $\mathcal{P}$-scheme can be constructed from a $\mathcal{P}'$-scheme, where $\mathcal{P}'$ is some subgroup system over $G$.
\end{itemize}
The first three of them are introduced in Section~\ref{sec_common_crq}. We use them to prove Lemma~\ref{lem_aggregation}, as promised in Chapter~\ref{chap_constructnum}.

In Section~\ref{sec_induction}, we discuss the {\em induction} of $\mathcal{P}$-schemes. 
For $G'\subseteq G$ and a subgroup system $\mathcal{P}$ over $G$, this operation produces a $\mathcal{P}$-scheme from a $\mathcal{P}'$-scheme, where $\mathcal{P}'$ is a certain subgroup system over $G'$. We apply this operation in Section~\ref{sec_schconj} to establish reductions among a family of conjectures concerning $\mathcal{P}$-schemes, whose resolution would imply that polynomial factoring over finite fields can be solved in deterministic polynomial time under GRH if an irreducible lifted polynomial with a special Galois group is given. See below for a more detailed discussion on these conjectures.

The rest of the above list is discussed in Section~\ref{sec_extension}--\ref{sec_dwproduct}. In particular, we discuss {\em primitivity} of homogeneous $m$-schemes in Section~\ref{sec_primitiveorbit}. By exploiting the connection between homogeneous primitive orbit $m$-schemes and primitive permutation groups, we prove that for $m\geq 3$, every antisymmetric homogeneous orbit $m$-scheme on a finite set $S$ where $|S|>1$ has a matching. Previously this was  known for $m\geq 4$, as proved in \citep{IKS09}.

\paragraph{Schemes conjectures.}
The work \citep{IKS09} proposed a conjecture on $m$-schemes called  the {\em schemes conjecture}. 
\begin{conj*}[schemes conjecture]\index{schemes conjecture}
There exists a constant $m\in \N^+$ such that  every antisymmetric homogeneous $m$-scheme on a finite set $S$ where $|S|>1$ has a matching.
\end{conj*}

 Assuming this conjecture (and GRH), polynomial factorization over finite fields can be solved in deterministic polynomial time, as shown in \citep{IKS09}. We reprove this result in Section~\ref{sec_schconj} using a $\mathcal{P}$-scheme algorithm.  
 
 For each family  $\mathcal{G}$ of finite permutation groups, we also formulate an  analogous conjecture, called the {\em schemes conjecture for $\mathcal{G}$}, in terms of the notation $d(G)$ introduced in Definition~\ref{defi_dg}.
\begin{conj*}[schemes conjecture for $\mathcal{G}$]\index{schemes conjecture!for permutation groups}
There exists a constant $m\in \N^+$ such that $d(G)\leq m$ for all $G\in\mathcal{G}$.
\end{conj*}
We show that assuming this conjecture (and GRH), a polynomial $f$ over finite fields can be factorized in deterministic polynomial time if we are also given an irreducible lifted polynomial $\tilde{f}$ of $f$ whose Galois group  is in $\mathcal{G}$ (as a permutation group  on the set of roots of $\tilde{f}$).

  Using induction of $\mathcal{P}$-schemes, we establish reductions among these conjectures for various families $\mathcal{G}$, so that the schemes conjecture for  $\mathcal{G}$ reduces to that for $\mathcal{G}'$ if the permutation groups in $\mathcal{G}$ are ``less complex'' than those in $\mathcal{G}'$. In particular, all these conjectures reduce to the one for the family of symmetric groups, and the latter turns out to be equivalent to (a slight relaxation of) the original schemes conjecture. 
In summary, the schemes conjectures for various families of finite permutation groups form a hierarchy of relaxations of the original schemes conjecture.
  
 Therefore, in order to prove the original schemes conjecture, it is necessary to prove our analogous conjectures for families of less complex permutation groups. On the other hand, one may hope that progress on the latter would shed some light on the original conjecture.
 We will follow this approach in subsequent chapters and prove some nontrivial results.
  
\section{Basic operations on \texorpdfstring{$\mathcal{P}$-schemes}{P-schemes}}\label{sec_common_crq}

In this section, we introduce some basic operations on $\mathcal{P}$-schemes, including inverse right translation, restriction, and passing to quotient groups. We then use them to prove Lemma~\ref{lem_aggregation}.

\paragraph{Inverse right translation of $\mathcal{P}$-schemes.}\index{inverse right translation!of $\mathcal{P}$-schemes} Let $\mathcal{P}$ be a subgroup system over a finite group $G$. 
For each $H\in\mathcal{P}$, the group $G$ acts on $H\backslash G$ by inverse right translation $\prescript{g}{}{Hh}=Hhg^{-1}$. This action induces an action of $G$ on the set of partitions of $H\backslash G$, defined by $\prescript{g}{}{P}=\{\prescript{g}{}{B}: B\in P\}$ for a partition $P$ of $H\backslash G$. Then $G$ also acts on the set of $\mathcal{P}$-collections by inverse right translation:

\begin{defi}
The action of $G$ on the set of $\mathcal{P}$-collections by inverse right translation is defined as follows: for a $\mathcal{P}$-collection $\mathcal{C}=\{C_H:H\in\mathcal{P}\}$ and $g\in G$, define $\prescript{g}{}{\mathcal{C}}=\{\prescript{g}{}{C_H}:H\in\mathcal{P}\}$.
\end{defi}

\begin{lem}\label{lem_invtrans}
For a $\mathcal{P}$-scheme $\mathcal{C}$ and $g\in G$, the $\mathcal{P}$-collection $\prescript{g}{}{\mathcal{C}}$ is also a $\mathcal{P}$-scheme. Moreover, if $\mathcal{C}$ is antisymmetric (resp. strongly antisymmetric), so is $\prescript{g}{}{\mathcal{C}}$.
\end{lem}

\begin{proof}
This follows in a straightforward manner from $G$-equivariance of projections and conjugations (see Lemma~\ref{lem_maps}).
\end{proof}

So $G$ also acts on the set of $\mathcal{P}$-schemes by inverse right translation, which preserves antisymmetry and strong antisymmetry.

\paragraph{Restriction to a subgroup.}

We define the {\em restriction} of a subgroup system $\mathcal{P}$ over $G$ and that of $\mathcal{P}$-collections to a subgroup of $G$.

\begin{defi}[restriction]\label{defi_res}
Let $\mathcal{P}$ be a subgroup system over a finite group $G$. For a subgroup $G'$ of $G$, define 
\[
\mathcal{P}|_{G'}:=\{H\in\mathcal{P}: H\subseteq G'\},
\]
which is a subgroup system over $G'$, called the {\em restriction}\index{restriction!of a subgroup system} of $\mathcal{P}$ to $G'$.

Let $\mathcal{C}=\{C_H: H\in\mathcal{P}\}$ be a $\mathcal{P}$-collection. For $H\in\mathcal{P}|_{G'}$, regard $H\backslash G'$ as a subset of $H\backslash G$ in the obvious way. Then the partition $C_H$ of $H\backslash G$ restricts to a partition of $H\backslash G'$, denoted by $C_H|_{G'}$. Define 
\[
\mathcal{C}|_{G'}:=\{C_H|_{G'}: H\in\mathcal{P}|_{G'}\}
\]
which is a  $\mathcal{P}|_{G'}$-collection, called the {\em restriction}\index{restriction!of a $\mathcal{P}$-collection} of $\mathcal{C}$ to $G'$.
\end{defi}
\nomenclature[e1a]{$\mathcal{P}\vert_{H}$}{restriction of a subgroup system $\mathcal{P}$ to $H$}
\nomenclature[e1b]{$\mathcal{C}\vert_{H}$}{restriction of a $\mathcal{P}$-collection $\mathcal{C}$ to $H$}

Next we show that when $\mathcal{C}$ is $\mathcal{P}$-scheme, its restriction $\mathcal{C}|_{G'}$ to a subgroup $G'$ is a $\mathcal{P}|_{G'}$-scheme. Moreover, antisymmetry and strong antisymmetry are preserved by restriction.

\begin{lem}\label{lem_res}
Let $\mathcal{P}$ be a subgroup system over a finite group $G$. For a subgroup $G'$ of $G$ and a $\mathcal{P}$-scheme $\mathcal{C}$,
the restriction $\mathcal{C}|_{G'}$ is a $\mathcal{P}|_{G'}$-scheme. Moreover, if $\mathcal{C}$ is antisymmetric (resp. strongly antisymmetric), so is $\mathcal{C}|_{G'}$.
\end{lem}

\begin{proof}
We have projections $\pi_{H,H'}$ and conjugations $c_{H,g}$ defined between coset spaces $H\backslash G$ for various subgroups $H\subseteq G$. And we also have projections and conjugations between coset spaces $H\backslash G'$ for  $H\subseteq G'$. We use $\pi'_{H,H'}$ and $c'_{H,g}$ for the latter maps to distinguish them from the former.

For each $H\in \mathcal{P}|_{G'}$, we have a projection $\pi_{H,G'}: H\backslash G\to G'\backslash G$.
This allows us to partition $H\backslash G$  into ``fibers'' of  $\pi_{H,G'}$, i.e., preimages of elements in $G'\backslash G$:
\[
H\backslash G=\coprod_{y\in G'\backslash G} \pi_{H,G'}^{-1}(y).
\]
We say $x\in H\backslash G$ is in the $y$-fiber if $\pi_{H,G'}(x)=y$, and $y$ is called the index of $x$. Note that the subset $H\backslash G'\subseteq H\backslash G$ is precisely the $y$-fiber with $y=G'e\in G'\backslash G$.

Consider $H,H'\in  \mathcal{P}|_{G'}$ and a map $\tau: H\backslash G\to H'\backslash G$ that is either a projection $\pi_{H,H'}$, or a conjugation $c_{H,g}$ for some $g\in G'$ satisfying $H'=gHg^{-1}$. We claim $\pi_{H,G'}=\pi_{H',G'}\circ\tau$, i.e., the map $\tau$ preserves the indices of elements. This can be checked directly:
for $Hh\in H\backslash G$, we have $\pi_{H,G'}(Hh)=G'h$. If $\tau=\pi_{H,H'}$, we have $\pi_{H',G'}\circ\tau(Hh)=\pi_{H',G'}(H'h)=G'h$. And if $\tau=c_{H,g}$ with $g\in G'$, we have $\pi_{H',G'}\circ\tau(Hh)=\pi_{H',G'}(H'gh)=G'gh=G'h$. So the claim holds.

This means the map $\tau$ is also fibered over $G'\backslash G$ such that its ``$y$-fiber'' $\tau_y:=\tau|_{\pi_{H,G'}^{-1}(y)}$ maps the $y$-fiber of $H\backslash G$ to the $y$-fiber of $H'\backslash G$. Setting $y=G'e$ gives us the map $\tau_y: H\backslash G'\to H'\backslash G'$ that is either the projection $\pi'_{H,H'}$, or the conjugation $c'_{H,g}$. 

From this observation it is easy to see that compatibility,  invariance, and regularity of $\mathcal{C}|_{G'}$ follows from the corresponding properties of $\mathcal{C}$: fix $y=G'e$. Assume to the contrary that $\mathcal{C}|_{G'}$ does not satisfy compatibility. Then some projection $\tau_y=\pi'_{H,H'}$ maps two elements in the same block of $C_{H}|_{G'}$ into different blocks of $C_{H'}|_{G'}$. But then $\tau=\pi_{H,H'}$ also maps these two elements that are in the same block of $C_H$ into different blocks of $C_{H'}$, contradicting compatibility of $\mathcal{C}$. Invariance is proved in the same way except that we consider  conjugations instead of projections. For regularity, note that for each projection $\pi'_{H,H'}: H\backslash G'\to H'\backslash G'$ and blocks $B\in C_{H}|_{G'}$, $B'\in C_{H'}|_{G'}$, we have $B=\tilde{B}\cap (H\backslash G')$, $B'=\tilde{B}'\cap (H\backslash G')$ where $\tilde{B}\in C_H$, $\tilde{B}'\in C_{H'}$. And for $z\in B'$ we have $\pi'^{-1}_{H,H'}(z)\cap B=\pi^{-1}_{H,H'}(z)\cap \tilde{B}\cap (H\backslash G')=\pi^{-1}_{H,H'}(z)\cap \tilde{B}$. Regularity of $\mathcal{C}|_{G'}$ then follows from regularity of $\mathcal{C}$.

 Now assume  $\mathcal{C}|_{G'}$ is not antisymmetric. Then for some $H\in \mathcal{P}|_{G'}$ and $g\in N_{G'}(H)$, the map $c'_{H,g}$ restricts to a nontrivial permutation of some block $B\in C_{H}|_{G'}$. Then we have $g\in N_G(H)$ and $c_{H,g}$ restricts to a nontrivial permutation of  $\tilde{B}$, where $\tilde{B}$ is the block of $C_{H}$ satisfying $\tilde{B}\cap (H\backslash G')=B$. So $\mathcal{C}$ is not antisymmetric.
 
 Finally, assume $\mathcal{C}|_{G'}$ is not strongly antisymmetric. Then there exists a nontrivial permutation $\tau$ of a block $B\in C_{H}|_{G'}$ for some subgroup $H\in \mathcal{P}|_{G'}$ such that $\tau$ is a composition of maps $\sigma_i: B_{i-1}\to B_i$, $i=1\dots,k$, where each $B_i$ is a block of $C_{H_i}|_{G'}$, $H_i\in \mathcal{P}|_{G'}$, and $\sigma_i$ is of the form $c'_{H_{i-1}, g}|_{B_{i-1}}$ (where $g\in G'$), $\pi'_{H_{i-1}, H_i}|_{B_{i-1}}$, or  $(\pi'_{H_{i}, H_{i-1}}|_{B_{i}})^{-1}$ (see Definition~\ref{defi_strongasym}). Each block $B_i$ is of the form $\tilde{B}_i\cap (H_i\backslash G')$ for some $\tilde{B}_i\in C_{H_i}$. In the case that $\sigma_i$ is of the form $(\pi'_{H_{i}, H_{i-1}}|_{B_{i}})^{-1}$, we know $|\pi'^{-1}_{H_{i}, H_{i-1}}(z)\cap B_{i}|=|\pi^{-1}_{H_{i},H_{i-1}}(z)\cap \tilde{B}_{i}|=1$ for all $z\in B_{i-1}$. So $(\pi_{H_{i},H_{i-1}}|_{B_{i}})^{-1}$ is well defined. Then $\tau=\tilde{\tau}|_B$ for the nontrivial permutation $\tau=\sigma_k\cdots\circ \sigma_1$ of the block $\tilde{B}=\tilde{B}_0\in C_H$, where each map $\tilde{\sigma}_i$  is of the form $c_{H_{i-1}, g}|_{\tilde{B}_{i-1}}$, $\pi_{H_{i-1}, H_i}|_{\tilde{B}_{i-1}}$, or  $(\pi_{H_{i}, H_{i-1}}|_{\tilde{B}_{i}})^{-1}$. So $\mathcal{C}$ is not strongly antisymmetric.
\end{proof}

Next we describe the analogue of  Definition~\ref{defi_res} for $m$-schemes.

\begin{defi}\label{defi_mres}
Let $\Pi=\{P_1,\dots,P_m\}$ be an $m$-scheme on a finite set $S$.
For $(x_1,\dots,x_k)\in S^{(k)}$ where $k<m$, define the   $(m-k)$-collection 
\[
\Pi|_{x_1,\dots,x_k}:=\{P'_1,\dots,P'_{m-k}\}
\]
on the set $S-\{x_1,\dots,x_k\}$, where $P'_i$ is the partition of $S^{(i)}$ such that two elements $(y_1,\dots,y_i), (y'_1,\dots,y'_i)$ are in the same block of $S^{(i)}$ iff $(x_1,\dots,x_k, y_1,\dots,y_i)$ and $(x_1,\dots,x_k, y'_1,\dots,y'_i)$ are in the same block of $S^{(i+k)}$.
\end{defi}
\nomenclature[e1c]{$\Pi\vert_{x_1,\dots,x_k}$}{See Definition~\ref{defi_mres}}

We also  have the analogue of Lemma~\ref{lem_res} for $m$-schemes.

\begin{lem}\label{lem_mres}
The $(m-k)$-collection $\Pi|_{x_1,\dots,x_k}$ in Definition~\ref{defi_res} is an $(m-k)$-scheme. 
Moreover, if $\Pi$ is antisymmetric (resp. strongly antisymmetric), so is $\Pi|_{x_1,\dots,x_k}$.
And if $\Pi$ does not have a matching, neither does $\Pi|_{x_1,\dots,x_k}$.
\end{lem}

The proof is straightforward by definition. Indeed, if we view $\Pi$ as a $\mathcal{P}$-scheme via Definition~\ref{defi_ptom} and Definition~\ref{defi_mtop}, where $\mathcal{P}$ is the system of stabilizers of depth $m$ with respect to the natural action of $G=\sym(S)$ on $S$.
Then $\Pi|_{x_1,\dots,x_k}$ is simply the restriction of this $\mathcal{P}$-scheme to the subgroup $G_{x_1,\dots,x_k}$.
We leave the details to the reader.

\paragraph{Passing to quotient groups.}

Let $G$ be a finite group and let $N$ be a normal in G. Write $\bar{G}$ for $G/N$ and $\phi$ for the quotient map $G\to \bar{G}$.

For a subgroup $H\subseteq \bar{G}$,  the group $G$ acts on $H\backslash\bar{G}$ by inverse right translation (through its quotient group $\bar{G}$). The stabilizer of $He\in H\backslash\bar{G}$ is $\phi^{-1}(H)$. So by Lemma~\ref{lem_equivaction}, we have an equivalence between the action of $G$ on $ H\backslash\bar{G}$ and that on $\phi^{-1}(H)\backslash G$, given by the bijection $\lambda_{He}:  H\backslash\bar{G}\to \phi^{-1}(H)\backslash G$  sending $H\phi(g)$ to $\phi^{-1}(H)g$ for $g\in G$.
%Denote the inverse of this map by $\phi_{H}$, which sends  $\pi^{-1}(H)g$ to $H\pi(g)$ for $g\in G$. We show that 

Let $\mathcal{P}$ be a subgroup system  over $\bar{G}$. Define $\tilde{\mathcal{P}}=\{\phi^{-1}(H): H\in \mathcal{P}\}$,
which is a subgroup system over $G$. By identifying $H\backslash\bar{G}$ with $\phi^{-1}(H)\backslash G$ via $\lambda_{He}$ for $H\in\mathcal{P}$, we see that  a $\mathcal{P}$-scheme over $\bar{G}$ is equivalent to a $\tilde{\mathcal{P}}$-scheme over $G$. This is made formal by the following lemma.

\begin{lem}\label{lem_quoidentify}
Let $\mathcal{P}$ and $\tilde{\mathcal{P}}$ be as above.
For a $\mathcal{P}$-collection $\mathcal{C}=\{C_H: H\in\mathcal{P}\}$, define the $\tilde{\mathcal{P}}$-collection
$\mathcal{C}'=\{C'_{\phi^{-1}(H)}: H\in \mathcal{P}\}$ by choosing
\[
C'_{\phi^{-1}(H)}=\{\lambda_{He}(B): B\in C_H\}.
\]
Then $\mathcal{C}\mapsto \mathcal{C}'$ is  a one-to-one correspondence between $\mathcal{P}$-schemes over $\bar{G}$ and $\tilde{\mathcal{P}}$-schemes over $G$. Moreover, $\mathcal{C}$ is antisymmetric (resp. strongly antisymmetric) iff  $\mathcal{C}'$ is antisymmetric (resp. strongly antisymmetric).
And $\mathcal{C}$ is homogeneous (resp. discrete) on a subgroup $H\in\mathcal{P}$ iff $\mathcal{C}'$ is homogeneous (resp. discrete) on $\phi^{-1}(H)$. 
\end{lem}
\begin{proof}
We check that the maps $\lambda_{He}$ commute with conjugations and projections: write $\pi_{H,H'}$ and $c_{H,g}$ for  conjugations and projections between coset spaces of $\bar{G}$ and write $\pi'_{H,H'}$ and $c'_{H,g}$ for those between coset spaces of $G$. Then we always have 
\[
\lambda_{H'e}\circ\pi_{H,H'}=\pi'_{\phi^{-1}(H),\phi^{-1}(H')}\circ \lambda_{He}
\]
for $H,H'\in\mathcal{P}$, $H\subseteq H'$, and
\[
\lambda_{H'e}\circ c_{H,\phi(g)}=c'_{\phi^{-1}(H),g}\circ \lambda_{He}.
\]
for $H,H'\in\mathcal{P}$, $g\in G$, $H'=\phi(g)H\phi(g)^{-1}$.
Also note that the maps $\lambda_{He}$ are bijections. The lemma then follows easily by definition.
\end{proof}

We conclude this section by proving Lemma~\ref{lem_aggregation} using the results developed above. 
First we prove the following lemma.
\begin{lem}\label{lem_chainreduction}
Let $k\in \N^+$ and $G_k\subseteq G_{k-1}\subseteq\dots\subseteq G_1\subseteq G_0$ be a chain of finite groups.
Let $\mathcal{P}$ be a subgroup system over $G_0$.
We have:
\begin{enumerate}
\item If  for all $i\in [k]$, all strongly antisymmetric $\mathcal{P}|_{G_{i-1}}$-schemes are discrete on $G_i$, then all strongly antisymmetric $\mathcal{P}$-schemes are discrete on $G_k$. 
\item  If  for some $i\in [k]$, all strongly antisymmetric $\mathcal{P}|_{G_{i-1}}$-schemes are inhomogeneous on $G_i$, then all strongly antisymmetric $\mathcal{P}$-schemes are inhomogeneous on $G_k$.  
\end{enumerate} 
The same holds if strong antisymmetry is replaced by antisymmetry.
\end{lem}

\begin{proof}
Assume that there exists a strongly antisymmetric $\mathcal{P}$-scheme $\mathcal{C}=\{C_H: H\in\mathcal{P}\}$ that is not discrete on $G_k$. 
Then there exist two different elements $x,x'\in G_k\backslash G$ lying in the same block of $C_{G_k}$.
Pick the greatest integer $i\in [k]$ satisfying $\pi_{G_k,G_{i-1}}(x)=\pi_{G_k,G_{i-1}}(x')$.
Such $i$ exists as $\pi_{G_k,G_0}(x)=\pi_{G_k,G_0}(x')$.  
Let $y=\pi_{G_k,G_{i}}(x)$ and $y'=\pi_{G_k,G_{i}}(x')$.
Then (1) $y\neq y'$ by maximality of $i$ and the fact that $x\neq x'$,
(2) $y, y'$ are in the same block of $C_{G_{i}}$ by compatibility of $\mathcal{C}$ and the fact that $x,x'$ are in the same block of $C_{G_k}$, 
and (3) $\pi_{G_{i},G_{i-1}}(y)=\pi_{G_{i},G_{i-1}}(y')$ since $\pi_{G_{i},G_{i-1}}(y)=\pi_{G_k,G_{i-1}}(x)$ and $\pi_{G_{i},G_{i-1}}(y')=\pi_{G_k,G_{i-1}}(x')$. 

Suppose  $\pi_{G_{i},G_{i-1}}(y)=\pi_{G_{i},G_{i-1}}(y')=G_{i-1} g$. By replacing $\mathcal{C}$ with $\prescript{g}{}{\mathcal{C}}$ (with respect to the action of $G_k$ on the set of $\mathcal{P}$-schemes by inverse right translation) and applying Lemma~\ref{lem_invtrans}, we may assume $G_{i-1}g=G_{i-1}e$. Then we can write $y=G_{i}h$ and $y'=G_{i}h'$ for some $h,h'\in G_{i-1}$.
By Lemma~\ref{lem_res}, the restriction $\mathcal{C}|_{G_{i-1}}=\{C_H|_{G_{i-1}}: H\in\mathcal{P}|_{G_{i-1}}\}$  is a strongly antisymmetric $\mathcal{P}|_{G_{i-1}}$-scheme. As $y, y'$ are  in the same block of $C_{G_{i}}$, they are also in the same block of $C_{G_{i}}|_{G_{i-1}}$.
As $y\neq y'$, we know $\mathcal{C}|_{G_{i-1}}$ is not discrete on $G_{i}$. This proves the first claim of the lemma.

%Define $\tilde{P}_i:=\{\pi^{-1}_i(H): H\in\mathcal{P}_i\}$ which is a subgroup system over $G_{i-1}$ and contains $G_{i}=\pi_i^{-1}(G_i/N_i)$.
%By definition, we have  $\tilde{P}_i \subseteq \mathcal{P}|_{G_{i-1}}$.
%So  from $\mathcal{C}|_{G_{i-1}}$ we obtain a strongly antisymmetric  $\tilde{P}_i$-scheme that is not discrete on $G_{i}$.
%Then by Lemma~\ref{lem_quoidentify}, there exists a strongly antisymmetric $\mathcal{P}_i$-scheme that is not discrete on $G_i/N_i$. This proves the first claim of the lemma.

For the second claim, assume to the contrary that it does not hold.
Choose $i\in [k]$ such all strongly antisymmetric $\mathcal{P}|_{G_{i-1}}$-schemes are inhomogeneous on $G_i$.
Let  $\mathcal{C}=\{C_H: H\in\mathcal{P}\}$  be a strongly antisymmetric  $\mathcal{P}$-scheme that is homogeneous on $G_k$. By compatibility, we know  $\mathcal{C}$ is  homogeneous on $G_i$.
Then $\mathcal{C}|_{G_{i-1}}$ is also homogeneous on $G_i$. It is also strongly antisymmetric by Lemma~\ref{lem_res}, which contradicts the assumption.

The proof for antisymmetry is the same.
\end{proof}

Now we are ready to prove Lemma~\ref{lem_aggregation}. For convenience, we restate the lemma.

\begin{lem}
Let $k\in \N^+$ and $G_k\subseteq G_{k-1}\subseteq\dots\subseteq G_1\subseteq G_0$ be a chain of finite groups.
For $i\in [k]$, let $N_i$ be a subgroup of $G_i$ that is normal in $G_{i-1}$, $\pi_i: G_{i-1}\to G_{i-1}/N_i$ be the corresponding quotient map, and $\mathcal{P}_i$ be a subgroup system over $G_{i-1}/N_i$ that contains $G_i/N_i$. Define
\[
\mathcal{P}=\{g\pi_i^{-1}(H)g^{-1}: 1\leq i\leq k, H\in \mathcal{P}_i, g\in G_0\},
\]
which is a subgroup system over $G_0$ and contains $\pi_i^{-1}(G_i/N_i)=G_i$ for all $i\in [k]$.
Then we have
\begin{enumerate}
\item If  for all $i\in [k]$, all strongly antisymmetric $\mathcal{P}_i$-schemes are discrete on $G_i/N_i$, then all strongly antisymmetric $\mathcal{P}$-schemes are discrete on $G_k$. 
\item  If  for some $i\in [k]$, all strongly antisymmetric $\mathcal{P}_i$-schemes are inhomogeneous on $G_i/N_i$, then all strongly antisymmetric $\mathcal{P}$-schemes are inhomogeneous on $G_k$. 
\end{enumerate} 
The same holds if strong antisymmetry is replaced by antisymmetry.
\end{lem}
\begin{proof}
Fix $i\in [k]$.
By Lemma~\ref{lem_quoidentify} and the definition of $\mathcal{P}$, if all strongly antisymmetric $\mathcal{P}_i$-schemes are discrete (resp. inhomogeneous) on $G_i/N_i$, then all strongly antisymmetric $\mathcal{P}|_{G_{i-1}}$-schemes are discrete (resp. inhomogeneous) on $G_i$. The same holds if strong antisymmetry is replaced by antisymmetry. The lemma now follows from Lemma~\ref{lem_chainreduction}.
\end{proof}

\section{Induction of \texorpdfstring{$\mathcal{P}$-schemes}{P-schemes}}\label{sec_induction}

Let $G$ be a finite group and let $G'$ be a subgroup of $G$. Let $\mathcal{P}$ be a subgroup system over $G$ 
and let
\[
\mathcal{P}'=\{G'\cap H : H\in \mathcal{P}\},
\]
which is a subgroup system over $G'$. In this section, we show that every $\mathcal{P}'$-scheme  induces a $\mathcal{P}$-scheme in a way that preserves antisymmetry and strong antisymmetry. 
To achieve it, we need the following lemma.

\begin{lem}\label{lem_induction_isom}
Given $g_1,\dots,g_k\in G$ such that $\{g_1^{-1},\dots,g_k^{-1}\}$ is a complete set of representatives of $H\backslash G/G'$, there exists a bijection
\[
\phi: \coprod_{i=1}^k (G'\cap g_i H g_i^{-1}) \backslash G' \to H\backslash G
\]
defined as follows:
For $g\in G$, define the map
\[
\phi_{H,g}: (G'\cap g H g^{-1}) \backslash G' \to H\backslash G
\]
sending $(G'\cap g H g^{-1}) h$ to $Hg^{-1}h$ for $h\in G'$.
The maps $\phi_{H,g}$ are well defined. For $i\in [k]$, the restriction of $\phi$ to  $(G'\cap g_i H g_i^{-1}) \backslash G'$ is $\phi_{H,g_i}$.
\end{lem}

\begin{proof}
Consider the action of $G'$ on $H\backslash G$ by inverse right translation.
For $i\in [k]$, let $O_i=\{Hg_i^{-1}g^{-1}: g\in G'\}$ be the $G'$-orbits of $Hg_i^{-1}\in H\backslash G$. 
Then $\{O_1,\dots,O_k\}$ is the partition of $H\backslash G$ into the $G'$-orbits, i.e., $H\backslash G=\coprod_{i=1}^k O_i$.
Fix $i\in [k]$. The stabilizer of $Hg_i^{-1}$ is $G'\cap g_iHg_i^{-1}$. So by Lemma~\ref{lem_equivaction}, we have an equivalence of actions of $G'$
\[
\lambda_{Hg_i^{-1}}: O_i\to (G'\cap g_i H g_i^{-1}) \backslash G'
\] 
sending $\prescript{h}{}{(Hg_i^{-1})}=Hg_i^{-1}h^{-1}$ to $(G'\cap g_i H g_i^{-1})h^{-1}$ for $h\in G'$.
The inverse of this map is exactly $\phi_{H,g_i}$. As we are allowed to choose $g_i$ to be any $g\in G$, all the maps $\phi_{H,g}$ are well defined.
\end{proof}

For each $H\in\mathcal{P}$, the subgroups $G'\cap g H g^{-1}$ are  in $\mathcal{P}'$ for all  $g\in G$. By Lemma~\ref{lem_induction_isom}, we can combine partitions of $G'\cap g_i H g_i^{-1}\backslash G'$, $i=1,\dots,k$, into a partition of $H\backslash G$. This leads to the following definition.

\begin{defi}[induction]\label{defi_pschemeind}
Let $G$, $G'$, $\mathcal{P}$ and $\mathcal{P}'$ be as above. Let $\mathcal{C}'=\{C'_H: H\in\mathcal{P}'\}$ be a $\mathcal{P}'$-scheme. For $H\in\mathcal{P}$, choose   $g_1,\dots,g_k\in G$ such that $\{g_1^{-1},\dots,g_k^{-1}\}$ is a complete set of representatives of $H\backslash G/G'$. Define the partition $C_H$ of $H\backslash G$ by
\[
C_H=\left\{\phi_{H,g_i}(B): i\in [k], B\in C'_{G'\cap g_i H g_i^{-1}}\right\},
\]
where the maps $\phi_{H,g_i}$ are as in Lemma~\ref{lem_induction_isom}, i.e., each $\phi_{H,g_i}$ sends $(G'\cap g_i H g_i^{-1})h$ to $Hg_i^{-1}h$ for $h\in G'$.  Define the $\mathcal{P}$-collection  $\mathcal{C}=\{C_H: H\in\mathcal{P}\}$, called the {\em induction}\index{induction of a $\mathcal{P}$-scheme} of $\mathcal{C}'$ to $\mathcal{P}$.
\end{defi}

 The $\mathcal{P}$-collection  $\mathcal{C}$ constructed as above is indeed a $\mathcal{P}$-scheme:

\begin{thm}\label{thm_pschmind}
The $\mathcal{P}$-collection  $\mathcal{C}$ in Definition~\ref{defi_pschemeind} is a well defined  $\mathcal{P}$-scheme, which does not depend on the choices of the elements $g_i$. Moreover, if $\mathcal{C}'$ is antisymmetric (resp. strongly antisymmetric), so is $\mathcal{C}$.
\end{thm}

\begin{proof}
 Fix $H\in\mathcal{P}$. It follows from Lemma~\ref{lem_induction_isom} that $C_H$ is indeed a partition of $H\backslash G$. We need to show that $C_H$ is independent of the choices of the elements $g_1,\dots,g_k$. So consider $g'_1,\dots,g'_k\in G$ such that $\{g'^{-1}_1,\dots,g'^{-1}_k\}$ is a complete set of representatives of $H\backslash G/G'$ as well. We want to show
 \[
 C_H=\left\{\phi_{H,g'_i}(B): i\in [k], B\in C'_{G'\cap g'_i H g'^{-1}_i}\right\}.
 \]
As the right hand side is also a partition of $H\backslash G$, it suffices to show that $\phi_{H,g'_i}(B)\in C_H$ for $i\in [k]$ and $B\in  C'_{G'\cap g'_i H g'^{-1}_i}$. Fix $i$ and $B$. Choose $j\in [k]$ such that $Hg_j^{-1}G'=Hg'^{-1}_iG'$. And choose $g\in G'$ such that $Hg^{-1}_j=Hg'^{-1}_ig^{-1}$. We have the conjugation
\[
c_{G'\cap g'_i H g'^{-1}_i, g}: (G'\cap g'_i H g'^{-1}_i)\backslash G' \to (G'\cap g_j H g_j^{-1})\backslash G'
\]
sending $(G'\cap g'_i H g'^{-1}_i)h$ to $(G'\cap g_j H g_j^{-1})gh$ for $h\in G'$. By invariance of $\mathcal{C}'$, the set $c_{G'\cap g'_i H g'^{-1}_i, g}(B)$ is a block of $C'_{G'\cap g_j H g^{-1}_j}$. So $\phi_{H,g_j}\circ c_{G'\cap g'_i H g'^{-1}_i, g}(B)$ is a block of $C_H$.
We claim 
\[
\phi_{H,g_j}\circ c_{G'\cap g'_i H g'^{-1}_i, g}=\phi_{H,g'_i},
\]
which holds since
\begin{align*}
\phi_{H,g_j}\circ c_{G'\cap g'_i H g'^{-1}_i, g}((G'\cap g'_i H g'^{-1}_i)h)&=\phi_{H,g_j}((G'\cap g_j H g_j^{-1})gh)
=Hg_j^{-1}gh\\
&=Hg'^{-1}_i h
=\phi_{H,g'_i}((G'\cap g'_i H g'^{-1}_i)h)
\end{align*}
for $h\in G'$. It follows that $\phi_{H,g'_i}(B)\in C_H$, as desired.  So $C_H$ does not depend on the choices of $g_1,\dots,g_k$.

Next we prove that $\mathcal{C}$ is a $\mathcal{P}$-scheme. To prove compatibility, consider $H,H'\in\mathcal{P}$ satisfying $H\subseteq H'$. For $g\in G$, the following diagram commutes:
\[
\begin{tikzcd}[column sep=4cm]
(G'\cap g H g^{-1}) \backslash G' \arrow[r, "\pi_{G'\cap g H g^{-1}, G'\cap g H' g^{-1}}"] \arrow[d, "\phi_{H,g}"']
& (G'\cap g H' g^{-1})\backslash G'  \arrow[d, "\phi_{H',g}"] \\
H\backslash G  \arrow[r,  "\pi_{H, H'}"]
& H'\backslash G\nospacepunct{.}
\end{tikzcd}
\]
For $B\in C_H$, we want to show that $\pi_{H, H'}(B)$ is contained in a block of $C_{H'}$.
Note
\[
\pi_{H, H'}(B)=\pi_{H, H'}\circ \phi_{H,g}(\tilde{B})=\phi_{H',g}(y)\circ \pi_{G'\cap g H g^{-1}, G'\cap g H' g^{-1}}(\tilde{B}).
\]
Here $\pi_{G'\cap g H g^{-1}, G'\cap g H' g^{-1}}(\tilde{B})$ is contained in a block of $C'_{G'\cap g H' g^{-1}}$ by compatibility of $\mathcal{C}'$, and hence $\pi_{H,H'}(B)$ is contained in a block of $C_{H'}$. It follows that $\mathcal{C}$ is compatible.

For regularity, consider $H,H'$ as above and $B\in C_H$. Choose $B'\in C_{H'}$ containing $\pi_{H,H'}(B)$.
We claim that $\pi_{H,H'}|_B: B\to B'$ has constant degree,  i.e., the number of preimages $|(\pi_{H,H'}|_B)^{-1}(y)|$ is independent of the choices of $y\in B'$. Choose $g\in G$ and $\tilde{B}\in C'_{G'\cap g H g^{-1}}$ such that $B=\phi_{H,g}(\tilde{B})$. Let $\tilde{B}'=\pi_{G'\cap g H g^{-1}, G'\cap g H' g^{-1}}(\tilde{B})$. Then $B'=\phi_{H',g}(\tilde{B}')$. By regularity of $\mathcal{C}'$, the map $\pi_{G'\cap g H g^{-1}, G'\cap g H' g^{-1}}|_{\tilde{B}}: \tilde{B}\to \tilde{B}'$ has constant degree. The claim follows by noting that $\phi_{H,g}|_{\tilde{B}}:\tilde{B}\to B$ and  $\phi_{H',g}|_{\tilde{B}'}:\tilde{B}'\to B'$ are bijective. So $\mathcal{C}$ is regular. 

For invariance, consider $H,H'\in\mathcal{P}$ and $h\in G$ satisfying $H'=hHh^{-1}$. For $g\in G$, we have $G'\cap g H g^{-1}=G'\cap gh^{-1} H' (gh^{-1})^{-1}$, and the following diagram commutes
\[
\begin{tikzcd}[column sep=large]
(G'\cap g H g^{-1}) \backslash G' \arrow[r, "\mathrm{id}"] \arrow[d, "\phi_{H,g}"']
& (G'\cap g H g^{-1})\backslash G'  \arrow[d, "\phi_{H',gh^{-1}}"] \\
H\backslash G  \arrow[r,  "c_{H, h}"]
& H'\backslash G\nospacepunct{,}
\end{tikzcd}
\]
where $\mathrm{id}$ denotes the identity map. It follows that $c_{H,h}$ maps blocks of $C_H$ to blocks of $C_{H'}$. So $\mathcal{C}$ is invariant.

 Now assume $\mathcal{C}$ is not strongly antisymmetric and we prove that $\mathcal{C}'$ is not either. By definition, there exists a nontrivial permutation $\tau=\sigma_k\circ\cdots\circ\sigma_1$ of a block $B\in C_H$ for some $H\in\mathcal{P}$ such that each $\sigma_i: B_{i-1}\to B_i$ is a map of the form $\pi_{H_{i-1}, H_i}|_{B_{i-1}}$,  $(\pi_{H_{i}, H_{i-1}}|_{B_{i}})^{-1}$,  or $c_{H_{i-1}, h}|_{B_{i-1}}$, and $B_i\in C_{H_i}$, $H_i\in \mathcal{P}$, $B=B_0=B_k$, $H=H_0=H_k$ (see Definition~\ref{defi_strongasym}). 
 
 By the two diagrams above, we can choose $g_i\in G$ and $\tilde{B}_i\in C'_{G'\cap g_i H g_i^{-1}}$ for $0\leq i\leq k$, and choose $\tilde{\sigma}_i: \tilde{B}_{i-1}\to \tilde{B}_i$
of the form $\pi_{G'\cap g_{i-1} H_{i-1} g_{i-1}^{-1}, G'\cap g_i H_i g_i^{-1}}|_{\tilde{B}_{i-1}}$, 
$(\pi_{G'\cap g_i H_i g_i^{-1}, G'\cap g_{i-1} H_{i-1} g_{i-1}^{-1}}|_{\tilde{B}_{i}})^{-1}$, 
or the identity map on $\tilde{B}_i$ for $i\in [k]$, such that $\phi_{H_{i}, g_{i}}(\tilde{B}_i)=B_i$ and 
$\sigma_i\circ \phi_{H_{i-1}, g_{i-1}}|_{\tilde{B}_{i-1}}=\phi_{H_{i}, g_{i}}|_{\tilde{B}_{i}}\circ \tilde{\sigma}_i$
 for $i\in [k]$.\footnote{For the case that $\sigma_i=(\pi_{H_{i}, H_{i-1}}|_{B_{i}})^{-1}$, we choose $\tilde{\sigma}_i=(\pi_{G'\cap g_i H_i g_i^{-1}, G'\cap g_{i-1} H_{i-1} g_{i-1}^{-1}}|_{\tilde{B}_{i}})^{-1}$, which is well defined since $\phi_{H_{i-1},g_{i-1}}$ and $\phi_{H_i,g_i}$ are bijective.}
Define $\tilde{\tau}:=\tilde{\sigma}_k\circ\cdots\circ\tilde{\sigma}_1$ which is a map from $\tilde{B}_0$ to $\tilde{B}_k$. Then the following diagram commutes.
\[
\begin{tikzcd}[column sep=large]
\tilde{B}_0 \arrow[r, "\tilde{\tau}"] \arrow[d, "\phi_{H,g_0}|_{\tilde{B}_0}"']
& \tilde{B}_k  \arrow[d, "\phi_{H,g_k}|_{\tilde{B}_k}"] \\
B  \arrow[r,  "\tau"]
& B 
\end{tikzcd}
\]
We have $Hg_0^{-1}G'=Hg_k^{-1}G'$, since otherwise the image of $\phi_{H,g_0}$ and that of $\phi_{H,g_k}$ would be disjoint (see Lemma~\ref{lem_induction_isom}). So $H g_0^{-1}=Hg_k^{-1}g^{-1}$ for some $g\in G'$. The first part of the proof shows that $\phi_{H,g_0}\circ c_{G'\cap g_k H g_k^{-1},g}=\phi_{H,g_k}$.
By composing $\tilde{\tau}$ with $c_{G'\cap g_k H g_k^{-1},g}$, we may assume $g_k=g_0$ and $\tilde{B}_k=\tilde{B}_0$. Then as $\tau$ is a nontrivial permutation of $B$ and $\phi_{H,g_0}|_{\tilde{B}_0}: \tilde{B}_0\to B$ is bijective, we know $\tilde{\tau}$ is a nontrivial permutation of $\tilde{B}_0$. So $\mathcal{C}$ is not strongly antisymmetric.
 
 The proof for antisymmetry is the same except that we only consider maps $\tau$ that are conjugations.
\end{proof}

\begin{cor}\label{cor_inductionsubgroupgel}
Let $G,G',\mathcal{P},\mathcal{P}'$ be as above and let $H$ be a subgroup in $\mathcal{P}$.
\begin{enumerate}
\item Suppose  all antisymmetric $\mathcal{P}$-schemes are discrete on $H$. Then all antisymmetric $\mathcal{P}'$-schemes are discrete on $G'\cap g H g^{-1}$ for all $g\in G$.
\item Suppose all antisymmetric $\mathcal{P}$-schemes are inhomogeneous on $H$, and $G'$ acts transitively on $H\backslash G$ by inverse right translation. Then all antisymmetric $\mathcal{P}'$-schemes are inhomogeneous on $G'\cap g H g^{-1}$ for all $g\in G$.
\end{enumerate}
The same claims hold if antisymmetry is replaced with strong antisymmetry.
\end{cor}

\begin{proof}
We prove the claims by contrapositive.
For the first claim, suppose $\mathcal{C}'=\{C'_{H'}: H'\in\mathcal{P}'\}$ is an antisymmetric $\mathcal{P}'$-scheme that is not discrete on $G'\cap g H g^{-1}$ for some $g\in G$. Choose $B\in C'_{G'\cap g H g^{-1}}$ that is not a singleton.  By Theorem~\ref{defi_pschemeind}, the induced $\mathcal{P}$-scheme $\mathcal{C}=\{C_{H'}:H'\in\mathcal{P}\}$ is  antisymmetric. Moreover, we know $\mathcal{C}$ is not discrete on $H$ since  the block $\phi_{H,g}(B)\in C_{H}$  is not a singleton.

For the second claim, suppose $\mathcal{C}'=\{C'_{H'}: H'\in\mathcal{P}'\}$ is an antisymmetric $\mathcal{P}'$-scheme that is homogeneous on $G'\cap g H g^{-1}$ for some $g\in G$.   By Theorem~\ref{defi_pschemeind}, the induced $\mathcal{P}$-scheme $\mathcal{C}=\{C_{H'}:H'\in\mathcal{P}\}$ is antisymmetric. 
As $G'$ acts transitively on $H\backslash G$, the double coset space $H\backslash G/G'$ has only one double coset $Hg^{-1}G'$, which implies that $\phi_{H,g}: G'\cap g H g^{-1}\backslash G\to H\backslash G$ is surjective.  As  $\mathcal{C}'$ is homogeneous on $G'\cap g H g^{-1}$, we know $\mathcal{C}$ is homogeneous on $H$ .

The proof for strong antisymmetry is the same.
\end{proof}

Now let $S$ be a finite $G$-set and let $G'$ be a subgroup of $G$. Fix $m\in\N^+$ and let $\mathcal{P}=\{G_T: 1\leq |T|\leq m\}$ be the system of stabilizers of depth $m$ with respect to the action of $G$ on $S$. 
Then $\mathcal{P}'$ is exactly the system of stabilizers of depth $m$ with respect to the action of $G'$ on $S$ restricted from that of $G$. Therefore we have:

\begin{cor}\label{cor_inductionsubgroup}
Let $G$ be a finite group acting on a finite set $S$, $G'$ a subgroup of $G$, and $m\in \N^+$. Let $\mathcal{P}$ (resp. $\mathcal{P}'$) be the system of stabilizers of depth $m$ over $G$ (resp. $G'$) with respect to the action of $G$ (resp. $G'$) on $S$. 
\begin{enumerate}
\item Suppose  all antisymmetric $\mathcal{P}$-schemes are discrete on $G_{x}$ for all $x\in S$. Then all  antisymmetric $\mathcal{P}'$-schemes are discrete on $G'_x$ for all $x\in S$.
\item Suppose all antisymmetric $\mathcal{P}$-schemes are inhomogeneous on $G_{x_0}$ for some $x_0\in S$, and $G'$ acts transitively on $S$. Then all antisymmetric $\mathcal{P}'$-schemes are inhomogeneous on $G'_x$ for all $x\in S$.
\end{enumerate}
The same claims hold if antisymmetry is replaced with strong antisymmetry.  
\end{cor}

In particular, we see  $d(G)$ and $d'(G)$ (cf. Definition~\ref{defi_dg}) are monotone with respect to inclusion of permutation groups:
\begin{cor}\label{cor_monotonein}
Let  $G$ be a finite permutation group on a finite set $S$, and let $G'$ be a subgroup of $G$ on $S$. Then $d(G')\leq d(G)$ and $d'(G')\leq d'(G)$.
\end{cor}

%the properties in Corollary~\ref{cor_inductionsubgroup} for an arbitrary finite group $G$ acting on $S$ are subsumed by those for the full symmetric group $\sym(S)$.
%
%\begin{cor}\label{cor_symworst}
%Let  $G$ be a finite group acting on a finite set $S$, and $m\in \N^+$. Let $\mathcal{P}$ (resp. $\mathcal{P}'$) be the system of stabilizers of depth $m$ over $\sym(S)$ (resp. $G$) with respect to the natural action of $\sym(S)$ (resp. the action of $G$) on $S$. 
%\begin{enumerate}
%\item Suppose  all antisymmetric $\mathcal{P}$-schemes are discrete on $\sym(S)_{x_0}$ for some $x_0\in S$. Then all  antisymmetric $\mathcal{P}'$-schemes are discrete on $G_x$ for all $x\in S$.
%\item Suppose all antisymmetric $\mathcal{P}$-schemes are inhomogeneous on $\sym(S)_{x_0}$ for some $x_0\in S$ and in addition $G$ acts transitively on $S$. Then all antisymmetric $\mathcal{P}'$-schemes are inhomogeneous on $G_x$ for all $x\in S$.
%\end{enumerate}
%The same claims hold if antisymmetry is replaced with strong antisymmetry.
%\end{cor}
%
%\begin{proof}
%By Lemma~\ref{lem_quoidentify}, we may assume $G$ acts faithfully on $S$ and hence regard $G$ as a subgroup of $\sym(S)$.
%The claims then follow from Corollary~\ref{cor_inductionsubgroup}. 
%\end{proof}

%We will see applications of the above corollaries in the next section.

We also mention the following variant of Corollary~\ref{cor_inductionsubgroup}, which allows $G'\subseteq G$ to act on a proper subset of $S$.
\begin{cor}\label{cor_inductionsubgroupvar}
Let $G$ be a finite group acting on a finite set $S$, $G'$ a subgroup of $G$, and $m\in \N^+$.
Let $T$ a subset of $S$ such that the action of $G'$ on $S$ fixes $T$ setwisely and $S-T$ pointwisely.
Let $\mathcal{P}$ (resp. $\mathcal{P}'$) be the system of stabilizers of depth $m$ over $G$ (resp. $G'$) with respect to the action of $G$ (resp. $G'$) on $S$ (resp. $T$). 
Suppose  all antisymmetric $\mathcal{P}$-schemes are discrete on $G_{x}$ for all $x\in S$. Then all  antisymmetric $\mathcal{P}'$-schemes are discrete on $G'_x$ for all $x\in T$.
The same claims hold if antisymmetry is replaced with strong antisymmetry.
\end{cor}
\begin{proof}
If $S=T$, the claim holds by Corollary~\ref{cor_inductionsubgroup}. So assume $S\neq T$.
Let $\mathcal{P}''$ be the system of stabilizers of depth $m$ over $G'$ with respect to the action of $G'$ on $S$. Then $\mathcal{P}''=\mathcal{P}'\cup \{G\}$. A $\mathcal{P}'$-scheme $\mathcal{C}$ always extends to a $\mathcal{P}''$-scheme $\mathcal{C}':=\mathcal{C}\cup \{C_G\}$, where $C_G$ is the only partition of the singleton $G\backslash G$, and such an extension clearly preserves antisymmetry and strong antisymmetry.
The claim then follows from  Corollary~\ref{cor_inductionsubgroup}.
\end{proof}

%\begin{rem}
% In \citepp{IKS09}, a deterministic polynomial factoring algorithm was developed based on the notion of $m$-schemes, which unlike our algorithms, does not employ a lifted polynomial $\tilde{f}$. This algorithm may be viewed as a $\mathcal{P}$-scheme algorithm that assumes the worst case $G=\sym(S)$, since in this case a $\mathcal{P}$-scheme (where $\mathcal{P}$ is as in Corollary~\ref{cor_symworst})  is essentially equivalent to an $m$-scheme (see Theorem~\ref{thm_mandp}).
%\end{rem}

\section{Schemes conjectures} \label{sec_schconj}

We investigate the following conjecture proposed in \citep{IKS09}, known as the {\em schemes conjecture}. 

\begin{conj}[schemes conjecture]\label{conj_schconj}\index{schemes conjecture}
There exists a constant $m\in \N^+$ such that  every antisymmetric homogeneous $m$-scheme on a finite set $S$ where $|S|>1$ has a matching.
\end{conj}

It was shown in \citep{IKS09} that this conjecture is true for orbit $m$-schemes with $m=4$. We improve this result  in Section~\ref{sec_primitiveorbit} by showing that one can even choose $m=3$. For general $m$-schemes, antisymmetric homogeneous $m$-schemes with no matching do exist for $m=1,2,3$  (see Section~\ref{sec_3anti}) but no counterexamples are known for $m\geq 4$.

The following theorem was proved in \citep{IKS09}.

\begin{thm}\label{thm_polyalgconj}
Assuming GRH and the schemes conjecture, there exists a deterministic polynomial-time algorithm that computes the complete factorization of a given polynomial $f(X)\in\F_q[X]$ over a finite field $\F_q$. 
\end{thm}

We reprove this theorem using the machinery of $\mathcal{P}$-schemes. First note that by Lemma~\ref{lem_antmatching}, an $m$-scheme with a matching is not strongly antisymmetric. So we can replace the schemes conjecture by  the following variant, which is implied by the original one.

\begin{conj}\label{conj_variantconj}
There exists a constant $m\in \N^+$ such that every strongly antisymmetric $m$-scheme on a finite set $S$ where $|S|>1$ is inhomogeneous.
\end{conj}

We also need the following simple lemma whose proof is  deferred to Section~\ref{sec_mschemeres}.  
It shows that inhomogeneity in Conjecture~\ref{conj_variantconj} can be replaced by discreteness.

\begin{lem}\label{lem_equivdisinhom}
Suppose there exists a strongly antisymmetric $m$-scheme on a finite set $S$ that is not discrete, where $m\in\N^+$ and $|S|>1$. Then for some finite set $T$ satisfying $1<|T|\leq |S|$, there exists a strongly antisymmetric homogeneous $m$-scheme on $T$.
\end{lem}

Now we complete the proof of Theorem~\ref{thm_polyalgconj}.

\begin{proof}[Proof of Theorem~\ref{thm_polyalgconj}]
First assume  that $\F_q=\F_p$ is a prime field and $f$ is square-free and completely reducible over $\F_p$. 
Fix the constant $m\in\N^+$ as guaranteed by the schemes conjecture, and let $n=\deg(f)$.
The algorithm first lifts $f$ to $\tilde{f}(X)\in\Z[X]$ of degree $n$ such that  all coefficients of $\tilde{f}$ are between zero and $p$. 
We can assume $\tilde{f}$ is irreducible over $\Q$ using the factoring algorithm for rational polynomials \citep{LLL82}. 
Let $S$ be the set of roots of $\tilde{f}$ in its splitting field. The Galois group $\gal(\tilde{f}/\Q)$ of $\tilde{f}$ is then a permutation group on $S$. 

Run the $\mathcal{P}$-scheme algorithm in Chapter~\ref{chap_alg_prime} that we used to prove  Corollary~\ref{cor_algdgbound}.
By Corollary~\ref{cor_algdgbound}, it suffices to prove $d(\gal(\tilde{f}/\Q))\leq m$. Assume to the contrary that $d(\gal(\tilde{f}/\Q))>m$. By Corollary~\ref{cor_monotonein}, we have $d(\sym(S))>m$, where $\sym(S)$ acts naturally on $S$.
Then by Lemma~\ref{lem_ptom}, there exists a strongly antisymmetric non-discrete $m$-scheme on $S$.
By Lemma~\ref{lem_equivdisinhom}, for some finite set $T$ satisfying $|T|>1$,  there exists a strongly antisymmetric homogeneous $m$-scheme on $T$. But this is a contradiction to Conjecture~\ref{conj_variantconj} and hence to the schemes conjecture.

For general  $f$ and $\F_q$, we either reduce to the previous case using Berlekamp's reduction \citep{Ber70} and  square-free factorization \citep{Yun76, Knu98}, or run the generalized $\mathcal{P}$-scheme algorithm in Chapter~\ref{chap_alg_general} and apply Corollary~\ref{cor_algdgboundg} instead.
\end{proof}

\paragraph{Schemes conjectures for a family of permutation groups.}\index{schemes conjecture!for permutation groups}

In the proof of Theorem~\ref{thm_polyalgconj}, we reduce to the case of the full symmetric group $\sym(S)$ and then apply the schemes conjecture. On the other hand, if the Galois group $G$ is ``less complex'' than $\sym(S)$,  we expect that the schemes conjecture can be replaced with a more moderate assumption. Formalizing this intuition leads to a hierarchy of conjectures, which we explain now.

Let $\mathcal{G}$ be a family of finite permutation groups. We formulate a conjecture for $\mathcal{G}$ as follows.

\begin{conj}[schemes conjecture for $\mathcal{G}$]\label{conf_schconjg}
There exists a constant $m\in \N^+$ such that $d(G)\leq m$ for all $G\in\mathcal{G}$.
\end{conj}

By Corollary~\ref {cor_algdgbound} and Corollary~\ref{cor_algdgboundg}, assuming this conjecture (and GRH) guarantees a polynomial-time factoring algorithm for the case that the Galois group $G$ is in $\mathcal{G}$ as a permutation group:

\begin{thm}\label{thm_polyalgconjg}
Assuming GRH and the schemes conjecture for $\mathcal{G}$, there exists a deterministic polynomial-time algorithm that given a polynomial $f(X)\in\F_q[X]$ and an irreducible\footnote{The assumption that $\tilde{f}$ is irreducible is not necessary, and can be avoided by adapting Lemma~\ref{lem_compstabsysg}. We omit the details.} lifted polynomial $\tilde{f}$ of $f$, computes the complete factorization of $f$ over $\F_q$, provided that the Galois group of $\tilde{f}$, as a permutation group  on the set of roots of $\tilde{f}$, is permutation isomorphic to some group in $\mathcal{G}$.
\end{thm}
%
%\begin{proof}
%Let $\mathcal{P}$ be the system of stabilizers of depth $m$ over $G$ with respect to the action of $G$ on the set of roots of $\tilde{f}$, where $m\in \N^+$ is the integer as guaranteed by Conjecture~\ref{conf_schconjg}.
%For the case that $\F_q=\F_p$ is a prime field and $f$ is square-free and completely reducible over $\F_p$,
%run the algorithm in Section~\ref{sec_algputtogether} on $(f,\tilde{f})$ where the collection of fields $\mathcal{F}$ corresponds to $\mathcal{P}'$ and is constructed using Lemma~\ref{lem_compstab}.
%  The claim then follows from Theorem~\ref{thm_algmain2formal}.
%  For general  $f$ and $\F_q$,  run the algorithm in Section X and apply Theorem X instead.
%\end{proof}

There exist reductions among these schemes conjectures defined for various families $\mathcal{G}$. To formulate them, we need the following notation: for two families $\mathcal{G}$ and $\mathcal{G}'$, write $\mathcal{G}\preceq \mathcal{G}'$ if any permutation group $G\in\mathcal{G}$  is permutation isomorphic to a subgroup of some permutation group $G'\in\mathcal{G}'$ (where action of this subgroup is restricted from that of $G'$).
Then we have

\begin{thm}\label{thm_conjred}
The schemes conjecture for $\mathcal{G}$ is implied by that for $\mathcal{G}'$ if $\mathcal{G}\preceq\mathcal{G}'$.
\end{thm}

\begin{proof}
This follows directly from Corollary~\ref{cor_monotonein}.
\end{proof}

In particular, all these conjectures are subsumed by that for the family of symmetric groups $\{\sym(n): n\in \N^+\}$, where each symmetric group $\sym(n)$ acts naturally on $[n]$. The latter is equivalent to Conjecture~\ref{conj_variantconj} by the connection between $m$-schemes and $\mathcal{P}$-schemes (see Theorem~\ref{thm_mandp}). 

Therefore, the conjectures for different families of finite permutation groups form a hierarchy, partially ordered by the relation $\preceq$, and Conjecture~\ref{conj_variantconj}  is the most difficult one.
One possible approach to the schemes conjecture is first relaxing it to those for simpler permutation groups which may be easier to prove. We will prove results in the same spirit in subsequent chapters.

Finally, we note that the schemes conjecture hold for the family of primitive solvable permutation groups, or more generally for primitive permutation groups $G$ not involving $\alt(d)$ (i.e., $\alt(d)$ is not isomorphic to a subquotient of $G$), where $d$ is a constant.

\begin{thm}
The schemes conjecture for $\mathcal{G}$ is true if $\mathcal{G}$ is the family of primitive solvable permutation groups, or the family of primitive permutation groups $G$ not involving $\alt(d)$, where $d\in\N^+$ is a constant.
\end{thm}
\begin{proof}
Let $G$ be a primitive permutation group.
Seress \citep{Se96} proved $b(G)\leq 4$ when $G$ is solvable.
More generally, it was shown in \citep{GSS98} that there exists a function $g(\cdot)$ such that $b(G)\leq g(d)$ if $G$ does not involve $\alt(d)$. The theorem then follows from Lemma~\ref{lem_antibasebound}.
\end{proof}

\begin{rem} The schemes conjectures in this section are formulated in terms of discreteness of $\mathcal{P}$-schemes and are used for complete factorization. One can also formulated conjectures in terms of inhomogeneity and use them for proper factorization. We leave the details to the reader. To establish reductions between these conjectures (in terms of inhomogeneity rather than discreteness), one needs to restrict to families of {\em transitive} permutation groups as transitivity is required in Corollary~\ref{cor_inductionsubgroup}.  
\end{rem}

\section{Extension to the closure of a subgroup system}\label{sec_extension} 

Suppose $\mathcal{P},\mathcal{P}'$ are subgroup systems over a finite  group $G$ and $\mathcal{P}\subseteq \mathcal{P}'$. We can construct a $\mathcal{P}$-scheme  from a $\mathcal{P}'$-scheme by simply discarding the partitions of $H\backslash G$ for $H\in\mathcal{P}'-\mathcal{P}$. Conversely, we want to know if a $\mathcal{P}$-scheme can be extended to a $\mathcal{P}'$-scheme.
In this section, we show that this is possible in some cases by formulating the notion of the {\em closure} $\mathcal{P}_\mathrm{cl}$ of a subgroup system $\mathcal{P}$ and proving that $\mathcal{P}$-scheme can always be extended to a $\mathcal{P}_\mathrm{cl}$-scheme. As an application, we prove Lemma~\ref{lem_extsym} and Lemma~\ref{lem_extgl} as promised before.

\begin{defi}[closure]\label{defi_inherit}\index{closure}
Let $\mathcal{P}$ be  a subgroup system over a group $G$. Denote by $\mathcal{P}_\mathrm{cl}$  the set of subgroups $H$ of $G$ satisfying the following conditions:
\begin{enumerate}
\item $\mathcal{P}$ contains a subgroup $H'\subseteq H$, and the set of such subgroups has a unique maximal element (with respect to inclusion), denoted by $u_\mathcal{P}(H)$, or simply $u(H)$ when there is no confusion. 
%\item $H\subseteq N_G(u(H))$.
\item $u(H)$ is a normal subgroup of $H$.
\end{enumerate}
%This defines a map $u: \mathcal{P}'\to\mathcal{P}$.
Then $\mathcal{P}_\mathrm{cl}$ is a subgroup system\footnote{It is easy to see that $\mathcal{P}_\mathrm{cl}$ is closed under conjugation in $G$, so it is indeed a subgroup system over $G$.} over $G$ containing $\mathcal{P}$, called the {\em closure} of $\mathcal{P}$.
\end{defi}
\nomenclature[e1d]{$\mathcal{P}_\mathrm{cl}$}{closure of a subgroup system $\mathcal{P}$}

The usage of the term {\em closure} is justified by the obvious fact $\mathcal{P}\subseteq \mathcal{P}_\mathrm{cl}$ and the next lemma.
\begin{lem}
$(\mathcal{P}_\mathrm{cl})_\mathrm{cl}=\mathcal{P}_\mathrm{cl}$.
\end{lem}

\begin{proof}
Consider $H\in (\mathcal{P}_\mathrm{cl})_\mathrm{cl}$. Write $H'=u_{\mathcal{P}_\mathrm{cl}}(H)$ and $H''=u_{\mathcal{P}}(H')$. We show that $H\in \mathcal{P}_\mathrm{cl}$ and $u_\mathcal{P}(H)=H''$.

We first verify that $H''$ is normal in $H$.
By definition, we know $H'$ is normal in $H$.
% and $H''$ is normal in $H'$.
Then for any $g\in H$, we have
\[
H''=u_\mathcal{P}(H')=u_\mathcal{P}(gH'g^{-1})=gu_\mathcal{P}(H')g^{-1}=gH''g^{-1}.
\]
So $H''$ is normal in $H$.

Next we show that $H''$ is the unique maximal element in $\mathcal{P}$ subject to $H''\subseteq H$. Assume to the contrary that there exists an element $U\subsetneq H''$ in $\mathcal{P}\subseteq\mathcal{P}_\mathrm{cl}$ that is a subgroup of $H$. 
%We may assume $U$ is maximal in $\mathcal{P}$ subject to $U\subseteq H$.
As $H'$ is the unique maximal element in $\mathcal{P}_\mathrm{cl}$ subject to $H'\subseteq H$, we have $U\subseteq H'$.
Furthermore, as $H''$ is the unique maximal element in $\mathcal{P}$ subject to $H''\subseteq H'$, we have $U\subseteq H''$, contradicting the assumption $U\subsetneq H''$. 
%So $U$ is a proper subgroup of $H''\in \mathcal{P}$, contradicting its maximality in $\mathcal{P}$. 

By definition, we have  $H\in \mathcal{P}_\mathrm{cl}$ and $u_\mathcal{P}(H)=H''$.
\end{proof}

We show that a $\mathcal{P}$-scheme can always be extended to a $\mathcal{P}_\mathrm{cl}$-scheme where antisymmetry and strong antisymmetry are preserved.

 \begin{lem}\label{lem_ext}
Let $\mathcal{P}$ be a subgroup system over a group $G$ and let $\mathcal{C}=\{C_H: H\in\mathcal{P}\}$ be a $\mathcal{P}$-scheme. 
There exists a unique $\mathcal{P}_\mathrm{cl}$-scheme $\mathcal{C}'=\{C'_{H}: H\in \mathcal{P}_\mathrm{cl}\}$ extending $\mathcal{C}$ (i.e.,  $C'_H=C_H$ for $H\in\mathcal{P}$), given by
\[
C'_{H}=\{\pi_{u(H),H}(B): B\in C_{u(H)}\}.
\] 
Moreover, if $\mathcal{C}$ is antisymmetric (resp. strongly antisymmetric), so is $\mathcal{C}'$.
And $\mathcal{C}'$ is not discrete on $H\in  \mathcal{P}_\mathrm{cl}$ 
if $\mathcal{C}$ is antisymmetric and not discrete on $u(H)$.
 \end{lem} 
 
 \begin{proof}
We have $u(H)\in\mathcal{P}\subseteq \mathcal{P}_{\mathrm{cl}}$ for $H\in\mathcal{P}_{\mathrm{cl}}$. It follows from Lemma~\ref{lem_min} that $\mathcal{C}'$ as defined above is the only possible one extending $\mathcal{C}$.
 
 Then we check that $\mathcal{C}'$ is indeed well defined, i.e., for $H\in \mathcal{P}_\mathrm{cl}$, the set $C'_{H}=\{\pi_{u(H),H}(B): B\in C_{u(H)}\}$ is indeed a partition of $H\backslash G$. For two blocks $B_1,B_2\in C_{u(H)}$, we prove that $\pi_{u(H),H}(B_1)$ and $\pi_{u(H),H}(B_2)$ are either identical or disjoint.
Suppose  there exist $u(H)g_1\in B_1$ and $u(H)g_2\in B_2$ satisfying $\pi_{u(H),H}(u(H)g_1)=\pi_{u(H),H}(u(H)g_2)$, i.e., $Hg_1=Hg_2$. Then $g_2g_1^{-1}\in H\subseteq N_G(u(H))$. Note that $c_{u(H),g_2g_1^{-1}}(u(H)g_1)=u(H)g_2$. So by invariance of $\mathcal{C}$, we have $c_{u(H),g_2g_1^{-1}}(B_1)=B_2$. 
Then by Lemma~\ref{lem_maps}, we have 
\[
\pi_{u(H),H}(B_2)=\pi_{u(H),H}\circ c_{u(H),g_2g_1^{-1}}(B_1)=c_{H,g_2g_1^{-1}}\circ \pi_{u(H),H}(B_1)=\pi_{u(H),H}(B_1)
\]
as desired. So $\mathcal{C}'$ is well defined. Moreover,  we have $u(H)=H$ for  $H\in\mathcal{P}$. It follows that $\mathcal{C}'$ does extend $\mathcal{C}$.

Next we show that $\mathcal{C}'$ is a $\mathcal{P}_\mathrm{cl}$-scheme.
For  $H,H'\in \mathcal{P}_\mathrm{cl}$ with $H\subseteq H'$, we have $u(H)\subseteq H'$ and hence $u(H)\subseteq u(H')$ by the unique maximality of $u(H')$. By transitivity of projections  (see Lemma~\ref{lem_maps}), the following diagram commutes:
\[
\begin{tikzcd}[column sep=large]
u(H)\backslash G \arrow[r, "\pi_{u(H), u(H')}"] \arrow[d, "\pi_{u(H),H}"']
& u(H')\backslash G \arrow[d, "\pi_{u(H'),H'}"] \\
H\backslash G  \arrow[r,  "\pi_{H, H'}"]
& H'\backslash G 
\end{tikzcd}
\]
To show compatibility, consider $y,y'\in H\backslash G$ lying in the same block $B\in C'_H$. Choose $\tilde{B}\in C_{u(H)}$ satisfying $\pi_{u(H),H}(\tilde{B})=B$ and choose $x,x'\in \tilde{B}$ satisfying $\pi_{u(H),H}(x)=y$, $\pi_{u(H),H}(x')=y'$. By compatibility of $\mathcal{C}$, the elements $\pi_{u(H), u(H')}(x)$ and $\pi_{u(H), u(H')}(x')$ lie in the same block of $C_{u(H')}$. 
Then $\pi_{u(H'),H'}\circ \pi_{u(H), u(H')}$ maps $x$ and $x'$ into the same block of $C'_{H'}$ by the definition of $\mathcal{C}'$.
By commutativity of the diagram above and the facts  $\pi_{u(H),H}(x)=y$, $\pi_{u(H),H}(x')=y'$, we see that $\pi_{H, H'}(y)$ and $\pi_{H, H'}(y')$ lie in the same block of $C'_{H'}$. So $\mathcal{C}'$ is compatible.

For regularity, let $B$ be a block of $C'_H$. Then $\pi_{H,H'}(B)$ is contained in a unique block $B'$ of $C'_{H'}$ by compatibility of $\mathcal{C}'$.
Lift $B$ to a block $\tilde{B}\in C_{u(H)}$ along $\pi_{u(H),H}$, and let $\tilde{B}'=\pi_{u(H),u(H')}(\tilde{B})\in C_{u(H')}$.  By regularity of $\mathcal{C}$, the map $\pi_{u(H),u(H')}|_{\tilde{B}}:\tilde{B}\to\tilde{B}'$ has constant degree, i.e., the number of preimages $|(\pi_{u(H),u(H')}|_{\tilde{B}})^{-1}(y)|$ is independent of the choices of $y\in \tilde{B}'$.  We show that $\pi_{u(H),H}|_{\tilde{B}}$ (and similarly $\pi_{u(H)',H'}|_{\tilde{B}'}$) also has constant degree. Consider $y,y'\in B$. As $\pi_{u(H), H}(\tilde{B})=B$, there exists $x,x'\in\tilde{B}$ satisfying $\pi_{u(H),H}(x)=y$ and $\pi_{u(H),H}(x')=y'$. Note that all the elements in $(\pi_{u(H),H}|_{\tilde{B}})^{-1}(y)$ (resp. $(\pi_{u(H),H}|_{\tilde{B}})^{-1}(y')$) are of the form $c_{u(H),g}(x)$ (resp. $c_{u(H),g}(x')$) for some $g\in H$ since $H\subseteq N_G(u(H))$. And we have $c_{u(H),g}(x)\in\tilde{B}$ iff $c_{u(H),g}(x')\in\tilde{B}$   for $g\in H$ by invariance of $\mathcal{C}$.
It follows that $|(\pi_{u(H),H}|_{\tilde{B}})^{-1}(y)|=|(\pi_{u(H),H}|_{\tilde{B}})^{-1}(y')|$.
So $\pi_{u(H),H}|_{\tilde{B}}$ (and similarly $\pi_{u(H)',H'}|_{\tilde{B}'}$) has constant degree. Then $\pi_{H,H'}|_B$ also has constant degree by the commutativity of the diagram above. So $\mathcal{C}'$ is regular.

For invariance, note that for $H,H'\in \mathcal{P}_\mathrm{cl}$ with $H'=gHg^{-1}$, we have $u(H')=g u(H)g^{-1}$. And the following diagram commutes by Lemma~\ref{lem_maps}:
\[
\begin{tikzcd}[column sep=large]
u(H)\backslash G \arrow[r, "c_{u(H), g}"] \arrow[d, "\pi_{u(H),H}"']
& u(H')\backslash G \arrow[d, "\pi_{u(H'),H'}"] \\
H\backslash G  \arrow[r,  "c_{H, g}"]
& H'\backslash G 
\end{tikzcd}
\]
For a block $B$ of $C'_H$, lift it to a block $\tilde{B}$ of $C_{u(H)}$. Then $c_{H,g}(B)=\pi_{u(H'),H'}\circ c_{u(H),g}(\tilde{B})$ by the commutativity of the diagram above. Note that  $c_{u(H),g}(\tilde{B})$ is a block of $C_{u(H')}$ by invariance of $\mathcal{C}$. So $c_{H,g}(B)$ is a block of $C'_{H'}$ by definition. Therefore $\mathcal{C}'$ is invariant.

 Now assume $\mathcal{C}'$ is not strongly antisymmetric and we prove that $\mathcal{C}$ is not either. By definition, there exists a nontrivial permutation $\tau=\sigma_k\circ\cdots\circ\sigma_1$ of a block $B\in C'_H$ for some $H\in\mathcal{P}_\mathrm{cl}$ such that each $\sigma_i: B_{i-1}\to B_i$ is a map of the form $c_{H_{i-1}, g}|_{B_{i-1}}$, $\pi_{H_{i-1}, H_i}|_{B_{i-1}}$, or  $(\pi_{H_{i}, H_{i-1}}|_{B_{i}})^{-1}$, and $B_i\in C'_{H_i}$, $H_i\in \mathcal{P}_\mathrm{cl}$, $B=B_0=B_k$, $H=H_0=H_k$ (see Definition~\ref{defi_strongasym}). 
 By the two diagrams above, we can lift each $B_i$ to $\tilde{B}_i\in  C_{u(H_i)}$ for $0\leq i\leq k$ and lift each $\sigma_i$ to a map $\tilde{\sigma}_i: \tilde{B}_{i-1}\to\tilde{B}_i$ of the form  $c_{u(H_{i-1}), g}|_{\tilde{B}_{i-1}}$, $\pi_{u(H_{i-1}), u(H_i)}|_{\tilde{B}_{i-1}}$, or  $(\pi_{u(H_{i}), u(H_{i-1})}|_{\tilde{B}_{i}})^{-1}$ respectively, i.e., $\pi_{u(H_{i}), H_{i}}(\tilde{B}_i)=B_i$ and $\sigma_i\circ \pi_{u(H_{i-1}), H_{i-1}}|_{\tilde{B}_{i-1}}=\pi_{u(H_{i}), H_{i}}|_{\tilde{B}_{i}}\circ \tilde{\sigma}_i$.\footnote{For the case that $\sigma_i=(\pi_{H_{i}, H_{i-1}}|_{B_{i}})^{-1}$, we lift $\pi_{H_{i}, H_{i-1}}|_{B_{i}}$ to $\pi_{u(H_{i}), u(H_{i-1})}|_{\tilde{B}_{i}}$. As $\mathcal{C}$ is antisymmetric, both $\pi_{u(H_{i-1}), H_{i-1}}|_{\tilde{B}_{i-1}}$ and $\pi_{u(H_{i}), H_{i}}|_{\tilde{B}_{i}}$ are bijective. So $\pi_{u(H_{i}), u(H_{i-1})}|_{\tilde{B}_{i}}$ is also bijective and its inverse is well defined.}
 Then $\tilde{\tau}:=\tilde{\sigma}_k\circ\cdots\circ\tilde{\sigma}_1$ is a map from $\tilde{B}_0$ to $\tilde{B}_k$ lifting $\tau$.
  Note that  $\pi_{u(H),H}(\tilde{B}_0)=\pi_{u(H),H}(\tilde{B}_k)=B$.
 So $c_{u(H),g}(\tilde{B}_k)=\tilde{B}_0$ for some $g\in H$. By composing $\tilde{\tau}$ with $c_{u(H),g}$ (and noting that $c_{H,g}$ is the identity map), we may assume $\tilde{B}_k=\tilde{B}_0$. So $\tilde{\tau}$ is a permutation of $\tilde{B}_0$. Moreover $\tilde{\tau}$ is nontrivial since it lifts $\tau$. So $\mathcal{C}$ is not strongly antisymmetric.
  The proof for antisymmetry is the same except that we only consider maps $\tau$ that are conjugations. 
  
  Finally, to prove the last claim, assume $\mathcal{C}$ is antisymmetric and $\mathcal{C}'$ is discrete on $H\in \mathcal{P}_\mathrm{cl}$. We prove that $\mathcal{C}$ is discrete on $u(H)$. Consider distinct elements $x,x'\in u(H)\backslash G$ and let $y=\pi_{u(H),H}(x)$, $y'=\pi_{u(H),H}(x')$. If $y\neq y'$, they are in different blocks of $C'_H$ and hence $x,x'$ are in different blocks of $C_{u(H)}$ by the definition of $C'_H$. So assume $y=y'$. 
Then $x=u(H)g$, $x'=u(H)g'$ for some $g,g'\in G$ satisfying $Hg=Hg'$, i.e., $g'g^{-1}\in H\subseteq N_G(u(H))$.  As $x'=c_{u(H),g'g^{-1}}(x)$, the elements $x$ and $x'$ are in different blocks of $C_{u(H)}$ by antisymmetry of $\mathcal{C}$. So $\mathcal{C}$ is discrete on $u(H)$, as desired.
 \end{proof}

Recall that for a subgroup system $\mathcal{P}$ over a finite group $G$, we let $\mathcal{P}_+=\{H: H'\subseteq H\subseteq N_G(H'), H'\in\mathcal{P}\}$ which is also a  subgroup system over $G$ (see Section~\ref{sec_consother}). Clearly $\mathcal{P}_\mathrm{cl}\subseteq \mathcal{P}_+$.
We show that equality holds if $\mathcal{P}$ is {\em join-closed}.\index{join-closed}

 \begin{lem}\label{lem_joinclosure}
 Let $\mathcal{P}$ be a subgroup system that is join-closed, i.e., $\langle H, H'\rangle\in\mathcal{P}$ for all $H,H'\in \mathcal{P}$.  Then $\mathcal{P}_\mathrm{cl}=\mathcal{P}_+$.
 \end{lem}
\begin{proof}
Consider $H\in \mathcal{P}_+$. We prove $H\in \mathcal{P}_\mathrm{cl}$ by verifying the conditions in Definition~\ref{defi_inherit}. 

Choose a maximal element $H'\in\mathcal{P}$ subject to $H'\subseteq H$. Such an element exists by the definition of $\mathcal{P}_+$.
 We first show that $H'$ is unique.
Assume to the contrary that there exists  another maximal element $H''\subseteq H$ in $\mathcal{P}$ different from $H'$. Then $\langle H',H''\rangle\supsetneq H'$ is also a subgroup of $H$ and lies in $\mathcal{P}$ by join-closedness, contradicting maximality of $H'$. So $H'$ is unique.

Next we prove $H'$ is normal in $H$. Assume to the contrary that  there exists $g\in H$ such that $g H'g^{-1}\neq H'$. As $g H'g^{-1}\subseteq gHg^{-1}=H$ and  $g H'g^{-1}\in \mathcal{P}$, the join $\langle H', g H' g^{-1}\rangle\supsetneq H'$ is also a subgroup of $H$ and lies in $\mathcal{P}$ by join-closedness, again contradicting maximality of $H'$.
\end{proof}

As an application, we consider a system of stabilizers with respect to the natural action of a symmetric group or an alternating group. 

\begin{lem}\label{lem_joinclosedsys}
Let $S$ be a finite $G$-set where $G$ is $\sym(S)$ or $\alt(S)$ acting  naturally on $S$.
Let $\mathcal{P}=\mathcal{P}_m$ be the corresponding system of stabilizers of depth $m$, where $m<|S|/2$.
Then  $\mathcal{P}':=\mathcal{P}\cup\{G\}$ is join-closed.
\end{lem}

\begin{proof}
Note  $\mathcal{P}'=\{G_T: 0\leq T\leq m\}$.
Let $T$ and $T'$ be subsets of $S$ of cardinality at most $m$.
We show that $\langle G_T, G_{T'}\rangle\in\mathcal{P}'$.
Obviously we have $\langle G_T, G_{T'} \rangle\subseteq G_{T\cap T'}$.

First assume $G=\sym(S)$. 
We have $G_T\cong \sym(S-T)$, $G_{T'}\cong \sym(S-T')$  and $G_{T\cap T'}\cong\sym(S-(T\cap T'))$ by restricting to the subsets $S-T$, $S-T'$ and $S-(T\cap T')$ respectively. The group $\sym(S-(T\cap T'))$
 is generated by transpositions $(x~y)$ with $x,y\in S-(T\cap T')$. We claim that every such $(x~y)$ is contained in $\langle G_T, G_{T'}\rangle$. 
This is obvious if $x$ and $y$ are both in $S-T$ or $S-T'$. So we assume $x\in T- T'$ and $y\in T'- T$. 
As $m<|S|/2$, the set $S-(T\cup T')$ is not empty. Pick $z\in S-(T\cup T')$. Then $(x~y)=(y~z) (x~z)(y~z)^{-1}\in \langle G_T, G_{T'} \rangle$ since $y,z\in S-T$ and $x,z\in S-T'$. So $\langle G_T,G_{T'}\rangle=G_{T\cap T'}\in\mathcal{P}'$.

Next assume $G=\alt(S)$. If $|S|\leq 4$, one can directly verify that $\langle G_T, G_{T'}\rangle$ equals $G$, $G_T$ or $G_{T'}$.
So assume $|S|\geq 5$. Note that $G_{T\cap T'}\cong\alt(S-(T\cap T'))$ is generated by $3$-cycles $(x~y~z)$ with $x,y,z\in S-(T\cap T')$.  We claim that every such $(x~y~z)$ is contained in $\langle G_T, G_{T'}\rangle$.  This is obvious if $x,y,z$ are all in $S-T$ or $S-T'$. So we assume $x,y\in T- T'$ and $z\in T'- T$ (the other cases are symmetric).  Pick $w\in S-(T\cup T')$ and let $(w~z~u)$ be a $3$-cycle for some $u\in S-T-\{z,w\}$. Then $(x~y~z)=(w~z~u)(x~y~w)(w~z~u)^{-1}\in\langle G_T,G_{T'}\rangle$ since $w,z,u\in S-T$ and $x,y,w\in S-T'$. So again $\langle G_T,G_{T'}\rangle=G_{T\cap T'}\in\mathcal{P}'$.
\end{proof}

\begin{cor}\label{cor_symequal}
Let $S$ be a finite $G$-set where $G$ is $\sym(S)$ or $\alt(S)$ acting  naturally on $S$.
Let $\mathcal{P}=\mathcal{P}_m$ be the corresponding system of stabilizers of depth $m$, where $m<|S|/2$.
Then $\mathcal{P}_\mathrm{cl}=\mathcal{P}_+$.
\end{cor}
\begin{proof}
 Let $\mathcal{P}'=\mathcal{P}\cup \{G\}$.
Then by Lemma~\ref{lem_joinclosedsys}, we have $\mathcal{P}_+\subseteq \mathcal{P}'_+=\mathcal{P}' _\mathrm{cl}=\mathcal{P}_\mathrm{cl}\cup \{G\}$. If $G\in\mathcal{P}$, we have $\mathcal{P}_\mathrm{cl}\cup \{G\}=\mathcal{P}_\mathrm{cl}$ and hence $\mathcal{P}_+\subseteq \mathcal{P}_\mathrm{cl}$.
On the other hand, if $G\not\in\mathcal{P}$, none of the groups in $\mathcal{P}$ is normal in $G$, and hence $G\not\in \mathcal{P}_+$.
So we still have $\mathcal{P}_+\subseteq \mathcal{P}_\mathrm{cl}$.
\end{proof}

\begin{rem}
The condition $m<|S|/2$ is necessary:  suppose $|S|\geq 6$ is even and let $m=|S|/2$. Partition $S$ into $S_1$ and $S_2$ of the same cardinality $m$. When $G=\sym(S)$ (resp. $G=\alt(S)$), the subgroup $\langle G_{S_1},G_{S_2}\rangle$ is the product of two copies of the symmetric group (resp. alternating group) of degree $m$. It is a proper subgroup of $G$ but stabilizes no element of $S$. Therefore $\langle G_{S_1},G_{S_2}\rangle\not\in\mathcal{P}_m\cup\{G\}$. Indeed, we have $\langle G_{S_1},G_{S_2}\rangle\in (\mathcal{P}_m)_+ - (\mathcal{P}_m)_\mathrm{cl}$ since $G_{S_1}\subseteq \langle G_{S_1},G_{S_2}\rangle\subseteq N_G(G_{S_1})$ whereas both $G_{S_1}$ and $G_{S_2}$ are maximal among subgroups of  $\langle G_{S_1},G_{S_2}\rangle$ in $\mathcal{P}_m$. 
\end{rem}

Lemma~\ref{lem_extsym} now follows from Lemma~\ref{lem_ext} and Corollary~\ref{cor_symequal}.

We also consider the case  $G=\gl(V)$ with the natural action on a vector space $V$.

\begin{lem}\label{lem_glequal}
Let $V$ be a finite dimensional vector space  over a finite field $F$.
Let $\mathcal{P}=\mathcal{P}_m$ be the  system of stabilizers of depth $m$ with respect to the natural action of $G:=\gl(V)$ on $S:=V-\{0\}$, where $m<\dim_F V$. Then $\mathcal{P}_\mathrm{cl}=\mathcal{P}_+$.
\end{lem}
\begin{proof}
Consider $H\in  \mathcal{P}_+$ and we prove that $H\in \mathcal{P}_\mathrm{cl}$.
Choose $H'\in\mathcal{P}$ such that $H'\subseteq H\subseteq N_G(H')$.
It suffices to show that $H'$ is the unique maximal element in $\mathcal{P}$ subject to $H'\subseteq H$.
Assume to the contrary that there exists another maximal element $H''\subseteq H$  in $\mathcal{P}$.
As $m<\dim_F V$, we have $H'=G_{V'}$ and $H''=G_{V''}$ for some proper linear subspaces $V', V''$ of $V$. 
As $H''\not\subseteq H'$, we have $V'\not\subseteq V''$. Also note that $V-(V'\cup V'')\neq \emptyset$ since 
\[
|V'\cup V''|=|V'|+|V''|-|V'\cap V''|<2|V|/|F|\leq |V|.
\] 
Pick $v\in V'-V''$ and $v'\in V-(V'\cup V'')$. Choose $g\in H''=G_{V''}$ sending $v$ to $v'$ which is possible since $v,v'\not\in V''$. As $g\in H''\subseteq H\subseteq N_G(H')=N_G(G_{V'})$, we have $\prescript{g}{}{V'}=V'$. But $\prescript{g}{}{v}=v'\not\in V'$, and we get a contradiction.
\end{proof}

Lemma~\ref{lem_extsym} now follows from Lemma~\ref{lem_ext} and Lemma~\ref{lem_glequal}.

\section{Restricting to a subset}\label{sec_mschemeres}

Suppose $\Pi=\{P_1,\dots,P_m\}$ is an $m$-scheme on a finite set $S$ and $T$ is a subset of $S$. Then we can restrict $\Pi$ to $T$ and obtain an $m$-collection on $T$, denote by $\Pi\|_T$.\footnote{It should not be confused with the notation $\Pi|_{x_1,\dots,x_k}$ in Definition~\ref{defi_mres}, which is an $(m-k)$-scheme on $S-\{x_1,\dots,x_k\}$.} In this section, we investigate this operation  and use it to prove Lemma~\ref{lem_equivdisinhom} in Section~\ref{sec_induction}. We also discuss its generalization for $\mathcal{P}$-schemes, where $\mathcal{P}$ is a system of stabilizers.

\begin{defi}\label{defi_mschemeres}\index{restriction!of an $m$-collection to a subset}
Let $\Pi=\{P_1,\dots, P_m\}$ be an $m$-collection on a finite set $S$, where $m\in\N^+$. For a subset $T$ of $S$, define the $m$-collection $\Pi\|_T:=\{P'_1,\dots,P'_m\}$ on $T$, where $P'_k:=P_k|_{T^{(k)}}$ is the restriction of $P_k$ to $T^{(k)}\subseteq S^{(k)}$ for  $k\in [m]$. 
\end{defi}
\nomenclature[e1e]{$\Pi\Vert_T$}{restriction of an $m$-collection $\Pi$ to a subset $T$}

\begin{lem}\label{lem_mschemeres}
 Suppose $\Pi=\{P_1,\dots,P_m\}$ is an $m$-scheme  on $S$ and $T\subseteq S$ is a disjoin union of blocks in $P_1$. Then $\Pi\|_T$ is also an $m$-scheme. Moreover, if $\Pi$ is antisymmetric (resp. strongly antisymmetric), so is $\Pi\|_T$. And if $\Pi$ does not have a matching, neither does $\Pi\|_T$.
\end{lem}

\begin{proof}
By compatibility of $\Pi$, for $k\in [m]$ and $B\in P_k$, either $B\subseteq T^{(k)}$ or $B\cap T^{(k)}=\emptyset$, and hence $T^{(k)}$ is a disjoint union of blocks of $P_k$. Then the various properties of $\Pi\|_T$ (compatibility, regularity, etc.) follow from those of $\Pi$ in a straightforward manner.
\end{proof}

In particular, suppose $\Pi=\{P_1,\dots, P_m\}$ is a strongly antisymmetric $m$-scheme on $S$ that is not discrete. Let $T$ be a block of $P_1$ such that $|T|>1$. Then $\Pi\|_T$ is a strongly antisymmetric homogeneous $m$-scheme on $T$. Lemma~\ref{lem_equivdisinhom} now follows.

Next we discuss the analogue of Lemma~\ref{lem_mschemeres} for $\mathcal{P}$-schemes. 
Let $G$ be a finite group acting on a finite set $S$. Let $\mathcal{P}=\mathcal{P}_{m}$ be the corresponding system of stabilizers of depth $m$ over $G$ for some $m\in \N^+$. By Lemma~\ref{lem_equivaction}, for $x\in S$, we have an equivalence of group actions
\[
\lambda_{x}: Gx \to G_x\backslash G
\]
between the action of $G$ on the $G$-orbit $Gx$  and that on $G_x\backslash G$ by inverse right translation. It sends $\prescript{g}{}{x}$ to $G_x g^{-1}$ for $g\in G$.

\begin{defi}\label{defi_pschemeresset}
Let $m$,  $G$ and $\mathcal{P}$ be as above. Let $\mathcal{C}=\{C_H: H\in \mathcal{P}\}$ be a $\mathcal{P}$-scheme.
Let $T$ be a subset of $S$ such that for $z\in T$, the set $\lambda_z(Gz\cap T)$ is a disjoint union of blocks in $C_{G_z}$.
Moreover, define $G'$ to be the setwise stabilizer $G_{\{T\}}$ and suppose it satisfies the following conditions:
\begin{enumerate}
\item For $U,U'\subseteq T$ satisfying $1\leq |U|, |U'|\leq m$ and $G'_U\subseteq G'_{U'}$, we have $G_U\subseteq G_{U'}$.
\item For $k\in [m]$ and $x\in T^{(k)}$, we have $G'x=Gx\cap T^{(k)}$.
\end{enumerate}
Let $\mathcal{P}'$ be the system of stabilizers of depth $m$ over $G'$ with respect to the action of $G'$ on $T$ (restricted from the action of $G$ on $S$). We define a $\mathcal{P}'$-collection $\mathcal{C}'=\{C'_H: H\in\mathcal{P}'\}$ as follows:

For $H\in\mathcal{P}'$, choose a nonempty subset $U\subseteq T$ of cardinality at most $m$ such that $H=G'_U$. Identify $G'_U\backslash G'$ with a subset of $G_U\backslash G$ via the injective map $i_U: G'_U\backslash G'\hookrightarrow G_U\backslash G$ sending $G'_Ug$ to $G_Ug$ for $g\in G'$.\footnote{This is indeed a well defined injective map by  Lemma~\ref{lem_equivaction}. Let $G'$ act on $G_U\backslash G$ by inverse right translation and let $O$ be the $G'$-orbit of $G_Ue$. The stabilizers of $G_Ue$ is $G'_U$. So we have a bijection $O\to G'_U\backslash G'$ whose inverse (composed with $O\hookrightarrow G_U\backslash G$) is $i_U$.}
Then define $C'_{H}$ to be the restriction of $C_{G_U}$ to $G'_U\backslash G'$.
\end{defi}

The assumption that  $\lambda_z(Gz\cap T)$ is a disjoint union of blocks in $C_{G_z}$ for all $z\in T$ is the analogue of the assumption in Lemma~\ref{lem_mschemeres} that $T$ is a disjoint union of blocks in $P_1$. If $G$ acts transitively on $S$, we have $Gz=S$, in which case this assumption is equivalent to that  $\lambda_z(T)$ is a disjoint union of blocks in $C_{G_z}$ for some $z\in T$. Note that we also need two additional conditions on $G'$. They are satisfied in the following important cases.

\begin{exmp}
Suppose $G$ is the full symmetric group $\sym(S)$ acting naturally on $S$.   The image of the permutation representation $G'\to \sym(T)$ is   $\sym(T)$. In this case the  two conditions in Definition~\ref{defi_pschemeresset} are satisfied for any subset $T$ of $S$ whose cardinality greater than $m+1$.\footnote{The first condition does not hold for $|T|
\leq m+1$: if $U,U'\subseteq T$ are different subsets of cardinality $|T|-1$, we have $G'_U=G'_{U'}=G'_T$, but $G_U\neq G_{U'}$ unless $T=S$.}
Indeed, if we view the $\mathcal{P}$-scheme $\mathcal{C}$ as an $m$-scheme by Theorem~\ref{thm_mandp}, the construction of $\mathcal{C}'$ from $\mathcal{C}$ is precisely the restriction of an $m$-scheme to the subset $T$ (see Definition~\ref{defi_mschemeres}).
\end{exmp}

\begin{exmp}
Suppose $S=V-\{0\}$ where $V$ is a finite dimensional vector space over a finite field $F$. Let $G$ be the general linear group $\gl(V)$ acting naturally on $S$. Let $T=V'-\{0\}\subseteq S$ where $V'$ is a linear subspace of $V$. The image of the permutation representation $G'\to \sym(T)$ is isomorphic to $\gl(V')$. It is easy to verify that in this case the two conditions in Definition~\ref{defi_pschemeresset} are also satisfied.
\end{exmp}

We prove the following generalization of Lemma~\ref{lem_mschemeres}.

\begin{lem}
The $\mathcal{P}'$-collection  $\mathcal{C}'$ is a well defined $\mathcal{P}'$-scheme. Moreover, if $\mathcal{C}$ is antisymmetric (resp. strongly antisymmetric), so is $\mathcal{C}'$.
\end{lem}

\begin{proof}
In  Definition~\ref{defi_pschemeresset} we define each $C'_H$ by picking $U\subseteq T$ of cardinality at most $m$ satisfying $H=G'_U$. Here the group $G_U$  and the map $i_U$ do not depend on the choice of $U$ by the first condition in Definition~\ref{defi_pschemeresset}. So $\mathcal{C}'$ is well defined.

For $H, H'\in\mathcal{P}'$ satisfying $H\subseteq H'$, we pick nonempty subsets $U,U'\subseteq T$ of cardinality at most $m$ such that $H=G'_U$ and $H'=G'_{U'}$. Then $G_U\subseteq G_{U'}$ by the first condition in Definition~\ref{defi_pschemeresset}. And the following diagram commutes.
\[
\begin{tikzcd}[column sep=large]
G'_U\backslash G' \arrow[r, "\pi_{G'_U, G'_{U'}}"] \arrow[d, "i_U"']
& G'_{U'}\backslash G' \arrow[d, "i_{U'}"] \\
G_U\backslash G  \arrow[r,  "\pi_{G_U, G_{U'}}"]
& G_{U'}\backslash G 
\end{tikzcd}
\]
For $H, H'\in\mathcal{P}'$ and $g\in G'$ satisfying $H'=gHg^{-1}$, we pick a nonempty subset $U\subseteq T$ of cardinality at most $m$ such that $H=G'_U$, and let $U'=\prescript{g}{}{U}\subseteq T$. Then $H'=G'_{U'}=gG'_Ug^{-1}$ and $G_{U'}=gG_Ug^{-1}$. And the following diagram commutes.
\[
\begin{tikzcd}[column sep=large]
G'_U\backslash G' \arrow[r, "c_{G'_U, g}"] \arrow[d, "i_U"']
& G'_{U'}\backslash G' \arrow[d, "i_{U'}"] \\
G_U\backslash G  \arrow[r,  "c_{G_U, g}"]
& G_{U'}\backslash G 
\end{tikzcd}
\]
Let  $U$ be a nonempty subset of $T$ of cardinality at most $m$. We claim $i_U$ maps each block of $C'_{G'_U}$ to a block of $C_{G_U}$. The rest of the proof focuses on this claim. Combining it with the two diagrams above, we can derive the various properties of $\mathcal{C}'$ (compatibility, regularity, invariance, antisymmetry and strong antisymmetry) from the corresponding properties of $\mathcal{C}$ in a straightforward manner.

Let $B$ be a block of $C'_{G_U}$  and  $B'$ be the block of $C_{G_U}$ containing $i_U(B)$. Assume to the contrary that $i_U(B)\neq B'$. Choose $G_Ug^{-1}, G_Ug'^{-1}\in G_U\backslash G$, represented by $g^{-1},g'^{-1}\in G$ respectively, such that $G_Ug^{-1}\in i_U(B)$ and $G_Ug'^{-1}\in B'-i_U(B)$. We may assume $g\in G'$ and hence $\prescript{g}{}{z}\in \prescript{g}{}{T}=T$ for all $z\in T$. Also note $B=i_U^{-1}(B')$  by construction. So from $G_Ug'^{-1}\in B'- i_U(B)$ we know $G_Ug'^{-1}\not\in i_U(G'_U\backslash G')$.

Assume there exists $z\in U$ such that $\prescript{g'}{}{z}\not\in T$. As $G_Ug^{-1}$ and $G_Ug'^{-1}$ are in the same block $B'$ of $C_{G_U}$, by compatibility of $\mathcal{C}$ we know $\pi_{G_U,G_z}(G_Ug^{-1})=G_z g^{-1}$ and $\pi_{G_U, G_z}(G_Ug'^{-1})=G_z g'^{-1}$ are in the same block of $C_{G_z}$.
On the other hand, we have $G_z g^{-1}=\lambda_z(\prescript{g}{}{z})\in\lambda_z(Gz\cap T)$ and $G_z g'^{-1}=\lambda_z(\prescript{g'}{}{z})\not\in\lambda_z(Gz\cap T)$ since $\prescript{g}{}{z}\in G z\cap T$, $\prescript{g'}{}{z}\not\in T$ and $\lambda_z: Gz\to G_z\backslash G$ is a bijection. But this contradicts the assumption that $\lambda_z(Gz\cap T)$ is a disjoint union of blocks of $C_{G_z}$.

Now assume $\prescript{g'}{}{z}\in T$ for all $z\in U$. Suppose $U=\{x_1,\dots,x_k\}$, where $x_i$ are distinct and ordered in an arbitrary way. Let $x=(x_1,\dots,x_k)\in T^{(k)}$.  Then $\prescript{g'}{}{x}$ is in $Gx\cap T^{(k)}$ and hence  in $G'x$ by the second condition in Definition~\ref{defi_pschemeresset}. So $\prescript{g'}{}{x}=\prescript{g''}{}{x}$ for some $g''\in G'$. Then $g'^{-1}g''\in G_x=G_U$. So $G_Ug'^{-1}=G_Ug''^{-1}=i_U(G'_Ug''^{-1})$,   contradicting the fact  $G_U g'^{-1}\not\in i_U(G'_U\backslash G')$ above.
 This proves the claim that $i_U$ maps each block of $C'_{G'_U}$ to a block of $C_{G_U}$. 
\end{proof}

\section{Primitivity of homogeneous \texorpdfstring{$m$-schemes}{m-schemes}} \label{sec_primitiveorbit} The notion of {\em primitivity} is important for permutation groups as well as association schemes. In this section, we extend it to homogeneous $m$-schemes. As an application, we show that every antisymmetric homogeneous orbit $m$-scheme on a finite set $S$ has a matching if  $|S|>1$ and $m\geq 3$.  

\begin{defi}[primitivity]\label{defi_mschemeprimitive}
Let $\Pi=\{P_1,\dots,P_m\}$ be a homogeneous $m$-scheme  on a finite set $S$. For $B\in P_2$, denote by $G_B$ the simple graph\footnote{A {\em simple graph} is an undirected graph without loops or multiple edges.} on the vertex set $S$ such that there exists an edge between two distinct vertices $u, v$ iff $(u,v)$ or $(v,u)$ is in $B$.
We say $\Pi$ is {\em primitive}\index{primitive!$m$-scheme} if $G_B$ is connected for all $B\in P_2$. Otherwise $\Pi$ is {\em imprimitive}\index{imprimitive!$m$-scheme}.
\end{defi}

 The reader familiar with {\em primitivity of association schemes}\index{primitive!association scheme} (see, e.g., \citep{CGS78}) may recognize that when $m\geq 3$,  Definition~\ref{defi_mschemeprimitive} simply defines $\Pi=\{P_1,\dots, P_m\}$ to be primitive iff  $P(\Pi')$ is primitive, where $\Pi'$ denotes the homogeneous $3$-scheme $\{P_1, P_2, P_3\}$ and $P(\Pi')$ is the corresponding association scheme (see Definition~\ref{defi_3sch}). 

\begin{rem}
Our definition of primitivity coincides with the notion of {\em primitivity at level 2} introduced in the full version of \citep{IKS09}. 
The same paper also generalizes the notion of primitivity to higher levels. We will not discuss their generalization in this thesis, but refer the interested reader to \citep{IKS09} for further details. 
\end{rem}

\paragraph{Restricting to a connected component.} We note that restricting a homogeneous $m$-scheme to a connected component yields another homogeneous $m$-scheme:

\begin{lem}\label{lem_mprimitive}
Let $\Pi=\{P_1,\dots, P_m\}$ be a homogeneous $m$-scheme on a finite set $S$ where $m\geq 3$. For each $B\in P_2$ and a connected component $T\subseteq S$ of $G_B$, the $m$-collection $\Pi\|_T$ (see Definition~\ref{defi_mschemeres}) is a homogeneous $m$-scheme on $T$. Moreover, if $\Pi$ is antisymmetric (resp. strongly antisymmetric), then so is $\Pi\|_T$. And if $\Pi$ has no matching, then neither does $\Pi\|_T$.
\end{lem}
\begin{proof}
Let $T\subseteq S$ be as in the lemma. It is well known that there exist blocks $B_1,\dots,B_k\in P_2$ such that the union of these blocks and $1_S=\{(x,x):x\in S\}$ yields an equivalence relation $\sim$ on $S$, and $T$ is one of its equivalence classes (see, e.g., \citep{CGS78}).

  For $k\in [m]$, define the equivalence relation $\sim_k$ on $S^{(k)}$ such that $(x_1,\dots,x_k)\sim_k (y_1\dots,y_k)$ iff $x_i\sim y_i$ for all $i\in [k]$.
These equivalence relations are respected by the maps  $\pi^k_i$ and $c^k_g$.
%For $k\in \{2,\dots,m\}$, $x,y\in S^{(k)}$, and a map $\tau: S^{(k)}\to S^{(k-1)}$ of the form $\pi^k_i$ or $c^k_g$, we have $x\sim_k y$ iff $\tau(x)\sim_{k-1} \tau(y)$. 
The various properties of $\Pi\|_T$ then follow from the corresponding properties of $\Pi$ in a straightforward manner.
\end{proof}

%Recall that strong antisymmetry of an $m$-scheme $\Pi$ strengthens the property that $\Pi$ has no matching (see Definition~\ref{defi_matching} and Lemma~\ref{lem_antmatching}).
%The same argument in the proof of Lemma~\ref{lem_mprimitive} also applies to the latter property and yields the following lemma, which is needed later.

\paragraph{Primitivity of homogeneous orbit $m$-schemes.} 

The next lemma states that primitivity of homogeneous orbit $m$-schemes is equivalent to primitivity of the associated  permutation group.

\begin{lem}\label{lem_orbitprimitive}
A homogeneous  orbit $m$-scheme  on a finite set $S$ associated with $K\subseteq \sym(S)$ is primitive iff $K$ is a primitive permutation group on $S$.
\end{lem}
\begin{proof}
Let $\Pi=\{P_1,\dots,P_m\}$ be a  homogeneous  orbit $m$-scheme associated with a group $K\subseteq \sym(S)$. Then $K$ acts transitively on $S$.
The graphs $G_B$ for $B\in P_2$ are known as the non-diagonal (undirected) {\em orbital graphs}\index{orbital graph}.
The lemma then follows from Definition~\ref{defi_mschemeprimitive} and the well known fact that a transitive permutation group is primitive iff every non-diagonal orbital graph is connected \citep{Hig67}.
\end{proof}

%Suppose $\Pi$ is a homogeneous orbit $m$-scheme on a finite set $S$ associated with a group $K\subseteq \sym(S)$ that acts transitively on $S$. In this case $\Pi$ is primitive iff $K$ is a primitive permutation group on $S$. This follows directly from the well known fact that the association scheme on $S$ whose blocks are $K$-orbits\footnote{An association scheme whose blocks are orbits under some transitive group action is known as a {\em Schurian} association scheme.}  is primitive iff $K$ is a primitive permutation group on $S$. For completeness, we give a proof as follows.

In general, we can obtain a primitive orbit $m$-scheme from a possibly imprimitive one by restricting to a minimal set that is a connected component:
%Recall that for  a permutation group $K$ on a finite set $S$, a set of imprimitivity of $K$ is  a subset $T$ of $S$ satisfying $\prescript{g}{}{T}\cap T=\emptyset$ or $\prescript{g}{}{T}=T$ for all $g\in K$.
% We say a set of imprimitivity $T$ is {\em minimal} if  $|T|>1$ and every set of imprimitivity $T'\subsetneq T$ is a singleton.  

\begin{lem}\label{lem_resmorbit}
Let $\Pi=\{P_1,\dots,P_m\}$ be a homogeneous  orbit $m$-scheme  on $S$ associated with $K\subseteq \sym(S)$, where $|S|>1$.
Let $T$ be a minimal subset of $S$ such that $T$ is a connected component of $G_B$ for some $B\in P_2$.
Let $K'$ be the image of the permutation representation $K_{\{T\}}\to \sym(T)$.
Then $\Pi\|_T$ is a primitive homogeneous orbit $m$-scheme on $T$, and is the orbit $m$-scheme associated with  $K'$.
\end{lem}

\begin{proof}
As already noted, for  $B\in P_2$ and any connected component $T'$ of $G_B$, there exist blocks $B_1,\dots,B_k\in P_2$ such that the union of these blocks and $1_S=\{(x,x):x\in S\}$ yields an equivalence relation   on $S$ where $T'$ is an equivalence class \citep{CGS78}. Primitivity of $\Pi$ then follows from minimality of $T$.

Choose $B\in P_2$ such that $T$ is a connected component of $G_B$.
Note that for $g\in K$ and $(u,v)\in S^{(2)}$, the edge $(u,v)$ is in $G_B$ iff $(\prescript{g}{}{u}, \prescript{g}{}{v})$ is in $G_B$. So for $g\in K$, the set $\prescript{g}{}{T}$ is a connected component of $G_B$. It follows that  $T$ is a set of imprimitivity of $K$, i.e., $\prescript{g}{}{T}\cap T=\emptyset$ or $\prescript{g}{}{T}=T$ for all $g\in K$.

Consider $k\in [m]$ and $x,y\in T^{(k)}$ in the same block of $P_k|_{T^{(k)}}\in \Pi\|_T$.  There exists $g\in K$ sending $x$ to $y$. As  $T$ is a set of imprimitivity of $K$, we have $\prescript{g}{}{T}=T$ and hence $g\in K'$.
So  $\Pi\|_T$ is  the orbit $m$-scheme on $T$ associated with  $K'$.
\end{proof}

\paragraph{Antisymmetric homogeneous orbit $m$-schemes for $m\geq 3$.} As an application, we prove that for $m\geq 3$, an antisymmetric homogeneous orbit $m$-scheme $\Pi$ on a finite set $S$ where $|S|>1$ always has a matching. In particular, it is not strongly antisymmetric by  Lemma~\ref{lem_antmatching}. The same claim for $m\geq 4$ was proved in \citep{IKS09}. Note that strongly antisymmetric homogeneous orbit $m$-schemes on sets $S$ where $|S|>1$ do exist for $m=1$ and $m=2$ (see Section~\ref{sec_3anti}).

We need the following result from finite group theory.

\begin{lem} \label{lem_primitivesol}
Let $G$ be a primitive solvable permutation group on a finite set $S$. The set $S$ can be identified with a 
finite dimensional vector space $V$ over a finite field $F$ such that $G$ acts on it as a subgroup of the {\em general affine group}\index{general!affine group}
\[
\agl(V)=\{\phi_{g,u}: g\in\gl(V), u\in V\},
\]
where $\phi_{g,u}$ sends $x\in V$ to $\prescript{g}{}{x}+u$. Moreover, the group $G$ contains the translation $\phi_{e, u}:x\mapsto x+u$ for all $u\in V$.
\end{lem}
\nomenclature[e1f]{$\agl(V)$}{general affine group on $V$}

 See \citep[Section~\RN{1}.4]{Su76} for its proof. We have

\begin{thm}\label{thm_nonexistence3sch}
Let $\Pi=\{P_1,\dots,P_m\}$ be an antisymmetric homogeneous orbit $m$-scheme  on a finite set $S$ associated with a group $K\subseteq \sym(S)$, where $m\geq 3$ and $|S|>1$. Then $\Pi$ has a matching. 
\end{thm}

\begin{proof}
We may assume $m=3$. Assume to the contrary that $\Pi$ has no matching.
Let $T$ be  a minimal subset of $S$ such that $T$ is a connected component of $G_B$ for some $B\in P_2$.
Let $K'$ be the image of the permutation representation $K_{\{T\}}\to \sym(T)$.
By Lemma~\ref{lem_mprimitive} and Lemma~\ref{lem_resmorbit}, the $m$-scheme $\Pi\|_T$ is the orbit $m$-scheme on $T$ associated with $K'$ which is antisymmetric,  homogeneous, primitive and has no matching. 
By replacing $\Pi$ with $\Pi\|_T$, $S$ with $T$, and $K$ with $K'$, we may assume $\Pi$ is primitive. Then $K$ is a primitive permutation group on $S$ by Lemma~\ref{lem_orbitprimitive}. 
Also note that $|K|$ is odd by Lemma~\ref{lem_antisym_orbitm}. It follows by the {\em Odd Order Theorem}\index{Odd Order Theorem} \citep{FT63} that $K$ is solvable. We conclude that $K$ is a primitive solvable permutation group on $S$ of odd order.

By Lemma~\ref{lem_primitivesol}, we can identify $S$ with a finite dimensional vector space $V$ over a finite field $F$, and $K$ with a subgroup of $\agl(V)$  acting on $V$ that contains all the translations $\phi_{e,u},u\in V$. Moreover, we have $\mathrm{char}(F)\neq 2$ since $|K|$ is odd. 

Choose $v\in V-\{0\}$. Let $x=(0,v,2v)\in S^{(3)}$, $y=\pi^3_3(x)=(0,v)\in S^{(2)}$ and $z=\pi^3_1(x)=(v,2v)\in S^{(2)}$. Let $B=Kx \in P_3$, $B'=\pi^3_3(B)=Ky\in P_2$ and $B''=\pi^3_1(B)=Kz\in P_2$. We claim that $B$ together with the maps $\pi^3_3|_B:B\to B'$, $\pi^3_1|_B:B\to B''$ is a matching of $\Pi$, which contradicts the assumption. To see this, note that the translation $\phi_{e,v}: x\mapsto x+v$ is in $K$ and sends $y$ to $z$. So $B'=B''$.
We also need to prove $|B|=|B'|$. By the orbit-stabilizer stabilizer theorem, it suffices to show $K_x=K_y$, which holds since $2v$ lies on the affine line spanned by $0$ and $v$, and $K$ acts affine linearly on $V$. The claim follows.
\end{proof}

\begin{rem}
The first half of our proof basically follows  \citep{IKS09} which reduces to the case that $K$ is primitive solvable. In \citep{IKS09}, the proof is completed by a result of Seress \citep{Se96} that bounds the minimal base size of primitive solvable permutation groups of odd order. This result allows them to prove the theorem for $m\geq 4$. We substitute it with the more elementary fact in Lemma~\ref{lem_primitivesol}, and use the above argument to prove the theorem for $m\geq 3$.
\end{rem}

\section{Direct products and wreath products}\label{sec_dwproduct}

We describe two more techniques of constructing new $\mathcal{P}$-schemes (resp. $m$-schemes) from old ones, namely the direct product and the wreath product. They extend the direct product and the wreath product of association schemes (see, e.g., \citep{SS98}).   As  an application, we show that either the schemes conjecture (Conjecture~\ref{conj_schconj}) is true, or there exist infinitely many counterexamples.

\paragraph{Direct products.}

Suppose $\mathcal{P}$ and $\mathcal{P}'$ are subgroup systems over finite groups $G$ and $G'$ respectively. Define 
\[
\mathcal{P}\times\mathcal{P}':=\{H\times H': H\in \mathcal{P}, H'\in \mathcal{P}'\}\]
 which is a subgroup system over $G\times G'$. For $H\in\mathcal{P}$ and $H'\in\mathcal{P}'$, we have a bijection
 \[
 \phi_{H,H'}: H\backslash G \times H'\backslash G' \to (H\times H')\backslash (G\times G')
 \]
 sending $(Hg, H'g')$ to $(H\times H')(g,g')$ for $g\in G$ and $g'\in G'$. Then we define the direct product of a $\mathcal{P}$-collection and a $\mathcal{P}'$-collection as follows.
 
\begin{defi}\label{defi_productpscheme}
For a  $\mathcal{P}$-collection $\mathcal{C}=\{C_H: H\in\mathcal{P}\}$ and a $\mathcal{P}'$-collection $\mathcal{C}'=\{C_H': H\in\mathcal{P}'\}$, define the $(\mathcal{P}\times\mathcal{P}')$-collection $\mathcal{C}\times \mathcal{C}'=\{C''_{H\times H'}: H\times H'\in\mathcal{P}\times \mathcal{P}'\}$ by
\[
C''_{H\times H'}=\{\phi_{H,H'}(B\times B'): B\in C_H, B'\in C'_{H'}\},
\]
called the {\em direct product}\index{direct product!of $\mathcal{P}$-collections} of $\mathcal{C}$ and $\mathcal{C'}$.
\end{defi}

We have

\begin{lem}\label{lem_productpscheme}
The direct product $\mathcal{C}\times \mathcal{C}'$ is a $(\mathcal{P}\times\mathcal{P}')$-scheme if $\mathcal{C}$ is a $\mathcal{P}$-scheme and $\mathcal{C}'$ is a $\mathcal{P}'$-scheme. Moreover, if $C$ and $C'$ are antisymmetric (resp. strongly antisymmetric), so is  $\mathcal{C}\times \mathcal{C}'$.
\end{lem}
\begin{proof}
Write $\pi_{H,H'}$ (resp. $\pi'_{H, H'}$, $\pi''_{H,H'}$) for a projection between coset spaces of subgroups in $G$ (resp. $G'$, $G\times G'$). Similarly write $c_{H,g}$ (resp. $c'_{H, g}$, $c''_{H,g}$) for a conjugation between coset spaces of subgroups in $G$ (resp. $G'$, $G\times G'$). For $H=H_1\times H_2, H'=H'_1\times H'_2\in\mathcal{P}\times\mathcal{P}'$ satisfying $H\subseteq H'$, we have $H_1\subseteq H_1'$, $H_2\subseteq H_2'$ and 
\[
\pi''_{H_1\times H_2,H_1'\times H_2'}\circ \phi_{H_1,H_2}(x,y)=\phi_{H_1',H_2'}(\pi_{H_1,H_1'}(x),\pi'_{H_2,H_2'}(y))
\]
 for all $x\in H_1\backslash G$ and $y\in H_2\backslash G'$. Similarly, for $H_1\times H_2\in\mathcal{P}\times\mathcal{P}'$ and $(g,g')\in G\times G'$, we have  
\[
c''_{H_1\times H_2,(g,g')}\circ \phi_{H_1,H_2}(x,y)=\phi_{gH_1g^{-1},g'H_2g'^{-1}}(c_{H_1,g}(x),c'_{H_2,g'}(y))
\]
 for all $x\in H_1\backslash G$ and $y\in H_2\backslash G'$.  The various properties of $\mathcal{C}\times\mathcal{C}'$ (compatibility, regularity, invariance, antisymmetry, and strong antisymmetry) then follow from those of $\mathcal{C}$ and $\mathcal{C}'$ in a straightforward manner.
\end{proof}

Similarly, we define the direct product of $m$-schemes:

\begin{defi}\label{defi_productmscheme}\index{direct product!of $m$-schemes}
Let $\Pi=\{P_1,\dots,P_m\}$ and $\Pi'=\{P'_1,\dots,P'_m\}$ be $m$-schemes on finite sets $S$ and $S'$ respectively, where $m\in \N^+$.
Define the $m$-collection $\Pi\times \Pi'=\{P''_1,\dots,P''_m\}$ on $S\times S'$ in the following way:
for $k\in [m]$, two elements $z=((x_1,y_1),\dots,(x_k,y_k)), z'=((x'_1,y'_1),\dots,(x'_k,y'_k))\in (S\times S')^{(k)}$ are in the same block of $P''_k$ iff the following conditions are satisfied:
\begin{enumerate}
\item For $i,j\in [k]$, it holds that $x_i=x_j$ iff $x'_i=x'_j$, and $y_i=y_j$ iff $y'_i=y'_j$. 
\item Omit a minimal subset $T$ of coordinates in $[k]$ such that all $x_i$ are distinct, and so are all $x'_i$. Let $k'=k-|T|$. Suppose the remaining $x$-coordinates of $z$ and $z'$ are $x_{i_1},\dots,x_{i_{k'}}$ and $x'_{i_1},\dots,x'_{i_{k'}}$ respectively. Then $(x_{i_1},\dots,x_{i_{k'}})$ and $(x'_{i_1},\dots,x'_{i_{k'}})$ are in the same block of $P_{k'}$.\footnote{The order of these coordinates does not matter by invariance of $\Pi$. Under the previous condition, the choice of $T$ does not matter either.}
\item The previous condition holds with $x$-coordinates replaced by $y$-coordinates and $P_{k'}$ replaced by $P'_{k'}$.
\end{enumerate}
\end{defi}

We have the following analogue of Lemma~\ref{lem_productpscheme} whose proof is left to the reader.
\begin{lem}
The $m$-collection $\Pi\times\Pi'$ is an $m$-scheme on $S\times S'$. Moreover, if $\Pi$ and $\Pi'$ are antisymmetric (resp. strongly antisymmetric), so is $\Pi\times \Pi'$. And if $\Pi$  and $\Pi'$ have no matching, neither does $\Pi\times \Pi'$.
\end{lem}

\begin{rem}
The connection between Definition~\ref{defi_productpscheme} and Definition~\ref{defi_productmscheme} is as follows. Given $m\in \N^+$, let $\mathcal{P}$ (resp. $\mathcal{P}'$, $\mathcal{P}''$) be the system of stabilizers of depth $m$ over $G=\sym(S)$ (resp. $G'=\sym(S')$, $G''=\sym(S\times S')$) with respect to the natural action of $G$ on $S$ (resp. $G'$ on $S'$, $G''$ on $S\times S'$).
 Let $\tilde{\mathcal{P}}$ be the system of stabilizers of depth $m$ with respect to the {\em product action} of $G\times G'$ on $S\times S'$.\footnote{The product action is defined by $\prescript{(g,g')}{}{(x,x')}=(\prescript{g}{}{x},\prescript{g'}{}{x'})$ for $(g,g')\in G\times G'$ and $(x,x')\in S\times S'$.}
 Then $\tilde{\mathcal{P}}\subseteq \mathcal{P}\times\mathcal{P}'$.\footnote{To see this, note that for a subset $U\subseteq S\times S'$ whose projections to $S$ and $S'$ are $U_1$ and $U_2$, respectively, we have $(G\times G')_U=G_{U_1}\times G'_{U_2}$.}
So we obtain a $\tilde{\mathcal{P}}$-scheme $\tilde{\mathcal{C}}$ from $\mathcal{C}\times\mathcal{C}'$. Using induction of $\tilde{\mathcal{P}}$-schemes, we obtain a $\mathcal{P}''$-scheme $\mathcal{C}''$ (see Definition~\ref{defi_pschemeind}). 
Using the connection between $m$-schemes and $\mathcal{P}$-schemes (see Theorem~\ref{thm_mandp}), we see that the construction of $\mathcal{C}''$ from $\mathcal{C}$ and $\mathcal{C}'$ corresponds to a construction of an $m$-scheme on $S\times S'$ from those on $S$ and $S'$. This is exactly Definition~\ref{defi_productmscheme}.
\end{rem}
 
 It is obvious that the direct product  also preserves homogeneity and discreteness. 
By taking iterated direct products, we can construct infinitely many antisymmetric homogeneous $m$-schemes with no matching if there exists a single one. As  an application, we know that either the schemes conjecture (Conjecture~\ref{conj_schconj}) is true, or there exist infinitely many counterexamples.\footnote{This claim also holds for the variant of the schemes conjecture (Conjecture~\ref{conj_variantconj}) for the same reason.}

\begin{cor}\label{cor_counterexampleinf}
For any $m\in \N^+$, there exist either infinitely many  antisymmetric homogeneous $m$-schemes with no matching or none.
\end{cor}

\paragraph{Wreath products.} There exists another operation of $\mathcal{P}$-schemes and $m$-schemes called the {\em wreath product}. While this operation is interesting on its own, we do not need it anywhere else in this thesis, except that it provides an alternative proof of Corollary~\ref{cor_counterexampleinf}. For this reason, we only give the definitions as well as  the statements, and leave the proofs to the reader.

 We first define the wreath product of groups.

\begin{defi}\label{defi_wrgroup}
Let $G$ and $G'$ be groups and let $\Omega$ be a $G'$-set. Let $G^\Omega$ be the group consisting of all the functions $f: \Omega\to G$. Its group operation is defined by $(ff')(x)=f(x)f'(x)$.
Define the {\em wreath product}\index{wreath product!of groups} $G\wr G'$ as the group consisting of all the pairs $(f,g)\in G^\Omega\times G'$, with its group operation defined by
\[
(f,g)(f',g')=(f\cdot \prescript{g}{}{f'}, g g') 
\]
for $(f,g),(f',g')\in G\wr G'$, where $\prescript{g}{}{f'}: \Omega \to G$ sends $x\in\Omega$ to $f'(\prescript{g^{-1}}{}{x})$. In other words, the group $G\wr G'$ is the semidirect product $G\rtimes_\varphi G'$ where $\varphi:G'\to \aut(G^\Omega)$ sends $g\in G'$ to the automorphism $f\mapsto \prescript{g}{}{f}$ of $G^\Omega$. For convenience, we identify $G^\Omega$ and $G'$ with subgroups of $G\wr G'$ and write $(f,g)\in G\wr G'$ as $fg$.
\end{defi}
\nomenclature[e1g]{$G\wr G'$}{wreath product of groups $G$ and $G'$}

Use the following notations: let $G$ and $G'$ be finite groups and let $\Omega$ be a finite $G'$-set. For a family $\mathcal{H}=\{H_{x}: x\in \Omega\}$ of subgroups of $G$ indexed by $\Omega$ and a subgroup $H'$ of $G'$ satisfying the following condition:
\begin{equation}\label{eq_wp}
H_x=G \text{ for all } x\in \Omega \text{ not fixed by } G',
\end{equation}
write $\mathcal{H} \wr H'$ for the subset
\[
\{fg: f(x)\in H_x \text{ for all } x\in\Omega, g\in H'\}
\]
of $G\wr G$, which is a subgroup of $G\wr G$ by \eqref{eq_wp}.
Suppose $\mathcal{P}$ and $\mathcal{P}'$ are subgroup systems over finite groups $G$ and $G'$ respectively. 
Define 
$\mathcal{P}\wr\mathcal{P}'$ to be the poset of subgroups of $G\wr G'$ consisting of the subgroups $\mathcal{H} \wr H'$ for all $\mathcal{H}=\{H_x\in \mathcal{P}:x\in\Omega\}$ and $H'\in\mathcal{P}'$ satisfying \eqref{eq_wp}.
Then $\mathcal{P}\wr\mathcal{P}'$ is a subgroup system over $G\wr G'$. 

For  $\mathcal{H}=\{H_x\in \mathcal{P}:x\in\Omega\}$ and $H'\in\mathcal{P}'$ satisfying \eqref{eq_wp}, we have a bijection
 \[
 \phi_{\mathcal{H},H'}: \left(\prod_{x\in\Omega} H_x\backslash G\right) \times H'\backslash G' \to (\mathcal{H} \wr H')\backslash (G\wr G')
 \]
 defined as follows: for $f\in \prod_{x\in\Omega} H_x\backslash G$ whose $x$-factor is $f_x\in H_x\backslash G$, pick $g_x\in G$ such that $f_x=H_x g_x$. Then define $f': \Omega\to G$ sending $x\in\Omega$ to $g_x$.
Define $\phi_{\mathcal{H},H'}$ such that it sends $(f,Hg')$ to $(\mathcal{H}\wr H')f'g'$ for $g'\in G'$. It can be shown that $\phi_{\mathcal{H},H'}$ is a well defined bijection.
 Finally, we define the wreath product of a $\mathcal{P}$-collection and a $\mathcal{P}'$-collection as follows.
 
\begin{defi}\label{defi_wppscheme}
For a  $\mathcal{P}$-collection $\mathcal{C}=\{C_H: H\in\mathcal{P}\}$ and a $\mathcal{P}'$-collection $\mathcal{C}'=\{C_H': H\in\mathcal{P}'\}$, define the $(\mathcal{P}\wr\mathcal{P}')$-collection $\mathcal{C}\wr \mathcal{C}'=\{C''_{\mathcal{H}\wr H'}: \mathcal{H}\wr H'\in\mathcal{P}\wr\mathcal{P}'\}$ by
\[
C''_{\mathcal{H}\wr H'}=\left\{\phi_{\mathcal{H},H'}\left(\left(\prod_{x\in\Omega}B_x\right)\times B'\right): \text{$B_x\in C_{H_x}$ for $x\in\Omega$}, B'\in C'_{H'}\right\},
\]
where  $\mathcal{H}=\{H_x:x\in\Omega\}$.
We call $\mathcal{C}\wr\mathcal{C}'$ the {\em wreath product}\index{wreath product!of $\mathcal{P}$-schemes} of $\mathcal{C}$ and $\mathcal{C'}$.
\end{defi}

We have
\begin{lem}\label{lem_wppscheme}
The wreath product $\mathcal{C}\wr \mathcal{C}'$ is a $(\mathcal{P}\wr\mathcal{P}')$-scheme if $\mathcal{C}$ is a $\mathcal{P}$-scheme and $\mathcal{C}'$ is a $\mathcal{P}'$-scheme. Moreover, if $C$ and $C'$ are antisymmetric (resp. strongly antisymmetric), then so is  $\mathcal{C}\wr \mathcal{C}'$.
\end{lem}

Similarly, we define the wreath product of $m$-schemes:

\begin{defi}\label{defi_wpmscheme}\index{wreath product!of $m$-schemes}
Let $\Pi=\{P_1,\dots,P_m\}$ and $\Pi'=\{P'_1,\dots,P'_m\}$ be $m$-schemes on finite sets $S$ and $S'$ respectively, where $m\in \N^+$.
Define the $m$-collection $\Pi\wr \Pi'=\{P''_1,\dots,P''_m\}$ on $S\times S'$ in the following way:
for $k\in [m]$, two elements $z=((x_1,y_1),\dots,(x_k,y_k)), z'=((x'_1,y'_1),\dots,(x'_k,y'_k))\in (S\times S')^{(k)}$ are in the same block of $P''_k$ iff the following conditions are satisfied:
\begin{enumerate}
\item For $i,j\in [k]$, it holds that $y_i=y_j$ iff $y'_i=y'_j$. 
\item For $i\in [k]$, let $T_i$ be set of indices $j\in[k]$ satisfying $y_i=y_j$. Suppose $T_i=\{i_1,\dots,i_\ell\}$, ordered in an arbitrary way. Then $(x_{i_1},\dots,x_{i_\ell})$ and $(x'_{i_1},\dots,x'_{i_\ell})$ are in the same block of $P_\ell$.
\item Omit a minimal subset $T$ of coordinates in $[k]$ such that all $y_i$ are distinct. Let $k'=k-|T|$. Suppose the remaining $y$-coordinates of $z$ and $z'$ are $y_{i_1},\dots,y_{i_{k'}}$ and $y'_{i_1},\dots,y'_{i_{k'}}$ respectively. Then $(y_{i_1},\dots,y_{i_{k'}})$ and $(y'_{i_1},\dots,y'_{i_{k'}})$ are in the same block of $P'_{k'}$.
\end{enumerate}
\end{defi}

We have the following analogue of Lemma~\ref{lem_wppscheme}.
\begin{lem}
The $m$-collection $\Pi\wr\Pi'$ is an $m$-scheme on $S\times S'$. Moreover, if $\Pi$ and $\Pi'$ are antisymmetric (resp. strongly antisymmetric), then so is $\Pi\wr \Pi'$.  And if $\Pi$  and $\Pi'$ have no matching, then neither does $\Pi\wr \Pi'$.
\end{lem}

\begin{rem}
The connection between Definition~\ref{defi_wppscheme} and Definition~\ref{defi_wpmscheme} is as follows. Given $m\in \N^+$, let $\mathcal{P}$ (resp. $\mathcal{P}'$, $\mathcal{P}''$) be the system of stabilizers of depth $m$ over $G=\sym(S)$ (resp. $G'=\sym(S')$, $G''=\sym(S\times S')$) with respect to the natural action of $G$ on $S$ (resp. $G'$ on $S'$, $G''$ on $S\times S'$).
 Let $\tilde{\mathcal{P}}$ be the system of stabilizers of depth $m$ with respect to the {\em imprimitive wreath product action}\index{imprimitive wreath product action} of $G\wr G'$ on $S\times S'$.\footnote{The imprimitive wreath product action is defined by $\prescript{(f,g)}{}{(x,x')}=(\prescript{f(\prescript{g}{}{x'})}{}{x},\prescript{g}{}{x'})$ for $(f,g)\in G\wr G'$ and $(x,x')\in S\times S'$.}
 Then $\tilde{\mathcal{P}}\subseteq \mathcal{P}\wr\mathcal{P}'$.\footnote{To see this, consider a subset $U\subseteq S\times S'$. For $x'\in S'$, let $U_{x'}=\{x\in S: (x,x')\in S'\}$ and $H_{x'}=G_{U_{x'}}$. Let $\mathcal{H}=\{H_{x'}: x'\in S'\}$ and let $U'$ be the projection of $U$ to $S'$. 
Then $(G\wr G')_U=\mathcal{H}\wr G'_{U'}$. Moreover, if $x'\in S'$ is not fixed by $G'_{U'}$, then  $x'\not\in U'$ and hence $U_{x'}=\emptyset$, which implies $H_{x'}=G$.
}
So we obtain a $\tilde{\mathcal{P}}$-scheme $\tilde{\mathcal{C}}$ from $\mathcal{C}\wr\mathcal{C}'$. Using induction of $\tilde{\mathcal{P}}$-schemes, we obtain a $\mathcal{P}''$-scheme $\mathcal{C}''$ (see Definition~\ref{defi_pschemeind}). 
Using the connection between $m$-schemes and $\mathcal{P}$-schemes (see Theorem~\ref{thm_mandp}), we see that the construction of $\mathcal{C}''$ from $\mathcal{C}$ and $\mathcal{C}'$ corresponds to a construction of an $m$-scheme on $S\times S'$ from those on $S$ and $S'$. This is exactly Definition~\ref{defi_wpmscheme}.
\end{rem}

\chapter{Symmetric groups and linear groups}\label{chap_sym}

Let $G$ be a finite permutation group. 
Motivated by the $\mathcal{P}$-scheme algorithms developed in Chapter~\ref{chap_alg_prime} and Chapter~\ref{chap_alg_general}, we are interested in the problem of bounding the integer $d(G)$, introduced in Definition~\ref{defi_dg}.

In this chapter, we study this problem for symmetric groups and linear groups with various special group actions.

\paragraph{Symmetric groups.}

%Consider the symmetric group $G=\sym(S)$ acting naturally on a finite set $S$.
%, the problem of bounding $d(G)$ is essentially equivalent to bounding the smallest $m\in\N^+$ such that all strongly antisymmetric $m$-scheme on $S$ are discrete. 
For convenience, we introduce the following notation:

\begin{defi}
For $n\in\N^+$, define $d_{\sym}(n):=d(G)$, where $G$ is the symmetric group $\sym(S)$ acting naturally on a finite set $S$  of cardinality $n$.\footnote{Clearly $d_{\sym}(n)$  only depends on $n$ but not on $S$.}
\end{defi}
\nomenclature[f1a]{$d_{\sym}(n)$}{alias for $d(G)$ where $G=\sym(S)$ acts naturally on $S$, $\lvert S\rvert=n$}

Note that $d_{\sym}(n)$ is nondecreasing in $n$ by Corollary~\ref{cor_inductionsubgroupvar}.
The best known general upper bound for $d_{\sym}(n)$ is 
\[
d_{\sym}(n)\leq \left(\frac{2}{\log 12}\right) \log n+O(1),
\]
proven in \citep{Gua09,Aro13} in different notations, based on the work of \citep{Evd94, IKS09}. In Section~\ref{sec_symnatural}, we review this result and interpret it as a result about $\mathcal{P}$-schemes.  
%Based on their arguments, we also provide a lower bound for the cardinality of the largest block in a $\mathcal{P}$-scheme (see Lemma~\ref{lem_lbblocksize}).

In Section~\ref{sec_symstandard}, we study the more general action of $\sym(S)$ on the set of $k$-subsets of $S$, where $1\leq k\leq |S|$,  and that on (an orbit of) the set of partitions of $S$. These actions are called the {\em standard action}\index{standard action} of symmetric groups, and play an important role in the study of minimal base sizes of primitive permutation groups (see, e.g., \citep{LS99}). Our results for these group actions will be used in Chapter~\ref{chap_primitive}.

\paragraph{Linear groups.} Let $V$ be a vector space of dimension $n\in\N^+$ over a finite field $\F_q$. We have the {\em general linear group}\index{general!linear group} $\gl(V)$ consisting of all the invertible linear transformations of $V$ over $\F_q$. It is a subgroup of the {\em general semilinear group}\index{general!semilinear group} $\gammal(V)$, which consists of all the invertible {\em semilinear transformations}\index{semilinear transformation} of $V$. Here we say a map $\phi: V\to V$ is a {\em semilinear transformation} of $V$ if
$\phi(x+y)=\phi(x)+\phi(y)$ and
$\phi(c x)=\tau_{\phi}(c)\phi(x)$
hold for all $x,y\in V$ and $c\in \F_q$, where $\tau_\phi$ is an automorphism of the field $\F_q$.
We have the {\em natural action}\index{natural action of linear groups} of $\gl(V)$ and that of $\gammal(V)$ on $V-\{0\}$, defined in the obvious way.

Denote by $\proj V$ the {\em projective space}\index{projective!space} associated with $V$, i.e.,
$\proj V$ is the set of equivalence classes of  $V-\{0\}$
where $x,y\in V-\{0\}$ are equivalent iff $x=c y$ for some $c\in\F_q^\times$.
Define the {\em projective linear group}\index{projective!linear group} $\pgl(V):=\gl(V)/\F_q^\times$ and the {\em projective semilinear group}\index{projective!semilinear group} $\pgammal(V):=\gammal(V)/\F_q^{\times}$, where $\F_q^\times$ is identified with the subgroup of the {\em scalar linear transformations}\index{scalar linear transformation} of $\gl(V)$ (resp. $\gammal(V)$) so that $c\in\F_q^{\times}$ sends $x\in V$ to $c x$.
The natural action of $\gl(V)$ (resp. $\gammal(V)$) on $V-\{0\}$ induces an action of $\pgl(V)$ (resp. $\pgammal(V)$) on $\proj V$, called the {\em natural action} of $\pgl(V)$ (resp.  $\pgammal(V)$) on $\proj V$. Finally, when $V=\F_q^n$, we also use the notations $\gl_n(q)$, $\gammal_n(q)$, $\pgl_n(q)$, and $\pgammal_n(q)$.
\nomenclature[f1b]{$\gl(V)$, $\gl_n(q)$}{general linear group}
\nomenclature[f1c]{$\gammal(V)$, $\gammal_n(q)$}{general semilinear group}
\nomenclature[f1d]{$\pgl(V)$, $\pgl_n(q)$}{projective linear group}
\nomenclature[f1e]{$\pgammal(V)$, $\pgammal_n(q)$}{projective semilinear group}

We call the above groups $\gl(V)$, $\gammal(V)$, $\pgl(V)$, and $\pgammal(V)$ {\em linear groups}\index{linear group}. In Section~\ref{sec_lingroup}, we investigate $d(G)$ for the natural action of a linear group $G$. 
For convenience, we introduce the following notations:
\begin{defi}\label{defi_dlinear}
 Let $V$ be a vector space of dimension $n\in\N^+$ over a finite field $\F_q$.
  Define $d_{\gl}(n,q):=d(G)$, where $G$ is the permutation group $\gl(V)$ acting naturally on $V-\{0\}$.
 Similarly define $d_{\gammal}(n,q)$, $d_{\pgl}(n,q)$, and $d_{\pgammal}(n,q)$ by choosing $G$ to be the permutation group $\gammal(V)$, $\pgl(V)$, $\pgammal(V)$ acting naturally on $V-\{0\}$, $\proj V$, $\proj V$, respectively.\footnote{Clearly these definitions only depend on $n$ and $q$ but not on $V$.}
\end{defi}
We show that the problems of bounding $d_{\gl}(n,q)$ $d_{\gammal}(n,q)$, $d_{\pgl}(n,q)$, and $d_{\pgammal}(n,q)$ are all equivalent: an upper bound $f(n,q)$ for any one of them implies an upper bound $f(n,q)+O(1)$ for the others.
So it suffices to investigate just one of them. 
\nomenclature[f1f]{$d_{\gl}(n,q)$, $d_{\gammal}(n,q)$, $d_{\pgl}(n,q)$, $d_{\pgammal}(n,q)$}{See Definition~\ref{defi_dlinear}}

Finally, we prove a bound 
\[
d_{\gl}(n,q)\leq \left(\frac{\log q}{\log q + (\log 12)/4 }\right) n+O(1),
\]
slightly improving the trivial bounds.

%In addition, we also investigate the {\em subspace actions} of these groups.

\paragraph{Self-reduction.} The results in Section~\ref{sec_symstandard} and Section~\ref{sec_lingroup} require  a technique called {\em self-reduction of discreteness}, which we introduce in Section~\ref{sec_selfreduction}. It reduces discreteness of a strongly antisymmetric $\mathcal{P}$-scheme to discreteness of its restrictions to stabilizer subgroups. 
In many cases, such a reduction greatly simplifies the problem.
Our results in  Chapter~\ref{chap_primitive} also rely heavily on this technique.

\section{The natural action of a symmetric group}\label{sec_symnatural}

%By Theorem~\ref{thm_mandp}, the problem of bounding $d_\sym(n)$ is essentially equivalent to bounding the smallest $m\in\N^+$ such that all strongly antisymmetric $m$-scheme on $[n]$ are discrete. 
We introduce the following notations about $m$-schemes:
\begin{defi}\label{defi_funmn}
For $n\in \N^+$, let $m(n)$ (resp. $m'(n)$) be the smallest positive integer such that any non-discrete antisymmetric $m(n)$-scheme (resp. $m'(n)$-scheme) on $[n]$ has a matching (resp. is not strongly antisymmetric).
\end{defi}
It is easy to see that $m(n)$ and $m'(n)$ are nondecreasing in $n$.
We also have $d_\sym(n)\leq m'(n)\leq m(n)$ by Lemma~\ref{lem_ptom} and Lemma~\ref{lem_antmatching}.
\nomenclature[f1g]{$m(n)$, $m'(n)$}{See Definition~\ref{defi_funmn}}

It was proven \citep{Gua09} and independently in \citep{Aro13} that $m(n)\leq \left(\frac{2}{\log 12}\right) \log n+O(1)$.
We review the proof of this bound, starting from the following lemma:

\begin{lem}\label{lem_sizeshrink2}
Let $\Pi=\{P_1,\dots,P_m\}$ be an antisymmetric $m$-scheme on a finite set $S$ where $m\geq 3$. Suppose $B\in P_1$ satisfies $|B|\geq 3$. Let $x$ be an element of $B$ so that $\Pi|_x=\{P'_1,\dots, P'_{m-1}\}$ is an $(m-1)$-scheme on $S-\{x\}$ (see Definition~\ref{defi_mres}). Then at least one of the two conditions is satisfied.
\begin{enumerate}
\item There exists $B'\in P'_1$ contained in $B$ satisfying $|B'|\leq (|B|-1)/4$.
\item There exist distinct elements $y,z\in B-\{x\}$ such that for the $(m-2)$-scheme $\Pi|_{x,y}=\{P''_1,\dots,P''_{m-2}\}$ on $S-\{x,y\}$, the block $B''$ of $P''_1$ containing $z$ satisfies $|B''|\leq (|B|+1)/12$. 
Furthermore, $(x,y)$, $(y,z)$, and $(z,x)$ are in the same block of $P_2$.
\end{enumerate}
\end{lem}

\begin{proof}
By replacing $\Pi$ with $\Pi\|_B$, we may assume $\Pi$ is homogeneous and  $S=B$.
By antisymmetry, we know $|P_2|$ is even. If $|P_2|\geq 4$, there exists $B_1\in P_2$ of cardinality at most $|B|(|B|-1)/4$. Let $B':=\{y\in B: (x,y)\in B_1\}$. Then $B'$ is a block of $P'_1$ by definition, and its cardinality is $|B_1|/|B|\leq (|B|-1)/4$ by regularity of $\Pi$. And the first condition is met.

So assume $|P_2|=2$. Then $P_2$ contains two blocks $B_1$ and $B_2$ of the same cardinality $|B|(|B|-1)/2$. Choose $y\in B-\{x\}$ such that $(x,y)\in B_1$. Such an element $y$ exists by regularity and homogeneity of $\Pi$.
By Lemma~\ref{lem_3sch} and Lemma~\ref{lem_3schsym}, we have an antisymmetric association scheme $P(\Pi)=P_2\cup\{1_B\}$ that has three blocks. 
By Lemma~\ref{lem_antiasso}, the number of elements $z\in B-\{x,y\}$ satisfying $(y,z),(z,x)\in B_1$ is precisely $(|B|+1)/4>0$. The cardinality of the set $T:=\{(a,b,c): (a,b),(b,c),(c,a)\in B_1\}$ is then $|B_1|\cdot (|B|+1)/4$.  Choose $z\in B-\{x,y\}$ such that $(x,y,z)\in T$. 
Let $B'_1$, $B'_2$, and $B'_3$ be the blocks of $P_3$ containing $(x,y,z)$, $(y,z,x)$ and $(z,x,y)$ respectively, which are all subsets of $T$. They have the same cardinality by invariance of $\Pi$, and are distinct by antisymmetry of $\Pi$. So $|B'_1|\leq |T|/3=|B_1|\cdot (|B|+1)/12$. By regularity of $\Pi$, the cardinality of the set $\{u\in S-\{x,y\}:(x,y,u)\in B'_1\}$ is $|B'_1|/|B_1|\leq (|B|+1)/12$, and this set is exactly the block $B''$ of $P''_1$ containing $z$ by definition. So the second condition is satisfied.
\end{proof}

Lemma~\ref{lem_sizeshrink2} implies the following recursive relation:
\begin{lem}\label{lem_recrel}
For $n\geq 3$,
\[
m(n)\leq \max\left\{m\left(\frac{n-1}{4}\right)+1, m\left(\frac{n+1}{12}\right)+2\right\}.
\]
The  inequality also holds for $m'(\cdot)$ in replaced of $m(\cdot)$.
\end{lem}
\begin{proof}
Let $\Pi=\{P_1,\dots,P_m\}$ be a non-discrete antisymmetric $m$-scheme on a finite set $S$ of cardinality $n$, where $m\geq 3$. Also assume 
\[
m\geq \max\left\{m\left(\frac{n-1}{4}\right)+1, m\left(\frac{n+1}{12}\right)+2\right\}.
\]
We want to show that $\Pi$ has a matching.

Choose $B\in P_1$ such that $|B|>1$. 
Let $x$ be an element of $B$ and suppose $\Pi|_x=\{P'_1,\dots, P'_{m-1}\}$.  Then $\Pi|_x$ is an antisymmetric $(m-1)$-scheme on $S-\{x\}$.
Note that  $\Pi\|_B$ is a homogeneous antisymmetric $m$-scheme on $B$ by Lemma~\ref{lem_mschemeres}, which implies $|B|\geq 3$. 
Then either of the two conditions in Lemma~\ref{lem_sizeshrink2} is satisfied.

If the first condition is satisfied, there exists $B'\in P'_1$ contained in $B$ satisfying $|B'|\leq (|B|-1)/4\leq (n-1)/4$.
If $|B'|>1$, we see $(\Pi|_x)\|_{B'}$ is a non-discrete antisymmetric $(m-1)$-scheme on $B'$. It has a  matching since $m-1\geq m((n-1)/4)\geq m(|B'|)$. So $\Pi$ also has a matching by Lemma~\ref{lem_mres} and Lemma~\ref{lem_mschemeres}.
On the other hand, if $|B'|=1$, we let $y$ be the unique element in $B'$ and let $B_1$ be the block of $P_2$ containing $(x,y)$. Note that $|B'|=|B_1|/|B|$, which implies $|B_1|=|B|$. As $x,y\in B$, we have $\pi^2_1(B_1)=\pi^2_2(B_1)=B$. Then $B_1$ is a matching of $\Pi$.

Next assume the second condition is satisfied. So there exist distinct elements $y,z\in B-\{x\}$ such that for the $(m-2)$-scheme $\Pi|_{x,y}=\{P''_1,\dots,P''_{m-2}\}$ on $S-\{x,y\}$, the cardinality of the block $B''$ of $P''_1$ containing $z$ is at most $(|B|+1)/12\leq (n+1)/12$.
 Furthermore, $(x,y)$, $(y,z)$, and $(z,x)$ are in the same block $B_0$ of $P_2$. 
If $|B''|>1$, we see $(\Pi|_{x,y})\|_{B''}$ is a non-discrete antisymmetric $(m-2)$-scheme on $B''$. It has a  matching since $m-2\geq m((n+1)/12)\geq m(|B''|)$. So $\Pi$ also has a matching by Lemma~\ref{lem_mres} and Lemma~\ref{lem_mschemeres}.
On the other hand, if $|B''|=1$, we let $B_0'$ be the block of $P_3$ containing $(x,y,z)$.
We have $\pi^3_1(B_0')=\pi^3_3(B_0')=B_0$ since $(x,y),(y,z)\in B_0$. 
Also note that $|B''|=|B_0'|/|B_0|$, which implies $|B_0|=|B_0'|$. 
So $B_0'$ is a matching of $\Pi$.

This proves the inequality for $m(\cdot)$.
The proof for $m'(\cdot)$ is similar, and we leave it to the reader.
\end{proof}

%\begin{lem}\label{lem_sizeshrink}
%Let $\Pi=\{P_1,\dots,P_m\}$ be an antisymmetric $m$-scheme on a finite set $S$ where $m\geq 2$. Let $B\in P_1$ and $x\in B$, so that $\Pi|_x=\{P'_1,\dots, P'_{m-1}\}$ is an $(m-1)$-scheme on $S-\{x\}$. Any block $B'$ of the partition $P'_1$ contained in $B$ satisfies $|B'|\leq (|B|-1)/2$. Moreover, these blocks form a  partition of $B-\{x\}$. 
%\end{lem}
%
%\begin{proof}
%By definition, two elements $y,y'\in S-\{x\}$ are in the same block of $P'_1$ iff $(x,y)$ and $(x,y')$ are in the same block of $P_2$, which holds only if $y$ and $y'$ are both in $B-\{x\}$ or $S-B$ by compatibility of $\Pi$. So the blocks $B'\in P'_1$ contained in $B$ form a partition of $B-\{x\}$. For the other claim, assume $B\neq \{x\}$ since otherwise the claim is trivial. Fix $B'\in P'_1$ contained in $B$ and choose $y\in B'$. Let $B_1$ and $B_2$ be the blocks of $P_2$ containing $(x,y)$ and $(y,x)$ respectively. Then $|B_1|=|B_2|$ by invariance of $\Pi$ and $B_1\neq B_2$ by antisymmetry of $\Pi$. As $x,y\in B$, we have $B_1,B_2\subseteq B^{(2)}$ by compatibility of $\Pi$. So $|B_1|\leq |B^{(2)}|/2=|B|(|B|-1)/2$. By definition, we have $B'=\{z\in S-\{x\}: (x,z)\in B_1\}$, and hence the map $z\mapsto (x,z)$ is a bijection from $B'$ to $(\pi^2_2)^{-1}(x)\cap B_1$. So $|B'|=|(\pi^2_2)^{-1}(x)\cap B_1|$. The later equals $|B_1|/|B|$ by regularity of $\Pi$. So we have
%\[
%|B'|=|B_1|/|B|\leq (|B|-1)/2
%\]
%as desired.
%\end{proof}

\begin{thm}[\citep{Gua09, Aro13}]\label{thm_symbound2}
For all $n\in\N^+$,
\[
m(n)\leq \left(\frac{2}{\log 12}\right) \log n+O(1).
\]
More generally,  an antisymmetric $m$-scheme $\Pi=\{P_1,\dots,P_m\}$ on a finite set $S$ always has a matching if  $P_1$ has a block $B$ of cardinality $k>1$ and $m\geq  m(k)$. In particular it holds for sufficiently large $m=\left(\frac{2}{\log 12}\right) \log k+O(1)$.  
\end{thm}

\begin{proof}
Note $m(1)=1$ and $m(2)=2$. The first claim then follows from Lemma~\ref{lem_recrel} and a simple induction. 
The second claim follows by considering  $\Pi\|_B$ and applying  Lemma~\ref{lem_mschemeres}.
\end{proof}

Theorem~\ref{thm_symbound2} implies a bound for $d_\mathrm{Sym}(n)$, and also a bound for $d(G)$ by  Corollary~\ref{cor_monotonein}, where $G$ is an arbitrary permutation group on a set of cardinality $n$:

\begin{cor}\label{cor_ubsysp}
Let $G$ be a permutation group on a set of cardinality $n\in\N^+$. Then $d(G)\leq d_\mathrm{Sym}(n)\leq \left(\frac{2}{\log 12}\right) \log n+O(1)$.
\end{cor}
%\begin{proof}
%The first inequality follows from Corollary~\ref{cor_monotonein} whereas the second follows from Corollary~\ref{cor_symbound2}, Lemma~\ref{lem_ptom} and Lemma~\ref{lem_antmatching}.
%\end{proof}

We conclude this section with the following technical lemma, which is used later in the proof of Theorem~\ref{thm_slightimp}.

\begin{lem}\label{lem_lbblocksize}
Let $G$ be a permutation group  on a finite set $S$, and let $\mathcal{P}$ be the corresponding system of stabilizers of depth $m$ where $1\leq m\leq |S|$. Let $\mathcal{C}=\{C_H: H\in\mathcal{P}\}$ be a strongly antisymmetric $\mathcal{P}$-scheme. Suppose   $\mathcal{C}$ is non-discrete on $G_x$ for some $x\in S$.
Then there exists $(x_1,\dots,x_m)\in S^{(m)}$ such that $C_{G_{x_1,\dots,x_m}}$ has a block of cardinality at least $2^{\left(\frac{\log 12}{4}\right)m^2-O(m)}$.
\end{lem}

\begin{proof}
Let $\mathcal{P}'$ be the system of stabilizers of depth $m$ with respect to the natural action of $G':=\sym(S)$ on $S$. 
Let $\mathcal{C}'=\{C'_H: H\in\mathcal{P}\}$ be the induction of $\mathcal{C}$ to $\mathcal{P}'$ (see Definition~\ref{defi_pschemeind}), which is strongly antisymmetric by Lemma~\ref{thm_pschmind} and is non-discrete on $G'_x$ for $x\in S$ in the lemma since $\mathcal{C}$ is non-discrete on $G_x$.
Assume the lemma holds for $\sym(S)$, $\mathcal{P}'$, and an $m$-tuple $(y_1,\dots,y_m)\in S^{(m)}$, i.e., there exists $B'\in C'_{G'_{y_1,\dots,y_m}}$ of cardinality at least $2^{\left(\frac{\log 12}{4}\right)m^2-O(m)}$. By Definition~\ref{defi_pschemeind}, we know $B'$ is of the form
$\phi_{G'_{y_1,\dots,y_m}, g}(B)$, where $g\in G'$, $\phi_{G'_{y_1,\dots,y_m},g}$ is an injection from $(G\cap gG'_{y_1,\dots,y_m}g^{-1})\backslash G$ to $G'_{y_1,\dots,y_m}\backslash G'$, and $B$ is a block of $C_{G\cap gG'_{y_1,\dots,y_m}g^{-1}}$. Let $x_i=\prescript{g}{}{y_i}$ for $i\in [m]$. Then $G\cap gG'_{y_1,\dots,y_m}g^{-1}=G_{x_1,\dots,x_m}$.
So $(x_1,\dots,x_m)$ and $B\in C_{G_{x_1,\dots,x_m}}$ satisfy the condition in the lemma.

Thus we may assume $G=\sym(S)$ and it acts naturally on $S$. By Lemma~\ref{defi_ptom}, it suffices to show that for any non-discrete strongly antisymmetric $m$-scheme $\Pi=\{P_1,\dots, P_m\}$ on $S$, the partition $P_m$ has a block of cardinality at least $2^{\left(\frac{\log 12}{4}\right)m^2-cm}$, where $c=O(1)$. We prove this claim by induction on $m$. The case $m=1$ is trivial. For $m>1$, assume the claim for $m'<m$. Let $B_0$ be a block of $P_1$ of cardinality $k>1$. By Theorem~\ref{thm_symbound2}, we have $m\leq \left(\frac{2}{\log 12}\right)\log k+c'$ for some $c'=O(1)$, or equivalently $k\geq 2^{\frac{\log 12}{2}(m-c')}$. Choose $x\in B_0$ and consider the $(m-1)$-scheme $\Pi$-scheme $\Pi':=\Pi|_x=\{P'_1,\dots,P'_{m-1}\}$ on $S-\{x\}$.  It is strongly antisymmetric by Lemma~\ref{lem_mres}. Let $B_1$ be a block of $P'_1$ contained in $B_0$, which exists by compatibility of $\Pi$ and the fact $k>1$. 
If $|B_1|=1$, we have seen in the proof of  Lemma~\ref{lem_recrel} that $\Pi$ has matching, contradicting the assumption that $\Pi$ is strongly antisymmetric.
So $|B_1|>1$. By  Lemma~\ref{lem_mschemeres}, the homogeneous $(m-1)$-scheme $\Pi'\|_{B_1}=\{P''_1,\dots,P''_{m-1}\}$ on $B_1$ is strongly antisymmetric. By the induction hypothesis, the partition $P''_{m-1}$ has a block $B'\subseteq B_1^{(m-1)}$ of cardinality at least $2^{\left(\frac{\log 12}{4}\right)(m-1)^2-c(m-1)}$. And $B'$ is also a block of $P'_{m-1}\in \Pi'$ by definition and compatibility of $\Pi'$.
Then $P_m\in \Pi$ has a block $B$ containing $(x,x_1,\dots,x_{m-1})$ for all $(x_1,\dots,x_{m-1})\in B'$.
By regularity of $\Pi$, we have 
\[
|B|=|B_0||B'|\geq 2^{\frac{\log 12}{2}(m-c')}\cdot 2^{\left(\frac{\log 12}{4}\right)(m-1)^2-c(m-1)}\geq 2^{\left(\frac{\log 12}{4}\right)m^2-cm}
\]
for sufficiently large $c=O(1)$.
\end{proof}

\section{Self-reduction of discreteness}\label{sec_selfreduction}\index{self-reduction of discreteness} 

In this section, we prove a ``self-reduction'' lemma, which states that discreteness of a strongly antisymmetric $\mathcal{P}$-scheme is implied by discreteness of its restrictions to stabilizer subgroups.

We need the following technical lemma.

\begin{lem}\label{lem_selfred0}
Suppose $G$ is a finite group, $\mathcal{P}$ is a subgroup system over $G$, and  $\mathcal{C}=\{C_H: H\in\mathcal{P}\}$ is a  $\mathcal{P}$-scheme. Suppose $H_0, H_1, H_2$ are subgroups in $\mathcal{P}$ such that $H_0\subseteq H_1\cap H_2$ and $\mathcal{C}|_{H_1}$, $\mathcal{C}|_{H_2}$ are both discrete on $H_0$. For $i=0,1,2$, let $B_i$ be the block of $C_{H_i}$ containing $H_i e\in H_i\backslash G$.  
Then $(\pi_{H_0,H_2})|_{B_0} \circ(\pi_{H_0, H_1}|_{B_0})^{-1}$ is a well-defined bijection from $B_1$ to $B_2$ sending $H_1 e\in H_1\backslash G$ to $H_2 e\in H_2\backslash G$.
\end{lem}

\begin{proof}
Note that $\pi_{H_0, H_1}|_{B_0}$ is a surjective map from $B_0$ to $B_1$ sending $H_0 e$ to $H_1 e$, and $\pi_{H_0, H_2}|_{B_0}$ is a surjective map from $B_0$ to $B_2$ sending $H_0 e$ to $H_2 e$. So it suffices to prove that these two maps are injective. The set $B_0\cap (H_0\backslash H_1)$ contains $H_0 e$ and is a block of $C_{H_0}|_{H_1}\in\mathcal{C}|_{H_1}$ by the definition of restriction (Definition~\ref{defi_res}). By discreteness of $\mathcal{C}|_{H_1}$ on $H_0$, this set is just the singleton $\{H_0e\}$. 
On the other hand, the set $H_0\backslash H_1\subseteq H_0\backslash G$  is precisely the preimage of $H_1 e$ under 
$\pi_{H_0, H_1}: H_0\backslash G\to H_1\backslash G$. 
 So $B_0\cap  (H_0\backslash H_1)$ is the preimage of $H_1 e$ under  $\pi_{H_0, H_1}|_{B_0}: B_0\to B_1$. By regularity of $\mathcal{C}$, the map $\pi_{H_0, H_1}|_{B_0}$  is injective.
Similarly $\pi_{H_0, H_2}|_{B_0}$   is also injective.
\end{proof}

The bijection in Lemma~\ref{lem_selfred0} can be used to separate elements in a strongly antisymmetric $\mathcal{P}$-scheme:

\begin{lem}\label{lem_selfred}
Let $G$ be a finite group acting on a finite set $S$, and let $x\in S$. Let $\mathcal{P}$ be a subgroup system over $G$ such that $G_x\in\mathcal{P}$, and let $\mathcal{C}=\{C_H: H\in\mathcal{P}\}$ be a  $\mathcal{P}$-scheme.
Suppose $y=\prescript{g}{}{x}$ and $z=\prescript{g'}{}{x}$ in $S$ satisfy 
(1) $G_{y}, G_{z}, G_{y,z}\in\mathcal{P}$ and (2) $\mathcal{C}|_{G_{y}}$ and $\mathcal{C}|_{G_{z}}$ are both discrete on $G_{y,z}$.
Then there exists a bijection between blocks of $C_{G_x}$ sending $G_x g^{-1}$ to $G_x g'^{-1}$ that can be written as a  composition of conjugations, projections and their inverses between blocks of  $C_{G_x}$,  $C_{G_{y}}$, $C_{G_{z}}$ and $C_{G_{y,z}}$.
%\footnote{Note that $G_x g^{-1}$ and $G_x g'^{-1}$ only depend on $y$ and $z$. In fact they are the images of $y$ and $z$ under the map $\lambda_x: S\to G_x\backslash G$ in Lemma~\ref{lem_equivaction} respectively.}
In particular, if $\mathcal{C}$ is strongly antisymmetric, then $G_x g^{-1}$ and  $G_x g'^{-1}$ are in different blocks of $C_{G_x}$.
\end{lem}

\begin{proof}
Let $B_0$ (resp. $B_1$, $B_2$) be the block of $C_{G_{y,z}}$ (resp. $C_{G_{y}}$, $C_{G_{z}}$) containing $G_{y,z} e$ (resp. $G_{y}e$, $G_{z}e$). By Lemma~\ref{lem_selfred0}, the map $\pi_{G_{y,z}, G_{z}}|_{B_0}\circ (\pi_{G_{y,z}, G_{y}}|_{B_0})^{-1}$ is a bijection  from $B_1$ to $B_2$ sending $G_y e$ to $G_z e$.
Let $B_1'$ and $B_2'$ be the blocks of $C_{G_x}$ containing  $G_x g^{-1}$ and $G_x g'^{-1}$ respectively. 
We have the conjugations $c_{G_x, g}|_{B_1'}: B_1'\to B_1$ sending $G_x g^{-1}$ to $G_{y} e$
and $c_{G_{z},g'^{-1}}|_{B_2}: B_2\to B_2'$ sending $G_{z} e$ to $G_x g'^{-1}$.
Then the map
\[
c_{G_{z},g'^{-1}}|_{B_2}\circ \pi_{G_{y,z}, G_{z}}|_{B_0}\circ (\pi_{G_{y,z}, G_{y}}|_{B_0})^{-1}\circ c_{G_x, g}|_{B_1'}
\]
is a bijection from $B_1'$ to $B_2'$ sending $G_x g^{-1}$ to $G_x g'^{-1}$.
\end{proof}

%We choose $H=G_{\prescript{g}{}{x},\prescript{g'}{}{x}}$ in most applications of Lemma~\ref{lem_selfred}. 

This provides a way of proving discreteness of a strongly antisymmetric $\mathcal{P}$-scheme using discreteness of   its restrictions to stabilizers.  For example, if $\mathcal{C}$ is strongly antisymmetric and the conditions in Lemma~\ref{lem_selfred} hold for all pairs $(y,z)\in G x\times G x$, then $\mathcal{C}$ is discrete on $G_x$. 
In fact,  we only need to verify the conditions for a subset of pairs that form a connected graph:
%This is the key idea we use to bound $m$ for the actions of $\sym(S)$ on sets of $k$-subsets or partitions.

\begin{lem}[self-reduction lemma]\label{lem_selfredspan}
Let $G$ be a  finite group acting on a finite set $S$, and let $x\in S$. Let $\mathcal{P}$ be a subgroup system over $G$ such that $G_x\in\mathcal{P}$, and let $\mathcal{C}=\{C_H: H\in\mathcal{P}\}$ be a strongly antisymmetric $\mathcal{P}$-scheme.
Suppose $R$ is a subset of $S\times S$ satisfying the following conditions:
\begin{enumerate}
\item For all $(y,z)\in R$, it holds that (1) $G_{y}, G_{z}, G_{y,z}\in\mathcal{P}$ and (2) $\mathcal{C}|_{G_{y}}$ and $\mathcal{C}|_{G_{z}}$ are both discrete on $G_{y,z}$.
\item Let $\mathcal{G}_R$ be the undirected graph on $S$ such that $\{y,z\}$ is an edge iff $(y,z)\in R$ or $(z,y)\in R$. Then $G x$ is contained in a connected component of $\mathcal{G}_R$ (in particular, this condition is satisfied if $\mathcal{G}_R$ is connected).
\end{enumerate} 
Then  $\mathcal{C}$ is discrete on $G_x$.
\end{lem}
\begin{proof}
For $y\in S$, denote by $B_y$ the block of $C_{G_y}$ containing $G_y e\in G_y\backslash G$.
For $(y,z)\in S\times S$, write $y\sim z$ if there exists a bijection $\tau: B_y\to B_z$ sending $G_y e$ to $G_z e$ 
 such that $\tau$ is a composition of  
maps of the form  $\pi_{H,H'}|_B$ or $(\pi_{H,H'}|_B)^{-1}$ (where $H, H\in\mathcal{P}$ and $B$ is block of $C_H$).
Then $\sim$ is an equivalence relation on $S$.
By the first condition and Lemma~\ref{lem_selfred0}, we have $y\sim z$ for all $(y,z)\in R$.
And by the second condition, we have $y\sim z$ for all $(y,z)\in G x \times G x$.

Consider any $g,g'\in G$ and  let $y=\prescript{g}{}{x},z=\prescript{g'}{}{x}\in Gx$. Let  $\tau: B_y\to B_z$ be a bijection sending $G_y e$ to $G_z e$ as above. Let $B$ and $B'$ be the blocks of $C_{G_x}$ containing  $G_x g^{-1}$ and $G_x g'^{-1}$ respectively. 
We have the conjugations $c_{G_x, g}|_{B}: B\to B_y$ sending $G_x g^{-1}$ to $G_{y} e$
and $c_{G_{z},g'^{-1}}|_{B_z}: B_z\to B'$ sending $G_{z} e$ to $G_x g'^{-1}$.
Then the map
\[
c_{G_{z},g'^{-1}}|_{B_z}\circ \tau \circ c_{G_x, g}|_{B}
\]
is a bijection from $B$ to $B'$ sending $G_x g^{-1}$ to $G_x g'^{-1}$.
In particular, if $G_x g^{-1}\neq G_x g'^{-1}$, then $B\neq B'$ by strong antisymmetry of $\mathcal{C}$.
As $g,g'\in G$ are arbitrary, we know $\mathcal{C}$ is discrete on $G_x$.
\end{proof}

%\begin{exmp}
%Suppose $G=\sym(S)$ acts transitively on a finite set $S$. We say two elements $x,y$ of a $G$-set $S$ are {\em adjacent} if there exists a transposition of $S$ sending $x$ to $y$.
%Let $R$ be the set of all pairs $(x,y)\in S\times S$ where $x,y$ are adjacent. 
%As $\sym(S)$ is generated by transpositions,  the graph $\mathcal{G}_R$ in Lemma~\ref{lem_selfredspan} is connected.
%\end{exmp}

\section{The actions of symmetric groups on \texorpdfstring{$k$-subsets}{k-subsets} or partitions}\label{sec_symstandard}\index{standard action}

Let $S$ be a finite set of cardinality $n$, and let $G=\sym(S)$.
For $k\in [n]$, the natural action of $G$ on $S$ induces a (transitive) action of $G$  on the set of $k$-subsets (i.e., subsets of cardinality $k$) of $S$. Similarly, it induces an action of $G$ on (an orbit of) the set of partitions $P$ of $S$,  given by $\prescript{g}{}{P}:=\{\prescript{g}{}{B}: B\in P\}$.

In these cases, we expect to have a bound $d(G)=O(\log n)$ as we have in Section~\ref{sec_symnatural}.
Let $S'$ be the underlying set on which $G$ acts.
The naive approach  is to embed $G$ in $\sym(S')$ and apply Corollary~\ref{cor_ubsysp}.
In general, however, the cardinality of $S'$ is much larger than $n$. For example,  we have $|S'|=\binom{n}{k}$  for the action of $G$ on the set of $k$-subsets of $S$, and hence Corollary~\ref{cor_ubsysp} only implies the bound $d(G)=O(\log |S'|)=O(k\log n)$. The same problem exists for the action of $G$ on an orbit of the set of partitions of $S$, in which case $|S'|$ is the number of partitions of $S$ into subsets with prescribed cardinalities.

In this section, we extend the result in Section~\ref{sec_symnatural} and show that in the above cases, we  have $d(G)\leq d_{\sym}(n)+O(1)=O(\log n)$. 
In fact, we prove  more general criteria for a subgroup system $\mathcal{P}$ over $G$ (or more generally, over a subgroup $H\subseteq G$) to have the property that all strongly antisymmetric $\mathcal{P}$-schemes are discrete on all $x\in S'$.\footnote{To derive $d(G)\leq d_{\sym}(n)+O(1)$, we only need the case $H=G$.   The more general setting $H\subseteq G$ is needed for applications in Chapter~\ref{chap_primitive}.}
It is possible to design a subgroup system $\mathcal{P}$   of complexity $|S'|^{O(1)} n^{O(\log n)}$  that satisfies these criteria. An algorithm of constructing the corresponding collection of number fields will be given in Chapter~\ref{chap_primitive}.
%There exist  subgroup systems $\mathcal{P}$ satisfying these criteria whose complexity $c(\mathcal{P})$ is bounded by $|S'|^{O(1)} n^{O(\log n)}$.

\paragraph{The action of $\sym(S)$ on the set of $k$-subsets of $S$.}

Suppose $S$ is a finite set of cardinality $n$  and consider the action of $G=\sym(S)$ on the set $S'$ of $k$-subsets of $S$. 
We say two elements $x,y\in S'$ are {\em adjacent} if there exists $g\in G$ sending $x$ to $y$ and $g$ is a transposition (i.e. 2-cycle) on $S$.
The following technical lemma is needed:

\begin{lem}\label{lem_orbitbsub}
For  all adjacent $x,y\in S$ and $z\in G_{x}y$ adjacent to $y$, it holds that $|G_{x,y}z|\leq n$.
\end{lem}
\begin{proof}
Choose $h\in G_x$ that sends $y$ to $z$.
As $x$ and $y$ are adjacent, we know $x=\prescript{h}{}{x}$ and $z=\prescript{h}{}{y}$ are also adjacent.
Let $u=x\cap y$ (as the intersection of two $k$-subsets).
As $x$ and $y$ are adjacent, there exist distinct elements $a,b\in S$ such that $x=u\cup\{a\}$ and $y=u\cup\{b\}$. 
Then $G_{x,y}$ fixes $u$ setwisely as well as $a,b$.
If $b\not\in z$, we have $z=u\cup\{c\}$ for some $c\in S$ since $y$ and $z$ are adjacent.
In this case, as $G_{x,y}$ fixes the subset $u$ of $z$ of cardinality $k-1$ setwisely, we have $|G_{x,y}z|\leq |S|=n$, as desired.
Next assume $b\in z$. Since $x$ and $z$ are adjacent, we have $z=(x-\{b'\})\cup\{b\}$ for some $b'\in x$. As $G_{x,y}$ fixes $x$ setwisely as well as $a,b\in S$, the elements in $G_{x,y}z$ are of the form $(x-\{b''\})\cup\{b\}$ where $b''\in x$. In this case we have $|G_{x,y}z|\leq |x|=k\leq n$.
\end{proof}
 
 We state a criterion for a subgroup system $\mathcal{P}$  over a subgroup $H\subseteq G$ to have the property that all strongly antisymmetric $\mathcal{P}$-schemes are discrete on all $x\in S'$. 
 
\begin{thm}\label{thm_tuples}
Let $G$, $n$, and $S'$ be as above, and let $H$ be a subgroup of $G$. Suppose $\mathcal{P}$ is a subgroup system over $H$ satisfying the following conditions:
\begin{enumerate}
\item $H_x, H_{x,y}\in \mathcal{P}$ for all $x,y\in S'$.
\item $H_{\{x,y\}\cup T}\in \mathcal{P}$ for all $x,y,z\in S'$ and $T\subseteq H_{x,y} z$ satisfying $|H_{x,y} z|\leq n$ and $1\leq |T|\leq  d_{\sym}(n)$.
\end{enumerate}
 Then all strongly antisymmetric $\mathcal{P}$-schemes are discrete on $H_x$ for all $x\in S'$.
% In particular, this is the case if $\mathcal{P}=\mathcal{P}_{S,m}$ where $m\geq m_{\sym}(d)+2$.
\end{thm}

\begin{proof}
Let $\mathcal{C}=\{C_H: H\in\mathcal{P}\}$ be a strongly antisymmetric $\mathcal{P}$-scheme.
We want to prove that $\mathcal{C}$ is discrete  on $H_x$ for all $x\in S'$.
As $G$ is generated by transpositions on $S$, by Lemma~\ref{lem_selfredspan}, we just need to verify for all adjacent $x,y\in S'$ that (1) $H_x,H_y,H_{x,y}\in\mathcal{P}$, and (2)  $\mathcal{C}|_{H_x}$ and  $\mathcal{C}|_{H_y}$ are discrete on $H_{x,y}$.
Fix adjacent $x,y\in S'$.
Note that (1) follows from the first condition in the theorem.

 So it remains to prove that $\mathcal{C}|_{H_x}$ is discrete on $H_{x,y}$ (the claim for $\mathcal{C}|_{H_y}$ is symmetric). This is trivial if $x=y$. So assume $x\neq y$.
%Next apply Lemma~\ref{lem_selfredspan} to the action of $G_x$ on $G_{x}y$: 
we claim that for all $z,w\in H_x y\subseteq G_{x}y$, there exists a sequence of elements $u_0,\dots,u_t\in G_{x}y$ such that $u_0=z$, $u_t=w$, and $u_{i-1}, u_i$ are adjacent for $i\in [t]$.  
This follows from the fact that $G_x\cong \sym(x) \times \sym(S-x)$ is generated by transpositions on $S$.
Let $\mathcal{P}':=\mathcal{P}|_{H_x}$ and $\mathcal{C}':=\mathcal{C}|_{H_x}$. 
By Lemma~\ref{lem_selfredspan} and the previous claim,
it remains to show that for all adjacent $z,w\in G_{x}y$, it holds that (a) $(H_x)_z, (H_x)_w, (H_x)_{z,w}\in\mathcal{P}'$, or equivalently $H_{x,z}, H_{x,w}, H_{x,z,w}\in\mathcal{P}$, and (b) $\mathcal{C}'|_{(H_x)_z}$ and $\mathcal{C}'|_{(H_x)_w}$ are discrete on $(H_x)_{z,w}$.
Fix such $z,w\in G_{x}y$.
Note that $z$ and $w$ are adjacent to $x$ since $y$ is adjacent to $x$.
%\footnote{As $y$ is adjacent to $x$, an element $\prescript{g}{}{y}\in G_{x}y$ is adjacent to $\prescript{g}{}{x}=x$ for all $g\in G_x$.} 
It follows from Lemma~\ref{lem_orbitbsub} that $|H_{x,z}w|\leq |G_{x,z}w|\leq n$.
 Then (a) follows from the two conditions in the theorem.

It remains to show that $\mathcal{C}'|_{(H_x)_z}$ is  discrete on $(H_x)_{z,w}$ (the claim for $\mathcal{C}'|_{(H_x)_w}$ is symmetric). 
Let $\mathcal{P}'':=\mathcal{P}'|_{(H_x)_z}$.
%By Lemma~\ref{lem_orbitbsub}, we have $|H_{x,z}w|\leq n$.
By the second condition of the theorem and the fact $|H_{x,z}w|\leq n$, we have $H_{\{x,z\}\cup T}\in \mathcal{P}''$ for all $T\subseteq H_{x,z}w$ satisfying $1\leq |T|\leq d_\sym(n)$. 
This means that $\mathcal{P}''$ contains the system of stabilizers of depth $d_\sym(n)$ with respect to the action of $H_{x,z}$ on $H_{x,z} w$. By Corollary~\ref{cor_monotonein} and the fact $|H_{x,z}w|\leq n$, we see all strongly antisymmetric $\mathcal{P}''$-schemes are discrete on $(H_{x,z})_w=(H_x)_{z,w}$. Finally, note that $\mathcal{C}'|_{(H_x)_z}$ is strongly antisymmetric by Lemma~\ref{lem_mres}, and hence discrete on $(H_x)_{z,w}$, as desired.
\end{proof}

Choosing $H=G$ in Theorem~\ref{thm_tuples}, we obtain

\begin{cor}
$d(G)\leq d_{\sym}(n)+2$.
\end{cor}

\paragraph{The action of $\sym(S)$ on the set of partitions of $S$.}

Suppose $S$ is a finite set of cardinality $n$  and consider the action of $G=\sym(S)$ on an orbit $S'$ of the set of partitions of $S$.
We prove an analogue of Theorem~\ref{thm_tuples} for this case.
The following notations are needed: 
again, we call two elements $x,y\in S'$  {\em adjacent} if there exists $g\in G$ sending $x$ to $y$ and $g$ is a transposition   on $S$.
For $x,y,z\in S'$,   write $y\sim_x z$ if there exists $g\in G_x$ sending $y$ to $z$ such that either (1) $g$ is a transposition on $S$ fixing all the blocks of $x$ setwisely, or (2) $x-y\neq x-z$, and $g$ exchanges two blocks of $x$ while fixing the other blocks of $x$ pointwisely. 

We also need the following technical lemma:
\begin{lem}\label{lem_orbitbpar}
For  all adjacent $x,y\in S'$ and $z\in G_{x}y$ satisfying $y\sim_x z$, it holds that $|G_{x,y}z|\leq 4n$.
\end{lem}
\begin{proof}
We may assume $x\neq y$.
As $x$ and $y$ are adjacent, there exists a transposition $(a~b)$ of $S$ sending $x$ to $y$ where $a\in B_1$, $b\in B_2$ and $B_1$, $B_2$ are distinct blocks of $x$. 
So we have 
\begin{equation}\label{eq_yvalue}
y=(x-\{B_1,B_2\})\cup  \{(B_1-\{a\})\cup \{b\}, (B_2-\{b\})\cup\{a\}\}.
\end{equation}
Fix $h\in G_x$ sending $y$ to $z$ such that either (1) $h$ is a transposition on $S$ fixing all the blocks of $x$ setwisely, or (2) $x-z\neq x-y$ and $h$ exchanges two blocks of $x$ while fixing the other blocks of $x$ pointwisely. 
We claim that in either case, $h$ fixes at least one of $B_1$ and $B_2$ pointwisely. This is obvious in Case (1). And in Case (2), if $h$  fixes neither $B_1$ nor $B_2$ pointwisely, it exchanges $B_1$ and $B_2$. But then we have $x-y=x-z=\{B_1,B_2\}$, contradicting the assumption.

So assume $h$ fixes $B_1$ pointwisely (the other case is symmetric). 
%Then $y|_{B_1}=z|_{B_1}=\{B_1-\{a\},\{a\}\}$.
Consider arbitrary $w=\prescript{g}{}{z}\in G_{x,y}z$ where $g\in G_{x,y}$.
We have
\[
w=\prescript{gh(a~b)}{}{x}=\prescript{gh(a~b)(gh)^{-1}}{}{x}=\prescript{(a'~b')}{}{x},
\]
where $a'=\prescript{gh}{}{a}$ and $b'=\prescript{gh}{}{b}$.
%As $g$ fixes $x$ and $y$, it also fixes $x-y=\{B_1,B_2\}$ setwisely. 
So $w$ is determined by  the pair $(a',b')$.

There are at most $n$ choices of $b'\in S$.
Now consider the number of choices of $a'$.
Note that $a'=\prescript{gh}{}{a}=\prescript{g}{}{a}$ since $h$ fixes $B_1$ pointwisely.
As $\{a\}\in y|_{B_1}$, we see $\{a'\}\in \prescript{g}{}{y}|_{\prescript{g}{}{B_1}}=y|_{\prescript{g}{}{B_1}}$.
As $g$ fixes $x$ and $y$, it  fixes $x-y=\{B_1,B_2\}$ setwisely. 
So $\prescript{g}{}{B_1}\in \{B_1,B_2\}$. It follows that $\{a'\}$ is in $y|_{B_1}$ or $y|_{B_2}$.
By \eqref{eq_yvalue}, we see $\{a'\}$ equals $\{a\}$, $\{b\}$, $B_1-\{a\}$ or $B_2-\{b\}$.
So the number of choices of $a'$ is at most four.
Therefore $|G_{x,y}z|\leq 4n$.
\end{proof}

 We have following criterion for a subgroup system $\mathcal{P}$  over  a subgroup $H\subseteq G$ to have the property that all strongly antisymmetric $\mathcal{P}$-schemes are discrete on all $x\in S'$.

\begin{thm}\label{thm_partitions}
Let $G$, $n$, and $S'$ be as above, and let $H$ be a subgroup of $G$.  Suppose $\mathcal{P}$ is a subgroup system over $H$ satisfying the following conditions:
\begin{enumerate}
\item $H_x, H_{x,y}\in \mathcal{P}$ for all $x,y\in S'$.
\item $H_{\{x,y\}\cup T}\in \mathcal{P}$ for all $x,y,z\in S'$ and $T\subseteq H_{x,y} z$ satisfying $|H_{x,y} z|\leq 4n$ and $1\leq |T|\leq  d_{\sym}(4n)$.
\end{enumerate}
 Then all strongly antisymmetric $\mathcal{P}$-schemes are discrete on $H_x$ for all $x\in S'$.
\end{thm}

\begin{proof}
Let $\mathcal{C}=\{C_H: H\in\mathcal{P}\}$ be a strongly antisymmetric $\mathcal{P}$-scheme.
We want to prove that $\mathcal{C}$ is discrete  on $H_x$ for all $x\in S'$.
As $G$ is generated by transpositions on $S$, by Lemma~\ref{lem_selfredspan}, we just need to verify for all adjacent $x,y\in S'$ that (1) $H_x,H_y,H_{x,y}\in\mathcal{P}$, and (2)  $\mathcal{C}|_{H_x}$ and  $\mathcal{C}|_{H_y}$ are discrete on $H_{x,y}$.
Fix adjacent $x,y\in S'$.
Note that (1) follows from the first condition in the theorem.
 So it remains to prove that $\mathcal{C}|_{H_x}$ is discrete on $H_{x,y}$ (the claim for $\mathcal{C}|_{H_y}$ is symmetric). 
%Next we apply Lemma~\ref{lem_selfredspan} to the action of $H_y$ on $H_{y}z$: 
This obviously holds if $x=y$. So assume $x\neq y$. 

We claim that for all $z,w\in H_{x}y\subseteq G_{x}y$, there exists a sequence of elements $u_0,\dots,u_t\in G_{x}y$ such that $u_0=z$, $u_t=w$ and $u_{i-1}\sim_x u_i$ for $i\in [t]$.  
To see this, note that we can choose distinct elements $u_0,\dots,u_t$ such that $u_0=z$, $u_t=w$, and  for $i\in [t]$, $u_{i-1}$ is sent to $u_i$ by some $g_i\in G_x$ such that $g_i$ is in either of the following two cases:
\begin{enumerate}
\item $g_i$ is  a transposition on $S$ fixing the blocks of $x$ setwisely, or
\item $g_i$  exchanges two blocks of $x$ while fixing the other blocks of $x$ pointwisely. 
\end{enumerate}
This is because $G_x$ is generated by such permutations $g_i$.
Furthermore, if $g_i$ is in the latter case, we may assume $x-u_{i-1}\neq x-u_i$. To see this, note that 
 $u_{i-1}$ and $u_i$ are adjacent to $x$ since $y$ is adjacent to $x$. 
So there exist transpositions $(a~b)$ and $(a'~b')$  on $S$ sending $x$ to $u_{i-1}$ and $u_i$ respectively.
Choose $B_1,B_2,B'_1,B'_2\in x$ such that $a\in B_1$, $b\in B_2$, $a'\in B'_1$ and $b'\in B'_2$.
Then $x-u_{i-1}=\{B_1,B_2\}$ and $x-u_i=\{B'_1,B'_2\}$.
Suppose $x-u_{i-1}=x-u_i$. Then by exchanging $a'$ with $b'$ and $B'_1$ with $B'_2$ if necessary, we may assume  $a,a'\in B_1$ and $b,b'\in B_2$.
So we have
\[
u_{i-1}=x-\{B_1,B_2\}\cup \{(B_1-\{a\})\cup \{b\}, (B_2-\{b\})\cup\{a\}\}
\]
 and
 \[
u_{i}=x-\{B_1,B_2\}\cup \{(B_1-\{a'\})\cup \{b'\}, (B_2-\{b'\})\cup\{a'\}\}.
\]
   If $a=a'$. Then $u_i=\prescript{(b~b')}{}{u_{i-1}}$ and we may replace $g_i$ by $(b~b')\in G_x$ which is in the first case above. Similarly, if $b=b'$, we may replace $g_i$ by $(a~a')\in G_x$. Finally, if $a\neq a'$ and $b\neq b'$, we insert $u'_i=\prescript{(a~a')}{}{u_{i-1}}$ into the sequence between $u_{i-1}$ and $u_i$, so that  $u_i=\prescript{(b~b')}{}{u'_i}$. 
It follows that we may always assume $u_{i-1}\sim_x u_i$ for all $i\in [t]$.
 
Let $\mathcal{P}':=\mathcal{P}|_{H_x}$ and $\mathcal{C}':=\mathcal{C}|_{H_x}$. 
By Lemma~\ref{lem_selfredspan} and the previous paragraph,
it suffices to show, for all $z,w\in G_{x}y$ satisfying $z\sim_x w$,  that (a) $H_{x,z}, H_{x,w}, H_{x,z,w}\in\mathcal{P}'$, and (b) $\mathcal{C}'|_{(H_x)_z}$ and $\mathcal{C}'|_{(H_x)_w}$ are discrete on $(H_x)_{z,w}$.
%As $z,w\in G_{x}y$ are also adjacent to $x$, we have $H_{x,z}, H_{x,w}\in\mathcal{P}$. 
Fix such $z,w\in G_{x}y$.
Note that $z$ and $w$ are adjacent to $x$ since $y$ is adjacent to $x$.
%\footnote{As $y$ is adjacent to $x$, an element $\prescript{g}{}{y}\in G_{x}y$ is adjacent to $\prescript{g}{}{x}=x$ for all $g\in G_x$.} 
It follows from Lemma~\ref{lem_orbitbpar} that $|H_{x,z}w|\subseteq|G_{x,z}w|\leq 4n$.
 Then (a) follows from the two conditions in the theorem.

It remains to show that $\mathcal{C}'|_{(H_x)_z}$ is  discrete on $(H_x)_{z,w}$ (the claim for $\mathcal{C}'|_{(H_x)_w}$ is symmetric). 
Let $\mathcal{P}'':=\mathcal{P}'|_{(H_x)_z}$.
%By Lemma~\ref{lem_orbitbpar}, we have $|H_{x,z}w|\leq 4n$.
By the second condition of the theorem and the fact $|H_{x,z}w|\leq 4n$, we have $H_{\{x,z\}\cup T}\in \mathcal{P}''$ for all $T\subseteq H_{x,z}w$ satisfying $1\leq |T|\leq d_\sym(4n)$. 
This means that $\mathcal{P}''$ contains the system of stabilizers of depth $d_\sym(4n)$ with respect to the action of $H_{x,z}$ on $H_{x,z} w$. By Corollary~\ref{cor_monotonein} and the fact $|H_{x,z}w|\leq 4n$, we see all strongly antisymmetric $\mathcal{P}''$-schemes are discrete on $(H_x)_{z,w}$. Finally note that $\mathcal{C}'|_{(H_x)_z}$ is strongly antisymmetric by Lemma~\ref{lem_mres}, and hence is discrete on $(H_x)_{z,w}$, as desired.
\end{proof}

Note  $d_{\sym}(4n)\leq d_{\sym}(n)+O(1)$   by Lemma~\ref {lem_recrel} and Theorem~\ref{thm_mandp}. 
Choosing $H=G$ in Theorem~\ref{thm_partitions}, we obtain
\begin{cor}
$d(G)\leq d_{\sym}(4n)+2\leq d_{\sym}(n)+O(1)$.
\end{cor}

\section{The natural actions of linear groups}\label{sec_lingroup}\index{natural action of linear groups}

In this section,  we show that $d_{\gl}(n,q)$, $d_{\gammal}(n,q)$, $d_{\pgl}(n,q)$, $d_{\pgammal}(n,q)$ are equal up to an additive constant. In addition, we prove an upper bound for $d_{\gl}(n,q)$, slightly improving the trivial bounds.

\paragraph{Equivalence between various linear groups.}

We have the following theorem.
\begin{thm}\label{thm_equivlin}
For  $f_1,f_2\in\{d_{\gl}, d_{\gammal}, d_{\pgl}, d_{\pgammal}\}$, there exists a constant $c\in\N$ such that
$f_1(n,q)\leq f_2(n,q) + c$ holds for all $n\in\N^+$ and prime powers $q$. And if $f_2=d_{\gl}$, choosing $c=6$ suffices.
\end{thm}

We break Theorem~\ref{thm_equivlin} into six inequalities, corresponding to the the arrows in the following diagram.
\[
\begin{tikzcd}
\gl(V) \arrow[Leftrightarrow]{r} \arrow[Leftrightarrow]{d}
& \gammal(V)   \\
\pgl(V)  \arrow[Leftrightarrow]{r}
& \pgammal(V) 
\end{tikzcd}
\]

Fix $n\in\N^+$, a prime power $q$ and a vector space $V$ of dimension $n$ over $\F_q$ from now on.
By Lemma~\ref{cor_monotonein} and the facts $\gl_n(q)\subseteq\gammal_n(q)$ and $\pgl_n(q)\subseteq\pgammal_n(q)$, we have

\begin{lem}\label{lem_gltogammal}
$d_{\gl}(n,q)\leq d_{\gammal}(n,q)$ and $d_{\pgl}(n,q)\leq d_{\pgammal}(n,q)$.
\end{lem}

In the other direction, we have 
\begin{lem}\label{lem_gammaltogl}
$d_{\gammal}(n,q)\leq d_{\gl}(n,q)+2$. 
\end{lem}
\begin{proof}
Let $G=\gammal(V)$, $S=V-\{0\}$, and $m=d_{\gl}(n,q)+2\geq 3$. Let $\mathcal{P}$ be the system of stabilizers of depth $m$ over $G$ with respect to the natural action of $G$ on $S$. 
Let $\mathcal{C}$ be a strongly antisymmetric $\mathcal{P}$-scheme.
Fix $x\in S$.
We want to show that $\mathcal{C}$ is discrete on $G_x$. Let $\alpha$ be an element in $\F_q^\times$ not contained in any proper subfield of $\F_q$. By Lemma~\ref{lem_min}, it suffices to show that $\mathcal{C}$ is discrete on $G_{x,\alpha x}$.
Let $G$ act diagonally on $S\times S$ and let $O$ be the $G$-orbit of $(x,\alpha x)$.
By Lemma~\ref{lem_selfredspan}, it suffices to prove, for all $u,v\in O$,  that (1)
$G_u,G_v,G_{u,v}\in \mathcal{P}$ and  (2) $\mathcal{C}|_{G_u}$ and $\mathcal{C}|_{G_v}$ are discrete on $G_{u,v}$.

Fix $u,v\in O$.
Note we have $u=(y,\beta y)$ and $v=(z,\gamma z)$ for some $y,z\in S$ and $\beta,\gamma\in\F_q^\times$. And $\beta,\gamma$ are not contained in any proper subfield of $\F_q$.
Let $g\in G_u$.
Then $g$ sends $\beta y$ to $\tau_{g}(\beta) y=\beta y$ where $\tau_{g}$ is the automorphism of $\F_q^\times$ determined by $g$. So  $\beta$ is in the subfield fixed by the cyclic group generated by $\tau_{g}$. As $\beta$ is not in any proper subfield of $\F_q$, we conclude that $\tau_{g}$ is the identity.  So $G_u\subseteq \gl(V)$.
We have $G_{u}=G_{y,\beta y}$, $G_{v}=G_{z,\gamma z}$, and $G_{u,v}=G_{y,\beta y, z,\gamma z}=G_{y,\beta y, z}$.
As $m\geq 3$, these subgroups are all in $\mathcal{P}$.

It remains to prove that $\mathcal{C}|_{G_u}$ is discrete on $G_{u,v}$ (the claim for  $\mathcal{C}|_{G_v}$ is symmetric).
Let $\mathcal{P}'$ be the system of stabilizers of depth $m-2$ over $G_u$ with respect to the action of $G_u$ on $S$. 
As $G_u\subseteq \gl(V)$, we have  $d(G_u)\leq d_{\gl}(n,q) =m-2$ by Corollary~\ref{cor_monotonein}. So all strongly antisymmetric $\mathcal{P}'$-schemes are discrete on $(G_u)_z=G_{u,v}$.
Also note $\mathcal{P}'\subseteq \mathcal{P}|_{G_u}$ since $G_{u}=G_{y,\beta y}$. It follows that all strongly antisymmetric $\mathcal{P}|_{G_u}$-schemes are discrete on $G_{u,v}$.
As $\mathcal{C}|_{G_u}$ is strongly antisymmetric by Lemma~\ref{lem_res}, it is discrete on $G_{u,v}$, as desired.
\end{proof}

Similarly, we have
\begin{lem}\label{lem_pgammaltopgl}
$d_{\pgammal}(n,q)\leq d_{\pgl}(n,q)+4$. 
\end{lem}
\begin{proof}

Let $G=\pgammal(V)$, $S=\proj V$, and $m=d_{\gl}(n,q)+4\geq 5$. Let $\mathcal{P}$ be the system of stabilizers of depth $m$ over $G$ with respect to the natural action of $G$ on $S$. 
Let $\mathcal{C}$ be a strongly antisymmetric $\mathcal{P}$-scheme.
By Lemma~\ref{lem_selfredspan}, it suffices to prove, for all $(w,w')\in S^{(2)}$, that $\mathcal{C}|_{G_{w}}$  is discrete on $G_{w,w'}$. Fix $(w,w')\in S^{(2)}$. Again by Lemma~\ref{lem_selfredspan}, it suffices to prove, for all $(x,x')\in (G_w w')^{(2)}$,
that $\mathcal{C}|_{G_{w,x}}$  is discrete on $G_{w,x,x'}$
 (note $G_{w,x},G_{w,x'},G_{w,x,x'}\in\mathcal{P}$  since $m\geq 3$). 
 
 Fix $(x,x')\in (G_w w')^{(2)}$.
Choose  representatives  $\tilde{w},\tilde{x},\tilde{x}'\in V-\{0\}$  of $w$, $x$ and $x'$ respectively.
Note that $\tilde{w}$, $\tilde{x}$ and $\tilde{x}'$ are pairwise linearly independent over $\F_q$ since $w, x, x'$ are distinct.
Let $\alpha$ be an element in $\F_q^\times$ not contained in any proper subfield  of $\F_q$.
Define $\tilde{y}=\tilde{w}+\alpha \tilde{x}\neq 0$ and let $y$ be the element in $S$ represented by $\tilde{y}$.
%By regularity of $\mathcal{C}|_{G_{w,x}}$, it suffices to prove that $\mathcal{C}|_{G_{w,x}}$ is discrete on $G_{w,x,x',y}$.
Consider the diagonal action of $G_{w,x}$ on $S^2$ and let $O$ be the orbit of $(x',y)$ under this action.
We have $(G_{w,x})_{(x',y)}=G_{w,x,x',y}\in\mathcal{P}|_{G_x}$ since $m\geq 4$.
By  Lemma~\ref{lem_min}, it suffices to prove  that $\mathcal{C}|_{G_{w,x}}$ is discrete on $(G_{w,x})_{(x',y)}$. Let $G'=G_{w,x}$. Applying Lemma~\ref{lem_selfredspan} to the action of $G'$ on $O$, we see that it suffices to prove for all $u,v\in O$ that (1) $G'_u, G'_v, G'_{u,v}\in\mathcal{P}|_{G'}$ and (2) $\mathcal{C}|_{G'_u}$ and $\mathcal{C}|_{G'_v}$ are discrete on $G'_{u,v}$.

Fix $u=(x'_1,y_1)=\prescript{g_1}{}{(x',y)}$ and $v=(x'_2,y_2)=\prescript{g_2}{}{(x',y)}$ in $O$, where $g_1,g_2\in G'$. 
Lift $g_1$ to $\tilde{g}_1\in\gammal(V)$. As $g_1\in G_x$, we have $\prescript{\tilde{g}_1}{}{\tilde{x}}=c \tilde{x}$ for unique $c\in\F_q^\times$.
Define $\tilde{x}'_1=\prescript{\tilde{g}_1}{}{\tilde{x}'}$ and $\tilde{y}_1=\prescript{\tilde{g}_1}{}{\tilde{y}}$ so that they are representatives of $x'_1$ and $y_1$ respectively.
Consider arbitrary $g\in G'_u=G_{w,x,x'_1,y_1}$. 
We claim $g\in\pgl(V)$. To see this, lift $g$ to $\tilde{g}\in \gammal(V)$.
Note that $\tilde{y}_1=\prescript{\tilde{g}_1}{}{(\tilde{w}+\alpha\tilde{x})}=\prescript{\tilde{g}_1}{}{\tilde{w}}+\alpha_1\prescript{\tilde{g}_1}{}{\tilde{x}}$ where $\alpha_1=\tau_{\tilde{g}_1}(\alpha)$ and $\tau_{\tilde{g}_1}$ is the automorphism of $\F_q^\times$ determined by $\tilde{g}_1$. Here $\prescript{\tilde{g}_1}{}{\tilde{w}}$ and $\prescript{\tilde{g}_1}{}{\tilde{x}}$ are collinear with $\tilde{w}$ and $\tilde{x}$ respectively since $g_1\in G_{w,x}$. And   $\prescript{\tilde{g}_1}{}{\tilde{w}}$, $\prescript{\tilde{g}_1}{}{\tilde{x}}$  are linearly independent over $\F_q$ since $\tilde{w}$ and $\tilde{x}$ are linearly independent. As $g\in G_{w,x,x',y_1}$, we see that $\tilde{g}$ scales $\prescript{\tilde{g}_1}{}{\tilde{w}}$, $\prescript{\tilde{g}_1}{}{\tilde{x}}$  and $\tilde{y}_1=\prescript{\tilde{g}_1}{}{\tilde{w}}+\alpha_1\prescript{\tilde{g}_1}{}{\tilde{x}}$. Therefore $\tau_{\tilde{g}}(\alpha_1)=\alpha_1$, where $\tau_{\tilde{g}}$ is the automorphism of $\F_q^\times$ determined by $\tilde{g}$. 
 So $\alpha_1$ is in the subfield fixed by the cyclic group generated by $\tau_{\tilde{g}}$.  But $\alpha_1=\tau_{\tilde{g}_1}(\alpha)$ is not in any proper subfield of $\F_q$. It follows  that $\tau_{\tilde{g}}$ is the identity.  So we have $\tilde{g}\in \gl(V)$ and hence 
 $g\in \pgl(V)$. 
 We conclude that $G'_u=\pgl(V)_{w,x,x'_1,y_1}$, and similarly $G'_v=\pgl(V)_{w,x,x'_2,y_2}$.
 Moreover, observe that $\tilde{g}$ above scales $\tilde{w}$ and $\tilde{x}$ by the same factor since $g$ fixes $y_1$.
 So it also scales any vector in the span of $\tilde{w}$ and $\tilde{x}$ over $\F_q$. We know $\tilde{y}_1$ is in this span and by the same argument, so is $\tilde{y}_2$. So $g$ fixes $y_2$.
This shows $G'_{u,v}=\pgl(V)_{w,x,x'_1,y_1,x'_2}$.
 We then have $G'_u, G'_v, G'_{u,v}\in\mathcal{P}|_{G'}$ since $m\geq 5$.

It remains to prove that $\mathcal{C}|_{G'_u}$ is discrete on  $G'_{u,v}$  (the claim for $\mathcal{C}|_{G'_v}$  is symmetric). 
Let $\mathcal{P}'$ be the system of stabilizers of depth $m-4$ over $G'_u$ with respect to the action of $G'_u$ on $S$. 
As $G'_u\subseteq \pgl(V)$, we have  $d(G'_u)\leq d_{\pgl}(n,q) =m-4$ by Corollary~\ref{cor_monotonein}. So all strongly antisymmetric $\mathcal{P}'$-schemes are discrete on $(G'_u)_{x'_2}=G'_{u,v}$.
Also note $\mathcal{P}'\subseteq \mathcal{P}|_{G'_u}$ since $G'_u=G_{w,x,x'_1,y_1}$. 
It follows that all strongly antisymmetric $\mathcal{P}|_{G'_u}$-schemes are discrete on $G'_{u,v}$.
As $\mathcal{C}|_{G'_u}$ is strongly antisymmetric by Lemma~\ref{lem_res}, it is discrete on $G'_{u,v}$, as desired.
\end{proof}

It remains to show the equivalence between $\gl(V)$ and $\pgl(V)$. To achieve this, we need a lemma about pointwise stabilizers of the natural action of $\pgl(V)$  on $\proj V$. Let $T$ be a subset of $\proj V$. 
For each $x\in T$, choose a representative $\tilde{x}\in V-\{0\}$. Call a subset of $T$ {\em dependent} if the corresponding set of representatives are linearly dependent over $\F_q$. Clearly, this definition does not depend on the choices of the representatives. Define the relation $\sim_T$ on $T$ such that $x\sim_T y$ iff there exists a minimal dependent subset of $T$ containing both $x$ and $y$. It is easy to show that this is an equivalence relation.\footnote{To prove transitivity of $\sim_T$, consider $x,y,z\in T$  such that $x\sim_T y$ and $y\sim_T z$. Then $x$ and $y$ (resp. $y$ and $z$) are in a dependent subset $T_1$ (resp. $T_2$) of $T$. Then $T_1\cup T_2$ is a dependent set. We obtain a minimal dependent set containing $x$ and $z$ by removing elements in $T_1\cup T_2-\{x,z\}$.} So it defines a partition of $T$ into the equivalence classes.\footnote{In the language of matroid theory, the dependent sets define a {\em matroid} on $T$, and the equivalent classes are known as the {\em connected components} of this matroid.}

Let $\pi$ denote the quotient map $\gl(V)\to\pgl(V)$. We have
\begin{lem}\label{lem_pglstabilizer}
Suppose $T$ is a subset of $\proj V$ and $T_1,\dots,T_k\subseteq T$ are the equivalent classes with respect to $\sim_T$.
For $i\in [k]$, let $V_i$ be the subspace of $V$ spanned by (the representatives of) the elements in $T_i$.
Then $g\in\gl(V)$ is sent to an element of $\pgl(V)_T$ under $\pi$ iff $g$ restricts to a scalar linear transformation on each $V_i$. 
\end{lem}
\begin{proof}
Suppose $g\in \gl(V)$ restricts to a scalar linear transformation on each $V_i$. Then obviously $\pi(g)$ fixes each $T_i$ pointwisely. So $\pi(g)\in \pgl(V)_T$. Conversely, suppose $\pi(g)\in \pgl(V)_T$. Then for every $x\in T$ and its representative $\tilde{x}\in V-\{0\}$, there exists a unique scalar $c_x\in\F_q^\times$ such that $\prescript{g}{}{\tilde{x}}=c_x\tilde{x}$. We need to show that for $x,y$ in the same equivalence class $T_i$, it holds that $c_x=c_y$. By definition, there exists a minimal dependent subset of $T$ containing both $x$ and $y$. So we can write 
\[
\tilde{x}=\sum_{\tilde{v}\in I} c_{v} \tilde{v},\qquad c_{v}\in \F_q^\times~\text{for all}~\tilde{v}\in I,
\]
where   $I$ is a finite set of linearly independent vectors $\tilde{v}\in V_i$, each $\tilde{v}$ represents an  element $v\in T_i$, and $\tilde{y}\in I$. As $\tilde{x}$ and all $\tilde{v}\in I$ are scaled by $g$, they are scaled by the same factor. So $c_x=c_y$, as desired. 
\end{proof}

In one direction, we have
\begin{lem}\label{lem_gltopgl}
$d_{\gl}(n,q)\leq d_{\pgl}(n,q)$. 
\end{lem}

\begin{proof}
Assume $n>1$ as otherwise $d_{\gl}(n,q)=d_{\pgl}(n,q)=1$.
Fix $m\in\N^+$ and let $\mathcal{P}$ (resp. $\mathcal{P}'$) be the system of stabilizers of depth $m$ over $\gl(V)$ (resp. $\pgl(V)$) with respect to the natural action of $\gl(V)$ on $V-\{0\}$ (resp. $\pgl(V)$ on $\proj V$).
Fix $x\in \proj V$ and let $\tilde{x}$ be a representative of $\tilde{x}$ in $V-\{0\}$.
Suppose $\mathcal{C}$ is a strongly antisymmetric $\mathcal{P}$-scheme that is not discrete on $\gl(V)_{\tilde{x}}$.
We prove that there exists a strongly antisymmetric $\mathcal{P}'$-scheme that is not discrete on $\pgl(V)_x$.

Define $\mathcal{P}''=\{\pi^{-1}(H): H\in\mathcal{P}'\}$ which is a subgroup system over $\gl(V)$.  
 We claim $\mathcal{P}''\subseteq  \mathcal{P}_{\mathrm{cl}}$ (see Definition~\ref{defi_inherit}).
Consider $H=\pgl(V)_T\in\mathcal{P}'$, where $T\subseteq \proj V$ satisfies $1\leq |T|\leq m$. Let $T_1,\dots, T_k\subseteq T$ be the equivalence classes with respect to $\sim_T$. For $i\in [k]$,  let $V_i$ be the subspace of $V$ spanned by (the representatives of) the elements in $T_i$. And let $W$ be the subspace of $V$ spanned by  (the representatives of) those in $T$, i.e., $W=\sum_{i=1}^k V_i$. Let $H':=\gl(V)_{W}$. Note that $H'=\gl(V)_B$ for any basis $B$ of $W$ over $\F_q$,  and $\dim_{\F_q} W\leq |T|=m$. So $H'\in\mathcal{P}$.
We claim $H'=u_\mathcal{P}(\pi^{-1}(H))$ and $\pi^{-1}(H)\subseteq N_{\gl(V)}(H')$.

By Lemma~\ref{lem_pglstabilizer}, the group $\pi^{-1}(H)$  consists of $g\in\gl(V)$ that  restricts to a scalar linear transformation on each $V_i$. So $\pi^{-1}(H)$ fixes $W$ setwisely. Therefore we have $H'\subseteq \pi^{-1}(H)\subseteq N_{\gl(V)}(H')$. Suppose  $H''$ is another subgroup in $\mathcal{P}$ contained in $\pi^{-1}(H)$.   It has the form $\gl(V)_{W'}$ where $W'$ is a subspace of $V$. If $W\not\subseteq W'$, there exists a representative $\tilde{y}\in V-\{0\}$ of some $y\in T$ such that $\tilde{y}\not\in W'$. Then there exists $g\in \gl(V)$ that fixes $W'$ pointwisely but sends $\tilde{y}$ to a vector $\tilde{y}'$ such that $\tilde{y}$ and $\tilde{y}'$ are not collinear. Such an element $g$ is in $H''$ but not in $\pi^{-1}(H)$, contradicting the assumption  $H''\subseteq \pi^{-1}(H)$. So $W\subseteq W'$ and hence $H''\subseteq H'$. Therefore $H'$ is the unique maximal subgroup in $\mathcal{P}$ contained in $\pi^{-1}(H)$, i.e., $H'=u_\mathcal{P}(\pi^{-1}(H))$. By definition, we have $\pi^{-1}(H)\in \mathcal{P}_{\mathrm{cl}}$. So $\mathcal{P}''\subseteq  \mathcal{P}_{\mathrm{cl}}$.

Note that  $\gl(V)_{\tilde{x}}=u_\mathcal{P}(\pi^{-1}(\pgl(V)_x))$.
By Lemma~\ref{lem_ext}, the existence of  $\mathcal{C}$ implies that there exists a strongly antisymmetric $\mathcal{P}_{\mathrm{cl}}$-scheme that is not discrete on $\pi^{-1}(\pgl(V)_x)$. As $\mathcal{P}''\subseteq  \mathcal{P}_{\mathrm{cl}}$,  there also exists a strongly antisymmetric $\mathcal{P}''$-scheme that is not discrete on $\pi^{-1}(\pgl(V)_x)$. 
Finally, by Lemma~\ref{lem_quoidentify}, there exists a strongly antisymmetric $\mathcal{P}'$-scheme that is not discrete on $\pgl(V)_x$, as desired.
\end{proof}

In the other direction, we have
\begin{lem}\label{lem_pgltogl}
$d_{\pgl}(n,q)\leq d_{\gl}(n,q)+2$. 
\end{lem}

\begin{proof}
Fix $m\in\N^+$ and let $\mathcal{P}$ (resp. $\mathcal{P}'$) be the system of stabilizers of depth $m+2$ (resp. $m$) over $\pgl(V)$ (resp. $\gl(V)$) with respect to the natural action of $\pgl(V)$ on $\proj V$ (resp. $\gl(V)$ on $V-\{0\}$).
Suppose there exists a strongly antisymmetric $\mathcal{P}$-scheme  $\mathcal{C}$ that is not discrete on $\pgl(V)_{x}$ for some $x\in \proj V$, i.e., $d_{\pgl}(n,q)>m+2$.
We prove that there exists a strongly antisymmetric $\mathcal{P}'$-scheme that is not discrete on $\gl(V)_{y}$ for some $y\in V-\{0\}$, i.e., $d_{\gl}(n,q)>m$.

By Lemma~\ref{lem_res} and Lemma~\ref{lem_selfredspan}, there exists $u,v\in\proj V$ such that the $\mathcal{P}|_{\pgl(V)_{u}}$-collection $\mathcal{C}|_{\pgl(V)_{u}}$ is a strongly antisymmetric  $\mathcal{P}|_{\pgl(V)_{u}}$-scheme and is not discrete on $\pgl(V)_{u,v}$. Let $\tilde{u}$ be a representative of $u$ in $V-\{0\}$. 
The map $\pi:\gl(V)\to\pgl(V)$ restricts to a map $\pi|_{\gl(V)_{\tilde{u}}}:\gl(V)_{\tilde{u}}\to \pgl(V)_{u}$.
The latter map is surjective (and in fact bijective) since  every element in $\pgl(V)_{u}$ can be lifted to an element in $\gl(V)_{\tilde{u}}$. Define $\mathcal{P}'':=\{(\pi|_{\gl(V)_{\tilde{u}}})^{-1}(H): H\in \mathcal{P}|_{\pgl(V)_{u}}\}$, which is a subgroup system over $\gl(V)_{\tilde{u}}$.
By Lemma~\ref{lem_quoidentify}, there exists a strongly antisymmetric $\mathcal{P}''$-scheme $\mathcal{C}'$ that is not discrete on $(\pi|_{\gl(V)_{\tilde{u}}})^{-1}(\pgl(V)_{u,v})$.

Let $\tilde{\mathcal{P}}$ be the system of stabilizers of depth $m$ over $\gl(V)_{\tilde{u}}$ with respect to the action of $\gl(V)_{\tilde{u}}$ on $V-\{0\}$ restricted from that of $\gl(V)$. We claim that $\tilde{\mathcal{P}}\subseteq\mathcal{P}''$. Consider arbitrary $H\in \tilde{\mathcal{P}}$. It has the form $H=\gl(V)_{\{\tilde{u}\}\cup T}$, where $1\leq |T|\leq m$.
Let $W$ be the subspace of $V$ spanned by $\tilde{u}$  and the elements in $T$. 
Extend $\{\tilde{u}\}$ to an $\F_q$-basis  $B=\{\tilde{u},x_1,\dots,x_k\}$ of $W$. Then $k\leq m$.
Let $w=\tilde{u}+x_1+\dots+x_k\in V-\{0\}$. Let $B'=B\cup\{w\}$ and let $\bar{B}'$ be the subset of $\proj V$ consisting of the elements represented by those in $B'$. Then $|\bar{B}'|\leq m+2$ and $u\in \bar{B}'$. So $\pgl(V)_{\bar{B}'}\in \mathcal{P}|_{\pgl(V)_{u}}$.
As $B$ is a basis of $W$, the set $\bar{B}'$ is a minimal dependent subset   and hence is the only equivalence class with respect to $\sim_{\bar{B}'}$. So $\pi^{-1}(\pgl(V)_{\bar{B}'})$ consists of the elements in $\gl(V)$ that restricts to scalar linear transformations on $W$. Then $(\pi|_{\gl(V)_{\tilde{u}}})^{-1}(\pgl(V)_{\bar{B}'})$ consists of the elements in $\gl(V)$ that fixes $W$ pointwisely, i.e., $(\pi|_{\gl(V)_{\tilde{u}}})^{-1}(\pgl(V)_{\bar{B}'})=\gl(V)_{\{\tilde{u}\}\cup T}=H$. By definition, we have $H\in\mathcal{P}''$. So $\tilde{\mathcal{P}}\subseteq\mathcal{P}''$.

Recall that $\mathcal{C}'$ is a strongly antisymmetric $\mathcal{P}''$-scheme that is not discrete on the subgroup $(\pi|_{\gl(V)_{\tilde{u}}})^{-1}(\pgl(V)_{u,v})$.
Let $\tilde{v}$ be a representative of $v$ in $V-\{0\}$. Then $\gl(V)_{\tilde{u},\tilde{v}}=(\gl(V)_{\tilde{u}})_{\tilde{v}}\in \tilde{\mathcal{P}}\subseteq\mathcal{P}''$.
Note $\gl(V)_{\tilde{u},\tilde{v}}\subseteq (\pi|_{\gl(V)_{\tilde{u}}})^{-1}(\pgl(V)_{u,v})$. By  Lemma~\ref{lem_min}, we know $\mathcal{C}'$ is not discrete on $\gl(V)_{\tilde{u},\tilde{v}}$.
As $\tilde{\mathcal{P}}\subseteq\mathcal{P}''$, there exists a strongly antisymmetric $\tilde{\mathcal{P}}$-scheme that is not discrete on  $\gl(V)_{\tilde{u},\tilde{v}}=(\gl(V)_{\tilde{u}})_{\tilde{v}}$. 
By Corollary~\ref{cor_inductionsubgroup}, there exists a strongly antisymmetric $\mathcal{P}'$-scheme that is not discrete on  $\gl(V)_{\tilde{v}}$, as desired. 
\end{proof}

Theorem~\ref{thm_equivlin} now follows from Lemma~\ref{lem_gltogammal}, Lemma~\ref{lem_gammaltogl}, Lemma~\ref{lem_pgammaltopgl}, Lemma~\ref{lem_gltopgl}, and Lemma~\ref{lem_pgltogl}.

\paragraph{Upper bounds for $d_{\gl}(n,q)$.}

It is easy to see that we have two upper bounds for $d_{\gl}(n,q)$:
\begin{enumerate}
\item $d_{\gl}(n,q)\leq \left(\frac{2}{\log 12}\right) \log (q^n-1)+O(1)= \left(\frac{2\log q}{\log 12}\right)  n+O(1)$. This follows from Corollary~\ref{cor_ubsysp}.
\item $d_{\gl}(n,q)\leq n$. This follows from Lemma~\ref{lem_antibasebound} and the fact that the natural action of $\gl_n(q)$ has a base of size $n$.
\end{enumerate}

The first bound is asymptotically better if $q\in \{2,3\}$. Otherwise the second one is better. Now we prove another upper bound that slightly improves both of the two bounds  above.

\begin{thm}\label{thm_slightimp}
$d_{\gl}(n,q)\leq \left(\frac{\log q}{\log q + (\log 12)/4 }\right) n+O(1)$.
\end{thm}

\begin{proof} 
Let $G=\gl_n(q)$ and $S=\F_q^n-\{0\}$.
Fix a positive integer $m\leq n$.
Let $\mathcal{P}$ be the system of stabilizers of depth $m$ with respect to the natural action of $G$ on $S$.
Suppose there exists  a strongly antisymmetric $\mathcal{P}$-scheme $\mathcal{C}=\{C_H: H\in\mathcal{P}\}$ that is not discrete on $G_x$ for some $x\in S$. 
We prove that  $m\leq\left(\frac{\log q}{\log q + (\log 12)/4 }\right) n+O(1)$.

By Lemma~\ref{lem_lbblocksize}, there exists a subset $T=\{x_1,\dots,x_m\}\subseteq S$ of cardinality $m$ such that $C_{G_T}$ has a block $B$ of cardinality at least $2^{\left(\frac{\log 12}{4}\right)m^2-O(m)}$.
We claim that the elements $x_i$ in $T$ may be assumed to be  linearly independent: if they are not, replace $T$ by a set $T'$ of cardinality $m$ such that (1) the elements in $T'$ are linearly independent, and (2) the subspace spanned by $T'$ contains the one spanned by $T$. Then replace $B$ with a block $B'$ of $C_{G_{T'}}$ such that $\pi_{G_{T'},G_T}(B')=B$. We have $|B'|\geq |B|$. 
This proves the claim.

Note that $N_G(G_T)$ is the setwise stabilizer of subspace spanned by $T$. Therefore 
$N_G(G_T)/G_T\cong \gl_m(q)$.
By antisymmetry, the group $N_G(G_T)/G_T$ acts semiregularly on the set of blocks of $C_{G_T}$. 
So we have 
\[
|G_T\backslash G|\geq |N_G(G_T)/G_T|\cdot |B|\geq 2^{\left(\frac{\log 12}{4}\right)m^2-O(m)}\cdot \prod_{i=0}^{m-1}(q^m-q^i).
\]
On the other hand, note that $G_T$ is the stabilizer of $u:=(x_1,\dots,x_m)\in S^{(m)}$ under the diagonal action of $G$ on $S^{(m)}$. By the orbit-stabilizer theorem, we have $|G_T\backslash G|=|G u|$, which is the number of $m$-tuples of linearly independent vectors in $V-\{0\}$.
Therefore $|G_T\backslash G|=\prod_{i=0}^{m-1} (q^n-q^i)$. So we have 
\[
2^{\left(\frac{\log 12}{4}\right)m^2-O(m)}\cdot \prod_{i=0}^{m-1}(q^m-q^i)\leq \prod_{i=0}^{m-1} (q^n-q^i).
\]
Solving the inequality yields the desired bound.
\end{proof}

As $q\geq 2$, we have $\log q + (\log 12)/4 < (\log 12)/2$. So  Theorem~\ref{thm_slightimp}
is indeed an improvement of the bound $d_{\gl}(n,q)\leq  \left(\frac{2\log q}{\log 12}\right)  n+O(1)$ above. 

%\section{Subspace actions of linear groups} 

\chapter{Groups with restricted noncyclic composition factors}\label{chap_primitive}

In this chapter, we consider the problem of factoring a polynomial $f(X)\in\F_q[X]$ using a lifted polynomial $\tilde{f}$ where the Galois group of $\tilde{f}$ has {\em restricted noncyclic composition factors}.

\paragraph{Simple groups, composition factors, and CFSG.}   To formally state our result, we first review some definitions and facts in group theory.
A {\em simple group}\index{simple group} is a nontrivial group whose only   normal subgroups are the trivial group and the group itself. 
A {\em composition series}\index{composition series} of a group $G$ is a finite chain of subgroups
\[
\{e\}=H_0\subseteq H_1\subseteq\cdots\subseteq H_k=G
\]
 such that  for every $i\in [k]$, $H_{i-1}$ is a maximal normal subgroup of $H_i$, so that $H_i/H_{i-1}$ is simple.
Such a series always exists when $G$ is finite. 
The  groups $H_i/H_{i-1}$ are called the {\em composition factors}\index{composition factor} of $G$.
 It is a consequence of the {\em Jordan-H\"{o}lder theorem}\index{Jordan-H\"{o}lder theorem} that the set of the composition factors of $G$ does not depend on the choice of composition series (see, e.g., \citep{Lan02}).

Now suppose $G$ is a finite group.  
%Denote by $\mathrm{Cf}(G)$ the set of composition factors of $G$. 
The composition factors of $G$ are finite simple groups, which are classified by the {\em classification of finite simple groups} (CFSG):
%\footnote{Isomorphic groups are regarded as the same element in $\mathrm{Cf}(G)$.}
\begin{thm}[classification of finite simple groups]\index{CFSG}\index{classification of finite simple groups|see{CFSG}}
A finite simple group is  isomorphic to one of the following groups: a cyclic group of prime order, an  alternating group $\alt(n)$ ($n\geq 5$), a classical group\index{classical group}, an exceptional group of Lie type\index{exceptional group of Lie type}, or one of the 26 sporadic simple groups\index{sporadic simple group}. 
\end{thm}

See, e.g., \citep{Gor13}. We do not describe these families of finite simple groups, except mentioning that a finite simple group is a classical group if it has one of the following forms (see, e.g., \citep{KL90}):
\[
\mathrm{PSL}_n(q),\enskip \mathrm{PSU}_n(q),\enskip \mathrm{PSp}_n(q)~\text{($n$ even)},\enskip \mathrm{P}\Omega_n^{\pm}(q)~\text{($n$ even)},\enskip \Omega_n(q)~\text{($n$ odd)}.
\]
%The parameter $n$ is called the {\em rank}\index{rank!of a classical group} of these classical groups.
We denote by $k(G)$ the maximum degree of the  alternating groups that appear as noncyclic composition factors of $G$, and let $k(G)=1$ if such alternating groups do not exist.
Similarly, denote by $r(G)$ the  maximum order of the   classical  groups that appear as noncyclic composition factors of $G$, and let $r(G)=1$ if such classical groups do not exist.

\paragraph{Main result.}
Let $\F_q$, $A_0$ and $K_0$ be as in Chapter~\ref{chap_alg_general}.
 The main result of this chapter is a GRH-based deterministic algorithm that factorizes $f(X)\in\F_q[X]$ using a lifted polynomial $\tilde{f}(X)\in A_0[X]$, such that the running time of the algorithm is controlled by $k(G)$ and $r(G)$, where $G=\gal(\tilde{f}/K_0)$ is the Galois group of $\tilde{f}$ over $K_0$.

\begin{thm} \label{thm_compbound}
Under GRH, there exists a  deterministic  algorithm that, given a polynomial $f(X)\in\F_q[X]$  of degree $n\in\N^+$ and a  lifted polynomial $\tilde{f}(X)\in A_0[X]$ of $f$  with the Galois group $G:=\gal(\tilde{f}/K_0)$ over $K_0$, computes the complete factorization of $f$ over $\F_q$ in time polynomial in $n$, $\log q$, $k(G)^{\log k(G)}$ and $r(G)$.
\end{thm}

For $k\in\N^+$, denote by $\Gamma_k$ the family of finite groups   whose noncyclic composition factors are all  isomorphic to subgroups of $\sym(k)$.
 %These families play a significant role in graph isomorphism testing \cite{Luk82, Mi83}, asymptotic group theory \cite{BCP82, Pyb93, PS97} and computational group theory \cite{Luk93, Ser03}. 
It is known that a classical group $H$ is isomorphic to a subgroup of $\sym(k)$ only if $|H|=k^{O(\log k)}$ \cite{Coo78}. Therefore
%by Theorem~\ref{thm_compbound} and the fact $d_{\sym}(n)=O(\log n)$ (see Corollary~\ref{cor_ubsysp}), 
we have

\begin{thm}\label{thm_gamma}
Under GRH, there exists a  deterministic  algorithm that, given a  polynomial $f(X)\in\F_q[X]$  of degree $n\in\N^+$ and a  lifted polynomial $\tilde{f}(X)\in A_0[X]$ of $f$, computes the complete factorization of $f$ over $\F_q$ in time polynomial in $n$, $\log q$ and $k^{\log k}$,  where $k$ is the smallest positive integer satisfying $\gal(\tilde{f}/K_0)\in \Gamma_k$.
 In particular, the algorithm runs in polynomial time if $k=2^{O(\sqrt{\log n})}$.
\end{thm}

\bgroup
\def\arraystretch{1.2}
\begin{table}[hbt]
\centering
\caption{Known deterministic polynomial-time factoring algorithms for $\Gamma_k$} \label{tab_comp}
    \begin{tabular}{| c| c | }
    \hline
    $k$ & Reference  \\ \hhline{|==|}
    $4$ & \cite{Evd92}  \\ \hline
    $O(1)$ & \cite{Evd92} + \cite{BCP82}  \\ \hline
    $2^{O(\sqrt{\log n})}$ & {\textbf{Our result}}  \\ \hline
    $n$ & Goal \\ \hline
    \end{tabular}
\end{table}
\egroup

By Theorem~\ref{thm_gamma}, we have a deterministic polynomial-time algorithm that given $\tilde{f}(X)$, completely factorizes $f(X)$ under GRH , provided that $\gal(\tilde{f}/K_0)\in \Gamma_k$ for some $k=2^{O(\sqrt{\log n})}$ (note that achieving $k=n$ would fully resolve the problem of deterministic polynomial factoring under GRH).
Previously, such an algorithm was known only  for bounded $k$: for $k\leq 4$ this follows directly from the deterministic polynomial-time factoring algorithm for solvable Galois groups \cite{Evd92} (see Theorem~\ref{thm_algsolvable} and Theorem~\ref{thm_algsolvableg}). For $k=O(1)$, it follows from the proof in \cite{Evd92} together with the bound  in \cite{BCP82} for the orders of primitive permutation groups.
See Table~\ref{tab_comp} for a summary.
% It is really the power of $\mathcal{P}$-schemes that allows us to push $d$ to $2^{O(\sqrt{\log n})}$. 

\paragraph{Overview of the proof.}
We prove Theorem~\ref{thm_compbound} using the generalized $\mathcal{P}$-scheme algorithm in Chapter~\ref{chap_alg_general}. 
If the input polynomial $f$ is assumed to satisfy Condition~\ref{cond_spoly}, we may also use the simpler algorithm in Chapter~\ref{chap_alg_prime}.
These algorithms reduce the problem of factoring $f$ to the one of constructing a collection of (relative) number fields such that the associated subgroup system $\mathcal{P}$ has the property that all strongly antisymmetric $\mathcal{P}$-schemes are discrete on a certain subgroup (see Theorem~\ref{thm_algmain2formal} and Theorem~\ref{thm_algmain2formalg}). 

We further reduce the latter problem to the case that the Galois group $\gal(\tilde{f}/K_0)$ is a primitive permutation group on the set of roots of $\tilde{f}$, using  Theorem~\ref{thm_primitivereduction} and some facts from group theory.
Next we consider the following special kind of subgroup systems.
\begin{defi}\label{defi_subgroupsysgn}
Let $G$ be a finite permutation group on a finite set $S$.
For $N\in\N^+$, define the subgroup system $\mathcal{P}_{G,N}$ over $G$ by
\[
\mathcal{P}_{G,N}:=\left\{G_{U\cup U'}: 
\begin{array}{c}
\emptyset\neq U\subseteq S,~ x\in S,~ U'\subseteq G_U x \\
|S|^{|U|}, |G_U x|^{|U'|}\leq N
\end{array}
\right\}.
\]
\end{defi}
\nomenclature[g1a]{$\mathcal{P}_{G,N}$}{See Definition~\ref{defi_subgroupsysgn}}

We prove a sufficient condition for a subgroup system  $\mathcal{P}_{G,N}$ over a primitive permutation group to have the desired property:

\begin{thm}\label{thm_primcriterion}
Let $G$ be a primitive permutation group on a finite set $S$.
For sufficiently large $N=\mathrm{poly}(k(G)^{d_{\sym}(k(G))},r(G),|S|)\geq |S|$, all strongly antisymmetric $\mathcal{P}_{G,N}$-schemes are discrete on $G_x\in \mathcal{P}_{G,N}$ for all $x\in S$.
\end{thm}

It is easy to see that the complexity $c(\mathcal{P}_{G,N})$ of $\mathcal{P}_{G,N}$ is polynomial in $N$.
We modify the algorithm in Lemma~\ref{lem_compstabsysg} to construct a collection of (relative) number fields in time polynomial in $n$, $\log q$ and $c(\mathcal{P}_{G,N})$ such that the associated subgroup system is precisely $\mathcal{P}_{G,N}$. Theorem~\ref{thm_compbound} then  follows from Theorem~\ref{thm_primcriterion}.

Finally, to prove Theorem~\ref{thm_primcriterion}, we apply  the {\em O'Nan-Scott theorem} \citep{LPS88} in permutation group theory, which states that a finite primitive permutation group is in exactly one of the following five categories: almost simple type,  affine type,  diagonal type,  product type, and twisted wreath type. We  prove Theorem~\ref{thm_primcriterion} by verifying it in these five cases  separately.

%The exact formulation is given in Section~\ref{sec_onanscott}.

\paragraph{Outline of the chapter.} In Section~\ref{sec_partialstab}, we derive Theorem~\ref{thm_compbound} from Theorem~\ref{thm_primcriterion} using an algorithm that constructs the collection of (relative) number fields corresponding to $\mathcal{P}_{G,N}$. The rest of the chapter focuses on the proof of  Theorem~\ref{thm_primcriterion}: Section~\ref{sec_onanscott} describes the O'Nan-Scott theorem \citep{LPS88} and  the five categories of primitive permutation groups.
In Sections~\ref{sec_prim_as}--\ref{sec_prim_prod}, we prove Theorem~\ref{thm_primcriterion} for primitive permutation groups of almost simple type, affine type,  diagonal type and  product type respectively. We also  address twisted wreath type at the end of Section~\ref{sec_prim_prod} by reducing to the case of product type using an argument in \citep{Pra90}. Finally, we discuss possible directions for future research  in Section~\ref{sec_future}.

\section{Proof of the main theorem}\label{sec_partialstab}

We start by describing an algorithm $\mathtt{SubgroupSystem}$ that computes a $(K_0,g)$-subfield system  $\mathcal{F}$ given a number field $K_0$, an integer $N\in\N^+$, and a polynomial $g(X)\in K_0[X]$ irreducible over $K_0$, such that the subgroup system associated with $\mathcal{F}$ is exactly $\mathcal{P}_{G,N}$, where $G=\gal(g/K_0)$.

The pseudocode is given in Algorithm~\ref{alg_syspn}.
First compute the greatest integer $d\in\{0,\dots,\deg(g)\}$ subject to $\deg(g)^{d}\leq N$.
Run the algorithm $\mathtt{Stabilizers}$ in  Lemma~\ref{lem_compstabsysg} on the input  $(K_0, d, g)$
to obtain a $(K_0,g)$-subfield system $\mathcal{F}'$, and let $\mathcal{F}=\mathcal{F}'$.
Next enumerate $K\in\mathcal{F}'$ and the  irreducible factors $g_0$ of $g$ over $K$.
For each $(K,g_0)$,  let $d'$ be the greatest integer in $\{0,\dots,\deg(g_0)\}$ subject to $\deg(g_0)^{d'}\leq N$,  run the algorithm $\mathtt{Stabilizers}$  on the input  $(K, d', g_0)$ to obtain a $(K,g_0)$-subfield system $\mathcal{F}''$, and
add the fields in $\mathcal{F}''$ to $\mathcal{F}$.\footnote{We add a field to $\mathcal{F}$ only if it is non-isomorphic to all fields in $\mathcal{F}$ over $K_0$, so that the fields in $\mathcal{F}$ are always mutually non-isomorphic over $K_0$.}
 
\begin{algorithm}[htbp]
\caption{$\mathtt{SubgroupSystem}$}\label{alg_syspn}
\begin{algorithmic}[1]
\INPUT  number field $K_0$, $N\in\N^+$, and $g(X)\in K_0[X]$ irreducible over $K_0$ 
\OUTPUT \strut  $(K_0,g)$-subfield system  $\mathcal{F}$ 
\State $d\gets\max\{i\in\N: 0\leq i\leq \deg(g), \deg(g)^{i}\leq N\}$
\State run $\mathtt{Stabilizers}$ on $(K_0, d, g)$ to compute a $(K_0,g)$-subfield system $\mathcal{F}'$  
\State $\mathcal{F}\gets \mathcal{F}'$ 
  \For{$K\in\mathcal{F}'$}
      \State factorize $g$ over $K$
       \For{ irreducible factor $g_0$ of $g$ over $K$}
           \State $d'\gets\max\{i\in\N: 0\leq i\leq \deg(g_0), \deg(g_0)^{i}\leq N\}$
           \State run $\mathtt{Stabilizers}$ on $(K, d', g_0)$ to compute a $(K,g_0)$-subfield system $\mathcal{F}''$
           \For{$K'\in \mathcal{F}''$}
                 \State compute a relative number field $\tilde{K}'$ over $K_0$ such that $\tilde{K}'\cong_{K_0} K'$
                 \If{$\tilde{K}'$ is non-isomorphic to all fields in $\mathcal{F}$ over $K_0$}
                        \State $\mathcal{F}\gets\mathcal{F}\cup\{\tilde{K}'\}$
                 \EndIf    
           \EndFor
       \EndFor 
 \EndFor
\State \Return $\mathcal{F}$
\end{algorithmic}
\end{algorithm} 

The following lemma states that the subgroup system associated with $\mathcal{F}$ is precisely $\mathcal{P}_{G,N}$.

\begin{lem}\label{lem_syspn}
Given a number field $K_0$, an integer $N\in\N^+$, and a polynomial $g(X)\in K_0[X]$ irreducible over $K_0$, the algorithm $\mathtt{SubgroupSystem}$ computes a $(K_0,g)$-subfield system $\mathcal{F}$,
such that the subgroup system   associated with $\mathcal{F}$ is  precisely $\mathcal{P}_{G,N}$ over $G:=\gal(g/K_0)$, where $G$ is regarded as a permutation group on the set of roots of $g$ in the splitting field of $g$ over $K_0$.
Moreover, the algorithm runs in time polynomial in $c(\mathcal{P}_{G,N})=N^{O(1)}$ and the size of the input.
\end{lem}

\begin{proof}
Let $S$ be the set of roots of $g$ in the splitting field of $g$ over $K_0$.
Let $d=\max\{i\in\N: 0\leq i\leq \deg(g), \deg(g)^{i}\leq N\}$.
By definition, the subgroup system $\mathcal{P}_{G,N}$ consists of the pointwise stabilizers $G_{U\cup U'}$, such that $U$ is a nonempty subset of $S$ of cardinality at most $d$, and $U'$ is a subset of a $G_U$-orbit $O\subseteq S$ satisfying $|O|^{|U'|}\leq N$. 

Note that when we fix $U'=\emptyset$, the groups $G_{U\cup U'}=G_U$ are precisely those in the system of stabilizers of depth $d$ with respect to the action of $G$ on $S$.  We construct the corresponding fields by  running the algorithm $\mathtt{Stabilizers}$ on $(K_0, d, g)$.

Next consider the groups  $G_{U\cup U'}$ where $U'\neq \emptyset$.
We enumerate $K=L^{G_U}$ and the irreducible factor $g_0$ of $g$ over $K$. By Galois theory, the set of roots of $g_0$ is a $G_U$-orbit $O\subseteq S$. 
Let $d'=\max\{i\in\N: 0\leq i\leq \deg(g_0), \deg(g_0)^{i}\leq N\}$.
We run the algorithm $\mathtt{Stabilizers}$ on $(K, d', g_0)$ to  construct the fields corresponding to the subgroups $(G_U)_{U'}=G_{U\cup U'}$, where $U'\subseteq O$ and $1\leq |U'|\leq d'$. Moreover, all the groups $G_U$ and the $G_U$-orbits in $S$ are enumerated.
%\footnote{Note that we may ignore the $G_U$-orbits of cardinality one: if $O=\{x\}$ is a $G_U$-orbit, then $(G_{U})_x=G_U$.} 
It follows that  the subgroup system   associated with $\mathcal{F}$ is  precisely $\mathcal{P}_{G,N}$.

Finally, the fact $c(\mathcal{P}_{G,N})=N^{O(1)}$ and the claim about the running time follow from Lemma~\ref{lem_complexitystab} and Lemma~\ref{lem_compstabsysg}. 
\end{proof}

We also need the following lemma, which states that restricting to a subgroup does not increase the quantities $k(G)^{\log k(G)}$ and $r(G)$ by much.

\begin{lem}\label{lem_subquo}
Let $G$ be a permutation group on a finite set $S$, and let $G'$ be a subquotient of $G$.
% (i.e., a quotient group of a subgroup) 
Then $k(G')^{\log k(G')}$ and $r(G')$ are  polynomial in $k(G)^{\log k(G)}$, $r(G)$ and $|S|$.
\end{lem}
\begin{proof}
Let $H'$ be a noncyclic composition factor $G'$.
% that is either an alternating group or a classical group.
Then $H'$ is isomorphic to a subquotient of a noncyclic composition factor $H$ of $G$, i.e., there exists a subgroup $H''$ of $H$ and a normal subgroup $N$ of $H''$ such that $H'\cong H''/N$. Fix such $H$, $H''$ and $N$. 
 We want to prove (1) if $H'$ is an alternating group $\alt(k')$, then $k'^{\log k'}$ is polynomial in $k(G)^{\log k(G)}$, $r(G)$ and $|S|$, and (2) if $H'$ is a classical group, then $|H'|$ is    polynomial in $k(G)^{\log k(G)}$, $r(G)$ and $|S|$.
 
By CFSG, the group $H$ is either an alternating group or a group of Lie type (i.e. a classical group or an exceptional group of Lie type).
First assume $H$ is an alternating group of degree $k\leq k(G)$. If $H'$ is also an alternating group, its degree $k'$ is obviously bounded by $k$. So $k'^{\log k'}\leq k(G)^{\log k(G)}$. Now consider the case that $H'$ is a classical group of the form $\mathrm{PSL}_n(q)$, $\mathrm{PSU}_n(q)$, $\mathrm{PSp}_n(q)$, $\mathrm{P}\Omega_n^{\pm}(q)$, or $\Omega_n(q)$ over a finite field $\F_q$ for some $n\in\N^+$.
We have $|H'|=q^{\Theta(n^2)}$.
  Denote by $\mu(T)$ the minimal degree of a faithful permutation representation of a finite group $T$. 
It was proven in \citep{KP00} that if $\bar{T}$ is a quotient group of $T$ with no nontrivial abelian normal subgroup, then $\mu(\bar{T})\leq\mu(T)$. As $H'\cong H''/N$ is simple and noncyclic, we have
\[
\mu(H')=\mu(H''/N)\leq \mu(H'') \leq \mu(H)\leq k.
\]
On the other hand, it was shown in \citep{Coo78} that $\mu(H')=q^{\Theta(n)}$ (see also \citep[Table~5.2.A]{KL90}).
So we have $n=O(\log k/\log q)$ and  
\[
|H'|=q^{\Theta(n^2)}= k^{O(\log k/\log q)}= k(G)^{O(\log k(G))}.
\]

Next assume $H$ is a group of Lie type over a finite field $\F_q$, and has Lie rank\index{Lie rank} $\ell$.\footnote{Each finite simple group of Lie type has an associated {\em Lie rank}. See, e.g., \citep[Section~5.1]{KL90}.}
Then $|H|=q^{\Theta(\ell^2)}$ \citep[Table~5.1.A]{KL90}. It is also known that $H$ has a faithful  projective linear representation $H\hookrightarrow \pgl_d(\bar{\F}_q)$  of degree $d=O(\ell)$, where $\bar{\F}_q$ is the algebraic closure of $\F_q$ (see \citep[Proposition~5.4.13]{KL90}). As $H$ is finite, this also holds for some finite field $F$ in place of $ \bar{\F}_q$. Identify $H''\subseteq H$ with a subgroup of $\pgl_d(F)$.
%, and let $\tilde{H}''$ be the preimage of $H''$ in $\gl_d(F)$.
Then  $H'\cong H''/N$ is %isomorphic to  a quotient group of $\tilde{H}''$, 
a subquotient of $\pgl_d(F)$, and hence also a subquotient of $\gl_d(F)$.
Choose the largest $s\in\N^+$ such that $H'$ has a subquotient isomorphic to  $\alt(s)$. Then $\alt(s)$ is  isomorphic to a subquotient of $\gl_d(F)$. 
On the other hand, it is known that  $\alt(s)$ has a finite preimage in $\gl_d(F)$ only if $s=O(d)$ \citep[Theorem~5.7A]{DM96}.
So we have $s=O(d)=O(\ell)$.

Suppose $H$ is a classical group.  If $H'=\alt(s)$, we have $s^{\log s}= \ell^{O(\log \ell)}=|H|^{O(1)}=r(G)^{O(1)}$. And if $H'$ is a classical group, we have the obvious bound $|H'|\leq |H|\leq r(G)$.

Finally, suppose $H$ is an exceptional group of Lie type. Then $s=O(\ell)=O(1)$.  In the case $H'=\alt(s)$, we have $s^{\log s}=O(1)$. 
So assume  $H'$ is a classical group of the form $\mathrm{PSL}_n(q')$, $\mathrm{PSU}_n(q')$, $\mathrm{PSp}_n(q')$, $\mathrm{P}\Omega_n^{\pm}(q')$ or $\Omega_n(q')$ over a finite field $\F_q$ for some $n\in\N^+$.
It is easy to see that $H'$ has a subquotient isomorphic to an alternating group of degree $\Omega(n)$ (see, e.g., \citep[Proposition~16.4.4]{LS12}). So $s=\Omega(n)$, which implies $n=O(1)$. 
Then $\mu(H')=q'^{\Theta(n)}=q'^{\Theta(n^2)}=|H'|^{\Theta(1)}$.
On the other hand, we see above that $\mu(H')\leq \mu(H)$ since $H'$ is a subquotient of $H$ and is a noncyclic simple group.  For the same reason, we have $\mu(H)\leq \mu(G)\leq |S|$. It follows that $|H'|=|S|^{O(1)}$.
\end{proof}

Now we are ready to prove Theorem~\ref{thm_compbound} under the assumption of Theorem~\ref{thm_primcriterion}.

\begin{proof}[Proof of Theorem~\ref{thm_compbound}]
The first step is to reduce to the case that $\tilde{f}$ is irreducible over $K_0$, as in Chapter~\ref{chap_alg_general}:
Let $p=\mathrm{char}(\F_q)$.  
Using Lemma~\ref{lem_subprob}, we compute an integer $D$ satisfying $D\equiv 1\pmod{p}$ and a factorization of $D\cdot\tilde{f}$ into irreducible factors $\tilde{f}_1,\dots,\tilde{f}_k\in A_0[X]$ over $K_0$. 
Then we have $f(X)=\prod_{i=1}^k \tilde{\psi}_0(f_i)(X)$.
 The Galois groups $\gal(\tilde{f}_i(X)/K_0)$ are quotient groups of $G=\gal(\tilde{f}/K_0)$.
 So the set of the composition factors of each $\tilde{f}_i$ is a subset of that of $G$.
By replacing $\tilde{f}(X)$ with $\tilde{f}_i(X)$ and $f(X)$ with $\tilde{\psi}_0(f_i)\in\F_q[X]$ for each $i\in [k]$, we reduce to the case that $\tilde{f}$ is irreducible over $K_0$.

Choose sufficiently large  $N=\mathrm{poly}(k(G)^{\log k(G)},r(G),\deg(f))\geq \deg(f)$.  Assume for a moment that the value of $N$ is known to the algorithm. 
First consider the case that $G$ acts primitively on the set of roots of $\tilde{f}$.
We  compute a $(K_0,\tilde{f})$-subfield system $\mathcal{F}$ using the algorithm $\mathtt{SubgroupSystem}$ above.
By Lemma~\ref{lem_syspn}, the associated subgroup system $\mathcal{P}$ over $G$ equals $\mathcal{P}_{G,N}$.
Then by Theorem~\ref{thm_primcriterion} and the fact $d_{\sym}(k(G))=O(\log k(G))$ (see Corollary~\ref{cor_ubsysp}),  all strongly antisymmetric $\mathcal{P}$-schemes are discrete on $G_x\in \mathcal{P}$ for all roots $x$ of $\tilde{f}$.

Now consider the general case, where the action of $G$ may be imprimitive. We run the algorithm $\mathtt{GeneralAction}$ in Theorem~\ref{thm_primitivereduction} to compute $\mathcal{F}$, as well as  a tower of relative number fields $K_{0} \subseteq K_{1}\subseteq \dots \subseteq K_{k-1}\subseteq K_k$ over $K_0$ and $g_i(X)\in K_{i-1}[X]$ for $i\in [k]$, such that
\begin{enumerate}
\item $K_i$ is isomorphic to $K_{i-1}[X]/(g_i(X))$ over $K_{i-1}$, and
\item the Galois group $G_i:=\gal(L_i/K_{i-1})$ acts primitively on the set of roots of $g_i$ in $L_i$, where $L_i$ is the Galois closure of $K_i/K_{i-1}$.
\end{enumerate}
We implement the algorithm $\mathtt{PrimitiveAction}$ required in Theorem~\ref{thm_primitivereduction} using the algorithm $\mathtt{SubgroupSystem}$. The latter has an extra parameter $N$, which is chosen as above.
For $i\in [k]$, let  $\mathcal{F}_i$ be the  $(K_{i-1},g_i)$-subfield system computed by $\mathtt{SubgroupSystem}$ on the input $(K_{i-1}, N, g_i)$, and let $\mathcal{P}_i$ be the associated subgroup system over $G_i$.
Note that the groups $G_i$ are subquotients of $G$. Then by  Theorem~\ref{thm_primcriterion}, Lemma~\ref{lem_syspn}, and Lemma~\ref{lem_subquo},  for all $i\in [k]$,   
 all strongly antisymmetric $\mathcal{P}_i$-schemes are discrete on $(G_i)_x$ for all roots $x$ of $g_i$,  provided that $N=\mathrm{poly}(k(G)^{\log k(G)},r(G),\deg(f))$ is sufficiently large. In this case, by Theorem~\ref{thm_primitivereduction}, all strongly antisymmetric $\mathcal{P}$-schemes are discrete   on $G_x\in \mathcal{P}$ for all roots $x$ of $\tilde{f}$, where $\mathcal{P}$ is the subgroup system associated with $\mathcal{F}$.

Finally, we run the generalized $\mathcal{P}$-scheme algorithm in Chapter~\ref{chap_alg_general} using the $(K_0,\tilde{f})$-subfield system $\mathcal{F}$ computed above, so that $\tilde{f}$ is completely factorized by  Theorem~\ref{thm_algmain2formal}. 
If  $f$  satisfies Condition~\ref{cond_spoly}, we may also use the simpler algorithm in Chapter~\ref{chap_alg_prime} and apply Theorem~\ref{thm_algmain2formal} instead.

The above algorithm  assumes that the value of a sufficiently large integer $N$ is known. We may avoid this assumption by running the algorithm multiple times, where $N$ is initially a constant and is doubled each time, until $f$ is completely factorized. It only causes an extra factor of $O(\log N)$  in the running time.
\end{proof}

\section{The O'Nan-Scott theorem for finite primitive permutation groups}\label{sec_onanscott}

The O'Nan-Scott theorem for finite primitive permutation groups \citep{LPS88} is one of the most influential theorems in permutation group theory. In this section, we describe this theorem and the related definitions.

We start with the notion of the {\em socle} of a finite group: 
 \begin{defi}[socle]
The {\em socle}\index{socle} of a finite group $G$, denoted by $\soc(G)$, is the subgroup  generated by  the minimal (nontrivial) normal subgroups of $G$.
\end{defi}
\nomenclature[g1b]{$\soc(G)$}{socle of a finite group $G$}
 
Next we define the five categories of finite primitive permutation groups appeared in the O'Nan-Scott theorem.
 
\paragraph{Almost simple type.}

Let $T$ be a noncyclic finite simple group, so that its center $Z(T)$ is trivial. We identify $T$ with the inner automorphism group $\inn(T)\subseteq\aut(T)$ via the isomorphism 
sending $g\in T$ to the conjugation $h\mapsto ghg^{-1}$ (this map is indeed an isomorphism since its kernel equals $Z(T)$ and hence is trivial). 
%The image of $T$ in $\aut(T)$ is known as the {\em inner automorphism group}, denote by $\inn(T)$.

We say a finite group is {\em almost simple}\index{almost simple group} if it is isomorphic to a group $G$ satisfying $T\subseteq G\subseteq \aut(T)$ for some noncyclic finite  simple group $T$.  
It is known that  in this case $T=\soc(G)$ holds.

A finite permutation group of {\em almost simple type} is simply a finite  primitive permutation group that is  also almost simple as an abstract group:

\begin{defi}[almost simple type]
A finite permutation group is said to be of {\em almost simple type}\index{almost simple type} if it is primitive and almost simple.
\end{defi}

\paragraph{Affine type.} Finite permutation groups {\em of affine type} are primitive groups arising as  subgroups of general affine    groups that contain all the translations:

\begin{defi}[affine type] \label{defi_affine}
A finite permutation group is said to be of {\em affine type}\index{affine type} if it is primitive and permutation isomorphic to a subgroup  $G$ of a general affine group $\agl(V)$ acting naturally on a finite-dimensional vector space $V$ over a prime field $\F_p$, and $G$ contains the subgroup of translations $V^\sharp:=\{x\mapsto x+u: u\in V\}\subseteq \agl(V)$.
\end{defi}

For example, Lemma~\ref{lem_primitivesol} states that finite primitive solvable permutation groups are of affine type.

\paragraph{Diagonal type.}

Let $T$ be a noncyclic finite simple group and let $k\geq 2$ be an integer.
%Regard $T$ as a normal subgroup of $\aut(T)$ as we did in the case of permutation groups of almost simple type.
Consider the subgroup $A$ of $\aut(T)^k$, defined by
\[
A:=\{(a_1,\dots,a_k)\in \aut(T)^k: a_i\inn(T)=a_j\inn(T) ~\text{for all}~i,j\in [k] \}.
\]
The group $\sym(k)$ acts on $A$ by permuting the $k$ coordinates, sending $(a_1,\dots,a_k)\in A$ to $(a_{\pi^{-1}(1)},\dots,a_{\pi^{-1}(k)})$. So we can form the semidirect product
\[
W:=A\rtimes \sym(k).
\]
Also define the subgroups $M, D\subseteq W$ by 
\[
M:=\inn(T)^k\subseteq A\subseteq W
\]
 and 
\[
D:=\{(a,\dots,a) \pi: a\in \aut(T), \pi\in \sym(k)\}\subseteq W.
\]
Then $W$ acts on the right coset space $D\backslash W$ by inverse right translation.
Permutation groups of diagonal type arise as subgroups of $W$:
\begin{defi}[diagonal type]\label{defi_primdiag}
A finite permutation group is said to be of {\em diagonal type}\index{diagonal type} if it is primitive and is permutation isomorphic to a group $G$ satisfying $M\subseteq G\subseteq W$ acting on $D\backslash W$ by inverse right translation, where $D,M,W$ are as above. 
%determined by a noncyclic finite simple group $T$ and an integer $k\geq 2$.
\end{defi}

\begin{exmp}[holomorph of a noncyclic finite simple group]\label{exmp_diag}
Let $T$ be a noncyclic finite simple group. We may form the semidirect product $\mathrm{Hol}(T):=T\rtimes\aut(T)$ with respect to the natural action of $\aut(T)$ on $T$. The group $\hol(T)$ is called the {\em holomorph}\index{holomorph} of $T$.
By identifying $T$ (as a set) with the left coset space $\hol(T)/\aut(T)$ via the bijection $T\to \hol(T)/\aut(T)$ sending $g\in T$ to $g \aut(T)$, we see that the action of $\hol(T)$ on $\hol(T)/\aut(T)$ by left translation is equivalent to its action on the set $T$ defined by $\prescript{hg}{}{h'}=h\prescript{g}{}{h'}$ for $h,h'\in T$, $g\in \aut(T)$. By the following lemma, this is an example of  finite primitive permutation groups of diagonal type.
\nomenclature[g1c]{$\hol(G)$}{holomorph of a group $G$}

\begin{lem}\label{lem_holom}
$\hol(T)$ is a finite primitive permutation group of diagonal type on $T$.\footnote{Lemma~\ref{lem_holom} holds more generally for any group $G$ satisfying $T\rtimes\inn(T)\subseteq G\subseteq \hol(T)$. In an alternative formulation of the O'Nan-Scott theorem, such a group $G$ is said to have type HS (holomorph of a simple group). See, e.g., \citep{PLN97}. We do not use this notation in the thesis.}
\end{lem}

\begin{proof}
The action of $\hol(T)$ on $T$ is obviously transitive. It is faithful since $\hol(T)_e=\aut(T)$ acts faithfully on $T$. To prove that  $\hol(T)$ is primitive, we want to show that $\aut(T)$ is maximal in $\hol(T)$. Consider any group $G$ satisfying $\aut(T)\subseteq G\subseteq \hol(T)$. The kernel of $G$ under the quotient map $\hol(T)\to \aut(T)$ is a normal subgroup of $T$, and hence is either $\{e\}$ or $T$. So $G$ equals either $\aut(T)$ or $\hol(T)$. Therefore $\hol(T)$ acts primitively on $T$.

Now define the groups $D, M, W, A$ as above with respect to $T$ and $k=2$.
For $g\in T$, denote by $\tau_g\in\inn(T)$ the conjugation by $g$ which sends $x\in T$ to $gxg^{-1}$.
Define the map $\rho:\hol(T)\to A$ via $\rho(gh)=(\tau_g h, h)$ for $g\in T$, $h\in\aut(T)$. It is straightforward to check that $\rho$ is a well defined injective group homomorphism and $M\subseteq \rho(\hol(T))$. The action of $A$ on $D\backslash W$ by inverse right translation thus induces an action of $\hol(T)$ on $D\backslash W$, which is transitive since $M\subseteq \rho(\hol(T))$.
The stabilizer of $De\in D\backslash W$ with respect to this action is $\rho^{-1}(D\cap A)=\aut(T)\subseteq\hol(T)$, which is exactly the stabilizer of $e\in T$ with respect to the action of $\hol(T)$ on $T$.  By Lemma~\ref{lem_equivaction}, the action of $\hol(T)$ on $T$ and that on $D\backslash W$ are equivalent. The lemma follows by definition.
\end{proof}

\end{exmp}

\paragraph{Product type.} Let $H$ be a primitive permutation group on a finite set $\Gamma$ of almost simple type or diagonal type. 
Let $k\geq 2$ be an integer. Define the wreath product
\[
W:=H \wr \sym(k)= H^k\rtimes\sym(k),
\]
where $\sym(k)$ permutes the $k$ factors of $H^k$.
The group $W$ has a natural {\em primitive wreath product action}\index{primitive wreath product action} on $\Gamma^k$ where  $H^k$ acts coordinatewise and $\sym(k)$ permutes the coordinates. Also define 
\[
M:=\soc(H)^k\subseteq W.
\]
Permutation groups of product type arise as subgroups of $W$:
\begin{defi}[product type]\label{defi_prodprim}
A finite permutation group is said to be of {\em product type}\index{product type} if it is primitive and is permutation isomorphic to a group $G$ satisfying $M\subseteq G\subseteq W$ acting on $\Gamma^k$ via the primitive wreath product action, where $M,W,\Gamma,k$ are as above.
% determined by a primitive permutation group $H$ of almost simple type and an integer $k\geq 2$.
\end{defi}

\paragraph{Twisted wreath type.}

Let $T$ be a noncyclic finite simple group.
Let $P$ be a transitive permutation group on $[k]$ where $k\geq 2$. 
Denote by $\map(P, T)$ the set of the maps from $P$ to $T$.
Suppose  $\varphi: P_1\to \aut(T)$ is a group homomorphism from the stabilizer $P_1$ of $1\in [k]$ to $\aut(T)$. 
Define
\[
B:=\{f\in \map(P, T): f(pq^{-1})=\prescript{\varphi(q)}{}{(f(p))}~\text{for all}~p\in P, q\in P_1\},
\]
which is a group under coordinatewise multiplication. The group $P$ acts on $B$ via $(\prescript{p}{}{f})(px)=f(x)$, or equivalently
\[
(\prescript{p}{}{f})(x)=f(p^{-1}x) \qquad \text{for all}~p,x\in P, f\in B.
\]
It is easy to check that this is a well defined action.\footnote{For example, the map $\prescript{p}{}{f}$ is indeed in $B$ for $p\in P$ and $f\in B$ since
$(\prescript{p}{}{f})(p'q^{-1})=f(p^{-1}p'q^{-1})=\prescript{\varphi(q)}{}{(f(p^{-1}p'))}=\prescript{\varphi(q)}{}{((\prescript{p}{}{f})(p'))}$
for all $p'\in P$ and $q\in P_1$.}
So we can form the semidirect product $G:=B\rtimes P$ with respect to this action. The group $G$ is also called the {\em twisted wreath product} with respect to the data $(T, P, \varphi)$, denoted by $T\twr_{\varphi} P$ \citep{Neu63, DM96}. 
\nomenclature[g1d]{$\map(S, T)$}{set of all maps from the set $S$ to the set $T$}
\nomenclature[g1e]{$T\twr_{\varphi} P$}{twisted wreath product with respect to the data $(T, P, \varphi)$}

Finite  permutation groups of twisted wreath type are defined as follows.
\begin{defi}[twisted wreath type]\label{defi_twrprim}
A finite permutation group is said to be of {\em twisted wreath type}\index{twisted wreath type} if it is primitive and is permutation isomorphic to a group $G=T\twr_{\varphi} P$ acting on the left coset space $G/P$ via left translation, where $T$, $P$, and $\varphi$ are as above.
\end{defi}

\paragraph{The O'Nan-Scott theorem.} Now we are ready to state the O'Nan-Scott theorem for finite primitive permutation groups \citep{LPS88}.

\begin{thm}[O'Nan-Scott theorem]\index{O'Nan-Scott theorem}
A finite primitive permutation group is of exactly one of the following types: almost simple type, affine type, diagonal type, product type, and twisted wreath type.
\end{thm}

\paragraph{Schreier conjecture.}
We conclude this section by mentioning the fact that the outer automorphism group of every finite simple group is solvable. This is known as the {\em Schreier conjecture}, and is now known to be true as a result of CFSG. See, e.g., \citep{DM96}.
\begin{thm}\label{thm_schreier}\index{Schreier conjecture}
The outer automorphism group $\out(T)$ of every finite simple group $T$ is solvable.
\end{thm}

\section{Almost simple type}\label{sec_prim_as}\index{almost simple type}

In this section, we prove Theorem~\ref{thm_primcriterion} for finite primitive permutation groups of almost simple type.
Our proof is based on the work on the minimal base sizes of  such permutation groups, including the work on Pyber's base size conjecture, and the constant bounds for non-standard actions.

\paragraph{Pyber's base size conjecture.}
Recall that a base of a permutation group $G$ on a finite set $S$ is a subset $T\subseteq S$ satisfying $G_T=\{e\}$, and  the minimal base size $b(G)$ is the minimum cardinality of a base  of $G$. By the orbit-stabilizer theorem, we have the lower bound $b(G)\geq \log|G|/\log |S|$. {\em Pyber's base size conjecture} \citep{Pyb93} asserts that this is asymptotically tight if $G$ is primitive:
\begin{conj}[Pyber's base size conjecture]\index{Pyber's base size conjecture}
Let $G$ be a finite primitive permutation group on a finite set $S$. Then $b(G)=\Theta(\log|G|/\log |S|)$.
\end{conj}

There has been extensive work on Pyber's  conjecture \citep{Se96, GM98,  GSS98, LS02, Ben05, Faw13, LS14, BS15}.
Recently, Duyan, Halasi, and Mar{\'o}ti announced a proof of this conjecture \citep{DHM16}.

We only need the special case of the conjecture for almost simple type, which is verified in \citep{Ben05}.

\begin{thm}[ \citep{Ben05}]\label{thmpybas}
Let $G$ be a finite primitive permutation group of almost simple type on a finite set $S$. Then $b(G)=\Theta(\log|G|/\log |S|)$.
\end{thm}

\paragraph{Bounds for non-standard actions.}\index{standard action}

We also need a result on non-standard actions of primitive permutation groups of almost simple type.
Recall that an action of a symmetric group $\sym(n)$ is {\em standard} if  it is equivalent to the action on the set of $k$-subsets of $[n]$ for some $k\in [n]$, or the action on an orbit of the set of partitions of $[n]$, induced from the natural action of $\sym(n)$ on $[n]$ (see Chapter~\ref{chap_sym}). And we say an action of  $\alt(n)$ on a finite set $S$ is standard if it is restricted from a standard action of $\sym(n)$ on $S$. Analogously, one can define standard actions of a classical group which, roughly speaking, are actions that permute subspaces (or pairs of subspaces of complementary dimension) of the natural module. See \citep{LS99, Bur07} for the rigorous definition. Finally, an action of  a primitive permutation group  of almost simple type is {\em  non-standard}\index{non-standard action} if it is not a standard action.

It was conjectured in \citep{Cam92, CK93} that the minimal base sizes of non-standard actions are bounded by an absolute constant $c\in\N$. This conjecture was proved by Liebeck and  Shalev \citep{LS99}.\footnote{In addition, a chain of  papers \citep{Bur07, BLS09, BOW10, BGS11} shows that the minimum possible value of the constant $c$ is 7.} We state the following weaker form of this result, where we do not distinguish standard and non-standard actions of classical groups. This weaker form is sufficient for our goal.

\begin{thm}\label{thm_asbase}
Let $G$ be a finite primitive permutation group $G$ of almost simple type, and let $T=\soc(G)$. Then one of the following holds:
\begin{enumerate}
\item $G$ is permutation isomorphic to a symmetric group or an alternating group with a standard action. 
\item $T$ is a classical simple group.
\item $b(G)\leq c$, where $c\in\N$ is some absolute constant.
\end{enumerate}
\end{thm}

See \citep[Theorem~1.3]{LS99} for the original statement.

\paragraph{Proof of Theorem~\ref{thm_primcriterion} for almost simple type.}
Now we are ready to prove  Theorem~\ref{thm_primcriterion} for a primitive permutation group $G$ of almost simple type. In fact, we prove it in the following  general form which applies to   any subgroup $H\subseteq G$.

\begin{lem}\label{lem_sysas}
Let $G$ be a primitive permutation group of almost simple type on a finite set $S$, and let $H$ be a subgroup of $G$ on $S$.
 Then for sufficiently large $N=\mathrm{poly}(k(G)^{d_{\sym}(k(G))},r(G),|S|)\geq |S|$, all strongly antisymmetric $\mathcal{P}_{H,N}$-schemes are discrete on $H_x\in \mathcal{P}_{H,N}$ for all $x\in S$.
\end{lem}

\begin{proof}
Let $T=\soc(G)$ and $\mathcal{P}=\mathcal{P}_{H,N}$. Consider the three cases in Theorem~\ref{thm_asbase}. First assume $G$ is permutation isomorphic to a symmetric group $\sym(k)$ or an alternating group $\alt(k)$ with a standard action.
Note $k\leq k(G)$.
  We have $H_{x,y}\in \mathcal{P}$ for all $x,y\in S$ provided $N\geq |S|^2$. We also have $H_{\{x,y\}\cup U}\in\mathcal{P}$ for all $x,y,z\in S$ and $U\subseteq H_{x,y}z$ satisfying $|H_{x,y}z|\leq k$ and $1\leq |U|\leq d_{\sym}(k)$, provided that $N=k(G)^{\Omega(d_\sym(k(G)))}$ is sufficiently large. The lemma holds by Theorem~\ref{thm_tuples} in this case.
  
Next assume $T$ is a classical simple group. Then $|T|\leq r(G)$.
It is also known by CFSG that $|\out(T)|=O(\log |T|)$ (see \citep{CCNP85}) and hence $|G|\leq |\aut(T)|=|T|^{O(1)}=r(G)^{O(1)}$.
  By Lemma~\ref{lem_antibasebound} and Theorem~\ref{thmpybas}, we have
\[
d(H)\leq b(H)\leq b(G)=\Theta(\log |G|/\log |S|) 
\]
and hence $|S|^{d(G)}=|G|^{\Theta(1)}=r(G)^{O(1)}$.  It follows that for sufficiently large $N=r(G)^{\Omega(1)}$, the subgroup system $\mathcal{P}$ contains the system of stabilizers of depth $d(H)$. So the lemma also holds in this case.

Finally, in the last case of Theorem~\ref{thm_asbase}, we have  $d(H)\leq b(H)\leq b(G)\leq c$, and the lemma holds for $N\geq |S|^c$.
\end{proof}

Choosing $H=G$ in Lemma~\ref{lem_sysas}, we have 
\begin{cor}\label{cor_verias}
Theorem~\ref{thm_primcriterion} holds for finite primitive permutation groups of almost simple type.
\end{cor}

\section{Affine type}\index{affine type}

In this section, we prove Theorem~\ref{thm_primcriterion} for finite primitive permutation groups of affine type.
The following definitions are needed. 

\begin{defi}[irreducible / primitive linear group]
A group $H\subseteq \gl(V)$ is said to be an {\em irreducible linear group}\index{irreducible linear group} on $V$ if $H$ does {\em not} fixes any subspace $W\subseteq V$ other than $\{0\}$ and $V$.
%, otherwise a {reducible linear group}. 
And  $H\subseteq \gl(V)$ is said to be a {\em primitive linear group}\index{primitive!linear group} on $V$ if it is an irreducible linear group, and $V$ {\em cannot} be written as a direct sum $V=\bigoplus_{i=1}^k V_i$ such that $k>1$ and $H$ permutes the direct summands $V_i$.
\end{defi}

%Denote by $V^\sharp:=\{x\mapsto x+u: u\in V\}\subseteq G$ the subgroup of translations. And let by $G_0\subseteq G$  the stabilizer of the origin $0\in V$, which is a subgroup of $\gl(V)$. 

The following fact is well known (see, e.g., \citep[Section~\RN{1}.4]{Su76}).

\begin{lem}\label{lem_affinesemi}
Let $G$ be a finite primitive permutation group $G$ of affine type on a vector space $V$ over a prime field $\F_p$. Then the stabilizer $G_0\subseteq \gl(V)$ of the origin $0\in V$ is an irreducible linear group on $V$.
\end{lem}

We prove Theorem~\ref{thm_primcriterion} for   affine type by studying the stabilizer of the  origin.
In the following, we first discuss  the case that this stabilizer is a primitive linear group (over $\F_p$), and then the case of (possibly imprimitive) irreducible linear groups. 

%The group $G_0$ is irreducible but may not be a primitive linear group on $V$. We proceed by first reducing to the case of  (subgroups of) a primitive linear groups and then applying a structure theorem in \citep{LS14} on primitive linear groups.

\paragraph{Primitive linear groups.}

 Our analysis  is based on the work \citep{LS02, LS14} on bases of primitive linear groups.
% which verifies part of Pyber's base size conjecture.
 We start with the following definitions.
 
 \begin{defi}[fully deleted permutation module \citep{KL90}]\label{defi_delmod}\index{fully deleted permutation module}
 Fix  a finite field $\F_q$ and $k\in\N^+$.
 Define \begin{align*}
  E(k,q)&:=\{(a,\dots,a): a\in \F_q\}\subseteq \F_q^k,\\
 M(k,q)&:=\{(a_1,\dots,a_k)\in \F_q^k: a_1+\dots+a_k=0\},\\
 U(k,q)&:=M(k,q)/(M(k,q)\cap E(k,q)).
 \end{align*}
  Let $\sym(k)$ act on $\F_q^k$ by permuting the $k$ coordinates, which induces an action on $U(k,q)$.
%The faithful linear representation of $\sym(k)$ on $U(k,q)$ identifies $\sym(k)$ with a subgroup of $\gl(U(k,q))$.
We call $U(k,q)$ the {\em fully deleted permutation module} for $\sym(k)$ over $\F_q$.
 \end{defi}
\nomenclature[g1f]{$U(k,q)$}{fully deleted permutation module for $\sym(k)$ over $\F_q$}
 
 \begin{defi}[tensor product of linear groups]\label{defi_tenlin}\index{tensor product of linear groups}
 Let $V_1,\dots,V_k$ be  vector spaces over a finite field $\F_q$. Let $G_1,\dots,G_k$ be finite groups where $G_i\subseteq \gl(V_i)$ for $i\in [k]$. Define an action of $G_1\times\dots\times G_k$ on the tensor product $U:=V_1\otimes\dots\otimes V_k$ (over $\F_q$) by letting
 \[
 \prescript{(g_1,\dots,g_k)}{}{a_1\otimes\dots\otimes a_k}=\prescript{g_1}{}{a_1}\otimes\dots\otimes\prescript{g_k}{}{a_k}
 \]
and extending to all tensors multilinearly. This gives a linear representation $\rho: G_1\times\dots\times G_k\to \gl(U)$.
Write $g_1\otimes\dots\otimes g_k$ for $\rho(g_1,\dots,g_k)\in \gl(U)$.
%\footnote{In general, such an expression is not unique. For example, we have $e\otimes c=c\otimes e$ for $c\in\F_q^\times$.}
 And write $G_1\otimes\dots\otimes G_k$ for $\rho(G_1\times\dots\times G_k)\subseteq \gl(U)$, called the {\em tensor product} of $G_1,\dots,G_k$ (over $\F_q$).
 \end{defi}
\nomenclature[g1g]{$G_1\otimes\dots\otimes G_k$}{tensor product of the linear groups $G_i$}
\nomenclature[g1h]{$g_1\otimes\dots\otimes g_k$}{image of $(g_1,\dots,g_k)$ in $G_1\otimes\dots\otimes G_k$}

We need the following structure theorem  in \citep{LS14} on primitive linear groups. 
See \citep[Theorem~1]{LS14} for a more detailed statement.

\begin{thm}[\citep{LS14}]\label{thm_primlinstr}
Let $p$ be a prime number, $V$ a finite-dimensional vector space over $\F_p$,
and $G$  a  primitive linear group on  $V$. Choose the largest power $q$ of $p$ such that $V$ can be identified with a vector space  $V(q)$ over $\F_q$ and $G\subseteq\gammal(V(q))$.
Let $H:=G\cap \gl(V(q))$ act on $V(q)$.
Then there exists an absolute  constant $C\in\N^+$ such that either $b(H)\leq C$, or $V(q)$ can be identified with a tensor product over $\F_q$
\[
V(q)= \bigotimes_{i=1}^s U(k_i, q) \otimes W_0\otimes \bigotimes_{j=1}^t W_j,
\]
where $k_i\geq 5$\footnote{The condition $k_i\geq 5$ is implicit in \citep{LS14}. If $k_i<5$, we may always remove the factor $U(k_i,q)$ by replacing $W_0$ with $U(k_i,q)\otimes W_0$ (see \citep[Lemma~3.3]{LS02}).} and $U(k_i,q)$ is the fully deleted permutation module for $\sym(k_i)$ over $\F_q$ for $i\in[s]$, and $W_j$ is  vector space of dimension $d_j\in\N^+$ over $\F_q$ for $0\leq j\leq t$. Moreover, in the latter case, the group $H$ is a subgroup of 
\[
\bigotimes_{i=1}^s \sym(k_i) \otimes D_0\otimes  \bigotimes_{j=1}^t D_j
\]
acting on $V(q)$ that satisfies the following conditions:
\begin{enumerate}
\item For $i\in [s]$, the group $\sym(k_i)$ acts faithfully on $U(k_i,q)$ (see Definition~\ref{defi_delmod}).\footnote{We regard $\sym(k)$ as a subgroup of $\gl(U(k,q))$ via the faithful linear representation $\sym(k)\hookrightarrow \gl(U(k,q))$.}
\item $D_0\subseteq\gl(W_0)$ acts on $W_0$ and $b(D_0)\leq C$.
\item For $j\in [t]$, the group $D_j$ acting on $W_j$ is the normalizer in $\gl(W_j)$ of one of the quasisimple classical groups\index{quasisimple group} $\mathrm{SL}_{d_j}(q_j), \mathrm{SU}_{d_j}(q_j^{1/2}), \mathrm{Sp}_{d_j}(q_j), \Omega_{d_j}(q_j)\subseteq \gl_{d_j}(q_j)$.\footnote{
For the definitions of these classical groups, see, e.g., \citep{KL90, Asc00}. A group $G$ is {\em quasisimple} if it equals its commutator subgroup and its inner automorphism group is simple, or equivalently, if it is a {\em perfect central extension} of a simple group  \citep{Asc00}.
}
Here $\F_{q_j}$ is a subfield of $\F_q$, and we identify $\gl_{d_j}(q_j)$ with a subgroup $\gl(W_j')\subseteq \gl(W_j)$ for some  vector space $W_j'\subseteq W_j$ over $\F_{q_j}$ by fixing an $\F_{q_j}$-basis of $W_j'$ that is also an $\F_q$-basis of $W_j$.

%\[
%\gl(W_j')=\gl(W_j)_{\{W_j'\}}\subseteq \gl(W_j)
%\]
%where $W_j'\subseteq W_j$ is a vector space over $\F_{q_j}$ spanned by a fixed $\F_q$-basis of $W_j$ over $\F_{q_j}$.
\item $H$ contains the group $\bigotimes_{i=1}^s \alt(k_i) \otimes \{e\} \otimes \bigotimes_{j=1}^t D_j^{(\infty)}$, where $D_j^{(\infty)}$ denotes the last term in the derived series of $D_j$.
\end{enumerate}
\end{thm}

The following lemma implies that the group $D_j$ in Definition~\ref{thm_primlinstr} for each $j\in [t]$ is a subgroup of $\F_q^\times \gl_{d_j}(q_j)$. For its proof, see \citep[Proposition~4.5.1]{KL90}.
\begin{lem}\label{lem_normsub}
Suppose $\F_{q_0}\subseteq\F_q$,  and  $G\subseteq \gl_d(q)$ is one of the quasisimple classical groups $\mathrm{SL}_{d}(q_0), \mathrm{SU}_{d}(q_0^{1/2}),  \mathrm{Sp}_{d}(q_0), \Omega_{d}(q_0)\subseteq\gl_d(q_0)\subseteq\gl_d(q)$.
Then $N_{\gl_d(q)}(G)\subseteq \F_q^\times \gl_d(q_0)$.
\end{lem}

For convenience, we also make the following definition.

\begin{defi}[primary tensor]\label{defi_eletensor}\index{primary tensor}
Use the notations in Theorem~\ref{thm_primlinstr} and assume $b(G)>C$. So $W$ is identified with the tensor product
\[
 \bigotimes_{i=1}^s U(k_i, q) \otimes W_0\otimes \bigotimes_{j=1}^t W_j
\]
over $\F_q$ by Theorem~\ref{thm_primlinstr}. We say an element $x\in V-\{0\}$ is a {\em primary tensor} if 
$x$ is a pure tensor, i.e., $x= u_1\otimes \dots\otimes u_s\otimes w_0\otimes w_1\otimes\dots\otimes w_t$, where $u_i\in U(k_i, q)$ for $i\in [s]$ and $w_j\in W_j$ for $0\leq j\leq t$, and in addition,
\begin{enumerate}
\item for $i\in [s]$, $u_i\in  U(k_i, q)$ is represented by an element in $M(k_i,q)\subseteq\F_{q_i}^{k_i}$ (see Definition~\ref{defi_delmod})  that has exactly two nonzero coordinates, and
\item for $j\in [t]$, $w_j\in W_j$ has the form $w_j=c w_j'$ where $c\in\F_q^\times$ and $w_j'\in W'_j$ (see Definition~\ref{thm_primlinstr}).
\end{enumerate}
In addition, for two primary tensors $x,y\in V-\{0\}$, we write $x\sim y$ if $x$ and $y$ can be written as tensor products of vectors satisfying the above conditions and they differ at no more than one vector $u_i$ or $w_j$. In other words, we can write  
\[
x=u_1\otimes \dots\otimes u_s\otimes w_0\otimes\dots\otimes w_t
\] and either
\[
y=u_1\otimes \dots\otimes u_{i-1}\otimes u'_i\otimes u_{i+1} \otimes\dots\otimes u_s\otimes w_0\otimes\dots\otimes w_t
\]
for some $i\in [s]$ and $u'_i\in U(k_i,q)$, or
\[
y=u_1\otimes  \dots\otimes u_s\otimes w_0\otimes \dots\otimes  w_{j-1}\otimes w'_j\otimes w_{j+1}\otimes\dots\otimes w_t
\]
for some $0\leq j\leq t$ and $w_j\in W_j$, so that the vectors $u_i$ (resp. $u'_i$) and $w_j$ (resp. $w'_j$) satisfy the above defining conditions of  primary tensors.
\end{defi}

Note that in Definition~\ref{defi_eletensor}, a vector space $M(k_i,q)$ is spanned by vectors with exactly two nonzero coordinates, and   $W_j$ is spanned by vectors in $W'_j$ over $\F_q$.
So any $x\in V$ can be written as  a finite sum of primary tensors. Also note that for any two  primary tensors $x,y\in V-\{0\}$, there exists a finite sequence of primary tensors $x_0,\dots,x_k\in V-\{0\}$ such that $x_0=x$, $x_k=y$, and $x_{i-1}\sim x_i$ for all $i\in [k]$.

Now we are ready to prove the following analogue of Theorem~\ref{thm_primcriterion} for subgroups of primitive linear groups over $\F_p$.

\begin{lem}\label{lem_prilindisc}
%Use the notations in Theorem~\ref{thm_primlinstr}.
Let $G$ be a primitive linear group on  a vector space $V$ over $\F_p$ as in Theorem~\ref{thm_primlinstr}, and let $G'$ be a subgroup of $G$ on $V$.
Then for sufficiently large $N=\mathrm{poly}(r(G),|V|)\geq |V|$, all strongly antisymmetric $\mathcal{P}_{G',N}$-schemes are discrete on $G'_x\in \mathcal{P}_{G',N}$ for all $x\in V$.
\end{lem}

\begin{proof}
Use the notations in Theorem~\ref{thm_primlinstr}.
Fix $\alpha\in \F_q^\times$ that does not lie in any proper subfield of $\F_q$.
First assume $b(H)\leq C$. Let $B\subseteq V$ be a base of $H$ of cardinality at most $C$.
Pick a nonzero element $z\in B$. Then $B\cap \{\alpha z\}$ is a base of $G$ since $G_{z,\alpha z}\subseteq G\cap \gl(V(q))=H$. So $d(G)\leq b(G)\leq C+1$.
Then for $N\geq |V|^{C+1}$, all strongly antisymmetric $\mathcal{P}_{G',N}$-schemes are discrete on $G'_x\in \mathcal{P}_{G',N}$ for all $x\in V$, as desired.

So assume $b(H)>C$. Then we have $V(q)= \bigotimes_{i=1}^s U(k_i, q) \otimes W_0\otimes \bigotimes_{j=1}^t W_j$
 and $H\subseteq \bigotimes_{i=1}^s \sym(k_i) \otimes D_0\otimes  \bigotimes_{j=1}^t D_j$ as in Theorem~\ref{thm_primlinstr}.
Let $\mathcal{P}=\mathcal{P}_{G',N}$ and let $\mathcal{C}$ be a strongly antisymmetric $\mathcal{P}$-scheme.
Let $N\geq |V|^4$ so that $G'_{x,y,z,w}\in \mathcal{P}$ for all $x,y,z,w\in V$.
Fix $x\in V$. We want to prove that $\mathcal{C}$ is discrete on $G'_x$. By Lemma~\ref{lem_min}, it suffices to prove that
$\mathcal{C}$ is discrete on $G'_{x,\alpha x}$.

Consider the diagonal action of $G'$ on $V\times V$, and let $O$ be the $G'$-orbit of $(x, \alpha x)$.
The elements in $O$ are of the form $(y,\beta y)$, where $y\in V$ and $\beta\in\F_q^\times$ is a conjugate of $\alpha$, i.e., $\beta=\prescript{g}{}{\alpha}$ for some $g\in\gal(\F_q/\F_p)$.
Also note that for any distinct $y,z\in V$, the difference $z-y$ can be written as a finite sum of primary tensors.
By  Lemma~\ref{lem_selfredspan}, it suffices to prove, for all distinct $y,z\in V$ whose difference $z-y$ is a primary tensor  and conjugates $\beta,\gamma\in\F_q^\times$ of $\alpha$, that $\mathcal{C}|_{G'_{y, \beta y}}$ is discrete on $G'_{y,\beta y,z, \gamma z}=G'_{y,\beta y,z}$.
Fix such $y,z,\beta,\gamma$.

Let $H'=G'\cap H$. Then $G'_{y,\beta y}=H'_y$ and $G'_{y,\beta y,z}=H'_{y,z}=H'_{y,z-y}$. So we want to prove that  $\mathcal{C}|_{H'_y}$ is discrete on $H'_{y,z-y}$. Note that  every element in the $H'_y$-orbit of $z-y$ is  a primary tensor.
As noted after Definition~\ref{defi_eletensor}, for any two primary tensors $u,v\in V-\{0\}$, there exists a finite sequence of primary tensors $x_1,\dots,x_k \in V-\{0\}$ such that $x_1=u$, $x_k=v$, and $x_{i-1}\sim x_i$ for all $i\in [k]$. Again by Lemma~\ref{lem_selfredspan}, it suffices to prove, for all primary tensors $u, v\in V-\{0\}$ satisfying $u\sim v$, that $\mathcal{C}|_{H'_{y,u}}$ is discrete on $H'_{y,u,v}$ (note $H'_{y,u}=G'_{y,\beta y,u}, H'_{y,u,v}=G'_{y,\beta y,u,v}\in\mathcal{P}$). 
Fix such $u,v$.
 Suppose $u= u_1\otimes \dots\otimes u_s\otimes w_0\otimes\dots\otimes w_t$ where $u_i\in U(k_i, q)$ for $i\in [s]$ and $w_j\in W_j$ for $0\leq j\leq t$  satisfy the conditions in Definition~\ref{defi_eletensor}.
 
 First consider the case that $v$ has the form
 \[
 v=u_1\otimes\dots \otimes u_{r-1}\otimes u'_r\otimes u_{r+1} \otimes \dots\otimes u_s\otimes w_0\otimes\dots\otimes w_t,
 \]
 where  $r\in [s]$ and $u'_r\in  U(k_r, q)$ is represented by a vector  $\tilde{u}'_r\in M(k_r,q)$ with exactly two nonzero coordinates.
 Let $n:=|H'_{y,u} v|$. We prove a bound on $n$. Consider an element $g\in H'_{y,u}$. By Theorem~\ref{thm_primlinstr}, we may write $g=g_1\otimes\dots \otimes g_s\otimes h_0\otimes \dots\otimes h_t$ where $g_i\in\sym(k_i)$ for $i\in [s]$ and $h_j\in D_j$ for $0\leq j\leq t$. As $g$ fixes $u$, we know $g_r\in\sym(k_r)$ sends $u_r$ to $c u_r$ for some $c\in\F_q^\times$.
As $k_r\geq 5$ and $\tilde{u}_r$ has exactly two nonzero coordinates, either $g_r$ fixes $\tilde{u}_r$ and $c=1$, or $g_r$ swaps the two nonzero coordinates of $\tilde{u}_r$ and $c=-1$.
From $\prescript{g}{}{u}=u$, it is easy to see that 
\[
\prescript{g}{}{v}=c^{-1} (u_1\otimes\dots \otimes u_{r-1}\otimes \prescript{g_r}{}{u'_r}\otimes u_{r+1} \dots\otimes u_s\otimes w_0 \otimes\dots\otimes w_t).
\]
So $c$ and $ \prescript{g_r}{}{u'_r}$ determine $\prescript{g}{}{v}$. 
The  number of possible values of $ \prescript{g_r}{}{u'_r}$ is bounded by $|\sym(k_r) u'_r|\leq |\sym(k_r) \tilde{u}'_r|\leq k_r^2$. It follows that $n=|H'_{y,u} v|\leq 2k_r^2$. 
Also note that $|V|\geq |U(k_r,q)|\geq q^{k_r-2}$ and hence $k_r=O(\log|V|)$. 
Let $N\geq n^{d_\sym(n)}=|V|^{O(1)}$. Then $\mathcal{P}|_{H'_{y,u}}$  contains the system of stabilizers of depth $d_\sym(n)$ with respect to the action of $H'_{y,u}$ on $H'_{y,u} v$.
So $\mathcal{C}|_{H'_{y,u}}$ is discrete on $H'_{y,u,v}$, as desired.

Next consider the case  that $v$ has the form
\[
 v=u_1\otimes\dots \otimes u_s\otimes w'_0\otimes w_1 \otimes \dots\otimes w_t
\]
for some $w'_0\in  W_r$.
Let $B\subseteq W_0$ be a base of $D_0$ of cardinality at most $C$, which exists by Theorem~\ref{thm_primlinstr}.
For any subset $T$ of $W_0$, define
\[
\tilde{T}:=\{u_1\otimes\dots \otimes u_s \otimes a\otimes w_1 \otimes \dots\otimes w_t: a\in T\}\subseteq V.
\]
Consider $g=g_1\otimes\dots \otimes g_s\otimes h_0\otimes \dots\otimes h_t\in (H'_{y,u})_{\tilde{B}}$ where $g_i\in\sym(k_i)$ for $i\in [s]$ and $h_j\in D_j$ for $0\leq j\leq t$.
As $g$ fixes every element in $\tilde{B}$, we see $h_0\in D_0$ scales every element in $B$ by the same factor $c\in\F_q^\times$.
Then $c^{-1}h_0\in (D_0)_B=\{e\}$ and hence $h_0=c$. Therefore $h_0$ scales every element in $W_0$ by the factor $c$.
So $g\in (H'_{y,u})_{\tilde{W}_0}$.
It follows that $(H'_{y,u})_{\tilde{B}}= (H'_{y,u})_{\tilde{W}_0}$. 
Also note that $H'_{y,u}$ fixes $\tilde{W}_0$ setwisely. So $(H'_{y,u})_{\tilde{W}_0}$ is normal in $H'_{y,u}$. 
Let $N\geq |V|^{C+3}$ so that $(H'_{y,u})_{\tilde{W}_0}=H_{\{y,\beta y,u\}\cup \tilde{B}}\in\mathcal{P}$. 
 By antisymmetry of $\mathcal{C}|_{H'_{y,u}}$, we  know $\mathcal{C}|_{H'_{y,u}}$ is discrete on $(H'_{y,u})_{\tilde{W}_0}$. 
 By Lemma~\ref{lem_min},  it is also discrete on $H'_{y,u,v}\supseteq (H'_{y,u})_{\tilde{W}_0}$, as desired.

Finally, consider the case  that $v$ has the form
 \[
 v=u_1\otimes\dots \otimes u_s\otimes w_0\otimes\dots \otimes w_{r-1}\otimes w'_r\otimes w_{r+1} \otimes \dots\otimes w_t
 \]
 for some $0\leq r\leq  t$ and $w'_r\in  W_r$ such that $w'_r=c_0 w''_r$ for some $c_0\in\F_q^\times$ and $w''_r\in W'_r$.
 We claim that $|H'_{y,u} v|\leq q_r^{d_r}$.  To see this, consider  $g=g_1\otimes\dots \otimes g_s\otimes h_0\otimes \dots\otimes h_t\in H'_{y,u}$
where $g_i\in\sym(k_i)$ for $i\in [s]$ and $h_j\in D_j$ for $0\leq j\leq t$.
By Lemma~\ref{lem_normsub}, we have $h_r\in \F_q^\times \gl(W'_r)$.
As $g$ fixes $u$, we have $\prescript{h_r}{}{w_r}=c_1 w_r$ for some $c_1\in\F_q^\times$. Then  it is easy to see that
\begin{align*}
\prescript{g}{}{v}&=c_1^{-1}(u_1\otimes\dots \otimes u_s\otimes w_0\otimes\dots\otimes w_{r-1}\otimes \prescript{  h_r}{}{w'_r}\otimes w_{r+1} \otimes \dots \otimes  w_t)\\
&=u_1\otimes\dots \otimes u_s\otimes w_0\otimes\dots\otimes w_{r-1}\otimes \prescript{c_1^{-1}  h_r}{}{w'_r}\otimes w_{r+1} \otimes \dots \otimes  w_t.
\end{align*}
As $h_r\in \F_q^\times \gl(W'_r)$, we may write $h_r=c_2 h'_r$ for some $c_2\in\F_q^\times$ and $h'_r\in \gl(W'_r)$. Note that $h'_r=c_2^{-1}h_r\in \gl(W'_r)\subseteq \gl(W_r)$ sends $w_r$ to $c_2^{-1}c_1 w_r$.
Then $c_2^{-1} c_1\in\F_{q_r}^\times$.
Therefore $c_1^{-1} h_r=c_1^{-1}c_2 h'_r\in \gl(W'_r)$. 
%Also recall that $w'_r=$ for some $c_0\in\F_q^\times$ and $w''_r\in W'_r$ by assumption.
It follows that 
\[
|H'_{y,u} v|\leq |\gl(W'_r) w'_r|=|\gl(W'_r) w''_r|\leq |W'_r|=q_r^{d_r}
\]
as claimed.  Let $V'\subseteq V$ be the vector space over $\F_q$ spanned by the elements in $H'_{y,u} v$.
Let $B\subseteq H'_{y,u} v$ be an $\F_q$-basis of $V'$.
Then $|B|=\dim_{\F_q} V'\leq \dim_{\F_q} W_r=d_r$.
Note $q_r^{d_r^2}=|D_r|^{O(1)}=r(H)^{O(1)}=r(G)^{O(1)}$.
Let $N\geq q_r^{d_r^2}\geq |H'_{y,u} v|^{d_r}$, so that  $\mathcal{P}|_{H'_{y,u}}$  contains the system of stabilizers of depth $d_r$ with respect to the action of $H'_{y,u}$ on $H'_{y,u} v$.
Then $(H'_{y,u})_{V'}=(H'_{y,u})_{B}\in\mathcal{P}$.
Note that $H'_{y,u}$ fixes $V'$ setwisely, and hence $(H'_{y,u})_{V'}$ is normal in $H'_{y,u}$. 
By antisymmetry of $\mathcal{C}|_{H'_{y,u}}$, we  know $\mathcal{C}|_{H'_{y,u}}$ is discrete on $(H'_{y,u})_{V'}$.  
 By Lemma~\ref{lem_min},  it is also discrete on $H'_{y,u,v}\supseteq (H'_{y,u})_{V'}$, as desired.
\end{proof}

\paragraph{Irreducible linear groups.}

Next we extend Lemma~\ref{lem_prilindisc} to irreducible linear groups over $\F_p$.
For a group $G\subseteq \gl(V)$ and a subspace $W\subseteq V$, the setwise stabilizer $G_{\{W\}}$ acts on $W$, which gives a linear representation $\pi_W: G_{\{W\}}\to \gl(W)$. Write $G|_W$ for its image $\pi_W(G_{\{W\}})\subseteq \gl(W)$. 

We need the following lemma, whose proof can be found in, e.g., \citep[Section~\RN{4}.15]{Su76}.
 
\begin{lem}\label{lem_primlin}
Let $G$ be an irreducible linear group on a finite-dimensional vector space $V\neq \{0\}$.
Then there exists a nonzero subspace $W\subseteq V$ such that $G|_W$ is a primitive linear group on $W$, and $G$ permutes the subspaces in the set $\{\prescript{g}{}{W}: g\in G\}$.
% Moreover,  for any such $W$, we have a transitive action of $G_0$ on $S_W$ where $g\in G_0$ sends $\prescript{h}{}{W}\in S_W$ to $\prescript{gh}{}{W}$.
\end{lem}

We have the following generalization of Lemma~\ref{lem_prilindisc}.
\begin{lem}\label{lem_redprimlin}
%Use the notations in Theorem~\ref{thm_primlinstr}.
Let $G$ be an irreducible linear group on  a vector space $V$ over $\F_p$, and let $G'$ be a subgroup of $G$ on $V$.
Then for sufficiently large $N=\mathrm{poly}(r(G),|V|)\geq |V|$, all strongly antisymmetric $\mathcal{P}_{G',N}$-schemes are discrete on $G'_x\in \mathcal{P}_{G',N}$ for all $x\in V$.
\end{lem}

\begin{proof}
Assume $V\neq \{0\}$ as otherwise the claim is trivial. By Lemma~\ref{lem_primlin}, we may choose a nonzero subspace $W\subseteq V$ such that $G|_W$ is a primitive linear group on $W$, and $G$ permutes the subspaces in the set $S_W:=\{\prescript{g}{}{W}: g\in G\}$. 
Note $|S_W|=\log |V|/\log |W|=O(\log |V|)$.
We claim $r(G|_W)=\mathrm{poly}(r(G), |V|)$. To see this, consider a classical group $H$ that is a composition factor of $G_{\{W\}}$. 
The group $G$ permutes the subspaces in $S_W$, which gives a permutation representation $\rho:G\to \sym(S_W)$.
Then $H$ is either a composition factor of $\rho(G_{\{W\}})$ or that of $\ker(\rho)\cap G_{\{W\}}=\ker(\rho)$. 
In the former case, the group $H$ is a subquotient of $\sym(S_W)$. And Lemma~\ref{lem_subquo} implies that $|H|=r(H)$ is polynomial in $|S_W|^{\log |S_W|}=|V|^{O(1)}$.
In the latter case, we have $|H|\leq r(G)$ since $\ker(\rho)\unlhd G$.
So in either case, we have $r(G|_W)\leq r(G_{\{W\}})=\mathrm{poly}(r(G), |V|)$.

Let $\mathcal{P}=\mathcal{P}_{G',N}$. Let $N\geq |V|^3$ so that $G'_{x,y,z}\in\mathcal{P}$ for all $x,y,z\in V$.
Suppose $\mathcal{C}=\{C_H: H\in\mathcal{P}\}$ is a strongly antisymmetric $\mathcal{P}$-scheme.
We want to show that all strongly antisymmetric $\mathcal{P}$-schemes are discrete on $G'_x$ for all $x\in V$.
Note that for any $x,y\in V$, we may choose a sequence of elements $z_0,\dots,z_t\in S$ such that $z_0=x$, $z_t=y$, and for all $i\in [t]$, the vector $z_i-z_{i-1}$ is in $\prescript{g}{}{W}$ for some $g\in G$.
By Lemma~\ref{lem_selfredspan}, it suffices to prove, for all $x,y\in V$ and $g\in G$ satisfying $x-y\in \prescript{g}{}{W}$,  that  $\mathcal{C}|_{G'_x}$ is discrete on $G'_{x,y}$.
Fix such $x,y\in V$ and $g\in G$.

%Fix $x\in V$. We want to prove that $\mathcal{C}$ is discrete on $G'_x$. This is obvious if $x=0$ since $G'_0=G_0$. So assume $x\neq 0$.

Let $z=y-x$.
Note that $G'_{x,y}=G'_{x,z}$.
Every element in $G'_x z$ is in a subspace  $\prescript{g'}{}{W}$ for some $g'\in G$.
Consider distinct $u,v\in G'_x z$ that are in the same subspace  $\prescript{g'}{}{W}$. 
Pick $h,h'\in G'_x$ such that $u=\prescript{h}{}{z}$ and $v=\prescript{h'}{}{z}$.
We claim that $G'_{x,y} h^{-1}, G'_{x,y} h'^{-1}\in G'_{x,y}\backslash G'_x$ are in different blocks of $C_{G'_{x,y}}|_{G'_x}$. By Lemma~\ref{lem_selfred}, it suffices to show that $\mathcal{C}|_{G'_{x,u}}$ and $\mathcal{C}|_{G'_{x,v}}$ are discrete on $G'_{x,u,v}$.
We only prove it for $\mathcal{C}|_{G'_{x,u}}$ as the claim for $\mathcal{C}|_{G'_{x,v}}$ is symmetric.

Note that $G'_{x,u}$  is a subgroup of $G_{\{\prescript{g'}{}{W}\}}$ since $u\in \prescript{g'}{}{W}$ and $G$ permutes the subspaces in the set $S_W$. Define 
\[
\mathcal{P}':=\{(G'_{x,u}|_{\prescript{g'}{}{W}})_B: B\subseteq \prescript{g'}{}{W}, (G'_{x,u})_B\in\mathcal{P}\},
\]
 which is a subgroup system  over $G'_{x,u}|_{\prescript{g'}{}{W}}$.
By Lemma~\ref{lem_quoidentify}, it suffices to show that all strongly antisymmetric $\mathcal{P}'$-schemes are discrete on $(G'_{x,u}|_{\prescript{g'}{}{W}})_v$.
Let $N':=\lfloor N/|V|^2\rfloor$. Then $\mathcal{P}'$ contains the subgroup system $\mathcal{P}_{G'_{x,u}|_{\prescript{g'}{}{W}}, N'}$ with respect to the faithful action of $G'_{x,u}|_{\prescript{g'}{}{W}}$ on $\prescript{g'}{}{W}$.
Note that $G'_{x,u}|_{\prescript{g'}{}{W}}$ is a subgroup of $G|_{\prescript{g'}{}{W}}$ and the latter is a primitive linear group since  $G|_W$ is a primitive linear group.  
Also note $G|_{\prescript{g'}{}{W}}\cong G|_W$ and hence $r(G|_{\prescript{g'}{}{W}})=r(G|_W)=\mathrm{poly}(r(G), |V|)$.
Applying Lemma~\ref{lem_prilindisc} to $G'_{x,u}|_{\prescript{g'}{}{W}}\subseteq G|_{\prescript{g'}{}{W}}$, we see that all strongly antisymmetric $\mathcal{P}_{G'_{x,u}|_{\prescript{g'}{}{W}}, N'}$-schemes are discrete on $(G'_{x,u}|_{\prescript{g'}{}{W}})_v$, and hence  all strongly antisymmetric $\mathcal{P}'$-schemes are discrete on $(G'_{x,u}|_{\prescript{g'}{}{W}})_v$, as desired. This proves the claim that $G'_{x,y} h^{-1}$ and $G'_{x,y} h'^{-1}$ are in different blocks of $C_{G'_{x,y}}|_{G'_x}$ given that $u=\prescript{h}{}{z}$ and $v=\prescript{h'}{}{z}$ are distinct elements in the same subspace  $\prescript{g'}{}{W}$.

Consider an arbitrary block $\{G'_{x,y} g_1^{-1},\dots,G'_{x,y} g_s^{-1}\}$ of  $C_{G'_{x,y}}|_{G'_x}$ of cardinality $s\in\N^+$. By the claim just proved, the elements $\prescript{g_1}{}{z},\dots,\prescript{g_s}{}{z}$ are in distinct subspaces in the set $S_W$.
So $s\leq |S_W|=O(\log |V|)$.
Therefore we have $m(s)=O(\log s)=O(\log \log |V|)$ by Theorem~\ref{thm_symbound2} (see Definition~\ref{defi_funmn} for the definition of $m(\cdot)$).
Choose the largest $m\in\N$ satisfying $|G'_{x}z|^{m}\leq N$.
By definition, the subgroup system $\mathcal{P}|_{G'_{x}}$ contains the system of stabilizers of depth $m$ over $G'_x$ (with respect to the 
action of $G'_x$ on $G'_x z$).
Lemma~\ref{lem_ptom} and Theorem~\ref{thm_pschmind} then imply the existence of a strongly antisymmetric $m$-scheme $\Pi=\{P_1,\dots, P_m\}$ on $G'_{x}z$ 
such that $P_1$ has a block of cardinality $s$.
Note  $|G'_{x}z|\leq |S_W|\cdot |W|$.
And we have $|S_W|^{m(s)}=(\log |V|)^{O(\log \log |V|)}=|V|^{O(1)}$ and $|W|^{m(s)}=|W|^{O(\log |S_W|)}=|V|^{O(1)}$.
Then for  sufficiently large $N=|V|^{\Omega(1)}$, we have 
\[
|G'_{x}z|^{m(s)}\leq (|S_W|\cdot |W|)^{m(s)}\leq N,
\]
and hence $m\geq  m(s)$.
Theorem~\ref{thm_symbound2} then forces $s=1$.
So $\mathcal{C}|_{G'_x}$ is discrete on $G'_{x,z}=G'_{x,y}$, as desired.
\end{proof}

Now we are ready to prove  Theorem~\ref{thm_primcriterion} for finite primitive permutation groups of affine type.
\begin{lem}
Theorem~\ref{thm_primcriterion} holds for finite primitive permutation groups of affine type.
\end{lem}
\begin{proof}
Let $G$ be a finite primitive permutation groups of affine type on a vector space $V$ over a prime field $\F_p$.
Then the stabilizer $G_0\subseteq \gl(V)$ of the origin $0\subseteq V$ is an irreducible linear group by Lemma~\ref{lem_affinesemi}.
Let  $V^\sharp\subseteq G$ be the group of translations. Then $G\cong V^\sharp \rtimes G_0$ and hence  $r(G_0)\leq r(G)$.

Let $\mathcal{C}$ be a strongly antisymmetric $\mathcal{P}_{G, N}$-scheme. 
By  Lemma~\ref{lem_selfredspan}, it suffices to prove for all $x,y\in V$ that $\mathcal{C}|_{G_{x}}$ is discrete on $G_{x,y}$. Fix such $x,y\in V$. 
By invariance of $\mathcal{C}$ and the fact that $G$ acts transitively on $V$, we may assume $x=0$. 
So we want to show that $\mathcal{C}|_{G_{0}}$ is discrete on $G_{0,y}$. 
This follows from  Lemma~\ref{lem_redprimlin} applied to the irreducible linear group $G_0$   on $V$ and the subgroup system $\mathcal{P}_{G, N}|_{G_0}$ over $G_0$.
\end{proof}

\section{Diagonal type}\index{diagonal type}

In this section, we verify Theorem~\ref{thm_primcriterion} for a finite primitive permutation group $G$ of diagonal type.
By Definition~\ref{defi_primdiag}, we may assume $G$ is a permutation group satisfying $M\subseteq G\subseteq W$ and acting on a set  $S:=D\backslash W$ by inverse right translation, where
\begin{align*}
A&=\{(a_1,\dots,a_k)\in \aut(T)^k: a_i\inn(T)=a_j\inn(T) ~\text{for all}~i,j\in [k] \},\\
W&=A\rtimes \sym(k),\\
M&=\inn(T)^k\subseteq A\subseteq W,\\
D&=\{(a,\dots,a) \pi: a\in \aut(T), \pi\in \sym(k)\}\subseteq W
\end{align*}
for a noncyclic finite simple group $T$ and an integer $k\geq 2$.
The cardinality of $S$ is $|W|/|D|=|T|^{k-1}$.

Let $x_0$ denote the element $De\in S$, so that $G_{x_0}=D\cap G$. 
It is a consequence of CFSG that every finite simple group is generated by at most two elements (\citep{AG84}). 
So we can choose $r,s\in \inn(T)-\{e\}$ that generate $\inn(T)\cong T$.
For $g\in \inn(T)$, define $a_g:=(g,e,\dots,e)\in M\subseteq G$.
We have the following lemma.

\begin{lem}\label{lem_setu}
For $U=\{x_0,\prescript{a_r}{}{x_0},\prescript{a_s}{}{x_0}, \prescript{a_{rs}}{}{x_0}\}$, it holds that $W_U=\sym(k)_1$.
% and $N_{G_{x_0}}(G_U)\supseteq \{(a,\dots,a)\pi\in G_{x_0}: a\in\aut(T), \pi\in \sym(k)_1\}$.
\end{lem}

\begin{proof}
%It is easy to see that $(a,\dots,a)\pi\in G_{x_0}$ normalizes $\sym(k)_1\cap G$ for $a\in \aut(T)$ and $\pi\in \sym(k)_1$. So it suffices to verify that $G_U=\sym(k)_1\cap G$. 
Note that 
\[
W_U=D\cap a_r D a_r^{-1}\cap a_s D a_s^{-1}\cap a_{rs} D a_{rs}^{-1}
\] from which it is straightforward to see $\sym(k)_1\subseteq W_U$. 

For the other direction, consider $g=(a,\dots,a)\pi\in D\subseteq W_U$, where $a\in \aut(T)$ and $\pi\in \sym(k)$.
%We want to show $g\in\sym(k)_1$.
We have
\begin{equation}\label{eq_eq1diag}
a_r^{-1}g a_r=a_r^{-1} (a,\dots,a)\pi a_r=a_r^{-1} (a,\dots,a) \prescript{\pi}{}{a_r}\pi \in D
\end{equation}
since $a_r^{-1}W_U a_r\subseteq  D$.

First assume $k>2$. 
 Suppose $\pi$ sends $1$ to $i\in [k]$.
 Note that all coordinates of $a_r$ (resp.  $\prescript{\pi}{}{a_r}$) are identity except that the first (resp. $i$th) coordinate is $r\neq e$.
 As $k>2$ and $a_r^{-1} (a,\dots,a) \prescript{\pi}{}{a_r}\pi \in D$, we must have $i=1$ and $r^{-1} a r=a$.
 So $\pi\in\sym(k)_1$.
The same argument using the fact  $a_s^{-1}W_U a_s\subseteq D$ implies  $s^{-1} a s=a$.
Then $a$ commutes with $\langle r,s\rangle=\inn(T)$.
 Note that  the isomorphism $T\cong \inn(T)$ sending $h\in T$ to the inner automorphism $x\mapsto hxh^{-1}$ is an equivalence between the action of $\aut(T)$ on $T$ and that on $\inn(T)$ by conjugation. So $a$ fixes $T$ pointwisely, which implies $a=e$.
Then we have $g=\pi\in \sym(k)_1$, as desired.

Next assume $k=2$.  
If $\pi=e$, we have $a_r^{-1}g a_r=(r^{-1}a r, a)\in D$ by \eqref{eq_eq1diag}.
So $r^{-1} a r=a$, and the same argument  using the fact  $a_s^{-1}W_U a_s\subseteq D$ implies  $s^{-1} a s=a$.
Again we conclude that $a$ commutes with $\langle r,s\rangle=\inn(T)$, which implies $a=e\in\sym(k)_1$.
Now consider the case $\pi\neq e$, i.e., $\pi=(1~2)\in\sym(2)$.
Note that the proof for the previous case $\pi=e$ shows $W_U\cap A=\{e\}$. Therefore 
\[
|W_U|=[W_U:W_U\cap A]\leq [W:A]=|\sym(k)|=2.
\]
  The lemma is trivial if $|W_U|=1$. 
  So assume $|W_U|=2$. Then $W_U=\{e, g\}$, where $g=(a,a)\pi$ is as above.
  By \eqref{eq_eq1diag}, we have $(r^{-1}a, a r)\pi\in D$. 
 So $ara^{-1} =r^{-1}$. 
 The same argument  using the facts  $a_s^{-1}W_U a_s\subseteq D$  and $a_{rs}^{-1}W_U a_{rs}\subseteq D$ implies
 $asa^{-1}=s^{-1}$ and $arsa^{-1}=(rs)^{-1}=s^{-1}r^{-1}$.
 On the other hand, we have $arsa^{-1}=(ara^{-1})(asa^{-1})=r^{-1}s^{-1}$. So $r$ commutes with $s$.
 Then $T=\langle r,s\rangle$ is abelian, contradicting the assumption that $T$ is a noncyclic finite simple group. 
\end{proof}

We prove Theorem~\ref{thm_primcriterion} for a finite  primitive permutation group $G$ of diagonal type in the following  general form that applies to any subgroup $H\subseteq G$.
\begin{lem}\label{lem_cridiag}
Let $G$ be a finite primitive permutation group of diagonal type  on $S=D\backslash W$ as above, and let $H$ be a subgroup of $G$ on $S$.
 Then for sufficiently large $N=\mathrm{poly}(|S|)\geq |S|$, all strongly antisymmetric $\mathcal{P}_{H,N}$-schemes are discrete on $H_x\in \mathcal{P}_{H,N}$ for all $x\in S$.
\end{lem}
\begin{proof}
Let $\mathcal{P}=\mathcal{P}_{H,N}$. By choosing $N\geq |S|^2$, we may assume $H_{x,y}\in\mathcal{P}$ for all $x,y\in S$. 
Let $\mathcal{C}=\{C_{H'}: H'\in\mathcal{P}\}$ be a strongly antisymmetric $\mathcal{P}$-scheme.
Define $Z$ to be the set of elements $g\in M=\inn(T)^k$ such that  $g$ has exactly one coordinate different from the identity.
%$g\in W=A\rtimes \sym(k)$ satisfying either of the following two conditions:
%\begin{enumerate}
%\item $g$ is a transposition in $\sym(k)$.
%\item $g\in A\subseteq \aut(T)^k$ and all but one coordinate of $g$ equal the identity.
%\end{enumerate}
 Note that $g\in Z$ iff $g^{-1}\in Z$, and the elements in $Z$ generate $M$.
Also note that $M$ acts transitively on $S$.
Then by Lemma~\ref{lem_selfredspan}, it suffices to show that for all $x\in S$ and $g\in Z$,  
the $\mathcal{P}$-scheme $\mathcal{C}|_{H_x}$ is discrete on $H_{x,\prescript{g}{}{x}}\in\mathcal{P}|_{H_x}$.
Fix $x\in S$ and $g\in Z$.
As $M$ acts transitively on $S$, there exists $h\in M$ sending $x$ to $x_0$.
Let $y:=\prescript{h}{}{(\prescript{g}{}{x})}=\prescript{hgh^{-1}}{}{x_0}$.
By invariance of $\mathcal{C}$, it suffices to show that $\mathcal{C}|_{H_{x_0}}$ is discrete on $H_{x_0,y}\in\mathcal{P}|_{H_{x_0}}$.

Let $g':=hgh^{-1}$, so that $y=\prescript{g'}{}{x_0}$. Note $g'\in Z$. Suppose the $i$th coordinate of $g'$ is different from the identity. Choose $U=\{x_0,\prescript{a_r}{}{x_0},\prescript{a_s}{}{x_0}, \prescript{a_{rs}}{}{x_0}\}$ as in Lemma~\ref{lem_setu}, and let $U':=\prescript{(1~i)}{}{U}$. 
As $(1~i) \subseteq D$ fixes $x_0$, we have $x_0\in U'$.
% and $|U'|\leq 2|U|-1\leq 7$. 
By Lemma~\ref{lem_setu}, we have 
\begin{equation}\label{eq_eq2diag}
H_{U'}=\sym(k)_{i}\cap H=\sym(k)_{i}\cap H_{x_0}.
\end{equation}
Note  $g'\in N_G(\sym(k)_{i})$, and hence 
\[
\sym(k)_{i}=g'\sym(k)_{i}g'^{-1}\subseteq g'Dg'^{-1}= g'W_{x_0}g'^{-1}=W_y.
\] 
So $H_{U'}\subseteq H_{x_0,y}$. 
We  have $H_{U'}\in\mathcal{P}$ provided that $N\geq |S|^{|U'|}=|S|^{O(1)}$.
By Lemma~\ref{lem_min}, it suffices to prove that $\mathcal{C}|_{H_{x_0}}$ is discrete on $H_{U'}$.
By Lemma~\ref{lem_selfredspan}, it suffices to show, for all $h,h'\in H_{x_0}$, 
that (1) $H_{\prescript{h}{}{U'}\cup \prescript{h'}{}{U'}}\in \mathcal{P}|_{H_{x_0}}$ 
and (2) $\mathcal{C}|_{H_{\prescript{h}{}{U'}}}$ is discrete on $H_{\prescript{h}{}{U'}\cup \prescript{h'}{}{U'}}$. 
Fix $h,h'\in H_{x_0}$. 
%Note that both $\prescript{h}{}{U'}$ and $\prescript{h'}{}{U'}$ contain $x_0$.
%So $|\prescript{h}{}{U'}\cup \prescript{h'}{}{U'}|\leq 2|U'|-1\leq 13$. 
We have $H_{\prescript{h}{}{U'}\cup \prescript{h'}{}{U'}}\in  \mathcal{P}|_{H_{x_0}}$ provided that $N\geq |S|^{|\prescript{h}{}{U'}\cup \prescript{h'}{}{U'}|}=|S|^{O(1)}$.  

So  it remains to prove that $\mathcal{C}|_{H_{\prescript{h}{}{U'}}}$ is discrete on $H_{\prescript{h}{}{U'}\cup \prescript{h'}{}{U'}}$. 
Write $h=b\pi$ and $h'=b'\pi'$ where $b,b'\in A$ and $\pi,\pi'\in\sym(k)$.
As $h\in H_{x_0}\subseteq D$, the $k$ coordinates of $b$ are equal.
So $b$ commutes with $\sym(k)$.
%As $\pi$, $\pi'$, $h$ and $h'$ all fix $x_0$, so do $b$ and $b'$.
By \eqref{eq_eq2diag}, we have
\[
H_{\prescript{h}{}{U'}}=h(\sym(k)_{i}\cap  H_{x_0})h^{-1}=\sym(k)_{\prescript{\pi}{}{i}}\cap H_{x_0}.
\]
Similarly,  we have $H_{\prescript{h'}{}{U'}}=\sym(k)_{\prescript{\pi'}{}{i}}\cap H_{x_0}$ and 
\[
H_{\prescript{h}{}{U'}\cup \prescript{h'}{}{U'}}=\sym(k)_{\prescript{\pi}{}{i},\prescript{\pi'}{}{i}}\cap H_{x_0}.
\]
Let $n:=[H_{\prescript{h}{}{U'}}: H_{\prescript{h}{}{U'}\cup \prescript{h'}{}{U'}}]$. We have
\[
n\leq [\sym(k)_{\prescript{\pi}{}{i}}:\sym(k)_{\prescript{\pi}{}{i},\prescript{\pi'}{}{i}}]\leq k.
\]
Also note $k=\log |S|/\log |T|+1=O(\log |S|)$.
Consider the action of $H_{\prescript{h}{}{U'}}$ on $H_{\prescript{h}{}{U'}\cup \prescript{h'}{}{U'}}\backslash H_{\prescript{h}{}{U'}}$ by inverse right translation.
Each  one-point stabilizer with respect to this action is a pointwise stabilizer of a set $S'\subseteq S$ of cardinality at most $|U'|=O(1)$.
Choose sufficiently large $N\geq n^{|U'|d_\sym(n)}=k^{O(\log k)}=|S|^{O(1)}$ so that $\mathcal{P}|_{H_{\prescript{h}{}{U'}}}$ contains the system of stabilizers of depth $d_\sym(n)$ with respect to this action.
Then all strongly antisymmetric $\mathcal{P}|_{H_{\prescript{h}{}{U'}}}$-schemes, including $\mathcal{C}|_{H_{\prescript{h}{}{U'}}}$, are discrete on $H_{\prescript{h}{}{U'}\cup \prescript{h'}{}{U'}}$, as desired. 
\end{proof}

Choosing $H=G$ in Lemma~\ref{lem_cridiag}, we have 
\begin{cor}\label{cor_cridiag}
Theorem~\ref{thm_primcriterion} holds for finite primitive permutation groups of diagonal type.
\end{cor}

\section{Product type and twisted wreath  type}\label{sec_prim_prod}

In this section, we verify Theorem~\ref{thm_primcriterion} for finite primitive permutation groups of product type and those of twisted wreath type.

\paragraph{Product type.}\index{product type}
Suppose $G$ is a finite primitive permutation group of product type.
By Definition~\ref{defi_prodprim}, there exist an integer $k\geq 2$ and a primitive permutation group $H$ on a finite set $\Gamma$ that is of almost simple type or diagonal type  such that $G$ is a subgroup of $W:=H \wr \sym(k)= H^k\rtimes\sym(k)$ acting on $S:=\Gamma^k$, and $M:=\soc(H)^k\subseteq W$ is a subgroup of $G$.

We prove Theorem~\ref{thm_primcriterion} for a finite  primitive permutation group $G$ of product type  in the following  general form that applies to   any subgroup $G'\subseteq G$.
\begin{lem}\label{lem_criprod}
 Let $G$ be a finite primitive permutation group of product type  on $S$ as above.
 Let $G'$ be a subgroup of $G$ on $S$.
 Then for sufficiently large $N=\mathrm{poly}(k(G)^{d_{\sym}(k(G))},r(G),|S|)\geq |S|$, all strongly antisymmetric $\mathcal{P}_{G',N}$-schemes are discrete on $G'_x\in \mathcal{P}_{G',N}$ for all $x\in S$.
%Suppose $\mathcal{P}$ is a subgroup system over $G'$ satisfying\todo{Prove it for subgroups as well.}
%\begin{enumerate}
%\item $G_U\in\mathcal{P}$ for $U\in S$ satisfying $1\leq |U|\leq 3$.
%\item for all $x,y\in S$ that differ at exactly one coordinate, the subgroup system $\mathcal{P}|_{G_{x,y}}$ satisfies the conditions in Theorem~\ref{thm_almsimpcri} and those in Theorem~\ref{thm_cridiag} (with respect to the action of $G_{x,y}$ on any $G_{x,y}$-orbit in $S$).
%\item for all $x,y\in S$ that differ at exactly one coordinate, and subsets $U\subseteq G_x y$ of cardinality at most $\rceil\log k\rceil +1$, it holds that $G_U\in\mathcal{P}$.
%\end{enumerate}
%Then for all $x\in S$, all strongly antisymmetric $\mathcal{P}$-schemes are discrete on $G_x\in \mathcal{P}$.
\end{lem}

\begin{proof}
Let $\mathcal{P}=\mathcal{P}_{G',N}$.
Choose $N\geq |S|^3$ so that $G'_{x,y,z}\in \mathcal{P}$ for all $x,y,z\in S$.
Suppose $\mathcal{C}=\{C_H: H\in\mathcal{P}\}$ is a strongly antisymmetric $\mathcal{P}$-scheme.
Fix $x\in S$. We prove that $\mathcal{C}$ is discrete on $G'_x$. Note that for any $y,z\in S$, we may choose a sequence of elements $y_0,\dots,y_t\in S$ such that $y_0=y$, $y_t=z$, and for all $i\in [t]$, the elements $y_{i-1}, y_i\in S=\Gamma^k$ differ at exactly one coordinate. By Lemma~\ref{lem_selfredspan}, it suffices to prove, for all $y,z\in S$ differing at exactly one coordinate, that $\mathcal{C}|_{G'_y}$ is discrete on $G'_{y,z}$.
Fix such $y,z\in S$.
Also note that all elements in $G'_{y}z$ differ from $y$ at exactly one coordinate. 
In particular, we have $|G'_{y}z|\leq k|\Gamma|$.

Consider $u,v\in G'_{y}z$ differing from $y$ at the same coordinate whose index is denoted by $i\in [k]$.
Pick $g,g'\in G'_{y}$ such that $u=\prescript{g}{}{z}$ and $v=\prescript{g'}{}{z}$.
We claim that $G'_{y,z}g^{-1}, G'_{y,z}g'^{-1}\in G'_{y,z}\backslash G'_y$ are in different blocks of $C_{G'_{y,z}}|_{G'_y}\in\mathcal{P}|_{G'_y}$. By Lemma~\ref{lem_mres} and Lemma~\ref{lem_selfred}, it suffices to verify that  $\mathcal{C}|_{G'_{y,u}}$ and $\mathcal{C}|_{G'_{y,v}}$ are discrete on $G'_{y,u,v}$.
We only prove it for $\mathcal{C}|_{G'_{y,u}}$  since the claim for $\mathcal{C}|_{G'_{y,v}}$ is symmetric.
Note that $\mathcal{C}|_{G'_{y,u}}$ is a strongly antisymmetric   $\mathcal{P}|_{G'_{y,u}}$-scheme by Lemma~\ref{lem_mres}.
We show  that in fact all strongly antisymmetric $\mathcal{P}|_{G'_{y,u}}$-schemes are discrete on $G'_{y,u,v}$.
As $G'_{y,u}$ fixes $y$ and $u$ which differ at the $i$th coordinate, the image of $G'_{y,u}$ under the quotient map  $H \wr \sym(k)\to \sym(k)$ is contained in $\sym(k)_i$. Define 
\[
P:=\{(g_1,\dots,g_k)\pi\in  G'_{y,u}: g_i=e\}\subseteq G'_{y,u}.
\]
Then $P$ is a normal subgroup of $G'_{y,u}$. 
Suppose $v=(v_1,\dots,v_k)\in S=\Gamma^k$.
Define 
\[
S':=\{(v_1,\dots,v_{i-1},v'_i,v_{i+1},\dots,v_k)\in S: v'_i\in \Gamma\}.
\]
%We regard $Q:=G'_{y,u}/P$ as a subgroup of $H$ via the map $(g_1,\dots,g_k)\pi P \mapsto g_i$.
%Then $G'_{y,u}$ is the direct product of its subgroups $P$ and $Q$.
The action of $G'_{y,u}$ on $S$ restricts to an action on $S'$ which factors through $\bar{G}:=G'_{y,u}/P$.
And the action of $\bar{G}$ on $S'$ is permutation isomorphic to $H'$ on $\Gamma$, where $H'\subseteq H$ is defined by 
\[
H':=\{g\in H: (g_1,\dots,g_k)\pi\in G'_{y,u},~ g_i=g\}.
\]
Let $N'=\lfloor N/|S|^2\rfloor$. 
Note $\mathcal{P}_{G'_{y,u}, N'}\subseteq \mathcal{P}|_{G'_{y,u}}$. 
By Lemma~\ref{lem_quoidentify}, we just need to prove that all strongly antisymmetric $\mathcal{P}_{\bar{G}, N'}$-schemes are discrete on $\bar{G}_v$. Equivalently, we want to prove all strongly antisymmetric $\mathcal{P}_{H', N'}$-schemes (defined with respect to the action of $H'$ on $\Gamma$) are discrete on $H'_{v_i}$. 
Note $H'\subseteq H$ where $H$ is a primitive permutation group of almost simple type or diagonal type on $\Gamma$.
If $H$ is of almost simple type, we have $k(H)=k(\soc(H))\leq k(G)$ by Theorem~\ref{thm_schreier} and similarly  $r(H)=r(\soc(H))\leq r(G)$.
It follows from Lemma~\ref{lem_sysas} that all strongly antisymmetric $\mathcal{P}_{H', N'}$-schemes are discrete on $H'_{v_i}$ for sufficiently large  $N=\mathrm{poly}(k(G)^{d_{\sym}(k(G))},r(G),|S|)$. 
If $H$ is of diagonal type,  then we apply Lemma~\ref{lem_sysas} instead to conclude that all strongly antisymmetric $\mathcal{P}_{H', N'}$-schemes are discrete on $H'_{v_i}$ for sufficiently large  $N=\mathrm{poly}(|\Gamma|)$. 
So  $\mathcal{C}|_{G'_{y,u}}$ is discrete on $G'_{y,u,v}$. Therefore $G'_{y,z}g^{-1}$ and $G'_{y,z}g'^{-1}$ are in different blocks of $C_{G'_{y,z}}|_{G'_y}$, as claimed.

Consider an arbitrary block $\{G'_{y,z}g_1^{-1},\dots,G'_{y,z}g_s^{-1}\}$ of  $C_{G'_{y,z}}|_{G'_y}$ of cardinality $s\in\N^+$. By the claim just proved, the elements $\prescript{g_1}{}{z},\dots,\prescript{g_s}{}{z}$ differ from $y$ at distinct coordinates. So $s\leq k$.
Then $m(s)=O(\log s)=O(\log k)$ by Theorem~\ref{thm_symbound2} (see Definition~\ref{defi_funmn} for the definition of $m(\cdot)$).
Choose the largest $m\in\N$ satisfying $|G'_{y}z|^{m}\leq N$.
By definition, the subgroup system $\mathcal{P}|_{G'_{y}}$ contains the system of stabilizers of depth $m$ over $G'_y$ (with respect to the 
action of $G'_y$ on $G'_y z$).
Lemma~\ref{lem_ptom} and Theorem~\ref{thm_pschmind} then imply the existence of a strongly antisymmetric $m$-scheme $\Pi=\{P_1,\dots, P_m\}$ on $G'_{y}z$ 
such that $P_1$ has a block of cardinality $s$.
Note  $|G'_{y}z|\leq k|\Gamma|$, $|S|=|\Gamma|^k$, and $k=\log|S|/\log|\Gamma|\leq \log |S|$.
Then for  sufficiently large $N=|S|^{\Omega(1)}$, we have 
\[
|G'_{y}z|^{m(s)}=(k|\Gamma|)^{O(\log k)}\leq N
\]
and hence $m\geq  m(s)$. Theorem~\ref{thm_symbound2} then forces $s=1$.
So $\mathcal{C}|_{G'_y}$ is discrete on $G'_{y,z}$, as desired.
\end{proof}

Choosing $G'=G$ in Lemma~\ref{lem_criprod}, we have
\begin{cor}\label{cor_veriprod}
Theorem~\ref{thm_primcriterion} holds for finite primitive permutation groups of  product type.
\end{cor}

\paragraph{Twisted wreath type.}\index{twisted wreath type}

Suppose $G$ is a finite primitive permutation group of twisted wreath type.
By Definition~\ref{defi_twrprim}, we may assume $G=B\rtimes P$ acting on $S:=G/P$ by left translation, where
\begin{itemize}
\item $T$ is a noncyclic finite simple group,
\item $P\subseteq \sym(k)$ is a transitive permutation group  on $[k]$ for some integer $k\geq 2$,
\item $\varphi$ is a  group homomorphism from $P_1$ to $\aut(T)$,
\item $B$ is the group $\{f\in \map(P, T): f(pq^{-1})=\prescript{\varphi(q)}{}{(f(p))}~\text{for all}~p\in P, q\in P_1\}$ under coordinatewise multiplication, and
\item $P$ acts on $B$ via $(\prescript{p}{}{f})(x)=f(p^{-1}x)$ for $p,x\in P$, $f\in B$.
\end{itemize}

It turns out that $G$ can be embedded in a finite primitive permutation group of product type on $S$.
This is explained in \citep[Section~3.6]{Pra90}. We provide a detailed proof of this fact.
\begin{lem}\label{lem_groupemb}
 Let $G$ be a finite primitive permutation group of  twisted wreath type  on $S=G/P$ as above.
Then $G$ is permutation isomorphic to a subgroup of a finite primitive permutation group $\hol(T)\wr P$ of product type on $S$.
\end{lem}
\begin{proof}
Identifying $S=G/P$ with the set $B$ via the bijection $B\to G/P$  sending $g\in B$ to $gP\in G/P$, we may regard $G=B\rtimes P$ as a permutation group on the set $B$ where $B\subseteq G$ acts on $B$ by left translation and $P\subseteq G$ acts by  $(\prescript{p}{}{f})(x)=f(p^{-1}x)$ for $p,x\in P$, $f\in B$.
Pick $g_1,\dots,g_k\in P$  such that $\prescript{g_i}{}{1}=i\in [k]$.
Then  $g_1,\dots,g_k$  form a complete set of representatives of $P/ P_1$.
We further regard $G$ as a permutation group on $T^k$ by identifying the set $B$ with $T^k$ via the bijection $B\to T^k$ sending $f\in B$ to $(f(g_1),\dots,f(g_k))\in T^k$.

The holomorph $\hol(T)$ of $T$ is a primitive permutation group of diagonal type on $T$ where the action is defined by $\prescript{hg}{}{h'}=h\prescript{g}{}{h'}$ for $h,h'\in T$ and $g\in\aut(T)$ (see Example~\ref{exmp_diag} and Lemma~\ref{lem_holom}).
Denote by $G'$ the wreath product $\hol(T)\wr P$ acting faithfully on the set $T^k$ by the primitive wreath product action, i.e., $\hol(T)^k$ acts on $T^k$ coordinatewisely and $P\subseteq\sym(k)$ permutes the $k$ coordinates. We claim that $G$ is permutation isomorphic to  a subgroup of  $G'$ on $T^k$. To see this, note that a permutation $f\in B\subseteq G$ of $T^k$ is the same as the permutation $(f(g_1),\dots,f(g_k))\in T^k\unlhd \hol(T)^k\unlhd G'$. Now consider $\pi\in P\subseteq G$ and we show that it is also a permutation in $G'$.
For  $i\in [k]$,  the permutations $\pi^{-1}g_i$ and $g_{\prescript{\pi^{-1}}{}{i}}$ of $[k]$ both send $1$ to $\prescript{\pi^{-1}}{}{i}$, and hence
 $\pi^{-1}g_iP_1=g_{\prescript{\pi^{-1}}{}{i}}P_1$.
  %Note that the map $hP_1\mapsto g^{-1}hP_1$ for $hP_1\in P/P_1$ is a permutation of $P/P_1$. 
  So we can choose  $h_1,\dots,h_k\in P_1$ such that $\pi^{-1}g_i=g_{\prescript{\pi^{-1}}{}{i}} h_i^{-1}$ holds for all $i\in [k]$. We claim that $\pi\in P\subseteq G$, as a permutation  of $T^k$, equals $(\varphi(h_1),\dots,\varphi(h_k))\pi\in G'$. This is because for $f\in B$, we have
\begin{align*}
\prescript{(\varphi(h_1),\dots,\varphi(h_k))\pi}{}{(f(g_1),\dots,f(g_k))}
&=\prescript{(\varphi(h_1),\dots,\varphi(h_k))}{}{\left(f\left(g_{\prescript{\pi^{-1}}{}{1}}\right),\dots,f\left(g_{\prescript{\pi^{-1}}{}{k}}\right)\right)}\\
&=\left(f\left(g_{\prescript{\pi^{-1}}{}{1}}h_1^{-1}\right),\dots,f\left(g_{\prescript{\pi^{-1}}{}{k}}h_k^{-1}\right)\right)\\
&=(f(\pi^{-1}g_1),\dots,f(\pi^{-1}g_k))\\
&=((\prescript{\pi}{}{f})(g_1),\dots,(\prescript{\pi}{}{f})(g_1))\\
&=\prescript{\pi}{}{(f(g_1)\dots,f(g_k))}.
\end{align*}
Here $\pi$ in the last  equation acts as an element of $G$, whereas $(\varphi(h_1),\dots,\varphi(h_k))\pi$ is an element of $G'$.
It follows that $G=B\rtimes P$ is permutation isomorphic to a subgroup of $G'$ on $T^k$. As $G$ acts primitively on $T^k$, so does $G'$. By definition, the group $G'$ is a finite primitive permutation group of product type on $T^k$. The lemma follows.
\end{proof}

For the groups $G=T\twr_{\varphi} P$  and $G'=\hol(T)\wr P$ in Lemma~\ref{lem_groupemb}, we have $k(G)=k(G')$ and $r(G)=r(G')$ by Theorem~\ref{thm_schreier}. 
Then by Lemma~\ref{lem_criprod} and Lemma~\ref{lem_groupemb}, we have

\begin{cor}\label{cor_veritwist}
Theorem~\ref{thm_primcriterion} holds for finite primitive permutation groups of twisted wreath type.
\end{cor}

\section{Future research} \label{sec_future}

In this section, we suggest some possible directions for future research.

\paragraph{Dependence on classical groups.}

As we have shown, the running time of the factoring algorithm in this chapter is controlled by the alternating groups and the classical groups among the composition factors of the Galois group. Nevertheless, the exact relation between the running time and the classical groups is not fully investigated. The bound we use for classical  simple groups is simply the group order $r(G)$, and  a natural problem is  to improve this bound. In the case of the natural action of a general linear group $G:=\gl_n(q)$ on $S:=\F_q^n-\{0\}$, it yields the bound $r(G)=|\mathrm{PSL}_n(q)|=q^{O(n^2)}$. Note that Corollary~\ref{cor_algdgbound} or Corollary~\ref{cor_algdgboundg} gives the same bound $|S|^{d(\gl_n(q))}=|S|^{d_\gl(n,q)}=q^{O(n^2)}$ if we use the trivial $O(n)$ bound  for $d_\gl(n,q)$ (see Section~\ref{sec_lingroup}).
This observation suggests that proving $d_\gl(n,q)=o(n)$ is possibly the first step towards a faster factoring algorithm for classical  groups.

\paragraph{Factoring algorithms and $\mathcal{P}$-schemes for various permutation groups.}

The main results 
%(Theorem~\ref{thm_compbound} and Theorem~\ref{thm_gamma})
 of this chapter demonstrate that the problem of deterministic polynomial factoring may be much easier when the Galois group has a relatively simple structure. In particular, the results are obtained for Galois groups with restricted composition factors. It is an interesting problem to see if similar results can be obtained for other families of permutation groups  under possibly different restrictions. 
 %Techniques from permutation group theory or general finite group theory may be useful for research along this line.
 
A related problem is proving the schemes conjectures (Conjecture~\ref{conf_schconjg}) for more general families of  permutation groups. As we  observed in Section~\ref{sec_schconj},  proving these conjectures for various permutation groups are intermediate steps towards proving the  original schemes conjecture in \cite{IKS09}.  

\paragraph{Connections with association schemes.}

Another approach is to exploit the connections between our notion of $\mathcal{P}$-schemes and association schemes.
For example, by drawing connections  between $m$-schemes \citep{IKS09} and  association schemes, 
the work \citep{AIKS14} gave a factoring algorithm that finds a nontrivial factor of a reducible polynomial $f(X)\in \F_q[X]$ of prime degree $n$ in time  $\mathrm{poly}(\log q, n^{r+\log \ell})$ under GRH,  provided that $n-1$ has an $r$-smooth divisor $s$ satisfying $s\geq\sqrt{n/\ell}+1$. We have shown that $\mathcal{P}$-schemes generalize $m$-schemes, in the sense that
an $m$-schemes is essentially a $\mathcal{P}$-scheme with  $\mathcal{P}$ chosen to be  the system of stabilizers of depth $m$ over a multiply transitive group (see Theorem~\ref{thm_mandp}). Thus it is a curious question if the theory of association schemes can find more applications in deterministic polynomial factoring within our framework of $\mathcal{P}$-schemes.
% In the other direction, we wonder if our generalization of association schemes and $m$-schemes  

\printbibliography[heading=bibintoc]

\appendix

\chapter{A unifying definition of \texorpdfstring{$\mathcal{P}$-schemes}{P-schemes}}\label{chap_unify}

We present an alternative ring-theoretic definition of $\mathcal{P}$-schemes, such that the three defining properties (compatibility, invariance, regularity) are given in a unifying way. 

\paragraph{Ring $\ind^G K$.} Let $G$ be a finite group and $K$ be an arbitrary field of characteristic zero. Define $\ind^G K$ to be the set of all the functions $\phi:G\to K$. 
%\nomenclature[9a]{$\ind^G K$}{set of all functions $\phi:G\to K$}
We make it into a commutative ring by defining addition and multiplication entry-wisely.
Let $G$ act on it by $(\prescript{g}{}{\phi})(gh)=\phi(h)$, or equivalently
\[
(\prescript{g}{}{\phi})(h)=\phi(g^{-1}h)
\]
for $g,h\in G$ and $\phi\in\ind^G K$.
%\footnote{The notation $\ind^G K$ is chosen as it is the $G$-module induced from the trivial $H$-module $K$ where $H=\{e\}$.}

For a subgroup $H\subseteq G$, the subring $(\ind^G K)^H$ of $H$-invariant elements consists of functions $\phi:G\to K$ taking a constant value on each right coset in $H\backslash G$. So $(\ind^G K)^H$ is  identified with the commutative ring consisting of all the functions from $H\backslash G$ to $K$ where  addition and multiplication are defined entry-wisely.

We define inclusions, conjugations and trace maps between $(\ind^G K)^H$ for various subgroups $H\subseteq G$:
\begin{itemize}
\item (inclusion) for $H\subseteq H'\subseteq G$, the ring $(\ind^G K)^{H'}$ is a subring of  $(\ind^G K)^{H}$. Define the map $i_{H,H'}: (\ind^G K)^{H'} \hookrightarrow (\ind^G K)^{H}$ to be the natural inclusion.
\item (conjugation) for $g\in G$ and $H'=gHg^{-1}$,  define $c^*_{H,g}: (\ind^G K)^{H'} \to\linebreak[0](\ind^G K)^{H}$ to be the map sending $\phi$ to $\prescript{g^{-1}}{}{\phi}$.
\item (trace map)  for $H\subseteq H'$, define $\tr_{H,H'}: (\ind^G K)^{H} \to (\ind^G K)^{H'}$ to be the map sending $\phi$ to $\sum_{gH\in H'/H}\prescript{g}{}{\phi}$.
\end{itemize}
%\nomenclature[9b]{$i_{H,H'}$}{inclusion $(\ind^G K)^{H'} \hookrightarrow (\ind^G K)^{H}$}
%\nomenclature[9c]{$c^*_{H,g}$}{conjugation from  $(\ind^G K)^{gHg^{-1}}$ to $(\ind^G K)^{H}$}
%\nomenclature[9d]{$\tr_{H,H'}$}{trace map from $(\ind^G K)^{H}$ to $(\ind^G K)^{H'}$}
Note that trace maps are indeed well defined: as $\phi\in (\ind^G K)^{H}$ is fixed by $H$, the function $\prescript{g}{}{\phi}$ depends only on the left coset $gH$, and the image $\tr_{H,H'}(\phi)$ does lie in $(\ind^G K)^{H'}$, since for $h\in H'$ we have
\[
\prescript{h}{}{\tr_{H,H'}(\phi)}=
\prescript{h}{}{\left(\sum_{gH\in H'/H}\prescript{g}{}{\phi}\right)}
=\sum_{gH\in H'/H}\prescript{hg}{}{\phi}
=\sum_{gH\in H'/H}\prescript{g}{}{\phi}
=\tr_{H,H'}(\phi).
\]
The third equality  holds since if $g$ ranges over a complete set of representatives for $H'/H$, so does $hg$.

\paragraph{Subring $R_P$ associated with a partition $P$.} For a subgroup $H\subseteq G$ and  a partition $P$ of $H\backslash G$, define $R_P$ as the subring of $(\ind^G K)^H$ consisting of functions $\phi:H\backslash G\to K$ taking a constant value on each block $B$ of $P$.
%\nomenclature[9e]{$R_P$}{subring consisting of functions taking a constant value on each block $B$ of $P$}

The connection between these subrings and $\mathcal{P}$-schemes is described by the following theorem.

\begin{thm}\label{thm_unify}
For a $\mathcal{P}$-collection  $\mathcal{C}=\{C_H: H\in\mathcal{P}\}$,  
\begin{itemize}
\item $\mathcal{C}$ is compatible iff $i_{H,H'}(R_{C_{H'}})\subseteq R_{C_H}$ holds for all $H,H'\in\mathcal{P}$ with $H\subseteq H'$,
\item  $\mathcal{C}$ is invariant iff $c^*_{H,g}(R_{C_{H'}})\subseteq R_{C_H}$ holds for all $H,H'\in\mathcal{P}$, $g\in G$ with $H'=gHg^{-1}$, and
\item $\mathcal{C}$ is regular iff $\tr_{H,H'}(R_{C_{H}})\subseteq R_{C_{H'}}$ holds for all $H,H'\in\mathcal{P}$ with $H\subseteq H'$.
\end{itemize}
\end{thm}

\begin{proof}
Make every ring $(\ind^G K)^H$ as well as $R_{C_H}$ into a $K$-algebra by defining scalar multiplication of $K$ entry-wisely.
Note that maps $i_{H,H'}$, $c^*_{H,g}$ and $\tr_{H,H'}$ are $K$-linear. For $H\in G$ and $B\in C_H$, define the function $\delta_B: H\backslash G\to K$ by
\[
\delta_B(x)=\begin{cases}
1 & x\in B,\\
0 & x\not\in B.
\end{cases}
\]
Then $R_{C_H}$ is spanned by the functions $\delta_B$ over $K$ where $B\in C_H$.
%\nomenclature[9f]{$\delta_B$}{function whose value is one for $x\in B$ and zero for $x\not\in B$}
So by $K$-linearity, we have $i_{H,H'}(R_{C_{H'}})\subseteq R_{C_H}$  iff  $i_{H,H'}(\delta_B)\in R_{C_H}$ for all $B\in C_{H'}$, and similar claims hold for $c^*_{H,g}$ and $\tr_{H,H'}$.

Suppose $\mathcal{C}$ is compatible. Fix $H,H'\in\mathcal{P}$ with $H\subseteq H'$, and we check  $i_{H,H'}(\delta_B)\in R_{C_H}$ for all $B\in C_{H'}$, i.e., the function $i_{H,H'}(\delta_B)$ takes a constant value on each block of $C_H$  for all $B\in C_{H'}$. 
By definition, its value at $Hh\in H\backslash G$ equals $\delta_B(H'h)=\delta_B(\pi_{H,H'}(Hh))$, which equals one if $\pi_{H,H'}(Hh)\in B$ and zero otherwise. The claim follows by compatibility of $\mathcal{C}$.

Conversely, assume $\mathcal{C}$ is not compatible, i.e., for some $H,H'\in\mathcal{P}$ with $H\subseteq H'$, $B\in C_H$, $B'\in C_{H'}$ and elements $Hh,Hh'\in B$, we have $H'h=\pi_{H,H'}(Hh)\in B'$ but $H'h'=\pi_{H,H'}(Hh')\not\in B'$. We show that $i_{H,H'}(\delta_{B'})\not\in R_{C_H}$. By definition, we have $(i_{H,H'}(\delta_{B'}))(Hh)=\delta_{B'}(H'h)=1$ but $(i_{H,H'}(\delta_{B'}))(Hh')=\delta_{B'}(H'h')=0$. So the value of $i_{H,H'}(\delta_{B'})$ is not a constant on the block $B$. Therefore $i_{H,H'}(\delta_{B'})\not\in R_{C_H}$.

The proof for invariance is similar. Suppose $\mathcal{C}$ is invariant. Fix $H,H'\in\mathcal{P}$, $g\in G$ with $H'=gHg^{-1}$, and we check  $c^*_{H,g}(\delta_B)\in R_{C_H}$ for all $B\in C_{H'}$, i.e., the function $c^*_{H,g}(\delta_B)$ takes a constant value on each block of $C_H$  for all $B\in C_{H'}$. 
By definition, we have $c^*_{H,g}(\delta_B)=\prescript{g^{-1}}{}{\delta_B}$ with respect to the action of $G$ on $\ind^G K$ defined at the beginning, where $\delta_B$ is regarded as an element of $\ind^G K$.
Then for $Hh\in H\backslash G$, we have 
\[
(c^*_{H,g}(\delta_B))(Hh)=(\prescript{g^{-1}}{}{\delta_B})(h)=\delta_B(gh)=\delta_B(H'gh)=\delta_B(c_{H,g}(Hh))
\]
which equals one if $c_{H,g}(Hh)\in B$ and zero otherwise. The claim follows by invariance of $\mathcal{C}$.

Conversely, assume $\mathcal{C}$ is not invariant, i.e., for some   for $H,H'\in\mathcal{P}$, $g\in G$ with $H'=gHg^{-1}$, $B\in C_H$, $B'\in C_{H'}$ and elements $Hh,Hh'\in B$, we have $H'gh=c_{H,g}(Hh)\in B'$ but $H'gh'=c_{H,g}(Hh')\not\in B'$. We show that $c_{H,g}^*(\delta_{B'})\not\in R_{C_H}$. By definition, we have $(c_{H,g}^*(\delta_{B'}))(Hh)=\delta_{B'}(H'gh)=1$ but $(c_{H,g}^*(\delta_{B'}))(Hh')=\delta_{B'}(H'gh')=0$. So the value of $c_{H,g}^*(\delta_{B'})$ is not a constant on the block $B$. Therefore $c_{H,g}^*(\delta_{B'})\not\in R_{C_H}$.

Now suppose $\mathcal{C}$ is regular. Fix $H,H'\in\mathcal{P}$ with $H\subseteq H'$, and we check  $\tr_{H,H'}(\delta_B)\in R_{C_{H'}}$ for all $B\in C_{H}$, i.e., the map $\tr_{H,H'}(\delta_B)$ takes a constant value on each block of $C_{H'}$  for all $B\in C_{H}$. By definition, we have $\tr_{H,H'}(\delta_B)=\sum_{gH\in H'/H}\prescript{g}{}{\delta_B}$ with respect to the action of $G$ on $\ind^G K$ defined at the beginning, where $\delta_B$ is regarded as an element of $\ind^G K$. Then for $H'h\in H'\backslash G$, we have 
\begin{align*}
(\tr_{H,H'}(\delta_B))(H'h)&=\sum_{gH\in H'/H}(\prescript{g}{}{\delta_B})(h)=\sum_{Hg\in H\backslash H'}(\prescript{g^{-1}}{}{\delta_B})(h)
=\sum_{Hg\in H\backslash H'} \delta_B(gh)\\
&=|\{Hg\in H\backslash H': Hgh\in B\}|\\
&=|\{Hg\in H\backslash G: Hgh\in B, \pi_{H,H'}(Hgh)=H'h\}|\\
&=|\{Hg\in H\backslash G: Hg\in B, \pi_{H,H'}(Hg)=H'h\}|,
\end{align*}
which counts the number of elements in $B$ mapped to $H'h$ by $\pi_{H,H'}$.
By regularity, this value is a constant when $H'h$ ranges over a block of $C_{H'}$, as desired.

Conversely,  assume $\mathcal{C}$ is not regular, i.e., for some $H,H'\in\mathcal{P}$ with $H\subseteq H'$, $B\in C_H$, $B'\in C_{H'}$, and $H'h,H'h'\in B'$, the number of elements in $B$ mapped to $H'h$ is different from the number of those mapped to $H'h'$.
As shown in the previous paragraph, these two numbers are precisely $(\tr_{H,H'}(\delta_B))(H'h)$ and $(\tr_{H,H'}(\delta_B))(H'h')$ respectively. So the value of $\tr_{H,H'}(\delta_{B})$ is not a constant on the block $B'$. Therefore $\tr_{H,H'}(\delta_{B})\not\in R_{C_{H'}}$.
\end{proof}

 By Theorem~\ref{thm_unify}, we have the following alternative definition for $\mathcal{P}$-schemes, which is equivalent to the original one (Definition~\ref{defi_pscheme}).
\begin{defi}[$\mathcal{P}$-scheme, alternative definition]
A $\mathcal{P}$-collection $\mathcal{C}=\{C_H: H\in \mathcal{P}\}$ is a {\em $\mathcal{P}$-scheme}\index{Pscheme@$\mathcal{P}$-scheme} if it has the following properties:
\begin{itemize}
\item {\em (compatibility)}\index{compatibility!of a $\mathcal{P}$-collection} $i_{H,H'}(R_{C_{H'}})\subseteq R_{C_H}$ holds for all $H,H'\in\mathcal{P}$ with $H\subseteq H'$.
\item {\em (invariance)}\index{invariance!of a $\mathcal{P}$-collection}  $c^*_{H,g}(R_{C_{gHg^{-1}}})\subseteq R_{C_H}$ holds for all $H\in\mathcal{P}$ and $g\in G$.
\item {\em (regularity)}\index{regularity!of a $\mathcal{P}$-collection} $\tr_{H,H'}(R_{C_{H}})\subseteq R_{C_{H'}}$ holds for all $H,H'\in\mathcal{P}$ with $H\subseteq H'$.
\end{itemize}
\end{defi}

\begin{rem}
The reader familiar with the notion of {\em affine schemes}\index{affine scheme} \citep{Mum99} may recognize the right coset space $H\backslash G$  as (the underlying set of) the affine scheme associated with the commutative ring $(\ind^G K)^H$. More generally, each a partition $P$ of $H\backslash G$ determines a quotient set of $H\backslash G$  which is (the underlying set of) the affine scheme associated with the subring $R_P$. It is known that the language of affine schemes and that of commutative rings are equivalent.\footnote{Formally, this is known as the fact that  the category of affine schemes is anti-equivalent to the category of commutative rings. See, e.g., \citep[Section~\RN{2}.2, Corollary~1]{Mum99}.}
Theorem~\ref{thm_unify} is a manifestation of this equivalence.

Therefore in principle, statements and proofs about $\mathcal{P}$-schemes may be carried out either set-theoretically or ring-theoretically. We stick to the more elementary set-theoretic language in this thesis.
\end{rem}

\chapter{Proofs omitted from Chapter \texorpdfstring{\RN{3}}{3}}\label{chap_omitted}

This chapter contains proofs that are omitted from Chapter~\ref{chap_alg_prime}.

\lempi*

\begin{proof}
We first show that $P(I)$ and $I(P)$ are well defined. For $P(I)$ we note that $\prescript{g^{-1}}{}{(i_{K,L}(\delta))}$ depends only on the coset $Hg$, since $i_{K,L}(\delta)\in i_{K,L}(\bar{\ord}_K)$ is fixed by $H$. The relation  $\prescript{g^{-1}}{}{(i_{K,L}(\delta))}\equiv \prescript{g'^{-1}}{}{(i_{K,L}(\delta))}\pmod{\bar{\mathfrak{Q}}_0}$ for all $\delta\in I$ is obviously an equivalence relation on $H\backslash G$, and hence defines a partition of $H\backslash G$.

 For $I(P)$, we fix $B\subseteq H\backslash G$ and show that  $t:=\sum_{g\in G: Hg\in B}\prescript{g}{}{\delta_{\bar{\mathfrak{Q}}_0}}$ does lie in the image of $i_{K,L}$ so that $\delta_B=i_{K,L}^{-1}(t)$ is well defined.  
By Corollary~\ref{cor_idealcoset}, each coset $x=Hg$ corresponds to a maximal ideal
$\mathfrak{P}_x:=(\prescript{g}{}{\mathfrak{Q}_0}\cap \ord_K)/p\ord_K$ 
of $\bar{\ord}_K$.
By Lemma~\ref{lem_idealidem},  there exists a unique idempotent $\delta$ of $\bar{\ord}_K$ satisfying $\delta\equiv 1\pmod{\mathfrak{P}_x}$ for $x\in B$, and $\delta\equiv 0\pmod{\mathfrak{P}_x}$ for $x\not\in B$. 
%Also note that for any $g\in G$, we have $\prescript{g}{}{\mathfrak{Q}_0}\cap R_K=\mathfrak{P}_x$ where $x=Hg\mathcal{D}_{\mathfrak{Q}_0}$.
It follows that for  $g\in G$, the residue of $i_{K,L}(\delta)$ modulo $\prescript{g}{}{\bar{\mathfrak{Q}}_0}$ equals one  if $Hg\in B$ and zero otherwise.
The same holds for $t$ by definition: for  $g\in G$, the residue of $t$ modulo $\prescript{g}{}{\bar{\mathfrak{Q}}_0}$ equals one  if $Hg\in B$ and zero otherwise.
As all the maximal ideals of the semisimple ring $\bar{\ord}_L$ have the form $\prescript{g}{}{\bar{\mathfrak{Q}}_0}$ where $g\in G$, we have $t=i_{K,L}(\delta)$, as desired.
Furthermore, by choosing $B=H\backslash G$ and $t=i_{K,L}(1)=1$, we see that $\sum_{g\in G} \prescript{g}{}{\delta_{\bar{\mathfrak{Q}}_0}}=1$. It follows that $I(P)$ is a well defined idempotent decomposition of $\bar{\ord}_K$.

For the second claim, we first check that the sets $B_\delta$ form a partition of $H\backslash G$ and the map $\delta\mapsto B_\delta$ is injective. To see this, note that if an  element $Hg$ lies in both $B_{\delta}$ and $B_{\delta'}$ for distinct $\delta,\delta'\in I$, then $\prescript{g^{-1}}{}{(i_{K,L}(\delta))}\equiv\prescript{g^{-1}}{}{(i_{K,L}(\delta'))}\equiv 1\pmod{\bar{\mathfrak{Q}}_0}$ by definition. But then $\prescript{g^{-1}}{}{(i_{K,L}(\delta\delta'))}\equiv 1\pmod{\bar{\mathfrak{Q}}_0}$, contradicting the fact that $\delta\delta'=0$. So the sets $B_\delta$ are disjoint and the map $\delta\mapsto B_\delta$ is injective.
Furthermore, each $Hg\in H\backslash G$ lies in at least one set $B_\delta$ since 
\begin{equation}\label{eq_sum}
\sum_{\delta\in I}\prescript{g^{-1}}{}{(i_{K,L}(\delta))}\equiv\prescript{g^{-1}}{}{\left(i_{K,L}\left(\sum_{\delta\in I}\delta\right)\right)}\equiv 1\pmod{\bar{\mathfrak{Q}}_0}.
\end{equation}
So the sets $B_\delta$ form a partition of $H\backslash G$.

Fix $B\in P(I)$ and let $\delta=\delta_B=i_{K,L}^{-1}\left(\sum_{g\in G: Hg\in B}\prescript{g}{}{\delta_{\bar{\mathfrak{Q}}_0}}\right)$. It remains to verify that  $B_\delta=B$. For $Hh\in H\backslash G$, we have 
\[
\prescript{h^{-1}}{}{(i_{K,L}(\delta))}=\sum_{g\in G: Hg\in B}\prescript{h^{-1}g}{}{\delta_{\bar{\mathfrak{Q}}_0}}.
\]
Note that the residue of $\prescript{h^{-1}g}{}{\delta_{\bar{\mathfrak{Q}}_0}}$ modulo $\bar{\mathfrak{Q}}_0$ equals one if $h=g$, and zero otherwise.
% (see Corollary~\ref{cor_idealcoset}).
So the residue of $\prescript{h^{-1}}{}{(i_{K,L}(\delta))}$ modulo $\bar{\mathfrak{Q}}_0$ equals one if  $Hh\in B$ and zero otherwise. It follows by definition that $B_\delta=B$. 
\end{proof}

\lemcomputeresidue*
\begin{proof}
Let $d=[K:\Q]$. Suppose the structure constants of $K$ and $\ord_K'$ are given in the $\Q$-basis $B$ of $K$ and the $\Z$-basis $B'=\{x_1,\dots,x_d\}$ of $\ord_K'$ respectively. 
 Then we may assume the structure constants of $\bar{\ord}_{K}$ is given in the $\F_p$-basis $\{x_1+p\ord_K,\dots,x_d+p\ord_K\}$ of $\bar{\ord}_K$. The goal is computing the constants $c_1,\dots,c_d\in\F_p$ determined by
\begin{equation}\label{eqringhom}
\alpha+p\ord_K=\sum_{i=1}^{d} c_i(x_i+p\ord_{K}).
\end{equation}
Note that $B'$ is also a $\Q$-basis of $K$.
The change-of-basis matrix $M$ from $B'$ to $B$ is given by the inclusion $\ord_{K}'\hookrightarrow K$, whose entries are rational numbers of polynomial size. So the entries of $M^{-1}$ are also rational numbers of polynomial size. 
We apply $M^{-1}$ and write $\alpha$ in the basis $B'$:
 \[
 \alpha=\sum_{i=1}^{d} r_i x_i, \qquad r_i\in\Q.
 \]
 For $i\in [d]$, write $r_i$ in the form $a_i/b_i$ where $a_i,b_i$ are coprime integers and $b_i> 0$. Let $m$ be the least common multiple of all the denominators $b_i$. Then we have $m \alpha=\sum_{i=1}^{d} m r_i x_i$ with the coefficients $m r_i\in\Z$. So $m\alpha\in\ord_{K}'\subseteq\ord_{K}$. Passing to the quotient ring $\bar{\ord}_{K}$, we obtain
\[
 m \alpha+p\ord_K=\sum_{i=1}^{d} c'_i (x_i+p\ord_{K}), \qquad c'_i=m r_i\bmod p\in\F_p.
\]
Suppose $m=p^e m'$ where $e\in\N$, $m'\in\Z$ and $p\nmid m'$. We claim $e=0$. Assume to the contrary that $e>0$. For some $i_0\in [d]$, we have $p^e|b_{i_0}$ but $p^{e+1}\nmid b_{i_0}$. Then $p\nmid a_{i_0}$ since $a_{i_0},b_{i_0}$ are coprime. So $p\nmid m r_{i_0}$.
Then $c'_{i_0}\neq 0$ and hence $m \alpha +p\ord_K\neq 0$.
But as $\alpha +p\ord_K\in\bar{\ord}_K$, we have $m \alpha +p\ord_K\in m\bar{\ord}_{K}=0$, which is a contradiction. So $e=0$ and $p\nmid m$. Let $s$ be the multiplicative inverse of $m\bmod p\in \F_p$. We compute $s$ and let $c_i=s c'_i$ for $i\in [d]$, which satisfy \eqref{eqringhom}.
\end{proof}

\lemringhom*

\begin{proof}

Let $d=[K:\Q]$. Suppose the structure constants of $\ord_K'$ is given in the $\Z$-bases $\{x_1,\dots,x_d\}$ of $\ord'_K$. Then we may assume the structure constants of $\bar{\ord}_{K}$ is given in the $\F_p$-bases $\{x_1+p\ord,\dots,x_d+p\ord\}$ of $\bar{\ord}_K$.

For $i\in [d]$, we need to compute $\bar{\phi}(x_i+p\ord_K)\in \bar{\ord}_{K'}$. Note that $\bar{\phi}(x_i+p\ord_K)=\phi(x_i)+p\ord_{K'}$.
First compute $\phi(x_i)\in K'$ using the inclusion $\ord'_K\hookrightarrow K$ and the embedding $\phi: K\to K'$ given in the input.
Here $\phi(x_i)$ is actually in $\ord_{K'}$ since $x_i\in \ord'_K\subseteq \ord_K$.
Use the algorithm $\mathtt{ComputeResidue}$ to compute $\phi(x_i)+p\ord_{K'}\in  \bar{\ord}_{K'}$, and we are done.
\end{proof}

\lemfree*

\begin{proof}
We   maintain a submodule $N$ of $M$ that is free over $A$.
Initially $N$ equals $\{0\}$ and we iteratively enlarge it. Each time we pick $x\in M-N$ and check if the sum $N+Ax$ is a direct sum, i.e., if $N\cap Ax=\{0\}$. If so, we replace $N$ with $N+Ax$. Otherwise we find a nonzero element $y\in N\cap Ax$ and $a\in A$ satisfying $y=ax$, and return $a$.
Note that in the latter case, the element $a$ is indeed a zero divisor: otherwise $a$ would be invertible and hence $x=a^{-1}y$ is in $N$, contradicting the assumption $x\not\in N$.

If $N$ eventually becomes $M$, we conclude that $M$ is free over $A$, in which case we return zero. The algorithm clearly runs in polynomial time.
%
%The pseudo-code is given below. All steps can be implemented  by solving systems of linear equations over $\F_p$.
%
%\begin{algorithm}[htbp]
%\caption{ $\mathtt{FreeModuleTest}$}
%\begin{algorithmic}[1]
%\INPUT semisimple $\F_p$-algebra $A$ and finitely generated $A$-module $M$
%\OUTPUT zero divisor $a$ of $A$ which equals zero only when $M$ is a free $A$-module
%\State $N\gets \{0\}\subseteq M$
%\While{$N\neq M$}
%\State choose $x\in M-N$
%\If{$N\cap Ax\neq \{0\}$}
%    \State choose $y\in (N\cap Ax)-\{0\}$
%    \State choose $a\in A$ such that $ax=y$
%    \State \Return $a$
%\EndIf
%\State $N\gets N+Ax$ 
%\EndWhile
%\State \Return $0$
%\end{algorithmic}
%\end{algorithm}
%
%The loop in Line 2--10 iterates at most $\dim_{\F_p} M /\dim_{\F_p} A$ times. It follows that the algorithm runs in polynomial time. If zero is returned on Line 10, we have $N=M$ which is free over $A$. So assume that the algorithm returns $a$ on Line 7. The element $a$ is nonzero since $ax=y\neq 0$. It remains to prove that $a$ is a zero divisor of $A$. Assume it is not. As $A$ is semisimple, the element $a$ has a multiplicative inverse  $a^{-1}\in A$. But then $x=a^{-1}y\in N$, contradicting the choice of $x$.
\end{proof}

\lemzerodivisor*

\begin{proof}
We pick an element $\tilde{a}\in R$ lifting $a$, i.e., $\pi(\tilde{a})=a$.  Compute the ideal $(\tilde{a})$ of $R$ generated by $\tilde{a}$. As $R$ is semisimple, we have $(\tilde{a})=(\gamma')$ for some idempotent $\gamma'$ of $R$.
 Compute $\gamma'$ by solving a system of linear equations using the fact that $\gamma'$ is the unique element in $(\tilde{a})$ satisfying $\gamma'x=x$ for all $x\in (\tilde{a})$. Finally we replace $\gamma$ with $\gamma'\gamma$ and $(1-\gamma')\gamma$. It remains to show that $\gamma'\gamma \not\in \{0,\gamma\}$.

Note that $\pi(\gamma')\in\bar{R}$ generates the ideal $(a)$ of $\bar{R}$, and hence $\pi(\gamma')$ is also a nonzero zero divisor of $\bar{R}$. 
But $\pi(\gamma')=\gamma'+(1-\gamma)=\gamma'\gamma+(1-\gamma)$. So $\gamma'\gamma\neq 0$.
It also follows that $\gamma'\gamma\neq \gamma$ since otherwise we would have $\pi(\gamma')=\gamma'\gamma+(1-\gamma)=\gamma+(1-\gamma)=1+(1-\gamma)$, which is the unity of $\bar{R}$ and not a zero divisor.
\end{proof}

\lemauto*

To prove Lemma~\ref{lem_auto}, we need the following lemma.

\begin{lem}[\citep{Ron92, IKRS12}]\label{lem_iexp}
There exists an algorithm $\mathtt{IteratedExp}$ that, given a semisimple $\F_p$-algebra $A$, a prime number $\ell\neq p$, and elements $x,y$ in the multiplicative group $A^\times$ of order $n_x$ and $n_y$ respectively such that $n_x,n_y$ are powers of $\ell$ and $n_x\geq n_y$, returns a zero divisor of the form $x^k-y\in A$, $k\in\N$, in time polynomial in $\log |A|$ and $\ell$. In particular,  zero is returned only if $y$ is a power of $x$.
\end{lem}

\begin{proof}
The algorithm is as follows: try to find $k\in \{0,\dots,\ell-1\}$ such that $x^k-y$ is a zero divisor.
If such an integer $k$ is found, simply return $x^k-y$. Otherwise raise $x$ to its $\ell$th power and repeat.

To analyze the algorithm, note that there exists a maximal ideal $\mathfrak{m}$ of $A$ such that the order of $x+\mathfrak{m}\in (A/\mathfrak{m})^\times$ is $n_x$, and the order of $y+\mathfrak{m}\in (A/\mathfrak{m})^\times$, which we denote by $n'_y$, divides $n_y$. 
Then $x^{n_x/n'_y}+\mathfrak{m}$ and $y+\mathfrak{m}$ are both primitive $n'_y$-th roots of unity in $ (A/\mathfrak{m})^\times\cong\F_p$.
Then there exists $k\in\{0,\dots,\ell-1\}$ such that $x^{kn_x/n'_y}-y$ is in $\mathfrak{m}$ and hence  is a zero divisor. Such a zero divisor is guaranteed to be found when $x$ is raised to $x^{n_x/n'_y}$ (or earlier).
\end{proof}

\begin{proof}[Proof of Lemma~\ref{lem_auto}]
For $x,y\in A$  linearly independent over $\F_p$, at least one element in the set $\{y-cx: c\in\F_p\}$ is a nonzero zero divisor. If $p\leq \dim_{\F_p} A$, we can find such an element in polynomial time by choosing $x,y$ and enumerating $c$.
So assume $p>\dim_{\F_p} A$. 
In this case, the pseudocode of the algorithm  is given in Algorithm~\ref{alg_aut}.
Here $\mathrm{id}$ denotes the identity map on $A$. 

\begin{algorithm}[htbp]
\caption{ $\mathtt{Automorphism}$}\label{alg_aut}
\begin{algorithmic}[1]
\INPUT ring $A$ isomorphic to a finite product of $\F_p$, automorphism $\sigma\neq\mathrm{id}$ of $A$
\OUTPUT zero divisor $a\neq 0$ of $A$
\State $n\gets 1$
\Repeat
    \State find $z\in A$ satisfying $\sigma^n(z)\neq z$
    \If{$\sigma^n(z)-z$ is a zero divisor of $A$}
    \State \Return $\sigma^n(z)-z$
    \EndIf
\Until{$\sigma^n=\mathrm{id}$}
\State compute the least prime factor $\ell$ of $n$
\State $\sigma\gets \sigma^{n/\ell}$
\State compute $\F_{p^d}$, where $d$ is the smallest positive integer satisfying $\ell| p^d-1$
\State compute $A\otimes \F_{p^d}$ and the inclusion $i: A\hookrightarrow A\otimes \F_{p^d}$ sending $t\in A$ to $t\otimes 1$
\State compute the automorphism $\sigma\otimes 1$ of $A\otimes \F_{p^d}$ sending  $t\otimes u\in A$ to $\sigma(t)\otimes u$ 
\State pick an $\ell$th power non-residue $\gamma$ of $\F_{p^d}$
\State $\xi\gets \gamma^{(p^d-1)/\ell}$
\State compute a nonzero element $x\in A\otimes \F_{p^d}$ satisfying $(\sigma\otimes 1)(x)=\xi x$
 \If{$x$ is a not zero divisor of $A\otimes \F_{p^d}$}
\State $k\gets$ the largest factor of $p^d-1$ coprime to $\ell$
%\State compute the factor $k$ of $p^d-1$ such that $\ell\nmid k$ and $(p^d-1)/k$ is a power of $\ell$
\State \parbox[t]{\dimexpr\linewidth-\algorithmicindent}{call $\mathtt{IteratedExp}$ with the input $A\otimes \F_{p^d}$, $\ell$, $\gamma^k$, and $x^k$ to obtain a zero divisor $b$ of $A\otimes \F_{p^d}$\strut}
\State \strut $x\gets b$
\EndIf
\State choose $a\in A-\{0\}$ such that $i(a)$ is in the ideal $(x)$ of  $A\otimes \F_{p^d}$
\State \Return $a$
\end{algorithmic}
\end{algorithm}

The loop in Lines 2--6 of the algorithm computes the powers $\sigma^n$ of $\sigma$ for $n=1,2,\dots$ and tries to find $z\in A$ satisfying $\sigma^n(z)\neq z$. The loop exits either when such an element $z$ is found, or when the condition $\sigma^n=\mathrm{id}$ is satisfied.
In the former case, the algorithm returns the zero divisor $\sigma^n(z)-z\neq 0$, and in the latter case, the algorithm proceeds. Note that initially $n=1$ and we have $\sigma\neq \mathrm{id}$ by assumption. 

By assumption, we may identify $A$ with a product $\prod_{i=1}^m \F_p$ where $m=\dim_{\F_p} A$. For $i\in [m]$, let $\delta_i$ be the element of $A$ whose $i$th coordinate is one and the other components are zero. So $\delta_1,\dots,\delta_m$ are the primitive idempotents of $A$. The automorphism $\sigma$ of $A$ permutes these primitive idempotents, i.e., it is associated with a permutation $\pi$ of $[m]$ such that $\sigma(\delta_i)=\delta_{\pi(i)}$ for $i\in [m]$. By $\F_p$-linearity of $\sigma$ (which is automatic since $\F_p$ is a prime field), we know $\sigma$ sends $(x_1,\dots,x_m)\in A$ to $(x_{\pi^{-1}(1)},\dots,x_{\pi^{-1}(m)})$.

Let $H$ be the cyclic group generated by $\pi$, and it acts on $[m]$. Assume the $H$-orbits of $[m]$ do not have the same cardinality. We claim that in this case a zero divisor $\sigma^n(z)-z\neq 0$ is returned at Line $5$ for some $n\leq m$. To see this, suppose $O_1$ and $O_2$ are two $H$-orbits of distinct cardinalities $n_1,n_2\leq m$ respectively. We may assume $n_1\leq n_2$. Then $\sigma^{n_1}$ fixes all elements in $O_1$ but not all in $O_2$. So $\sigma^{n_1}\neq\mathrm{id}$. If the loop returns a (nonzero) zero divisor at Line 5 in the $n$th iteration for some $n<n_1$ then we are done. Otherwise, an element $z$  satisfying $\sigma^{n_1}(z)-z\neq 0$ is found at Line 3 in the $n_1$th iteration. 
Note that for any $i\in O_1$, the $i$th coordinate of $\sigma^{n_1}(z)-z$ is a zero, and hence $\sigma^{n_1}(z)-z$ is annihilated by $\delta_i$. It follows that $\sigma^{n_1}(z)-z$ is a zero divisor and is returned at Line 5.

So  assume all the $H$-orbits of $[m]$ have the same cardinality and the algorithm reaches Line 7. Then the order of $\sigma$ equals $n$.
Line 8 replaces $\sigma$ with its $(n/\ell)$th power where $\ell$ is the least prime factor of $n$. Then the order of $\sigma$ becomes the prime number $\ell$. Note that $\ell<p$ since $n\leq m< p$.

At Line 9, we compute the finite field $\F_{p^d}$ where $d$ is the smallest positive integer satisfying $\ell| p^d-1$. Equivalently, the integer $d$ is the (multiplicative) order of $p$ in the group $(\Z/\ell \Z)^\times$. So we have $d\leq |(\Z/\ell \Z)^\times|=\ell-1$. 
Under GRH (or Hypothesis ($*$) in the introduction), the field $\F_{p^d}$ can be computed in deterministic polynomial time. It is the smallest extension of $\F_p$ containing the primitive $\ell$th roots of unity. 

At Line 10, we compute the $\F_{p^d}$-algebra $A\otimes \F_{p^d}$ (where the tensor product is taken over $\F_p$) and the inclusion $i:A\hookrightarrow \F_{p^d}$ sending $t\in A$ to $t\otimes 1$. Suppose $\{b_1,\dots,b_m\}$ is an $\F_p$-basis of $A$ and $b_i b_j=\sum_{k=1}^m {c_{ijk}} b_k$ where $c_{ijk}\in\F_p$, then $A\otimes \F_{p^d}$ can be defined as an $\F_{p^d}$-algebra in the $\F_{p^d}$-basis  $\{b_1\otimes 1,\dots, b_m\otimes 1\}$ satisfying $(b_i\otimes 1)(b_j\otimes 1)=\sum_{k=1}^m {c_{ijk}} (b_k\otimes 1)$. It follows from the universal property of tensor products that this definition does not depend on the choice of the basis. See, e.g., \citep{AM69}.
In particular, identify $A$ with $\prod_{i=1}^m \F_p$ and then  $A\otimes \F_{p^d}$ is simply $\prod_{i=1}^m \F_{p^d}$.

At Line 11, we compute  $\F_{p^d}$-linear automorphism $\sigma\otimes 1$ of $A\otimes \F_{p^d}$ sending  $t\otimes u\in A$ to $\sigma(t)\otimes u$. It follows from the universal property of tensor products that such an automorphism exists and is unique.
At Line 12, we pick an $\ell$th power non-residue $\gamma$ of $\F_{p^d}$, which  be done in deterministic polynomial time under GRH (or Hypothesis ($*$) in the introduction). Then at Line 13, we compute $\xi=\gamma^{(p^d-1)/\ell}$, which is a primitive $\ell$th root of unity. 

At Line 14, we compute  a nonzero element $x\in A\otimes \F_{p^d}$ satisfying $(\sigma\otimes 1)(x)=\xi x$. 
We claim that such an element $x$ exists. To see this, note that as $\sigma$ has order $\ell$, the permutation $\pi$ of $[m]$   associated with $\sigma$ has an $\ell$-cycle $(i_1~i_2~\cdots~i_\ell)$. Then we can choose $x$ to be the element in $ A\otimes \F_{p^d}=\prod_{i=1}^m\F_{p^d}$ whose $i_j$th coordinate is $\xi^{-j}$ for $j\in[\ell]$ and remaining coordinates are zero. 

If  the element $x$ is a zero divisor of $A\otimes \F_{p^d}$, the preimage of the ideal $(x)$ of $A\otimes \F_{p^d}$ in $A$ under the map $i$ is strictly between $\{0\}$ and $A$.
In this case, we compute a nonzero element $a$ in it (or equivalently, an element $a$ satisfying $i(a)\in (x)$) at Line 19 and return it. Note that $a$ is guaranteed to be a zero divisor of $A$.

On the other hand, if $x$ is not a zero divisor of $A\otimes \F_{p^d}$, we replace it with a zero divisor $b\neq 0$ in Lines 16--18: 
suppose $p^d-1=k \ell^e$ where $k$ is coprime to $\ell$. We compute $k$ at Line 16. As $\gamma\in \F_{p^d}$ is an $\ell$th power non-residue, the order of $\gamma^k$ is $\ell^e=(p^d-1)/k$. As $x$ is not a zero divisor, we have $x^k\in (A\otimes \F_{p^d})^\times$ and its order divides $\ell^e$. 
Also note that $\sigma\otimes 1$ fixes $\gamma^k$ (by $\F_{p^d}$-linearity) and sends  $x^k$ to $\xi^k x^k\neq x^k$. So $x^k$ is not a power of $\gamma^k$.
By Lemma~\ref{lem_iexp}, a zero divisor $b\neq 0$  of $A\otimes \F_{p^d}$ is obtained at Line 17 and we assign its value to $x$. Then we obtain the zero divisor $a\neq 0$ of $A$ and return it at Line 19 as before. 
\end{proof}

\chapter{Proofs omitted from Chapter \texorpdfstring{\RN{5}}{5}}  \label{chap_omitted2}

This chapter contains proofs that are omitted from Chapter~\ref{chap_alg_general}.

\lemsubprob*
\begin{proof}
Factorize $\tilde{f}$ into irreducible factors $g_1,\dots,g_k$ over $K_0$ using polynomial factoring algorithms for number fields  \citep{Len83, Lan85}. Note that coefficients of each factor $g_i$ lie in $K_0=\Q[Y]/(\tilde{h}(Y))$ but not necessarily in $A_0=\Z[Y]/(\tilde{h}(Y))$.
Here a coefficient $\alpha\in K_0$ is represented by a unique polynomial $r_\alpha(Y)\in \Q[Y]$ of degree at most $\deg(\tilde{h})-1$ such that $\alpha=r_\alpha(Y)+(\tilde{h}(Y))$. And $\alpha\in A_0$ holds iff the coefficients of $r_\alpha(Y)$ are all integers.

For each factor $g_i$, use $r_\alpha$, where $\alpha$ ranges over coefficients of $g_i$, to compute the smallest $e_i\in\Z$ and $D_i\in\N^+$ coprime to $p$ such that all the coefficients of $p^{e_i}D_i g_i$ are in $A_0$. Compute an integer $D\in\N^+$ such that $D$ is a multiple of $\prod_{i=1}^k D_i$ and $D\equiv 1{\pmod p}$. Compute $\tilde{f}_i:=p^{e_i} D_i g_i$ for $i=2,\dots,k$ and $\tilde{f}_1=(p^{e_1}D/\prod_{i=2}^k D_i)g_1$. Then the polynomials $\tilde{f}_i(X)$ are all in $A_0[X]$.

It remains to show that the product of $\tilde{f}_i$ equals $D\cdot\tilde{f}$, which reduces to proving $\sum_{i=1}^k e_i=0$.
Note that for all $i\in [k]$, the polynomial $p^{e_i}D_i g_i(X)$ is in $A_0[X]$ but not in $pA_0[X]$, since otherwise we may replace $e_i$ with $e_i-1$, contradicting the minimality of $e_i$. The ideal $pA_0[X]$ is a prime ideal of $A_0[X]$, since $A_0[X]/pA_0[X]\cong \F_q[X]$ is an integral domain. Therefore
\[
\prod_{i=1}^k p^{e_i}D_i g_i(X)=\left(\prod_{i=1}^k p^{e_i}\right)\cdot \left(\prod_{i=1}^k D_i\right)\cdot \tilde{f}(X)
\]
is not in $pA_0[X]$ either. So $\sum_{i=1}^k e_i\leq 0$. Rewrite the equation above as
\[
\left(\prod_{i=1}^k p^{-e_i}\right)\cdot \left(\prod_{i=1}^k p^{e_i}D_i g_i(X)\right)= \left(\prod_{i=1}^k D_i\right)\cdot \tilde{f}(X).
\]
As $\tilde{\psi}_0(\tilde{f})\neq 0$, we have $\tilde{f}(X)\not\in pA_0[X]$. And the integers $D_i$ are coprime to $p$ and hence not in $pA_0[X]$ either. The equation above then implies $\sum_{i=1}^k e_i\geq 0$.
\end{proof}

\begin{rem}
An alternative way of proving  $\sum_{i=1}^k e_i=0$  is to consider the localization of $A_0$ at the prime ideal $pA_0$ and apply  Gauss Lemma (see \citep[Section~\RN{4}.2]{Lan02}). We leave the details to the reader.
\end{rem}

\lempig*

\begin{proof}
We first show that $P(I)$ and $I(P)$ are well defined. For $P(I)$ we note that $\prescript{g^{-1}}{}{(i_{K,L}(\delta))}\bmod{\mathfrak{Q}_0}$ depends only on the double coset $Hg\mathcal{D}_{\mathfrak{Q}_0}$ since $H$ fixes $i_{K,L}(\delta)\in i_{K,L}(R_K)$ and $\mathcal{D}_{\mathfrak{Q}_0}$ fixes any element modulo $\bar{\mathfrak{Q}}_0$. The relation  $\prescript{g^{-1}}{}{(i_{K,L}(\delta))}\equiv \prescript{g'^{-1}}{}{(i_{K,L}(\delta))}\pmod{\bar{\mathfrak{Q}}_0}$ for all $\delta\in I$ is obviously an equivalence relation on $H\backslash G/\mathcal{D}_{\mathfrak{Q}_0}$, and hence defines a partition of $H\backslash G/\mathcal{D}_{\mathfrak{Q}_0}$.

  For $I(P)$, we fix $B\subseteq H\backslash G/\mathcal{D}_{\mathfrak{Q}_0}$ and first show that 
  \[
  t:=\sum_{g\mathcal{D}_{\mathfrak{Q}_0}\in G/\mathcal{D}_{\mathfrak{Q}_0}: Hg\mathcal{D}_{\mathfrak{Q}_0}\in B}\prescript{g}{}{\delta_{\bar{\mathfrak{Q}}_0}}
  \]
   is well defined and does not depend on the choices of the representatives $g$.
  Note that for $h\in \mathcal{D}_{\mathfrak{Q}_0}$, the primitive idempotents $\delta_{\bar{\mathfrak{Q}}_0}$ and $\prescript{h}{}{\delta_{\bar{\mathfrak{Q}}_0}}$ correspond to the same maximal ideal $\bar{\mathfrak{Q}}_0=\prescript{h}{}{\bar{\mathfrak{Q}}_0}$ and hence are equal (see Lemma~\ref{lem_idealidem}).
So $\delta_{\bar{\mathfrak{Q}}_0}$ is fixed by $\mathcal{D}_{\mathfrak{Q}_0}$. It follows that $t$ is well defined.
  
Next we prove $t\in i_{K,L}(R_K)$ so that $\delta_B=i_{K,L}^{-1}(t)$ is well defined.  
By Lemma~\ref{lem_idealdoublecosetg}, each double coset $x=Hg\mathcal{D}_{\mathfrak{Q}_0}$ corresponds to a maximal ideal
\[
\mathfrak{P}_x:=\frac{(\prescript{g}{}{\mathfrak{Q}_0}\cap \ord_K)/p\ord_K}{\rad(\bar{\ord}_K)}\cap R_K
\] 
of $R_K$.
Let $\delta$ be the idempotent of $R_K$ satisfying $\delta\equiv 1\pmod{\mathfrak{P}_x}$ for $x\in B$, and $\delta\equiv 0\pmod{\mathfrak{P}_x}$ for $x\not\in B$ (see Lemma~\ref{lem_idealidem}). 
%Also note that for any $g\in G$, we have $\prescript{g}{}{\mathfrak{Q}_0}\cap R_K=\mathfrak{P}_x$ where $x=Hg\mathcal{D}_{\mathfrak{Q}_0}$.
It follows that $i_{K,L}(\delta)\equiv 1\pmod{\prescript{g}{}{\bar{\mathfrak{Q}}_0}}$ if $Hg\mathcal{D}_{\mathfrak{Q}_0}\in B$
and $i_{K,L}(\delta)\equiv 0\pmod{\prescript{g}{}{\bar{\mathfrak{Q}}_0}}$ if $Hg\mathcal{D}_{\mathfrak{Q}_0}\not\in B$.
By definition, we also have $t\equiv 1\pmod{\prescript{g}{}{\bar{\mathfrak{Q}}_0}}$ if $Hg\mathcal{D}_{\mathfrak{Q}_0}\in B$
and $t\equiv 0\pmod{\prescript{g}{}{\bar{\mathfrak{Q}}_0}}$ if $Hg\mathcal{D}_{\mathfrak{Q}_0}\not\in B$.
 So $t=i_{K,L}(\delta)$, as desired.
Furthermore, by choosing $B=H\backslash G/\mathcal{D}_{\mathfrak{Q}_0}$ and $t=i_{K,L}(1)=1$, we see that $\sum_{g\in G/\mathcal{D}_{\mathfrak{Q}_0}} \prescript{g}{}{\delta_{\bar{\mathfrak{Q}}_0}}=1$. It follows that $I(P)$ is a well defined idempotent decomposition of $R_K$.

For the second claim, we first check that the sets $B_\delta$ form a partition of $H\backslash G/\mathcal{D}_{\mathfrak{Q}_0}$ and the map $\delta\mapsto B_\delta$ is injective. To see this, note that if a double coset $Hg\mathcal{D}_{\mathfrak{Q}_0}$ lies in both $B_{\delta}$ and $B_{\delta'}$ for distinct $\delta,\delta'\in I$, then $\prescript{g^{-1}}{}{(i_{K,L}(\delta))}\equiv\prescript{g^{-1}}{}{(i_{K,L}(\delta'))}\equiv 1\pmod{\bar{\mathfrak{Q}}_0}$ by definition. But then $\prescript{g^{-1}}{}{(i_{K,L}(\delta\delta'))}\equiv 1\pmod{\bar{\mathfrak{Q}}_0}$, contradicting the fact that $\delta\delta'=0$. So the sets $B_\delta$ are disjoint and the map $\delta\mapsto B_\delta$ is injective.
Furthermore, each $Hg\mathcal{D}_{\mathfrak{Q}_0}\in H\backslash G/\mathcal{D}_{\mathfrak{Q}_0}$ lies in at least one set $B_\delta$ since 
\begin{equation}\label{eq_sumg}
\sum_{\delta\in I}\prescript{g^{-1}}{}{(i_{K,L}(\delta))}\equiv\prescript{g^{-1}}{}{\left(i_{K,L}\left(\sum_{\delta\in I}\delta\right)\right)}\equiv 1\pmod{\bar{\mathfrak{Q}}_0}.
\end{equation}
So the sets $B_\delta$ form a partition of $H\backslash G/\mathcal{D}_{\mathfrak{Q}_0}$.

Fix $B\in P(I)$ and let $\delta=\delta_B$. It remains to verify that $B_\delta=B$. 
For $h\in G$, we have 
\[
\prescript{h^{-1}}{}{(i_{K,L}(\delta))}=\sum_{g\mathcal{D}_{\mathfrak{Q}_0}\in G/\mathcal{D}_{\mathfrak{Q}_0}: Hg\mathcal{D}_{\mathfrak{Q}_0}\in B}\prescript{h^{-1}g}{}{\delta_{\bar{\mathfrak{Q}}_0}}.
\]
Note that the residue of $\prescript{h^{-1}g}{}{\delta_{\bar{\mathfrak{Q}}_0}}$ modulo $\bar{\mathfrak{Q}}_0$ equals one if $h\mathcal{D}_{\mathfrak{Q}_0}=g\mathcal{D}_{\mathfrak{Q}_0}$, and zero otherwise.
So the residue of $\prescript{h^{-1}}{}{(i_{K,L}(\delta))}$ modulo $\bar{\mathfrak{Q}}_0$ equals one if  $Hh\mathcal{D}_{\mathfrak{Q}_0}\in B$ and zero otherwise.
 It follows by definition that $B_\delta=B$. 
\end{proof}

\lemtransitivitybc*

\begin{proof}
Let $A=\bar{\ord}_K/\rad(\bar{\ord}_K)$ and $H=\langle \sigma_{K,i}\rangle$.
Equivalently, we want to prove that $H$ acts transitively on the set of the maximal ideals of $A_{K,i}/\mathfrak{m}A_{K,i}$, where the action is induced from that on $A_{K,i}$.

We have a short exact sequence
\[
0\to \mathfrak{m} \to A \to A/\mathfrak{m}\to 0,
\]
which by \citep[Proposition~2.18]{AM69} induces an exact sequence
\[
 \mathfrak{m}\otimes_{\F_q} \F_{q^i} \to A_{K,i} \to (A/\mathfrak{m})\otimes_{\F_q}\F_{q^i} \to 0.
\]
Also note that the image of $ \mathfrak{m}\otimes_{\F_q} \F_{q^i} $ in $A_{K,i}$ is $\mathfrak{m}A_{K,i}$.
Then we have
\[
A_{K,i}/\mathfrak{m}A_{K,i}\cong (A/\mathfrak{m})\otimes_{\F_q}\F_{q^i}.
\]
So we want to prove that $H$ acts transitively on the set of the maximal ideals of $(A/\mathfrak{m})\otimes_{\F_q}\F_{q^i}$.

Suppose   $\mathfrak{m}_1,\dots,\mathfrak{m}_k$ are maximal ideals of  $(A/\mathfrak{m})\otimes_{\F_q}\F_{q^i}$ that form an  $H$-orbit, and $\delta_1,\dots,\delta_k$ are the corresponding primitive idempotents. Define $t:=\sum_{i=1}^k \delta_i$ which is a nonzero idempotent fixed by $H$.
It suffices to prove $t=1$.

Note that we have the exact sequence 
\[
0\to (A/\mathfrak{m})^H\to A/\mathfrak{m}\xrightarrow{\tau} A/\mathfrak{m},
\]
where $\tau$ sends $x\in A$ to $x^q-x$. It induces a sequence
\[
0\to (A/\mathfrak{m})^H\otimes_{\F_q}\F_{q^i}\to (A/\mathfrak{m})\otimes_{\F_q}\F_{q^i}\xrightarrow{\tau'} (A/\mathfrak{m})\otimes_{\F_q}\F_{q^i},
\]
where $\tau'$ sends $x\in (A/\mathfrak{m})\otimes_{\F_q}\F_{q^i}$ to $\sigma_{K,i}(x)-x$.
This sequence is exact since $\F_{q^i}$ is a {\em flat} $\F_q$-module\index{flat module} (see, e.g., \citep[Proposition~2.19 and Exercise~2.4]{AM69}). 
So  we have 
\[
((A/\mathfrak{m})\otimes_{\F_q}\F_{q^i})^H\cong (A/\mathfrak{m})^H\otimes_{\F_q}\F_{q^i}\cong \F_q\otimes_{\F_q}\F_{q^i}\cong\F_{q^i}
\]
and the only nonzero idempotent it contains is $1$.
It follows that $t=1$, as desired.
\end{proof}

\lemcomputerings*

\begin{proof}
First use Corollary~\ref{lem_reltoord} to compute an ordinary number field $\tilde{K}$ isomorphic to $K$ and an isomorphism $\phi: K\to \tilde{K}$ in some $\Q$-basis of $K$. Apply Lemma~\ref{lem_residuering} to $\tilde{K}$ and $p$ to compute $\bar{\ord}_K$, $\ord'_K$ as well as the maps $\ord'_K\hookrightarrow \tilde{K}$ and $\ord'_K\to \bar{\ord}_K$.
Compose $\ord'_K\hookrightarrow \tilde{K}$ with $\phi^{-1}$ to obtain the map $\ord'_K\hookrightarrow K$. 

Next we compute an $\F_p$-basis $B=\{x_1,\dots,x_s\}$ of the radical $\rad(\bar{\ord}_K)\subseteq \bar{\ord}_K$ using Theorem~\ref{thm_comprad}. Extend $B$ to an $\F_p$-basis $B'=\{x_1,\dots,x_s,y_1,\dots,y_t\}$ of $\bar{\ord}_K$.
Compute $c_{ij}^k\in\F_p$ for $i,j\in [t]$, $k\in [s]$ and $d_{ij}^k\in\F_p$  for $i,j,k\in [t]$ such that 
\[
y_i y_j=\sum_{k=1}^s c_{ij}^k x_k + \sum_{k=1}^t d_{ij}^k y_k \quad\text{for}~i,j\in [t].
\]
 Then the structure constants of $\bar{\ord}_K/\rad(\bar{\ord}_K)$ are given by $d_{ij}^k$ in the $\F_p$-basis $\{y_1+\rad(\bar{\ord}_K),\dots, y_t+\rad(\bar{\ord}_K)\}$  of $\bar{\ord}_K/\rad(\bar{\ord}_K)$ since 
 \[
 (y_i+\rad(\bar{\ord}_K))(y_j+\rad(\bar{\ord}_K))=\sum_{k=1}^t d_{ij}^k y_k+\rad(\bar{\ord}_K)
 \]
 holds for $i,j\in [t]$. The map $\bar{\ord}_K\to \bar{\ord}_K/\rad(\bar{\ord}_K)$ is given in the basis $B'$ which sends each $x_i$ to zero and each $y_i$ to $y_i+\rad(\bar{\ord}_K)$.

Finally, we compute an $\F_p$-basis of $R_K$ in $\bar{\ord}_K/\rad(\bar{\ord}_K)$ by solving the system of $\F_p$-linear equations given by $x^p=x$. It also gives the inclusion $R_K\hookrightarrow \bar{\ord}_K/\rad(\bar{\ord}_K)$. The structure constants of $R_K$ can be computed from those of $\bar{\ord}_K/\rad(\bar{\ord}_K)$.
\end{proof}

\lemringhomg*

\begin{proof}
To compute the map $\bar{\phi}$, we identify $K$ and $K'$ with ordinary number fields and apply Lemma~\ref{lem_ringhom}:
compute isomorphisms $\tau: K\to \tilde{K}$ and $\tau': K'\to \tilde{K}'$ using Corollary~\ref{lem_reltoord} where $\tilde{K}$ and $\tilde{K}'$ are ordinary number fields. Compute the maps $\ord'_K\hookrightarrow \tilde{K}$, $\ord'_{K'}\hookrightarrow \tilde{K'}$ by composing $\ord'_K\hookrightarrow K$, $\ord'_{K'}\hookrightarrow K'$ with $\tau$ and $\tau'$ respectively.
And compute the field embedding $\phi'=\tau'\circ\phi\circ\tau^{-1}$ from $\tilde{K}$ to $\tilde{K}'$. Now use Lemma~\ref{lem_ringhom} to obtain the map $\bar{\phi}$.

The map  $\hat{\phi}: \bar{\ord}_K/\rad(\bar{\ord}_K)\to \bar{\ord}_{K'}/\rad(\bar{\ord}_{K'})$ induced from $\bar{\phi}$ sends $x+\rad(\bar{\ord}_K)\in \bar{\ord}_K/\rad(\bar{\ord}_K)$ to $\bar{\phi}(x)+\rad(\bar{\ord}_{K'})$. We can efficiently compute $\hat{\phi}$ from $\bar{\phi}$ since the quotient maps $\bar{\ord}_K\to \bar{\ord}_K/\rad(\bar{\ord}_K)$ and $\bar{\ord}_{K'}\to \bar{\ord}_{K'}/\rad(\bar{\ord}_{K'})$ are given.

  Finally,  we restrict $\hat{\phi}$ to $\hat{\phi}|_{R_K}: R_K\to  \bar{\ord}_{K'}/\rad(\bar{\ord}_{K'})$ using the given inclusion $R_K\hookrightarrow \bar{\ord}_K/\rad(\bar{\ord}_K)$. 
  Then compute $\tilde{\phi}: R_K\to R_{K'}$ from $\hat{\phi}|_{R_K}$ by lifting  along the given inclusion $R_{K'}\hookrightarrow \bar{\ord}_{K'}/\rad(\bar{\ord}_{K'})$. 
\end{proof}

{\renewcommand\footnote[1]{}\lemgenset*}

 \begin{proof}
 See Algorithm~\ref{alg_auxi} for the pseudocode of the subroutine.
It enumerates $K\in \mathcal{F}$, $\delta\in I_K$ and computes $e_\delta$, $f_\delta$, $s_\delta$ (if $e_\delta>1)$ and $t_\delta$ (if $f_\delta>1$).

 \begin{algorithm}[htbp]
\caption{$\mathtt{ComputeAdvice}$}\label{alg_auxi}
\begin{algorithmic}[1]
%\INPUT $\mathcal{I}=\{I_K:K\in\mathcal{F}\}$
 \For{\strut $K\in\mathcal{F}$}
     \For{$\delta\in I_K$}
     \State $J\gets  \{x\in\bar{\ord}_K: x+\rad(\bar{\ord}_K)\in (1-\delta)(\bar{\ord}_K/\rad(\bar{\ord}_K)) \}$
     \State compute $e_\delta$ as the smallest $i\in\N^+$ such that $J^i=J^{i+1}$
     \State compute $f_\delta$ as the smallest $i\in\N^+$ such that  $x\mapsto x^{q^i}$ fixes $\bar{\ord}_K/J$
     \If{$e_\delta>1$}
            \State find $s_\delta\in J-J^2$
            \State $U\gets$ the image of $\ann_{\bar{\ord}_K}(s_\delta^{e_\delta-1})$ in $\bar{\ord}_K/\rad(\bar{\ord}_K)$
            \State $U\gets U\cap \delta R_K$
            \State compute $\delta_0\in U$ satisfying $(1-\delta_0)U=\{0\}$
            \If {$\delta_0\delta\not\in\{0,\delta\}$}
                \State $I_K\gets I_K-\{\delta\}$
                \State $I_K\gets I_K\cup\{\delta_0\delta, (1-\delta_0)\delta\}$
                \State \Return
            \EndIf
      \EndIf
      \If{$f_\delta>1$}
        \State find a primitive $f_\delta$th root of unity $\xi\in \F_{q^{f_\delta}}$
        \State find nonzero $t_\delta\in \delta A_{K,f_\delta}$ satisfying $\sigma_{K,f_\delta}(t_\delta)=\xi t_\delta$
        \State $U\gets t_\delta A_{K,f_\delta}\cap R_K$
        \State find $\delta_0\in U$ satisfying $(1-\delta_0)U=\{0\}$
         \If {$\delta_0\delta\not\in\{0,\delta\}$}
                \State $I_K\gets I_K-\{\delta\}$
                \State $I_K\gets I_K\cup\{\delta_0\delta, (1-\delta_0)\delta\}$
                \State \Return
        \EndIf
      \EndIf
\EndFor  
\EndFor
\end{algorithmic}
\end{algorithm}

%When it is possible to refine $I_K$, we perform the refinment and return. 
 
 Fix  $K\in \mathcal{F}$ and $\delta\in I_K$. We compute the ideal $J$ of $\bar{\ord}_K$, which is defined to be the preimage of the ideal of $\bar{\ord}_K/\rad(\bar{\ord}_K)$ generated by $1-\delta$ (under the natural quotient map). Then $J$ is the product of the maximal ideals $\mathfrak{m}$ of $\bar{\ord}_K$ satisfying $\delta\equiv 1\pmod{\mathfrak{m}/\rad(\bar{\ord}_K)}$.
 Note that for any such $\mathfrak{m}$, we have 
 \[
\bar{\ord}_K\supsetneq\mathfrak{m}\supsetneq \mathfrak{m}^2\supsetneq \dots\supsetneq \mathfrak{m}^{e_\delta}=\mathfrak{m}^{e_\delta+1}
\]
and $\bar{\ord}_K/\mathfrak{m}\cong \F_{q^{f_\delta}}$.
So we can compute $e_\delta$ as the smallest positive integer $i$ such that $J^i=J^{i+1}$, and compute $f_\delta$ as the smallest positive integer $i$ such that the automorphism $x\mapsto x^{q^i}$ fixes $\bar{\ord}_K/J$.
 
 Suppose $e_\delta>1$. Choose $s_\delta$ to be an element in $J^2-J$. 
 So  we have $s_\delta\in\mathfrak{m}$ for all the maximal ideals $\mathfrak{m}$ of $\bar{\ord}_K$ satisfying $\delta\equiv 1\pmod{\mathfrak{m}/\rad(\bar{\ord}_K)}$,
 and $s_\delta\not\in \tilde{\mathfrak{m}}^2$  for some maximal  ideal $\tilde{\mathfrak{m}}$ of them.
 
Next compute the image of $\ann_{\bar{\ord}_K}(s_\delta^{e_\delta-1})$ in $\bar{\ord}_K/\rad(\bar{\ord}_K)$ and let $U$ be its intersection with $\delta R_K$, which is an ideal of $R_K$.  Choose an element $\delta_0$ in $U$ such that $(1-\delta_0) U=0$. Then $\delta_0$ is the unique idempotent of $R_K$ that generates $U$. If $\delta_0\delta\not\in\{0,\delta\}$, we use $\delta_0$ to properly refine $I_K$ and return.
 
 As $s_\delta\in\tilde{\mathfrak{m}}-\tilde{\mathfrak{m}}^2$, we have $s_\delta^{e_\delta-1}\in\tilde{\mathfrak{m}}^{e_\delta-1}-\tilde{\mathfrak{m}}^{e_\delta}$ and hence $\ann_{\bar{\ord}_K}(s_\delta^{e_\delta-1})\subseteq \tilde{\mathfrak{m}}$.
 So we have $\delta_0 \in U\subseteq  \tilde{\mathfrak{m}}/\rad(\bar{\ord}_K)$.
But we also have  $\delta\equiv 1\pmod{\tilde{\mathfrak{m}}/\rad(\bar{\ord}_K)}$. It follows that $\delta_0\delta\neq \delta$.
 
On the other hand, assume $s_\delta\in\tilde{\mathfrak{m}}'^2$ for some maximal  ideal $\tilde{\mathfrak{m}}'$ of $\bar{\ord}_K$ satisfying $\delta\equiv 1\pmod{\tilde{\mathfrak{m}}'/\rad(\bar{\ord}_K)}$. We claim $\delta_0\delta\neq 0$, in which case the subroutine properly refines $I_K$ and returns. To see this, note that $s_\delta^{e_\delta-1}\in \tilde{\mathfrak{m}}'^{2(e_\delta-1)}\subseteq \tilde{\mathfrak{m}}'^{e_\delta}$ since $2(e_\delta-1)\geq e_\delta$.
Then $\ann_{\bar{\ord}_K}(s_\delta^{e_\delta-1})\not\subseteq \tilde{\mathfrak{m}}'$. 
Let $\delta'$ be the idempotent of $\bar{\ord}_K/\rad(\bar{\ord}_K)$ that generates the image of $\ann_{\bar{\ord}_K}(s_\delta^{e_\delta-1})$ in $\bar{\ord}_K/\rad(\bar{\ord}_K)$. Then $\delta'\not\in \tilde{\mathfrak{m}}'/\rad(\bar{\ord}_K)$.
Note that $\delta_0=\delta\delta'\not\in \tilde{\mathfrak{m}}'/\rad(\bar{\ord}_K)$. So $\delta_0\delta=\delta_0\neq 0$, as desired.
%We conclude that either the condition that $s_\delta\in\mathfrak{m}-\mathfrak{m}^2$ for all the maximal ideals $\mathfrak{m}$ of $\bar{\ord}_K$ satisfying $\delta\equiv 1\pmod{\mathfrak{m}/\rad(\bar{\ord}_K)}$ is satisfied, or the subroutine properly refines $I_K$ and return.

Now suppose $f_\delta>1$. We pick a primitive $f_\delta$th root of unity $\xi$ in $\F_{q^{f_\delta}}$ which exists since $f_\delta$ divides $|\F_{q^{f_\delta}}^\times|=q^{f_\delta}-1$.\footnote{We use the fact that $f_\delta$ is coprime to $p$, which in turn relies on the assumption $p>\deg(f)$.}
 This step can be done efficiently assuming GRH. 
Choose  $t_\delta$ to be a nonzero element in $\delta A_{K,f_\delta}$ satisfying $\sigma_{K,f_\delta}(t_\delta)=\xi t_\delta$.
We claim that such an element always exists. To see this, note that the quotient map $A_{K,f_\delta}\to A_{K,f_\delta}/(1-\delta)$ is injective when restricting to $\delta A_{K,f_\delta}$. So it suffices to show that there exists a nonzero element $t\in  A_{K,f_\delta}/(1-\delta)$ satisfying $\sigma_{K,f_\delta}(t)+(1-\delta)=\xi t+(1-\delta)$.
This follows from the  argument used  in the proof of Lemma~\ref{lem_auto}.

Next compute  the ideal $U=t_\delta A_{K,f_\delta}\cap R_K$ of $R_K$, and choose an element $\delta_0$ in $U$ satisfying $(1-\delta_0) U=0$. Then $\delta_0$ is the unique idempotent of $R_K$ that generates $U$.  If $\delta_0\delta\not\in\{0,\delta\}$, we use $\delta_0$ to properly refine $I_K$ and return. 

Assume there exists a maximal ideal $\mathfrak{m}_0$ of $A_{K,f_\delta}$ satisfying $\delta\equiv 1\pmod{\mathfrak{m}_0}$ and $t_\delta\in\mathfrak{m}_0$. Then $t_\delta A_{K,f_\delta}\subseteq\mathfrak{m}_0$ and hence $U\subseteq\mathfrak{m}_0$.
As $\delta\not\in\mathfrak{m}_0$, we have $\delta_0\delta\neq \delta$. We claim $\delta_0\delta$ is nonzero, and hence the subroutine properly refines $I_K$ and returns. As $\delta_0\in t_\delta A_{K,f_\delta}\subseteq \delta A_{K,f_\delta}$. We have $\delta_0\delta=\delta_0$, which generates the ideal $U$ of $R_K$. So it suffices to prove $U\neq \{0\}$. As $t_\delta\neq 0$, there exists a maximal ideal $\mathfrak{m}$ of $A_{K,f_\delta}$ that does not contain $t_\delta$. Let $\mathfrak{m}'=\mathfrak{m}\cap R_K$, which is a maximal ideal of $R_K$. Let $\delta'$ be the primitive idempotent of $R_K$ corresponding to $\mathfrak{m}'$, i.e., $\delta'\equiv 1\pmod{\mathfrak{m}'}$ and $\delta'\equiv 0\pmod{\mathfrak{m}''}$ for all the maximal ideals $\mathfrak{m}''\neq \mathfrak{m}'$ of $R_K$. We claim $\delta'\in U$, or equivalently, $\delta'\in t_\delta A_{K,f_\delta}$. 
 For $i\in\Z$, we have $\sigma_{K,f_\delta}^{-i}(t_\delta)=\xi^{-i} t_\delta\not\in \mathfrak{m}$ and hence $t_\delta\not\in \sigma_{K,f_\delta}^i(\mathfrak{m})$. 
 By Lemma~\ref{lem_transitivitybc} and the choice of $\delta'$, the maximal ideals of $A_{K,f_\delta}$ not containing $\delta'$ are exactly those of the form $\sigma_{K,f_\delta}^i(\mathfrak{m})$, $i\in\Z$.
 It follows that $\delta'\in t_\delta A_{K,f_\delta}$, as desired.

The claim about the running time is straightforward.
 \end{proof}

\lemordpcolwd*
 
 \begin{proof}
First note that each element  $s_{\delta,H}$  (resp. $t_{\delta,H}$) is  fixed by $H$ and hence  $\prescript{g^{-1}}{}{s_{\delta,H}},\prescript{g'^{-1}}{}{s_{\delta,H}}$ (resp. $\prescript{g^{-1}}{}{t_{\delta,H}}$,  $\prescript{g'^{-1}}{}{t_{\delta,H}}$) only depend on the cosets $Hg$ and $Hg'$.
Fix $H\in\mathcal{P}$ and $K\in\mathcal{F}$ as in Definition~\ref{defi_ordpcollection}. Fix $B\in C_H$ and $g,g'\in G$ such that $Hg\mathcal{D}_{\mathfrak{Q}_0},Hg'\mathcal{D}_{\mathfrak{Q}_0}\in B$.
Let $\delta$ be the unique idempotent in $I_K$ such that  $\tilde{\tau}_H(\delta)=\delta_B$.
Then $e_\delta=e(B)$ and $f_\delta=f(B)$.

Suppose $e(\delta)>1$.
By Definition~\ref{defi_auxielem}, we have $s_{\delta,H}\in\mathfrak{m}-\mathfrak{m}^2$  for all the maximal ideals $\mathfrak{m}$ of $\bar{\ord}_{L^H}$ satisfying $\tilde{\tau}_H(\delta)\equiv 1\pmod{\mathfrak{m}/\rad(\bar{\ord}_{L^H})}$. Consider the maximal ideal 
\[
\mathfrak{m}_{g,H}=(\prescript{g}{}{\mathfrak{Q}_0}\cap \ord_{L^H})/p\ord_{L^H}
\]
of $\bar{\ord}_{L^H}$.  As $\tilde{\tau}_H(\delta)=\delta_B$ and $Hg\mathcal{D}_{\mathfrak{Q}_0}\in B$, we have 
\[
\tilde{\tau}_H(\delta)\equiv 1\pmod{\mathfrak{m}_{g,H}/\rad(\bar{\ord}_{L^H})}.
\] 
Therefore $s_{\delta,H}\in\mathfrak{m}_{g,H}-\mathfrak{m}_{g,H}^2$. 
So $s_{\delta,H}+\mathfrak{m}_{g,H}^2$ is a nonzero element in $\mathfrak{m}_{g,H}/\mathfrak{m}_{g,H}^2$.
Let $\mathfrak{m}_g$ be the maximal ideal $\prescript{g}{}{\mathfrak{Q}_0}/p\ord_{L}$ of $\bar{\ord}_L$, and let $k=e(\mathfrak{Q}_0)/e_\delta$.
Using the natural inclusion 
\[
\mathfrak{m}_{g,H}/\mathfrak{m}_{g,H}^2\hookrightarrow \mathfrak{m}_g^{k}/\mathfrak{m}_g^{k+1},
\]
 we see $s_{\delta,H}+\mathfrak{m}_g^{k+1}$ is a nonzero element  in $\mathfrak{m}_{g}^k/\mathfrak{m}_{g}^{k+1}$.
 Let $\mathfrak{m}_e:=\mathfrak{Q}_0/p\ord_L$, so that $\prescript{g}{}{\mathfrak{m}_e}=\mathfrak{m}_g$. 
 Then $\prescript{g^{-1}}{}{s_{\delta,H}}+\mathfrak{m}_e^{k+1}$ is a nonzero element in $\mathfrak{m}_{e}^k/\mathfrak{m}_{e}^{k+1}$. The same argument shows that $\prescript{g'^{-1}}{}{s_{\delta,H}}+\mathfrak{m}_e^k$ is a nonzero element in $\mathfrak{m}_{e}^k/\mathfrak{m}_{e}^{k+1}$ as well.
 As $\mathfrak{m}_{e}^k/\mathfrak{m}_{e}^{k+1}$ is an one-dimensional vector space over $\bar{\ord}_L/\mathfrak{m}_e\cong\kappa_{\mathfrak{Q}_0}$, we see that there exists a unique scalar $c\in\kappa_{\mathfrak{Q}_0}^\times$ satisfying
 \[
 \prescript{g^{-1}}{}{s_{\delta,H}}+\mathfrak{m}_e^{k+1}=c\cdot(\prescript{g'^{-1}}{}{s_{\delta,H}}+\mathfrak{m}_e^{k+1}).
 \]
 Note $\mathfrak{m}_e^{k+1}=(\mathfrak{Q}_{0}/p\ord_{L})^{e(\mathfrak{Q}_0)/e_\delta+1}$. We see that the second condition in Definition~\ref{defi_ordpcollection} is well defined.

Now suppose $f(\delta)>1$.
By Definition~\ref{defi_auxielem}, we have $t_{\delta,H}\not\in\mathfrak{m}$  for all the maximal ideals $\mathfrak{m}$ of $A_{L^H,f_\delta}$ satisfying $\tilde{\tau}_H(\delta)\equiv 1\pmod{\mathfrak{m}}$. As $\tilde{\tau}_H(\delta)=\delta_B$, $Hg\mathcal{D}_{\mathfrak{Q}_0}\in B$ and $\frac{\mathfrak{Q}_{0}/p\ord_{L}}{\rad(\bar{\ord}_{L})}\subseteq \mathfrak{m}_0$, we have
\[
\tilde{\tau}_H(\delta)\equiv 1\pmod{\prescript{g}{}{\mathfrak{m}_0}}.
\] 
So $t_{\delta,H}\not\in\prescript{g}{}{\mathfrak{m}_0}$. Then $\prescript{g^{-1}}{}{t_{\delta,H}}+\mathfrak{m}_0$ is a nonzero element in $A_{L,f_\delta}/\mathfrak{m}_0$. The same argument shows that $\prescript{g'^{-1}}{}{t_{\delta,H}}+\mathfrak{m}_0$ is a nonzero element in $A_{L,f_\delta}/\mathfrak{m}_0$ as well. It follows that  there exists a unique scalar $c\in (A_{L,f_\delta}/\mathfrak{m}_0)^\times$ satisfying
\begin{equation}\label{eq_cond3}
\prescript{g^{-1}}{}{t_{\delta,H}}+\mathfrak{m}_0=c \cdot (\prescript{g'^{-1}}{}{t_{\delta,H}}+\mathfrak{m}_0).
\end{equation}

We also check that Definition~\ref{defi_ordpcollection} is independent of the choice of $\mathfrak{m}_0$:
Let $\mathfrak{m}'_0$ be another  maximal ideal of $A_{L, f_\delta}$ containing $\frac{\mathfrak{Q}_{0}/p\ord_{L}}{\rad(\bar{\ord}_{L})}$.  By Lemma~\ref{lem_transitivitybc}, we have $\mathfrak{m}'_0=\sigma_{L,f_\delta}^i(\mathfrak{m}_0)$ for some  $i\in\Z$. Let $\sigma=\sigma_{L,f_\delta}^i$.
Then \eqref{eq_cond3} is equivalent to 
\begin{equation}\label{eq_cond3_2}
\sigma(\prescript{g^{-1}}{}{t_{\delta,H}})+\mathfrak{m}'_0=\sigma(c) \cdot \left(\sigma(\prescript{g'^{-1}}{}{t_{\delta,H}})+\mathfrak{m}'_0\right)
\end{equation}
where $\sigma(c)\in (A_{L,f_\delta}/\mathfrak{m}'_0)^\times$.
Also note that 
\[
\sigma(\prescript{g^{-1}}{}{t_{\delta,H}})=\prescript{g^{-1}}{}{(\sigma (t_{\delta,H}))}=\prescript{g^{-1}}{}{(\xi^i t_{\delta,H})}=\xi^i\prescript{g^{-1}}{}{t_{\delta,H}}.
\]
and similarly $\sigma(\prescript{g'^{-1}}{}{t_{\delta,H}})=\xi^i\prescript{g'^{-1}}{}{t_{\delta,H}}$.
Substituting them in \eqref{eq_cond3_2} and canceling $\xi^i+\mathfrak{m}'_0$ on both sides, we obtain
\[
\prescript{g^{-1}}{}{t_{\delta,H}}+\mathfrak{m}'_0=\sigma(c) \cdot (\prescript{g'^{-1}}{}{t_{\delta,H}}+\mathfrak{m}'_0).
\]
Note that $\sigma(c)$ and $c$ have the same order. We see that choosing $\mathfrak{m}'_0$ instead of $\mathfrak{m}_0$ does not affect the definition.

 Finally, it is easy to see that the  conditions in  Definition~\ref{defi_ordpcollection} are equivalence relations on $H\backslash G$. So they do define a partition $C_H$ on $H\backslash G$.
\end{proof}

%\lemexponentg*

\lembijtest*

  To prove Lemma~\ref{lem_bijtest}, we first prove the following lemma,  generalizing Lemma~\ref{lem_iexp}:

\begin{lem}\label{lem_exponentg}
There exists an algorithm $\mathtt{SplitByExp}$ that, given a semisimple $\F_p$-algebra $A$, $m\in\N^+$, and nonzero elements $x,y\in A$ satisfying the following conditions
\begin{itemize}
\item $x$ and $y$ generate the same ideal of $A$
\item  Let $n_x$ (resp. $n_y$) be the smallest positive integer such that $x^{n_x}$ (resp. $y^{n_y}$) is an idempotent. Then $n_y$ divides $n_x$ and all the prime factors of $n_x$ divide $m$
\end{itemize}
returns  an element $z=x^k-y\in A$ satisfying $zA\subsetneq xA$ in time polynomial in $m$ and $\log |A|$, where $k\in\N$.
\end{lem}

\begin{proof}
We find $k\in\N$ such that $x^k-y$ satisfies the requirement. Let $I=\ann_A(x)$. By replacing $A$, $x$, and $y$ with $A/I$, $x+I$ and $y+I$ respectively, we reduce to the case $x,y\in A^\times$, and the goal is to find $k\in\N$ such that $z=x^k-y$ is a zero divisor. 
In addition, we find the smallest $d\in\N^+$ such that the ideal $J$ generated by $\{x^{p^d}-x: x\in A\}$ is a proper ideal of $A$.
By replacing $A$ with $A/J$, we may assume $J=\{0\}$. Then $A$ is a finite product of copies of $\F_{p^d}$.

Enumerate the prime factors $\ell$ of $m$. For each $\ell$, compute $e_\ell\in\N$ and $f_\ell\in \N^+$ such that $p^{d}-1=\ell^{e_\ell}f_\ell$ and $f_\ell$ is coprime to $\ell$. Let $n_{x,\ell}$ and $n_{y,\ell}$ be the order of $x^{f_\ell}$ and $y^{f_\ell}$ respectively. Then $n_{x,\ell}, n_{y,\ell}$ are powers of $\ell$ and $n_{y,\ell}|n_{x,\ell}$.
Use the algorithm in Lemma~\ref{lem_iexp} (applied to $x^{f_\ell}$ and $y^{f_\ell}$) to compute $k_\ell\in\N$ such that $x^{k_\ell f_\ell}-y^{f_\ell}$ is a zero divisor. If $x^{k_\ell f_\ell}-y^{f_\ell}\neq 0$, we use Lemma~\ref{lem_zerodivisor} to find an idempotent $\gamma\not\in\{0,1\}$ of $A$ and solve the problem recursively on the quotient ring $A/(1-\gamma)$. So assume $x^{k_\ell f_\ell}=y^{f_\ell}$. Then the order of $x^{k_\ell}/y$ divides $f_\ell$ and hence is coprime to $\ell$.

Compute $k_\ell$ and $e_\ell$ for all the prime factors $\ell$ of $m$ as above. Use the extended Euclidean algorithm to find $k\in\N$ satisfying $k\equiv k_\ell\pmod{\ell^{e_\ell}}$ for all $\ell$. Then $k$ is the desired integer. 

We claim $x^k=y$. To see this, note that for each $\ell$, we have $x^k/y=(x^{k_\ell}/y)\cdot x^{t \ell^{e_\ell}}$ for some $t\in\Z$. As the orders of $x^{k_\ell}/y$ and $ x^{t \ell^{e_\ell}}$ are both coprime to $\ell$, so is the order of $x^k/y$. Therefore the order of $x^k/y$ is coprime to $m$. But the orders of $x^k$ and $y$ are only divisible by prime factors of $m$. So $x^k/y=1$, as desired.
\end{proof}

The pseudocode of the subroutine $\mathtt{SurjectivityTest}$ is given in Algorithm~\ref{alg_surjtest}. It enumerates $K\in F$ and $\delta\in I_K$. For each $K$ and $\delta$, a set $S$ of ideals of $A_{K,f_\delta}$ is computed. And for each $I\in S$, we find  $\delta_0\in I\cap R_K$ satisfying $(1-\delta_0)(I\cap R_K)=\{0\}$, which is the unique idempotent of $R_K$ that generates the ideal $I\cap R_K$ of $R_K$.\footnote{Here $R_K$ is regarded as a subring of $A_{K,f_\delta}$ via the inclusions $R_K\hookrightarrow \bar{\ord}_K/\rad(\bar{\ord}_K)$ and $\bar{\ord}_K/\rad(\bar{\ord}_K)\hookrightarrow A_{K,f_\delta}$.}
If $\delta_0\delta\not\in\{0,\delta\}$, we use $\delta_0$ to refine $I_K$ and return.

Fix $K\in F$ and $\delta\in I_K$. The corresponding set $S$ is computed as follows:
first assume $f_\delta>1$. We compute the largest factor $r$ of $q^{f_\delta}-1$ coprime to $f_\delta$, so that all the prime factors of $(q^{f_\delta}-1)/r$ divide $f_\delta$. Compute an element $\gamma\in \F_{q^{f_\delta}}^\times$ of order $(q^{f_\delta}-1)/r$, which can be done efficiently assuming GRH.\footnote{For example,  we can achieve this by computing an $\ell$th power non-residue $\gamma_\ell$ for each prime factor $\ell$ of $f_\delta$. By raising $\gamma_\ell$ to its $r_\ell$th power, where $r_\ell$ is the largest factor of $q^{f_\delta}-1$ coprime to $\ell$, we may assume the order of $\gamma_\ell$ is $(q^{f_\delta}-1)/r_\ell$.  Then let $\gamma$ be the product of all $\gamma_\ell$.} 
By the second condition in Definition~\ref{defi_auxielem}, the element $\delta t_\delta^r$ generates the ideal $\delta A_{K,f_\delta}$ of $ A_{K,f_\delta}$, and so does $\delta \gamma$. We call the subroutine $\mathtt{SplitByExp}$ in Lemma~\ref{lem_exponentg} on the input $(A_{K,f_\delta},f_\delta, \delta \gamma,\delta t_\delta^r)$ to obtain $x\in A_{K,f_\delta}$, and add the ideal $xA_{K,f_\delta}$  to $S$.

\begin{algorithm}[tbp]
\caption{$\mathtt{SurjectivityTest}$}\label{alg_surjtest}
\begin{algorithmic}[1]
 \For{\strut $K\in\mathcal{F}$}
     \For{$\delta\in I_K$}
        \State $S\gets\emptyset$
        \If{$f_\delta>1$}
            \State $r\gets$ the largest factor of $q^{f_\delta}-1$ coprime to $f_\delta$
            \State compute $\gamma\in \F_{q^{f_\delta}}^\times$ of order $(q^{f_\delta}-1)/r$
            \State call $\mathtt{SplitByExp}$  on $(A_{K,f_\delta},f_\delta, \delta \gamma,\delta t_\delta^r)$ to obtain  $x\in A_{K,f_\delta}$
            \State $S\gets S\cup\{xA_{K,f_\delta}\}$
        \EndIf
        \If{$e_\delta>1$}
             \State $J\gets$ the preimage of $\delta(\bar{\ord}_K/\rad(\bar{\ord}_K))$ in $\bar{\ord}_K$
             %\State $J\gets J^{[K:K_0]}$
             \State find $\delta'\in  \ann_{\bar{\ord}_K}(J^{e_\delta})$ satisfying $(1-\delta') \ann_{\bar{\ord}_K}(J^{e_\delta})=\{0\}$
             \State lift $\delta' s_\delta\in \bar{\ord}_K$ to $\tilde{s}\in \ord'_K$ 
             \State compute the image $s$ of $\tilde{s}^{e_\delta}/p$ in $\bar{\ord}_K/\rad(\bar{\ord}_K)$
             \State $r'\gets$ the largest factor of $q^{f_\delta}-1$ coprime to $e_\delta$
             \State compute $\mu\in \F_{q^{f_\delta}}^\times$ of order $(q^{f_\delta}-1)/r'$
             \State call $\mathtt{SplitByExp}$ on $(A_{K,f_\delta},e_\delta, \delta\mu, s^{r'})$ to obtain $y\in A_{K,f_\delta}$
             \State $S\gets S\cup \{yA_{K,f_\delta}\}$
             \If{$f_\delta>1$}
                \ForTo{$i$}{$0$}{$f_\delta-1$}
                     \State $S\gets S\cup\{yA_{K,f_\delta}+\sigma^i_{K,f_\delta}(x) A_{K,f_\delta}\}$
                \EndFor
             \EndIf
        \EndIf
        \For{$I\in S$}
             \State find $\delta_0\in I\cap R_K$ satisfying $(1-\delta_0)(I\cap R_K)=\{0\}$
             \If {$\delta_0\delta\not\in\{0,\delta\}$}
                 \State $I_K\gets I_K-\{\delta\}$
                 \State $I_K\gets I_K\cup\{\delta_0\delta, (1-\delta_0)\delta\}$
                 \State \Return
             \EndIf
        \EndFor
     \EndFor  
\EndFor
\end{algorithmic}
\end{algorithm}

Next  assume $e_\delta>1$. 
Compute the preimage $J$ of $(1-\delta)(\bar{\ord}_K/\rad(\bar{\ord}_K))$ under the quotient  map $\bar{\ord}_K\to \bar{\ord}_K/\rad(\bar{\ord}_K)$. 
Then $J$ is the product of the maximal ideals $\mathfrak{m}$ of $\bar{\ord}_K$ satisfying $\delta\equiv 1\pmod{\mathfrak{m}/\rad(\bar{\ord}_K)}$. 
%Raise $J$ to its $[K:K_0]$th power.  Then for any  maximal ideals $\mathfrak{m}$ of $\bar{\ord}_K$, we have $J\in \mathfrak{m}^k$ for all $k\in\N$ if $\delta\in \mathfrak{m}/\rad(\bar{\ord}_K)$ and $J\not\in \mathfrak{m}$ if $\delta\not \in \mathfrak{m}/\rad(\bar{\ord}_K)$. 
Find $\delta'\in  \ann_{\bar{\ord}_K}(J^{e_\delta})$ satisfying $(1-\delta') \ann_{\bar{\ord}_K}(J^{e_\delta})=\{0\}$, so that $\delta'$ is the unique idempotent of $\bar{\ord}_K$ generating $ \ann_{\bar{\ord}_K}(J^{e_\delta})$. Lift $\delta' s_\delta\in \bar{\ord}_K$ to $\tilde{s}\in \ord'_K$.

We claim $\tilde{s}^{e_\delta}\in p\ord_K$: this is equivalent to $(\delta' s_\delta)^{e_\delta}=\delta' s_\delta^{e_\delta}=0$. 
By the first condition in Definition~\ref{defi_auxielem},
we have $s_\delta^{e_\delta}\in\mathfrak{m}^{e_\delta}$ for all the maximal ideals $\mathfrak{m}$ of $\bar{\ord}_K$ satisfying $\delta\equiv 1\pmod{\mathfrak{m}/\rad(\bar{\ord}_K)}$.
And  by the definition of $J$, it holds that  $\delta'\in\mathfrak{m}^{k}$ for all the maximal ideals $\mathfrak{m}$ of $\bar{\ord}_K$ satisfying $\delta \in\mathfrak{m}/\rad(\bar{\ord}_K)$ and   $k\in \N$. It follows that $\delta' s_\delta^{e_\delta}=0$ and hence   $\tilde{s}^{e_\delta}\in p\ord_K$. 

Compute the image $s$ of $\tilde{s}^{e_\delta}/p\in \ord_K$ in $\bar{\ord}_K/\rad(\bar{\ord}_K)$.
This is done by first computing  $\tilde{s}^{e_\delta}+\ord_K\in \bar{\ord}_K$
using Lemma~\ref{lem_computeresidue} and then computing $s$ using the quotient map  $\bar{\ord}_K\to \bar{\ord}_K/\rad(\bar{\ord}_K)$.
Next  compute the largest factor $r'$ of $q^{f_\delta}-1$ coprime to $e_\delta$, so that all the prime factors of $(q^{f_\delta}-1)/r'$ divide $e_\delta$. Compute an element $\mu\in \F_{q^{f_\delta}}^\times$ of order $(q^{f_\delta}-1)/r'$, which can be done efficiently assuming GRH.
By the first condition in Definition~\ref{defi_auxielem}, the element $s^{r'}$ generates the ideal $\delta A_{K,f_\delta}$ of $ A_{K,f_\delta}$, and so does $\delta \mu$.\footnote{We let $A_{K,f_\delta}=\bar{\ord}_K/\rad(\bar{\ord}_K)$ if $f_\delta=1$.}
We call the subroutine $\mathtt{SplitByExp}$ on the input $(A_{K,f_\delta},e_\delta, \delta \mu,s^{r'})$ to obtain $y\in A_{K,f_\delta}$, and add the ideal $yA_{K,f_\delta}$  to $S$.
In addition, if $f_\delta>1$, we enumerate $i=0,1,\dots,f_\delta-1$ and for each $i$, we add the ideal  of $A_{K,f_\delta}$ generated by $y$ and $\sigma^i_{K,f_\delta}(x)$ to $S$, where $x\in A_{K,f_\delta}$ is computed in the case $f_\delta>1$ above. 

Now we prove Lemma~\ref{lem_bijtest}.

\begin{proof}[Proof of Lemma~\ref{lem_bijtest}]
Assume for some $H\in\mathcal{P}$, $B\in C_H$, and $\tilde{B}\in \tilde{C}_H$, the map $\pi_H: Hh\mapsto Hh\mathcal{D}_{\mathfrak{Q}_0}$ maps  $\tilde{B}$ to a proper subset of $B$. Let $K$ be the field in $\mathcal{F}$ isomorphic to $L^H$ over $K_0$.
Let $\delta$ be the idempotent in $I_K$ satisfying $\tilde{\tau}_H(\delta)=\delta_B$ (see Definition~\ref{defi_partitioncorg}).
We show that in the  corresponding iteration of the loop in Lines 3--27, we compute a set $S$ that contains an ideal $I$ of $A_{K,f_\delta}$ such that the unique idempotent $\delta_0\in R_K$  generating $I\cap R_K$ satisfies $\delta_0\delta\not\in\{0,\delta\}$. Consequently, some partition in $\mathcal{C}$ is properly refined.

Choose $g,g'\in G$ such that  $Hg\mathcal{D}_{\mathfrak{Q}_0} \in B-\pi_H(\tilde{B})$ and $Hg'\in \tilde{B}$. 
Let  $\mathfrak{m}_0$  be an arbitrary  maximal ideal of $A_{L, f_\delta}$ containing $\frac{\mathfrak{Q}_{0}/p\ord_{L}}{\rad(\bar{\ord}_{L})}$.  Fix $\sigma\in \mathcal{D}_{\mathfrak{Q}_0}$ whose image   in  $\gal(\kappa_{\mathfrak{Q}_0}/\bar{\ord}_{K_0})$ is the Frobenius automorphism $x\mapsto x^q$ over $\F_q$.

We necessarily have $f_\delta>1$ or $e_\delta>1$.
First assume $f_\delta>1$.  
  Let $\gamma\in \F_{q^{f_\delta}}^\times$ be of order $(q^{f_\delta}-1)/r$, where $r$ is the largest factor of $q^{f_\delta}-1$ coprime to $f_\delta$. 
Consider an element $x\in A_{L^H,f_\delta}$ of the form $x=(\delta_B\gamma)^k-\delta_B t_{\delta, H}^r=\delta_B(\gamma^k-t_{\delta, H}^r)$ such that $xA_{L^H,f_\delta}\subsetneq \delta_B A_{L^H,f_\delta}$, where $k\in \N$. 
Let $\delta_0$ be the unique idempotent  of $R_{L^H}$ generating $xA_{L^H,f_\delta}\cap R_{L^H}$. The assumption $xA_{L^H,f_\delta}\subsetneq \delta_B A_{L^H,f_\delta}$ implies  $\delta_0\delta_B=\delta_0$ and $\delta_0\neq \delta_B$.
If $\delta_0\neq 0$, by identifying $K$ with $L^H$ via the isomorphism $\tau_H: K\to L^H$, we see  the ideal  added to $S$ at Line 8 is used in Lines 24--26 to properly refine $I_K$.
So assume $\delta_0=0$, or equivalently $xA_{L^H,f_\delta}\cap R_{L^H}=\{0\}$. 

Consider arbitrary $h\in G$, and let $\delta_1$ be the primitive idempotent of $R_{L^H}$ corresponding to the maximal ideal $\prescript{h}{}{\bar{\mathfrak{Q}}_0}\cap R_{L^H}$.
Then a maximal ideal $\mathfrak{m}$  of $R_L$ satisfies $\delta_1\equiv 1\pmod{\mathfrak{m}}$ iff $\mathfrak{m}=\prescript{h'}{}{\bar{\mathfrak{Q}}_0}$ for some $h'\in Hh$. This follows from  Lemma~\ref{lem_idealdoublecosetg} and
the fact that $\mathcal{D}_{\mathfrak{Q}_0}$ fixes $\bar{\mathfrak{Q}}_0$ setwisely. 
So a maximal ideal $\mathfrak{m}'$  of $A_{L,f_\delta}$ satisfies $\delta_1\equiv 1\pmod{\mathfrak{m}'}$ iff $ \mathfrak{m}'\supseteq \prescript{h'}{}{\bar{\mathfrak{Q}}_0}$ for some $h'\in Hh$. As $xA_{L^H,f_\delta}\cap R_{L^H}=\{0\}$, we have $\delta_1\not\in xA_{L^H,f_\delta}$.
 So for some $h'\in Hh$ and a maximal ideal $ \mathfrak{m}'\supseteq \prescript{h'}{}{\bar{\mathfrak{Q}}_0}$ of $A_{L,f_\delta}$,
 we have  $x\in \mathfrak{m}'\cap A_{L^H,f_\delta}$, and hence $\prescript{h^{-1}}{}{x}=\prescript{h'^{-1}}{}{x} \in \prescript{h'^{-1}}{}{\mathfrak{m}'}\supseteq \bar{\mathfrak{Q}}_0$.
 By Lemma~\ref{lem_transitivitybc}, we have $\prescript{h'^{-1}}{}{\mathfrak{m}'}=\sigma_{L,f_\delta}^{-i_0}(\mathfrak{m}_0)$ for some $i_0\in \Z$.
Therefore 
\begin{equation}\label{eq_surjeq1}
 \sigma_{L,f_\delta}^{i_0}(\prescript{h^{-1}}{}{x})\in\mathfrak{m}_0.
 \end{equation}

 Suppose the element $h$ above satisfies $Hh\mathcal{D}_{\mathfrak{Q}_0}\in B$. Then $\sigma_{L,f_\delta}^{i_0}(\prescript{h^{-1}}{}{\delta_B})=\prescript{h^{-1}}{}{\delta_B}\equiv 1\pmod{\mathfrak{m}_0}$. As $x=\delta_B(\gamma^k-t_{\delta, H}^r)$, \eqref{eq_surjeq1} implies
 \[
\xi^{i_0 r}\prescript{h^{-1}}{}{(t_{\delta,H}^r)}=\prescript{h^{-1}}{}{(\sigma_{L,f_\delta}^{i_0}(t_{\delta,H}^r))}=  \sigma_{L,f_\delta}^{i_0}(\prescript{h^{-1}}{}{(t_{\delta,H}^r)})\equiv \gamma^k\pmod{\mathfrak{m}_0},
 \]
 where $\xi$ is the primitive $f_\delta$th root of unity satisfying $\sigma_{K,f_\delta}(t_{\delta})=\xi\cdot t_\delta$ as in  Definition~\ref{defi_auxielem}.
  Choosing $h$ to be  $g$ and $g'$ respectively and using the fact $r$ is coprime to $f_\delta$, we see that there exists an unique integer $i\in\{0,\dots,f_\delta-1\}$ satisfying
\begin{equation}\label{eq_surjeq2}
 \xi^{i r} \prescript{g^{-1}}{}{(t_{\delta,H}^r)}\equiv   \prescript{g'^{-1}}{}{(t_{\delta,H}^r)} \pmod{\mathfrak{m}_0}.
 \end{equation}
  As
 \[
 \xi^{ir} \prescript{g^{-1}}{}{(t_{\delta,H}^r)}\equiv  \prescript{g^{-1}}{}{( \sigma_{L,f_\delta}^{i}(t_{\delta,H}^r))} \equiv \sigma_{L,f_\delta}^{i}(\prescript{g^{-1}}{}{(t_{\delta,H}^r)})\equiv \prescript{\sigma^i g^{-1}}{}{(t_{\delta,H}^r)} \pmod{\mathfrak{m}_0},
 \]
 and $r$ is coprime to $f_\delta$, we see that $i$ is the unique integer in $\{0,\dots,f_\delta-1\}$ such that the order of the element $c$ in $(A_{L,f_\delta}/\mathfrak{m}_0)^\times$ satisfying
\[
\prescript{\sigma^i g^{-1}}{}{t_{\delta,H}}+\mathfrak{m}_0=c \cdot (\prescript{g'^{-1}}{}{t_{\delta,H}}+\mathfrak{m}_0) 
\]
 is coprime to $f_\delta$. So for all $\omega\in\mathcal{I}_{\mathfrak{Q}_0}$, the third condition in Definition~\ref{defi_ordpcollection} is    satisfied by $Hg\sigma^{-i}\omega^{-1}$ and $Hg'$, and is not satisfied by $Hg\sigma^{-i'}\omega^{-1}$ and $Hg'$ for $i'\in\{0,\dots,f_\delta-1\}-\{i\}$. In particular, if $e_\delta=1$, then $Hg\sigma^{-i}$ and $Hg'$ are in the same block $\tilde{B}$ by  Definition~\ref{defi_ordpcollection}, contradicting the assumption $Hg\mathcal{D}_{\mathfrak{Q}_0}\not\in\pi_H(\tilde{B})$. So the subroutine properly refines $I_K$ if $f_\delta>1$ and $e_\delta=1$.
 
 Next assume $e_\delta>1$. Consider the ideal $y A_{K,f_\delta}$ of $A_{K,f_\delta}$ added to $S$ at Line 18, and let $\delta_0$ be the idempotent of $R_K$ generating $y A_{K,f_\delta}\cap R_K$. Note $yA_{K,f_\delta}\subsetneq \delta A_{K,f_\delta}$.
 So $\delta_0\delta=\delta_0\neq \delta$. If $\delta\neq 0$, we properly refine $I_K$ using $\delta_0$ in Lines 24--26. So assume $\delta_0=0$, or equivalently $y A_{K,f_\delta}\cap R_K=\{0\}$. Using the isomorphism $\tau_H: K\to L^H$, we regard $y$ as an element of $A_{L^H,f_\delta}$. So the assumption becomes  $y A_{L^H,f_\delta}\cap R_{L^H}=\{0\}$.

 Let $c$ be the unique element in $\kappa_{\mathfrak{Q}_{0}}^\times$ satisfying
\begin{equation}\label{eq_surjeq25}
\prescript{\sigma^i g^{-1}}{}{s_{\delta,H}}+I= c\cdot  (\prescript{g'^{-1}}{}{s_{\delta,H}} +I),
 \end{equation}
 where $I=(\mathfrak{Q}_{0}/p\ord_{L})^{e(\mathfrak{Q}_0)/e_\delta+1}$ (see Definition~\ref{defi_ordpcollection}), and $i$ is the unique integer in $\{0,\dots,f_\delta-1\}$ satisfying  \eqref{eq_surjeq2} above (if $f_\delta=1$, we let $\sigma^i$ be the identity).
Then the element $\tilde{s}$ computed at Line 13 (regarded as an element of $\ord_{L}$) satisfies
 \[
 \prescript{ \sigma^i g^{-1}}{}{\tilde{s}}+\mathfrak{Q}_{0}^{e(\mathfrak{Q}_0)/e_\delta+1}=   c\cdot   (\prescript{g'^{-1}}{}{\tilde{s}} +\mathfrak{Q}_{0}^{e(\mathfrak{Q}_0)/e_\delta+1})
 \]
 and hence
  \[
 \prescript{\sigma^i  g^{-1}}{}{(\tilde{s}^{e_\delta})}+\mathfrak{Q}_{0}^{e(\mathfrak{Q}_0)+1}= c^{e_\delta}\cdot  (\prescript{g'^{-1}}{}{(\tilde{s}^{e_\delta})} +\mathfrak{Q}_{0}^{e(\mathfrak{Q}_0)+1}).
 \]
We have $p\in \mathfrak{Q}_{0}^{e(\mathfrak{Q}_0)}-\mathfrak{Q}_{0}^{e(\mathfrak{Q}_0)+1}$ and it is fixed by $G$.
So the element $s$ computed at Line 14 (regarded as an element of $\bar{\ord}_L/\rad(\bar{\ord}_L)$) satisfies
\begin{equation}\label{eq_surjeq3}
 \prescript{\sigma^i g^{-1}}{}{s}+\mathfrak{m}= c^{e_\delta}\cdot  (\prescript{g'^{-1}}{}{s} +\mathfrak{m}),
 \end{equation}
 where $\mathfrak{m}=\frac{\mathfrak{Q}_{0}/p\ord_L}{\rad(\bar{\ord}_L)}$.
 
  Fix a generator $\omega$ of $\mathcal{I}_{\mathfrak{Q}_0}$.  The proof of Lemma~\ref{lem_injsep} shows that 
\begin{equation}\label{eq_surjeq4}
 \prescript{\omega\sigma^i g^{-1}}{}{s_{\delta,H}}+I=c'(\prescript{\sigma^i g^{-1}}{}{s_{\delta,H}}+I)
 \end{equation}
for some primitive $e_\delta$th root of unity $c'\in \kappa_{\mathfrak{Q}_{0}}^\times$.
 
If $f_\delta=1$, we have $A_{L^H,f_\delta}=\bar{\ord}_{L^H}/\rad(\bar{\ord}_{L^H})$, and its maximal ideals correspond one-to-one to those of $R_{L^H}$. So $y A_{L^H,f_\delta}\cap R_{L^H}=\{0\}$ implies $y=0$.
 Note that $y$ is of the form $\delta_B\mu^\ell -s^{r'}$ where $\ell\in\N$, $r'$ is the largest factor of $q^{f_\delta}-1$ coprime to $e_\delta$, and $\mu$ is an element in $\F_{q^{f_\delta}}^\times$ of order $(q^{f_\delta}-1)/r'$. We have $\prescript{ g^{-1}}{}{\delta_B}, \prescript{g'^{-1}}{}{\delta_B}\equiv 1\pmod{\mathfrak{m}}$ since   $Hg\mathcal{D}_{\mathfrak{Q}_0},Hg'\mathcal{D}_{\mathfrak{Q}_0}\in B$.
 As $\prescript{ g^{-1}}{}{y}=\prescript{g'^{-1}}{}{y}=0$, we have $\prescript{ g^{-1}}{}{(s^{r'})}\equiv \prescript{g'^{-1}}{}{(s^{r'})}\pmod{\mathfrak{m}}$. Combining it with \eqref{eq_surjeq3}, we see $c^{r'}$ is an $e_\delta$th root of unity.
 On the other hand, we know $c'$ is  a  primitive $e_\delta$th root of unity, and so is $c'^{r'}$ since $r'$ is coprime to $e_\delta$.
 Therefore there exists $j\in \{0,\dots,e_\delta-1\}$ such that $(c'^{r'})^j c^{r'}=1$. Then the order of $c'^j c$ divides $r'$, and hence is coprime to $e_\delta$. On the other hand, by \eqref{eq_surjeq25} and \eqref{eq_surjeq4}, we have
 \[
\prescript{\omega^j\sigma^i g^{-1}}{}{s_{\delta,H}}+I= c'^j c\cdot  (\prescript{g'^{-1}}{}{s_{\delta,H}} +I).
 \]
So by Definition~\ref{defi_ordpcollection} and the fact $Hg'\in\tilde{B}$, we have $Hg\sigma^{-i}\omega^{-j}\in\tilde{B}$. But this is a contradiction to the assumption $Hg\mathcal{D}_{\mathfrak{Q}_0} \not\in \pi_H(\tilde{B})$.

Next consider the case $f_\delta>1$ (and $e_\delta>1$). 
Let $x,y\in A_{L^H,f_\delta}$ be as above.
We claim that there exists $i'\in\{0,\dots,f_\delta-1\}$ such that  
\begin{equation}\label{eq_proper}
yA_{L^H,f_\delta}+\sigma^{i'}_{L^H,f_\delta}(x)A_{L^H,f_\delta}\subsetneq \delta_B A_{L^H,f_\delta}.
\end{equation}
 To see this, choose a maximal ideal $\mathfrak{m}_1$ of $A_{L^H,f_\delta}$ containing $y$ but not $\delta_B$, which exists since $yA_{L^H,f_\delta}\subsetneq  \delta_B A_{L^H,f_\delta}$.
 Let $\mathfrak{m}'_1=\mathfrak{m}_1\cap R_{L^H}$. As $xA_{L^H,f_\delta}\cap R_{L^H}=\{0\}$, there exists a maximal ideal $\mathfrak{m}_2\supseteq \mathfrak{m}'_1$ of $A_{L^H,f_\delta}$ containing $x$.
By Lemma~\ref{lem_transitivitybc}, there exists $i'\in\Z$ such that 
$\sigma^{i'}_{L^H,f_\delta}(\mathfrak{m}_2)= \mathfrak{m}_1$ and hence
$\sigma^{i'}_{L^H,f_\delta}(x)\subseteq  \mathfrak{m}_1$.
As $\sigma^{f_\delta}_{L^H,f_\delta}$ fixes $\delta_B A_{L^H,f_\delta}$, we may assume $i'\in\{0,\dots,f_\delta-1\}$. As $\mathfrak{m}_1$ contains both $y$ and $\sigma^{i'}_{L^H,f_\delta}(x)$, but not $\delta_B$, the claim follows.

Let $I=yA_{L^H,f_\delta}+\sigma^{i'}_{L^H,f_\delta}(x)A_{L^H,f_\delta}\subsetneq \delta_B A_{L^H,f_\delta}$. Let $\delta_0$ be the idempotent of $R_K$ generating $I\cap R_K$. As $I\subsetneq \delta_B A_{L^H,f_\delta}$, we have $\delta_0\delta=\delta_0\neq \delta$.
If $\delta_0\neq 0$, we see it is used in Lines 24--26 to properly refine $I_K$. So assume $\delta_0=0$, or equivalently $I\cap R_K=\{0\}$.
Let $x'=\sigma^{i'}_{L^H,f_\delta}(x)$.
  Then there exists $i_1,i_2\in\Z$ such that
\[
 \sigma_{L,f_\delta}^{i_1}(\prescript{g^{-1}}{}{y}), \sigma_{L,f_\delta}^{i_1}(\prescript{g^{-1}}{}{x'}), \sigma_{L,f_\delta}^{i_2}(\prescript{g'^{-1}}{}{y}),  \sigma_{L,f_\delta}^{i_2}(\prescript{g'^{-1}}{}{x'})\in\mathfrak{m}_0.
\]
As $y=\delta_B\mu^\ell -s^{r'}$ and $x'=\sigma^{i'}_{L^H,f_\delta}(\gamma^k-t_{\delta, H}^r)$, we have
\begin{equation}\label{eq_surjeq5}
 \sigma_{L,f_\delta}^{i_1}\left(\prescript{g^{-1}}{}{(s^{r'})}\right)\equiv  \sigma_{L,f_\delta}^{i_2}\left(\prescript{g'^{-1}}{}{(s^{r'})}\right)\pmod{\mathfrak{m}_0}
 \end{equation}
 and
\begin{equation}\label{eq_surjeq6}
 \sigma_{L,f_\delta}^{i_1}\left(\prescript{g^{-1}}{}{(t_{\delta,H}^{r})}\right)\equiv  \sigma_{L,f_\delta}^{i_2}\left(\prescript{g'^{-1}}{}{(t_{\delta,H}^{r})}\right)\pmod{\mathfrak{m}_0}.
 \end{equation}
As  $\sigma_{L,f_\delta}(t_{\delta,H})=\xi\cdot t_{\delta,H}$ and $G$ commutes with $\sigma_{L,f_\delta}$, \eqref{eq_surjeq6} implies
 \[
  \xi^{(i_1-i_2)r} \prescript{g^{-1}}{}{(t_{\delta,H}^r)}\equiv   \prescript{g'^{-1}}{}{(t_{\delta,H}^r)} \pmod{\mathfrak{m}_0}.
 \]
 On the other hand,  we know $i$ is the unique integer in $\{0,\dots,f_\delta-1\}$ satisfying \eqref{eq_surjeq2}. So $i_1-i_2\equiv i\pmod{f_\delta}$.
 Let $s'=\sigma_{L,f_\delta}^{i_2}(s)$.
 Then by \eqref{eq_surjeq5}, Lemma~\ref{lem_congruencedi} and the fact that  $G$ commutes with $\sigma_{L,f_\delta}$, we have
 \[
  \prescript{\sigma^i g^{-1}}{}{(s'^{r'})}\equiv  \prescript{g'^{-1}}{}{(s'^{r'})}\pmod{\mathfrak{m}_0}.
 \]
 On the other hand, as  $\sigma_{L,f_\delta}$ fixes $\mathfrak{m}=\frac{\mathfrak{Q}_{0}/p\ord_L}{\rad(\bar{\ord}_L)}$ setwisely, \eqref{eq_surjeq3} implies
\[
 \prescript{\sigma^i g^{-1}}{}{s'}+\mathfrak{m}=\sigma_{L,f_\delta}^{i_2}(c^{e_\delta})\cdot  (\prescript{g'^{-1}}{}{s'} +\mathfrak{m}).
\]
  It follows that $\sigma_{L,f_\delta}^{i_2}(c)$ is an $e_\delta$th root of unity. So $c$ is also an $e_\delta$th root of unity, as in the case $e_\delta>1$, $f_\delta=1$.
The same proof  in the  case $e_\delta>1$, $f_\delta=1$ then shows that there exists $j\in\{0,\dots,e_\delta-1\}$ such that $Hg\sigma^{-i}\omega^{-j}\in\tilde{B}$, which contradicts the assumption $Hg\mathcal{D}_{\mathfrak{Q}_0} \not\in \pi_H(\tilde{B})$.
\end{proof}

\lemringhomtestg*

We need the following notation: suppose $K,K'$ are extensions of $K_0$ and $\phi: K'\hookrightarrow K$ is an embedding of $K'$ in $K$ over $K_0$.
Recall that $\phi$ induces a homomorphism of $\F_q$-algebras  $\hat{\phi}: \bar{\ord}_{K'}/\rad(\bar{\ord}_{K'})\to \bar{\ord}_{K}/\rad(\bar{\ord}_{K})$.
Also suppose $\psi$ is an embedding of $\F_{q^i}$ in $\F_{q^j}$ over $\F_q$ where $i,j\in \N^+$. Then $\hat{\phi}$ and $\psi$ determine a homomorphism of $\F_q$-algebras $A_{K',i}\to A_{K,j}$ sending $a\otimes b\in A_{K',i}$ to $\hat{\phi}(a)\otimes\psi(b)\in  A_{K,j}$ for $a\in \bar{\ord}_{K'}/\rad(\bar{\ord}_{K'})$ and $b\in \F_{q^i}$.
We denote this map by $\hat{\phi}\otimes_{\F_q} \psi$.
% In particular, when $\phi$ is the identity map on $K$ and $\psi$ is the automorphism $t\mapsto t^q$ of $\F_{q^i}$, we obtain an $\F_q$-linear automorphism of $A_{K,i}$, and we denote this automorphism by $\sigma_{K,i}$.

The pseudocode of the subroutine $\mathtt{RingHomTest}$ is given in Algorithm~\ref{alg_ringhom}. It enumerates  $(K,K')\in\mathcal{F}^2$, embeddings $\phi:K'\hookrightarrow K$ over $K_0$, and $(\delta, \delta')\in I_K\times I_{K'}$ such that $\tilde{\phi}(\delta')\delta=\delta$.
For each $(K,K',\phi,\delta,\delta')$, a set $S$ of ideals of $A_{K,f_\delta}$ is computed. And for each $I\in S$, we find  $\delta_0\in I\cap R_K$ satisfying $(1-\delta_0)(I\cap R_K)=\{0\}$, which is the unique idempotent of $R_K$ that generates the ideal $I\cap R_K$ of $R_K$. If $\delta_0\delta\not\in\{0,\delta\}$, we use $\delta_0$ to refine $I_K$ and return.

\begin{algorithm}[tbp]
\caption{$\mathtt{RingHomTest}$}\label{alg_ringhom}
\begin{algorithmic}[1]
 \For{\strut $(K,K')\in\mathcal{F}^2$}
     \For{embedding $\phi:K'\hookrightarrow K$ over $K_0$}
     \For{$(\delta, \delta')\in I_K\times I_{K'}$ satisfying $\tilde{\phi}(\delta')\delta=\delta$}
        \State $S\gets\emptyset$
        \If{$f_{\delta}>1$}
            \State $r\gets$ the largest factor of $q^{f_\delta}-1$ coprime to $f_\delta$
            \State compute $\gamma\in \F_{q^{f_\delta}}^\times$ of order $(q^{f_\delta}-1)/r$
            \State call $\mathtt{SplitByExp}$  on $(A_{K,f_\delta},f_\delta, \delta \gamma,\delta t_\delta^r)$ to obtain  $x\in A_{K,f_\delta}$
            \If{$f_{\delta'}>1$}
                \State compute an embedding $\psi: \F_{q^{f_{\delta'}}}\to \F_{q^{f_\delta}}$ over $\F_q$
                \State $t\gets (\hat{\phi}\otimes_{\F_q}\psi)(t_{\delta'})\in A_{K,f_\delta}$
                \State call $\mathtt{SplitByExp}$  on $(A_{K,f_\delta},f_\delta, \delta \gamma,\delta t^r)$ to obtain  $x'\in A_{K,f_\delta}$
                \ForTo{$i$}{$0$}{$f_\delta-1$}
                     \State $S\gets S\cup\{x'A_{K,f_\delta}+\sigma^i_{K,f_\delta}(x) A_{K,f_\delta}\}$
                \EndFor
            \EndIf
        \EndIf
        \If{$e_{\delta'}>1$}
             \State $J\gets$ the preimage of $(1-\delta)(\bar{\ord}_K/\rad(\bar{\ord}_K))$ in $\bar{\ord}_K$
             \State compute $u\in \ann_{\bar{\ord}_K}(J^{e_\delta})$ such that $u  \bar{\phi}(s_{\delta'}) - s_\delta^{e_{\delta}/e_{\delta'}} \in J^{e_{\delta}/e_{\delta'}+1}$
             \State \strut $\bar{u}\gets u+\rad(\bar{\ord}_K)\in \bar{\ord}_K/\rad(\bar{\ord}_K)$
             \State \strut $r'\gets$ the largest factor of $q^{f_\delta}-1$ coprime to $e_\delta$
             \State compute $\mu\in \F_{q^{f_\delta}}^\times$ of order $(q^{f_\delta}-1)/r'$
             \State call $\mathtt{SplitByExp}$ on $(A_{K,f_\delta}, e_\delta, \delta\mu, \bar{u}^{r'})$ to obtain $y\in A_{K,f_\delta}$
             \State $S\gets S\cup \{yA_{K,f_\delta}\}$
             \If{$f_\delta>1$}
                \ForTo{$i$}{$0$}{$f_\delta-1$}
                     \State $S\gets S\cup\{yA_{K,f_\delta}+\sigma^i_{K,f_\delta}(x) A_{K,f_\delta}\}$
                \EndFor
             \EndIf
        \EndIf
         \For{$I\in S$}
             \State find $\delta_0\in I\cap R_K$ satisfying $(1-\delta_0)(I\cap R_K)=\{0\}$
             \If {$\delta_0\delta\not\in\{0,\delta\}$}
                 \State $I_K\gets I_K-\{\delta\}$
                 \State $I_K\gets I_K\cup\{\delta_0\delta, (1-\delta_0)\delta\}$
                 \State \Return
             \EndIf
        \EndFor
     \EndFor
     \EndFor  
\EndFor
\end{algorithmic}
\end{algorithm}

Fix $(K,K',\phi,\delta,\delta')$. The corresponding set $S$ is computed as follows:
Note we have $f_{\delta'}|f_\delta$ and $e_{\delta'}|e_\delta$. 
First assume $f_{\delta}>1$. 
Compute the largest factor $r$ of $q^{f_\delta}-1$ coprime to $f_\delta$. Then compute an element $\gamma\in \F_{q^{f_\delta}}^\times$ of order $(q^{f_\delta}-1)/r$, which can be done efficiently assuming GRH.
Call the subroutine $\mathtt{SplitByExp}$ in Lemma~\ref{lem_exponentg} on $(A_{K,f_\delta},f_\delta, \delta \gamma,\delta t_\delta^r)$ to obtain $x\in A_{K,f_\delta}$.
Also perform the following computation if $f_{\delta'}>1$:  
compute an embedding $\psi: \F_{q^{f_{\delta'}}}\to \F_{q^{f_\delta}}$ over $\F_q$ deterministically in polynomial time using Lenstra's algorithm \citep{Len91}.
Compute $t=(\hat{\phi}\otimes_{\F_q}\psi)(t_{\delta'})\in A_{K,f_\delta}$.
By  Definition~\ref{defi_auxielem} and the fact $\tilde{\phi}(\delta')\delta=\delta$, we have $\delta t^r A_{K,f_\delta}=\delta A_{K,f_\delta}$.
Call the subroutine $\mathtt{SplitByExp}$ on $(A_{K,f_\delta},f_\delta, \delta \gamma,\delta t^r)$ to obtain $x'\in A_{K,f_\delta}$. Then add the ideal $x'A_{K,f_\delta}+\sigma^i_{K,f_\delta}(x) A_{K,f_\delta}$ to $S$ for all $i\in\{0,1,\dots,f_\delta-1\}$.

If $e_{\delta'}>1$, we perform the following computation: first compute the preimage $J$ of $(1-\delta)(\bar{\ord}_K/\rad(\bar{\ord}_K))$ under the quotient  map $\bar{\ord}_K\to \bar{\ord}_K/\rad(\bar{\ord}_K)$. 
Then $J$ is the product of the maximal ideals $\mathfrak{m}$ of $\bar{\ord}_K$ satisfying $\delta\equiv 1\pmod{\mathfrak{m}/\rad(\bar{\ord}_K)}$. 
Compute $u\in \ann_{\bar{\ord}_K}(J^{e_\delta})$ such that $u  \bar{\phi}(s_{\delta'}) - s_\delta^{e_{\delta}/e_{\delta'}} \in J^{e_{\delta}/e_{\delta'}+1}$.
We claim such $u$ exists: by the Chinese remainder theorem, it suffices to show, for each maximal ideal $\mathfrak{m}$ of $\bar{\ord}_K$ containing $J$, that  
\[
u \bar{\phi}(s_{\delta'}) \equiv s_\delta^{e_{\delta}/e_{\delta'}} \pmod{\mathfrak{m}^{e_{\delta}/e_{\delta'}+1}}
\]
 has a solution in $\bar{\ord}_K$.
Fix such $\mathfrak{m}$. We have $s_\delta\in \mathfrak{m}-\mathfrak{m}^{2}$ by Lemma~\ref{lem_genset} and hence $s_\delta^{e_{\delta}/e_{\delta'}}\in \mathfrak{m}^{e_{\delta}/e_{\delta'}}-\mathfrak{m}^{e_{\delta}/e_{\delta'}+1}$. 
Let $\mathfrak{m}'=\bar{\phi}^{-1}(\mathfrak{m})$.
By Lemma~\ref{lem_genset} and the fact  $\tilde{\phi}(\delta')\delta=\delta$, we have $s_{\delta'}\in \mathfrak{m}'-\mathfrak{m}'^{2}$ and hence $\bar{\phi}(s_{\delta'})\in \mathfrak{m}^{e_{\delta}/e_{\delta'}}-\mathfrak{m}^{e_{\delta}/e_{\delta'}+1}$.  The claim follows by noting  $\mathfrak{m}^{e_{\delta}/e_{\delta'}}/\mathfrak{m}^{e_{\delta}/e_{\delta'}+1}$ is an one-dimensional vector space over $\bar{\ord}_K/\mathfrak{m}$.
Next compute 
\[
\bar{u}:=u+\rad(\bar{\ord}_K)\in \bar{\ord}_K/\rad(\bar{\ord}_K).
\]
Compute the largest factor $r'$ of $q^{f_\delta}-1$ coprime to $e_\delta$, so that all the prime factors of $(q^{f_\delta}-1)/r'$ divide $e_\delta$. And compute an element $\mu\in \F_{q^{f_\delta}}^\times$ of order $(q^{f_\delta}-1)/r'$, which can be done efficiently assuming GRH.
Note that $\bar{u}^{r'}A_{K,f_\delta}=\delta A_{K,f_\delta}$.\footnote{Again, we let $A_{K,f_\delta}=\bar{\ord}_K/\rad(\bar{\ord}_K)$ if $f_\delta=1$.} Call the subroutine $\mathtt{SplitByExp}$ on the input $(A_{K,f_\delta},e_\delta, \delta \mu, \bar{u}^{r'})$ to obtain $y\in A_{K,f_\delta}$, and add the ideal $yA_{K,f_\delta}$  to $S$.
In addition, if $f_\delta>1$, we enumerate $i=0,1,\dots,f_\delta-1$, and for each $i$ we add the ideal  of $A_{K,f_\delta}$ generated by $y$ and $\sigma^i_{K,f_\delta}(x)$ to $S$, where $x\in A_{K,f_\delta}$ is computed in the case $f_\delta>1$ above. 

Now we prove Lemma~\ref{lem_ringhomtestg}.

\begin{proof}[Proof of Lemma~\ref{lem_ringhomtestg}]
Assume the algorithm does not properly refine any $I_K$.  We prove that $\tilde{\mathcal{C}}$ is compatible and invariant. Fix $H,H'\in\mathcal{P}$ and a map $\phi^*: H\backslash G\to H'\backslash G$ that is either a projection $\pi_{H, H'}$ (with $H\subseteq 
H'$) or a conjugation $c_{H,h}$ (with $H'=hHh^{-1}$).
Consider $g,g'\in G$ for which $Hg, Hg'\in H\backslash G$ are in the same block of $\tilde{C}_H$. We want to show that $\phi^*(Hg), \phi^*(Hg')\in H'\backslash G$ are in the same block of $\tilde{C}_{H'}$.

Let $B$ be the block of $C_H$ containing both $Hg\mathcal{D}_{\mathfrak{Q}_0}$ and $Hg'\mathcal{D}_{\mathfrak{Q}_0}$.
Let $\bar{\phi}^*:  H\backslash G/\mathcal{D}_{\mathfrak{Q}_0}\to H'\backslash G/\mathcal{D}_{\mathfrak{Q}_0}$ be the map $\pi_{H, H'}^{\mathcal{D}_{\mathfrak{Q}_0}}$ if $\phi^*=\pi_{H,H'}$, or $c_{H,h}^{\mathcal{D}_{\mathfrak{Q}_0}}$  if $\phi^*=c_{H,h}$.
As $\mathcal{C}$ is compatible and invariant, there exists $B'\in C_{H'}$ containing both $\bar{\phi}^*(Hg\mathcal{D}_{\mathfrak{Q}_0})$ and $\bar{\phi}^*(Hg'\mathcal{D}_{\mathfrak{Q}_0})$.
Let $K$ (resp. $K'$) be the field in $\mathcal{F}$ isomorphic to $L^H$ (resp. $L^{H'}$) over $K_0$.
Let $\delta$ (resp. $\delta'$) be the idempotent in $I_K$  (resp. $I_{K'}$) satisfying $\tilde{\tau}_H(\delta)=\delta_B$ (resp. $\tilde{\tau}_{H'}(\delta')=\delta_{B'}$). Let  $\mathfrak{m}_0$  be an arbitrary  maximal ideal of $A_{L, f_\delta}$ containing $\frac{\mathfrak{Q}_{0}/p\ord_{L}}{\rad(\bar{\ord}_{L})}$.
Fix an embedding $\psi: \F_{q^{f_{\delta'}}}\to \F_{q^{f_\delta}}$ over $\F_q$.
Let $\phi: L^{H'}\hookrightarrow L^H$ be the natural inclusion if $\phi^*=\pi_{H,H'}$, or the map $x\mapsto \prescript{h^{-1}}{}{x}$ if $\phi^*=c_{H,h}$.
Finally, let $s=\bar{\phi}(s_{\delta',H'})$ if  $e_{\delta'}>1$, and let $t= (\hat{\phi}\otimes_{\F_q}\psi)(t_{\delta', H'})$ if $f_{\delta'}>1$.

We claim that the following two conditions are satisfied:
\begin{enumerate}
\item  If $e_{\delta'}>1$, the order of the unique element $c$ in $\kappa_{\mathfrak{Q}_{0}}^\times$ satisfying
\[
\prescript{g^{-1}}{}{s}+I'= c\cdot  (\prescript{g'^{-1}}{}{s} +I')
\]
is coprime to $e_{\delta'}$, where $I'=(\mathfrak{Q}_{0}/p\ord_{L})^{e(\mathfrak{Q}_0)/e_{\delta'}+1}$.
\item    If $f_{\delta'}>1$, the order of the unique element $c$ in $(A_{L,f_\delta}/\mathfrak{m}_0)^\times$ satisfying
\[
\prescript{g^{-1}}{}{t}+\mathfrak{m}_0=c \cdot (\prescript{g'^{-1}}{}{t}+\mathfrak{m}_0) 
\]
 is coprime to $f_{\delta'}$.
\end{enumerate}
To see this claim implies that  $\phi^*(Hg)$ and $\phi^*(Hg')$ are in the same block of $\tilde{C}_{H'}$, pick $\bar{g}, \bar{g}'\in G$ such that $H'\bar{g}=H'\phi^*(Hg)$  and $H'\bar{g}'=H'\phi^*(Hg')$.  
Then  $c\in \kappa_{\mathfrak{Q}_{0}}^\times$ in the first condition is also the unique element satisfying
$\prescript{\bar{g}^{-1}}{}{s_{\delta',H'}}+I'= c\cdot  (\prescript{\bar{g}'^{-1}}{}{s_{\delta',H'} +I'})$.
And $c\in (A_{L,f_\delta}/\mathfrak{m}_0)^\times$  in the second condition is also the unique element satisfying
$\prescript{\bar{g}^{-1}}{}{t_{\delta', H'}}+\mathfrak{m}'_0=c \cdot (\prescript{\bar{g}'^{-1}}{}{t_{\delta', H'}}+\mathfrak{m}'_0)$, where $\mathfrak{m}_0'\supseteq \frac{\mathfrak{Q}_{0}/p\ord_{L}}{\rad(\bar{\ord}_{L})}$ is the preimage of $\mathfrak{m}_0$ under  $\id\otimes_{\F_q}\psi: A_{L,f_{\delta'}}\to A_{L,f_{\delta}}$, and $\id$ is the identity map on $\bar{\ord}_{L}/\rad(\bar{\ord}_{L})$. It follows by Definition~\ref{defi_ordpcollection} that $\phi^*(Hg)$ and $\phi^*(Hg')$ are in the same block assuming if these two conditions are satisfied.

The rest of the proof focuses on verifying the above two conditions. First assume $f_{\delta'}>1$. Suppose $x=(\delta_B\gamma)^k-\delta_B t_{\delta, H}^r$ and $x'=(\delta_B\gamma)^{k'}-\delta_B t^r$ satisfy $x A_{L^H,f_\delta}, x' A_{L^H,f_\delta} \subsetneq \delta_B A_{L^H,f_\delta}$, where $k,k'\in\N$.
Then there exists $i\in\{0,\dots,f_\delta-1\}$ such that 
\[
I_i:=x'A_{L^H,f_\delta}+\sigma^i_{L^H,f_\delta}(x) A_{L^H,f_\delta}\subsetneq \delta_B A_{L^H,f_\delta}.
\]
This follows from the same argument in the proof of Lemma~\ref{lem_bijtest} that shows the existence of $i'\in\{0,\dots,f_\delta-1\}$ satisfying \eqref{eq_proper}. We may also assume $I_i\cap R_{L^H}=\{0\}$: otherwise, by identifying $K$ with $L^H$ using the isomorphism $\tau_H$, we see the subroutine finds an idempotent $\delta_0$ of $R_K$ at Line 23 satisfying $\delta_0\delta\not\in\{0,\delta\}$, and  properly refines $I_K$.

 By Lemma~\ref{lem_transitivitybc} and the assumption $I_i\cap R_{L^H}=\{0\}$, we know there exist $i_1,i_2\in\Z$ such that
\begin{equation}\label{eq_ringhomeq1}
 \sigma_{L,f_\delta}^{i_1}(\prescript{g^{-1}}{}{x'}), \sigma_{L,f_\delta}^{i_1+i}(\prescript{g^{-1}}{}{x}), \sigma_{L,f_\delta}^{i_2}(\prescript{g'^{-1}}{}{x'}),  \sigma_{L,f_\delta}^{i_2+i}(\prescript{g'^{-1}}{}{x})\in\mathfrak{m}_0.
\end{equation}
By Definition~\ref{defi_auxielem}, there exist primitive $f_\delta$th roots of unity $\xi,\xi'\in\F_{q^{f_\delta}}$ satisfying $\sigma_{L,f_\delta}(t_{\delta,H})=\xi\cdot t_{\delta,H}$ and $\sigma_{L,f_\delta}(t)=\xi'\cdot t$.
As $x=\delta_B\gamma^k-\delta_B t_{\delta, H}^r$ and $x'=\delta_B\gamma^{k'}-\delta_B t^r$, \eqref{eq_ringhomeq1} implies
 \begin{equation}\label{eq_ringhomeq2}
\prescript{g^{-1}}{}{(t_{\delta,H}^r)}\equiv \xi^{(i_2-i_1) r}  \prescript{g'^{-1}}{}{(t_{\delta,H}^r)}\pmod{\mathfrak{m}_0}
\end{equation}
and
 \begin{equation}\label{eq_ringhomeq3}
\prescript{g^{-1}}{}{(t^r)}\equiv \xi^{(i_2-i_1) r}  \prescript{g'^{-1}}{}{(t^r)}\pmod{\mathfrak{m}_0}.
\end{equation}
 On the other hand, as $Hg\mathcal{D}_{\mathfrak{Q}_0}, Hg'\mathcal{D}_{\mathfrak{Q}_0}\in B$, we know from  Definition~\ref{defi_ordpcollection} that the order of the unique element $c\in (A_{L,f_\delta}/\mathfrak{m}_0)^\times$ satisfying
$\prescript{g^{-1}}{}{t_{\delta,H}}+\mathfrak{m}_0=c \cdot (\prescript{g'^{-1}}{}{t_{\delta,H}}+\mathfrak{m}_0)$
 is coprime to $f_\delta$.
As $r$ is coprime to $f_\delta$, we see from \eqref{eq_ringhomeq2} that $i_2-i_1$ is divisible by $f_\delta$.
Then  \eqref{eq_ringhomeq3} becomes $\prescript{g^{-1}}{}{(t^r)}\equiv   \prescript{g'^{-1}}{}{(t^r)}\pmod{\mathfrak{m}_0}$.
So the order of the unique element $c\in (A_{L,f_\delta}/\mathfrak{m}_0)^\times$ satisfying  $\prescript{g^{-1}}{}{t}+\mathfrak{m}_0=c \cdot (\prescript{g'^{-1}}{}{t}+\mathfrak{m}_0) $ is coprime to $f_\delta$, as desired.

Next assume $e_{\delta'}>1$.
%Suppose $u$ is an element in $\bar{\ord}_K$ such that (1) for all the maximal ideal $\mathfrak{m}$ of $\bar{\ord}_K$ satisfying $\delta\equiv 0\pmod{\mathfrak{m}/\rad(\bar{\ord}_K)}$, it holds that $u\in \mathfrak{m}$,  and (2) for all the maximal ideal $\mathfrak{m}$ of $\bar{\ord}_K$ satisfying $\delta\equiv 1\pmod{\mathfrak{m}/\rad(\bar{\ord}_K)}$, it holds that $u\bar{\phi}(s_{\delta',H'})-s_{\delta,H}^{e_\delta/e_{\delta'}}\in \mathfrak{m}^{e_\delta/e_{\delta'}+1}$.
 Let $\bar{u}$ be the element computed at Line 18 and regard it as an element of $\bar{\ord}_{L^H}/\rad(\bar{\ord}_{L^H})$ by identifying $K$ with $L^H$ using the isomorphism $\tau_H$.
Suppose $y=(\delta_B\mu)^{\ell}-\bar{u}^{r'}$ satisfies  $y A_{L^H,f_\delta} \subsetneq \delta_B A_{L^H,f_\delta}$, where $\ell\in\N$. We may assume $y A_{L^H,f_\delta}\cap R_{L^H}=\{0\}$, since otherwise the idempotent $\delta_0$ generating $y A_{L^H,f_\delta}\cap R_{L^H}$ satisfies $\delta_0\delta\not\in\{0,\delta\}$ and is used to properly refine $I_K$.
 
If $e_{\delta'}>1$ and $f_{\delta}=1$, the ring $A_{L^H,f_\delta}$ is just $\bar{\ord}_{L^H}/\rad(\bar{\ord}_{L^H})$, and we have $y=0$  in this case. So $\prescript{g^{-1}}{}{y}=\prescript{g'^{-1}}{}{y}=0$, which implies
 \begin{equation}\label{eq_ringhomeq35}
\prescript{g^{-1}}{}{(\bar{u}^{r'})}\equiv \prescript{g'^{-1}}{}{(\bar{u}^{r'})}\pmod{\frac{\mathfrak{Q}_{0}/p\ord_{L}}{\rad(\bar{\ord}_{L})}}.
\end{equation}
Let $c_1,c_2\in  \kappa_{\mathfrak{Q}_{0}}^\times$ be the residues of $\prescript{g^{-1}}{}{\bar{u}}$ and $\prescript{g'^{-1}}{}{\bar{u}}$ modulo $\frac{\mathfrak{Q}_{0}/p\ord_{L}}{\rad(\bar{\ord}_{L^H})}$ respectively.
Then $(c_2/c_1)^{r'}=1$. So the order of $c_2/c_1$ divides $r'$, which is coprime to $e_\delta$.
By Definition~\ref{defi_ordpcollection}, the order of the unique element $c\in \kappa_{\mathfrak{Q}_{0}}^\times$ satisfying
\[
\prescript{g^{-1}}{}{s_{\delta,H}}+I= c\cdot  (\prescript{g'^{-1}}{}{s_{\delta,H}}+I)
\]
is coprime to $e_{\delta}$ (and hence to $e_{\delta'}$), where $I=(\mathfrak{Q}_{0}/p\ord_{L})^{e(\mathfrak{Q}_0)/e_{\delta}+1}$.
Then we have
 \begin{equation}\label{eq_ringhomeq4}
\prescript{g^{-1}}{}{s_{\delta,H}^{e_\delta/e_{\delta'}}}+I'= c^{e_\delta/e_{\delta'}}\cdot  \left(\prescript{g'^{-1}}{}{s_{\delta,H}^{e_\delta/e_{\delta'}}}+I'\right)
\end{equation}
and the order of $c^{e_{\delta}/e_{\delta'}}$ is  coprime to $e_{\delta'}$.
By the definition of $\bar{u}$, we may rewrite \eqref{eq_ringhomeq4} as
\[
c_1(\prescript{g^{-1}}{}{s})+I'= c_2 c^{e_{\delta}/e_{\delta'}}\cdot  (\prescript{g'^{-1}}{}{s}+I').
\]
As the order of $c_2/c_1$ and that of $c^{e_{\delta}/e_{\delta'}}$ are coprime to $e_{\delta'}$, the second condition above is satisfied.

Finally, assume $e_{\delta'}>1$ and $f_{\delta}>1$. Then there exists $j\in\{0,\dots,f_\delta-1\}$ such that 
\[
I'_j:=yA_{L^H,f_\delta}+\sigma^i_{L^H,f_\delta}(x) A_{L^H,f_\delta}\subsetneq \delta_B A_{L^H,f_\delta},
\]
where  $x=\delta_B\gamma^k-\delta_B t_{\delta, H}^r$ is as above.
Again we may assume $I'_j\cap R_{L^H}=\{0\}$ since otherwise $I_K$ is properly refined. Then there exist $i_1,i_2\in\Z$ such that
%\begin{equation}\label{eq_ringhomeq1}
\[
 \sigma_{L,f_\delta}^{i_1}(\prescript{g^{-1}}{}{y}), \sigma_{L,f_\delta}^{i_1+j}(\prescript{g^{-1}}{}{x}), \sigma_{L,f_\delta}^{i_2}(\prescript{g'^{-1}}{}{y}),  \sigma_{L,f_\delta}^{i_2+j}(\prescript{g'^{-1}}{}{x})\in\mathfrak{m}_0.
 \]
%\end{equation}
 As  $x=\delta_B\gamma^k-\delta_B t_{\delta, H}^r$ and $\sigma_{L,f_\delta}(t_{\delta,H})=\xi\cdot t_{\delta,H}$, again we conclude that $i_2-i_1$ is divisible by $f_\delta$.
 As the order of $\sigma_{L^H,f_\delta}$ on $\delta_B A_{L^H,f_\delta}$ is $f_\delta$, we may assume $i_1=i_2$. 
As $y=(\delta_B\mu)^{\ell}-\bar{u}^{r'}$, we have
\[
\sigma_{L,f_\delta}^{i_1}\left(\prescript{g^{-1}}{}{(\bar{u}^{r'})}\right)\equiv \sigma_{L,f_\delta}^{i_1}\left(\prescript{g'^{-1}}{}{(\bar{u}^{r'})}\right)\pmod{\mathfrak{m}_0}.
\]
As $\sigma_{L,f_\delta}$ fixes every maximal ideal of $\bar{\ord}_{L}/\rad(\bar{\ord}_{L})$ setwisely, and 
\[
\mathfrak{m}_0\cap (\bar{\ord}_{L}/\rad(\bar{\ord}_{L}))=\frac{\mathfrak{Q}_{0}/p\ord_{L}}{\rad(\bar{\ord}_{L})},
\] 
we see   \eqref{eq_ringhomeq35} still holds. The rest of the proof is the same as in the case $e_{\delta'}>1$, $f_{\delta}=1$.
\end{proof}

\lemintegralitypschemeg*

\begin{proof}
Assume to the contrary that there exist $x\in S$, a $\mathcal{P}$-scheme of double cosets $\mathcal{C}=\{C_H: H\in\mathcal{P}\}$ with respect to $\mathcal{D}$ that is  homogeneous on  $G_x$,
 and an antisymmetric $(\mathcal{C},\mathcal{D})$-separated $\mathcal{P}$-scheme  $\tilde{\mathcal{C}}=\{\tilde{C}_H: H\in\mathcal{P}\}$. 
As  $G$ acts transitively on $S$, we know $\mathcal{C}$ is homogeneous on $G_x$ for all $x\in S$.

Fix $x_0\in S$ and consider the bijection $\lambda_{x_0}: S \to G_{x_0}\backslash G$ sending $\prescript{g}{}{x_0}$ to $G_{x_0} g^{-1}$.
It is an equivalence between the action of $G$ on $S$ and that on $G_{x_0}\backslash G$ by inverse right translation.
Let $B_0$ be a block of $\tilde{C}_{G_{x_0}}$ and define $T:=\lambda_{x_0}^{-1}(B_0)\subseteq S$. As $\mathcal{C}$ is homogeneous on $G_{x_0}$ and $\tilde{\mathcal{C}}$ is $(\mathcal{C},\mathcal{D})$-separated, we know $T$ is a complete set of representatives of the $\mathcal{D}$-orbits in $S$, and hence $|B_0|=|T|=k$. 

 The group $G$ acts diagonally on $S^{(\ell)}$. And $\sym(\ell)$ acts on $S^{(\ell)}$  by permuting the coordinates.
 As the two actions commute, we know $\sym(\ell)$ permutes the $G$-orbits in $S^{(\ell)}$. 
 Fix $z\in T^{(\ell)}$ and let $H_z$ be the subgroup of $\sym(\ell)$ fixing $G z$ setwisely.
 Using the bijection $\lambda_z: G z\to G_z\backslash G$, the action of $H_z$ on $Gz$ induces an action  on $G_z\backslash G$.
 In the proof of  Lemma~\ref{lem_integrality_pscheme}, we showed that the latter action induces a semiregular action on the set of the blocks of  $\tilde{C}_{G_{z}}$. 
 
 Let $U_z:=T^{(\ell)}\cap G z$. Suppose $z=(z_1,\dots,z_\ell)$.
 For $g\in G$, the element $\prescript{g}{}{z}$ is in $U_z$
 iff $\lambda_{x_0}(\prescript{g}{}{z_i})\in \lambda_{x_0}(T)=B_0$ for all $i\in [\ell]$.
 Fix $i\in [\ell]$ and choose $g_i\in G$ satisfying $\prescript{g_i}{}{x_0}=z_i$.
 Then $c_{x_0,g_i}: G_{x_0}\backslash G\to G_{z_i}\backslash G$ sends $B_0$ to a block $B_i\in \tilde{C}_{G_{z_i}}$.
 Also note that $c_{x_0,g_i}\circ\lambda_{x_0}=\lambda_{z_i}$. So $\lambda_{x_0}(\prescript{g}{}{z_i})\in B_0$ is equivalent to $\lambda_{z_i}(\prescript{g}{}{z_i})\in B_i$.
 As
 \[
 \lambda_{z_i}(\prescript{g}{}{z_i})=G_{z_i}g^{-1}=\pi_{G_z,G_{z_i}}(G_zg^{-1})=\pi_{G_z, G_{z_i}}\circ \lambda_z(\prescript{g}{}{z}),
 \]
 we see that $\lambda_z(U_z)$ consists of the elements $x\in G_z\backslash G$ satisfying $\pi_{G_z,G_{z_i}}(x)\in B_i$ for all $i\in [\ell]$. By compatibility of $\tilde{\mathcal{C}}$, the set $\lambda_{z}(U_z)$ is a disjoint union of blocks of $\tilde{C}_{G_z}$. Moreover, by regularity of $\tilde{\mathcal{C}}$, the cardinality of these blocks are all divisible by $|B_0|=k$.
 
 Note that the action of $H_z$ on $G z$ fixes the set $U_z$ setwisely. So the semiregular action of $H_z$ on the set of the blocks of $\tilde{C}_{G_z}$ restricts to a semiregular action on the subset of the blocks in $\lambda_z(U_z)$.
 By the previous paragraph, we know  $|U_z|$ is a multiple of $k|H_z|$.
 
 The set $T^{(\ell)}$ is a disjoint union of subsets of the form $U_z$ where $z\in T^{(\ell)}$.
 The group $\sym(\ell)$ permutes these subsets. 
  By the orbit-stabilizer theorem, each $\sym(\ell)$-orbit $O$ is a disjoint union of $|\sym(\ell)|/|H_z|$ subsets of the same cardinality $|U_z|$, where $z$ is an arbitrary element in $O$. So 
  \[
  |O|=\frac{|\sym(\ell)|}{|H_z|}\cdot {|U_z|}
  \]
  which is a multiple of $k\ell!$ by the previous paragraph. It follows that $|T^{(\ell)}|=k(k-1)\cdots(k-\ell+1)$ is a multiple of $k\ell!$.
But this is impossible since   none of the factors $k-1,\dots, k-\ell+1$ are divisible by the prime number $\ell$. 
 \end{proof}

\chapter{List of algorithms}\label{chap_algs}

\begin{table}[htb]
\centering
\begin{tabular}{| p{.50\textwidth} | p{.34\textwidth} |c|} 
\hline
Name & Reference & Page\\ \hhline{|===|}
$\mathtt {ComputeQuotientRing}$  & Lemma~\ref{lem_residuering} & \pageref{lem_residuering} \\ \hline 
$\mathtt {ComputeResidue}$ & Lemma~\ref{lem_computeresidue} & \pageref{lem_computeresidue}  \\ \hline 
$\mathtt {ComputeEmbeddings}$ & Lemma~\ref{lem_compembed} & \pageref{lem_compembed} \\ \hline
$\mathtt {ComputeRingHom}$ & Lemma~\ref{lem_ringhom} & \pageref{lem_ringhom} \\ \hline
$\mathtt {ExtractFactors}$ & Algorithm~\ref{alg_reduction},  Theorem~\ref{thm_extract} & \pageref{alg_reduction} \\ \hline
$\mathtt {ComputePscheme}$ & Algorithm~\ref {alg_mainalg},  Theorem~\ref {thm_comppscheme} & \pageref{alg_mainalg} \\ \hline
$\mathtt {CompatibilityAndInvarianceTest}$ & Algorithm~\ref {alg_citest},   Lemma~\ref {lem_citest} & \pageref{alg_citest} \\ \hline
$\mathtt {FreeModuleTest}$ & Lemma~\ref {lem_free} & \pageref{lem_free} \\ \hline 
$\mathtt {SplitByZeroDivisor}$ & Lemma~\ref {lem_zerodivisor} & \pageref{lem_zerodivisor}\\ \hline
$\mathtt {RegularityTest}$ & Algorithm~\ref {alg_reg},  Lemma~\ref {lem_rtest} & \pageref{alg_reg} \\ \hline
$\mathtt {StrongAntisymmetryTest}$ & Algorithm~\ref {alg_satest},  Lemma~\ref {lem_satest} & \pageref{alg_satest} \\ \hline
$\mathtt {PschemeAlgorithm}$ & Algorithm~\ref {alg_pschalg},  Theorem~\ref {thm_algmain2formal} & \pageref{alg_pschalg} \\ \hline
$\mathtt {Automorphism}$ & Algorithm~\ref {alg_aut},  Lemma~\ref {lem_auto} & \pageref{alg_aut} \\ \hline
\end{tabular}\caption{Algorithms and  subroutines in the $\mathcal{P}$-scheme algorithm} 
\end{table}

\begin{table}[htb]
\centering
\begin{tabular}{| p{.50\textwidth} | p{.34\textwidth} |c|} 
\hline
Name & Reference & Page\\ \hhline{|===|}
$\mathtt {AdjoinRoot}$ & Lemma~\ref {lem_comp} & \pageref{lem_comp} \\ \hline
$\mathtt {SplittingField}$ & Algorithm~\ref {alg_splittingfield}, Lemma~\ref {lem_compsplitgeneral} & \pageref{alg_splittingfield} \\ \hline
$\mathtt {Stabilizers}$ & Algorithm~\ref {alg_stabilizers}, Lemma~\ref {lem_compstabsysg} & \pageref{alg_stabilizers} \\ \hline
$\mathtt{Tower}$ & Theorem~\ref{thm_lm} & \pageref{thm_lm} \\ \hline
$\mathtt {GeneralAction}$ & Algorithm~\ref {alg_gaction}, Theorem~\ref {thm_primitivereduction} & \pageref{alg_gaction} \\ \hline
$\mathtt {SubgroupSystem}$ & Algorithm~\ref {alg_syspn}, Lemma~\ref {lem_syspn} & \pageref{alg_syspn} \\ \hline
\end{tabular}\caption{Algorithms for constructing number fields} 
\end{table}

\newpage

\begin{table}[htb]
\centering
\begin{tabular}{| p{.50\textwidth} | p{.34\textwidth} |c|} 
\hline
Name & Reference & Page\\ \hhline{|===|}
$\mathtt {ComputeRelEmbeddings}$ & Lemma~\ref {lem_comprelembed} & \pageref{lem_comprelembed} \\ \hline
$\mathtt {ComputeRings}$ & Lemma~\ref {lem_computerings} & \pageref {lem_computerings} \\ \hline
$\mathtt {ComputeRingHoms}$ & Lemma~\ref {lem_ringhomg} & \pageref{lem_ringhomg} \\ \hline 
$\mathtt {ExtractFactorsV2}$ & Algorithm~\ref {alg_reductiong}, Theorem~\ref {thm_extfactorredg} & \pageref{alg_reductiong} \\ \hline
$\mathtt {ComputeDoubleCosetPscheme}$\tablefootnote{The subroutine $\mathtt {ComputeDoubleCosetPscheme}$  is not  actually used in the generalized $\mathcal{P}$-scheme algorithm, but only serves as a preliminary version of $\mathtt {ComputeOrdinaryPscheme}$.} & Algorithm~\ref {alg_mainalgg}, Theorem~\ref {thm_maindc} & \pageref{alg_mainalgg} \\ \hline
$\mathtt {CompatibilityAndInvarianceTestV2}$ & Lemma~\ref {lem_citestg} & \pageref{lem_citestg} \\ \hline
$\mathtt {RegularityTestV2}$ & Lemma~\ref {lem_rtestg} & \pageref{lem_rtestg} \\ \hline
$\mathtt {StrongAntisymmetryTestV2}$ & Lemma~\ref {lem_satestg} & \pageref{lem_satestg} \\ \hline
$\mathtt {RamificationIndexTest}$ & Algorithm~\ref {alg_ramtest}, Lemma~\ref {lem_ramtest} & \pageref{alg_ramtest} \\ \hline
$\mathtt {InertiaDegreeTest}$ & Algorithm~\ref {alg_inetest}, Lemma~\ref {lem_inetest} & \pageref{alg_inetest} \\ \hline
$\mathtt {ComputeOrdinaryPscheme}$ & Algorithm~\ref {alg_mainalggv2}, Theorem~\ref {thm_comppschemeg} & \pageref{alg_mainalggv2} \\ \hline
$\mathtt {GeneralizedPschemeAlgorithm}$ & Algorithm~\ref {alg_genpscheme}, Theorem~\ref {thm_algmain2formalg} & \pageref{alg_genpscheme} \\ \hline
$\mathtt {ComputeAdvice}$ & Algorithm~\ref {alg_auxi}, Lemma~\ref {lem_genset} & \pageref{alg_auxi} \\ \hline
$\mathtt {SplitByExp}$ & Lemma~\ref {lem_exponentg} & \pageref{lem_exponentg} \\ \hline
$\mathtt {SurjectivityTest}$ & Algorithm~\ref {alg_surjtest}, Lemma~\ref {lem_bijtest} & \pageref{alg_surjtest} \\ \hline
$\mathtt {RingHomTest}$ & Algorithm~\ref {alg_ringhom}, Lemma~\ref {lem_ringhomtestg} & \pageref{alg_ringhom} \\ \hline
\end{tabular}\caption{Algorithms and subroutines in the generalized $\mathcal{P}$-scheme algorithm} 
\end{table}

\printnomenclature

\clearpage
\phantomsection
\addcontentsline{toc}{chapter}{\indexname}
\printindex

\end{document}